\documentclass[11pt,twoside]{cernrep}
\usepackage{graphicx,epsfig}
\usepackage{here,cite}
\usepackage{amsfonts}
\usepackage{amsmath}
\usepackage{amssymb}
\usepackage{longtable}
\usepackage{threeparttable}
\usepackage{comment}
\usepackage{minitoc} 
\usepackage{a4,epsfig}
\usepackage{rotating}

\newcommand{\bea}{\begin{eqnarray}}
\newcommand{\eea}{\end{eqnarray}}
\newcommand{\be}{\begin{equation}}
\newcommand{\ee}{\end{equation}}
\newcommand{\nn}{\nonumber}

\newcommand{\gsim}{\;\raisebox{-0.9ex}
           {$\textstyle\stackrel{\textstyle >}{\sim}$}\;}

\newcommand{\tanb}{\ensuremath{\tan\beta}}

\newcommand{\sQ}{\ensuremath{\tilde{Q}}}

\newcommand{\sQua}{\ensuremath{\tilde{q}}}

\newcommand{\sTop}{\ensuremath{\tilde{t}}} 
\newcommand{\stL}{\ensuremath{\tilde{t}_{\rm L}}}
\newcommand{\stR}{\ensuremath{\tilde{t}_{\rm R}}}
\newcommand{\stone}{\ensuremath{\tilde{t_1}}}
\newcommand{\sttwo}{\ensuremath{\tilde{t_2}}}
\newcommand{\sBot}{\ensuremath{\tilde{b}}}

\newcommand{\sEl}{\ensuremath{\tilde{e}}}

\newcommand{\sMu}{\ensuremath{\tilde{\mu}}}
\newcommand{\sNu}{\ensuremath{\tilde{\nu}}}
\newcommand{\sTau}{\ensuremath{\tilde{\tau}}}

\newcommand{\sGlu}{\ensuremath{\tilde{g}}}

\newcommand{\chiz}{\ensuremath{\tilde{\chi}^{0}}}

\newcommand{\chipm}{\ensuremath{\tilde{\chi}^{\pm}}}

\newcommand{\mtau}{\ensuremath{m_\tau}}

\begin{document}

\newpage
\thispagestyle{empty}

\bigskip

{\large
\begin{tabbing}
\hspace*{13cm} \= CERN-2004-005\\
\> 10 June 2004 \\
\> Physics Department \\
\> hep-ph/0412251
\end{tabbing}
}

\vspace*{2cm}

{\large\bf ORGANISATION EUROP\'EENNE POUR LA RECHERCHE NUCL\'EAIRE}

\vspace*{2mm}

{\LARGE\bf CERN} {\large\bf EUROPEAN ORGANIZATION FOR NUCLEAR RESEARCH}

\vspace*{3cm}

{\LARGE \begin{center}
{\bf PHYSICS AT THE CLIC MULTI-TeV LINEAR COLLIDER}
\end{center}}

\vspace{1cm}
{\large \begin{center}
{\bf Report of the CLIC Physics Working Group}
\end{center}}

\vspace{2cm}

{\Large \begin{center}
Editors: M.~Battaglia, A.~De~Roeck, J.~Ellis, D.~Schulte
\end{center}}

\vspace{7.5cm}

{\begin{center}
GENEVA \\
2004
\end{center}}

\newpage
\thispagestyle{empty}

\vspace*{4cm}

\vspace{-3cm}
{\large \begin{center}
{\bf CLIC Physics Working Group}
\end{center}}

\vspace{1cm}

\noindent
E.~Accomando (INFN, Torino), 
A.~Aranda (Univ. of Colima),
E.~Ateser (Kafkas Univ.),
C.~Balazs (ANL),
D.~Bardin (JINR, Dubna),
T.~Barklow (SLAC),
M.~Battaglia (LBL and UC Berkeley),
W.~Beenakker (Univ. of Nijmegen),
S.~Berge (Univ. of Hamburg),
G.~Blair (Royal Holloway College, Univ. of London),
E.~Boos (INP, Moscow), 
F.~Boudjema (LAPP, Annecy),
H.~Braun (CERN),
P.~Burikham (Univ. of Wisconsin),
H.~Burkhardt (CERN),
M.~Cacciari (Univ. Parma),
O.~\c{C}ak{\i}r (Univ. of Ankara),
A.K.~Ciftci (Univ. of Ankara),
R.~Ciftci (Gazi Univ., Ankara),
B.~Cox (Manchester Univ.), 
C.~Da Vi\'a (Brunel), 
A.~Datta (Univ. of Florida),
S.~De~Curtis (INFN and Univ. of Florence), 
A.~De~Roeck (CERN),
M.~Diehl (DESY),
A.~Djouadi (Montpellier),
D.~Dominici (Univ. of Florence),
J.~Ellis (CERN),
A.~Ferrari (Uppsala Univ.),
J.~Forshaw (Manchester Univ.),
A.~Frey (CERN),
G.~Giudice (CERN),
R.~Godbole (Bangalore),
M.~Gruwe (CERN), 
G.~Guignard (CERN), 
T.~Han (Univ. of Wisconsin),
S.~Heinemeyer (CERN), 
C.~Heusch (UC Santa Cruz),
J.~Hewett (SLAC),
S.~Jadach (INP, Krakow),
P.~Jarron (CERN), 
C.~Kenney (MBC, USA),
Z.~K{\i}rca (Osmangazi Univ.),
M.~Klasen (Univ. of Hamburg),
K.~Kong (Univ. of Florida),
M.~Kr\"amer (Univ. of Edinburgh),
S.~Kraml (HEPHY Vienna and CERN),
G.~Landsberg (Brown Univ.),
J.~Lorenzo Diaz-Cruz (BUAP, Puebla),
K.~Matchev (Univ. of Florida),
G.~Moortgat-Pick (Univ. of Durham),
M.~M\"uhlleitner (PSI, Villigen),
O.~Nachtmann (Univ. of Heidelberg),
F.~Nagel (Univ. of Heidelberg),
K.~Olive (Univ. of Minnesota),
G.~Pancheri (INFN, Frascati),
L.~Pape (CERN), 
S.~Parker (Univ. of Hawaii),
M.~Piccolo (LNF, Frascati),
W.~Porod (Univ. of Zurich),
E.~Recepoglu (Univ. of Ankara), 
P.~Richardson (Univ. of Durham), 
T.~Riemann (DESY-Zeuthen),
T.~Rizzo (SLAC), 
M.~Ronan (LBL, Berkeley),
C.~Royon (CEA, Saclay), 
L.~Salmi (HIP, Helsinki),
D.~Schulte (CERN), 
R.~Settles (MPI, Munich),
T.~Sjostrand (Lund Univ.),
M.~Spira (PSI, Villigen),
S.~Sultansoy (Gazi Univ., Ankara and IP Baku),
V.~Telnov (Novosibirsk, IYF),
D.~Treille (CERN), 
M.~Velasco (Northwestern Univ.),
C.~Verzegnassi (Univ. of Trieste),
G.~Weiglein (Univ. of Durham),
J.~Weng (CERN, Univ. of Karlsruhe),
T.~Wengler (CERN),
A.~Werthenbach (CERN), 
G.~Wilson (Univ. of Kansas),
I.~Wilson (CERN), 
F.~Zimmermann~(CERN)

\vspace*{7.5cm}

\begin{center}
CERN Scientific Information Service--1000--June 2004
\end{center}

\newpage
\renewcommand{\thepage}{\roman{page}}
\setcounter{page}{3}
\vglue4cm

\begin{center}
{\Large\bf Abstract}
\end{center}

\vspace*{1cm}

This report summarizes a study of the physics potential of the CLIC
$e^+e^-$ linear collider operating at centre-of-mass energies from 1~TeV
to 5~TeV with luminosity of the order of 10$^{35}$~cm$^{-2}$~s$^{-1}$.
First, the CLIC collider complex is surveyed, with emphasis on
aspects related to its physics capabilities, particularly the luminosity
and energy, and also possible polarization, $\gamma \gamma$ and $e^- e^-$
collisions. The next CLIC Test facility, CTF3, and its R\&D
programme are also reviewed. We then discuss aspects of experimentation at
CLIC, including backgrounds and experimental conditions, and present
a conceptual detector design used in the physics analyses,
most of which use the nominal CLIC centre-of-mass energy of 
3 TeV. CLIC
contributions to Higgs physics could include completing the profile of a
light Higgs boson by measuring rare decays and reconstructing the Higgs
potential, or discovering one or more heavy Higgs bosons, or probing CP
violation in the Higgs sector. Turning to physics beyond the Standard
Model, CLIC might be able to complete the supersymmetric spectrum
and make more precise measurements of sparticles detected previously at
the LHC or a lower-energy linear $e^+ e^-$ collider: $\gamma \gamma$
collisions and polarization would be particularly useful for these tasks.
CLIC would also have unique capabilities for probing other possible
extensions of the Standard Model, such as theories with extra dimensions
or new vector resonances, new contact interactions and models with strong
$WW$ scattering at high energies. In all the scenarios we have studied,
CLIC would provide significant fundamental physics information
beyond that available from the LHC and a lower-energy linear $e^+ e^-$
collider, as a result of its unique combination of high energy and
experimental precision.

\newpage
\thispagestyle{empty}
~

\newpage
\tableofcontents 

\newpage
\renewcommand{\thepage}{\arabic{page}}
\setcounter{page}{1}
\chapter{INTRODUCTION}
\label{chapter:one}

The energy range up to 100~GeV has been explored by the hadron--hadron
colliders at CERN and Fermilab, by the LEP $e^+ e^-$ collider and the
SLC, and by the $ep$ collider HERA. 
The next energy frontier is the range up to 1~TeV, which will first be
explored by the LHC. Just as $e^+ e^-$ colliders provided an essential
complement to hadron--hadron colliders in the 100~GeV energy range,
establishing beyond doubt the validity of the Standard Model, so we expect
that higher-energy $e^+ e^-$ colliders will be needed to help unravel the
TeV physics, to be unveiled by the LHC. They provide very clean experimental
environments and democratic production of all particles within the
accessible energy range, including those with only electroweak
interactions. These considerations motivate several projects for $e^+ e^-$
colliders in the TeV energy range, such as TESLA, the NLC and JLC. We
assume that at least one of these projects will start up during the
operation of the LHC. However, we do not expect that the full scope of
TeV-scale physics will then be exhausted, and we therefore believe that a
higher-energy $e^+ e^-$ collider will be needed.

The best candidate for new physics at the TeV scale is that associated 
with generating masses for elementary particles. This is expected to 
involve a Higgs boson, or something to replace it. The precision 
electroweak data from LEP and elsewhere rule out many alternatives to the 
single elementary Higgs boson predicted by the Standard Model, and suggest 
that it should weigh $\lappeq$~200~GeV. A single elementary Higgs boson is 
not thought to be sufficient by itself to explain the variety of the
different mass scales in physics. Many theories beyond the Standard
Model, such as those postulating supersymmetry, extra dimensions or
new strong interactions, predict the appearance of non-trivial new
dynamics at the TeV scale.

For example, supersymmetry predicts that every particle in the Standard
Model should be accompanied by a supersymmetric partner weighing 
$\lappeq$~1~TeV. Alternatively, theories with extra spatial dimensions
predict the 
appearance of new particle excitations or other structural phenomena at
the TeV scale. Finally, alternatives to an elementary Higgs boson, such as
new strong interactions, also predict many composite resonances and other
effects observable at the TeV energy scale. 

Whilst there is no direct evidence, there are various indirect 
experimental hints that there is indeed
new dynamics at the TeV scale. One is the above-mentioned agreement of 
precision
electroweak data with the Standard Model, {\it if} there is a relatively
light Higgs boson. Another is the agreement of the gauge couplings
measured at LEP and elsewhere with the predictions of simple grand unified
theories, {\it if} there is a threshold for new physics at the TeV scale,
such as supersymmetry. Another hint may be provided by the apparent
dominance of dark matter in the Universe, which may well consist of
massive, weakly-interacting particles, {\it in which case} they should
weigh $\lappeq$~1~TeV. Finally, we note that there {\it may be} a
discrepancy between the measurement of the anomalous magnetic moment of
the muon and the prediction of the Standard Model, which could only be
explained by new dynamics at the TeV scale.
\begin{figure}[!ht] 
  \begin{center}
    \epsfig{file=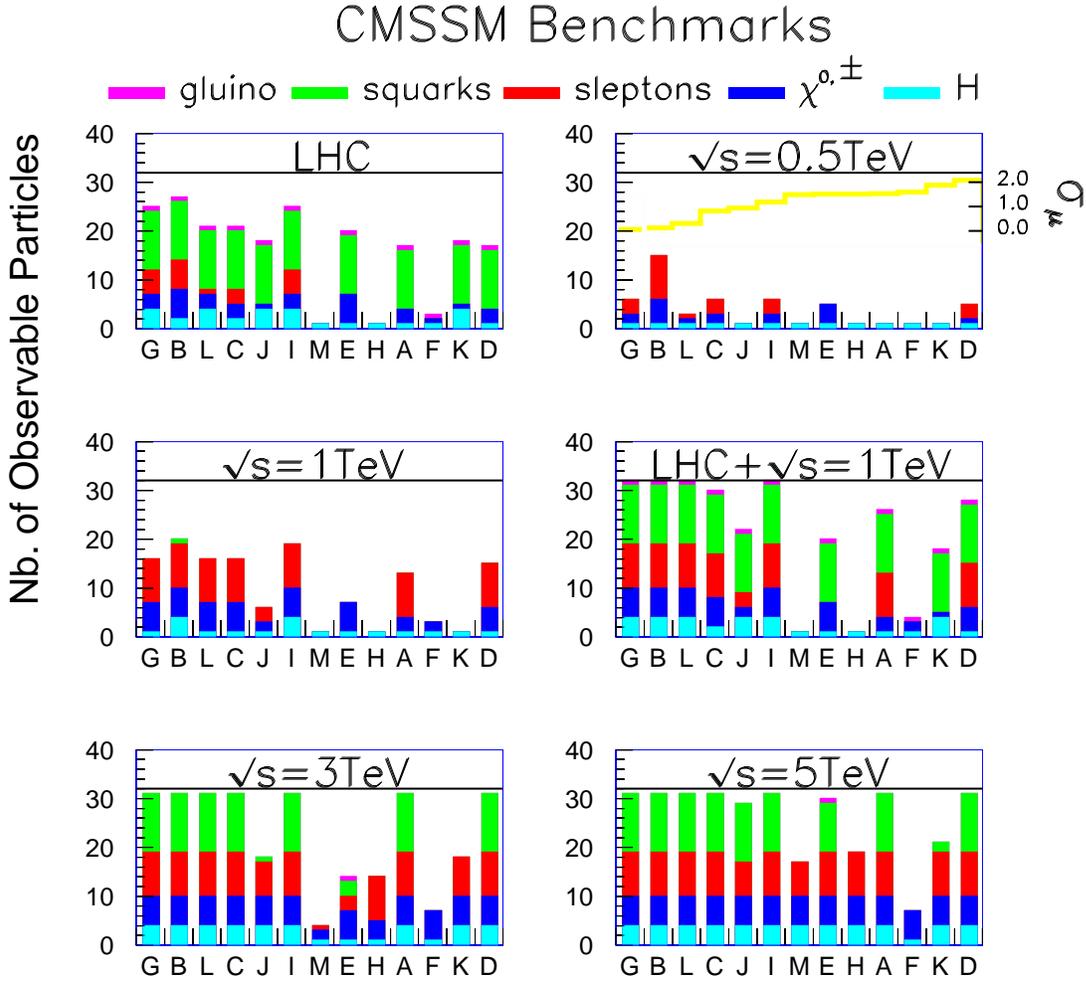,width=15cm}
    \caption{
Bar charts of the numbers of different sparticle species observable in a 
number of benchmark supersymmetric scenarios at 
different colliders, including the  LHC
and linear $e^+ e^-$ colliders with 
various centre-of-mass energies. The benchmark scenarios are ordered by 
their consistency with the most recent BNL measurement of $g_\mu - 2$ and 
are compatible with the WMAP data on cold dark matter density. We 
see that there are some scenarios where the LHC discovers only the 
lightest neutral supersymmetric Higgs boson. Lower-energy linear $e^+ 
e^-$ colliders largely complement the LHC by discovering or measuring 
better the lighter electroweakly-interacting sparticles. Detailed 
measurements of the squarks would, in many cases, be possible only at 
CLIC.}
    \label{fig:Manhattan}
  \end{center}  
\end{figure}

We expect that the clean experimental conditions at a TeV-scale linear
$e^+ e^-$ collider will enable many detailed measurements of this new
dynamics to be made. However, we also expect some aspects of TeV-scale
physics to require further study using a higher-energy $e^+ e^-$ collider.
For example, if there is a light Higgs boson, its properties will have been
studied at the  LHC
and the first $e^+ e^-$ collider, but one would wish to
verify the mechanism of electroweak symmetry breaking by measuring the
Higgs self-coupling associated with its effective potential, which would
be done better at a higher-energy $e^+ e^-$ collider. On the other hand,
if the Higgs boson is relatively heavy, measurements of its properties at
the LHC or a lower-energy $e^+ e^-$ collider will quite possibly have been
incomplete. As another example, if Nature has chosen supersymmetry, it is
quite likely that the LHC and the TeV-scale $e^+ e^-$ collider will not
have observed the complete sparticle spectrum, as seen in
Fig.~\ref{fig:Manhattan}. 
\begin{figure}[!ht] 
  \begin{center}
\hspace*{3.5cm}\epsfig{file=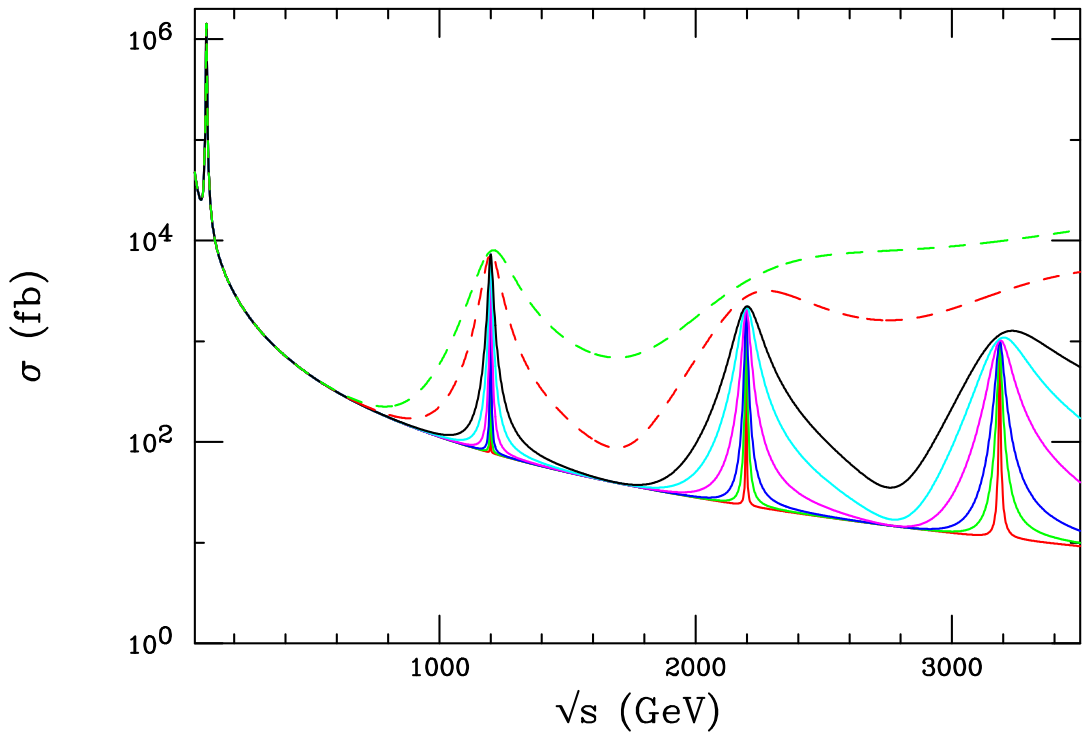,width=13cm,angle=90}

\vspace*{-11mm}

    \caption{An example of the dilepton spectrum that might be 
observed at the  LHC in some scenario for extra dimensions, including 
Kaluza--Klein excitations of the photon and $Z$ and their interferences.}
    \label{fig:extrad}
  \end{center}  
\end{figure}

Moreover, in many cases detailed measurements at
a higher-energy $e^+ e^-$ collider would be needed to complement previous
exploratory observations, e.g.\ of squark masses and mixing, or of heavier
charginos and neutralinos. Analogous examples of the possible
incompleteness of measurements at the LHC and the TeV-scale $e^+ e^-$
collider can be given in other scenarios for new physics, such as extra
dimensions, as discussed in later chapters of this report. Certainly a
multi-TeV linear $e^+ e^-$ collider would be able to distinguish smaller
extra dimensions than a sub-TeV machine.  Even if prior machines do
uncover extra dimensions, it would, for example, be fascinating to study
in detail at CLIC  a Kaluza--Klein excitation of the $Z$ boson that might
have been discovered at the LHC, as seen in Fig.~\ref{fig:extrad}.

For all the above reasons, we think that further progress in particle
physics will necessitate clean experiments at multi-TeV energies, such as
would be possible at a higher-energy $e^+ e^-$ collider like CLIC. 
This 
would, in particular, be the logical next step in CERN's vocation to study 
physics at the high-energy frontier. CERN and collaborating institutes 
have already made significant progress towards demonstrating the 
feasibility of this accelerator concept, whilst other projects for 
reaching multi-TeV energies, such as a $\mu^+ \mu^-$ collider or a very 
large hadron collider, seem to be more distant prospects.

Some exploratory studies of CLIC physics have already been made, but the
close integration of experiments at linear $e^+ e^-$ colliders with
the accelerator, particularly in the final-focus region, now mandate a
more detailed study, as described in this report.

Chapter 2 summarizes the design of the CLIC  accelerator, including the
overall design concept, its general parameters such as energy and
luminosity, the collision energy spread, the prospects for obtaining
polarized beams, and the option for a $\gamma\gamma$ collider.  
A crucial step has recently been demonstrated at the second CLIC 
test facility, namely the attainment of an accelerating gradient in excess of
150~MeV/m. Chapter 3 discusses experimental aspects, such as the
levels of experimental backgrounds expected, the specification of a
baseline detector, the luminosity measurement, and the simulation tools
available for experimental studies. 

Chapter 4 is devoted to Higgs physics, including the prospects for
measuring the triple-Higgs coupling for a relatively light Higgs boson,
heavy-Higgs studies, the possibility of observing CP violation in the
heavy-Higgs sector of the minimal supersymmetric extension of the Standard
Model (MSSM), and possible Higgs studies with a $\gamma\gamma$ collider.

Chapter 5 reviews possible studies of supersymmetry at CLIC, with
particular attention to certain benchmark MSSM scenarios where we
demonstrate their complementarity with studies at the  LHC
and a lower-energy 
$e^+ e^-$ collider. We discuss in particular possible precision 
measurements of sleptons, squarks, heavier charginos and neutralinos, and 
the possibility of gluino production in $\gamma\gamma$ collisions.

Other scenarios for new physics are presented in Chapter 6, including
direct and indirect observations of extra dimensions, black-hole
production, non-commutative theories, etc. Chapter 7 summarizes QCD
studies that would be possible at CLIC, in both $e^+ e^-$ and
$\gamma\gamma$ collisions.

Finally, Chapter 8 summarizes the conclusions of this study of the physics
accessible with CLIC.


\newpage
\chapter{ACCELERATOR ISSUES AND PARAMETERS}
\label{chapter:two}

\newcommand{\EPEM}{\mbox{e$^+$e$^-$}}

\newcommand{\EE}{\mbox{ee}}
\newcommand{\GG}{\mbox{$\gamma\gamma$}}
\newcommand{\GP}{\mbox{$\gamma$e$^+$}}
\newcommand{\GE}{\mbox{$\gamma$e}} 
\newcommand{\LGE}{\mbox{$L_{\GE}$}}
\newcommand{\LGG}{\mbox{$L_{\GG}$}}
\newcommand{\LEE}{\mbox{$L_{\EE}$}}

\newcommand{\MKM}{\mbox{$\mu$m}}
\newcommand{\CMS}{\mbox{cm$^{-2}$s$^{-1}$}}


\section{Overview of the CLIC Complex} 

The  CLIC (Compact Linear Collider) study aims at a multi-TeV,  
high-luminosity $e^+e^-$ linear collider. 
In order to reach high energies with a linear collider, a cost-effective
technology is of prime importance. In conventional linear accelerators, 
the RF power used to accelerate the main beam is generated by klystrons.
To achieve multi-TeV energies, high accelerating gradients are
necessary to limit the lengths of the two main linacs and hence
the cost. Such high gradients are easier to achieve at higher RF
frequencies since, for a given gradient, the peak power in the 
accelerating structure is smaller than at low frequencies. For this
reason, a frequency  
of 30~GHz has been chosen for CLIC to attain a gradient of 150~MV/m.
However, the production of highly efficient klystrons is
very difficult at high frequency. Even in the X-band at 11.5~GHz, a very
ambitious programme has been necessary at SLAC and KEK to develop
prototypes that come close to the required performance. At even higher
frequencies, the difficulties of building efficient high-power
klystrons are significantly larger. Instead, the CLIC  study is based
on the two-beam accelerator scheme. The RF power is  
extracted from a low-energy, high-current drive beam, which is
decelerated in power   
extraction and transfer structures of low impedance.
This power is then directly transferred into the high-impedance
structures of the main linac and used to accelerate the high-energy,
low-current main beam, which is   
later brought into collision. The two-beam approach offers a solution
that avoids the use of a large number of active RF elements,
e.g. klystrons or modulators, in the main  
linac. This potentially eliminates the need for a second tunnel.

In the  CLIC scheme, the drive beam is created and accelerated
at low frequency (0.937~GHz) where efficient klystrons can be realized
more easily. The frequency and intensity of the beam is then increased
in the chain of a delay loop and two combiner rings. This drive-beam
generation system can be installed at a central site, thus 
allowing easy access and replacement of the active RF elements.

\begin{figure}[htbp] 
\centerline{\includegraphics[width=13cm]{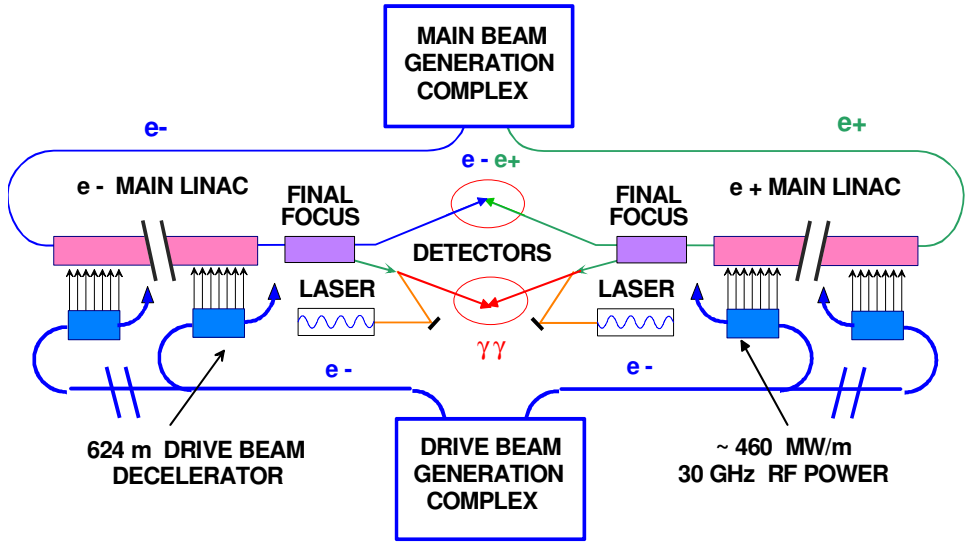}}
\caption{The schematics of the overall layout of the CLIC complex}
\label{fig1_1}
\end{figure}

The  CLIC design parameters have been optimized for a nominal
centre-of-mass energy $\sqrt{s}$~=~3~TeV with a luminosity of about
10$^{35}$~cm$^{-2}$s$^{-1}$~\cite{ref1_1}, but
the  CLIC concept allows its construction to be staged without
major modifications (see Fig.~\ref{fig1_1}). The possible
implementation of a lower-energy phase for physics 
would depend on the physics requirements at the time of construction. 
In principle, a first  CLIC stage could cover centre-of-mass
energies between $\sim$~0.1 and 0.5~TeV with a luminosity of
$\mathcal{L}$~=~10$^{33}$--10$^{34}$~cm$^{-2}$s$^{-1}$, providing an
interesting physics overlap with the LHC. This stage could then be
extended first to 1~TeV, with $\mathcal{L}$ above
10$^{34}$~cm$^{-2}$s$^{-1}$, and then to multi-TeV operation, with
$e^+e^-$ collisions at 3~TeV, which should break new physics  
ground. A final stage might reach a collision energy of 5~TeV or more.


The sketch of Fig.~\ref{fig1_1} gives an overall layout of the complex
with the linear decelerator units running parallel to the main
beam~\cite{ref1_2}. Each unit is 625~m long and decelerates a
low-energy, high-intensity $e^{-}$~beam, the drive-beam,  
which provides the RF power for each corresponding unit of the main 
linac through energy-extracting RF structures. Hence, there are no
active elements in the main tunnel. With a gradient of 150~MV/m, the
main beam is accelerated by $\sim$~70~GeV in each unit. Consequently,
the natural lowest value and step size of the colliding beam energy in
the centre of mass ($\sqrt{s}$) is $\sim$~140~GeV, though both can be
tuned by adjusting the drive-beam and decelerator. The nominal energy
of 3~TeV requires 2~$\times$~22~units, for a total two-linac length of
$\sim$~28~km. Each unit contains 500 power-extraction  
transfer structures (PETSs) feeding 1000 accelerating structures. 

The two-beam acceleration method of  CLIC ensures that the design
remains essentially independent of the final energy for all the major
subsystems, such as the main beam injectors, the damping rings, the
drive-beam generators, the RF power source, the main-linac and drive-beam 
decelerator units, as well as the beam delivery systems (BDSs).  
The  CLIC modularity is made easier by the fact that the complexes for
the generation of all the beams and the interaction point~(IP) are
located at a central position, where all power sources are also
concentrated. The main tunnel houses both linacs, the various 
beam transfer lines and, in its centre, the BDSs. 

This chapter summarizes the  CLIC two-beam study and discusses the 
interplay between the achievable energy and luminosity and the design
of the various accelerator system components, with emphasis on the
features most relevant to the CLIC physics performance. These systems
are the focus of a continuing research and development programme, in
particular for the high-gradient structures, the damping rings, the
vibration stabilization systems, and the beam delivery section. The
main-beam and the drive-beam parameters are summarized in
Table~\ref{tab:param}, for the nominal energy of 3~TeV as well as for
500~GeV, as an example for lower energies.

\begin{table}[t] 
\caption{Main CLIC machine parameters}
\label{tab:param}

\renewcommand{\arraystretch}{1.3} 
\begin{center}

\begin{tabular}{lcc}\hline \hline \\[-3mm]
\textbf{Collision energy \boldmath{$\sqrt{s}$} \,\, (TeV)} & 
\bf{0.5} & \textbf{3.0} \\[4mm]  

\hline \\[-3mm]
Design luminosity $\cal{L}$ (10$^{35}$~cm$^{-2}$s$^{-1}$) & 
0.2 & 0.8 \\
Linac repetition frequency (Hz) & 
200 & 100 \\
No.\ of ptcs./bunch $N$ (10$^{10}$) & 
0.4 & 0.4 \\
No.\ of bunches/pulse $n_b$ & 
154 & 154 \\
Bunch separation (ns) & 
0.67 & 0.67 \\
Bunch length ($\mu$m) & 
35 & 35 \\
Normalized emittance 
$\gamma \epsilon^*_x$/$\gamma \epsilon^*_y$ 
(m$\cdot$rad~$\times$~10$^{-6}$)&
2.0/0.01 & 0.68/0.01 \\
Beam size at collision 
$\sigma^*_x$/$\sigma^*_y$/$\sigma^*_z$ (nm/nm/$\mu$m))
$\hspace*{27mm}$ & 
202/12/35 & 60/0.7/35 \\
Energy spread $\Delta E/E$ (\%) &
0.25 & 0.35 \\
Crossing angle (mrad) & 
20 & 20 \\
Beamstrahlung $\delta_B$ (\%) &
4.4 & 21 \\
Beam power/beam (MW) & 
4.9 & 14.8 \\
Gradient unloaded/loaded (MV/m) & 
150 & 150 \\
Two-linac length (km) & 
5.0 & 28.0 \\
Beam delivery length (km) & 
5.2 & 5.2 \\
Final focus length (km) & 
1.1 & 1.1 \\
Total site length (km) & 
10.2 & 33.2 \\
Total AC power (MW) & 
175 & 410
\\[3mm] 
\hline \hline
\end{tabular}
\end{center}
\end{table}


\section{CLIC Energy and RF Technology Choice}

Linear, single-pass colliders are currently the most advanced concepts
for particle accelerators capable of reaching multi-TeV energies in
lepton collisions. At these energies, the choice of technology may be
narrowed down to high-frequency, normal-conducting cavities for the
reasons discussed above. Superconducting linac technology, such as
that proposed for the lower-energy TESLA collider, is limited  to
accelerating gradients of about 50~MV/m. At this value, the critical
magnetic field strength for superconductivity is reached on  
the cavity walls. This limit is fundamental and cannot be overcome, at
least in the present theoretical understanding of superconductivity.  
Using normal conducting linear accelerator technology, as employed for
the SLC and now proposed for the NLC/JLC projects, the achievable
accelerating gradients are considerably higher, in
principle. Gradients of 65~MV/m are now routinely achieved with
NLC/JLC test structures in the X~band at 11.5~GHz over long running
periods.  

In principle, higher RF frequencies facilitate higher field
strength. Furthermore in the high beamstrahlung regime the luminosity
increases with the RF frequency and is independent of the gradient.
Recent results at dedicated test facilities have shown, however, that
above $\simeq$~10~GHz the cavity geometry, material and surface
preparation become the predominant factors, determining the achievable
accelerating fields~\cite{gl_linac02}. The choice of frequency for
normal conducting linacs is therefore based on an optimization of
other aspects, such as beam dynamics, technical feasibility,  
power consumption, and investment costs.

The energy needed to establish a given accelerating field $E$ over a
given length scales with $\nu^2$, which can readily be understood from
the scaling of cavity dimensions with the frequency $\nu$. 
The time scale for the dissipation of this energy by resistive losses 
on the cavity walls in conjunction with the skin effect scales like
$\nu^{-3/2}$. Hence, the instantaneous power per unit length to
maintain a certain field strength scales as $\nu^{-1/2}$, favouring
higher frequencies. However, the small cavity dimensions at high
frequency lead to strong beam-induced transverse wakefields,
generating transverse instabilities. These limit the number of
particles that can be stored in a single bunch and hence the
luminosity. This effect can be offset by choosing a shorter bunch
length $\sigma_z$ at higher frequency, which increases the luminosity
for multi-TeV collisions. To take advantage of from this effect, the
horizontal beam size at the interaction point has to be decreased with
increasing frequency. This beam size has, however, a lower limit due to the
performance of damping rings and beam-delivery systems~\cite{lu_epac02}.

Considering these aspects together, it turns out that for an
accelerating field of 150~MV/m the overall cost as a function of
frequency has a rather flat minimum between 20~GHz and 30~GHz. For
lower frequencies the costs to supply the pulsed RF energy become
prohibitive and for higher frequencies the reduced bunch charge
precludes sufficient luminosity with present damping ring and
beam-delivery systems~\cite{oc_pac03}.

The RF power needed to establish an accelerating field of 150~MV/m is
about 100--200~MW/m at 30~GHz, with the precise value depending on the
details of the cavity geometry. While such power can only by supplied
over short pulses, it would be more favourable from the RF 
power-source point of view to supply a given amount of pulse energy in  
a longer pulse of lower power. To reconcile these conflicting
requirements, different RF pulse compression schemes have been
developed.  In the case of NLC/JLC, the compression is performed by
intermediate storage of long RF pulses in low loss waveguides. In the
CLIC design, the compression is achieved by accelerating a long drive
beam pulse in a low frequency (937~MHz) accelerator. This long pulse
is wound up in a combiner ring using a sophisticated injection scheme
based on RF dipoles. This scheme achieves multiplication of the bunch
repetition frequency and pulse compression simultaneously.

\begin{figure}[t] 
\begin{center}
\hspace*{-0.5cm} \epsfig{file=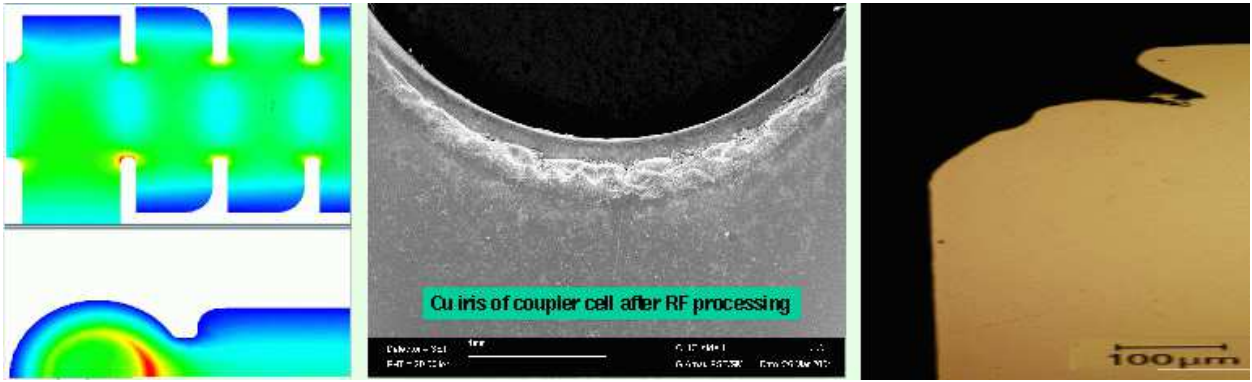,height=5.0cm}
\end{center}
\caption{Macro-photographs of the input coupler of a 30~GHz RF
copper structure, showing the  erosion damage subsequent to breakdown
in RF tests} 
\label{fig:erosion}
\end{figure}

\begin{figure}
\begin{center}
\epsfig{file=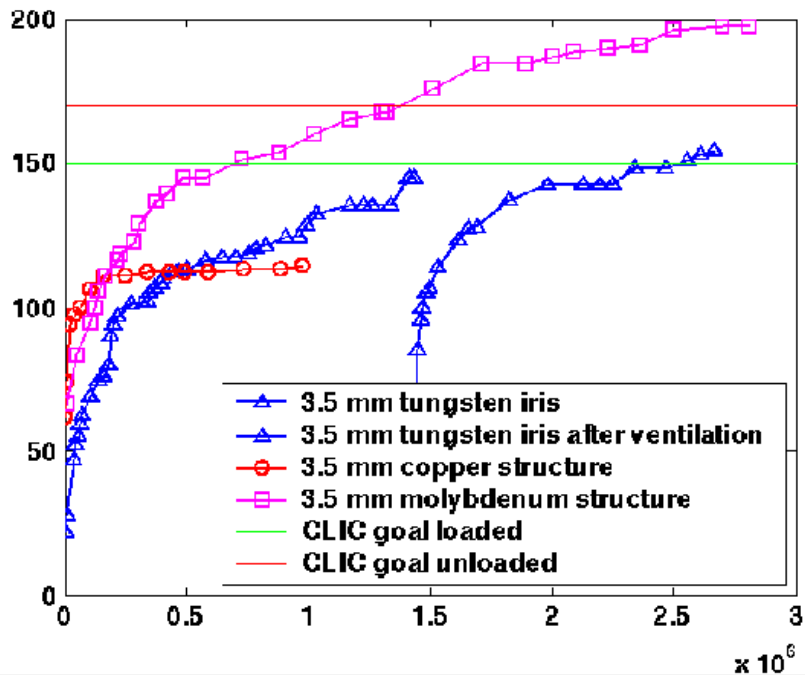,height=9.0cm}
\end{center}
\caption{Accelerating gradients obtained with 30 GHz structures of
different designs. The gradient measured in the first cell of the
structure is shown as a function of the number of applied RF pulses.}
\label{fig:gradient}
\end{figure}

The limiting factors to the achievable accelerating gradient are the
RF breakdown in the cavities and the related erosion of the cavity
surfaces. Erosion effects due to breakdown were observed in early
tests of a CLIC prototype structure, whose damaged iris 
is shown in~Fig.~\ref{fig:erosion}. The physics of these breakdowns
is currently not precisely known. It is generally believed that
field emission from the cavity walls triggering a runaway plasma
formation is the process responsible for it. Recent experiments in the
CLIC Test Facility 2 (CTF2) indicate that these effects can be
overcome, in normal operating conditions, by replacing the copper with  
either molybdenum or tungsten as material for the structure irises. 
Figure~\ref{fig:gradient} summarizes the results obtained with test
structures adopting this new configuration. The feasibility of
achieving gradients up to 193~MV/m has thus been
demonstrated~\cite{ww_pac03} in CTF2 for short RF pulses.  Tests with
pulses of nominal length will become possible in the  CLIC
Test Facility~3 (CTF~3).

Pulsed surface heating represents another potentially severe
limitation. Although the present CLIC RF~structure design attempts to
minimize this effect, a temperature rise still beyond that sustained
in present linacs is expected. A full understanding of its 
impact on the operation of the cavities will 
become possible once CTF3 provides
30~GHz RF pulses of the designed length and amplitude.

\section{CLIC Luminosity}

The luminosity ${\cal L}$ in a linear collider can be expressed as a function
of the effective transverse beam
sizes\footnote{In CLIC the colliding bunches will have significant
transverse tails and a much better focused core. To simplify the following
discussion, effective beam sizes are used. They give the sigmas of the Gaussian
distributions that would lead to the same luminosity and beam--beam interaction
at the collision point as the actual distributions~\cite{lu_epac02}.}
$\sigma_{x,y}$ at the interaction point (IP), the bunch
population $N$, the number of bunches $n_b$ per beam pulse and the number of
pulses per second $f_r$:
\begin{equation}
{\cal L}=H_D\frac{N^2}{4\pi\sigma_x\sigma_y}n_bf_r\,.
\end{equation}
Here, the luminosity enhancement factor $H_D$, which is usually in the range of
1--2, is due to the beam--beam interaction, which focuses the
$e^+e^-$ beams during collision. The equation can be expressed as a
function of the power consumption $P$ of the collider and of the total
power to beam power efficiency $\eta$,~obtaining: 
\begin{equation}
{\cal L}\propto H_D\frac{N}{\sigma_x}\frac{1}{\sigma_y}\eta P\,.
\end{equation}
The different parameters are not independent and the dependences can be quite
complex. However, three main fundamental limitations arise from the
factors $N/\sigma_x$, $\sigma_y$ and $\eta$, if the other parameters
are kept~fixed. 

\begin{itemize}
\item
The optimum ratio $N/\sigma_x$ is determined by the beam--beam
interaction\footnote{The relevant parameter is more precisely
$N/(\sigma_x+\sigma_y)$, but, in order to maximize luminosity and
simultaneously minimize beam--beam
effects, one normally has parameters with $\sigma_x\gg\sigma_y$. In this case
the important term is $N/\sigma_x$.}.
At large $N/\sigma_x$ the total luminosity is highest, but the colliding
particles strongly emit beamstrahlung during the collision. Hence the
luminosity spectrum will be degraded and the backgrounds higher.
\item
The value of $\sigma_y$ is, in the case of  CLIC, mainly limited by
the difficulty of creating such a small beam size and by the difficulty of
keeping two small beams in collision. The achievable $\sigma_y$ depends on the
bunch charge $N$.
\item
The efficiency of the beam acceleration $\eta$ mainly depends on the
RF technology chosen for the main linac and on the beam current, e.g. larger
$N$ leads to better efficiency.
\end{itemize}

\noindent
The above parameters are strongly coupled. An important example for a coupling
parameter is the bunch length $\sigma_z$. In a given main linac the bunch
length is a function of the bunch charge, larger $N$ requiring larger
$\sigma_z$. In turn, the optimum ratio $N/\sigma_x$ is a function of
$\sigma_z$. The achievable $\sigma_y$ also depends on $N$, since larger
$N$ and larger $\sigma_z$ lead to larger $\sigma_y$.

An additional limitation arises from the damping ring and the beam
delivery system.
\begin{itemize}
\item
For the nominal  CLIC parameters, a lower limit 
$\sigma_x\ge $~60~nm is currently found. The damping ring and the beam delivery
system contribute equally to this limit. It remains to be investigated
if this limit is fundamental. 
\end{itemize}

In the following the limitations for the three main factors that
determine the luminosity are presented. The trade-off between
luminosity and beamstrahlung at the interaction point is discussed
first. Then the issues related to achieving the small needed
$\sigma_y$ are detailed, with emphasis on the resulting luminosity. 

\subsection{Horizontal Beam Size and Bunch Charge}

A fundamental lower limit to the ratio of bunch charge and horizontal
beam size at the IP arises from the strong beam--beam
interaction. Because of this
effect, the beams are focused during the collision. While increasing
the luminosity ${\cal L}$, this gives rise to the emission of
beamstrahlung, with each beam particle typically emitting  
$O(1)$ photon. The beamstrahlung alters the beam particles' energies,
so that particles can collide at energies different from nominal and a wide
luminosity spectrum is delivered. In most physics investigations, only some
fraction of the luminosity ${\cal L}_1$ close to the nominal centre-of-mass
energy is of interest\footnote{The definition of which part of the
luminosity belongs to ${\cal L}_1$ depends on the experiment. For simplicity
one can assume
${\cal L}_1=\int_{(1-x)E_{{\rm c.m.},0}}^{E_{{\rm c.m.},0}}{\cal L}
(E_{\rm c.m.}) dE_{\rm c.m.}$, where $x\ll$~1. The precise value of
$x$ turns out not to be very important and we shall use~0.01.}.

\begin{figure}[t]
\begin{center}
\epsfig{file=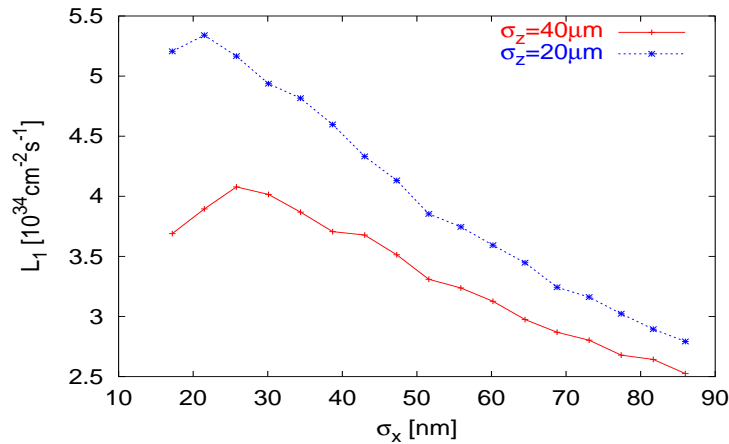,width=10.0cm,height=6.0cm}
\end{center}
\caption{The high-energy luminosity ${\cal L}_1$ as a function of the
horizontal beam size for two different bunch lengths and nominal bunch
charge. In both cases there exists a clear optimum. For the shorter bunch
length the total peak luminosity is higher since the beamstrahlung is
more suppressed.} 
\label{f:sx}
\end{figure}

If one keeps the other parameters constant, the beamstrahlung depends almost
completely on the ratio $N/\sigma_x$.
Decreasing $\sigma_x$ or increasing $N$ increases the total luminosity, but
also increases the beamstrahlung. Consequently the fraction of luminosity
close to the nominal energy ${\cal L}_1/{\cal L}$ decreases. For otherwise
fixed parameters an optimum $\sigma_x$ exists, which maximizes
${\cal L}_1$, see Fig.~\ref{f:sx}. However, the optimum $\sigma_x$ and
the maximum luminosity ${\cal L}_1$ depend on these other parameters.
As can be seen in the
figure, the use of a shorter bunch allows one to use a smaller horizontal beam
size and yields a higher luminosity even for the same transverse size.

However, it is not only the wish to maximize ${\cal L}_1$ that can
lead to a lower limit on $\sigma_x$. One may also require a certain
quality of the luminosity spectrum (e.g. for threshold studies) or
certain background conditions: at smaller $\sigma_x$ the background
levels will be higher. 

With the current damping ring and BDS designs it is found that one cannot
achieve horizontal beam sizes below about $\sigma_x\ge$~60~nm. If this limit
is fundamental, it will make it impossible to achieve the optimum $N/\sigma_x$
for small bunch charges, with the consequences discussed 
in~Section~\ref{r:sx} 

\subsection{Vertical Beam Size}
In order to achieve a small vertical beam size at the IP,
the vertical phase space occupied by the beam---the vertical emittance
$\epsilon_y$---must be small. The total effective beam size at the IP can be
expressed in a simplified way as a function of the total emittance and the
focal strength of the final-focus system (described by $\beta_y$):
\begin{equation}
\sigma_{y,{\rm eff}}\propto\sqrt{\beta_y\left(\epsilon_{y,{\rm DR}}
+\epsilon_{y,{\rm BC}}+\epsilon_{y,{\rm linac}}+\epsilon_{y,{\rm BDS}}
+\epsilon_{y,{\rm jitter}}\right)}\,.
\label{e:lumi}
\end{equation}
Consequently a number of challenges have to be met to achieve a small vertical
beam size.
\begin{itemize}
\item
First, a beam with a small emittance $\epsilon_{y,{\rm DR}}$ must be created in
the damping ring. The target for  CLIC is $\epsilon_{y,{\rm DR}}\le$~3~nm.
\item
This beam needs to be longitudinally compressed and transported to the main
linac with a small emittance growth $\epsilon_{y,{\rm BC}}$.
 The target is $\epsilon_{y,{\rm BC}}\le$~2~nm.
\item
The emittance growth during the acceleration in the main linac
$\epsilon_{{\rm linac}}$ has to be kept small.
 The target is $\epsilon_{y,{\rm linac}}\le$~10~nm.
\item
In the BDS the beam tails are scraped off and the beam
is focused to a very small spot size. This system must also lead to a
small emittance growth $\epsilon_{{\rm BDS}}$ and
at the same time achieve strong focusing, i.e. a small vertical
beta-function $\beta_y$. The target is 
$\epsilon_{y,{\rm BDS}}\le$~10~nm for the nominal $\beta_y$~=~70~$\mu$m.
\item
The beams need to collide. Dynamic effects in the whole accelerator lead to a
continuous motion of the beam trajectory, and this motion can be described by
a growth of the multi-pulse emittance $\epsilon_{y,{\rm jitter}}$. This growth
should be much smaller than the other contributions.
\end{itemize}
It is obvious that all the emittance contributions must be minimized to
achieve a small spot size and that further optimization of one value becomes
useless if the sum is dominated by some other contribution.
The different contributions are not independent, but for the sake of
simplicity, they are discussed separately in the following. For each
subsystem a design must first be developed, which in principle can
achieve the required 
performance; then the consequences of imperfect realizations of this design
must be considered and finally the effects of dynamic imperfections.

\subsubsection{Damping ring emittance}
The vertical emittance of the beam is large at production. Hence, it needs to
be reduced in a damping ring. The design value for the vertical emittance
after the damping ring is $\epsilon_{y,{\rm DR}}$~=~3~nm. The
possibility to achieve this is currently under
investigation. Simulations of different possible layouts of the ring
have so far not reached values better than 
$\epsilon_{y,{\rm DR}}$~=~9~nm~\cite{c:dr}. In addition, not all
effects in the ring have been studied yet, in particular the imperfections.
However, it is hoped that the design can be improved by a further
optimization that takes into account all the limiting physics effects
at the design stage.

\subsubsection{Bunch compressor emittance growth}

A design for the bunch compressor exists, but its performance has not been
completely evaluated~\cite{c:design}. In particular, the simulation of the
emittance growth due to coherent synchrotron radiation and imperfections
remains to be done. However, preliminary studies of the coherent synchrotron
radiation indicate that they remain acceptable~\cite{c:eric}.

\subsubsection{Main linac emittance growth}

The preservation of the emittance in the main linac is one of the
major challenges in a linear collider design. This is due to a large
extent to the so-called wakefields that the beam experiences when
passing the accelerating structures. The size of these wakefields is
strongly dependent on the chosen accelerating technology and
frequency. The design of the main linac has now reached a relatively  
mature state and some significant work has already been done to estimate and
minimize the effect of imperfections, which is the main issue.

Structure offsets from the nominal beam line induce a transverse
electric field, the wakefield, which induces a transverse kick on the
beam. This effect can be large, especially in a high-frequency linac.
The emittance growth due to imperfections can be tackled with
different countermeasures. 

\begin{itemize}
\item
First, the main linac lattice is designed to reduce the sensitivity to such 
imperfections. 
\item
Second, a sophisticated prealignment system using wires, lasers and
hydrostatic levelling devices is foreseen to position the elements in
CLIC with small errors to reduce the imperfections.
\item
Third, beam-based alignment will be used. Small remaining imperfections
are detected using the beam itself, and their effect on the beam is corrected.
Simulations predict that, after application of these
procedures, most of the remaining emittance growth is due to the wakefields
in the RF structures of the main linac\cite{c:emitt}.
\item
The accelerating structures are mounted on
movable girders and each of them incorporates a beam position monitor.
This allows one to correct their position with respect to the beam by direct
observation and minimization of the beam offset.
\item
Finally, so-called emittance tuning bumps are used. A few structures are moved
in order to minimize the emittance at the end of the linac. This globally
compensates the mean beam offset, which remains because of imperfect
measurement of the beam position in each structure.
\end{itemize}
The final emittance growth after these steps is about 1.5~nm and thus
significantly smaller than the target.

\begin{figure}[t]
\begin{center}
\begin{tabular}{c c}
\epsfig{file=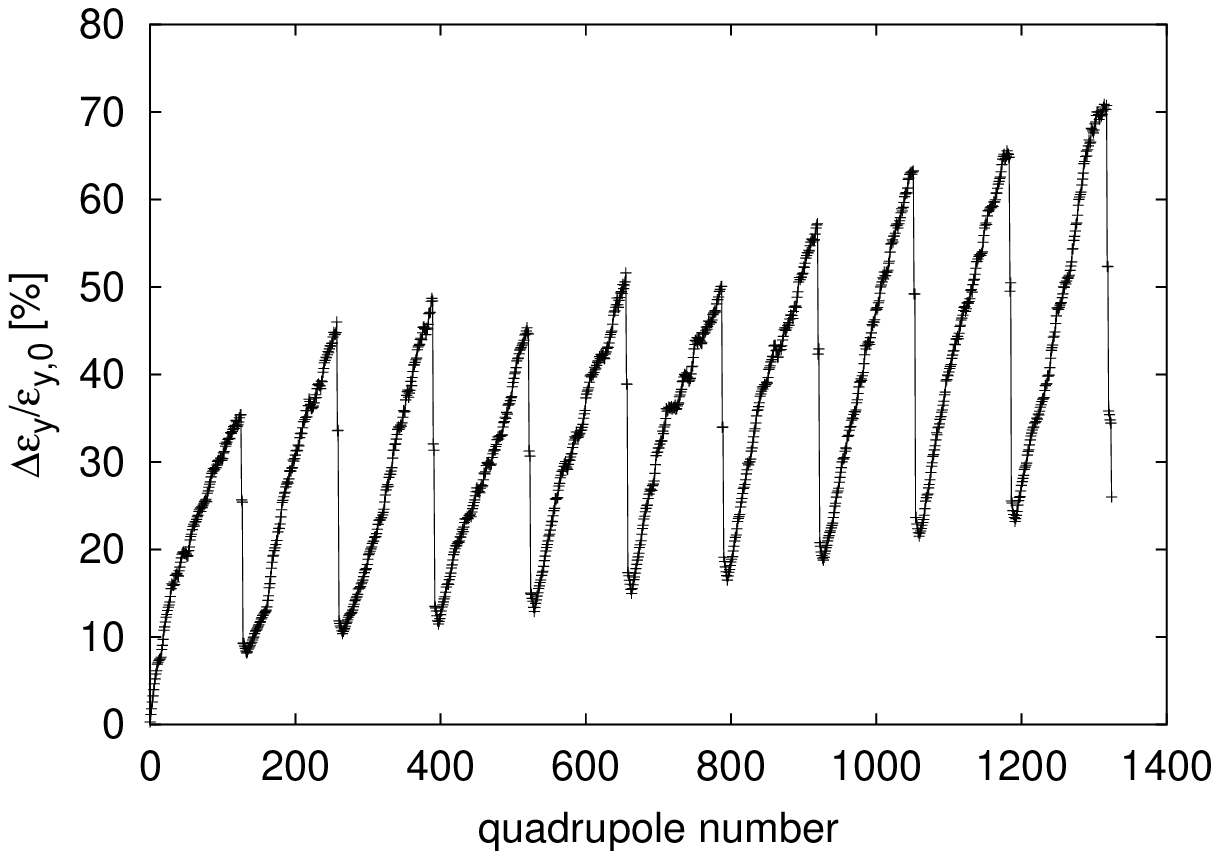,width=7.5cm,height=5.5cm} &
\epsfig{file=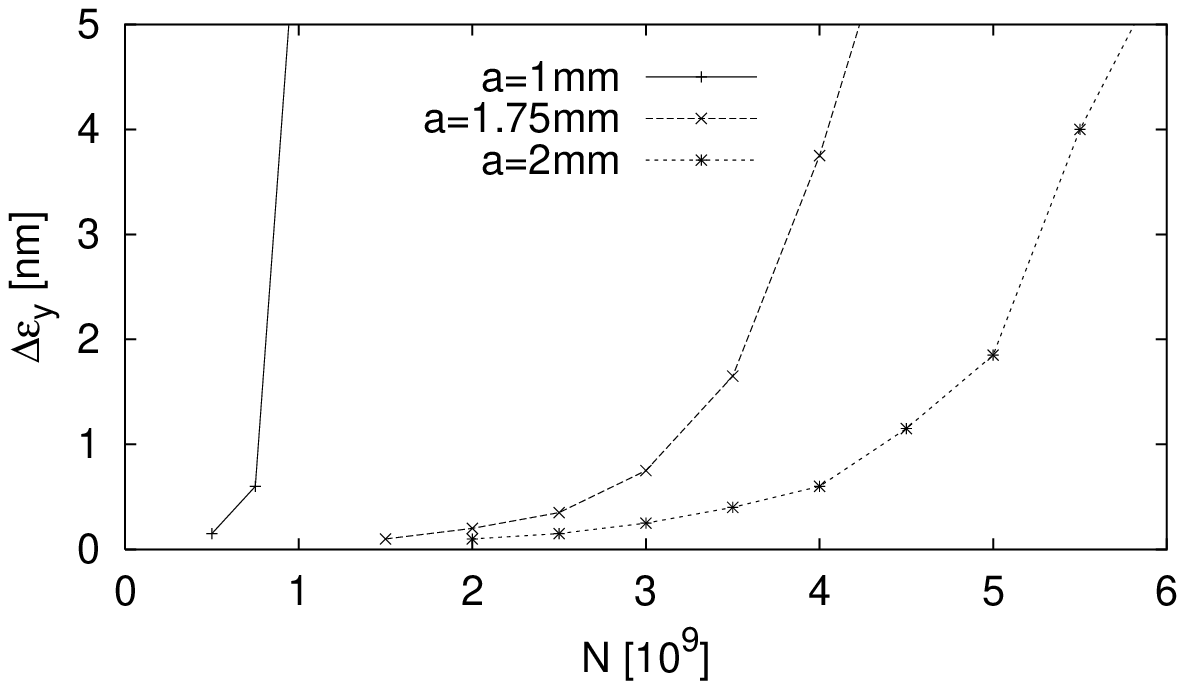,width=7.5cm,height=5.5cm} \\
\end{tabular}
\end{center}
\caption{The emittance growth $\epsilon_{{\rm linac}}$ in the main
linac as a function of the bunch charge $N$ for different structures}
\label{f:main}
\end{figure}

The dependence of the emittance growth on the bunch charge and structure can
be seen in~Fig.~\ref{f:main}\footnote{The beam that enters the linac has an
energy spread that leads to some emittance growth during acceleration; this
effect is neglected in the figure.}.
The structure with an iris radius $a$~=~2~mm
corresponds to the reference design of the accelerating structure.
As can be seen, the structures with larger values of
$a$ (the radius of the iris) allow larger bunch charges. However, it is
more difficult to achieve the required gradient in them.

\subsubsection{Beam delivery system emittance growth}
In the final focus system (FFS)
the beam is strongly focused, and consequently the system
has a tendency to be very chromatic. Since the beam has an energy spread,
one needs to reduce the chromaticity by a delicate system of cancelling
magnets; but some residual effect remains. Another problem arises from
the emission of synchrotron radiation in the magnets. While the resulting
stochastic energy change of the particles is smaller than the energy
difference between incoming particles, it can destroy the delicate
cancellation between different magnets. Because of these two effects there is
a lower limit to the achievable $\beta_y$, and $\epsilon_{y,{\rm BDS}}$ 
is not zero even for an error-free lattice. The current design
of the FFS achieves an effective vertical beam size 
of $\sigma_y \approx$~0.7~nm~\cite{c:ffs}, whereas the rms beam size
is much larger. If the two above-mentioned problems were not present,
the size would be $\sigma_y$~=~0.5~nm. The effect can be understood as
a doubling of the vertical emittance that enters the BDS,
corresponding to an emittance growth of 
$\epsilon_{y,{\rm BDS}}$~=~10~nm, though the actual dependence is more
complicated. This performance is better than the original target of
1~nm. It compensates approximately the fact that the effective
horizontal beam size is larger than the target. 

\subsubsection{Dynamic imperfections}
Dynamic effects finally limit $\sigma_{y,{\rm eff}}$ in two
ways. First, they make it more difficult to achieve a small emittance
in all the different subsystems. Second, 
they let the beams miss each other at
the IP; this does not change the size of a single bunch but the phase
space occupied by a number of consecutive bunch trains, as summarized in 
Eq.~(\ref{e:lumi}) by $\epsilon_{y,{\rm jitter}}$. A number of  
effects can lead to transverse jitter. Potentially important sources
are motion of the ground, vibration of quadrupoles in the beam line
due to cooling water, vibrations 
of the accelerating structures, and a number of other effects. Also very
important are variations of the RF amplitude and phase, which could for
example be induced by variation of the intensity or phase of the drive-beam
or by its transverse motion. The size of most of these effects remains to be
determined. However, some encouraging results have been achieved.
Preliminary tests show that feet stabilized with commercial 
supports using rubber pads and piezo-electric movers give results
that meet the  
requirements for the linac quadrupoles even in a noisy environment. 
A (non-optimized) quadrupole with flowing cooling water has been stabilized to
the required level for the main linac~\cite{c:quad}. Further reduction of
the vibration amplitudes by a factor 2--5 is being investigated for
the last final-focus doublets, which contribute predominantly to the
luminosity reduction. This clearly requires active stabilization,
optimized by the use of permanent magnets in order to  
reduce their weight.

Simulations of the luminosity in the presence of ground motion
as measured at different existing sites showed good performance for
motion levels measured at CERN and SLAC~\cite{c:ground}.

The effect of the jitter will be mitigated by the use of feedback in all
subsystems of the machine. Especially important will be the beam-position
feedback at the IP, which minimizes the offsets between the two beams. Such a
feedback, acting from train to train, has been studied at
500~GeV~\cite{c:ground}. A luminosity reduction could be almost completely
avoided if the
quadrupoles of the last doublet were stabilized and a quiet site (e.g. the
LEP tunnel) chosen. In a noisy site, significant luminosity loss can be
experienced. The possibility of an intra-pulse feedback, which has to respond
extremely fast since the pulse duration is short, has also been
investigated~\cite{c:fastfeed} and a substantial reduction of the luminosity
loss has been reached.

Further studies to determine the size of different dynamic effects,
their impact on the luminosity, and the possible counter measures
remain to be done. This was identified as  
an important R\&D issue for all future linear colliders~\cite{c:trc}.

\subsection{Efficiency and Luminosity}
\label{r:sx}
\begin{figure}[t] 
\begin{center}
\epsfbox{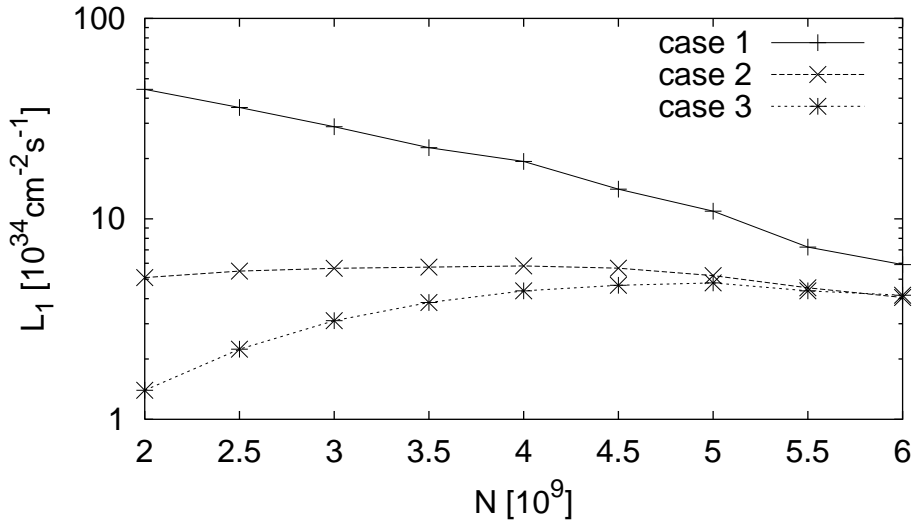}
\end{center}
\caption{The luminosity ${\cal L}_1$ as a function of the bunch
charge, under different assumptions, at $\sqrt{s}$~=3~TeV. In case~1 it
is assumed that the only source 
of vertical emittance is the main linac and that the horizontal beam size
can be optimized for maximum luminosity. In case~2, the other sources of
vertical emittance growth are taken into account. In case~3, the lower
limit of the horizontal beam size as given by the current reference design
of damping ring and BDS is taken into account.}
\label{f:simple}
\end{figure}

The efficiency of a future linear collider is affected by technical
limitations. The transformation of wall-plug power into RF power
in the klystrons is affected by  losses. Such inefficiencies can be improved
relatively independently of the main parameter choices. However, most RF
devices have reached a high level of maturity and large improvements
are not to be expected. 

Some efficiency limitations are, however, more complex, and arise from the
interplay of different collider parameters. An example is that,
for an otherwise unchanged design, a higher beam current will lead to higher
efficiency. A higher current can be achieved by increasing the bunch charge
$N$, which requires a longer bunch and leads to an increase of the wakefield
effects in the main linac and consequently of the vertical emittance 
$\epsilon_{y,{\rm linac}}$. In addition, the beamstrahlung will be more severe.
Taking into account the different limitations, one can thus determine
an optimum choice of $N$ giving the best compromise between efficiency and
vertical beam size and leading to maximum luminosity.
Figure~\ref{f:simple} 
illustrates this for the reference design. If
the only source of emittance growth were the linac, small bunch
charges would be favoured because the loss in efficiency is more than
compensated by the reduction of the emittance growth. Taking into
account the other sources of emittance growth, however, one  
finds an almost flat dependence with an optimum around
$N$~=~4~$\times$~10$^9$, the current reference bunch charge. For smaller bunch
charges, the loss in efficiency 
is slightly larger than the luminosity increase owing to the shorter bunch
and smaller linac emittance growth. At larger bunch charges the larger
emittance growth starts to dominate over the increased efficiency.
If one assumes, however, that a lower limit exists for the horizontal
beam size at the value of the current reference design, the luminosity
reduction at lower bunch charges is much stronger. In this case
the reference value of $N$~=~4~$\times$~10$^9$ is very close to the~optimum.

Another possibility to increase the beam current is to reduce the
bunch-to-bunch distance as much as possible. This requires that the wakefields
be sufficiently damped between the bunches to avoid a significant growth of
$\epsilon_{y,{\rm LET}}$. Considerable effort has gone and is still
going into the development of optimum damping
techniques~\cite{c:damp}, but large improvements are not to be expected.

\begin{figure}[t]
\begin{center}
\epsfbox{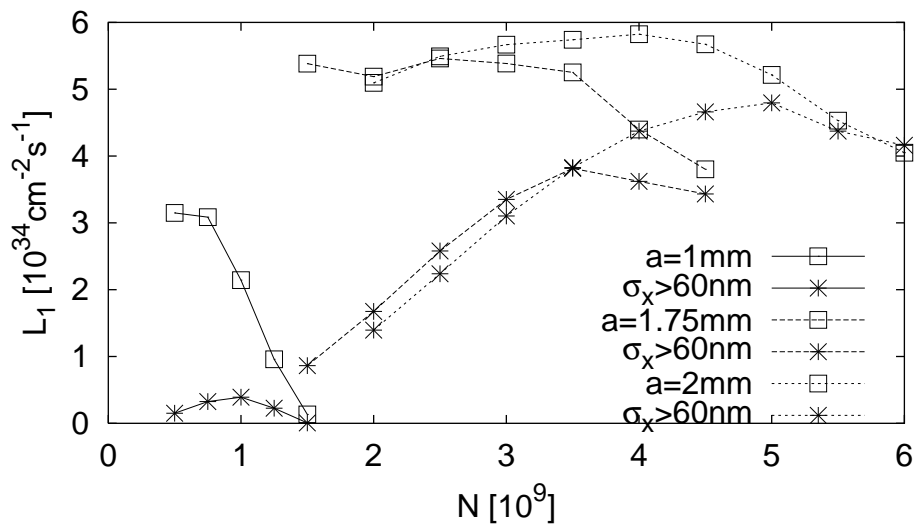}
\end{center}
\caption{The peak luminosity as a function of the bunch charge, for three
different structures}
\label{f:discuss}
\end{figure}

It is also possible to modify the design of the accelerating structures
in order to obtain a higher efficiency for a constant beam current, but again
this will increase the vertical emittance. An example of the luminosity with
different structures is shown in Fig.~\ref{f:discuss}, where the
variable $a$ is the radius of the structure iris. For small $a$ the
wakefields are larger, 
but it is easier in these structures to achieve a high gradient.
As can be seen in the plot, the luminosity ${\cal L}_1$ depends on the bunch
charge and on the assumption about the achievable $\sigma_x$.
If the $\sigma_x$ which is optimum for beamstrahlung can be used,
the structures differ by only a factor 2.
A delicate trade-off between several parameters is thus necessary to
determine an optimum machine parameter set.

\def\nix{
\subsection{Discussion}

\begin{itemize}
\item
Reduction of the vertical emittance produced in the damping ring.
\item
Reduction of the horizontal beam size at the interaction point or
increase of the bunch charge. The latter could be achieved by reducing
the RF frequency.
\end{itemize}
}

\section{The CLIC Energy Range}

The CLIC design aims at reaching multi-TeV centre-of-mass energies with high
luminosity. Studies of low emittance transfer and beam characteristics
for a luminosity of the order of 10$^{35}$~cm$^{-2}$s$^{-1}$
indicate that beam dilution and the sensitivity to vibrations in the
last doublet may limit the maximum $\sqrt{s}$ energy to $\sim$~5~TeV.  
This holds even when the wakefield effects of the 30~GHz structures
are controlled by a judicious choice of bunch length, charge, and
focusing strength. This limitation, for a 10$^{35}$ luminosity, 
comes mainly from the fact that the needed vertical geometric beam size at
the IP becomes critically small with respect 
to the estimated effects of
jitter and vibrations in the final-focus system. Therefore the  CLIC
design has been optimized for 3~TeV collision energy with a possible
upgrade path to 5~TeV, at constant luminosity. 
 
The injection system of the main beam remains essentially the same at
0.5~TeV and 3~TeV. However, while the klystrons of the injector linacs
have to provide the same peak power, the average power delivered is
lower at 3~TeV than at 0.5~TeV, since the repetition rate is two times
smaller. Considering the drive-beam generation, the characteristics of
each bunch train are the same, i.e. an energy of 2~GeV, an average
current of 147~A and a length of 130~ns, but the number of bunch
trains depends on $\sqrt{s}$. This means that the duration of the
initial long pulse accelerated by each drive-beam linac operating at
937~MHz differs and is proportional to the energy. The direct
consequence is an increase of the pulse length of the drive-beam
klystrons by a factor of 6. However, since the repetition rate is
correspondingly reduced from 200 to 100~Hz, the average power to be
provided by these klystrons increases only by a factor of 3 when
going from 0.5~TeV to 3~TeV and the same klystrons can be used at both
energies. The power consumption for accelerating the drive-beams
increases from $\sim$~106 MW at 0.5~TeV to $\sim$~319~MW~at~3~TeV.

The combiner rings remain unchanged while the repetition rate of the
RF deflectors is halved and their pulse is 6 times longer. Each
decelerator unit also remains the same, so that all the technical
problems related to the drive-beam control, RF power extraction and
transfer to the accelerating structures are identical, irrespective of  
the collision energy.
    
At 3 TeV, each linac contains 22 RF power source units, that is 22000
accelerating structures representing an active length of 11~km. With a
global cavity-filling factor of $\sim$~78\%, the total length of each
linac is $\sim$~14~km.  To keep the filling factor about constant
along the linac, the target values of the FODO focal length and
quadrupole spacing are scaled with $E^{1/2}$. For practical reasons,
however, the beam line consists of 12 sectors (5 at 500~GeV), each
with constant lattice cells and with matching insertions between
sectors. The total number of quadrupoles is 1324 per linac and their
length ranges from 0.5~m to 2.0~m from the start to the last
sector. The rms energy spread along the linac is about 0.55\% average
for BNS damping and decreases to  $\sim$~0.36\% at the linac end
(1\% full width).  

The beam delivery system has to be adjusted to the collision
energy. In particular, the design scaling and the bending angles are
different at 3~TeV and 0.5~TeV. The design has been optimized at
3~TeV, where it is most critical, and changing the energy by a large  
factor currently assumes some changes in the magnet positions, and in
the bend and quadrupole strengths. However, the 5.1~km total length of
the proposed system remains unchanged, as well as the 20~mrad crossing
angle. Calculations indicate an acceptable  
emittance growth in the presence of sextupole aberrations and Oide 
effects, provided that the last focusing quadrupole is properly
stabilized. The collimation efficiency remains to be checked through
numerical simulations. The optics, the collimator survival and the
control of wakefield effects are still being studied and improved. In
any case, the static luminosity optimization procedure needs further  
studies together with the time-dependent effects and their control via
feedbacks including a luminosity-related feedback.

The CLIC design allows one to increase the collider energy with the
number of two-beam units installed in each linac and the length of the
pulse required in each drive-beam  accelerator. As an illustration,
these correspond to 4 units with 17~$\mu$s, 22 units with 100 $\mu$s
and 37 units with 154 $\mu$s, for $\sqrt{s}=$ 500~GeV, 3~TeV and 5~TeV,  
respectively. These numbers correspond to a two-linac length of 5~km,
28~km and 46.5~km with total collider lengths of about 10~km, 33~km
and 51.5~km. A length of up to 40~km total is available at a site near
CERN, extending parallel to the Jura mountain range, in a molasse
comparable to that housing the  SPS and LHC tunnels. To get beyond
this length would require diging the tunnel in the limestone on one
end or crossing a 2~km-wide underground fault on the other end.  
In spite of the anticipated technical difficulties, this second
solution appears preferable as the additional cost would be limited
and this would open the possibility of extending the tunnel to a total
length of 52~km. The limitation is then set by the presence of a major
fault. With this extension, the tunnel length would be sufficient for
a collider capable of achieving 5~TeV with the proposed parameters. 

\section{Polarization Issues}

The linear collider physics potential is greatly enhanced
if the beams are polarized. The requirements
for CLIC are relaxed with respect to the NLC-II, JLC, or TESLA 
parameters, since at CLIC both the charge per bunch, and 
the average beam current are lower than in the lower-frequency,
lower-energy machines. Table~\ref{comp1} compares
the relevant CLIC parameters with those of the SLC and 
with a 1996 parameter set for NLC-II~\cite{nlczdr}.
%
\begin{table}[htbp]  
\caption{Comparison of electron source parameters achieved at the
SLC with those required for NLC-II~\cite{nlczdr} and CLIC}
\label{comp1}

\renewcommand{\arraystretch}{1.3} 
\begin{center}

\begin{tabular}{lccc}\hline \hline \\[-3mm]
\textbf{ Parameter } & 
$\hspace*{3mm}$ \textbf{SLC} $\hspace*{3mm}$ & 
$\hspace*{3mm}$ \textbf{NLC-II} $\hspace*{3mm}$ & 
$\hspace*{3mm}$ \textbf{CLIC} $\hspace*{3mm}$ \\[4mm]  

\hline \\[-3mm]
Bunch ch. (10$^{10}$~$e^-$) $\hspace*{15mm}$ & 7 & 2.8 &
0.4 \\
Total ch. (10$^{10}$~$e^-$) & 14 & 252 & 62 \\
Av. pulse current (A) & 0.4 & 3.2 & 1.0 \\
Pulse length (ns)  & 62 & 126 & 103 \\
Beam polarization  & $\sim$~80\% & $\sim$~80\% & $\sim$~80\% 
\\[3mm] 
\hline \hline
\end{tabular}
\end{center}
\end{table}


A polarized electron beam with 
about 80\% polarization can be produced by an SLC-type 
photoinjector~\cite{tang}. 
Though producing an intense polarized positron beam
is more difficult, Compton scattering off a
high-power laser beam may provide a source of positrons with
60\%--80\% polarization~\cite{omori,omori2}. 
Experimental R\&D and 
prototyping of a polarized positron source based on
Compton scattering is ongoing at KEK for the JLC project~\cite{hirose}.
This scheme is taking advantage of rapid advancements in laser~technology. 

\begin{figure}[!h] 
\begin{center}
\rotatebox{0}{\scalebox{1.2}{\includegraphics*{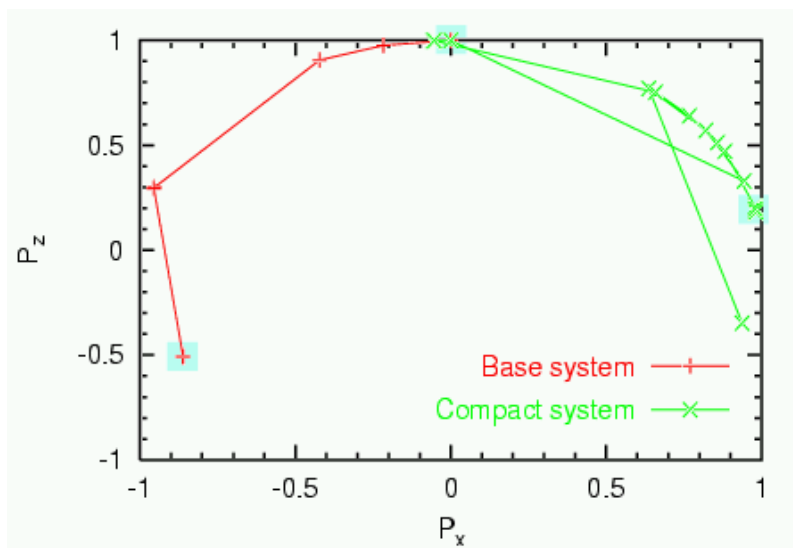}}}
\end{center}

\vspace*{-6mm}

\caption{Rotation of the polarization vector in the $x$--$z$ plane in
the nominal (`short') and an alternative (`base') CLIC final-focus
system at 3~TeV centre-of-mass energy.  
The initial ($P_x$~=~0, $P_z$~=~1, i.e. purely longitudinal) 
and final polarization values are indicated by underlaid
boxes.} 
\label{pol-xs}
\end{figure}

The geometry of the CLIC transfer lines and
the damping-ring energy are chosen so that the beam 
polarization is preserved, as was the case at the SLC. No  
significant depolarization is expected to
occur on the way to the collision point. 
We have demonstrated this
explicitly by spin tracking through two versions of
the  CLIC beam delivery system at the 3~TeV 
centre-of-mass energy~\cite{snowmasspol}.
However, the bending magnets of the beam delivery 
system rotate the polarization vector by about $\pi/2$ 
(see Fig.~\ref{pol-xs}) and the rotation angle changes with the 
beam energy. Complete control over the IP
spin orientation needs to be provided by an
orthogonal set of spin rotators, which can be installed
between the damping ring and the main linac.

During the beam--beam collision itself, because of  beamstrahlung and
the strong fields at 3~TeV, about 7\% of effective polarization will
be lost. About half of this loss is due to spin precession, the
other half to spin-flip radiation. The latter
is accompanied by a large energy change and thus does
not affect the luminosity-weighted polarization
at the nominal energy. 

In view of the fairly large depolarization in collision, 
the polarization should be measured both for the incoming and
for the spent beam. Therefore, we anticipate the installation
of two Compton polarimeters on either side 
of the detector. A measurement resolution of 
0.5\% for the incoming beam would be comparable to that
achieved at the SLC and expected for the other linear-collider
designs. Reaching a similar resolution for the highly
disrupted spent beam appears very challenging.

More details on polarization issues for CLIC at 3 TeV 
centre-of-mass energy can be found in~Ref.~\cite{snowmasspol}. 

\section{$\gamma \gamma$ Collisions at CLIC}

Gamma collider options have been considered in all linear collider studies.
The energy region of 0.5--1~TeV is particularly well suited for $\gamma\gamma$
collisions from a technical point of view: the wavelength of the
laser should be about 1~$\mu$m, i.e. in the region of the most
powerful solid-state lasers and collision effects do not restrict the
$\gamma\gamma$ luminosity~\cite{TEL2000,TESLATDR}.

In the multi-TeV energy region
the situation is more difficult: collision effects with coherent
$e^+ e^-$ pair creation in $\gamma\gamma$ collisions will be hard to
avoid and may restrict the luminosity. The optimum laser wavelength
increases proportionally with the energy. In addition, the required
laser flash energy increases because of non-linear Compton scattering. 
Options for a 3-TeV photon collider based on 4--6~$\mu$m wavelength
have been studied recently~\cite{TelnovBurkhardtGG2002}. We summarize here
the main results and give a tentative list of parameters and
luminosity spectra.

Parameters of a possible photon collider at CLIC with $2E_0$~=~3000~GeV
are listed in Table~\ref{tclic}. 
%
\begin{table}[!h] 
\caption{Possible parameters of the photon collider at
CLIC. Parameters of electron beams are the same as for $e^+ e^-$
collisions.}
\label{tclic}

\renewcommand{\arraystretch}{1.2} 
\begin{center}

\begin{tabular}{lc}\hline \hline \\[-3mm]
\textbf{\boldmath{ $2E_0$}} &
\textbf{3000 GeV} \\[4mm]  
\hline \\[-3mm]
$\lambda_L$\, [\MKM]/$x $ & 4.4 / 6.5 \\
$t_{L}$ $[\lambda_{{\rm scat}}]$ & 1 \\
$N$ / 10$^{10}$ & 0.4  \\  
$\sigma_{z}$\, [mm] & 0.03 \\  
$f_{rep}\times n_b$\, [kHz]& 15.4  \\
$\gamma \epsilon_{x/y}$/10$^{-6}$ [m$\cdot$rad] & 0.68 / 0.02  \\
$\beta_{x/y}$\, [mm] at IP& 8 / 0.15 \\
$\sigma_{x/y}$\, [nm] & 43 / 1  \\  
$b$\, [mm] & 3 \\
$L_{ee}$(geom)\, [10$^{34}$] \CMS\ & 4.5 \\  
$L_{\gamma\gamma}$ ($z>$~0.8~$z_{m, \gamma\gamma }$)\, [10$^{34}$] 
$\hspace*{25mm}$ & 0.45  \\
$L_{\gamma e}$ ($z>$~0.8~$z_{m,\gamma e}$)\, [10$^{34}$] & 0.9 \\
$L_{ee}$ ($z>$~0.65)\, [10$^{34}$]  & 0.6
\\[3mm] 
\hline \hline
\end{tabular}
\end{center}
\end{table}

\begin{figure}[t] 
  \center\includegraphics[scale=.9]{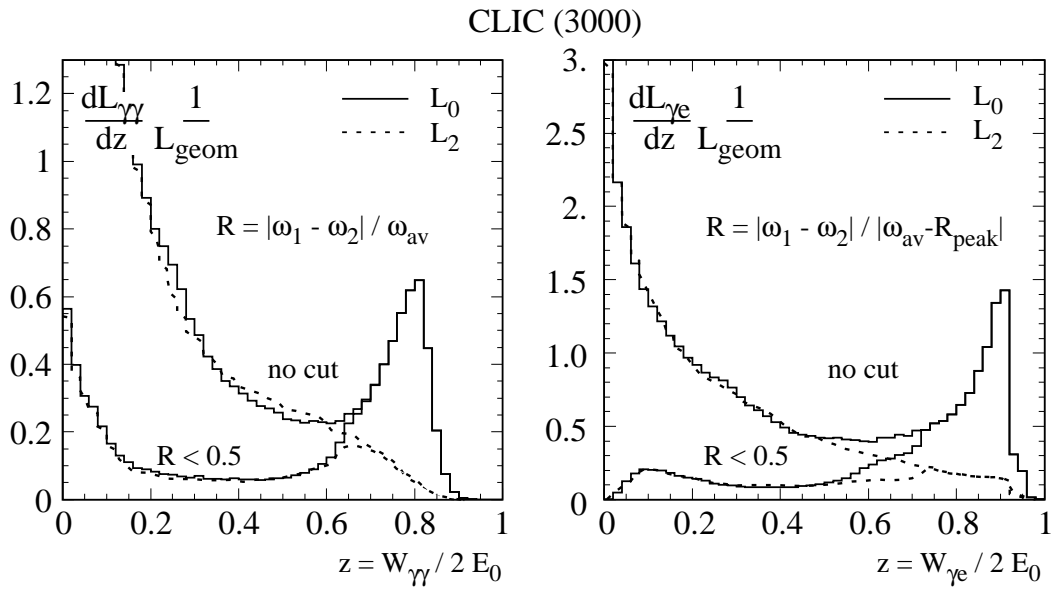}
\caption{$\gamma\gamma$ (left) and $\gamma e$ (right) luminosity
spectra at CLIC(3000). $L_0, L_2$ are the luminosities with the total
  helicity of two colliding photons in the case of $\gamma\gamma$
  collisions  (or the total helicity of the colliding photon and 
  electron in the case of $\gamma e$ collisions) equal to 0 and 2,
  respectively.} 
\label{e3022_telnov_fig2}
\end{figure}

%
\noindent
The electron beam parameters shown in
the table are the same as for $e^+ e^-$ collisions. As discussed
in~Ref.~\cite{TelnovBurkhardtGG2002}, this is somewhat conservative, and
there may be ways of decreasing electron beam sizes in collisions and
potentially increase the $\gamma\gamma$ luminosity by a factor of
about 3. The laser parameter 
$x$~=~6.5 approximately corresponds to the threshold for $e^+ e^-$ creation
for the non-linear parameter $\xi^2 \approx$~0.3. The corresponding
wavelength is 4.4 \MKM.  It is not clear at present which kind of
laser would be best suited for a photon collider at this wavelength.
Candidates are a gas CO laser, a free-electron laser, some solid-state 
laser or a parametric solid-state laser (the `short' wavelength
laser pulse is split in a non-linear laser medium into two beams with
longer wavelength). The luminosity spectra
obtained by a full simulation~\cite{TEL95} based on the parameters quoted here
are presented in~Fig.~\ref{e3022_telnov_fig2}. 

\section{$e^-e^-$ Collisions at CLIC}

While $e^-e^-$ collisions are considered an interesting option, not much
effort was made to study it in detail. Most subsystems used to provide a 
positron beam can also be used for an electron
beam, with minor modifications. The subsystems, which need larger changes,
e.g. the injector that produces the beam, are usually simpler for electrons.
The main concern is thus the beam--beam collision. In electron--positron
collisions the two beams focus each other while they will deflect each other in
electron--electron collisions. A preliminary study of the collision has been
performed~\cite{c:ee}.

\begin{figure}[t] 
  \begin{center}
\epsfig{file=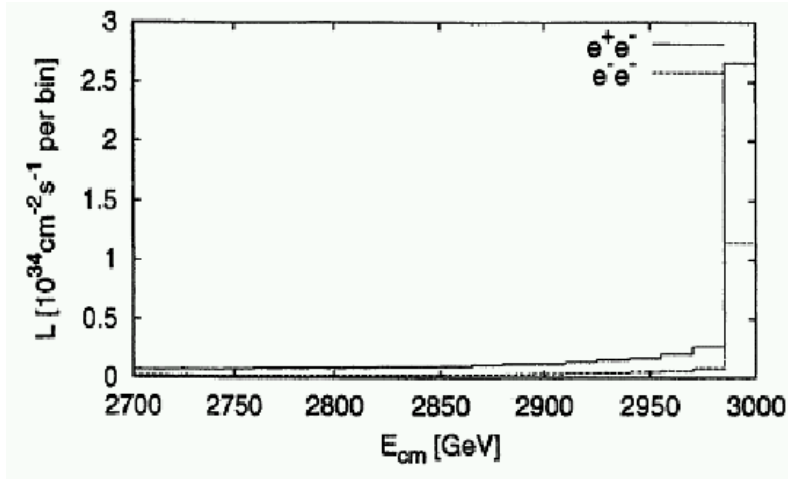,width=10.5cm}
    \caption{Absolute luminosity spectrum for the $e^+e^-$ and
      $e^-e^-$ cases. The bins have a width of 0.5\% of the
      center-of-mass energy.}
    \label{fig:1}
  \end{center}
\end{figure}

\begin{figure}[htbp!] 
  \begin{center}
\epsfig{file=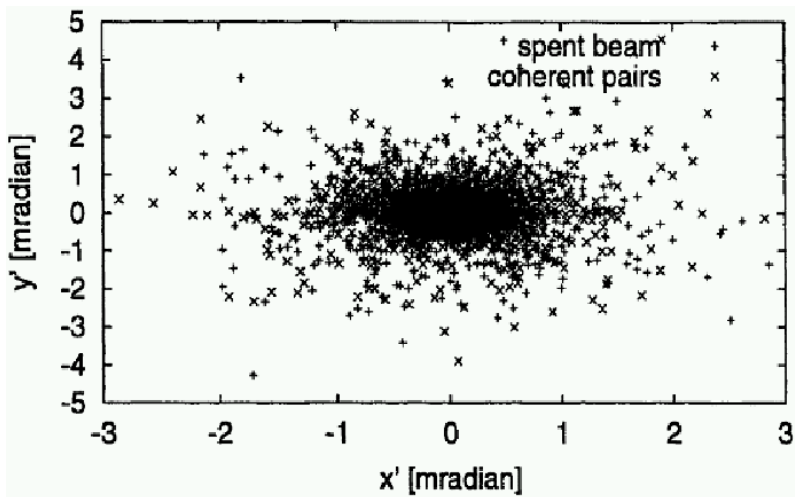,width=10.5cm}
    \caption{Angular distribution of the spent beam and the coherent
      pairs produced in the 3{TeV} collision}
    \label{fig:2}
  \end{center}
\end{figure}

The simulations show that the total luminosity is
reduced by a factor of roughly 4, but that the relative quality of the
luminosity spectrum is better in the $e^-e^-$ collisions. For the part of
the luminosity spectrum close to the nominal centre-of-mass energy, the
reduction in $e^-e^-$ mode is thus only a factor of about 2.5 compared
with $e^+e^-$. More remarkably, the background spectra of the $e^-e^-$
mode  have a minuscule lower-energy tail, as is clearly shown 
in~Fig.~\ref{fig:1}. 
Figure~\ref{fig:2} shows the spent-beam and coherent
pair production angular distributions, of major importance mainly for
detector configuration studies.
The number of beamstrahlung photons and coherent pairs is
slightly reduced. The incoherent pair and hadronic background are reduced by
a factor of 3 to~4. The angular distribution of the spent beam seems not to be
worse than the one from $e^+e^-$ collisions. A detector designed for the
latter should be perfectly capable of handling the $e^-e^-$ collisions as well.
%

\section{The CLIC Test Facility and Future R\&D}

The goals of the  CLIC scheme are ambitious, and require further R\&D
to demonstrate that they are indeed technically feasible. The basic principle 
of two-beam acceleration with 30 GHz accelerating structures has
already been demonstrated in CLIC Test Facilities 1 and 2 (CTF1 and
CTF2). The technical status of CLIC was recently evaluated by the 
International Linear Collider Technical Review Committee (ILC-TRC), which 
was nominated by the International Committee on Future Accelerators
(ICFA) in February 2001 to assess the current technical status of the
four electron--positron linear-collider designs in the various regions
of the world. The report~\cite{Loew} identified two groups of key issues 
for CLIC: (i)~those that were related specifically to CLIC
technology, and (ii)~those which were common to all linear collider
studies (such as the damping rings, the transport of low-emittance 
beams, the relative phase jitter of the beams, etc.). The CLIC study
is for the moment focusing its activities on the following five
CLIC-technology-related issues, which were given either an R1 
(R\&D needed for feasibility demonstration or an R2 (R\&D needed to
finalize design choices) rating by the ILC-TRC.

\begin{enumerate}
\item
Test of damped accelerating structures at design gradient and pulse 
length (R1)
\item
Validation of drive-beam generation scheme with a fully-loaded 
linac (R1)
\item
Design and test of damped ON/OFF power extraction structure (R1)
\item
Validation of beam stability and losses in the drive-beam 
decelerator, and design of \\
machine protection system (R2)
\item
Test of relevant two-beam linac subunit (R2).
\end{enumerate}

\begin{figure}[t] 
\begin{center}
\epsfig{file=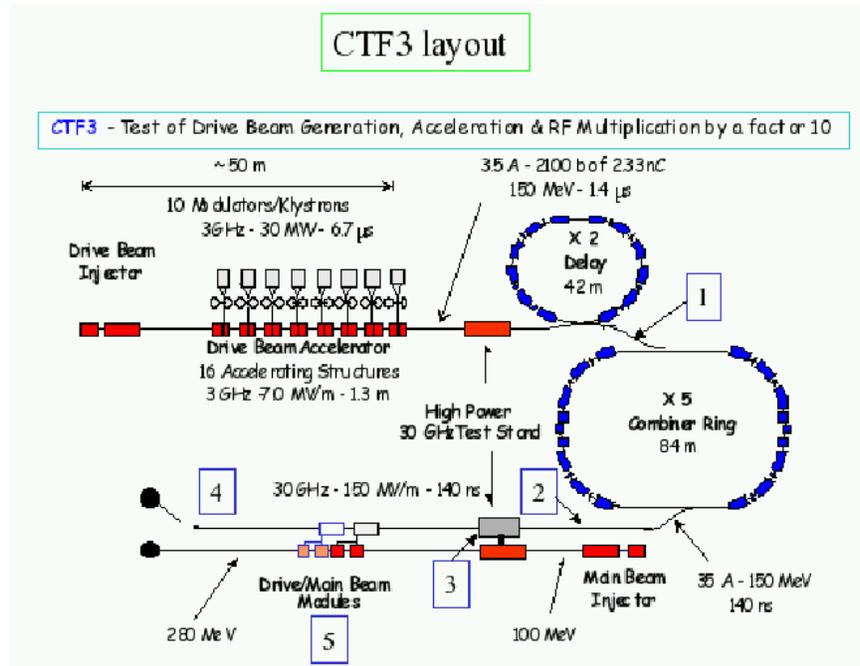,width=11.5cm}
\caption{Schematic layout of CTF3}
\label{fig:CTF3-1}
\end{center}
\end{figure}

Answers to these key R1 and R2 issues will be provided by the new CLIC
Test Facility (CTF3), which is being built to demonstrate the
technical feasibility of the key concepts of the novel CLIC RF power 
generation scheme, albeit on a smaller scale and re-using existing 
equipment, buildings and technical infrastructure that have become
available following the closure of LEP. A schematic layout of CTF3 is 
given in~Fig.~\ref{fig:CTF3-1}. CTF3 is being constructed in
collaboration  with INFN, LAL, Northwestern University (Illinois),
RAL, SLAC, and Uppsala University.  

The principal aim is to demonstrate the efficient CLIC-type production of 
short-pulse RF power at 30 GHz from 3-GHz long-pulse RF power. This involves
manipulations on intense electron beams in combiner rings using
transverse RF deflectors as required in the CLIC scheme~\cite{ctf3}.

The following are some significant details of the scheme. A 140-ns-long
 train of high-intensity electron bunches with a bunch spacing of
2~cm is created from a 1.4-$\mu$s continuous 
train of bunches spaced by 20~cm. The 2-cm spacing is required for an 
efficient generation of 30-GHz RF power. This is done by 
interleaving trains of bunches and is done in two stages. The first 
combination takes place in the delay loop, where every other 140-ns
slice of the 1.4-$\mu$s continuous train is sent round the 42-m (140-ns)
circumference of the loop before being interleaved with the 
following 140-ns slice. This results in a reduction in the bunch
spacing of a factor of 2 and an increase in the train intensity 
by a factor of 2. The second stage of combination --- this time by a 
factor 5 --- takes place in the combiner ring. After passing through
the delay loop, the 1.4-$\mu$s train from the linac is made up 
of five 140-ns pulses with bunches spaced by 10~cm, and five interspaced 
140-ns gaps. The combiner ring combines these five pulses into a
single 140-ns pulse using a novel system of beam interleaving. 

\begin{figure}
\begin{center}
\epsfig{file=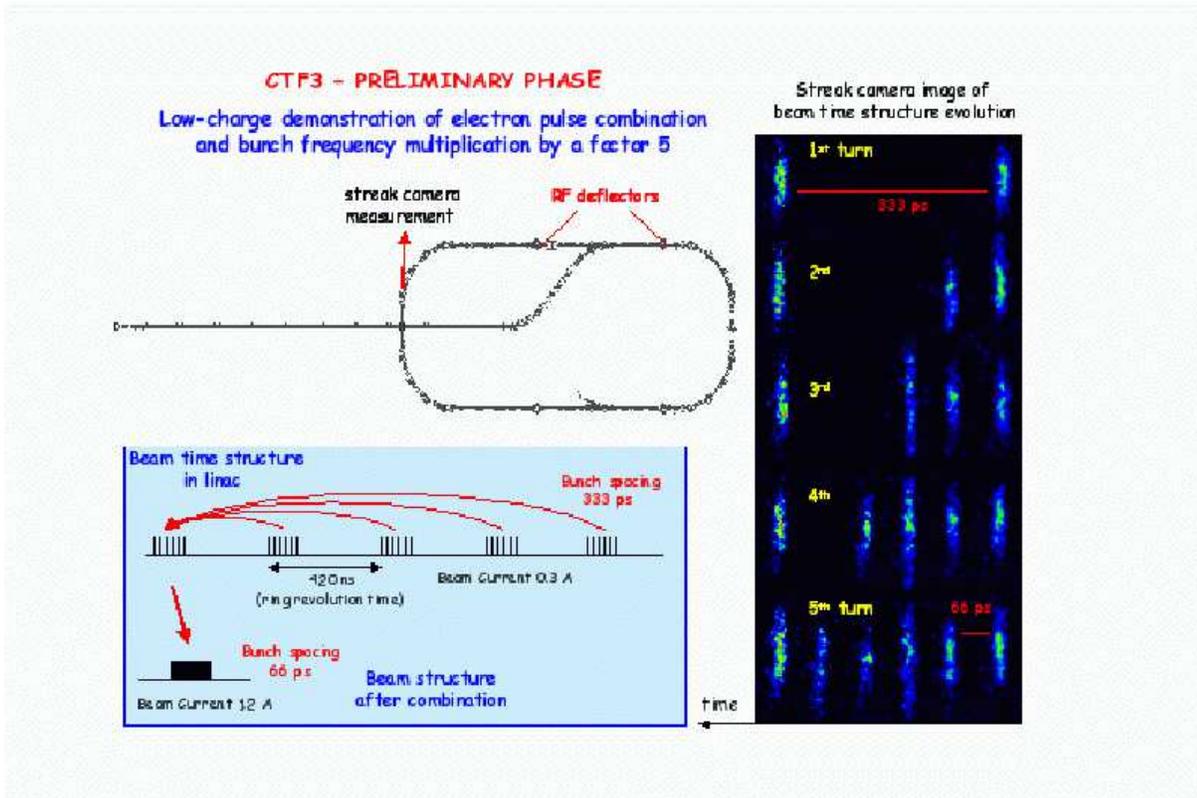,width=16cm}
\caption{Low-charge demonstration of electron pulse combination 
and bunch frequency multiplication by a factor 5 in CTF3}
\label{fig:CTF3-2}
\end{center}
\end{figure}

Progress CTF3 to date is as follows. Preliminary tests of bunch
interleaving with five trains using a new gun and a modified version of
the old LEP injector complex (LPI linac and EPA ring) at very low beam
current were successfully completed in November 2002, and the results are
summarized in Fig.~\ref{fig:CTF3-2}. This result confirms the basic
feasibility of the scheme. In December 2002 the old LPI linac was
dismantled and in June 2003 the new CTF3 injector was installed. A bunched
beam of the nominal current, pulse length and energy was obtained from the
injector for the first time in August 2003.

Good progress has also been made with CLIC machine studies. The 
following steps have been achieved within the 
30-GHz CLIC RF structure programme.

\begin{itemize}
\item[(i)] 
Peak accelerating gradients of almost 200 MV/m have been obtained
with short (16-ns) RF pulses with a 30-GHz molybdenum-iris
accelerating structure --- a comparative conditioning  
curve for three different materials is given in~Fig.~\ref{fig:gradient}.
\item[(ii)] 
A new fully-optimized design of the 30-GHz damped
accelerating structure has been made, with significantly lower
long-range transverse wakefields, which allow shorter bunch spacings
and hence shorter pulse lengths. 
\item[(iii)] 
A new RF design of the 30-GHz  power-generating structure has been
proposed, with the ability to turn the power ON and OFF.  
\item[(iv)] 
The CLIC  Stabilization Study Group has stabilized a prototype CLIC
quadrupole to the level of 0.5~nm using commercially available
equipment. Beam dynamics simulations of the main beams 
in the different parts of the machine have been integrated to give results 
with fully-consistent conditions.
\end{itemize}

\section{Summary}

We have summarized in this chapter the aims and status of the CLIC study,
with emphasis on features related to its physics performance. The
objective of a compact high-energy complex dictates the choice of a high
accelerating gradient, requiring a high-frequency accelerating structure
based on a two-beam approach. Accelerating structures capable of over
150~MV/m at 30~GHz have been operated successfully with short RF pulses in
CTF2.  Though optimized for a nominal centre-of-mass energy of 3~TeV
with a design luminosity of 
10$^{35}$~cm$^{-2}$s$^{-1}$, CLIC parameters for other energies between
0.5 and 5~TeV have also been proposed.

We have also discussed the principal machine characteristics that control
the achievable luminosity, including the horizontal beam size, the bunch
charge and the vertical beam size. The latter is constrained by the
emittance produced in the damping ring, and its subsequent growth in the
bunch compressor, the main linac and the beam delivery system. Further
studies of the damping ring are needed, as are further studies of the
effects of dynamical imperfections. In view of the small beam size, good
alignment and stability of the CLIC components are crucial, and seem
possible in a quiet site such as a tunnel through the molasse rock in the
neighbourhood of CERN.

The scaling of CLIC parameters with the centre-of-mass energy has been
discussed, and there are good prospects for polarized beams, $\gamma
\gamma$ and $e^- e^-$ collisions. Principal R\&D issues have been
identified by the CLIC team and the Loew Panel. The CLIC
 Test Facility 3
(CTF3) now under construction at CERN will address the most critical R1
issues, and should enable the technical feasibility of CLIC to be
established within a few years.

\mbox{}\hspace{-5cm}

\newpage
\thispagestyle{empty}
~

\newpage
\chapter{EXPERIMENTATION AT CLIC}
\label{chapter:three}
\newcommand{\qt}[1]{``{\em #1}''}

\newcommand{\defsum}{\raisebox{-0.13cm}{~\shortstack{\Large{$\Sigma$}
\\[-0.07cm] ${}_{m}$}}~}
\newcommand{\defsumg}{\raisebox{-0.13cm}{~\shortstack{\Large{$\Sigma$} 
\\[-0.07cm] ${}_{m, G}$}}~}

\newcommand{\Tb}{\tan\beta}
\newcommand{\sus}{{\tt SuSpect}}

The definition of the CLIC programme in the multi-TeV range still requires 
essential data; these will become available only after the first years of LHC 
operation and, possibly, also the results from $e^+e^-$ collisions at
lower energy.  
At present we have to envisage several possible scenarios for the
fundamental questions  
to be addressed by accelerator particle experiments after the LHC. 

It is therefore interesting to consider benchmark physics signatures
for assessing the impact of the accelerator characteristics on
experimentation and for defining the needs on the detector
response. Each physics signature may signal the manifestation of  
different physics scenarios, possibly beyond those we envisage
today. Nevertheless, the results of these studies should be generally
applicable also to those other processes,  
having similar characteristics.

While considering experimentation at a multi-TeV collider, it is also
interesting to verify to which extent extrapolations from experimental
techniques successfully developed at LEP, and subsequently extended in
the studies for a TeV-class linear collider, are still
applicable. This has important consequences on the  
requirements for the experimental conditions at the CLIC interaction region 
and for the definition of the CLIC physics potential.

Four main classes of physics signatures have been identified. These are:
{\it resonance scans}, {\it electro-weak fits}, {\it multijet final
states} and {\it missing energy and forward processes}. Their
sensitivities to the characteristics of the luminosity spectrum and the
underlying accelerator-induced backgrounds differ  
significantly. Results of detailed simulations of several physics processes 
representative of each of these classes of physics signatures are
discussed in the subsequent chapters. A physics matrix summarizing the
various processes studied in  
detail for CLIC, with their interdependence on  these classes of physics 
signatures and the aspects of the machine parameters, is given in 
Tables~\ref{tab:physprog} 
and~\ref{tab:physpar}. 

This chapter discusses the issues related to the experimental
conditions at CLIC, the conceptual design for the detector and the
software tools developed and used for simulation. 
%
\begin{table}[t] 
\caption{Physics signatures and CLIC physics programme: matrix of the 
simulated processes}
\label{tab:physprog}

\renewcommand{\arraystretch}{1.1} 
\begin{center}

\begin{tabular}{cccccc}
\hline \hline\\[-1mm]
$\hspace*{3.9mm}$ \textbf{Physics} $\hspace*{3.9mm}$ & 
$\hspace*{3.9mm}$ \textbf{Higgs} $\hspace*{3.9mm}$ & 
$\hspace*{3.9mm}$ \textbf{SUSY} $\hspace*{3.9mm}$ & 
$\hspace*{3.9mm}$ \textbf{SSB} $\hspace*{3.9mm}$ & 
$\hspace*{3.9mm}$ \textbf{New gauge} $\hspace*{3.9mm}$ & 
$\hspace*{3.9mm}$ \textbf{Extra} $\hspace*{3.9mm}$ \\ 
\textbf{signatures} & 
\textbf{sector} & & & 
\textbf{bosons} & 
\textbf{dimensions}  
\\[3mm]  \hline\\[-1mm]
Resonance scan & & $\tilde{\mu}$ &  D-BESS  & $Z'$ & KK \\ 
 & & thresholds & & & resonances \\[2mm] \hline \\[-2mm]
EW fits & & & & $\sigma_{f \bar f}, \, A^{f \bar f}_{\rm FB}$, 
&  $\sigma_{f \bar f}, \, A^{f \bar f}_{\rm FB}$, \\[4mm] \hline \\[-2mm]
Multijets     & $H^+H^-$ &   &   &           &  \\
              & $H^0A^0$ & &   &           &  \\        
              & $H^0 H^0 \nu \bar \nu$ &  &    &           &  
\\[2mm] \hline \\[-2mm]        
$E_{\rm miss}$, Fwd & $H^0e^+e^-$ & $\tilde \ell$ & $WW$ &  
               &  \\
               &       & $\chi^0_2$ & scattering &   &
\\[3mm] 
\hline \hline
\end{tabular}
\end{center}
\end{table}


\begin{table}[!h] 
\caption{Physics signatures and CLIC accelerator parameters}
\label{tab:physpar}

\renewcommand{\arraystretch}{1.1} 
\begin{center}

\begin{tabular}{cccccc}
\hline \hline\\[-2mm]
$\hspace*{3mm}$ \textbf{Physics} $\hspace*{3mm}$ & 
$\hspace*{3mm}$ \textbf{Beam-} $\hspace*{3mm}$ & 
$\hspace*{3mm}$ \textbf{Beam} $\hspace*{3mm}$ & 
$\hspace*{3mm}$ \boldmath{$e^+$} $\hspace*{3mm}$ & 
$\hspace*{3mm}$ \textbf{Pairs} $\hspace*{3mm}$ & 
$\hspace*{3mm}$ \boldmath{$\gamma \gamma$}~\textbf{bkgd} $\hspace*{3mm}$ \\ 
\textbf{signatures} & 
\textbf{strahlung} & 
\boldmath{$E$}~\textbf{spread} & 
\textbf{polarization} &  &  
\\[3mm]  \hline\\[-2mm]
Resonance scan & Stat. shape syst. &  Shape syst. & 
Couplings  & $\Gamma_{bb}, _{cc}, _{tt}$ &
$\Gamma_{bb}, _{cc}, _{tt}$ \\[3mm] \hline \\[-1mm]
EW fits        & Unfold boost & & Polarization & $b \bar b$, $c \bar c$ & 
$\cos \theta_{\rm min}$\\
               & & & measurement & tags & bkgd flavour      
\\[2mm] \hline \\[-2mm]
Multijets     & 5-C fit &  & & Tags for     & Fake jets  \\               
                  &   & &  & jet pairing  &  \\[2mm] \hline \\[-2mm]
$E_{\rm miss}$, Fwd & $\theta_{\rm miss}$ & & & Fwd tracking & 
$E_{\rm hem},\, E_T$  
\\[3mm] 
\hline \hline
\end{tabular}
\end{center}
\end{table}


\section{CLIC Luminosity}
\label{sec:3-luminosity}

In order to obtain the required high luminosity the beams need to have very
small transverse dimensions at the collision point of a linear collider.
This leads to strong beam--beam effects and subsequently to a smearing of the
luminosity spectrum and an increase of the background. The main beam parameters
and background numbers are summarized in Table~\ref{t:parameter}.
\begin{table}[!h]
\caption{The parameters of CLIC
at $E_{c.m.}$~=~500~GeV and $E_{c.m.}$~=~3~TeV.
For the latter energy two parameter sets are given: the first
one is used to evaluate the background. The second set has
been used for the ILC-TRC; it takes into account some recent findings of the
structure research and beam-delivery system. The first set is used for
background evaluation, as the second set is going to evolve soon, and
because it yields higher background levels, so that it presents a more
conservative view of the conditions. It should be noted that the beam
sizes $\sigma_x$ and $\sigma_y$ are determined by fitting the results
of full tracking with normal distributions. For background and luminosity the
full tracking is used, including non-linear effects.
$E_{c.m.}$: centre-of-mass energy,
{${\cal L}$}: {actual luminosity},
{${\cal L}_{0.99}$}: {luminosity with $E_{c.m.}>$~0.99~$E_{c.m.,0}$},
{$f_{\rm rep}$}: {repetition frequency},
{$N_{\rm b}$}: {number of bunches per train},
{$\Delta_{\rm b}$}: {distance between bunches},
{$N$}: {number of particles per bunch},
{$\sigma$}: {bunch dimensions at IP},
{$\epsilon$}: {normalized emittances},
{$\delta$}: {average energy loss},
{$n_\gamma$}: {number of photons per beam particle},
{$N_\perp$}: {number of particles from incoherent pair production,
produced with $p_\perp>$~20~MeV and $\theta>$~0.15},
{$N_{\rm hadr}$}: {number of hadronic events},
{$N_{\rm MJ}$}: {number of minijet pairs at $p_\perp>$~3.2~GeV/$c$;
for $E_{c.m.}$~=~3~TeV this has a significant theoretical uncertainty}.
}
\label{t:parameter}

\vspace*{5mm}

\renewcommand{\arraystretch}{1.35} 
\begin{center}

\begin{tabular}{clcccc}\hline \hline \\[-3mm]
 & $E_{c.m.}$ (TeV) &  & 0.5 & 3 & 3\\
 & ${\cal L}$ (10$^{34}$cm$^{-2}$s$^{-1}$) &  & 2.1&10.0&8.0 \\
 & ${\cal L}_{0.99}$ (10$^{34}$cm$^{-2}$s$^{-1}$) $\hspace*{12mm}$ & 
& 1.5 & 3.0 & 3.1 \\
 & $f_{\rm rep}$ (Hz) &  & 200 & 100 & 100 \\
 & $N_{\rm b}$  &  & 154 & 154 & 154\\
 & $\Delta_{\rm b}$ (ns) &  & 
$\hspace*{3mm}$ 0.67 $\hspace*{3mm}$ & 
$\hspace*{3mm}$ 0.67 $\hspace*{3mm}$ & 
$\hspace*{3mm}$ 0.67 $\hspace*{3mm}$ \\
 & ${N}$  (10$^{10}$)  &   & 0.4 & 0.4 & 0.4\\
 & ${\sigma_z}$  ($\mu$m)  &  &  35  &  30 & 35\\
 & ${\epsilon_x}$  ($\mu$m) &  &  2  &  0.68  & 0.68\\
 & ${\epsilon_y}$  ($\mu$m) &  &  0.01 & 0.02 & 0.01\\
 & ${\sigma_x^*}$  (nm) &   & 202 & 43 & $\approx$~60\\
 & ${\sigma_y^*}$  (nm) &  &  $\approx$~1.2  & 1 & $\approx$~0.7\\
 & ${\delta}$  (\%) &  & 4.4 & 31 & 21\\
 & ${n_\gamma}$    &  & 0.7 & 2.3 & 1.5\\
 & $N_\perp$  &   & 7.2 & 60 & 43\\
 & $N_{\rm hadr}$   &   &  0.07 & 4.05 & 2.3\\
 & $N_{\rm MJ}$   &   &  0.003 & 3.40 & 1.5
\\[3mm] 
\hline \hline
\end{tabular}
\end{center}
\end{table}


\subsection{Beam--Beam Interaction}

During collision in an electron--positron collider, the electromagnetic
fields of each beam accelerate the particles of the oncoming beam
toward its centre. In CLIC this effect is so strong that the particle
trajectories are significantly changed during the collision, leading to 
reduction of the transverse beam sizes, the so-called pinch effect. This
enhances the luminosity but since it bends  the particle trajectories it also
leads to the emission of beamstrahlung, which is comparable to synchrotron
radiation and reduces the particle energy. The average number of photons
emitted is of the order of 1, so the impact on the centre-of-mass
energy of colliding particles is somewhat comparable to initial-state
radiation.

The produced beamstrahlung photons also contribute to the production of
background. In the case of CLIC at high energies, the largest number of
particles is expected from the so-called coherent pair creation. In this
process
a real photon is converted into an electron--positron pair in the presence of
a strong electromagnetic field. The cross section for this process depends
exponentially on the field strength and the photon energy. It is therefore
very small at $E_{\rm c.m.}$~=~500~GeV but very 
important at~$E_{\rm c.m.}$~=~3~TeV.

Programs have been developed to simulate the pinch effect as well as the
production of beamstrahlung and the different sources of background. For our
estimates we use  GUINEAPIG~\cite{c:gp}.

For the old reference parameters, each particle emits on average 2.3 photons
per bunch crossing, see Table~\ref{t:parameter}. 
This corresponds to an average energy loss of about 30\%.
With the new parameters this is reduced to 1.5 photons per particle and a
loss of about 20\%. The number of coherent pairs is less than an order of
magnitude smaller than the number of beam particles. They will thus give rise
to some notable electron--electron and positron--positron luminosity.

\subsection{Luminosity Spectrum}

Not all the electron--positron collisions will take place at the nominal
centre-of-mass energy. Several sources for an energy reduction of the
initial-state particles exist. Each bunch has an initial energy spread of about
$\sigma_{E}/E\approx$~3~$\times$~10$^{-3}$, the bunch-to-bunch as well as
pulse-to-pulse energies vary and the beamstrahlung leads to energy loss during
the collision. The bunch-to-bunch as well as pulse-to-pulse energy variations
should be small, better than 0.1\% peak to peak. While the feasibility of
such a tight tolerance has been studied for the static bunch-to-bunch
variation~\cite{c:igor}, further studies remain to be performed. The main
sources of energy spread will, however, remain the single-bunch energy
spread and the beamstrahlung.
Most of the single-bunch energy spread is due to the single-bunch
beam loading in the main linac, i.e. the fact that a particle in the
head of a bunch   
extracts energy from the accelerating RF structure, so a particle in the tail
sees a reduced gradient.
In addition, because of the bunch length, not all particles
are accelerated at the same RF phase, which changes the gradient they
experience. 

In principle the energy spread in the main linac could be reduced somewhat
below the current reference value. This however would compromise the beam
stability and would thus probably imply a reduction in bunch charge and
consequently in luminosity.

\begin{figure}[htbp] %
\begin{center}
\begin{tabular}{c c}
\epsfig{file=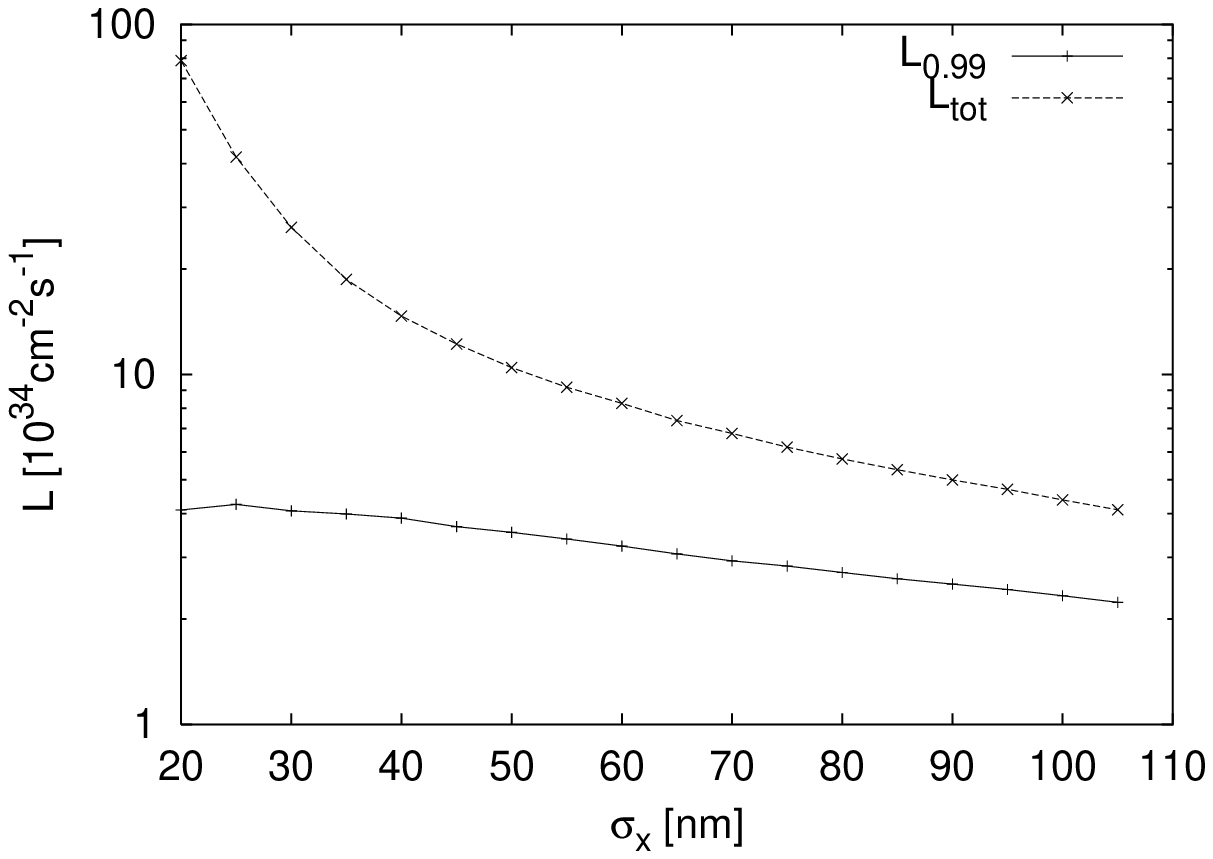,width=7.7cm,height=6.2cm} &
\epsfig{file=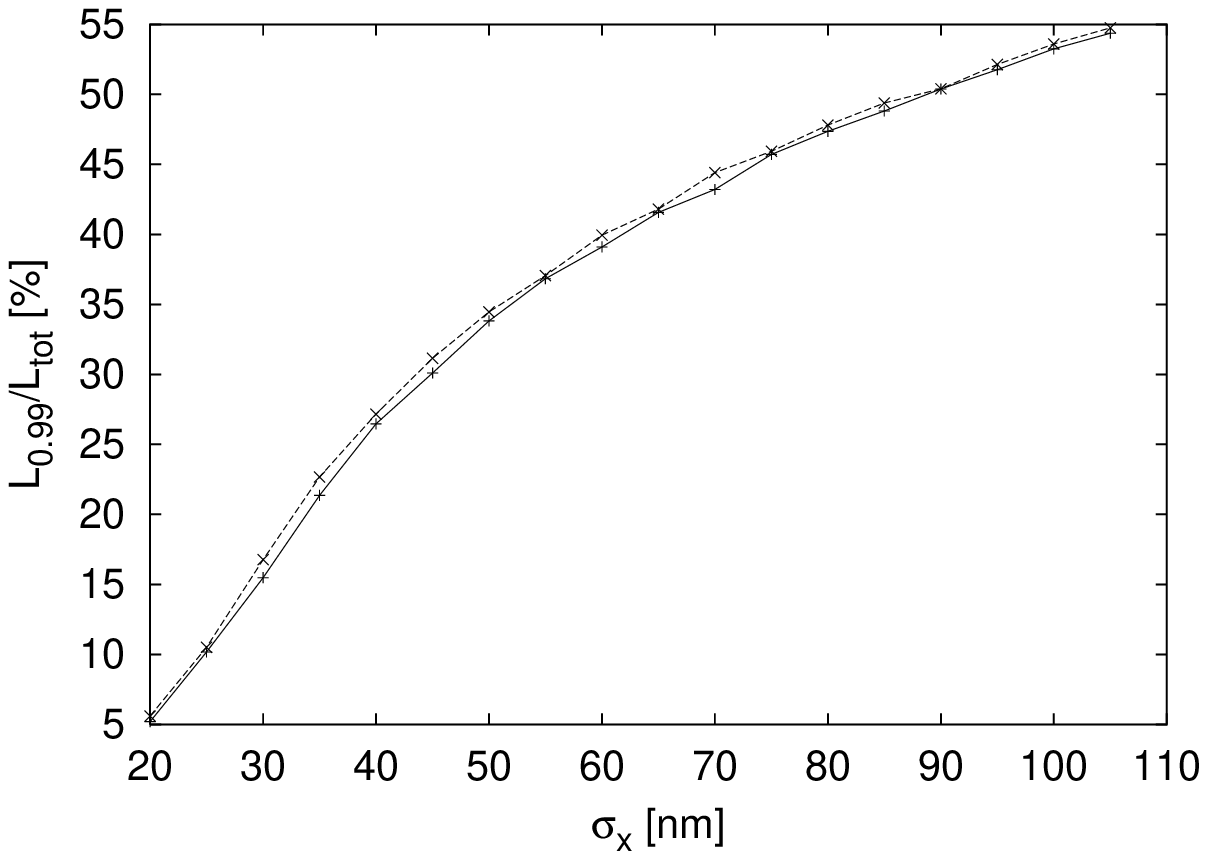,width=7.7cm,height=6.2cm} \\
\end{tabular}
\end{center}
\caption{Left: Total luminosity and the luminosity with 
$E_{c.m.}\ge$~0.99~$E_{c.m.,0}$ as a function of the horizontal beam size.
Right: The fraction of the luminosity with $E_{c.m.}\ge$~0.99~$E_{c.m.,0}$.}
\label{f:lumi}
\end{figure}

In the collision, beam particles lose energy because of beamstrahlung. This
limits the maximum luminosity that can be achieved close to the nominal
centre-of-mass energy. The lower-energy collisions can also compromise the
performance of experiments as they add to the background and make
cross section scans more difficult. For otherwise fixed parameters, the
beamstrahlung is a function of the horizontal beam size. A larger horizontal
beam size leads to the emission of fewer beamstrahlung photons and consequently
to a better luminosity spectrum. However, the total luminosity is reduced.
Figure~\ref{f:lumi} 
shows the luminosity for the nominal CLIC parameters
as a function of the horizontal beam size. On the right-hand side the total
luminosity above $\sqrt{s}\ge$~0.99~$E_{c.m.,0}$ is shown.

It should be noted that also the coherent pairs contribute to luminosity.
While they increase the $e^+e^-$ luminosity by some percent (mainly at
low centre-of-mass energies), they create $e^-e^-$ (and $e^+e^+$)
collisions, where an electron, from a coherent pair produced in the positron
beam, collides
 with the electron beam (and vice versa). Also there will occur a small
number of $e^-e^+$ collisions, 
where the initial-state particles come from
the wrong direction.

\section{Accelerator-Induced Backgounds and Experimental Conditions}

The characteristics of experimentation at CLIC will depend
significantly on the levels of backgrounds induced by the machine and
their impact on the accuracy in reconstructing the $e^+e^-$ collision
properties. Compared with the benign conditions  
experienced at LEP/SLC, and also with those anticipated for a 500~GeV linear 
collider, at the CLIC multi-TeV energies, $e^+e^-$ events will lose
part of their signature cleanliness, and resemble LHC collisions.
This is exemplified by the local track density, due to physics events
and backgrounds expected at CLIC, 
when compared with lower energy linear colliders and  LHC 
experiments (see Table~\ref{tab:tdens}). 
%
\begin{table}[htbp]  
\caption{Local track density on the innermost vertex tracker layer for
different LC designs and $\sqrt{s}$, compared with those expected at the LHC}
\label{tab:tdens}

\renewcommand{\arraystretch}{1.35} 
\begin{center}

\begin{tabular}{lccccc}\hline \hline \\[-4mm]
\textbf{LC} & 
$\hspace*{3mm}$ \boldmath{$\sqrt{s}$}~\textbf{(TeV)} $\hspace*{3mm}$ & 
$\hspace*{3mm}$ \textbf{{\it R} (cm)} $\hspace*{3mm}$ &  
$\hspace*{3mm}$ \boldmath\textbf{Hits mm$^{-2}$~BX$^{-1}$} $\hspace*{3mm}$ & 
$\hspace*{3mm}$ \boldmath\textbf{25 ns$^{-1}$} $\hspace*{3mm}$ & 
$\hspace*{3mm}$ \boldmath\textbf{train$^{-1}$} $\hspace*{3mm}$ 
\\[4mm]   

\hline \\[-3mm]
CLIC  & ~3.0 & 3.0 & 0.005  & 0.18 & 0.8 \\[3.5mm]
NLC   & ~0.5 & 1.2 & 0.100  & 1.80 & 9.5 \\
TESLA & ~0.8 & 1.5 & 0.050  & 0.05 & 225.0 \\ [3.5mm]
ATLAS & 14~ & 4.5 & 0.050  & 0.05 &      \\
ALICE & 5.5/n & 4.0 & 0.900  & 0.90 & 
\\[3mm] 
\hline \hline
\end{tabular}
\end{center}
\end{table}


The anticipated levels of backgrounds at CLIC also influence the
detector design. There are two main sources of backgrounds: those
arising from beam interactions, such as parallel muons from beam halo
and neutrons from the spent beam, and those from  
beam--beam effects, such as pair production and $\gamma \gamma \to $ hadrons. 
Beam dynamics at the interaction region also set constraints on the
detector design: the beam coupling to the detector solenoidal field
limits the strength of the magnetic field. 

\subsection{Beam Delivery System}

The beam delivery system (BDS) is the section
of beam line following the main linac
and extending through the interaction
region to the beam dump. Its task is
to transport and demagnify the beam,
to bring it into collision with a 
counter-propagating beam, and finally
to dispose of  the spent beam. 
A collimation system, which
provides a tolerable detector 
background and ensures machine 
protection against erroneous 
beam pulses, is also part of the BDS.

\begin{figure}[!h] 
\begin{center}
\rotatebox{-90}{\scalebox{.81}{\includegraphics*{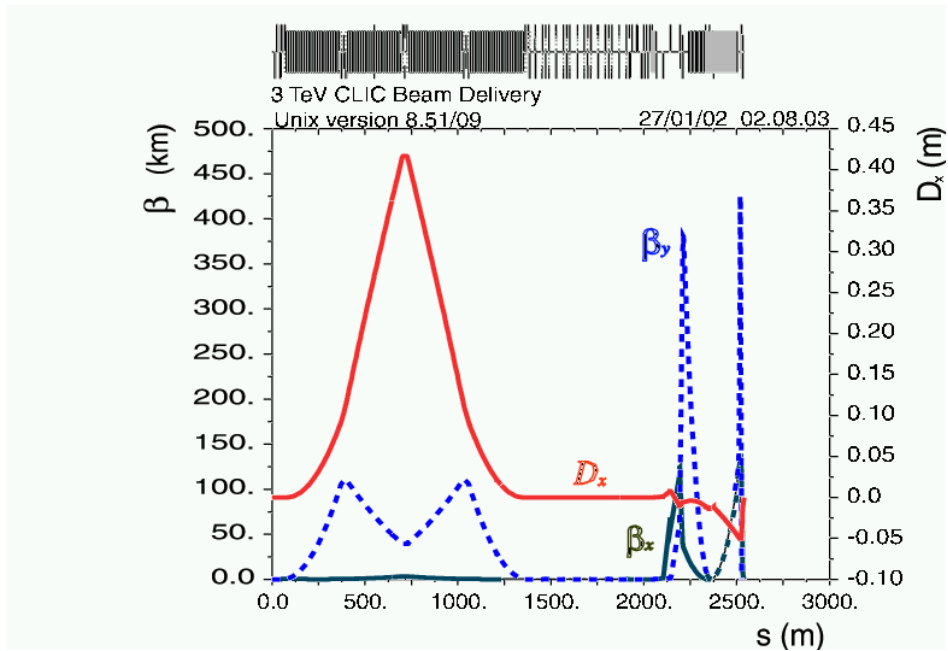}}}
\end{center}

\vspace*{-3mm}

\caption{Optics of the 3-TeV CLIC beam delivery system.
The collision point is on the right, at 2.6 km.}
\label{optics}
\end{figure}

The CLIC BDS is a modular design, consisting of
energy collimation, betatron collimation, final focus,
interaction region, and the exit line for the spent beam. 
Table~\ref{tab_bds1} 
lists the present design optics and beam parameters for the CLIC 
BDS at two different energies. Figure~\ref{optics} shows
the 3~TeV optics (from the end of the linac to the interaction point).
\begin{table}[t] 
\caption{Final-focus (FF), collimation system (CS), and beam parameters
at 3~TeV and 500~GeV cm energy. Emittance numbers refer to the
entrance of the BDS. The quoted spot sizes  refer to the rms values
obtained by particle tracking and are larger than the `effective' beam
sizes, which determine the luminosity.}
\label{tab_bds1}

\renewcommand{\arraystretch}{1.32} 
\begin{center}

\begin{tabular}{lccc}\hline \hline \\[-4mm]
\textbf{Parameter (unit)} & 
$\hspace*{4mm}$ \textbf{Symbol} $\hspace*{4mm}$ &
$\hspace*{4mm}$ \textbf{3~TeV} $\hspace*{4mm}$ &
$\hspace*{4mm}$ \textbf{500~GeV} $\hspace*{4mm}$ 
\\[4mm]   

\hline \\[-3mm]
FF length (km) & & 0.5 & 0.5 \\
CS length (km) & & 2.0 & 2.0 \\
BDS length (km) & &   2.5 & 2.5 \\ 
Hor.~emittance ($\mu$m) & 
$\gamma \epsilon_{x}$ & 0.68 & 2.0 \\
Vert.~emittance (nm) & 
$\gamma \epsilon_{y}$ &  10 & 10 \\
Hor.~beta function (mm) & 
$\beta_{x}^{\ast}$ & 6.0 & 10.0 \\
Vert.~beta function (mm) & 
$\beta_{y}^{\ast}$ & 0.07 &  0.05 \\
Effective spot size (nm) $\hspace*{15mm}$ & 
$\sigma_{x,y}^{\ast}$ & 65, 0.7 & 202, 1.2 \\
Bunch length ($\mu$m) & $\sigma_{z}^{\ast}$ & 35 & 35 \\
IP free length & $l^{\ast}$ & 4.3 & 4.3 \\
Crossing angle (mrad) & $\theta_{c}$ & 20 & 20 \\
Repetition rate (Hz) & $f_{\rm rep}$ & 100 & 200 \\
Luminosity ($10^{34}$ cm$^{-2}$s$^{-1}$) & $L_{0}$ & 8 & 2 
\\[3mm] 
\hline \hline
\end{tabular}
\end{center}
\end{table}


\begin{figure}[t] 
\begin{center}
\rotatebox{0}{\scalebox{.48}{\includegraphics*{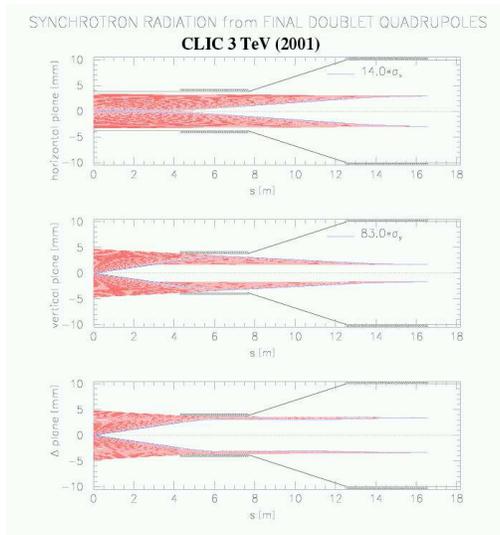}}}
\end{center}
\caption{Synchrotron radiation fans at 3~TeV with beam envelopes of 
14$\sigma_{x}$ and 83$\sigma_{y}$  
(Courtesy O.~Napoly \protect\cite{heacc01}).}
\label{srfan}
\end{figure}

The system length is kept constant, independently of the energy. Only
the sextupole 
strengths and bending angles are varied as the centre-of-mass energy is 
raised from 500~GeV to 3~TeV. This implies lateral displacements
of magnets by up to 10--20 cm. The vertical IP beta function
is squeezed down to values of 50--70~$\mu$m in order to optimize the
luminosity.  
These beta functions are still comfortably large compared
with the rms bunch length. In simulations, the target luminosity is 
reached at 3~TeV, and about twice the target value for 500~GeV. 
A solution for the design of the final quadrupole has been demonstrated,
based on permanent-magnet material~\cite{aleksa}.
Typical synchrotron radiation fans inside the two final quadrupoles
are depicted in Fig.~\ref{srfan}, 
for an envelope covering 14$\sigma_{x}$ and 83$\sigma_{y}$. The
requirement that synchrotron-radiation photons do no hit the
quadrupoles on the incoming side determines the collimation~depth.

Collimation efficiency, energy deposition by lost particles and
photons along the beam line, and muon background have been studied by various 
authors~\cite{droz,gb,hbu}, and were found to be acceptable.
Collimator wakefields and collimator survival
have been taken into account in the 
design optimization \cite{clicbds}.

A recent extended review of the present BDS designs for centre-of-mass
energies of 3~TeV and 500~GeV can be found in Ref.~\cite{clicbds}.
The final focus has also been 
discussed in~Refs.~\cite{heacc01,epac2002overview},  
and the collimation system in Refs.~\cite{epac2002overview,pac2001overview}.

\subsection{Muon Background}

The rate of muons, produced as secondary particles in the
collimation of high-energy (1.5~TeV) electrons', can be substantial and
requires a reliable simulation. 

First estimates for CLIC
have been obtained, based on the  MUBKG code developed for
TESLA~\cite{Sachwitz:1994xw} interfaced to  GEANT3 for the
simulation of the energy loss of the muons.

\begin{figure}[htbp] 
\centering
\includegraphics*[width=10cm]{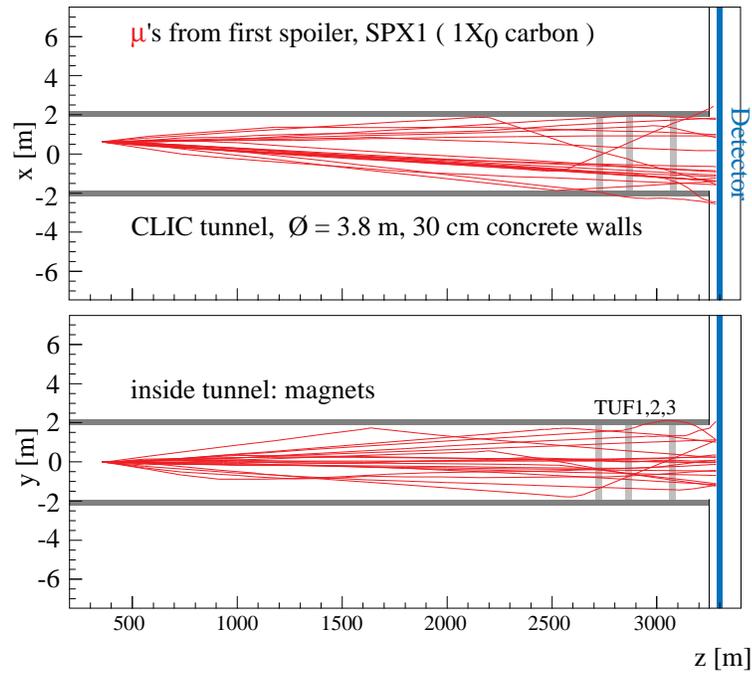}
\caption{Tracking of muons, produced at the first spoiler, through the
 beam delivery system up to the detector region.
Top is the horizontal and bottom the vertical plane.}
\label{spx1}
\end{figure}

Figure~\ref{spx1} shows some simulated tracks of muons, produced at the
first, horizontal spoiler (SPX1) and reaching the detector,
which is located at about 3~km.
The figure also shows the position of three optional, magnetized (2\,T) iron
`tunnel fillers',  each 10 or 30~m thick. They should be considered
as a first attempt at implementing a dedicated muon protection system,
as their properties and locations have not yet been optimized. 

\begin{figure}[htbp] 
\centering
\includegraphics*[width=10cm]{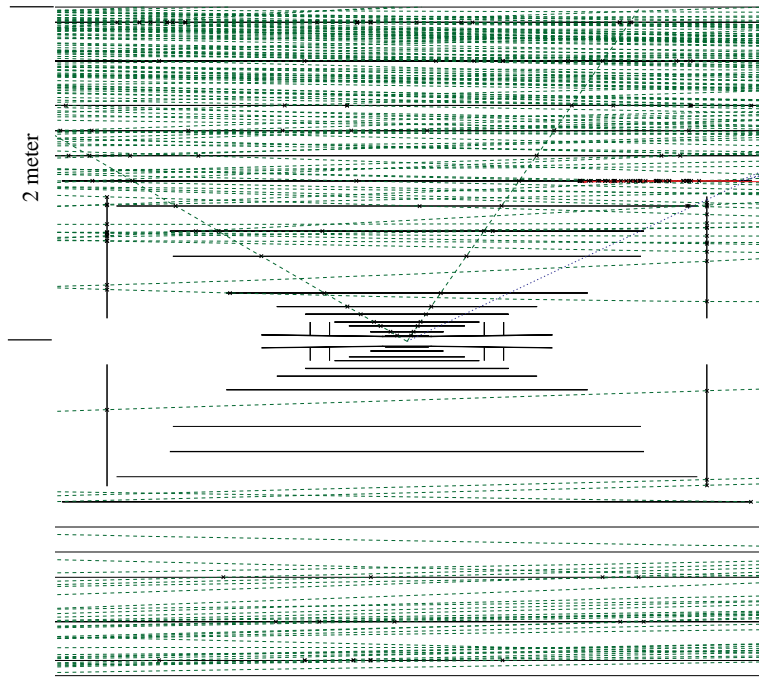}
\caption{Physics event 
($e^+ e^- \to \tilde{\mu}\tilde{\mu} \to \chi^0_1 \mu \chi^0_1 \mu$)
and muon background (in this case 1650 $\mu$ tracks) as seen in the
detector simulation} 
\label{evt_mu_smu}
\end{figure}

Tracks that reached the detector region were input to a GEANT3-based detector 
simulation. Figure~\ref{evt_mu_smu} shows the muon background
overlayed on a physics event in the  CLIC detector.

The simulation allows the prediction of the ratio $r_{e\mu}$ of beam
particles removed by the collimation system 
to the number of muons reaching the detector.
For an estimate of the muon flux in the detector, assumptions have to
be made about the fraction of halo electrons in the beam that will hit the 
collimators. The amount of beam particles in the tails is difficult to
predict, but it is generally expected to be small. 
Here, the fraction hitting the first spoiler was assumed to be
$f_{\rm tail}$~=~10$^{-3}$.  
With the parameters in Table~\ref{tab:MuFlux}, and
for the two (e$^+$ and e$^-$) beams in CLIC, we estimate that
2$N_{\rm e}\, N_{\rm b}\, f_{\rm tail}\, c_\mu
/r_{e\mu}\approx$~2.7~$\times$~10$^4$  
muons per bunch train crossing would reach the detector. With a muon
protection system of three tunnel fillers, their number could be
reduced to 4000 muons per train, or 26 per bunch crossing.
%
\begin{table}[t]
\caption{Parameters chosen to estimate the muon flux at the detector}
\label{tab:MuFlux}

\renewcommand{\arraystretch}{1.32} 
\begin{center}

\begin{tabular}{lcc}\hline \hline \\[-4mm]
\textbf{Parameter} & 
$\hspace*{5mm}$ \textbf{Symbol} $\hspace*{5mm}$ &
$\hspace*{5mm}$ \textbf{Value} $\hspace*{5mm}$ 
\\[4mm]   

\hline \\[-3mm]
Beam energy & $E$ & 1.5~TeV \\
Number of $e^+$, $e^-$ per bunch & $N_e$ & 4~$\times$~10$^9$ \\
Bunches per train & $N_{\rm b}$  & 154 \\
Fraction of tail particles & $f_{\rm tail}$ & 10$^{-3}$ \\
Secondaries and other processes $\hspace*{15mm}$ & $c_\mu$ & 2 \\
$e/\mu$ ratio without TUF & $r_{e\mu}$ & 9.2~$\times$~10$^4$ \\
$e/\mu$ ratio with TUF & $r_{e\mu}$ & 6.2~$\times$~10$^5$ 
\\[3mm] 
\hline \hline
\end{tabular}
\end{center}
\end{table}


The factor of $c_\mu$~=~2 is based on more complete simulations
for TESLA at 250~GeV; it accounts for muon production processes
that were not included in the GEANT3 simulation used here, such as
muon production from secondary photons in the cascade, 
$e^+ e^-$ annihilation 
and hadronic muon~production~\cite{Sachwitz:1994xw}.

\begin{figure}[!h]
\begin{center}
\begin{tabular}{c c}
\hspace*{-9mm}
\epsfig{file=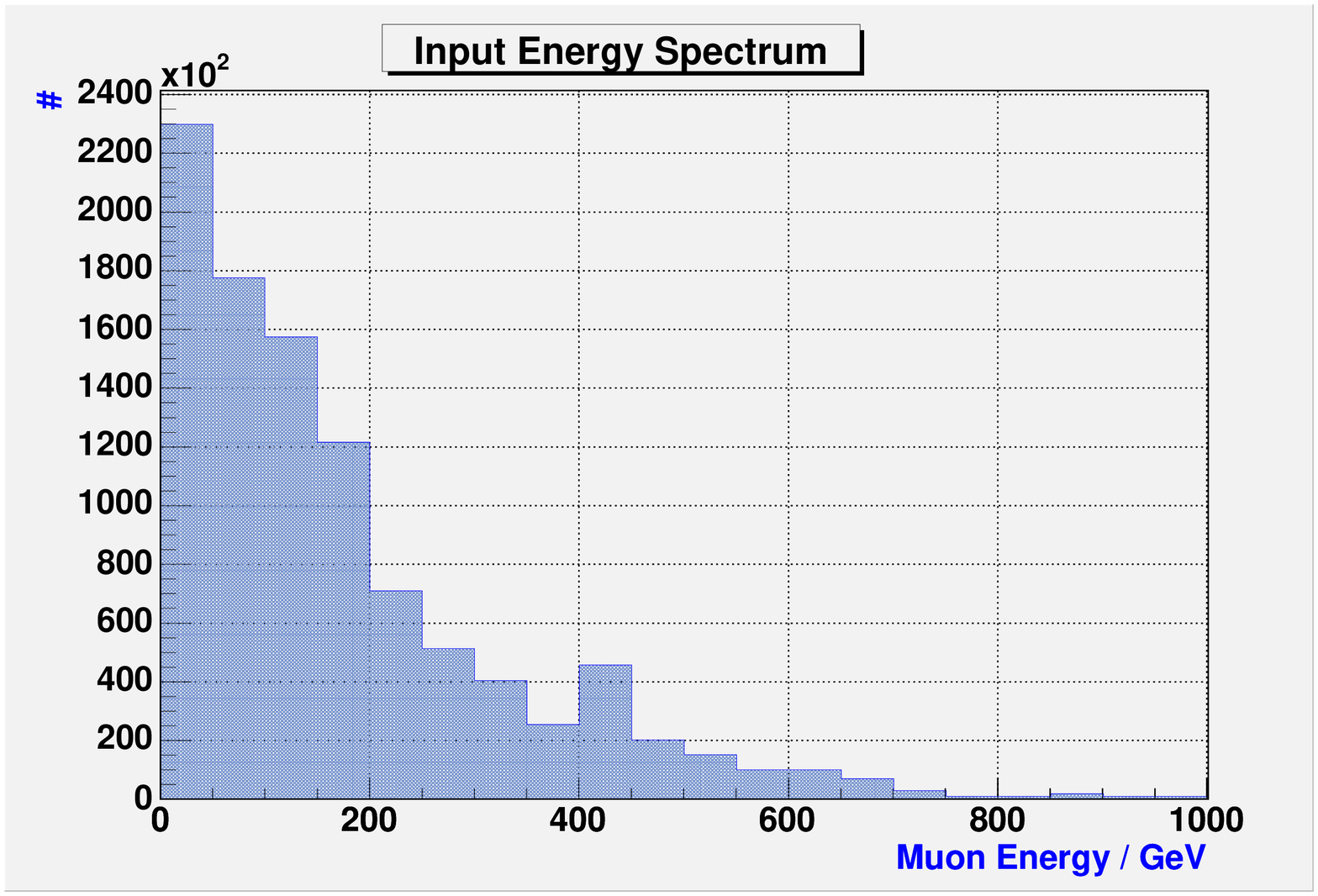,bbllx=0,bblly=0,bburx=580,bbury=700,
width=5.5cm,angle=-90}& \hspace*{7mm}
\epsfig{file=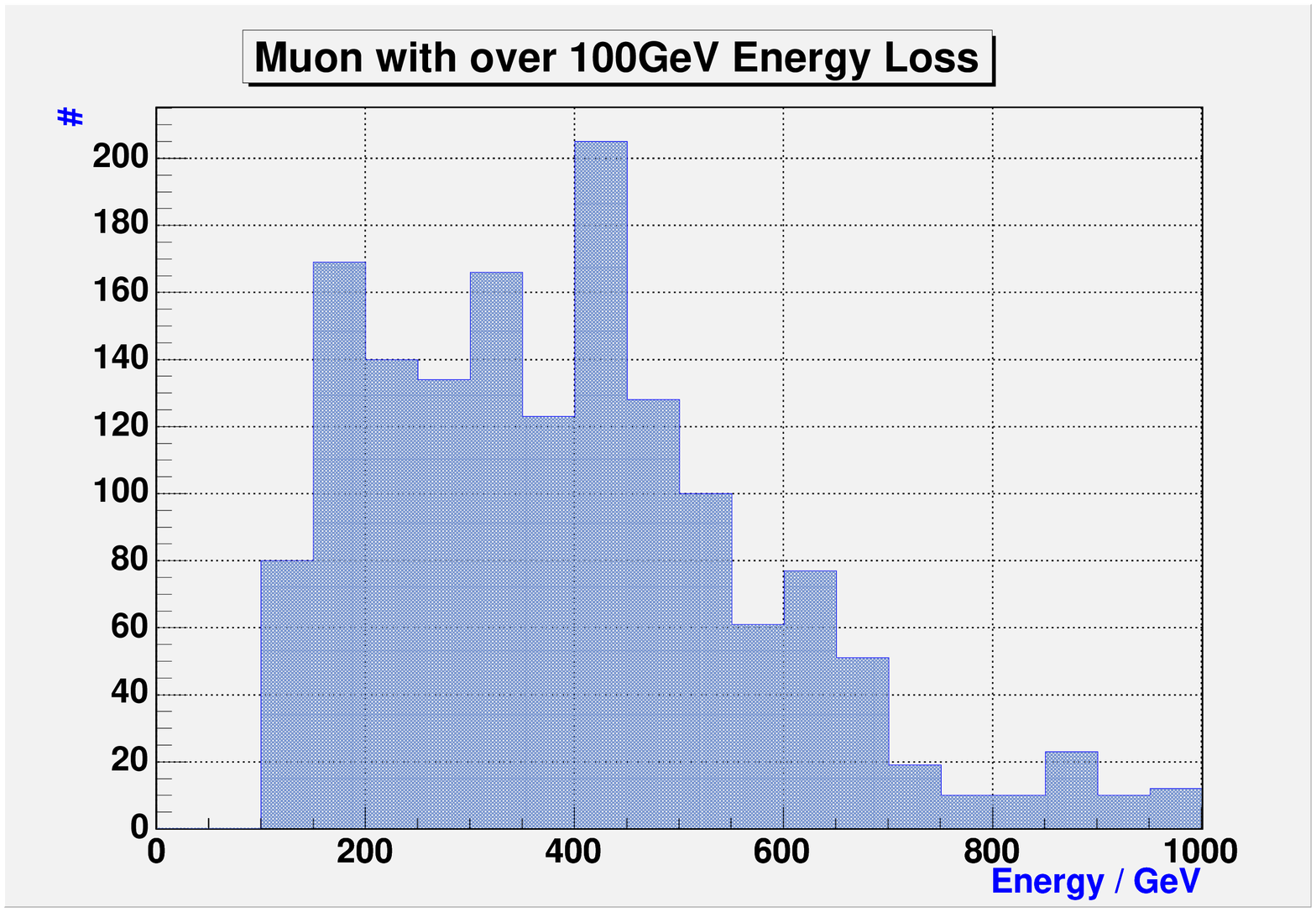,bbllx=0,bblly=0,bburx=580,bbury=700,
width=5.5cm,angle=-90}
\end{tabular}
\end{center}
\caption{Left: The energy spectrum of the background muons that hit the
detector surface; 
Right: The energy deposited in a block of iron of 8 interaction lengths}
\label{sec3:muonspec}
\end{figure}

While the rate is modest and will not deteriorate the tracker when
based on, say, silicon, the effect could be substantial for
calorimetric measurements 
due to catastrophic radiation events. GEANT4 was used  to estimate  
the amount of high energy showers that could be produced by the high 
energy muons. The expected  muon background spectrum
at the surface of the detector, shown in~Fig.~\ref{sec3:muonspec}~(left), 
from~Ref.~\cite{muons_battaglia}, was used to shoot muons into a block of
iron of 8 interaction lengths. This  corresponds to a typical endcap of a 
calorimeter. Using approximately 10$^6$ muons, the resulting spectrum of the
energy released in the block is shown in~Fig.~\ref{sec3:muonspec}~(right), for 
energy releases larger than 100~GeV. About 0.15\% of the muons of this
spectrum may leave a large energy shower in the calorimeter. Future studies
will need to develop tools to recognize these showers, including the 
investigation of calorimeter techniques.

More recently, a fully integrated approach using  GEANT4 has been
implemented. This 
includes the conversion of photons into a pair of muons in 
the presence of the fields of the
nucleus and the annihilation of high energy positrons with atomic 
electrons~\cite{MuPgen,AnnihiToMuPair}. The GEANT4 program has also
been extended to perform the tracking through the machine lattice and
materials in a combined, flexible manner. The simulation can then be
built from an existing machine description,  
allowing for background optimization during the machine design phase~\cite{gb}.

\subsection{Neutron Background}

No detailed study of the neutron flux expected from the
spent beam has been performed. Some estimate of the flux induced by
the hadrons produced in 
the beam--beam interaction is given in the next section. Regarding the spent
beam, a simple estimate of the neutron flux can be based on the giant resonance
production of neutrons. The low energy neutrons from this process tend to fly
in all spatial directions with an almost equal probability. The high energy
neutrons from other processes tend to move more in the direction of
the incoming particle beam and will thus fly away from the detector.

The total number of neutrons produced depends on the material of the beam dump.
With water, about one neutron is produced per 32~GeV of incoming
photon or electron energy. Neglecting  backscattering and shielding, the flux
at the interaction point would then be 4.6~$\times$~10$^{13}$~cm$^{-2}$ per
year (10$^7$~s) of operation, for a dump to IP distance of 100~m.
While a good fraction of the neutrons can be shielded, this shielding needs to
have a hole to let the spent beam pass. While the electron and positron beams
can in principle be bent in such a way 
that there remains no direct line of sight
between beam dump and detector, this is not possible for the photon beam.
The best (and probably  only) place where the vertex detector can be
shielded from the photon dump
is inside the detector. Such a mask can reduce the flux by about three orders
of magnitude~\cite{c:thesis}.
In the present case this would yield a flux of about
4.6~$\times$~10$^{10}$~cm$^{-2}$ per year. Increasing the distance 
between dump and interaction point would reduce this value
further. However, a careful, detailed study of this problem, including
backscattering of neutrons, remains to be done. 

\section{Beam--Beam Backgrounds and their Impact}

During the collision a number of particles are produced as background.
Beamstrahlung photons and coherent pairs have already been mentioned.
Their number is comparable to the number of beam particles, so they strongly
constrain the detector design.
Further electron--positron pairs are created by incoherent two-photon
production. Hadrons are produced in a similar fashion. The number of these
particles is small enough  not to modify the beam--beam
interaction any further, but they still have a significant impact on
the detector design.

\subsection{Coherent Pairs}

\begin{figure}[htbp] %
\begin{center}
\epsfig{file=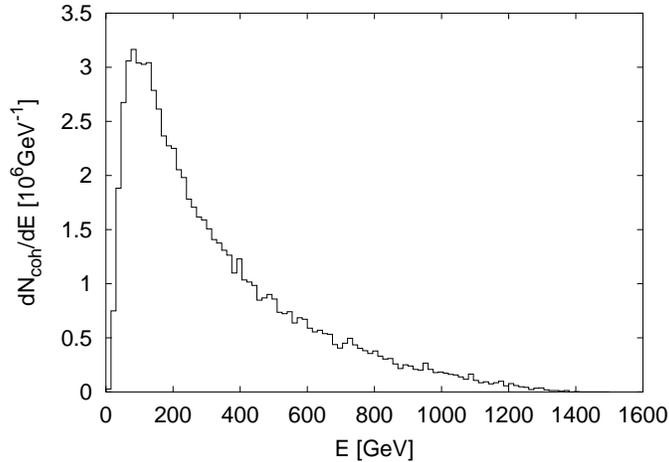,width=9.0cm}
\end{center}
\caption{The energy spectrum of the coherent pairs produced during the
collision}
\label{f:coherent}
\end{figure}

The number of coherent pairs is quite large: at $\sqrt{s}$~=~3~TeV about
7~$\times$~10$^8$ pairs are produced per bunch crossing. The spectrum of
the particles is shown in~Fig.~\ref{f:coherent}; it peaks at about
100~GeV and has a long tail 
toward high energies, almost reaching the full beam energy. At low energies
($\approx$~10~GeV) the production is strongly suppressed.

The particles initially have small angles with respect to
the beam axis. While a newly created electron that flies in the direction of
the electron beam is focused by the positron beam, a positron going in the
same direction is deflected away from the axis. Since the energy of a typical
coherent pair particle is lower than that of a beam particle, the angle after
the collision is significantly larger. The coherent pair production thus
significantly affects the aperture requirement 
for the spent beamline. 
Since the number of coherent pairs is very large, one has to avoid losing
even a small fraction of them in the detector region. In order to
achieve a reasonable 
statistics, a large number of particles thus has to be simulated, requiring
a significant amount of computing time.

\subsection{Spent Beam}

The spent beam consists mainly of the beam particles, the secondary
beamstrahlung photons and the coherent pairs. While the former two 
components are
confined to relatively small angles (of the order of 1  mrad), the
latter can reach larger angles. In order to avoid more losses in the detector
region, an exit aperture must be provided for these particles. 
Simulations with GUINEAPIG showed that an  aperture of about 10 mrad
around the axis of the spent beam is sufficient~\cite{c:qpext}.
Since space is needed around the axis of the incoming beam 
for the final-focus quadrupole, a total crossing angle
of about 20 mrad appears~reasonable.

\subsection{Incoherent Pairs}

The production of $e^+e^-$ pairs through two-photon processes can lead to
significant background at all energies. The main contributions arise from
$e e \to  ee e^+e^-$, $e\gamma \to e e^+e^-$,
and $\gamma\gamma \to e^+e^-$, where the photons are from
beamstrahlung. In the beam--beam simulations, the processes that include one
or two beam particles are calculated by replacing the particle with the 
equivalent photon spectra. This allows the effect of the beam size and
the strong beam fields onto the cross sections
to be taken into account.

\begin{figure}[htbp] %
\begin{center}
\epsfig{file=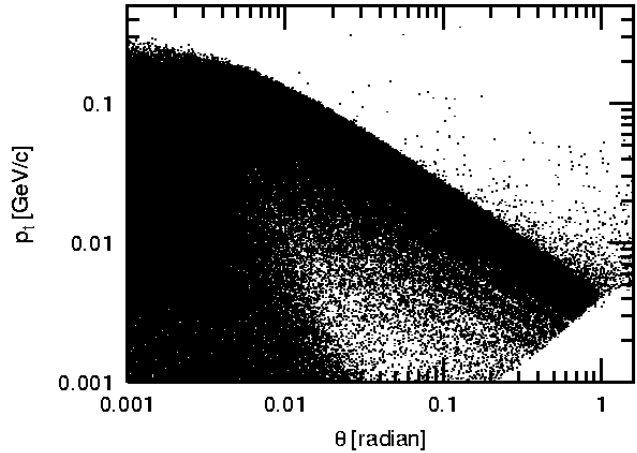,width=9cm}
\end{center}

\vspace*{-5mm}

\caption{Angle and transverse momentum of the incoherent pairs after the
collision. Only particles with an energy larger than 5 MeV were
tracked}
\label{f:scatter}
\end{figure}

It is important to track the produced particles through the fields of the
beams, since they can be strongly deflected. 
Figure~\ref{f:scatter} shows
particles after the collision. For the bulk of these, a clear
correlation is visible between the maximum particle angle and the
transverse momentum. These particles were produced at small angles and
obtained most of their transverse momentum from the deflection by the beams.
A few particles above this edge were produced with large angles and transverse
momenta. They can produce significant background in the vertex
detector.

\subsection{Hadronic Background and Resulting Neutrons}

\begin{figure}[!h] 
\hspace*{8mm} \hbox{
\epsfxsize=7cm
\epsfbox{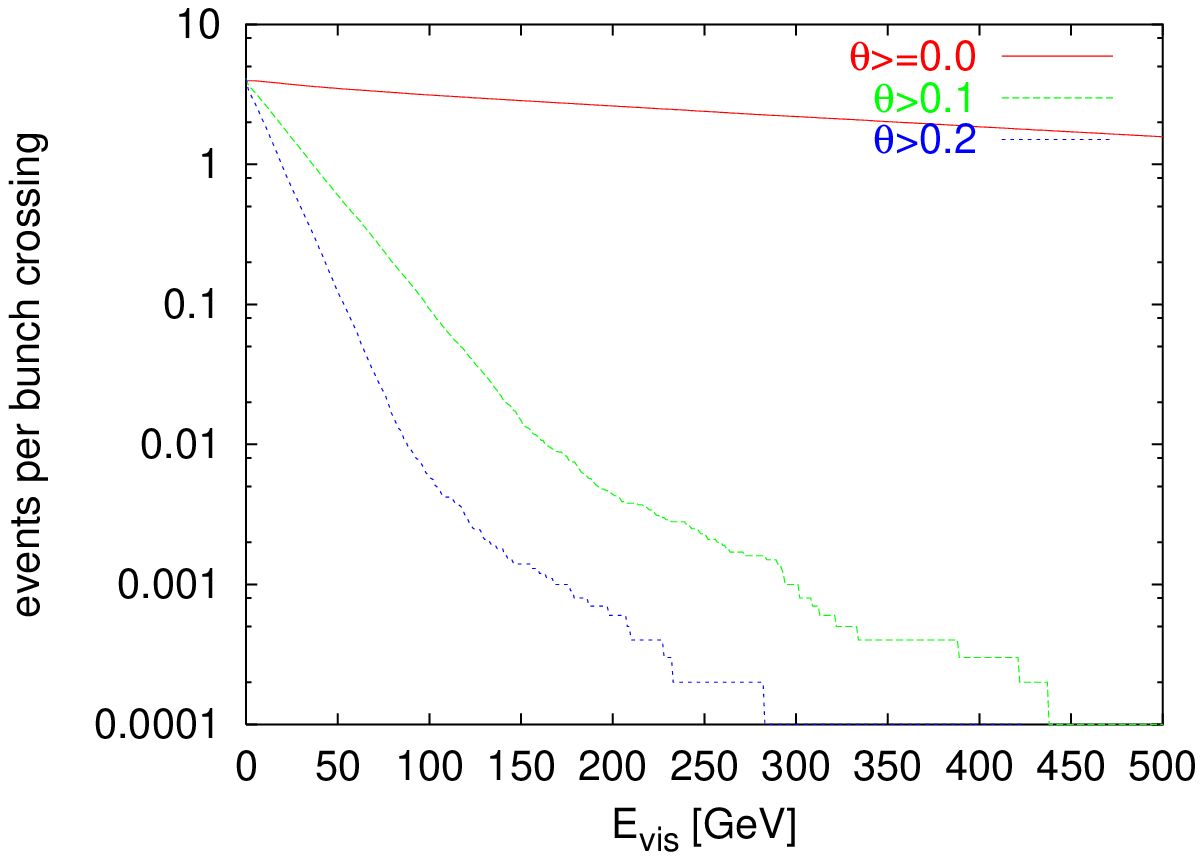}
\epsfxsize=7cm
\epsfbox{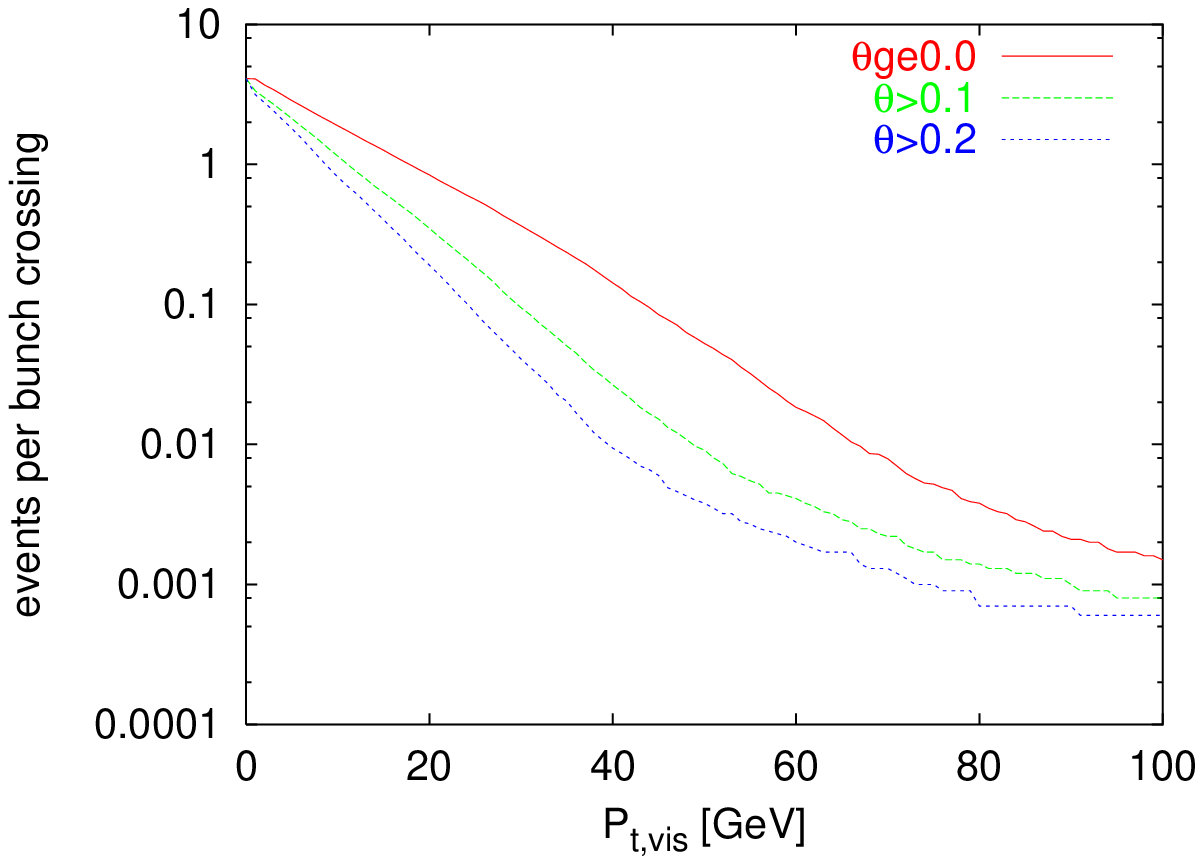}}
\caption{Left: The average number of hadronic events per bunch crossing
as a function of the visible energy above a given cut angle $\theta$.
Right: The number of events as a function of the transverse energy.}
\label{f:had}
\end{figure}

Two-photon collisions can also lead to the production of hadrons. The
cross section for this process is not very well established at higher
centre-of-mass energies, and measuring it will be an interesting
experiment at a future linear collider. In order to estimate the number of
these events, a simple parametrization of the cross section~\cite{c:hadcross}
is done with GUINEAPIG. For the old reference parameters, this
simulation yields 
about 4~events with a centre-of-mass energy above 5~GeV per bunch
crossing (Fig.~\ref{f:had}). For the new beam parameters this number
is reduced to 2.3~events per bunch crossing.
The final states of the hadronic background were simulated using  PYTHIA.
The average visible energy in the detector is about 90~GeV per bunch
crossing.
The hadronic background also causes charged hits in the vertex detector. The
highest density of these hits is 0.25~hits~mm$^{-2}$. While this is still
lower than the $\simeq$~1~hit~mm$^{-2}$ due to incoherent pairs, it is still 
a sizeable contribution. A database containing the hadronic
background of about 2000~bunch crossings has been made
available~\cite{c:hadesweb} and can be accessed  
using the HADES library (see Section~\ref{sec:3-hades})~\cite{c:hades}.

The secondary neutron flux from these hadronic events has also been simulated,
yielding a maximum flux of 3~$\times$~10$^9$~$n$~cm$^{-2}$ per year
(=~10$^7$~s) of operation~\cite{c:lcws2000}.
The flux is highest around the masks and smaller around the IP.

\subsection{Crossing Angle}

In CLIC the beams will collide with a full crossing of 20 mrad. Lower
limits of this crossing angle arise from the so-called multibunch kink
instability and the necessity to extract the spent beam from the detector with
minimal losses. An upper limit is imposed by the detector solenoidal  field,
which, in the presence of a crossing angle, affects the beams, leading to the
emission of synchrotron radiation and the subsequent increase of the beam 
spot size.

To avoid the luminosity reduction normally associated with collisions at a
crossing angle, the so-called crab-crossing cavities should be 
used. These cavities add a small horizontal kick to the beam particles, 
depending on their longitudinal position in the bunch. This scheme allows the 
bunches to collide head-on.

\subsubsection{Lower limit of the crossing angle}

During the beam collision a large number of electron--positron
pairs is created. These particles carry about $1\%$ of the total beam energy
and are deflected by the strong electromagnetic fields of the beams.
Consequently they can lead to large losses in the line for the spent beam.
While the detector is shielded against the resulting secondaries by a tungsten
mask, some of these are still of concern, in particular neutrons.
In addition, the sheer energy deposition may lead to problems. In particular, 
the tolerance for the position stability of the final quadrupoles,
which are inside the detector and must not move by more than
$\approx$ 0.2~nm, may be hard to meet. However,  providing an exit hole
lets all the particles through with an angle of less than 10~mrad
and solves the problem~\cite{c:qpext}. This can be achieved with a
crossing angle 
of at least 10~mrad. Since additional space is required for the
quadrupole around the incoming beam a reasonable lower limit is
$\theta_c \ge$~20~mrad.

Another issue is the kick that the outgoing beam applies on the incoming
one during the parasitic crossing inside the detector. This effect can be
reduced by shielding the incoming beam line 
from the outgoing one, but to
achieve this some material must be placed close to the IP. Another
possibility is to increase the crossing angle in order to reduce the
parasitic kicks. A sufficient suppression  
can be achieved with a crossing angle of $\theta_c$~=~20 mrad.

\subsubsection{Beam coupling to the detector field}

If the beams have an angle with respect to the detector solenoidal field, they
will travel on a helix toward the collision point. By choosing an appropriate
optics for the incoming beam, the associated effects can be compensated.
However, in the solenoidal field the particles will emit synchrotron radiation,
which modifies their energy and thus their trajectory slightly. As a
consequence the beam-spot size will increase in the vertical direction. This
effect can be reduced by decreasing either the detector solenoidal
field or the crossing angle. A preliminary study indicates that for
$B_z$~=~4~T a crossing angle of  
$\theta_c$~=~20 mrad is still acceptable~\cite{c:cross}. However, the
results depend significantly on the detector end fields, for which
only a simple model was used in the study. Therefore it will be
necessary to perform more precise simulations when a design  
for the detector solenoid becomes available.

\subsection{Impact on the Vertex Detector Design}

\begin{figure}[t] 
\begin{center}
\epsfig{file=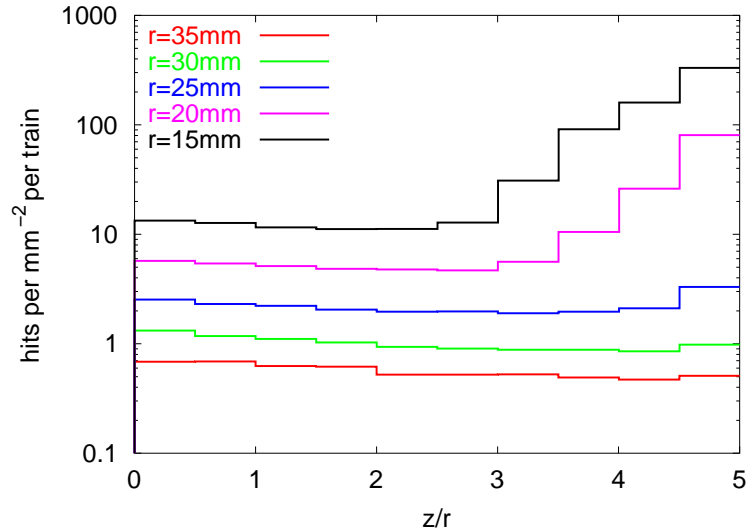,width=10.0cm}
\end{center}
\caption{Number of hits per train as function of the ratio of the
longitudinal position along the beam axis to the detector radius, for
different values of the radius}
\label{fig:nhits}
\end{figure}

A crucial constraint on the detector design arises from the hits of
incoherent pairs in the innermost vertex detector layer. An acceptable
density of hits from background on the innermost layer is of
$O(1)$~hits~mm$^{-2}$ per readout cycle. Beyond this level the pattern
recognition will be affected by the association of spurious hits to
particle tracks and by the creation of ghost tracks.  
To allow some safety margin a target of not more than 1~hit~ mm$^{-2}$
has been adopted. The number of hits in the vertex detector has been
obtained using a full GEANT simulation. Figure~\ref{fig:nhits} 
shows the expected hit density per beam pulse at different values of the
radius $r$ of the innermost vertex detector layer  
as a function of the longitudinal position $z$. At $r$~=~30~mm the hit
density reaches the limit of about 1 mm$^{-2}$, and this radius
has therefore been adopted in the detector design. A higher detector
solenoidal field could be used to further reduce~$r$. 
%
\begin{figure}[!h] 
\epsfxsize=13.5cm
\centerline{\epsfbox{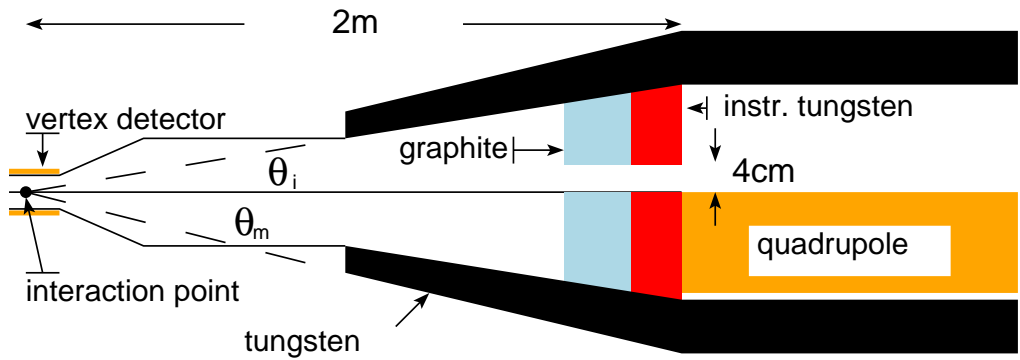}}
\caption{Sketch of the mask layout}
\label{f:mask}
\end{figure}

\subsection{Mask Design}

Another potentially important source of hits in the vertex detector 
is that of low-energy electrons and positrons from coherent pair
creation; these are backscattered in the final
quadrupoles and are then guided by the main solenoidal field back into
the vertex 
detector~\cite{c:thesis}. The total number of hits from this source can be
an order of magnitude larger than the direct hits. This effect can be
suppressed by using a mask that covers the side of the quadrupole
facing the detector. If the mask is covered with a low-$Z$ material,
backscattering can be almost completely suppressed;  graphite is
currently considered. Figure~\ref{f:mask} shows the full masking
system, of which the part discussed here 
consists of the graphite and instrumented tungsten layers.

Originally the mask has been developed for
a machine without a crossing angle~\cite{c:thesis} and it was found that it
worked excellently as soon as the
inner opening, through which the beam passes, is smaller than the vertex
detector. In the presence of a crossing angle the situation is more
complicated, as part of the vertex detector cannot be protected by the
mask. One therefore has to provide further masking downstream, to
prevent backscattering through the hole of the mask.

The inner mask also serves as a shield against neutrons, which are produced
by the spent beam and backscattered into the detector. It has been shown that
such a shield can reduce the neutron flux in the vertex detector by three
orders of magnitude~\cite{c:thesis}.
For this purpose, the opening in the mask needs to
be smaller than the vertex detector; the crossing angle does not
matter as the neutrons will mainly come antiparallel to the spent
beam. The further the mask is from the IP the larger the opening must
be to allow the passage of the spent beam. At the chosen distance of 2~m
between mask and IP, the opening
has a radius of 2 cm and satisfies both conditions.

\begin{figure}[t] 
\begin{center}
\epsfig{file=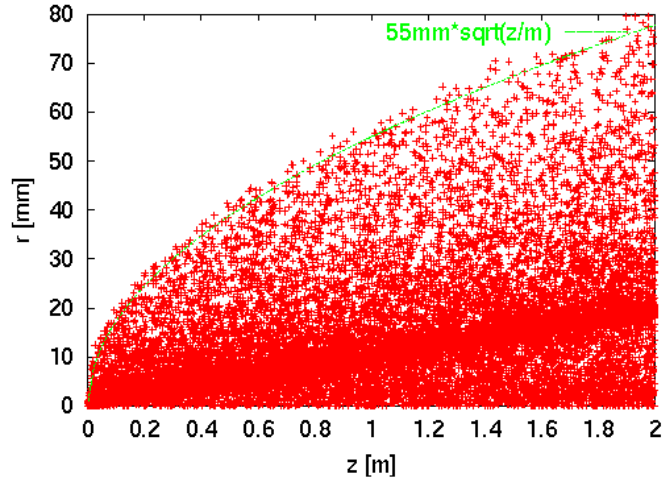,width=9.0cm}
\end{center}

\vspace*{-5mm}

\caption{The particles from incoherent pair production in the spent beam}
\label{f:aperture}
\end{figure}

Not all the coherently and incoherently produced pairs will pass the exit hole
for the spent beam. The fraction that is lost in the detector produces
secondary photons, which can cause severe background. A conical mask is
employed to shield the detector. Currently a simplified geometry is used for
this mask. The design is completely determined by the inner and outer opening
angle, $\theta_0$ and $\theta_1$, and the distance to the IP at which the mask
starts is chosen to be 1~m. The need to let the bulk of the
pairs enter the mask then fixes the inner opening angle in the detector
magnetic field of $B_z$~=~4~T. Figure~\ref{f:aperture} 
shows the distances 
from the detector axis reached by the incoherent pairs; they are similar for
the coherent ones. To allow some margin an inner opening angle
$\theta$~=~80~mrad is chosen. With an outer angle
$\theta_1$~=~120~mrad one can expect the 
mask to suppress the photons sufficiently~\cite{c:thesis}. Optimization of the
mask remains to be done, and in particular the instrumentation of this
area needs to be considered.

\section{The Detector: Concept and Techniques}

The concept of a detector for CLIC is greatly influenced by the
experience gained  
at LEP/SLC and by the technical solutions being adopted for the LHC. 
The extrapolation of these principles to a detector for energies up to
about 1~TeV, a  
factor of 5 more than at LEP, is already well advanced within the LC studies 
world-wide. As CLIC pushes the energy range up by another factor of 5, a major 
question arises on the validity of such an
extrapolation for multi-TeV $e^+e^-$ physics.

In order to appreciate the different characteristics of events from
multi-TeV $e^+e^-$ collisions, it is useful to consider that the jet
multiplicity will reach fourteen, with  
large charged and neutral particle multiplicity. The
$b$-hadrons will travel up to 20~cm, 
owing to the large boost in 
two-fermion events. The intense beam radiation will produce 
events boosted along the beam axis, and backgrounds will become of
primary importance with significant minijet and pair
production. There will be of the order of 2000  
photons per bunch crossing, and 25000 muons per train, 
traversing the detector.
%
\begin{table}[!h] 
\caption{Average reconstructed jet multiplicity in hadronic events at 
different $\sqrt{s}$ energies}
\label{tab:jmult}

\renewcommand{\arraystretch}{1.35} 
\begin{center}

\begin{tabular}{ccccccc}\hline \hline \\[-4mm]
$\hspace*{3.5mm}$ $\sqrt{s}$ (TeV) $\hspace*{3.5mm}$ &
$\hspace*{3.5mm}$ 0.09 $\hspace*{3.5mm}$ & 
$\hspace*{3.5mm}$ 0.20 $\hspace*{3.5mm}$ & 
$\hspace*{3.5mm}$ 0.5 $\hspace*{3.5mm}$ & 
$\hspace*{3.5mm}$ 0.8 $\hspace*{3.5mm}$ & 
$\hspace*{3.5mm}$ 3.0 $\hspace*{3.5mm}$ & 
$\hspace*{3.5mm}$ 5.0 $\hspace*{3.5mm}$ \\ 
$\langle N_{\rm Jets} \rangle$   & 2.8  & 4.2  & 4.8 & 5.3 & 6.4 & 6.7 
\\[3mm] 
\hline \hline
\end{tabular}
\end{center}
\end{table}


Studies of the detector concept for a 500~GeV LC have considered both
a small and a large detector. The first consists of a compact Si
tracker with high magnetic field ($B$~=~5--6~T). The large detector
relies on a continuous time projection chamber (TPC) main tracker, has
a lower magnetic field ($B$~=~3--4~T) and moves the calorimeter further  
away from the interaction region, which is advantageous for energy
flow reconstruction based on a fine granularity. 

At CLIC, the beam delivery system will constrain the solenoidal
magnetic field of the detector to below about 6~T. We consider here a model 
for a large detector, which inherits much of the solutions adopted in
the ECFA/DESY study for a 500--800~GeV LC, adapted to the experimental
conditions expected at~CLIC.

\subsection{Vertex Tracker}

A CLIC vertex tracker design may consist of a multilayered
detector with very high 3-D space resolution and stand-alone tracking
capabilities.
At CLIC, the anticipated background from $e^+e^-$ pairs produced in the 
interaction of the colliding beams will limit the approach to the interaction
region to about 3.0~cm, compared with $\simeq$~1.5~cm foreseen for the lower 
energy projects. This is partly compensated by the increase of the 
short-lived hadron decay length due to the larger boost. 
At $\sqrt{s}$~=~3~TeV, the average decay distance of a $B$ 
hadron is 9.0~cm, in the two-jet 
process $e^+e^- \to b \bar b$, and 2.5~cm in multiparton 
$e^+e^- \to H^+H^- \to t \bar b \bar t b \to W^+ b \bar b W^-
\bar b b$ decays. Because of the large boost and large hadronic multiplicity,
the local detector occupancy in $e^+e^- \to b \bar b$ events
is expected to  
increase, by  a factor of almost 10, to $>$~1 particle mm$^{-1}$ from 
$\sqrt{s}$~=~0.5~TeV to  $\sqrt{s}$~=~3.0~TeV.
This indicates the need to design a large vertex tracker based on
small-area pixel sensors, able to accurately reconstruct the trajectories of 
secondary particles originating few tens of centimetres away from the beam 
IP and contained in highly collimated hadronic jets. There are 
other background issues relevant to the conceptual design of a vertex tracker
for CLIC. These are the rate of $\gamma \gamma \to $ hadrons
events, estimated at 4.0~BX$^{-1}$, and the neutron flux, possibly of the
order of 10$^{10}$~1-MeV-equivalent~n~cm$^{-2}$~year$^{-1}$. The need to 
reduce the number of $\gamma \gamma$ events overlapped to an $e^+e^-$ 
interaction requires fast time stamping capabilities, while the neutron-induced
bulk damage has to be considered in terms of sensor efficiency reduction. 
Significant R\&D is under way for meeting the requirements of a~TeV-class 
$e^+e^-$ linear collider; further developments are expected when the issues 
related to the SLHC will be addressed. 

Emerging silicon sensor technologies and possible paths for the
upcoming R\&D are discussed in the next section.

\begin{figure}[!ht]
\begin{center}
\begin{tabular}{l r}
\epsfig{file=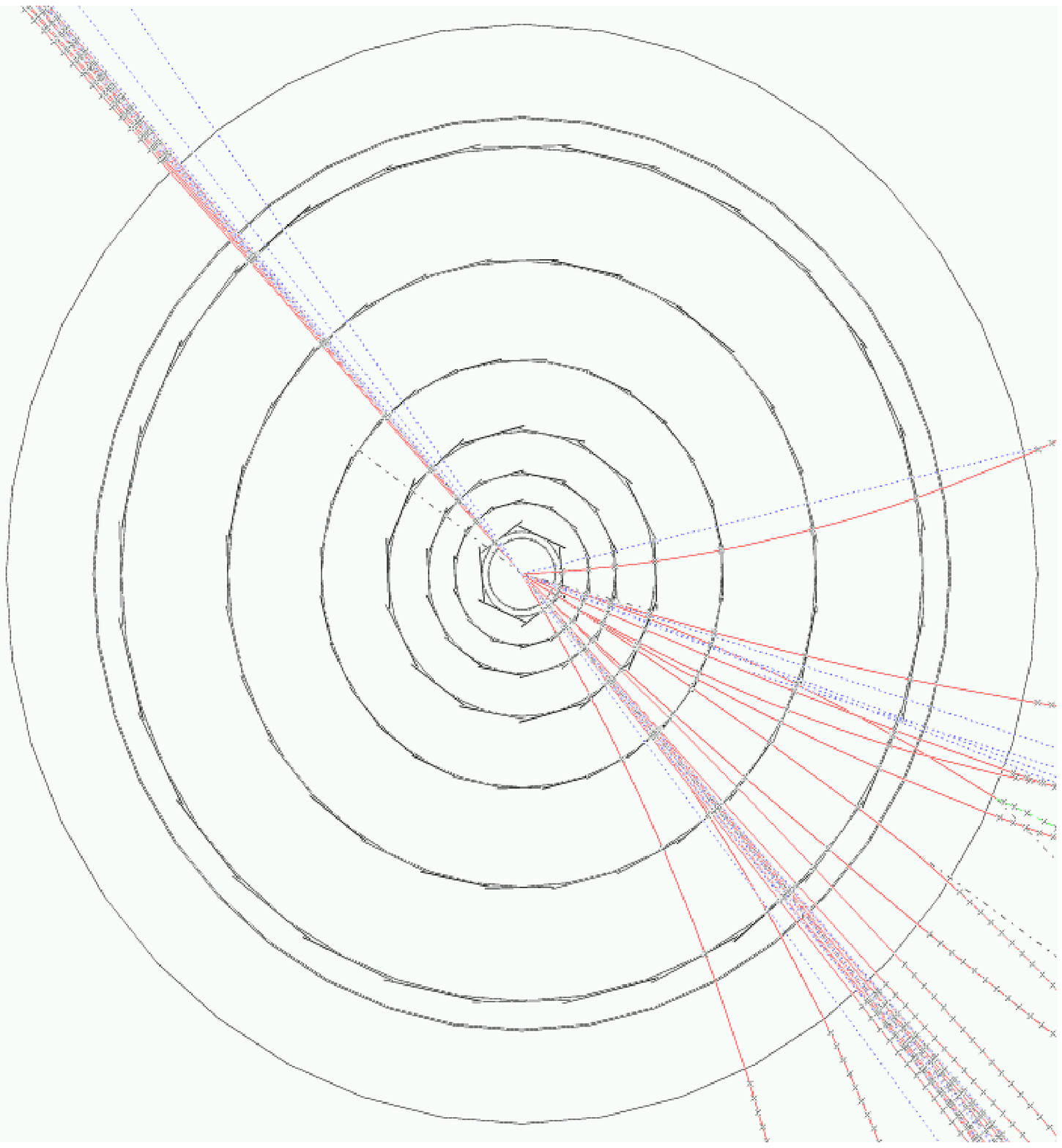,width=5.0cm,clip} & \hspace*{5mm}
\epsfig{file=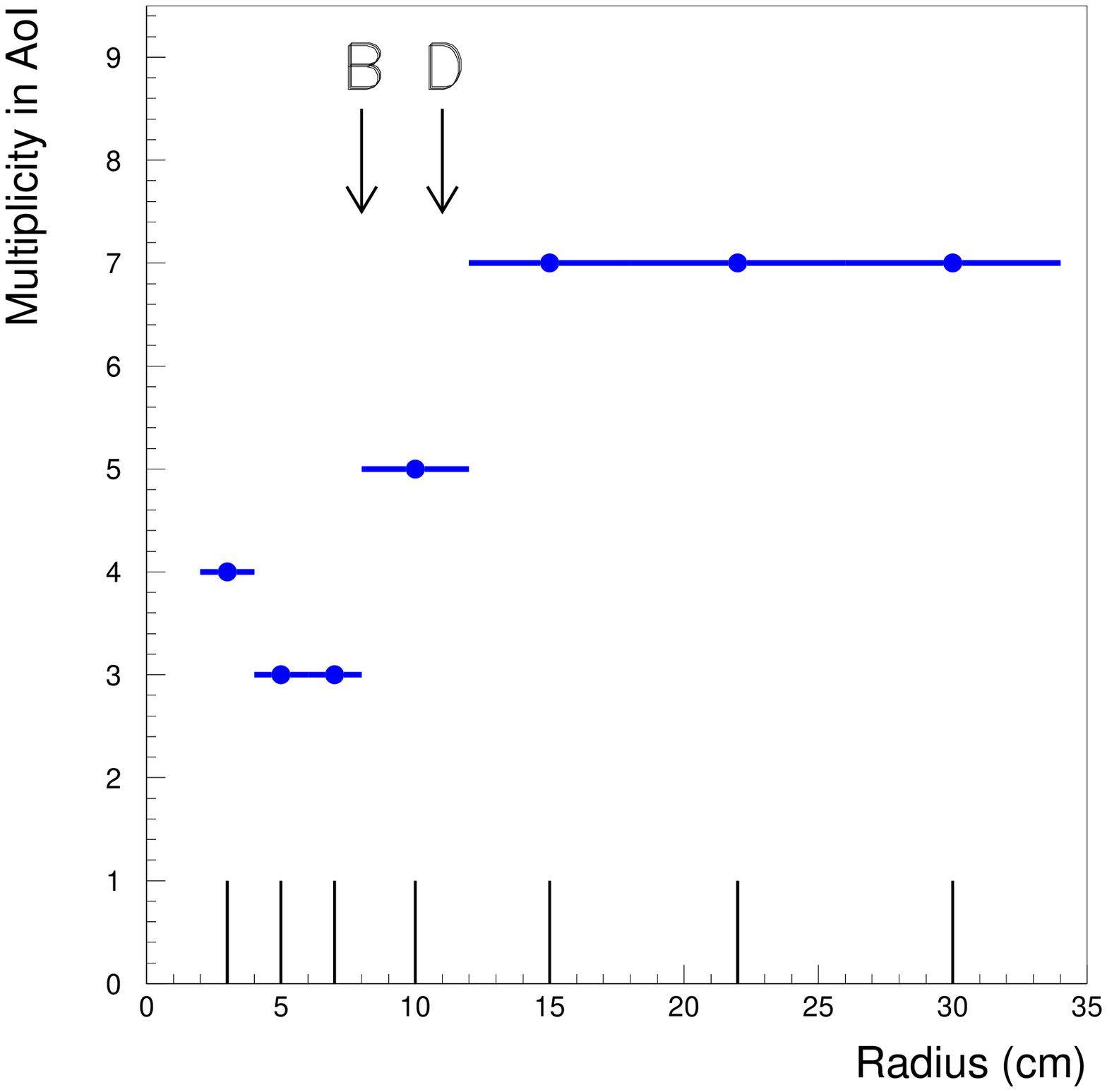,width=7.0cm,height=5.5cm,clip}\\
\end{tabular}
\caption{Display of a $e^+e^- \to b \bar b$ event
at $\sqrt{s}$~=~3~TeV (left) with the detected hit multiplicity steps from the 
cascade decay of a long-flying $B$ hadron (right)}
\label{fig:event}
\end{center}
\end{figure}

A detector consisting of seven concentric Si layers located from 3.0~cm
to 30~cm from the beam IP and based on pixel sensors,
with 20~ns time stamping and radiation hardness capabilities, demonstrated for 
their LHC applications can be considered as a baseline design for the
vertex tracker.  The layer spacing has been chosen to optimally sample
the heavy hadron decay length, resulting in a closer spacing for the
innermost layers (see Fig.~\ref{fig:event}).    

\subsection{Trends in Si Sensor Developments and Future R\&D}

Silicon sensors have been very successfully employed in collider
experiments for many  
years. They are based on well established technologies and 
their prices have dropped quite 
significantly over the years, making large-area 
trackers both feasible and affordable. 
Obvious advantages of Si pixel detectors are their accurate point
resolution, of the order of 10~$\mu$m, true 3D reconstruction of the
point of passage of the particle, 
and fast charge collection. The main drawback is certainly related to
the relatively  
high material burden represented by the present sensors and related read-out 
electronics. R\&D studies performed for the SLHC experiments to improve the 
radiation hardness of silicon devices have shown that fluences of several 
10$^{14}$ 1-MeV equivalent neutrons cm$^{-2}$
can be coped with, which is orders of 
magnitude higher than those expected at CLIC.

Tracking systems for future collider experiments will need very large area 
detectors with a macro-pad geometry (0.5~mm to 1~mm pixels). A large pixel 
granularity, with $<$~50~$\times$~50~$\mu$m$^2$ pixels, are required
for vertex detectors. The time structure of the colliding beams at
CLIC  and SLHC, high-multiplicity events, and the precise tracking
requirements will need time-stamping capabilities of~$O$(1--10~ns). 
These challenging requirements   
call both for  innovative detector technologies and readout 
electronics that profit from 
future nanoscale microelectronics technologies.

Based on the experience from the LEP silicon trackers and the past
R{\&}D work and production of prototype tracker systems, strip and
pixel, for the LHC experiments, some preliminary guidelines can be
drawn to direct the needed R{\&}D on Si detectors  
for future tracking systems. 
Identified issues in designing and constructing large-area Si tracking 
systems (with detector areas of tens to hundreds of square
metres) can be categorized as follows:

\begin{itemize}
\item \textit{Cost}: present Si detector technology, based on crystal
Si wafers, has  
reached an industrial maturity and the current production cost of about 
10~CHF/cm$^{2}$ cannot realistically be expected to decrease significantly.

\item \textit{Spatial resolution}: LHC hybrid pixel detectors 
currently have  a minimum 
pixel dimension of 50 $\mu$m, 
corresponding to the minimum pitch realistically feasible 
for the current bump bonding technology. The wafer thickness of 300
$\mu$m can be thinned down to 200 $\mu$m.

\item \textit{Charge collection speed}: the nominal charge collection time of 
a standard 
silicon detector is in the range of 10~ns to 20~ns.

\item \textit{Radiation hardness}: the R{\&}D work on radiation
hardness of crystal Si  
detectors done for the LHC experiments has shown that standard $n$-doped 
high-resistivity substrates reach the doping inversion at 
about 2 $~\times~$10$^{13}$~$n$/cm$^2$
and sustain a 
displacement damage at a fluence of few 10$^{14}$~$n$/cm$^{2}$. Above this 
limit, standard crystal Si detectors cannot be used.

\item \textit{Interconnections}: bonding between silicon detectors and
readout ASICs, wire bonding for strips and bump bonding for pixel detectors,
represents a substantial effort and expansive step in the construction
of silicon strip and silicon pixel detector modules. 

\item \textit{Cooling}: Si strip and pixel detectors 
in central LHC trackers operate 
at $-15^{\circ}$~C 
to control the leakage current and reverse annealing after radiation damage. 
This poses strong constraints on the design of cooling systems, substantially 
increasing the material budget.
\end{itemize}

Considering all these issues, it turns out that scaling up dimensions, 
improving spatial
precision, radiation hardness, charge collection speed of 
future silicon trackers is a very challenging task,
 without a technological breakthrough based on emerging solid-state
detector technologies, the 
cost issue  not being the least obstacle. Thinner standard crystal
Si (of order 100~$\mu$m) could in principle improve the detector 

\newpage

\begin{sidewaystable}
\caption{Comparison of the different pixel sensor technologies}
\label{tab1}

\vspace*{-17mm}


\vspace*{7mm}

\renewcommand{\arraystretch}{1.25} 
%
\large{
\begin{tabular}{lcccc}
 & & & & \\
\hline \hline\\[-4mm]
 & & & &  \\[-4mm]
\textbf{Properties} & 
\textbf{Standard planar crystal} &
\textbf{3-D silicon} &
\textbf{Monolithic CMOS pixel} &
\textbf{a-Si:H pixel detector} \\
 & \textbf{silicon} &  & \textbf{detector} & 
\\[2.5mm] \hline \\[-2mm]
Collection speed & 
10 ns & 
Short drift & 
Thermal drift & 
Short drift, high field \\
Electron transient t & 
20 ns &
$<$1 ns  & 
100~ns  & 
2~ns \\
Holes transient t & & 
1 ns & 
200~ns &
150~ns \\[3mm]
Thickness & 
300~$\mu$m & 
100--200~$\mu$m & 
2--8~$\mu$m & 
30--50~$\mu$m \\[3mm]
MIP charge signal & 
24 000 $e^-$ & 
10 000--20 000~$e^-$ & 
100--500 $e^-$ & 
1000--2000 $e^-$ \\[3mm]
Radiation hardness & 
3~$\times$~10$^{14}$  &  
At least 10$^{15}$ at +~20$^\circ$C & 
$<$~10$^{13}$, & 
$>$~5~$\times$~10$^{15}$, limit not known, \\
 &  &  & strong surface effects &  self-annealing by mobile H \\
Fluence (n/cm$^{2}$) & 
at --~20$^\circ$C &  &  & \\[3mm]
Operating temperature & 
--~20$^\circ$C, cryogenic & 
Room & 
Room & 
Room to 60$^\circ$C \\[3mm]
Manufacturing cost & 
High & 
High & 
Low & 
Low \\[3mm]
Field of applications & 
Microvertex detector &
Small detector area,  &
Microvertex detector,  &
Large--area detector, \\
 & tracker & fast timing, high  & low radiation level, &
macropad and microvertex, \\
 &  & radiation level & slow readout  & high radiation environment 
\\[5mm] 
\hline \hline
%
\end{tabular}
}
\end{sidewaystable}

\newpage

\noindent
speed and minimize 
voxel thickness for a better geometry. However, large-area 
thin-crystal detectors are 
impractical for outer-tracker systems 
and delicate for microvertex systems; furthermore, they do 
not solve the other issues mentioned above.

There are three novel solid-state detector 
technologies that are likely to have a 
significant impact on their design at future collider 
experiments. 
A comparison of the different pixel sensor technologies is given 
in~Table~\ref{tab1}.

\subsubsection{3D sensors}

\begin{figure}[htbp] 
\begin{center}
\begin{tabular}{c c}
\epsfig{file=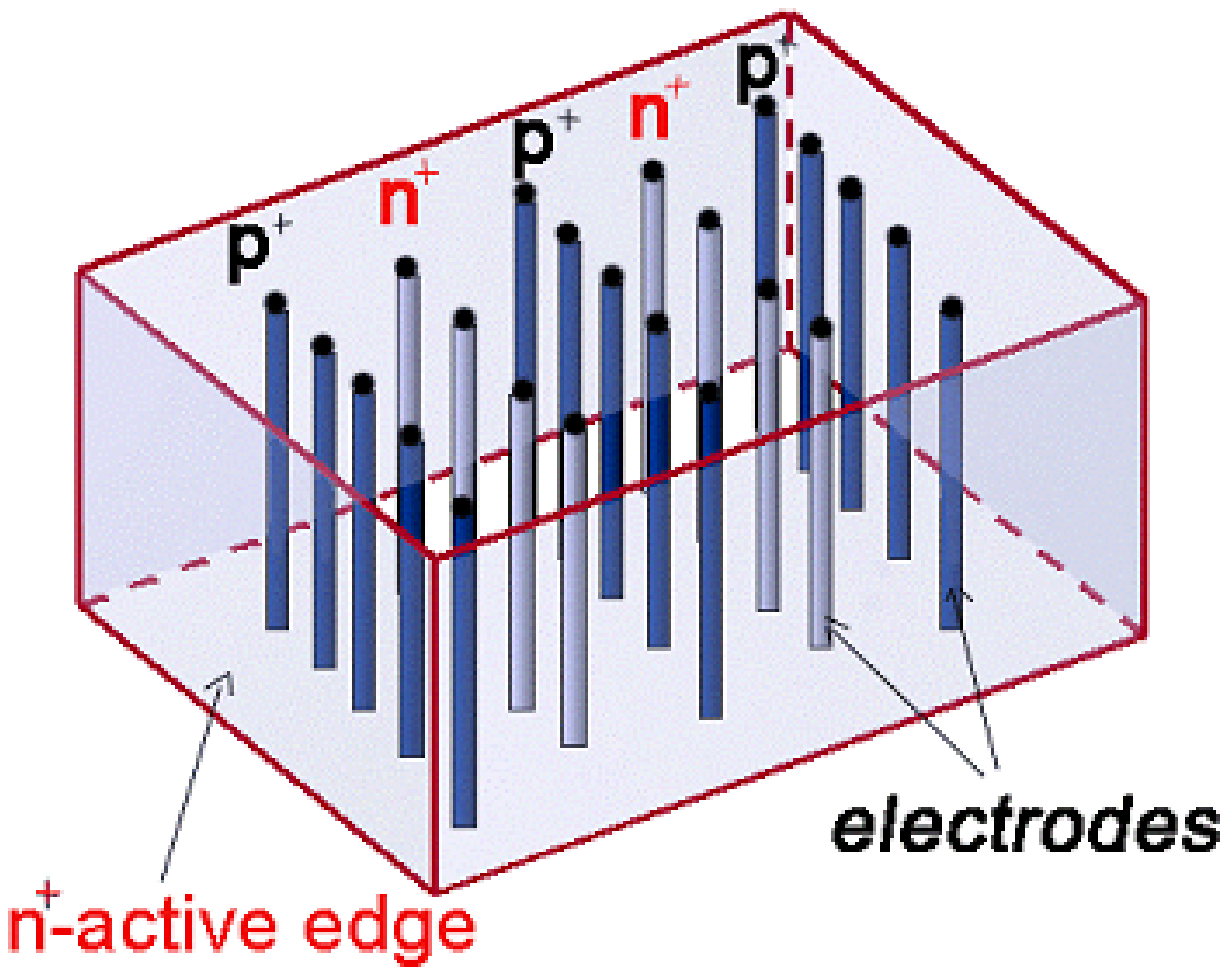,width=7cm} \hspace*{7mm}
\vspace*{3mm}\epsfig{file=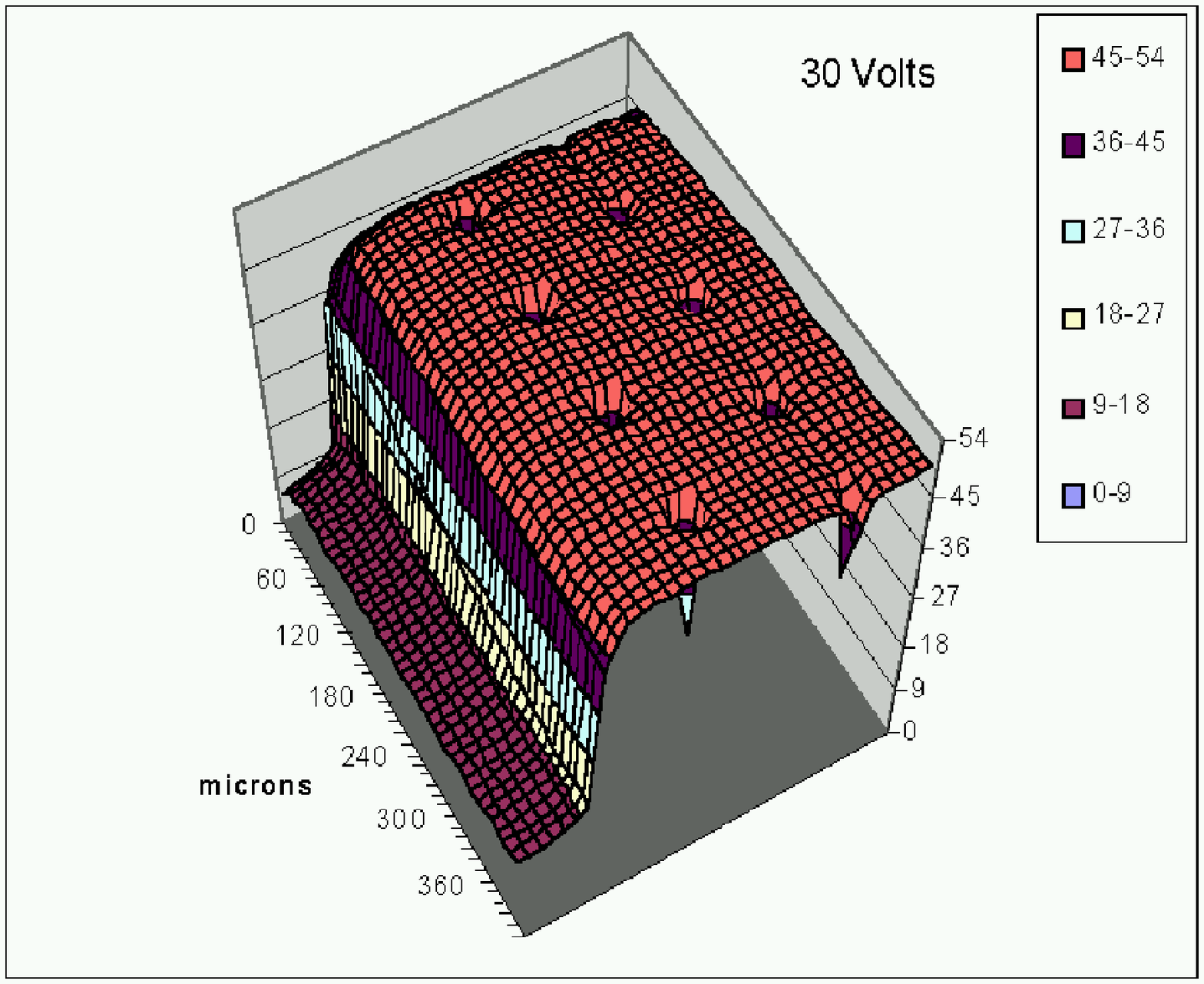,width=7cm} \\
\end{tabular}
\end{center}
\caption{Left: Sketch of a 3D detector where the p+ and n+ electrodes are
processed inside the silicon bulk. The edges are trench electrodes (active
edges) and surround the sides of the 3D device making the active volume
sensitive to a few microns from the physical edge.
Right: Scan of part of a 3D detector performed with a 12~keV  $X$--ray
beam. The edge signal turn-on is visible on the left. The vertical scale
represents the current pulse height in arbitrary units (also colour coded),
and the distance between electrodes is 100 microns.}
\label{3D1}
\end{figure}

3D detectors are a new generation of silicon sensors, which combine fast
readout signals, low depletion voltage and full sensitivity, independently
of the substrate thickness. This detector design was originally proposed
by S. Parker and C. Kenney~\cite{Parker:1997,Kenney:2001},
and is schematically represented in~Fig.~\ref{3D1} (left). 
The process involved in the fabrication of 3D devices is a
combination of traditional VLSI (Very Large System Integration) and Deep
Reactive Ion Etching (DRIE), which was developed for micromechanical
systems. DRIE allows microholes to be etched in silicon with a
thickness-to-diameter ratio as large as 20~:~1. 
In the 3D detectors
currently processed at Stanford, USA, by a collaboration involving
scientists from Brunel (UK), Hawaii, The Molecular Biology
Consortium (MBC), CERN, Stanford, ESRF and 
others, this technique is used
to etch holes as deep as several hundred microns, at distances as short as
50 microns from one another. These holes are  filled with
polycrystalline silicon doped with either Boron or Phosphorus, which is
then diffused into the surrounding single-crystal silicon to make the
detector electrodes. 
The silicon substrate used for this process can be either p-type or
n-type, and the crystal orientation is 
$<$~100~$>$, where 1,0,0 represent the
crystal plane coordinates. Silicon atoms line up in certain directions in
the crystal. $<$~100~$>$ corresponds 
to having a particular crystal plane at the
surface, and is preferred for a better surface quality. Once the
electrodes are filled, the polycrystalline silicon is removed from the
surfaces, and the dopant is diffused. The same process is then used to
fabricate `active edges', or all around trench electrodes doped with
either phosphorus or boron to properly complete the electric field lines.
The presence of active edges reduces the dead volume around the detector
to $<$~10~$\mu$m, as can be seen from the $X$-ray beam scan 
in~Fig.~\ref{3D1}~(right). 
From this figure, obtained using the 2~$\mu$m, 12-keV beam line at the
Advanced Light Source, Berkeley, it is possible to see the rapid turn
on response of the detector at the edge and the partial response of
the centre of the electrodes. 
%
Similar tests have been performed with
particle beams~\cite{totemtdr}. The
new etching machines and the new etching techniques that are becoming
available should allow one to reduce the electrode central part as much as
possible. When the etching process is completed, aluminium can be
deposited in a pattern that will depend on how the individual electrodes
are to be read out.

\begin{figure}[!t] 
\begin{center}
\begin{tabular}{c c}
\epsfig{file=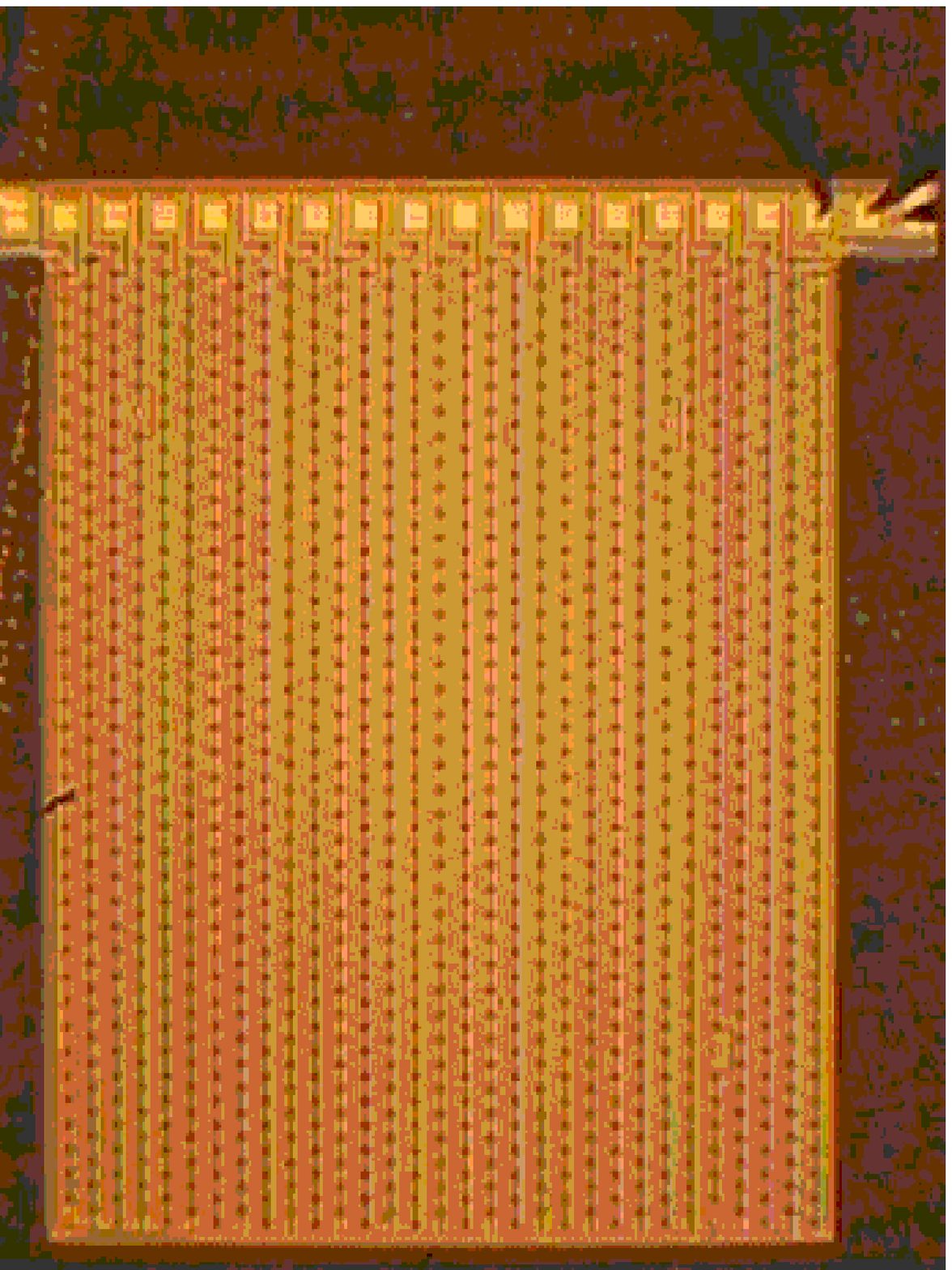,bbllx=0,bblly=0,
bburx=360,bbury=460,width=7cm} & \hspace*{5mm}
\epsfig{file=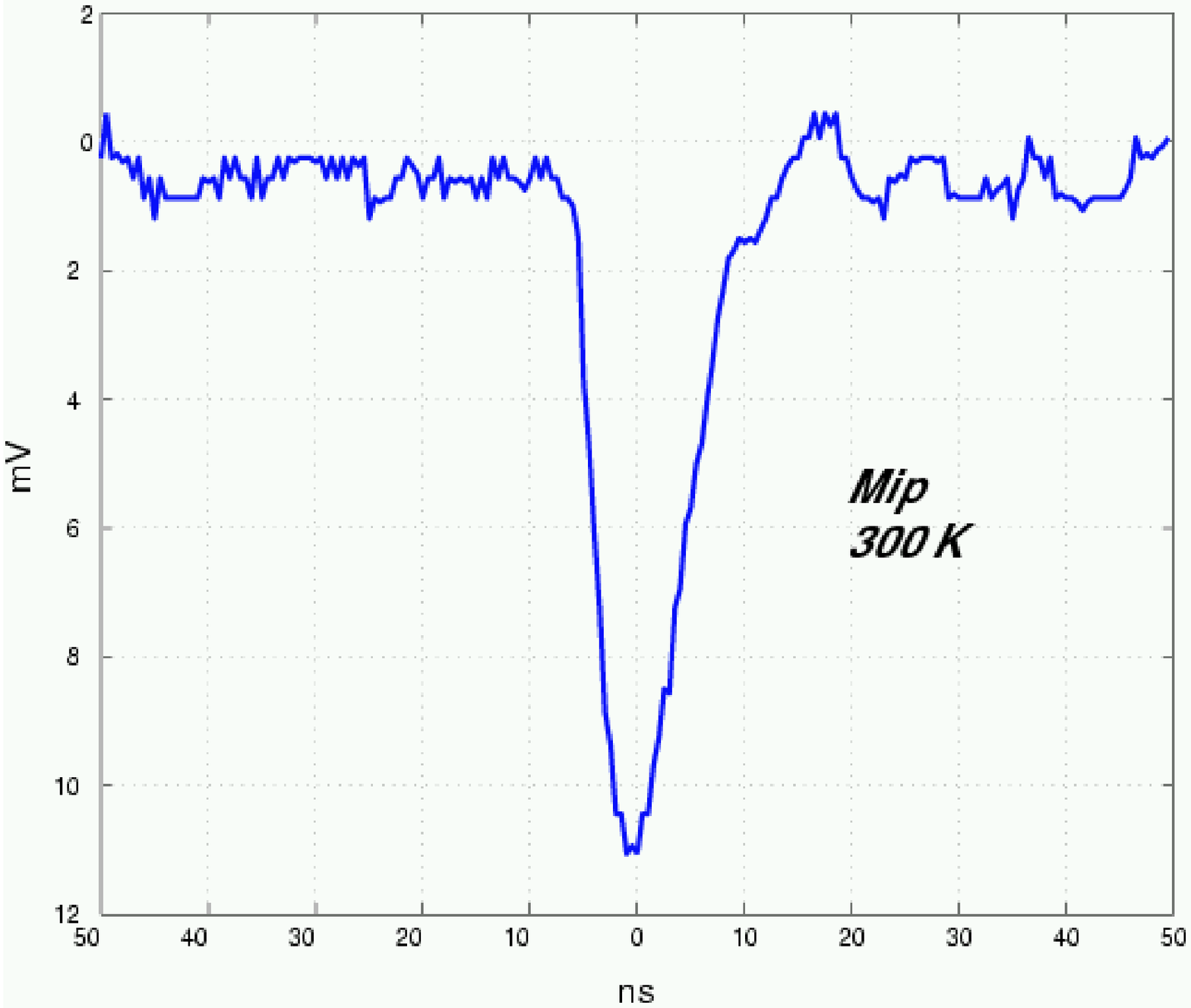,bbllx=0,bblly=0,
bburx=400,bbury=300,width=5.5cm} \\
\end{tabular}
\end{center}
\caption{Left: 
3D detector with microstrip readout configuration. The
electrodes and the aluminium strips, which tie the rows of 
${\rm p^+}$ and ${\rm n^+}$ electrodes, are clearly visible. 
The ${\rm p^+}$ strips end at the bonding pads (top). Right:
Oscilloscope traces of mips of 3D detectors at room temperature.
The rise time is 3.5 ns and is dominated by the readout electronics. The
electrode spacing is 100 microns.}
\label{3D2}
\end{figure}

A 3D detector, where all the electrodes have been connected together by an
aluminium microstrip, is shown in~Fig.~\ref{3D2}~(left). 
The oscilloscope trace of its
response to a minimum ionising particle is shown in~Fig.~\ref{3D2}
(right)~\cite{Davia:2003}. The fast
radiation-hard electronics used for this test, designed by the CERN
microelectronics group dominates the rise time of the pulse, which was
measured to be 3.5 ns and did not degrade 
after 1~$\times$~10$^{15}$ protons/cm$^2$. The
same device was tested at 130~K. At this temperature the pulse rise timed
improved to 1.5~ns while the full pulse width was measured to be less than
5~ns. The inter electrode spacing of these detectors was 100~mm. Devices
with shorter spacing have been processed and should have an intrinsic
charge collection time of $<$~1~ns~\cite{Parker:1997}. 
Tests with such devices and 0.13
microns electronics readout are in preparation.

\subsubsection{Monolithic Si sensors: MAPS}

The integration of detector and electronics is particularly crucial in 
high-resolution silicon pixel detectors for microvertex applications, where a 
very high density of pixel elements has to be achieved. The monolithic 
approach~\cite{Deputch:2002,Dulinski:2001} is based on Active Pixel Sensor 
(APS) technology and utilizes the epitaxial layer of the CMOS wafer as
detector substrate. 
The Monolithic Active Pixel Sensors (MAPS) detectors potentially 
provide unambiguous two-dimensional tracking, where the detecting element is 
inseparable from the readout electronics. In this approach, both the detector 
and the front-end readout electronics are integrated on the same 
silicon wafer, 
using standard CMOS fabrication processes. The development of MAPS detectors 
started with their use as photon detectors, in the visible band, 
where they are becoming increasingly 
popular at the expence of Charge Coupled Devices (CCDs). 
MAPS devices are detectors with a small-area readout section, yielding good 
noise performance. However, the charge collected originates from a thin 
active volume. In this approach, a thin, moderately doped, undepleted Si 
layer is used as radiation-sensitive volume, while the readout  
electronics is implemented on top of it. 
This particular structure makes the entire 
sensor surface sensitive (100{\%} fill factor), as required for 
particle tracking. The sensitive volume is usually an epitaxial layer, which 
is available in modern CMOS processes, where it is grown on a highly, usually 
${\rm p^{++}}$-type, doped substrate. 
The pixel readout electronics is placed in the ${\rm p}$-well. 
Because of the difference in doping levels (about 10$^3$), the epitaxial layer 
junctions with the ${\rm p}$-well and the substrate 
acts as reflective barriers, confining the charge carriers. 

The electron-hole collection is performed by thermal diffusion 
between ${\rm n}$-well 
electrodes. The charge-collection is then spread over several 
${\rm n}$-well collecting 
electrodes. But MAPS devices have a low tolerance to radiation 
effects. Ionizing 
radiation induces surface effects that deteriorate charge collection, and 
displacement damage decreases the minority carrier lifetime, thus rapidly 
deteriorating the charge collection efficiency. 
Only NMOS transistors can be used 
in pixel circuits, which represents an important limitation 
for the circuit design of the pixel cell.

\subsubsection{Hydrogenated amorphous Si on ASIC Technology}

A seed R{\&}D on the technology of deposition of hydrogenated amorphous 
silicon n.i.p layer on ASIC~\cite{Wyrsch:2003,Jarron:2003} has 
shown that this detector technology has a great potential to solve present 
limitations of standard Si crystal detector technology. 

\begin{figure}[t]
\begin{center}
\begin{tabular}{c c}
\epsfig{file=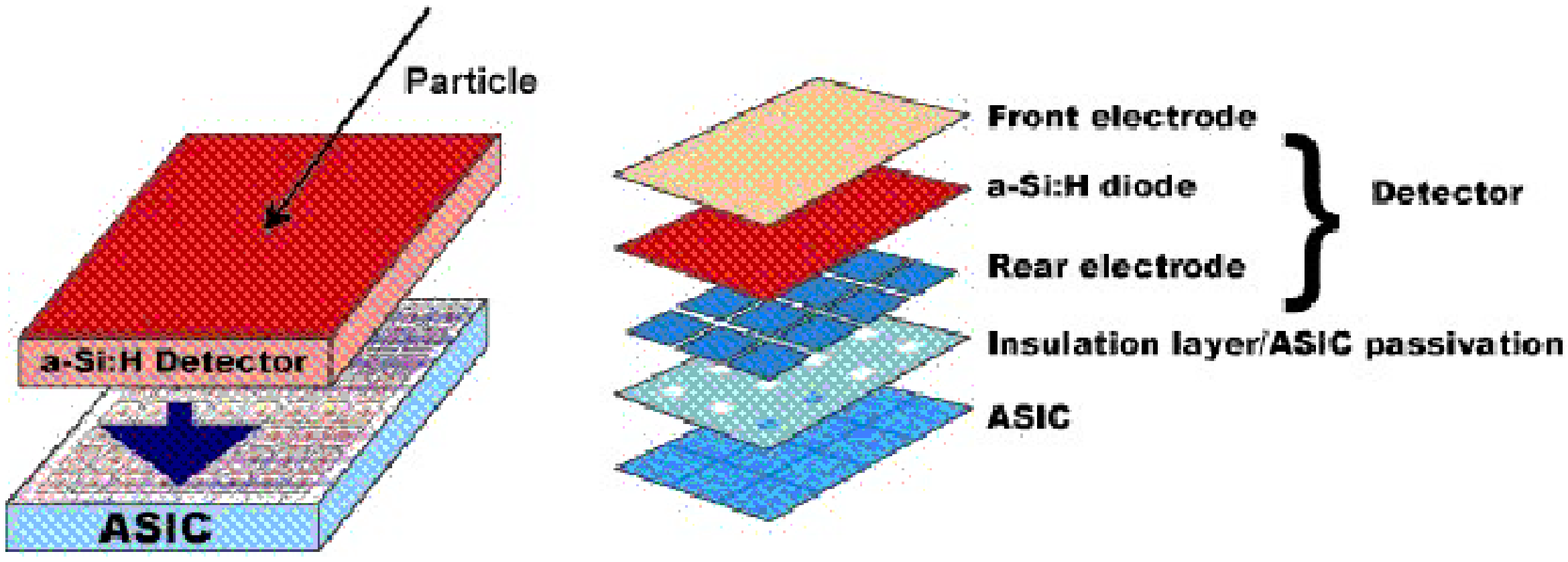,width=7.5cm,height=5.5cm} &
\epsfig{file=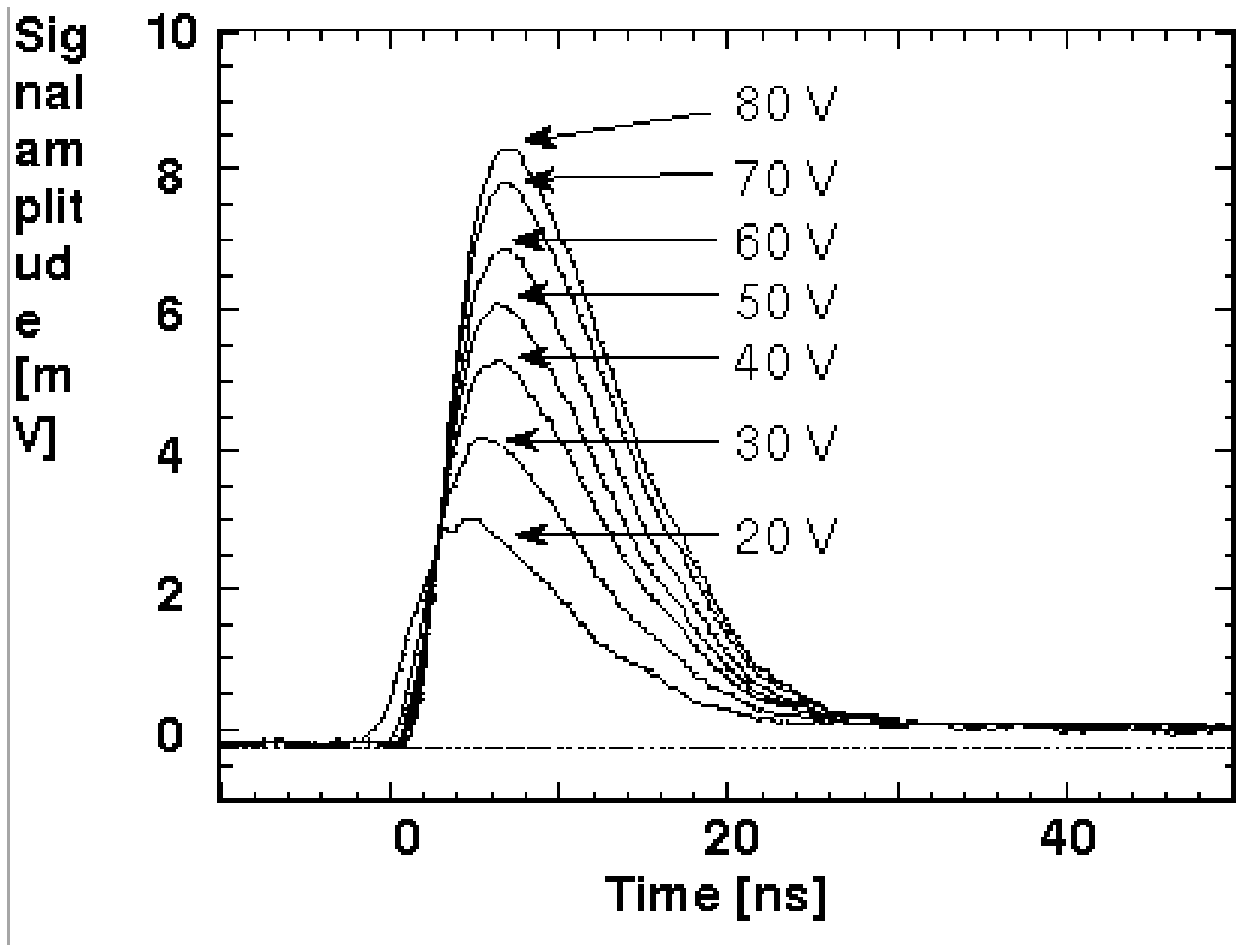,width=7.0cm,height=5.0cm} \\
\end{tabular}
\end{center}
\caption{Amorphous silicon detector. Left: a schematic of a hydrogenated 
amorphous Si detector on ASIC. Right: The time behaviour of the 
signal for different 
polarization voltages showing the intrinsically fast 
charge-collection properties.}
\label{fig:asi}
\end{figure}

One of the main advantages of such techniques 
is the potential cost reduction and the 
sensor integration with the readout electronics, 
as in pixel monolithic detectors. 
Compared with standard Si hybrid pixels, the deposition of a-Si:H on
ASIC greatly simplifies the pixel detector fabrication 
(see~Fig.~\ref{fig:asi}). 
The problem of interconnections to the electronics is also
solved. Another advantage  
is that a-Si:H semiconductor is extremely radiation-hard, 
thanks to a self-annealing 
process operated by mobile hydrogen (15\%) within the amorphous
tissue, and defects are 
continuously passivated. 
The intrinsic limit of radiation hardness is expected to be 
above 10$^{15}$~ptcs~cm$^{-2}$. 
An a-Si:H detector can also operate in a wide temperature 
range, from room temperature to up to $\simeq$~60$^\circ$~C, 
without affecting the charge 
collection efficiency. This feature represents a great 
simplification for the cooling 
systems, which are at present a bottleneck in the system 
design of the LHC trackers.

Preliminary results have demonstrated that charge collection 
is quite fast: 3~ns to 
5~ns for electron carriers (see Fig.~\ref{fig:asi}). 
High-energy electrons of~~$^{63}$Ni and $^{90}$Sr have been measured
with a short peaking time of 15~ns.  
However, a substantial R{\&}D effort is essential 
to master the deposition technology 
of a-Si:H on ASIC. Several technological issues have
to be mastered, in order to obtain 
an excellent a-Si:H, in particular a low dangling bond 
defect density, with a high 
deposition rate of 1 to 2~nm~s$^{-1}$. 
The need for a low defect density is essential to deplete 
a thick intrinsic layer, 50~$\mu$m with reasonable bias voltages. 

\subsubsection{Readout electronics, microelectronics and design effort}

The development of microelectronics will continue its pace towards nanoscaled 
transistor feature size. The particle physics community should monitor and 
understand how to profit from these technological advances, which have been 
a crucial basis for developing novel solid-state detector and 
radiation-hardened electronics for LHC applications. 
Low-power pixel readout architecture should be further developed in
adapting circuit design techniques to very low supply voltages and nanoscale
MOS gate~length.

\subsubsection{R{\&}D directions for Si sensors and associated electronics}

Emerging solid-state detector technologies need to be supported by an 
adequate R{\&}D effort. There are largely complementary: 
3-D detectors offer the 
fastest possible pixel response, pad, or strip detector 
with an excellent radiation 
hardness, the a-Si:H detector can be used to build 
large-area detector at low cost, 
and monolithic MAPS sensors offer a low material budget 
option for vertex trackers.

The 3-D detector technology is already developed at Stanford as a
small sensor,  
but a future fabrication of large area should be demonstrated, and 
transferred to industry. A study of its radiation hardness 
should be made to determine its maximum tolerance.
However, for detectors such as this one
a hybridization technique like bump bonding should 
still be used to form an hybrid detector module.

Fast readout techniques should be adapted to monolithic pixel MAPS, and MAPS 
operation should be demonstrated at wafer scales with a readout scheme 
compatible with the vertex tracker requirements 
of future collider experiments. 
Limitations in radiation hardness should be clearly assessed. The development 
design effort should be concentrated on three main axes. At the system level, 
possible implementation of on-chip, real-time hit-reconstruction readout 
architectures and data-transfer strategies are investigated. At the pixel 
level, the development should address the design of pixel readout circuitry 
allowing on-pixel integration of an amplifier and circuitry allowing double 
sampling operation leading to the effective zero suppression. Another issue
is the optimization of the charge-sensitive element oriented on the specific 
requirements and conditions met for a given tracking application, e.g. vertex 
detectors. This includes recently emerged designs with auto-reverse 
polarization of a charge-collecting diode and a new solution for  
charge-sensitive elements, featuring built-in signal amplification. A
conceptual design of this new element, realizing charge-to-current
conversion, called PhotoFET,  
is a charge-sensing element, in which ionization-generated 
charge carriers collected from the epitaxial layer directly 
modulate the channel current of a PMOS transistor implemented on top of an 
n-well. 

Sensors of the a-Si:H type have a significant potential, in particular 
for low-cost large-area 
detectors. However, the technology of 
amorphous silicon, though extremely well developed for photovoltaic solar 
panels and TFT screens, is still at an early stage in particle physics 
applications 
and needs a vigorous R{\&}D effort. In particular, the know-how to grow  
high-quality thick films of hydrogenated amorphous Si at high deposition 
rate needs to be acquired. 
This comes from the need of depositing a thick 
intrinsic a-Si:H 
layer, which is not usual for mainstream industrial applications. 
Also a good film 
quality, ensuring a maximum depletion layer thickness 
for good sensor sensitivity, 
crucially depends of the deposition of a film with a low defect density, a 
requirement that is not so stringent for other applications. 
Deposition of a-Si:H film on ASIC demands also the development
 of adapted lithography and planarization 
process. To demonstrate the full potential of the a-Si:H TFA technology it is 
mandatory to prove that good sensors can be manufactured 
on 8- or 12-inch wafers. 
This will require a substantial effort in circuit design 
and an improvement in the a-Si:H film deposition technology.

\subsection{\boldmath{$b$}-Tagging}

The physics studies at the multi-TeV frontier are expected 
to still largely rely on the detector ability to identify the flavour
of final-state fermions with high efficiency and purity.

A high resolution vertex tracker, located immediately outside the 
beam pipe, should
provide the track position and extrapolation accuracy needed to 
distinguish particles 
originated from secondary decays of short-lived hadrons, 
from those produced indirectly at the position of the colliding beams.

\begin{figure}[!h]
\begin{center}
\begin{tabular}{c c}
\epsfig{file=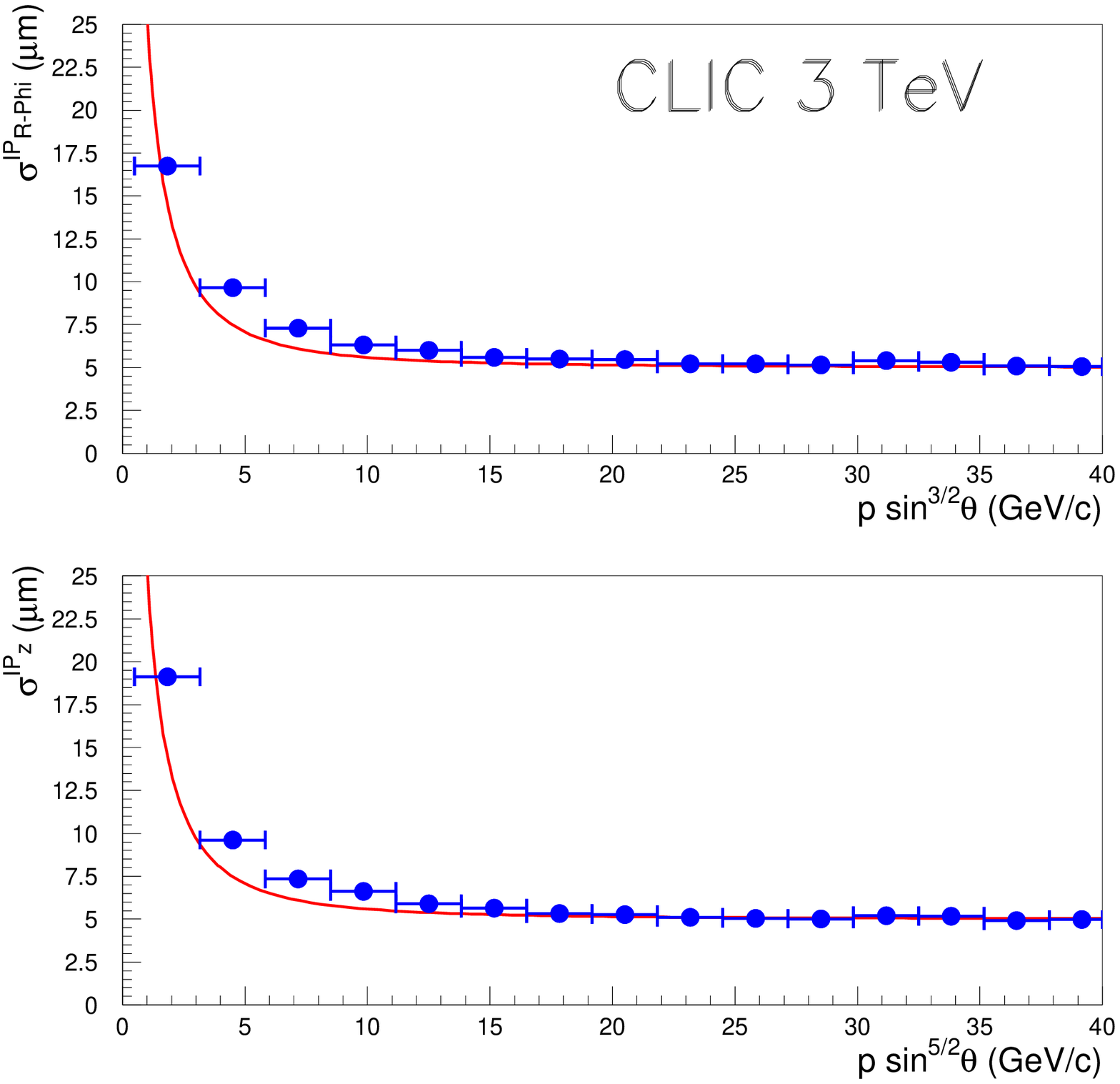,width=7cm,clip} & \hspace*{5mm}
\epsfig{file=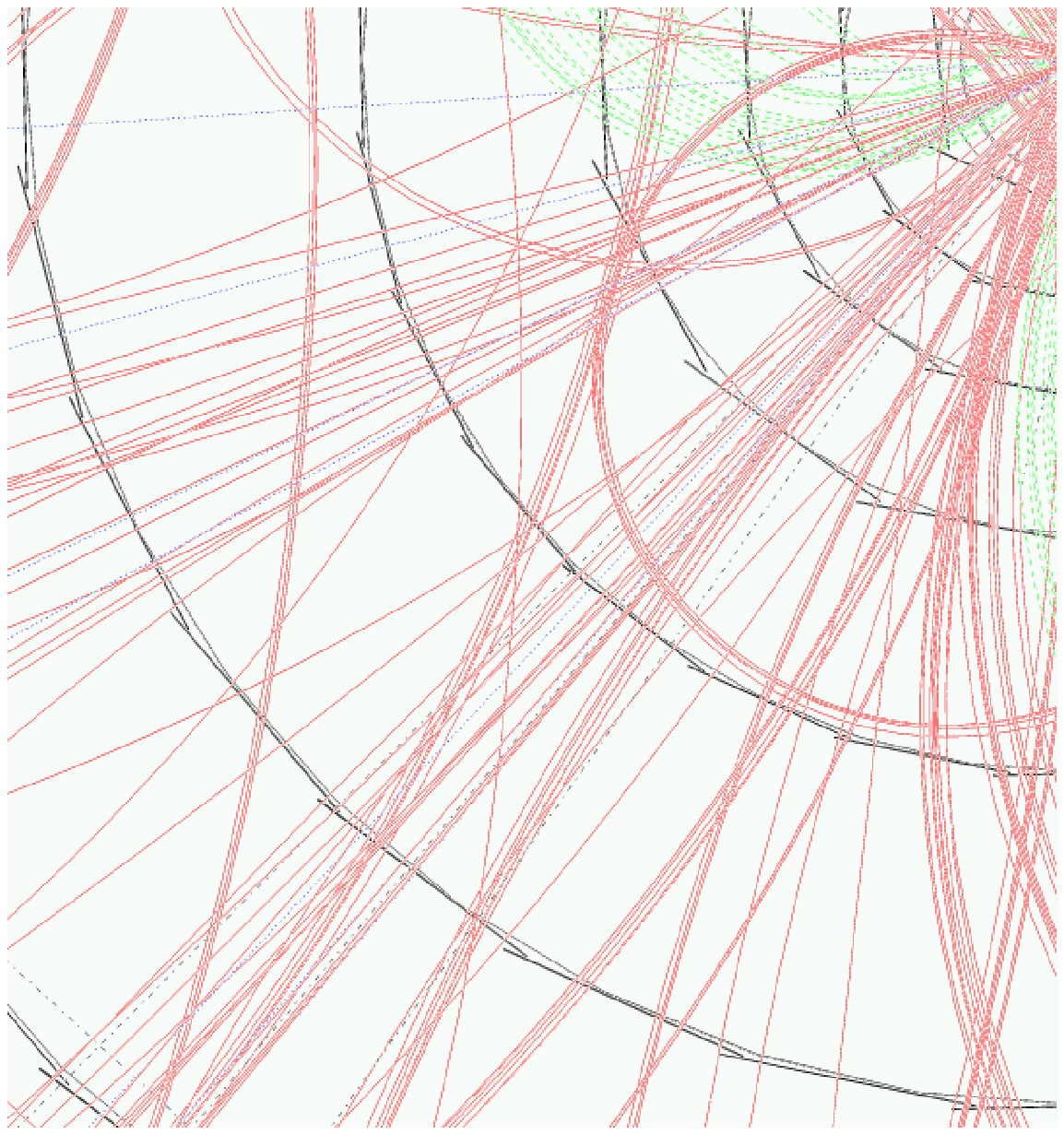,width=6.25cm,clip} \\
\end{tabular}


\caption{Track extrapolation resolution at
CLIC. Left: Impact parameter accuracy as function of 
$p_t$ for the $R$--$\phi$ (upper plot) and $R$--$z$ 
(lower plot) projections. Right: Display of a multijet event simulated with
GEANT3 showing secondary decays in~$b$~jets.}
\label{fig:ipres}
\end{center}
\end{figure}

At CLIC we envisage to propose a multilayered vertex tracker based on 
pixel Si sensors. The density of hits from the machine-induced backgrounds 
will limit the approach to the interaction region 
to a radius of about 3~cm. Assuming a 
rather modest single-point resolution of 5~$\mu$m, 
the achievable resolution on the 
point of closest approach to the colliding beam spot 
(impact parameter) is shown in~Fig.~\ref{fig:ipres}. 
Since the boost of the 
short-lived particles is large, this 
performance appears to be sufficient for effectively 
discriminating the secondary 
from the primary particles. Also, multiple-scattering 
effects are less severe here than at 
lower-energy colliders, since typically only 
10\% or less of the $b$ and $c$ decay
products have momenta below 3~GeV.

\vspace*{-4mm}

Pixel sensors of different technologies can be considered, from CCDs to 
the monolithic pixel sensors of new design currently under
development.  
Potentially interesting technologies and an outline of the needed R\&D are 
discussed later in this section.

Jet-flavour tagging, based on the combination of particle topology and 
kinematics, has succesfully been applied to experiments at
LEP and the SLC. 
The kinematics in multi-TeV $e^+e^-$ collisions suggests that the extensions 
of the reconstruction and tagging algorithms, pioneered at LEP and the
SLC and 
further developed for application at a lower-energy LC, may need to be 
reconsidered. The high jet collimation and large $b$ decay distance pose 
significant challenges to the track pattern recognition and reconstruction that
may affect the accuracy of $b$ and $c$ 
identification, relying only on secondary vertex 
and impact parameter tagging. It is thus interesting 
to consider new quark tagging 
techniques, which profit from the kinematics of multi-TeV $e^+e^-$ collisions.
A $b$-tagging algorithm based on the tag of the steps in particle multiplicity
originating from the heavy hadron decay along its 
flight path echoes a technique 
developed for charm photoproduction experiments~\cite{framm}. 
At CLIC, the signal of 
the production and decay of a $b$ or $c$ hadron can be 
obtained from an analysis of the 
number of hits recorded within a cone centred on the jet 
direction as a function of the 
radial position of the detector layer. In the tested implementation of the 
algorithm, a cone of half-aperture angle $\psi$~=~40~mrad, optimized to 
maximize the sensitivity in the presence of background and fragmentation 
particles, defines the area of interest (AoI) on each detector layer. 
The decay of a highly boosted 
short-lived hadron is characterized by a step in the number of hits recorded
in each AoI, corresponding to the additional charged multiplicity generated 
by the decay products. 

Since the average charged decay multiplicity of a 
beauty hadron is about 5.2 and that of a charm hadron 
2.3, the $B$-decay signature can consist of either one or two steps, 
depending on whether the charm decay length exceeds the vertex tracker layer 
spacing. 

The number of background hits in the AoI is estimated to be 0.7, 
constant with the radius, the increase of the AoI surface being compensated by 
the background track density reduction due to the detector solenoidal field.
This background density can be monitored, by sampling the region outside 
the AoI, 
and subtracted. The effect of significant fluctuations of the number of these 
background hits and of low-momentum curling tracks in the innermost layers can
also be removed. The typical multiplicity pattern of a tagged 
$B$ decay is shown in~Fig.~\ref{fig:event}. 

\begin{figure}[htbp] 
\begin{center}
\begin{tabular}{c c}
\epsfig{file=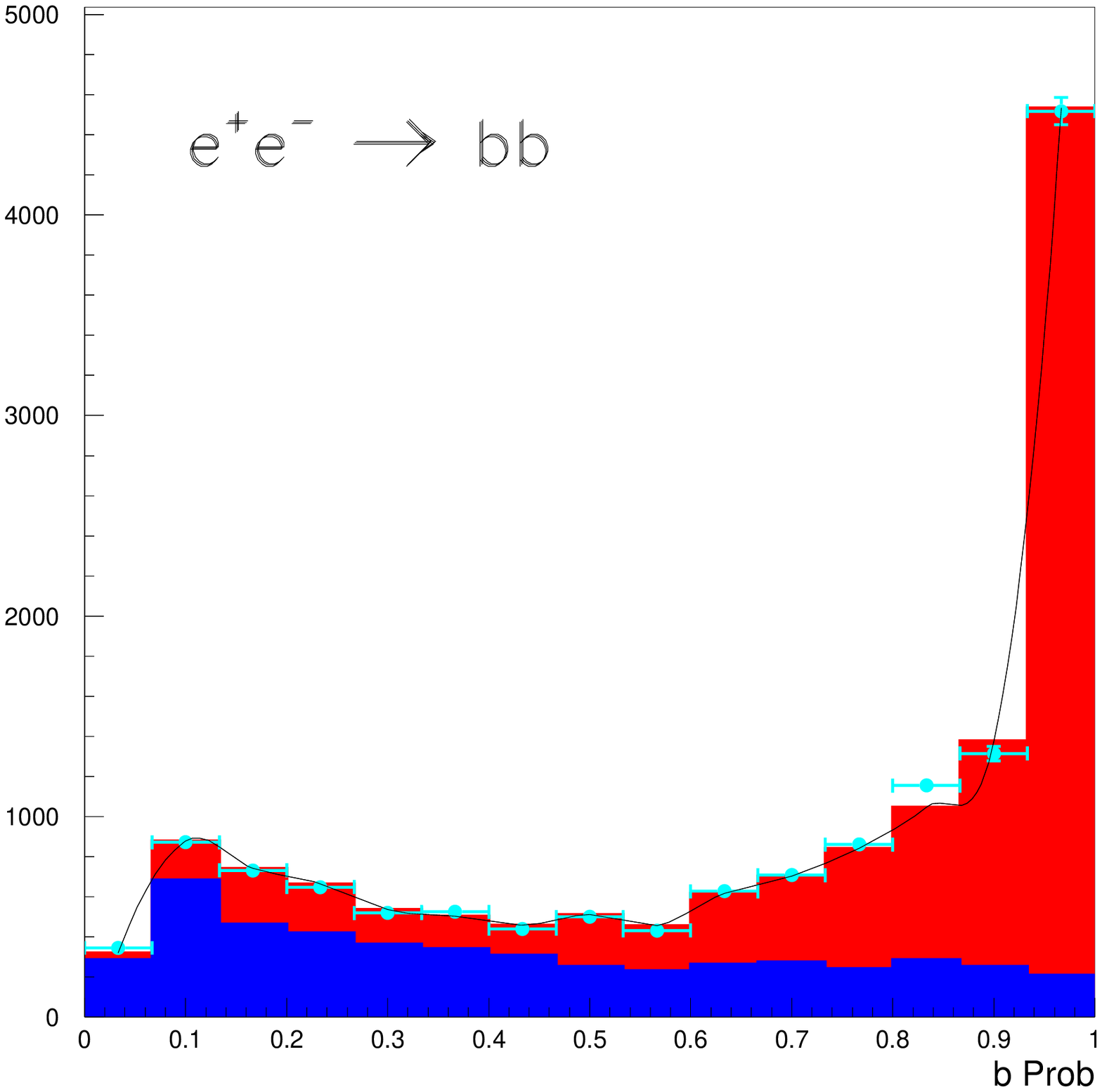,width=7.8cm,height=5.5cm} & 
\epsfig{file=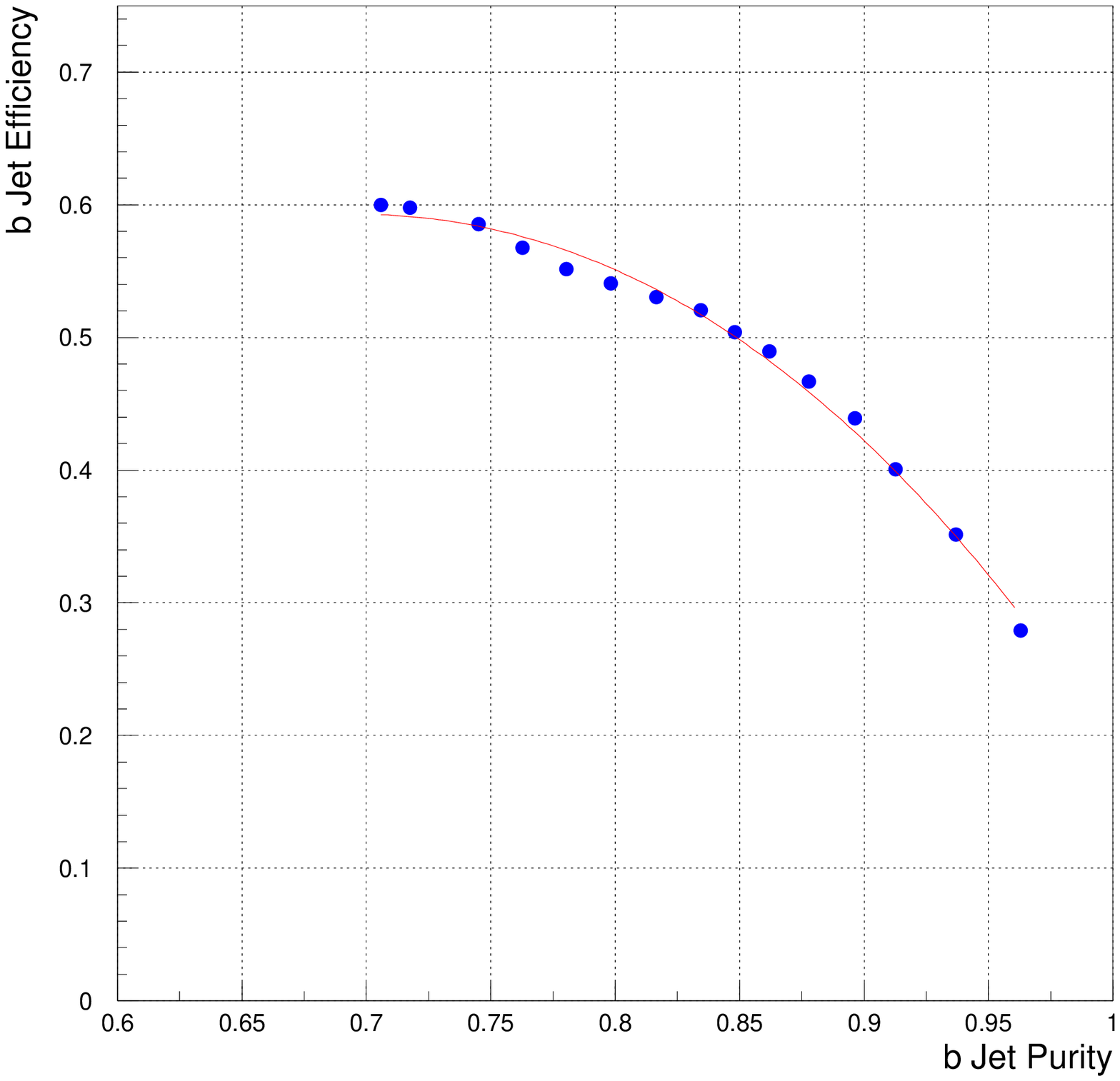,width=7.8cm,height=5.5cm}\\
\end{tabular}
\caption{The $b$ likelihood for jets in
$e^+e^- \to q \bar q$ events at $\sqrt{s}$~=~3~TeV (left) with a
multiplicity tag. The response for $b$ ($c$ and lighter) jets is shown in
light (dark) grey. The $b$-jet efficiency is given as a function of the purity
corresponding to different likelihood cut values (right).}
\label{fig:tag} 
\end{center}
\end{figure}

Jets with an upward multiplicity step at least larger 
than 1 and a total multiplicity 
increase larger than 2 have been considered. 
According to the simulation, 69\%, 29\% and
3\% of these jets are due to $b$, $c$ and light quarks respectively. The 
$b$ jets can be further discriminated using a $b$-likelihood based on the 
size, radial position and number of multiplicity steps, the fraction of the
jet energy and the invariant mass of the tracks originating at the detected 
multiplicity steps. The resulting likelihood for $b$ and lighter jets and the
$b$ efficiency and purity resulting from a cut on this discriminating variable
are shown in Fig.~\ref{fig:tag}. 

\begin{table}[!h] 
\caption{Average decay distance in space for $B$ hadrons 
at different $\sqrt{s}$} 
\label{table:bbdec}

\renewcommand{\arraystretch}{1.4} 
\begin{center}

\begin{tabular}{lccccc}\hline \hline \\[-4mm]
\boldmath{$\sqrt{s}$}~\textbf{(TeV)} $\hspace*{9mm}$ &
$\hspace*{3mm}$ 0.09 $\hspace*{3mm}$  & 
$\hspace*{3mm}$ 0.2 $\hspace*{3mm}$ & 
$\hspace*{3mm}$ 0.35 $\hspace*{3mm}$ & 
$\hspace*{3mm}$ 0.5 $\hspace*{3mm}$ & 
$\hspace*{10mm}$ 3.0 $\hspace*{3mm}$ \\ 
\textbf{Process} $\hspace*{6mm}$ &
$Z^0$ & $HZ$ & $HZ$ & $HZ$ & 
$\hspace*{1mm}$ {$H^+H^-$} $\hspace*{1mm}$ $|$ 
$\hspace*{1mm}$ $b \bar b$ $\hspace*{1mm}$ \\
\boldmath\textbf{$d_{\rm space}$~(cm)} & 
0.3 & 0.3 & 0.7 & 0.85 & 
$\hspace*{10mm}$ 2.5 $\hspace*{1mm}$ $|$ 
$\hspace*{1mm}$ 9.0 $\hspace*{1mm}$ 
\\[3mm] 
\hline \hline
\end{tabular}
\end{center}
\end{table}



\subsection{Main Tracker}

\begin{figure}[t] 
\begin{center}
\begin{tabular}{c c}
\epsfig{file=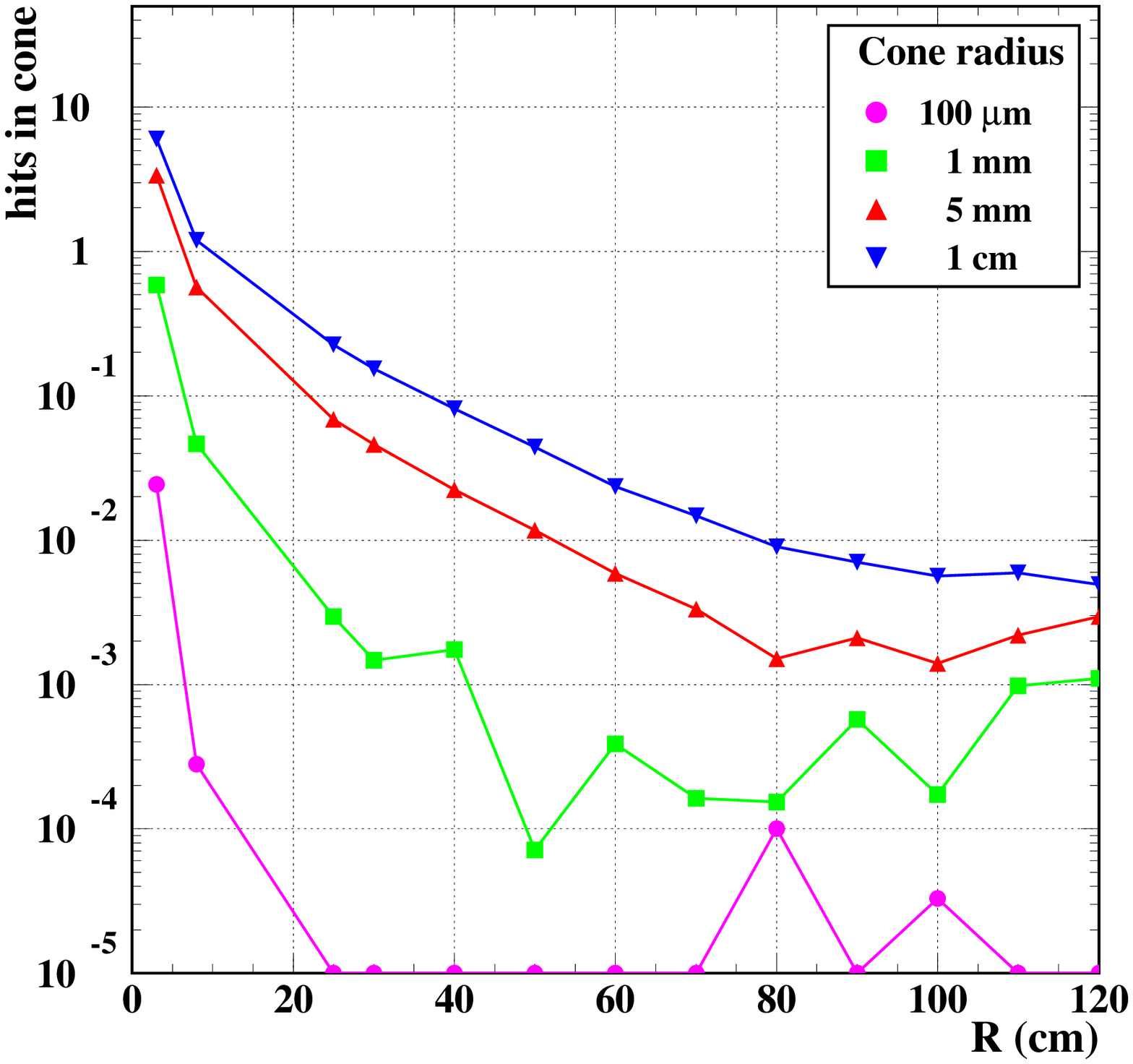,width=7.0cm} &
\epsfig{file=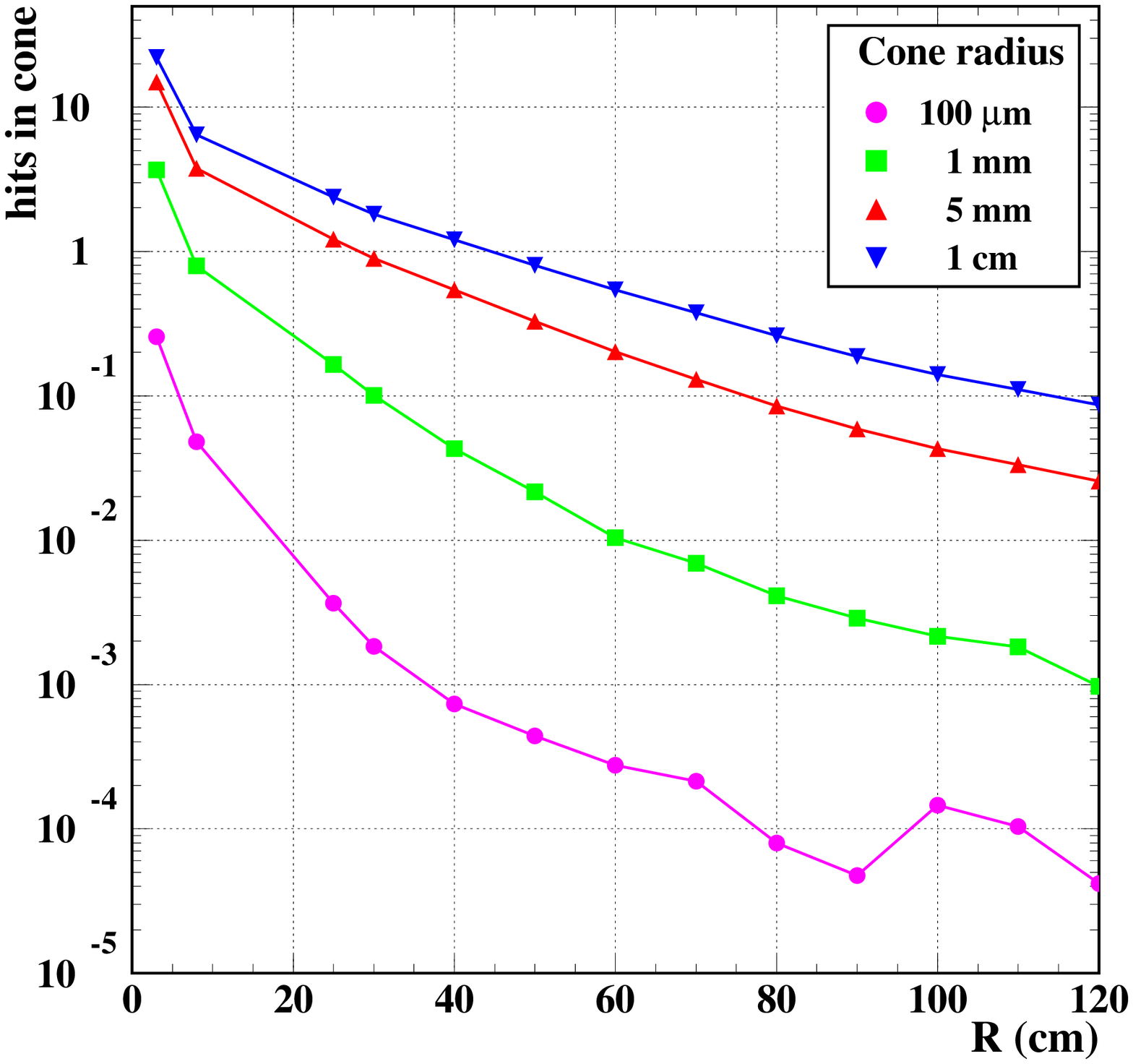,width=7.0cm} \\
\end{tabular}
\end{center}
\caption{Average number of additional hits within a cone of a given radius,
as a function of the radius of the tracking layer. Left: $\sqrt{s}~=~500$~GeV,
$B~=~4$~T; Right: $\sqrt{s}~=~3$~TeV, $B~=~6$~T.}
\label{fig:trkoccup}
\end{figure}

Possible concepts for the central tracking device are similar to those 
studied for the 500~GeV LC, a small silicon tracker surrounded 
by a large TPC, and for CMS at the 
LHC, a compact full Si tracker in an intense magnetic field. A TPC would 
offer some advantages by providing continuous hit information along 
the charged-particle 
trajectory, a large lever arm with a minimum material 
burden, and $dE/dx$ specific 
ionization information, useful for particle identification. However, local 
occupancies and two-track separation may represent serious 
limitations. The total 
occupancy of a TPC with a inner radius of 50~cm, which integrates 
a full train due to 
its maximum drift time of $\simeq $~50~$\mu$s is estimated 
to be a few per cent. However, 
the local track density in hadronic jets is extremely high 
due to the large boost. 
Figure~\ref{fig:trkoccup} shows the result of a GEANT3 simulation of 
$e^+e^- \to b \bar{b}$ events at $\sqrt{s}$~=~0.5~TeV and 3~TeV with 
a solenoidal filed 
of 4~T and 6~T, respectively. 

The average number of additional hits contained within a cone of given 
radius around a track hit is plotted as a function of the radial 
position from the beam line. In the case of 0.5~TeV centre-of-mass
energy, the probability to find additional hits within 1~cm distance
at a radius of 40~cm is below 10\%. However, in the case of multi-TeV
collisions, the jet collimation becomes so pronounced that for  
1~cm distance a 10\% probability for an additional hit is only
reached at radii exceeding 1~m. Hence, a solid-state tracker with its
intrinsic two-track resolution of 100~$\mu$m seems to be an
advantageous option for quality tracking at CLIC. 
The CMS experiment at the LHC will deploy an all-Si tracker with an active
surface of 230~m$^2$. While the physics at the LHC and at a 500~GeV LC
impose clear requirements on momentum resolution, from processes such
as $H \to ZZ \to 4\mu$ and $HZ \to X \mu^+\mu^-$, at multi-TeV
energies there are no obvious benchmark reaction  
striving for ultimate momentum resolution. A study of the process 
$e^+e^- \to \tilde{\mu}\tilde{\mu} \to \mu^+\mu^- \chi_0 \chi_0$,
where the $\tilde{\mu}$ smuon  
mass is determined from the edges of the muon momentum spectrum, has
shown that a momentum resolution comparable to that aimed at for the
500~GeV LC, i.e. 
$\delta p_{\rm T}/p_{\rm T}^2\le$~5~$\times$~10$^{-5}$~GeV$^{-1}$, 
seems adequate.
 
\begin{figure}[t]
\begin{center}
\begin{tabular}{c c}
\epsfig{file=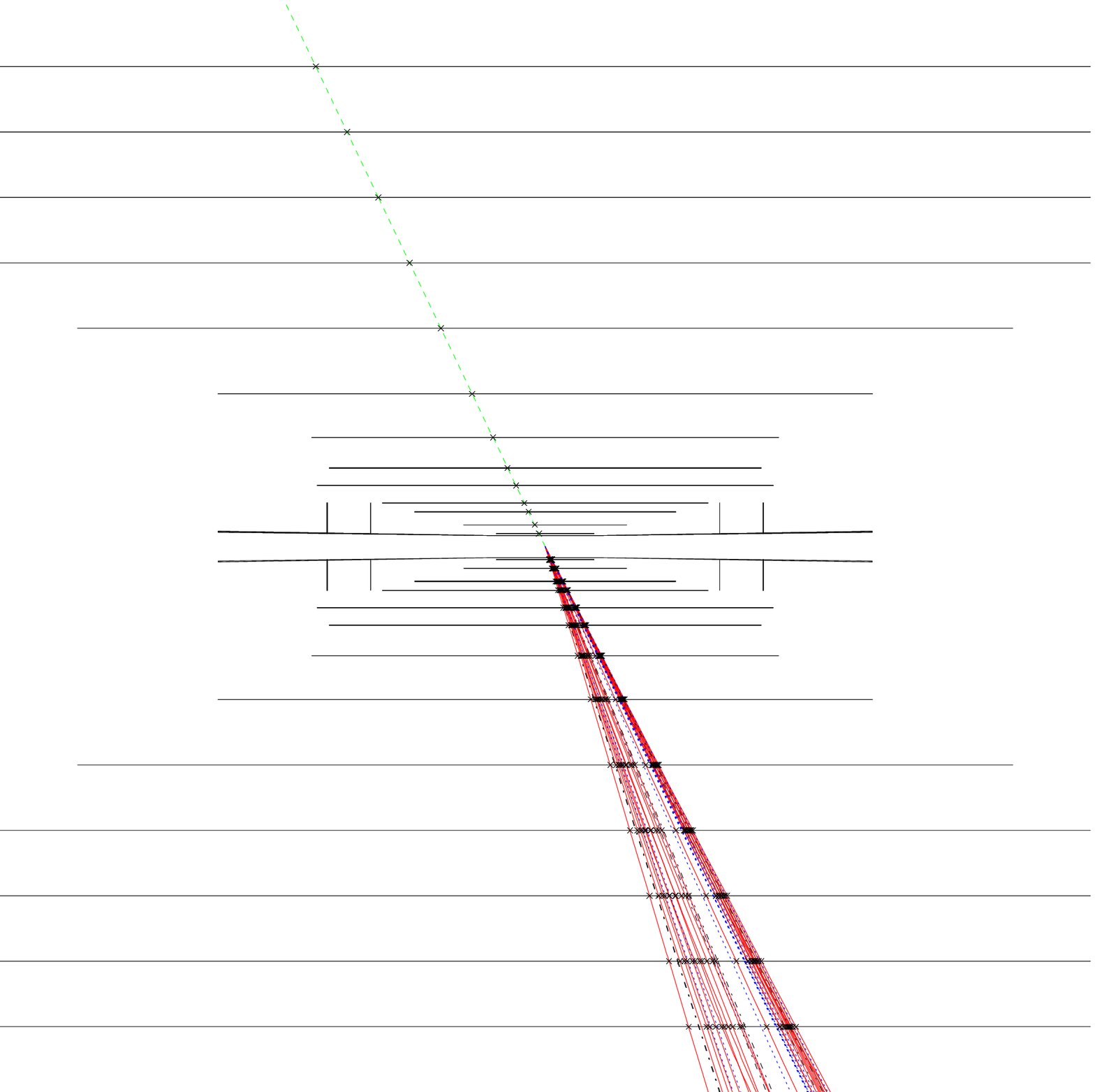,width=7.0cm} &
\epsfig{file=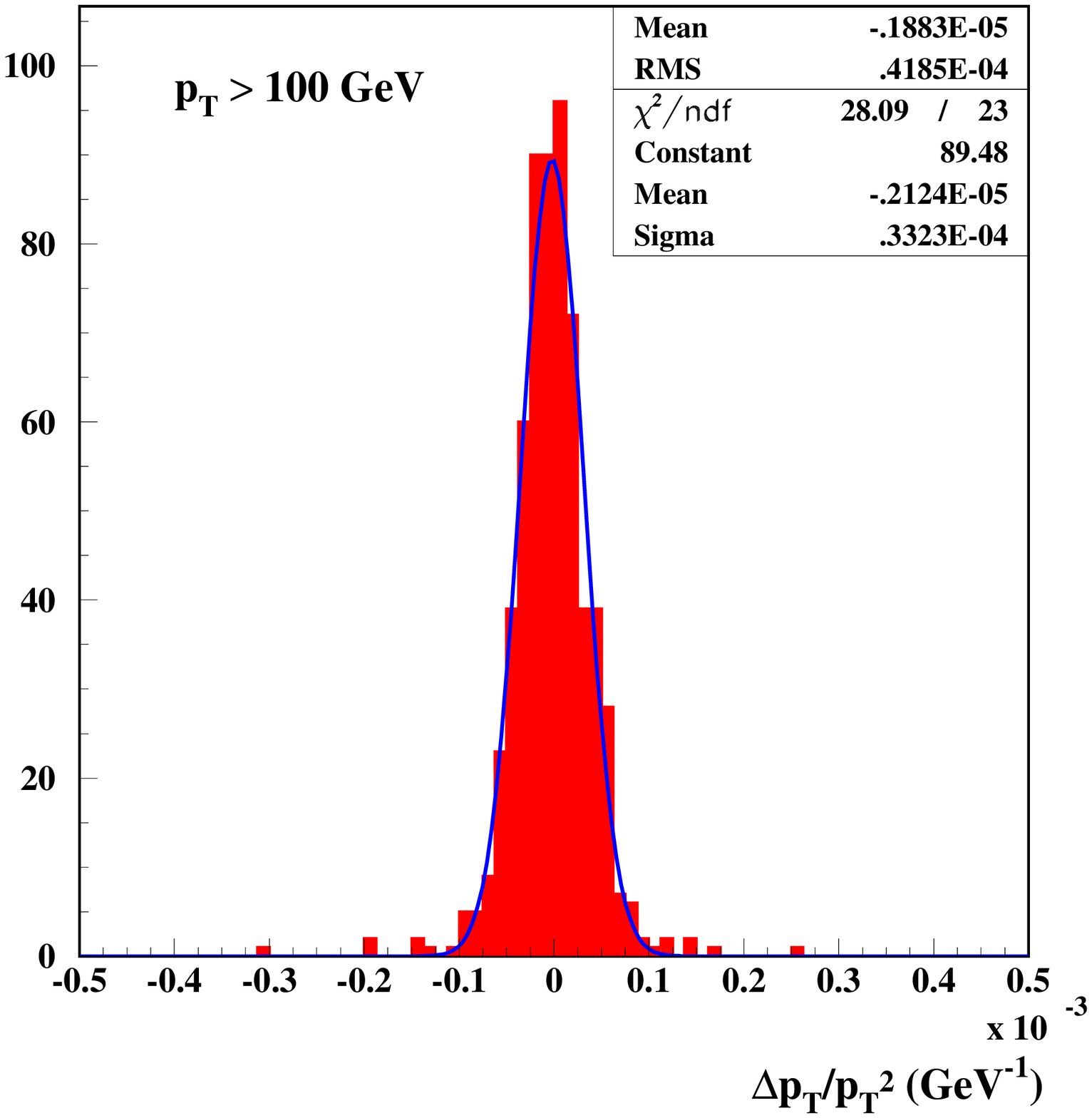,width=7.0cm} \\
\end{tabular}
\end{center}
\caption{The proposed Si tracker for the CLIC detector. Left: $R$--$z$
view of detector layout with a simulated event 
$e^+ e^- \to W^+ W^-$ at $\sqrt{s}$~=~3~TeV. Right: Momentum
resolution for tracks with $p_T >$~100~GeV obtained using the full
GEANT3 simulation.} 
\label{fig:dpresol}
\end{figure}

We consider here a central Si tracker consisting of eight concentric
layers located between 18~cm and 110~cm from the interaction point. A
model of this tracker has been implemented in GEANT3, assuming 0.65\%
of a radiation length per layer, and is shown 
in~Fig.~\ref{fig:dpresol}. Its performance has been studied with 
$b\bar{b}$ and $WW$ events. The targeted 
$\delta p_{\rm T}/p_{\rm T}^2\le$~5~$\times$~10$^{-5}$~GeV$^{-1}$ can be  
achieved as shown in~Fig.~\ref{fig:dpresol}.

\subsection{Calorimetry}

The quality of the reconstruction of jets at CLIC should be guiding the 
calorimetry concept. The reasons to achieve the best energy resolution
for jets are multiple. A classical example is the need to distinguish
hadronic decays of the $Z^0$ from those of the $W^{\pm}$, so as
to separate the two pair production  
fusion reactions $e^+e^- \to Z^0 Z^0 \nu \bar{\nu}$ and 
$e^+e^- \to W^+ W^- \nu \bar{\nu}$ in the search for possible strong
interactions among intermediate vector bosons. The ability to use
these hadronic final states besides their leptonic or mixed
counterparts would be a key asset in such studies. 

Many search channels also claim for a good two-jet mass
resolution. The possibility to identify a $Z^0$ or a Higgs boson in
the final states of supersymmetry is an important tool in deciphering
the nature of cascades. Having access to the hadronic modes, such  
as $h^0 \to b \bar{b}$, is required to ensure a sizeable event~yield.

Another example is the radion search, an object similar to the Higgs
bosons whose main decay channel is a pair of gluons. The little Higgs
scenario, to be identified, would require a precise measurement of its
$W^3_H$ boson modes into $\ell \bar{\ell}$, $q \bar{q}$ and $Z h$.

A first requirement from the calorimetry would be to avoid the usual weak 
points (of most systems conceived up to now), namely a break between
the electromagnetic sector and the hadronic part, either due to the
coil or to some other uninstrumented gap. Since any electromagnetic
system has about one interaction length, a large fraction  
of hadrons interact in its depth and the first interaction is poorly sampled.

\begin{figure}[t]
\begin{center}
\begin{tabular}{c c}
\epsfig{file=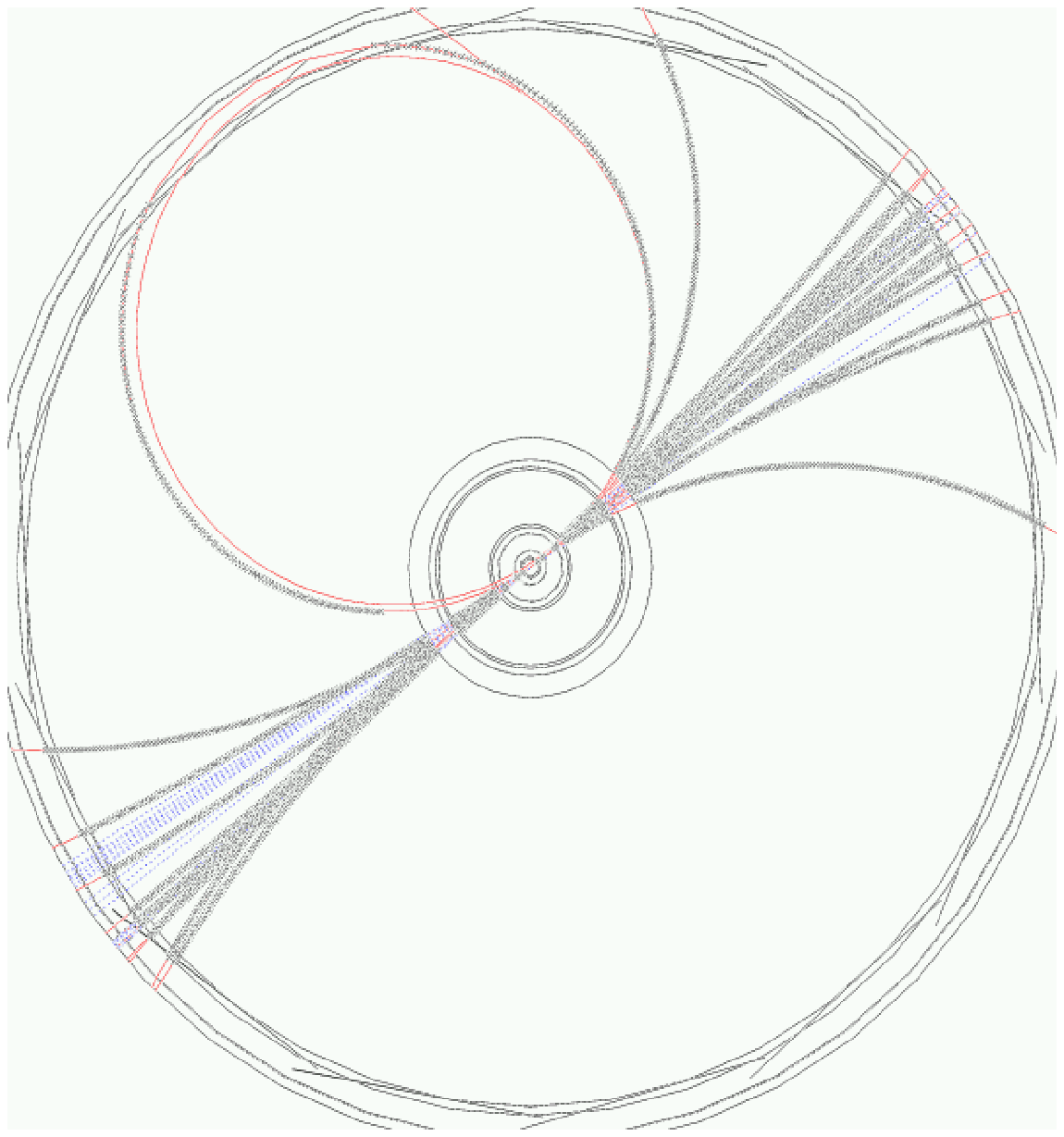,width=6.4cm} &
\hspace*{0.75cm}\epsfig{file=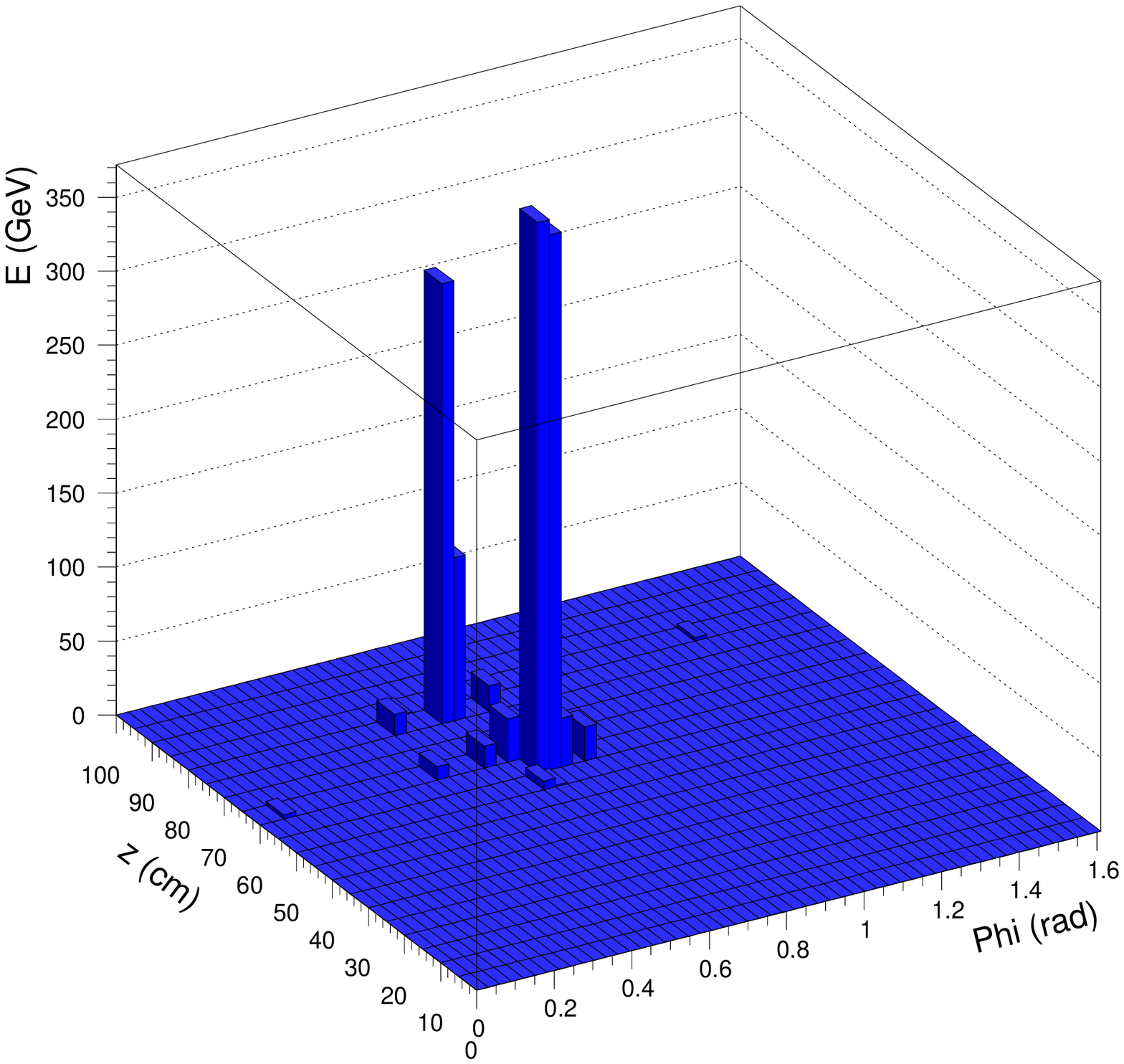,width=8cm} \\
\end{tabular}
\vspace*{2mm}
\end{center}
\caption{A $e^+e^- \to W^+W^-$ event at $\sqrt{s}$~=~3~TeV (left)
and the energy flowing at the entrance of the e.m. calorimeter located at a
radius of 170~cm, assuming a solenoidal field of 6~T and 5 $\times$ 5 cm$^2$
cell size (right)}
\label{fig:wwjj}
\end{figure}

It is likely that the way to approach the required performances at CLIC is to 
achieve the best possible energy-flow reconstruction.
The energy-flow technique was developed at LEP and has been adopted for the 
current LC detector studies. This technique combines tracking and calorimetric 
information in an optimal way to obtain the best possible resolution on
the energy of the produced partons. This requires an optimal distinction
between hadrons, neutral and charged, electrons, photons and
muons with a minimal double-counting. At CLIC the boost is large
enough to make intra-jet particle separation difficult,  
as exemplified by the energy-flow plot of Fig.~\ref{fig:wwjj}. 

Conventional wisdom and present experience then call, beside other
requirements, for high granularity, both transverse and
longitudinal, to disentangle neutral energy deposits from those
produced by charged particles, which are already  
accounted for by the more accurate tracker measurements. This requires
on the one hand a dense radiator material, with minimal Moliere radius,
and a finely segmented readout. On the other hand a sufficiently
large radius, with excellent matching between charged  
particle tracks, as seen in the main tracker and the calorimeter
information is needed. Beside having a minimal amount of material
along the tracker, this implies that the  
transition zone, involving the outermost end of the tracker and the
front part of the calorimeter should be particularly well studied, with
minimal thickness of the inert and minimal albedo effects. The problem
may be particularly acute in the forward region, because of the rates and
of the effect of interactions in the mask. 

Given the time scale and the rapid evolution of techniques, we are
still far from having a worked out scenario. On the other hand we
advise to define and set up, or join, a strong program of R\&D along
the lines defined in~\cite{teslatdr,brient,frey} aiming at a very
finely pixellated calorimeter, made of a dense tungsten radiator and a
very compact detecting medium, such as silicon. In its main lines this
activity should address the modelling of the transition from the main
tracker to the calorimeter, understand the merits of a digital
approach for the hadronic part with respect to a more classical
sampling approach,
and whether compensation should be used.

\subsection{Forward Region}

The small-angle environment at CLIC is very demanding for low-angle 
calorimetry: the region below 50~mrad, in particular, will be
difficult to instrument, both because of the high level of radiation
and particle density. At larger polar angles the situation improves
and we are led to believe that various types of detectors are  
suited for operating at polar angles above 100~mrad. Simulation
results show, however, that there still is a sizeable amount of energy
deposited by $\gamma \gamma$ background, between 100~mrad and 400~mrad.

The evaluation of the $\gamma \gamma \to {\mathrm{hadrons}}$ background was 
performed using the output of the GUI\-NEA\-PIG simulation. Backgrounds 
have then been tracked through the detector using the GEANT3 simulation up to 
scoring planes located at $\pm$150~cm downstream from the interaction region.

The results of the simulation show that the situation is manageable
down to 300~mrad for general hadronic events and down to about
100~mrad for identifying and measuring energetic electrons from Bhabha
events and from fusion processes such as  
$e^+e^- \to ZZ e^+e^-$. Assuming that the forward calorimetric
coverage is located  
at 200~cm from the interaction point, the $\gamma \gamma$ energy would
spread over an area of 20~cm radius. Using a small Moliere radius
material for the e.m. calorimeter, one should be able to confine
showers from high energy electrons to a transverse area about one
order of magnitude smaller than the figure reported above. 
Taking into account that the average energy per particle in $\gamma
\gamma$ events is $\simeq$~2.5~GeV, this contribution does not spoil
the energy measurement of a 250~GeV~electron.

\begin{figure}[t]
\centerline{\epsfig{file=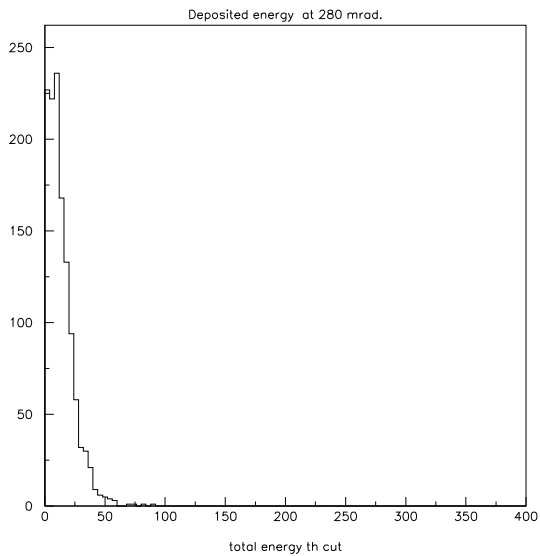,width=8.0cm} }
\caption{The average deposited energy for $\gamma\gamma$ event at angles
larger than  280~mrad}
\label{fig:forward}
\end{figure}

Even requiring a minimum angle close to 300~mrad, for total energy measurement 
purposes it is not feasible to integrate a full train. The average
deposited energy per $\gamma \gamma$ event at 280~mrad 
is shown~in~Fig.~\ref{fig:forward}. 

Timing capability should be implemented up to polar angles of about
400~mrad. Two strategies are available: either to incorporate 
time-stamping capabilities in the whole forward calorimeter or to break the
end-cap calorimeter in two sections.  
One with nsec-type time resolution and the other extending the same
technology as the main calorimeter. A specialized calorimeter at small
angle is better optimized and represents the option discussed here.

Small-angle calorimetry for luminosity measurement and beam condition
monitoring covers the angular region below 100~mrad. These devices
must have good properties for radiation hardness and time
resolution. Out of the various options  
for the active part of this type of calorimeters, one might use
diamond, quartz 
fibers, parallel-plate chambers, liquid scintillators; we focus
here on quartz fibers. Such detectors have been widely used in various
applications. R\&D has been carried out to address  
the main requirements. Energy resolutions better than 40\%/$\sqrt{E}$
have been  
routinely obtained. The output light is mostly due to Cherenkov effect by high 
energy particles in the shower. As for the timing characteristics, the
Cherenkov light is intrinsically fast and allows 
resolutions of the order of  1~nsec to be achieved. As noted before, a
requirement is for 
the low-angle calorimeter to have a very small  
Moliere radius. The choice of lead as radiator imposes small gaps for the 
quartz fibres, and using these as active medium improves on the 
already small Moliere radius achieved with lead. In fact the fibres
are sensitive to those high-energy particles in the shower which
travel close to the original primary trajectory. Such a phenomenon is
extremely important for hadronic showers. The  
shower radius has been found to shrink by a factor of 3
for the same radiator material. The effect is instead a factor of
about 2 for e.m.\ showers. The discrimination of high-energy
electrons benefits from this feature. Given the  
smaller radius, all edge effects will also be of lesser importance.
Using a volume ratio of 10:1 gives an equivalent Moliere radius
well below~1~cm. This can be further reduced by adopting tungsten as
the passive~radiator. 

\subsection{Summary of Detector Performance}

The parameters and response of the different detector components are 
summarized in~Table~\ref{tab:detperf}. The tracking performance has 
been validated using the full GEANT simulation in terms of both momentum 
resolution and track extrapolation accuracy. The calorimetric performances 
are derived from those assumed in the TESLA study, as they should not be 
significantly affected by the larger energy. The coverage is limited in 
the forward region by the beam delivery system and forward background.

\begin{table}[!h] 
\caption{Summary of detector performances}
\label{tab:detperf}

\renewcommand{\arraystretch}{1.7} 
\begin{center}

\begin{tabular}{lcc}\hline \hline \\[-4mm]
\textbf{Detector} &  & \textbf{CLIC studies} 
\\[4mm]   

\hline \\[-3mm]
Vertexing  &  $\hspace*{15mm}$ & 
$\delta(IP_{r\phi})= 
15~\mu{\rm m} \oplus {35~\mu{\rm m}~{\rm GeV}/c \over {p\sin^{3/2}\theta}}$ \\
 &  & $\delta(IP_{z})= 
15~\mu{\rm m}\oplus {35~\mu{\rm m}{\rm GeV}/c \over {p\sin^{5/2}\theta}} $ \\[3mm]
Solenoidal field &  & B~=~4.0~T \\
Tracking   &  & ${\delta p_t\over{{p_t}^2}}$~=~5.0~$\times$~10$^{-5}$
${\left (\frac{{\rm GeV}}{c} \right)}^{-1}$ \\
E.m. calorimeter &  & 
${\delta E \over E~ ({\rm GeV}) }$~=~0.10~${1\over \sqrt{E}}\oplus$~0.01 \\
Hadron calorimeter &  & 
${\delta E \over E~({\rm GeV})}$~=~0.50~${1\over \sqrt{E}}\oplus$~0.04 \\
$\mu$ detector &  & Instrumented Fe yoke 
$\frac{\delta p}{p}\simeq$~30\% at 100~GeV/$c$ \\
Energy flow  &   & 
${\delta E \over E~({\rm GeV})}\simeq$~0.3~$\frac{1}{\sqrt{E}}$\\
Coverage &  & $\left|\cos\theta\right|< $~0.98
\\[5mm] 
\hline \hline
\end{tabular}
\end{center}
\end{table}


\subsection{Luminosity Measurement and Energy Calibration}
\label{sec:3-lumimeas}

The intense beamstrahlung that smears significantly the luminosity
spectrum  will have to be accurately measured and unfolded from the
observed data, to relate them to theoretical predictions.  Accurate
determinations of both the absolute luminosity and the luminosity
spectrum are therefore crucial to preserve  the CLIC physics potential.
Bhabha scattering $e^+e^- \to e^+e^-$ represents a 
favourable reaction, with a cross section still sizeable beyond 1~TeV 
(9.4~pb at $\sqrt{s}$~=~3~TeV) and a simple, accurately measurable
final~state.  

\subsubsection{Luminosity determination with Bhabha scattering}

At LEP-1 a final luminosity precision of 0.07\% was achieved, by using
double-tags for the Bhabha process. The theoretical QED prediction for the 
Bhabha process at LEP was obtained using the BHLUMI Monte Carlo
program~\cite{bhlumi4:1996}. 
The theoretical uncertainty $\sigma_{\rm th}$ of the BHLUMI prediction was
estimated 
originally to be 0.1\%~\cite{th-95-38}, and later reduced to the level of 
0.07\%~\cite{YR-95-03-partIII,PL-bhabhaWG:1996,Ward:1998ht}.
The main contributions to the theoretical uncertainties at LEP-1 were
(a) the photonic second-order subleading correction 
$O(\alpha^2 L_e)$ where  
$L_e=\ln\frac{|t|}{m_e^2}$; (b) the hadronic vacuum polarization; and
(c) the $O(\alpha^2)$ light fermion-pair production. The
$s$-channel $Z$ contribution 
being only $\sim $~1\%, its contribution was well under control.
At LEP-1, the measured Bhabha rate had to be substantially larger than
that of the $s$-channel $Z$ at the resonance peak.
Hence, the polar-angle acceptance was pushed down to 25--50~mrad range,
corresponding to $\sqrt{|t|}$ of~1--2~GeV.

At CLIC, the angular range of Bhabha luminometer will need to be shifted to 
$\sim $~50--100~mrad due to background conditions.
At 3~TeV, the $t$-channel transfer becomes 75--150~GeV and the
$t$-channel $Z_t$ exchange can in principle be as important as that of
the $t$-channel photon exchange $\gamma_t$. The contribution from
hadronic vacuum polarization increases at higher transfers too. It is
thus important to estimate the magnitude of these theoretical 
uncertainties in the low-angle Bhabha (LABH) process at multi-TeV energies.

The contributions from photonic corrections will scale linearly with 
$L_e$, if some part of $O(\alpha^2 L_e)$ will still be missing
from the Monte Carlo 
generators (which is however unlikely by a realistic time for CLIC operation).
We therefore estimate the photonic uncertainty to increase, at most, by 
$\simeq $~30\% w.r.t. its LEP-1 value.
The $Z_t$ contribution was estimated by completely removing the $Z$ 
contribution. This changes the cross section by a few 0.1\% at
0.8~TeV and by 2--6\% at 3~TeV. 

\begin{figure}[htbp] 
\begin{tabular}{c c}
\epsfig{file=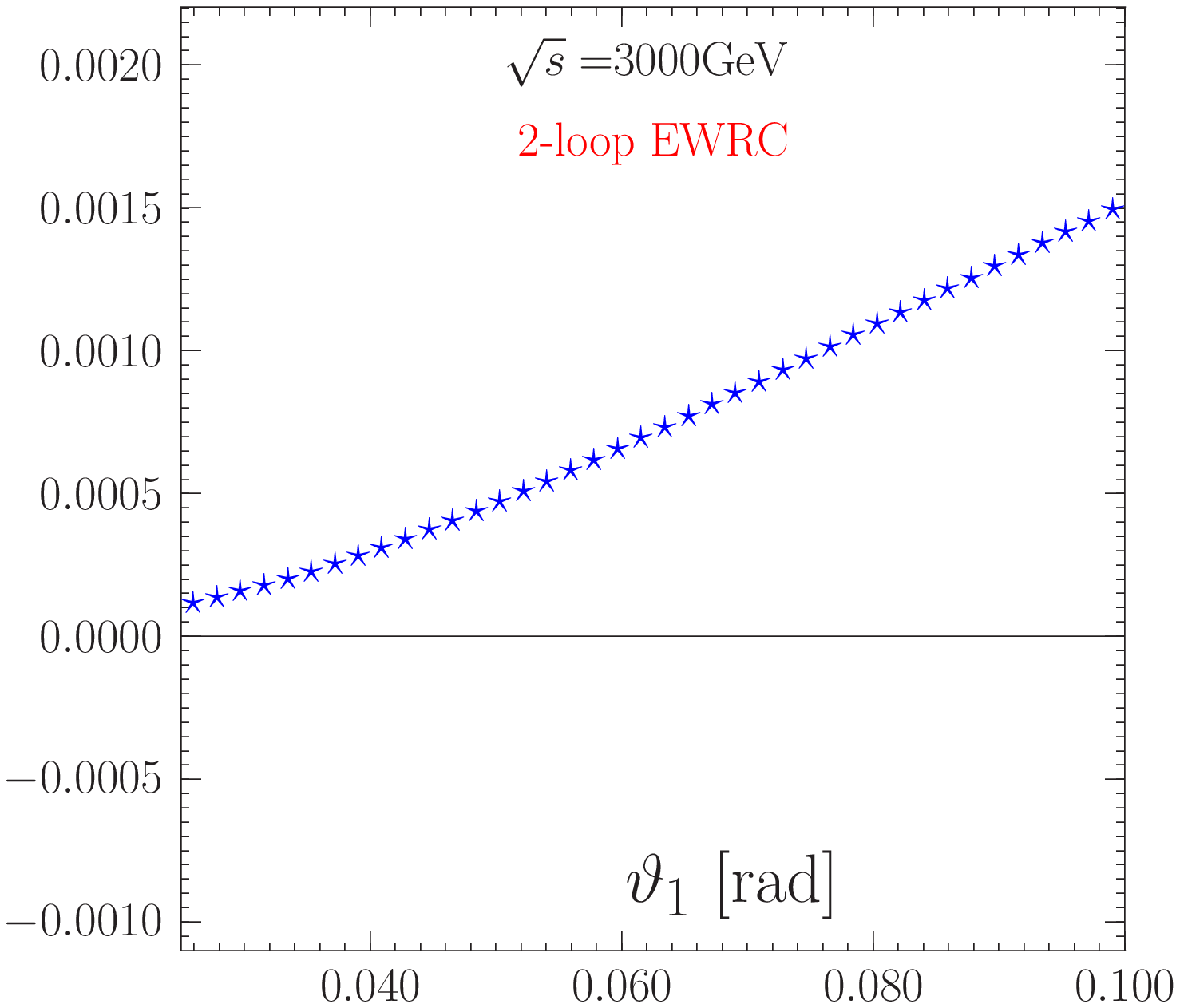,width=7.5cm} & \hspace*{5mm}
\epsfig{file=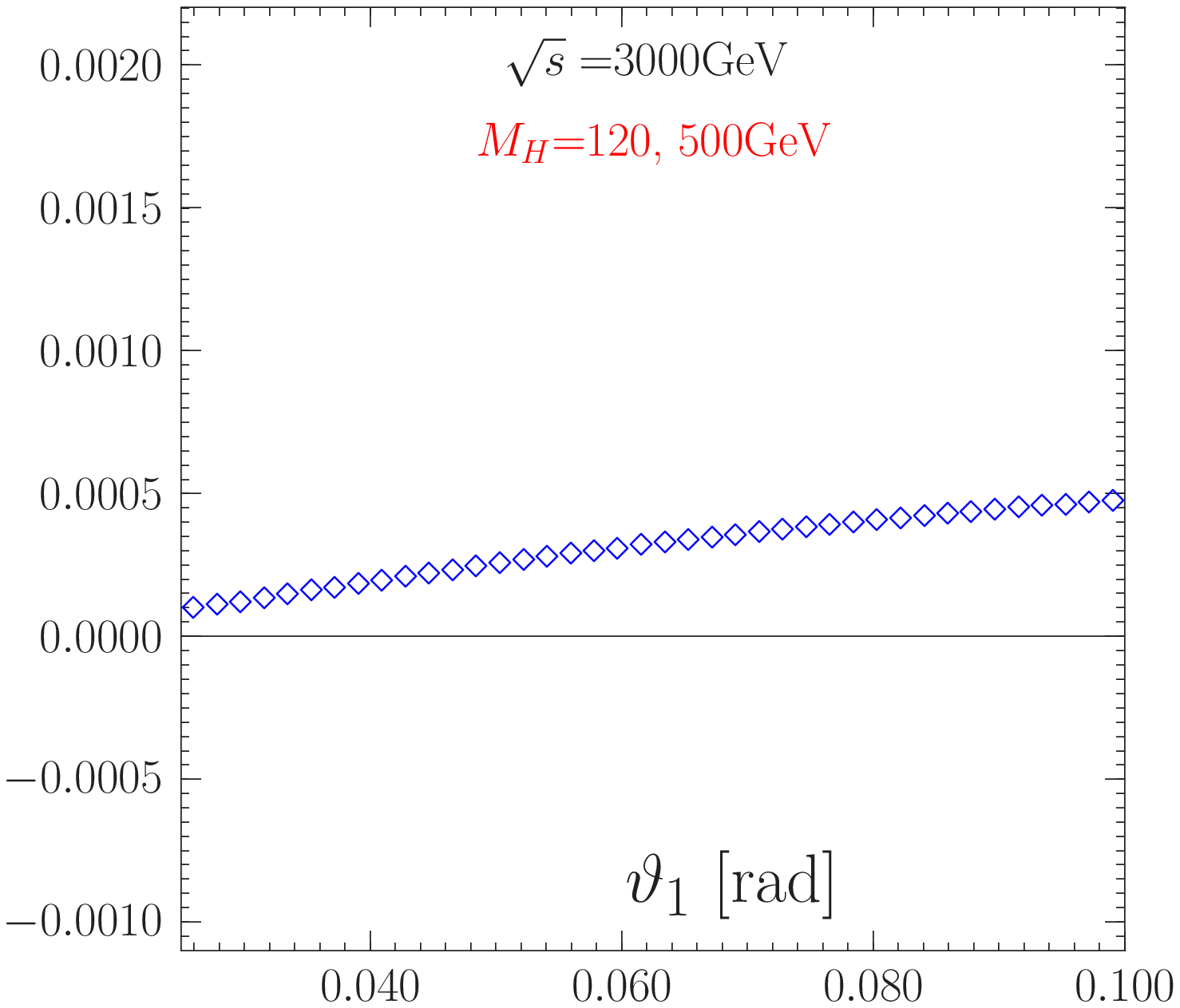,width=7.5cm} \\
\end{tabular}

\vspace*{6mm}

\caption{Change of Bhabha differential cross section due to uncontrolled
  2-loop corrections and due to Higgs mass error. Results were obtained 
  using the DIZET library of ZFITTER~\cite{Bardin:1989tq-orig,Bardin:1999yd}.}
\label{fig:ew}
\end{figure}

The theoretical uncertainty of the $O(\alpha)$ electroweak
corrections in  
the LABH process at CLIC have been estimated with the help of the DIZET EW 
library of  ZFITTER~\cite{Bardin:1989tq-orig,Bardin:1999yd}.
This was obtained by manipulating the non-leading $O(\alpha^2)$ EW 
corrections 
of $O(G_F^2 M_t^2 M_Z^2)$ of Degrassi et al., 
keeping $O(G_F^2 M_t^4)$ as accounted for. 
This is shown in~Fig.~\ref{fig:ew}, 
where the effect of change of $M_H$ from 120~GeV to 500~GeV is also given.
We estimate them to be 0.025\% at 0.8~TeV and 0.10\% at 3~TeV.
Changing $M_t$ from 165~GeV to 185~GeV has led to even smaller effect.
In summary, $\sigma_{\rm th} \simeq$~0.10\% of the LABH luminometer at CLIC
because of EW corrections emerges as a conservative estimate.

\vspace*{4mm}

The theoretical uncertainties of LABH due to hadronic vacuum
polarization (HVP), taking 1995 estimates of HVP 
of~Refs.\cite{burkhardt-pietrzyk:1995,eidelman-jegerlehner:1995}, 
was estimated as 0.03\% at LEP-2.
Using the same calculation of HVP we get 0.12\% at 3~TeV, i.e. larger
by a factor of 4. This means that HVP would become the dominant
component of theory systematics on the absolute luminosity at
CLIC. However, the recent improvements   
of HVP~\cite{Burkhardt:2001xp,Jegerlehner:2001ca} modifies the
situation substantially: according to preliminary estimates we get
a factor 2 reduction in the error due to HVP. 
By the time of CLIC operation, we hope for another factor 2 of improvement.

Summarizing, we think that a theoretical uncertainty 
$\sigma_{\rm th} \simeq $~0.1\% for the low-angle Bhabha process at
energies up to 3~TeV is~realistic. 

Since the LABH cross section depends as 1/$s$ on the total CMS energy,
in order to preserve the 0.1\% precision of LABH, a knowledge of the
absolute beam energy calibration to $\sim $~0.05\% accuracy is needed.
The energy spectrum of the luminosity also needs to be determined, which
can be obtained accurately using the acolinearity of Bhabha
scattering as discussed in the next section. 

\subsubsection{Bhabha scattering and $\sqrt{s'}$}

The reconstruction of the effective $e^+e^-$ energy $\sqrt{s'}$ from the 
acolinearity in large-angle Bhabha events has been proposed
for a lower energy linear collider~\cite{klaus:2000}, extending the
experience with 
the $\sqrt{s'}$ determination at LEP-2~\cite{sprime}. 

In the approximation where the energy lost before the $e^+e^-$ interaction is 
radiated in a single direction, the effective collision energy $\sqrt{s}$ can 
be related to the final state $e^+e^-$ acolinearity by 
\begin{displaymath}
\sqrt{s'} = \sqrt{s} \sqrt{1-2\frac{\sin (\theta_1+\theta_2)}
{\sin (\theta_1+\theta_2)-\sin \theta_1 -\sin \theta_2}}\,,
\end{displaymath}
where $\theta_1$ and $\theta_2$ are the angles of the final electron and the 
positron w.r.t.\ the photon direction. Therefore, with this assumption, 
the $\sqrt{s'}$ distribution can be measured by a determination of the $e^+$ 
and $e^-$ directions. At CLIC, there are two main processes 
leading to electron and positron energy loss, beamstrahlung (BS) and 
initial-state radiation (ISR). 
In order to infer the $\sqrt{s'}$ spectrum due to BS, it is
important that ISR can be reliably computed and unfolded from the measured  
distribution.  

A preliminary study has been performed by generating Bhabha events
with the BHLUMI~4.04 generator~\cite{bhlumi4:1996}. The CLIC electron
and positron energy spectra have been obtained from the result of the
beam simulation for the CLIC parameters at 1.5~TeV beam energy and the
effect of the beam energy spread in the Linac has been  
included in the form of a Gaussian smearing with a r.m.s. of 6~GeV.
The determination of $\sqrt{s'}$ has been based only on the electron and 
positron direction determination, assuming a tracking coverage down to 
$7^\circ$ in polar angle. While the calorimetric information may provide 
further important constraints, it needs to be validated by a full simulation,
accounting for the background conditions at small angles. Therefore, no 
attempt to reconstruct ISR photons has been made in this study. 
It is also important to measure the final-state-particle energy so as
to be able to disentangle the effect of the correlations in the
energies of the colliding particles, which are not taken into account  
in the approximation introduced. 

\begin{figure}[htbp] 
\begin{center}
\epsfig{file=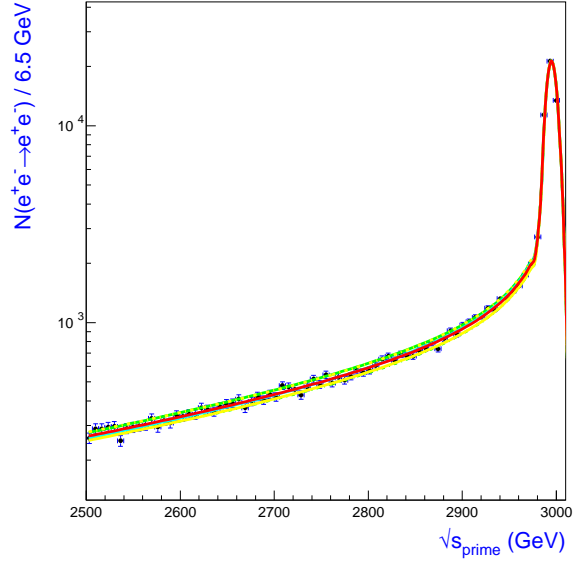,width=8.5cm}
\end{center}

\vspace*{-6mm}

\caption{The reconstructed luminosity spectrum for CLIC at 3~TeV 
(points with error bars). The continuous lines represent the fit
obtained using the first of the parametrizations discussed in the
text, for the fitted value and by  
varying the parameters within the $\pm$~2~$\sigma$~range.}
\label{fig:fit}
\end{figure}

The beamstrahlung spectrum has been parametrized using two models: 1) the 
modified Yokoya--Chen approximation~\cite{peskin99}: 
$e^{-N_{\gamma}} \big( \delta(x-1) + \frac{e^{-k(1-x)/x}}{x(1-x)} h(x)\big)$,
$h = f(\Upsilon)$ where $N_{\gamma}$ and $\Upsilon$ are treated as
free parameters,  
and 2) the CIRCE polynomial form: 
$a_0 \delta(1-x)+a_1x^{a_2}(1-x)^{a_3}$ were the free parameters 
are $a_0$, $a_2$, $a_3$, respectively. The fraction 
$F$ of events outside the 0.5\% of the nominal $\sqrt{s}$ energy has
also been left free and extracted from the reconstructed data.

The accuracy on the parameters has been obtained by performing a
likelihood fit to the reconstructed $\sqrt{s'}$ spectrum and the
uncertainty on the mean $\sqrt{s'}/\sqrt{s}$ has been extracted
accounting for correlations. The results are given 
in Table~\ref{tab:lyuba11} for an equivalent
luminosity $\int {\cal{L}}$~=~15~fb$^{-1}$, corresponding to 
$\simeq$ 3~days at the nominal luminosity of $10^{35}$~cm$^{-2}$s$^{-1}$.
The study has been performed for two different assumptions on the CLIC
parameters, denoted here as CLIC.01 and CLIC.02, corresponding to
different beamstrahlung spectra. 

\begin{table}[!b] 
\caption{The relative accuracy for the parameters in the 
Yokoya--Chen and CIRCE parametrization}
\label{tab:lyuba11}

\renewcommand{\arraystretch}{1.3} 
\begin{center}

\begin{tabular}{lccc}\hline \hline \\[-4mm]
\textbf{Parameter}  &  $\hspace*{7mm}$  & 
$\hspace*{8mm}$ \textbf{CLIC.01} $\hspace*{8mm}$  & 
$\hspace*{8mm}$ \textbf{CLIC.02} $\hspace*{8mm}$ 
\\[4mm]   

\hline \\[-3mm]
$\delta N_{\gamma} / N_{\gamma} $ &   & $\pm$ 0.044 & $\pm$ 0.084 \\
$\delta \Upsilon / \Upsilon$ &   & $\pm$ 0.019 & $\pm$ 0.018 
\\[2mm] \hline \\[-4mm]
$\delta \sqrt{s'} / \sqrt{s}$ &   & $\pm$~7.8~$\times$~10$^{-5}$ & 
$\pm$~5.3~$\times$~10$^{-5}$ \\[2mm] \hline \\[-4mm]
$\delta a_0 / a_0$ &   & $\pm$ 0.044 & $\pm$ 0.049 \\
$\delta a_2 / a_2$ &   & $\pm$ 0.089 & $\pm$ 0.058 \\   
$\delta a_3 / a_3$ &   & $\pm$ 0.018 & $\pm$ 0.021 
\\[2mm] \hline \\[-4mm]
$\delta \sqrt{s'} / \sqrt{s}$ &   & $\pm$~9.8~$\times$~10$^{-5}$ & 
$\pm$~7.2~$\times$~10$^{-5}$
\\[3mm] 
\hline \hline
\end{tabular}
\end{center}
\end{table}



The sensitivity to details of the beamstrahlung spectrum obtained in
this analysis needs to be further validated once the detector
resolution effects have been taken into account. However it has been
shown that the detector resolution can be made small with respect  
to the intrinsic beam energy spread. 

\section{Simulation Tools}

The influence of backgrounds on the extraction of physics signals 
cannot be forgotten. Tools have thus been developed to account for these 
effects by overlaying background particles to physics events and by simulating 
the $\sqrt{s}$ energy smearing effects due to the beamstrahlung. These
tools are interfaced to the simulation programs used in the physics
analyses discussed in the second part of this report.  The detector
response has been incorporated both through full GEANT3 simulation and
through a parametric smearing program. Both are based on the software
developed in the framework of the ECFA/DESY Workshops on Physics and
Detectors at a $e^+e^-$ Linear Collider, but have been customized to
describe the  CLIC specific geometry and detector concept.

\subsection{BDSIM}

A detailed simulation program  BDSIM~\cite{gb}
has been developed, based on GEANT4~\cite{Geant4}, 
to model the beam delivery system (BDS).
 BDSIM builds the BDS GEANT4 geometry directly from an
accelerator (MAD) optics file, which allows a rapid turn-around
as the BDS designs get updated. BDSIM 
incorporates efficient accelerator-style 
tracking based on transfer matrix techniques, together
with the standard shower generation
and physics processes of GEANT4.  
The program has been used to study 
electron/positron tracking, collimation, muon production 
~\cite{hb_epac,ds_epac}, and
the backgrounds associated with secondaries from showers
produced when primary beam or halo particles strike apertures
along the BDS.  BDSIM also allows for specialized
processes such as the tracking of
Compton-scattered electrons and photons from a laser-wire 
~\cite{gb_nanobeam}, 
including the effects of backgrounds
and limitations of beam line apertures.  
Such a study was first 
performed for the nominal long CLIC 1.5~TeV BDS~\cite{clicbds}.
BDSIM continues to be upgraded and now also incorporates
hadronic processes, including neutron production and tracking.
 GEANT4 is set up to allow for efficient weighting and biasing of
events and tracks, and  BDSIM will incorporate these techniques
to provide high-statistics background estimates for physics and detector
studies as well as for BDS optimization. It is also planned to interface
BDSIM directly to the next generation of detector simulation codes
as they are developed.

\subsection{GUINEAPIG}

The  GUINEAPIG code simulates the beam--beam interaction in a linear collider. 
It was originally developed for TESLA~\cite{c:thesis} and later 
extended to be able to simulate CLIC
 at high energies~\cite{c:qpext}. The program uses a
clouds-in-cells method to calculate the effect of the fields of one beam
on the incoming beam. A beam is typically represented by about 10$^5$
macroparticles. The emission of beamstrahlung is simulated by approximating
the particle trajectory locally with a circular motion and using the
Sokolov--Ternov spectrum~\cite{c:sk} for the radiation. The produced
photons are also 
tracked on the grid. The production of coherent pairs from the beamstrahlung
photons is implemented similarly to the beamstrahlung. It is also
possible to load  
an arbitrary distribution of particle from an input file, so as to be
able to take full 
advantage of the simulation of other components of the accelerator.

The production of incoherent pairs, bremsstrahlung and hadronic
background, is achieved using the Weiz\"acker--Williams approach. Each
beam particle is replaced by  
a number of virtual photons. In the collision these photons are
treated as real and 
the cross sections for $\gamma\gamma\to e^+e^-$ and
$\gamma\gamma\to hadrons$ are used. Secondary electrons and
positrons are tracked through the field of the beams. Beam-size effects
and those due to the strong field on the virtual photon spectrum can
also be taken into account~\cite{c:thesis}.

The produced particles and the beams can be saved after the simulation
and so used for further studies. Also the collision energies of
electrons, positrons and photons can be stored allowing the
reconstruction of the luminosity spectra, including all
correlations. The CALYPSO and 
HADES program libraries provide useful tools for interfacing with 
event generation and detector simulation programs, using the
GUINEAPIG outputs. 

\subsection{CALYPSO}

CALYPSO~\cite{c:calypsoweb} is a small Fortran library that gives
access to the luminosity output files from the GUINEAPIG
simulation. An initialization routine allows the file name to be
specified 
and the type of particles to be considered for each beam. Subsequent
calls of the main subroutine return the energies of colliding
particles. The use of the luminosity file allows the inclusion of the full
correlations between the colliding beams. Luminosity data files exist
for different energies and can be found in~Ref.~\cite{c:calypsoweb}.

There also exists a simple Fortran package that uses parametrizations 
to generate the
energies of colliding particles at $3{\,\rm~TeV}$. While the correlation is
only partially included and some simplifications are made for the
energy spread of the colliding beams, the program reproduces the
luminosity spectra quite accurately. This library needs no data files
but must be linked to the CERN library  CERNLIB. 

\subsection{HADES}
\label{sec:3-hades}

HADES consists of two parts. The first is a simple Fortan library that can 
write and retrieve background events to and from a file. Originally
designed for  
hadronic background, the library allows the generation of background
events with any 
generator and their storage in a file. With other routines, it is
easy to randomly 
choose a number of events from such a file and add to them an event generated
with  PYTHIA. HADES has been used in these studies also to load and overlay 
the muon background. The second part is a small program that uses PYTHIA to 
generate the hadronic background events from the initial photons provided by 
GUINEAPIG. Hadronic background files exist for CLIC
at~$E_{c.m.}$~=~3~~TeV and $E_{c.m.}$~=~5~TeV~\cite{c:hadesweb}.

\subsection{GEANT Simulation}

A full simulation of the CLIC tracking system has been developed,
based on GEANT3 and on the BRAHMS program set up for the TESLA study. In the 
CLIC version, the beam-pipe configuration and the tungsten-mask
geometry have been modified accordingly and a multilayered tracker
based on Si sensors has been implemented according to the  
conceptual design discussed above. The GEANT program has been interfaced with 
CALYPSO and  HADES to include the beamstrahlung effects and overlay 
backgrounds. The pair, $\gamma \gamma \to hadrons$ and
parallel muon backgrounds have been implemented and used in the
studies. The GEANT program has been used both to provide
parametrizations of the tracking performances used in the  
fast simulation and to carry out analyses based on the tracker only
such as the development of the $b$-tagging algorithm discussed above
and the smuon pair production. For this the track-reconstructed
parameters have been  
passed to the analysis code via the  VECP commons in the  VECSUB package, 
so that the same code could be used on the full and fast simulations.

\subsection{Parametric Detector Simulation}

The program  SIMDET simulates the reconstruction of $e^+e^-$ collision events, 
based on a parametrized detector response. This response was
originally tuned on that expected for the TESLA detector, according to
its full GEANT3 simulation. For its application to the CLIC studies,
several extensions have been implemented~\footnote{We thank the
European $e^+e^-$ Study Group for TESLA, who allowed us to use and develop
this program for CLIC.}. The response has been modified to reflect that
expected for a detector suitable for use at CLIC. The track momentum
resolution and extrapolation accuracy have been parametrized according
to the results of the full GEANT3 simulation of a discrete
multilayered Si tracker. Also the angular coverage has been modified,
while the calorimeter performances have been kept as for the TESLA detector. 

SIMDET provides a fast treatment of detector simulation and event 
reconstruction, while ensuring rather realistic performances. The
basic detector  components implemented are:

\newpage

\begin{itemize}
\item     Vertex tracker, 

\item     Main tracker and Forward tracker, 

\item     Electromagnetic calorimeter,

\item     Hadronic calorimeter 

\item     Low-angle tagger and low-angle luminosity calorimeter. 
\end{itemize}

The program applies Gaussian smearing to charged-particle momenta and
impact parameters with resolutions obtained from full simulation. The
calorimetric response is also  
treated parametrically. Pattern recognition is emulated by means of
cross-reference tables between generated particles and detector
response. An energy-flow algorithm defines the output of the
program. The track covariance matrix and $dE/dx$ information is also available.

SIMDET processes events generated in line with PYTHIA, and also events
stored in  
HepEvt format. As part of the CLIC extensions, an interface with partons 
generated with the  CompHep program has been set up. It is thus
possible to read  
CompHep event files and perform hadronization and short-lived particle decays 
in line. Also, the  CALYPSO and HADES interfaces have been added to 
read the luminosity spectrum and overlay backgrounds, respectively. In
the CLIC physics studies, the  GUINEAPIG luminosity files have been
used to describe the shape of the luminosity spectrum and to overlay the 
$\gamma\gamma \to hadrons$ and the parallel muon backgrounds.

After simulation, particles are classified, for the purpose of physics
analysis, as electrons, photons, muons, charged hadrons (taken as
pions), neutral hadrons (taken as long-lived kaons) and clusters of
unresolved particles. Detector coverage, tracking efficiencies, charge
mismeasurement and threshold energies for detector response are  
taken into account with user-adjustable values. These have been tuned
according to the CLIC detector concept discussed earlier on.

Best estimates for energy-flow objects have been stored using the  VECSUB 
conventions and used for the physics analyses.

\section{Generators and Physics Code}

\subsection{Monte Carlo Event Generators for CLIC}

Monte Carlo (MC) event generators are an essential tool in all  experimental
analyses. In this section we will briefly recall the key features of these
programs and describe those simulations which are available for CLIC
  physics.
Here we limit ourselves to the discussion of the physics event generators.
These programs must be interfaced to both detector simulations and beam  energy
spectra in order to give a complete simulation. In general the MC
event generation 
process can be divided into three  main phases\footnote{Those particles  which
decay before hadronization, for example, the top quark, are decayed before  the
hadronization phase and treated as a secondary hard process.}: (1) The  hard
process where the particles in the hard collision and their momenta are
generated, usually according to the leading-order matrix element. (2)  The
parton-shower phase where the coloured particles in the event are
perturbatively evolved from the hard scale of the collision to the  infrared
cut-off.  The emission of electromagnetic radiation from charged-particles  can
be handled in the same way. (3) A hadronization phase in which the partons 
left after the perturbative evolution are formed into the observed hadrons. 

Most generators fall into one of two classes:  general-purpose event
generators aim to perform the full simulation of the event starting
with the hard process  and 
finishing with the final-state hadrons; the second class of programs 
perform only the hard scattering part of the simulation and rely on one of the
general-purpose generators for the  rest of the simulation.

During the LEP era the most common practice was to rely on the  general-purpose
Monte Carlo programs for the description of hadronic final states
supplemented with more 
accurate parton-level generators interfaced to general-purpose event 
generators for specific processes, for example four-fermion production. At
future  linear colliders, and particularly CLIC,
this is even more important as we
wish to  study final states with higher multiplicities, for example six or even
eight fermions, which cannot be described by the general-purpose generators.

We will first discuss the general-purpose event generators. This is  followed
by a discussion of  the parton-level programs that are available to  calculate
particular final states and the automatic programs that are available to
calculate any final state.

\subsubsection{General-purpose event generators}

Historically the main general-purpose generators have been  
HERWIG~\cite{HERWIG},  ISAJET~\cite{ISAJET}, and 
PYTHIA~\cite{PYTHIA}.  While the general philosophy of these programs is
similar, the models used and approximations made are different.  In general, at
least for $e^+ e^-$ collisions, the  range of  hard production
processes available in the different generators is  very similar. All the
generators have a wide range of SM processes 
predicted by the MSSM, and some processes for other models.

The differences in the parton shower phase of the process are more  pronounced.
While  ISAJET still uses the original parton-shower algorithm which only 
resums collinear logarithms, both  HERWIG and  PYTHIA include the
effects of  soft logarithms via an angular-ordered parton shower in the case of
HERWIG and by  applying a veto to enforce angular ordering on a
virtuality-ordered shower in the case  of  PYTHIA. There is also a
separate program, ARIADNE\cite{Lonnblad:1992tz}, which implements the
dipole cascade and is interacted to  PYTHIA for  hadronization. %
Like the parton-shower phase the models used for the hadronization phase are
very different in the different programs.  ISAJET uses the Feynman-Field
independent fragmentation model while PYTHIA  uses the LUND string model
and  HERWIG the cluster hadronization model.

While these programs will continue to be used in the near future, a major 
programme is under way to produce a new generation of general-purpose event
generators in C++. The main aim of this programme is to provide  the
tools needed 
for the LHC. However, these tools will  be used for the next generation of
linear colliders. The only generator currently available in C++ that can
generate physics results is  APACIC++\cite{Kuhn:2000dk}; work is under
way to  rewrite both PYTHIA~\cite{Bertini:2000uh} and  
HERWIG~\cite{Gieseke:2002sg} in C++. These programs should be available in the
next few years and we  expect them to be the major tools for event 
generation~at~CLIC.

\subsubsection{Parton level programs}

Many programs are available to calculate an  individual
hard process, or some set of hard processes, and are interfaced to one  of  the
general-purpose generators, most often  PYTHIA, to perform the parton
shower and hadronization. In this brief review it is impossible to list all
these programs and discuss them.

As many of these programs were used by the LEP collaborations,  a detailed
discussion of both two-~\cite{Kobel:2000aw} and  four-~\cite{Grunewald:2000ju}
fermion generators can be found in the report of the LEP-II Monte Carlo
workshop~\cite{LEPII}. In addition there is also one program,  
LUSIFER~\cite{Dittmaier:2002ap}, for six-fermion processes. In practice the
computation of processes with six or more final-state particles is complex and
it seems likely that these calculations will be performed by the automatic
programs discussed in the next section.

\subsubsection{Automatic matrix element calculations}

There are an increasingly large number of programs available that  are
capable of calculating and integrating the matrix elements for large  numbers
of final-state particles automatically.

In general there are two important components of any such program. The first
step is the calculation  of the matrix element for a given  momentum
configuration. There are three different techniques in use to perform this step
of the calculation: (1) The matrix element squared can be evaluated
symbolically using traditional trace techniques. (2) A number of programs use
helicity amplitude techniques to  evaluate them. (3) There are techniques
\cite{Caravaglios:1995cd} which can be used to evaluate the matrix element
without using Feynman diagrams.

The second step of the process is to integrate the matrix element. There  are
two main techniques in use: One approach is to use adaptive integration
programs such as  VEGAS~\cite{Lepage:1980dq} to perform the integration.
A second approach is to use the knowledge of the matrix element to perform
multi-channel phase-space integration based on its peak structure.

In practice some programs combine these two approaches. In general, adaptive
programs such as  VEGAS~\cite{Lepage:1980dq} are ill-suited to the 
integration of functions which have complex peaked structures, such as 
multiparticle matrix elements, and therefore for most practical applications
multichannel integration techniques converge much faster.

There are a number of programs available which combine a variety of  these
techniques:

\begin{enumerate}
\item[---] 
{\sl AMEGIC++}~\cite{Krauss:2001iv} makes use of helicity amplitude
techniques to evaluate the matrix element together with efficient multichannel
phase-space integration to calculate the cross section.

\item[---] 
{\sl CompHep}~\cite{Pukhov:1999gg} is an automatic program for calculation
of the cross section for processes with up to eight  external
particles.\footnote{ CompHep can have up to six  final-state particles for
scattering processes and seven for decays.}  It uses the traditional trace
techniques to evaluate the  matrix element together with a modified adaptive
integrator to  compute the cross section.

\item[---] 
{\sl GRACE}~\cite{Ishikawa:1993qr} combines the calculation of matrix
element via helicity amplitude techniques and adaptive  integration to
calculate cross sections. Recently it has been used to  calculate the one-loop
corrections to Higgs production in $WW$ fusion~\cite{Belanger:2002ik}.

\item[---] 
{\sl HELAC/PHEGAS} uses the approach of~Ref.~\cite{Kanaki:2000ey}
which is  based 
on the Dyson--Schwinger equation together with  multichannel
integration~\cite{Papadopoulos:2000tt} to calculate  the cross section.

\item[---] 
{\sl MADGRAPH/MADEVENT}~\cite{Maltoni:2002qb} uses helicity amplitude
techniques for the matrix element together with an efficient multichannel
phase-space integrator  to compute the cross section.

\item[---] 
{\sl WHIZARD} is a multichannel integration package which can use  either
 CompHep, MADGRAPH or   OMEGA\footnote{  OMEGA uses the approach
of~Ref.~\cite{Caravaglios:1995cd} to evaluate  the matrix element but does not 
include any QCD processes.}~\cite{Moretti:2001zz} to calculate the matrix
elements.

\item[---] 
There are tools which are used to calculate loop processes 
\cite{LoopTools}:  FeynArts automatically generates the Feynman diagrams
for the  considered  process, while  FeynCalc evaluates analytically the
relevant  amplitudes.  Another package FormCalc can also be used for the
calculation of the  loop amplitudes with the numerical evaluation performed by
the program  LoopTools.  All these programs have been also used recently
to evaluate  the radiative  corrections to $e^+ e^- \to H \nu_e \bar{\nu}_e$ in
the~SM~\cite{Denner}.
\end{enumerate}

All of these programs apart from  HELAC/PHEGAS are publicly  available. At
present only  CompHep and  GRACE include supersymmetric processes,
although the both   MADGRAPH and  AMEGIC++ can be extended to add the
additional interactions which  are needed. In order  to simulate events these
programs needed to be interfaced to the general-purpose event generators. Most
of these programs are interfaced to  one of the  major general-purpose event
generators although the details vary from program~to~program.

\subsection{ Programs for Spectra, Decay and Production}

\subsubsection{Programs for the MSSM spectrum}

It is well known that in the unconstrained MSSM, it is a rather tedious  task
to deal with the basic parameters of the Lagrangian and to derive in an
exhaustive  manner their relationship with the particle masses  and  couplings.
This is mainly due to the fact that in  the MSSM, despite its minimality, 
there are more than one hundred new parameters.  Even if one
constrains the  model 
to have a viable phenomenology, there are still more than 20 free  parameters
left.  This large amount of input enters in the evaluation of the masses  of
$O$(30) SUSY particles and Higgs bosons as well as their  complicated
couplings, which involve several non-trivial aspects.  The situation  becomes
particularly difficult if one aims at rather precise calculations and  hence
attempts to include some refinements such as higher-order corrections,  which
for the calculation of a single parameter need the knowledge of a large  part
of the spectrum. 

However, there are well-motivated theoretical models where the soft
SUSY-breaking parameters obey a number of universal boundary conditions  at
the high (GUT) scale, leading to only a handful of basic parameters. This is
the case for instance of the minimal Supergravity model (mSUGRA) and  anomaly 
(AMSB) or gauge mediated (GMSB) SUSY-breaking models. However, this 
introduces another complication since the low-energy parameters are to be
obtained  through renormalization group equations (RGE) and the models should 
necessarily give radiative electroweak symmetry-breaking (EWSB). The
implementation of these two ingredients poses numerous non-trivial  technical
problems when done in an accurate way; this complication has to be added  to
the one from the calculation of the particle masses and couplings with 
radiative corrections (RC) which is still present.

To deal with the supersymmetric spectrum in all possible cases, one needs very
sophisticated programs to encode all the information and, eventually, to pass
it to other programs or Monte Carlo generators to  simulate the physical
properties of the new particles, decay branching ratios, production cross
sections at various colliders, etc. These programs  should have a high degree
of flexibility in the choice of the model and/or the  input parameters and an
adequate level of approximation at different stages, for instance in the
incorporation of the RGEs, the handling of the EWSB  and  the inclusion of
radiative corrections to (super)particle masses, which in  many cases can be
very important. They should also be reliable, quite fast to allow  for rapid
comprehensive scans of the parameter space, and simple enough to be  linked
with  other programs. 

There are four main public codes (there are also some private codes which  will
not be discussed here; see later) which make rather detailed calculations  of
the Supersymmetric particle spectrum in the MSSM or in constrained   scenarios
(mSUGRA, etc.): 

\begin{enumerate}
\item[---] 
{\sl ISASUSY}~\cite{ISAJET}, which is available since the early 90s  and is
implemented in the Monte Carlo generator  ISAJET;  it is the most widely
used for simulations in the MSSM.

\item[---] 
{\sl SuSpect}~\cite{SUSPECT}, a new version has been released very  recently
but a preliminary version of the program has existed  1998 and was described in
Ref.~\cite{GDR}. 

\item[---] 
{\sl SOFTSUSY}~\cite{SOFTSUSY}, a code written in object--oriented C++ but
which has a  Fortran interface for calculations in mSUGRA; it was released two
years  ago.  

\item[---] 
{\sl SPHENO}~\cite{SPHENO}, which was released this year and also calculates
the decay branching ratios and the production rates in $e^+e^-$
collisions of SUSY particles.
\end{enumerate}

The codes have different features in general, but they all incorporate the 
four main ingredients or requirements for any complete calculation of the SUSY 
spectrum: i) the RG evolution, ii) the implementation of radiative  EWSB,
iii) the calculation of the masses of the Higgs and SUSY  particles,
including the radiative corrections, and iv) the possibility of   performing
some checks of important theoretical  features and some experimental
constraints.

The radiative corrections are particularly important, as is well known, in  
the Higgs sector. Codes such as  ISASUSY and  SuSpect have  their 
own approximate calculations, but they are also linked with routines  which do 
a more sophisticated job. The main available routines for the Higgs  sector 
are:  Subhpole (SUBH)~\cite{SUBH} which calculates the leading  radiative
corrections in the effective potential approach with a two-loop RG
improvement;  HMSUSY (HHH)~\cite{HHH} which calculates the one-loop
corrections in the effective potential approach and includes the leading
two-loop standard QCD and EW corrections;    
FeynHiggsFast~\cite{FeynHiggsFast}, 
which calculates the  corrections in the Feynman
diagrammatic approach with the one and two-loop  QCD corrections at zero
momentum transfer (the version  FeynHiggs~\cite{FHF} has the full
one-loop corrections and is slower); and  BDSZ~\cite{BDSZ} gives the
leading one-loop corrections from the third-generation (s)fermion sector  as
well as the full $\alpha_s\lambda_t^2, \lambda_t^4$ and $\alpha_s  \lambda_b^2,
\lambda_t \lambda_b$ corrections at zero-momentum  transfer and also has the
corrections to the effective potential. The radiative corrections to the SUSY
spectrum are in general performed following Pierce, Bagger, Matchev and Zhang
\cite{PBMZ}. 

Detailed comparisons of these codes have been performed. The main  conclusion
is  that despite the different ways in which the various items
discussed above are  
implemented, they in general agree at the per cent level in large
parts of  the  
MSSM parameter space. Several more important differences occur, however, in 
some  areas of the parameter space, in particular in the high $\Tb$ and/or
focus point regions with large $m_0$ values, where the Yukawa couplings of top 
and bottom quarks  play an important role; see~Ref.~\cite{comp} for details.

\subsubsection{Decay and production}
 
For the production of SUSY and Higgs particles, the matrix elements of  most
important processes are already included in the MC generators. However, in 
some cases, specific-purpose programs which include some higher-order effects 
such as radiative corrections, spin correlations, width effects, etc. are very
useful. A non-exhaustive list of available public codes for Higgs and
sparticle  production in $e^+e^-$ collisions  including higher-order
effects is:  

\begin{enumerate}
\item[---] 
{\sl SUSYGEN}~\cite{SUSYGEN} for Higgs and sparticle production, also a  MC 
generator (see above).

\item[---] 
{\sl HZHA}~\cite{HZHA}: the most used Monte Carlo generator for Higgs 
production at LEP2. 

\item[---] 
{\sl HPROD}~\cite{HPROD}: a code which calculates the cross sections for SM
and MSSM Higgs production in $e^+e^-$ collisions in the main production
channels. 

\item[---] 
{\sl SPHENO}~\cite{SPHENO}: discussed earlier and which also calculates
the cross sections for SUSY particle production at $e^+e^-$ colliders. 

\item[---] 
Many four- or six- fermion production processes at $e^+e^-$ colliders as
discussed earlier; some of them have been discussed for the LEP--II MC
Workshop~\cite{LEPII}. 
\end{enumerate}

The decays of SUSY and Higgs particles can be rather complicated and it is 
important to determine them with good accuracy. There can be a large  number
of decay modes for some particles: simple two-body decays in which it is
important sometimes to include higher-order corrections (as is the case   for
Higgs bosons and strongly interacting sparticles), and rather complicated
many-body decay modes such as the three- or four- body decays of charginos, 
neutralinos and top squarks or important loop-induced decay modes. There  are
several  available codes doing this job with a different level of
sophistication: 

\begin{enumerate}
\item[---] 
{\sl ISASUSY}~\cite{ISAJET}: only tree-level two-body Higgs and  SUSY
decays (three-body for gauginos).  

\item[---] 
{\sl HDECAY}~\cite{HDECAY}: SM and MSSM Higgs decays with higher-order 
effects. 

\item[---] 
{\sl SDECAY}~\cite{SDECAY}: sparticle decays including higher-order  effects
(RC and multi-body). 

\item[---] 
{\sl SPHENO}~\cite{SPHENO}: discussed above and has two- and three-body SUSY 
particle decays. 
\end{enumerate}

\noindent Some decay routines are also included in the Monte Carlo
event  generators  
 SUSYGEN~\cite{SUSYGEN}, HZHA~\cite{HZHA}, PYTHIA
\cite{PYTHIA} and  HERWIG~\cite{HERWIG} with possible links to the
programs mentioned above. The treatment of SUSY production and decay in  the
various programs is very different. Both  ISAJET and  PYTHIA  assume that
the production and decay of the SUSY particles takes place independently, 
while  HERWIG and   SUSYGEN~\cite{SUSYGEN} include correlations
between the production and decay.  HERWIG includes these correlations in
all  processes~\cite{Richardson:2001df} while in   SUSYGEN they are only
included for some  processes and decays. Some  development in  this subject is
expected in the near future. 

\section{Standard Model Cross Sections}

In order to maximize the sensitivity to new physics signals, it is
important not only to consider radiative corrections to the process under
investigation, but also radiative corrections to so-called background
processes provided by the Standard Model. At multi-TeV energies, some
processes are of major interest not only for direct searches, but also as
backgrounds for new physics. This is especially true for gauge-boson and
fermion-pair production. Electroweak one-loop corrections are a central
building block in any precision study of fermion-pair production. The
complete set of electroweak contributions, including real hard-photon
corrections needs then to be 
calculated~\cite{Fleischer:2003kk,Dittmaier:2002ap}. One
novelty at CLIC is that electroweak radiative corrections become sizeable
at multi-TeV energies, due to the appearance of Sudakov logarithms. In
this section, we first describe the general status of these large
corrections, and then discuss the specific 
reaction~$e^+ e^- \to {\bar t} t$.

\subsection{Electroweak Sudakov Logarithms}

\def\sscr#1{\ensuremath { { \scriptscriptstyle #1}}}

At the next generation of linear colliders with centre-of-mass
energies in the  TeV range, well above the electroweak scale, one
enters the realm of large perturbative 
corrections. Even the effects arising from weak corrections are expected to be 
of the order of 10\% or
more~\cite{Kuroda:1991wn,Beenakker:1993tt,Ciafaloni:1998xg,Beccaria:1999fk,Layssac:2001ur,Denner:2000jv}, 
i.e., just as large as the well-known electromagnetic corrections. In 
order not
to jeopardize any of the high-precision studies at these high-energy colliders,
it is indispensable to improve the theoretical understanding of the 
radiative corrections in the weak sector of SM. 
In particular this will involve a careful analysis of effects beyond first 
order in the perturbative expansion in the (electromagnetic)  
coupling~$\,\alpha=e^2/(4\pi)$.

The dominant source of radiative corrections at~TeV-scale energies is given by 
logarithmically enhanced effects of the form $\,\alpha^n\log^k(m^2/s)\,$ 
for $\,k\le 2n$. These large logarithms arise when the finite particle
mass $m$,  
which is well below the collider energy $\sqrt{s}$, acts as a natural cut-off 
of a singularity. To give a numerical example, at a collider energy  
of~1\,TeV, logarithmically enhanced $W$-boson corrections of the form 
$\,\alpha\log^2(M_{_W}^2/s)\,$ and $\,\alpha\log(M_{_W}^2/s)\,$ amount to 
19\% and --4\%, respectively. A natural way of controlling the theoretical 
uncertainties therefore consists in a comprehensive study of these large 
logarithms, taking into account all possible sources. One class of large 
logarithms is of ultraviolet origin. They involve short-distance scales and are
controlled by the renormalization-group equations. The remaining large
logarithms are of soft and/or collinear origin. They involve long-distance 
scales and, based on QCD/QED experience, they are expected to possess specific
factorization properties. Here we review our present
understanding of the latter class of large-logarithmic effects.

The potentially most important electroweak corrections are the so-called 
Sudakov logarithms $\,\propto \alpha^n\log^{2n}(M^2/s)$, arising from 
the exchange of collinear-soft, effectively on-shell transverse gauge 
bosons~\cite{Sudakov:1956sw}. It should be noted, however, that for pure 
fermionic final states (numerical) cancellations can take place between 
leading and subleading logarithms~\cite{Kuhn:2001hz}. For on-shell bosons in 
the final state, the Sudakov logarithms in general tend to be 
dominant~\cite{Beenakker:1993tt,Denner:2000jv}.   

Over the last few years, various QCD-motivated methods have been applied 
to 
predict the electroweak Sudakov logarithms to all orders in perturbation 
theory~\cite{Ciafaloni:1999ub,Fadin:1999bq,Kuhn:1999nn}. The methods vary in 
the way that the QCD-motivated factorization and exponentiation properties are 
translated to the electroweak theory. This is caused by the fact that the 
electroweak theory is a spontaneously broken theory with two mass scales in 
the gauge-boson sector, whereas QCD is basically a single-scale theory. 
The main debate has therefore focused on the question to what extent
the SM behaves like an unbroken theory at high~energies. 

In~Ref.~\cite{Ciafaloni:2000rp} a first hint was given that the transition 
from QCD to electroweak theory does not come without surprises. It was shown 
that the Bloch--Nordsieck cancellation between virtual and real collinear-soft 
gauge-boson radiation~\cite{Bloch:1937pw} is violated for $W$ bosons in 
the SM 
as soon as initial- or final-state particles carry an explicit weak charge 
(isospin) and summation over the partners within an $SU(2)$ multiplet is not 
performed. In the case of final-state particles the event-selection procedure 
might (kinematically) favour one of the partners within the $SU(2)$ multiplet, 
leading to a degree of `isospin-exclusiveness'. In the initial state the 
situation is more radical. At an electron--positron collider the weak isospin 
of the initial state particles is fixed by the accelerator, in contrast to QCD 
where confinement forces average over initial colour at hadron colliders. 
As a result, the Bloch--Nordsieck theorem is in general violated for 
left-handed initial states, even for fully inclusive cross sections. If we 
start off with a pure left-handed electron beam, $W$-boson radiation will 
change 
this before the actual (hard) scattering process takes place into a mixture of 
left-handed electrons and neutrinos, with relative weights 
\begin{eqnarray}
  W_{e_L}   = \frac{1}{2}\,\Bigl\{1+\exp[-2L_W(s)]\Bigr\}
  \qquad\mbox{and}\qquad
  W_{\nu_e} = \frac{1}{2}\,\Bigl\{1-\exp[-2L_W(s)]\Bigr\}\,,
\end{eqnarray}
respectively. The Sudakov coefficient $L_W(s)$ is given by 
\begin{eqnarray}
  L_W(s) = \frac{\alpha}{4\pi\sin^2\theta_{\rm w}}\,\log^2(M_{_W}^2/s)\,,
\end{eqnarray}
with $\,\theta_{\rm w}\,$ being the weak mixing angle. The difference in the 
weights originates from the fact that the isospin-singlet component of the 
beam receives no Sudakov corrections, whereas the isospin-triplet component is 
subject to the Sudakov reduction factor $\,\exp[-2L_W(s)]$. The exponent of 
this reduction factor amounts to roughly $-0.13$ at $\sqrt{s}~=~1$\,TeV. 
This implies a 6\% neutrino component of the beam at that energy. 
Since the (hard) scattering cross section can be quite different for 
left-handed electrons and neutrinos, the corrections due to this electroweak 
phenomenon can be very large, exceeding the QCD corrections for energies in 
the~TeV range. They are such that at infinite energy the weak charges will 
become unobservable as asymptotic states~\cite{Ciafaloni:2000rp}, which implies
for instance an $SU(2)$ charge averaging (`pre-confinement') of the 
initial-state beams. 

With this in mind, explicit calculations of (virtual) Sudakov
corrections at the two-loop level have been performed to resolve any
ambiguity in the translation from QCD to the SM. In the first few
calculations, the focus was on pure fermionic 
processes, like $\,e^+e^- \to f\bar{f}\,$~\cite{Beenakker:2000kb} and  
fermion-pair production by an $\,SU(2)\times U(1)$-singlet 
source~\cite{Hori:2000tm,Melles:2000ed}. The survey was finally completed by 
an (Coulomb-gauge) analysis that covered all SM 
particles~\cite{Beenakker:2001kf}, including scalar particles as well as
transverse and longitudinal gauge bosons. By means of this explicit calculation
it was established to what extent the SM behaves like an unbroken theory at 
high energies.

It was observed that the SM behaves dynamically like an unbroken theory in the 
Sudakov limit, in spite of the fact that the explicit particle masses are 
needed at the kinematical (phase-space) level while calculating the Sudakov 
correction factors. For instance, a special version of the Equivalence Theorem
was obtained, which states that the longitudinal degrees of freedom of 
the massive gauge bosons can be {\it substituted} by the corresponding 
Goldstone-boson degrees of freedom without the need for finite compensation 
factors. As a result, the Sudakov form factors for longitudinal gauge bosons 
exhibit features that are typical for particles in the fundamental 
representation of $SU(2)$, whereas for the transverse gauge bosons the usual 
adjoint features are obtained. Moreover, in the transverse neutral gauge-boson 
sector the mass eigenstates decompose into the unbroken $SU(2)$ field $\,W^3\,$
and $U(1)$ field $\,B$, each multiplied by the corresponding Sudakov form 
factor. At the kinematical level, though, the large mass gap between the 
photon and the weak gauge bosons remains. 

These findings support the Ansatz made in Ref.~\cite{Fadin:1999bq}, where the 
Sudakov form factors were determined by postulating a generalized IR evolution 
equation for the SM in the unbroken phase. As a result,
the virtual Sudakov correction for an arbitrary on-shell external
particle with mass $m$, charge $Q$ and hypercharge $Y$, amounts to multiplying
the matrix element by an exponentiated external wave-function~factor 
\typeout{\theequation}
\begin{eqnarray}
Z^{1/2} = \exp\,(\delta Z^{\,(1)}/2)\,,
\end{eqnarray}
with
\begin{eqnarray}
  \delta Z^{\,(1)} &=& -\,\frac{\alpha}{4\pi}\,
       \left[\, \frac{C_2(R)}{\sin^2\theta_{\rm w}} 
            + \left( \frac{Y}{2\cos\theta_{\rm w}} \right)^{\!2}\,\right]\! 
       \log^2\left(\frac{M^2}{s}\right) 
\nonumber \\[1mm]
                   & & -\,\frac{\alpha}{4\pi}\,
       Q^{\,2}\,\Bigg[ \log^2\left(\frac{\lambda^2}{s}\right)
                     - \log^2\left(\frac{\lambda^2}{m^2}\right)
                     - \log^2\left(\frac{M^2}{s}\right) \Bigg]\,.
\label{deltaZ1}
\end{eqnarray}
Here $\lambda$ is the fictitious (infinitesimally small) mass of the photon, 
used as infrared regulator, and $M$ denotes the generic mass scale of the 
massive gauge bosons. The coefficient $\,C_2(R)\,$ is the $SU(2)$ Casimir 
operator of the particle. So, $\,C_2(R) = C^{\sscr{SU(2)}}_F$~=~3/4 for 
particles in the fundamental representation: the left-handed fermions 
($f_L/\bar{f}_R$), the physical Higgs boson ($H$) and the longitudinal gauge 
bosons ($W^{\pm}_L$ and $Z_L$) being equivalent to the Goldstone bosons 
$\phi^{\pm}$ and $\chi$. For the particles in the adjoint representation of 
$SU(2)$, i.e.\ the transverse $W$ bosons ($W^{\pm}_T$) and the $W^3$ components
of both the photon and the transverse $Z$ boson, one obtains 
$\,C_2(R) = C^{\sscr{SU(2)}}_A$~=~2. For the $SU(2)$ singlets, i.e., the 
right-handed fermions ($f_R/\bar{f}_L$) and the $B$ components of both the 
photon and the transverse $Z$ boson, the $SU(2)$ Casimir operator vanishes, 
$C_2(R)~=~0$. Note that the terms proportional to $Q^{\,2}$ in 
Eq.~(\ref{deltaZ1}) are the result of the mass gap between the photon
and the weak bosons. 

These Sudakov form factors apply, in principle in a universal way, to arbitrary
non-mass-suppressed electroweak processes at high energies. This universality
follows from the two defining conditions for the Sudakov corrections: the 
lowest-order matrix element of the process should not be mass suppressed to 
start with, and all kinematical invariants of the process other than the 
masses
the external particles should be of the order of $s$. We would like to 
stress, though, that the universality of the Sudakov form factors has to be 
interpreted with care. For an electroweak process like $\,e^+e^- \to 4f\,$ it 
is in general not correct to assume universality and merely calculate the 
Sudakov form factors for the external particles (i.e., the six fermions). 
Depending on the final state and the kinematical configuration, the process 
$\,e^+e^- \to 4f\,$ can be dominated by different near-resonance 
subprocesses~\cite{Grunewald:2000ju} like 
$\,e^+e^-\!\to W^+W^-\! \to 4f\ $ or $\ e^+e^-\!\to ZZ \to 4f$.
These subprocesses all have their own Sudakov correction factors, which can be 
determined by employing the so-called pole scheme~\cite{Stuart:1991xk} in the 
leading-pole approximation, which restricts the calculation to the on-shell 
residue belonging to the unstable particle that is close to its mass 
shell.
In that case, the Sudakov correction factor is given by the wave-function 
factors of the near-resonance intermediate particles rather than the four 
final-state fermions. The reason for this is that the invariant masses of 
those 
intermediate particles are close to being on-shell, and therefore {\it 
not\,} of 
the same order as $s$. The subsequent decays of the intermediate particles 
into 
the final-state fermions do not involve a large invariant mass, and will 
as 
such not give rise to Sudakov logarithms. In this way the Sudakov form factors 
for unstable particles, like the massive gauge bosons and the Higgs boson, can 
participate in the high-energy behaviour of reactions with exclusively stable 
particles in the final state. 

The observations for the virtual Sudakov corrections can be extended to 
real-emission processes in a relatively straightforward way. After all, since 
the Sudakov logarithms originate from the exchange of collinear-soft, 
effectively on-shell transverse gauge bosons, many of the features derived for 
the virtual corrections will be intimately related to properties of the 
corresponding real-emission processes. Two remarks are in order, though.
First of all, the Bloch--Nordsieck cancellation between virtual and real 
collinear-soft gauge-boson radiation can be violated in the SM, as was 
mentioned earlier. So, unlike in QED/QCD, the Sudakov corrections can show up 
in inclusive experimental observables. Secondly, unlike in QED/QCD, the masses
of the weak gauge bosons provide a physical cut-off for real $Z/W$-boson
emission. This means that, given a sufficiently good experimental resolution,
it is possible to construct exclusive experimental observables that do not 
receive contributions from real $Z/W$-boson radiation. Such observables thus 
provide direct access to the virtual Sudakov corrections.

Next we discuss the subleading logarithms.
For a complete understanding of the perturbative structure of large 
logarithmic correction factors, subleading logarithms originating from soft, 
collinear, or ultraviolet singularities cannot be 
ignored~\cite{Denner:2000jv,Kuhn:2001hz}. In the case of pure fermionic final 
states, for instance, the subleading effects can reach several per cent in the 
TeV energy range, giving rise to (numerical) cancellations between leading 
and subleading contributions~\cite{Kuhn:2001hz}. 

Over the last few years, some progress has been made in understanding and
predicting the subleading logarithms. In~Ref.~\cite{Ciafaloni:2001mu}, a first
step was made on the road towards the construction of a full set of collinear 
evolution equations in the electroweak theory. It was investigated how 
collinear logarithms factorize in a spontaneously broken gauge theory. 
As was to be expected on the basis of the aforementioned violation of the 
Bloch--Nordsieck theorem, a factorization pattern emerges that is qualitatively
different from the one in QCD. In order to deal with the explicit 
weak charges of the initial-state beams, extra splitting functions have to be 
introduced. These new splitting functions are infrared-sensitive, in the sense 
that they depend explicitly on the infrared cut-off provided by the 
symmetry-breaking scale. It is precisely this infrared sensitivity that 
gives 
rise to the observability of Sudakov logarithms in particular inclusive 
experimental quantities.

In~Ref.~\cite{Denner:2003wi}, finally, the (collinear-soft) angular-dependent, 
next-to-leading logarithms were determined at two-loop level. This explicit SM
calculation was found to agree with the exponentiation prescriptions proposed 
in~Refs.~\cite{Kuhn:2001hz,Kuhn:1999nn,Melles:2001dh}. These prescriptions are 
based on a symmetric $SU(2)\times U(1)$ theory matched with QED at the 
electroweak scale and involve a product of two exponentials.
The first exponential corresponds to the $SU(2)\times U(1)$-symmetric 
corrections, which are obtained by replacing the photon mass by the weak-boson 
mass scale $M$. The second exponential corresponds to the QED corrections, 
which are subtracted in such a way that they vanish when the photon mass 
becomes equal to $M$. The order of the two 
exponentials is such that the QED factor is the last one to act on the matrix 
element. In other words, the order respects the hierarchy of the generic 
scales: $\sqrt{s}$ for the hard-scattering matrix element, $M$ for the 
$SU(2)\times U(1)$-symmetric corrections, and $\lambda$ for the QED 
corrections. This is not surprising, since in a gauge-invariant 
effective-Lagrangian formulation of collinear and soft effects one would 
integrate out the momentum-scales exactly along this hierarchy.

\subsection{\boldmath{$e^+ e^- \to t\, \bar{t}$} Cross Sections}

The $e^+ e^- \to t \, \bar{t}$ reaction is of special
interest~\cite{Accomando:1997wt,Aguilar-Saavedra:2001rg2,Abe:2001wn}, due
to its larger cross section and a direct link of the top mass with the
coupling to the Higgs boson. This section discusses the process
{\ensuremath {e^+e^- \to t \bar{t} \,\,}} as an example of the
need to include electroweak radiative corrections to obtain reliable
predictions.

First we discuss the virtual one-loop corrections.
In lowest-order perturbation  theory, the process {\ensuremath {e^+e^- \to 
t \bar{t} \,\,}} can be
described by two Feynman diagrams. The radiative corrections are 
parametrized by means of form factors. 
Defining the following four matrix elements
\begin{eqnarray}
\label{eq:amplitude1}
{\cal M}_{1}^{ij} & = 
\left[ \bar{v}(p_4) \, \gamma^{\mu}\, {\bf G}^i \,  u(p_1) \, \right] \, { \times}  \, 
\left[ \bar{u}(-p_2) \, \gamma_{\mu}\, {\bf G}^j \,v(-p_3) \, \right] 
, ~~~i,j=1,5,  
\end{eqnarray}
with ${\bf G}^1 = { 1}$ and ${\bf G}^5 = { \gamma_{ 5}}$, 
the Born amplitude can be written in a compact form:
\begin{equation}
 \label{eqn:Mborn}
{\cal M}_{B} = {\cal M}_{\gamma} + {\cal M}_{Z } = 
{\sum}_{i,j=1,5}~{F_{1}^{ij,B}} \,\,  {\cal M}_{1}^{ij}  ,
\end{equation}
with
\begin{eqnarray}
 \label{eqn:bornFF}
 {F_1^{11,B}} & = &  v_e \, v_t \, 
\frac{e^2}{s - M_Z^2 +iM_Z\Gamma_Z}  ~+~ Q_e \, Q_t \, 
\frac{e^2}{s}{ ~~\equiv~~ F_1^{11,B,Z} + F_1^{11,B,\gamma}}\,, 
  \\
 {F_1^{15,B}} & =&  -  v_e \, a_t \, 
\frac{e^2}{s - M_Z^2 +iM_Z\Gamma_Z}\,,   \\
 {F_1^{51,B}} & =&  -  v_t \, a_e \, 
\frac{e^2}{s - M_Z^2 +iM_Z\Gamma_Z}\,,  \\
 {F_1^{55,B}} & = &   a_e \, a_t \, 
\frac{e^2}{s - M_Z^2 +iM_Z\Gamma_Z}\,.
\end{eqnarray}
Besides 
(\ref{eq:amplitude1}),  at 
one-loop level three further basic matrix-element structures are found (in 
the limit of vanishing electron mass): 
\begin{equation}
 \label{eqn:Mborna}
{\cal M}_{ \rm{1loop}} = 
\sum_{a=1}^4 {\sum}_{i,j=1,5}
~ {F_{ a}^{ij,\rm{1loop}}} \,\,  {\cal M}_{a}^{ij},  
\end{equation}
with 
\begin{eqnarray}
\label{eq:amplitudes1to4}
{\cal M}_{1}^{ij} & =  & 
\gamma^{\mu}\, {\bf G}^i \otimes \gamma_{\mu}\, {\bf G}^j,
\qquad
{\cal M}_{2}^{ij}  = {p\hspace{-0.43em}/}_2\, {\bf G}^i  \, \otimes \, {p\hspace{-0.43em}/}_4 \, {\bf G}^{ j}, \\
{\cal M}_{3}^{ij} & = & {p\hspace{-0.43em}/}_2\,{\bf G}^i \, \otimes \,  
{\bf G}^{j}\,, \qquad 
\hspace*{3mm} {\cal M}_{4}^{ij}  = \gamma^{\mu}\, {\bf G}^{i} \, \otimes \, 
\gamma_{\mu}\,{p\hspace{-0.43em}/}_4\, {\bf G}^{j}\,, 
\nonumber
\end{eqnarray}
and correspondingly there are sixteen scalar form factors $F_a^{ij}$ in 
total. 
The interferences of these matrix elements with the Born amplitude have
to be calculated. 
Only six of these interferences are independent, i.e., 
$ {\cal M}_{1}^{ij}$, ${{ \cal M}_3}^{11}$ and ${{ \cal M}_3}^{51}$.
In order to express the results compactly for possible later 
implementation into a 
full Monte Carlo program, the  virtual corrections are expressed  in terms 
of the six independent, modified, dimensionless form factors 
$\widehat{F}_1^{ij }, \widehat{F}_3^{11}, \widehat{F}_3^{{51}}$:    
\begin{eqnarray}
  \label{eq:eqf61}
\widehat{F}_1^{11} &=&  \Bigl[ F_1^{11} 
+ \frac{1}{4}(u-t)  ~F_2^{11} 
 -   \frac{1}{4}(u  +   t  +  2m_t^2)   ~F_2^{55} 
+ m_t~  (F_4^{55}-F_4^{11} ) \Bigr],
\\[2mm]
\widehat{F}_{ 1}^{ 15} &=& \Bigl[F_{ 1}^{ 15} 
- \frac{1}{4} (u+t-2m_t^2) ~F_{ 2}^{ 51} 
+ \frac{1}{4} (u-t) ~F_{ 2}^{ 15} \Bigr],
\\[2mm]
\widehat{F}_{ 1}^{ 51} &=& \Bigl[F_{ 1}^{ 51} 
+ \frac{1}{4} (u-t) ~F_{ 2}^{ 51}  
- \frac{1}{4} (u+t+2m_t^2)~F_{ 2}^{ 15}  
+m_t~(F_{ 4}^{ 15}-F_{ 4}^{ 51})\Bigr],
\\[2mm]
\widehat{F}_1^{55} &=&  \Bigl[F_1^{55} 
- \frac{1}{4} (u+t-2m_t^2   ) ~F_2^{11}
 +  \frac{1}{4} (u-t)~F_2^{55} \Bigr],
\\[2mm]
  \label{eq:eqf63}
\widehat{F}_3^{11}&=& \Bigl[F_3^{11}-F_4^{11}+F_4^{55}-m_t ~F_2^{55}\Bigr],
\\[2mm]
\widehat{F}_{ 3}^{ 51} &=& \Bigl[{F}_{ 3}^{ 51} +
F_{ 4}^{
15}-F_{ 4}^{ 51}-m_t ~F_{ 2}^{ 15} \Bigr]  \, .
\end{eqnarray}
The form factors $\widehat{F}_3^{11}$ and $\widehat{F}_{ 3}^{ 51}$ would
be modified explicitly, 
if the coupling of the top-quark deviates from the Standard Model 
prediction. The proportionality of the form factors to the mass of the 
produced quark ensures a particularly sensitive probe into physics beyond the 
Standard Model.

The resulting  cross-section formula is:
\begin{eqnarray}
  \label{eq:sigma6f}
  \frac{d\sigma}{d\cos\theta} &=& 
\frac{\pi \alpha^2}{2s} 
~c_t~ 
\beta
~2 \Re e \Bigl[ 
  (u^2+t^2+2m_t^2s)
\left(\bar{F}_1^{11} \bar{F}_1^{11,B*} +\bar{F}_1^{51} \bar{F}_1^{51,B*} \right)
\nonumber\\[2mm]
&&+~
  (u^2+t^2-2m_t^2s)
\left(\bar{F}_1^{15} \bar{F}_1^{15,B*} +\bar{F}_1^{55} \bar{F}_1^{55,B*} \right)
\nonumber\\[2mm]
&&
+~  (u^2-t^2)
\left( \bar{F}_1^{55} \bar{F}_1^{11,B*}+ \bar{F}_1^{15} \bar{F}_1^{15,B*}
  + \bar{F}_1^{51} \bar{F}_1^{51,B*}    + \bar{F}_1^{11} \bar{F}_1^{55,B*} \right)
\nonumber\\[2mm]
&&
+~ 2m_t(tu-m_t^4) \left( \bar{F}_3^{11}  \bar{F}_1^{11,B*} 
                       +\bar{F}_3^{51}  \bar{F}_1^{51,B*} \right)
\Bigr]\,,
\end{eqnarray}
where the dimensionless form factors are 
\begin{eqnarray}
\label{starbo} 
\bar{F}_1^{ij,B*}  &=& \frac{s}{e^2}~F_1^{ij,B*} , \hspace*{1cm}
 \bar{F}_{ a}^{ ij} = 
\frac{s}{e^2}~\left[
\frac{1}{2} \delta_{a,1} 
{F}_{ 1}^{ ij,B} +  
{ \frac{1}{16\pi^2}}~
\widehat {F}_{ a}^{ ij, \rm{1loop}}\right]\,,
\end{eqnarray}
\noindent
and  $c_t ~=~3$, $\alpha=e^2/4\pi$. 
The $\bar{F}_a^{ij}$ are defined so that  
double counting for the Born contributions ${F}_{1}^{ij,B}$ is
avoided.
The factor $1/(16\pi^2)$ is conventional.

Self-energy insertions, vertex and box diagrams as well as renormalization
lead to virtual corrections. A complete list of the contributing diagrams
can be found in~Ref.~\cite{Fleischer:2002rn}. By means of the package
DIANA~\cite{Tentyukov:1999is,Tentyukov:1999yq}, useful graphical
presentations of the diagrams and the input for subsequent FORM
manipulations are generated.  With the DIANA output (FORM input), two
independent calculations of the virtual form factors were performed, both
using the 't Hooft--Feynman gauge.

Both the ultraviolet (UV) and the infrared (IR) divergences are treated by
dimensional regularization, working in $d=4-2 \epsilon$ dimensions and
parametrizing the infinities as poles in $\epsilon$. For the infrared
divergences this is realized in the form of $ { P}_{IR}= -
\frac{1}{2\epsilon} + \frac{\gamma_E}{2} -\ln(2\sqrt{\pi})$. The UV
divergences have to be eliminated by renormalization on the amplitude
level, while the IR ones can only be eliminated on the cross-section level
by including the emission of soft photons.  For the IR divergences,
alternatively, a finite photon mass can be introduced, yielding a
logarithmic singularity in this mass.  Agreement to high precision was
achieved for the two approaches.  {Renormalization is performed closely
following~\cite{Fleischer:1981ub,Fleischer:2003kk,Dittmaier:2002ap}}, 
in the on-shell mass
scheme.

Next we discuss the real photonic radiative corrections.
The cross section for $e^+(p_4)e^-(p_1)\to t(p_2)\bar
t(p_3)\gamma(p)$ may be subdivided into gauge-invariant subsets of
initial-state radiation, final-state radiation and the interference
between them; the phase-space is five-dimensional. Basically, the approach
proposed in~Refs.~\cite{Bardin:1977qa,Passarino:1982zp} is followed
and extend to the massive fermion case.  The differential
bremsstrahlung cross section takes the form
\begin{eqnarray}
d\sigma = 
\frac{1}{(2\pi)^5}\frac{1}{2s\beta_0}|{\cal M}|^2\cdot
\frac{\pi}{16s}\mbox{d}\phi_\gamma\mbox{d}s'\mbox{d}V_2\mbox{d}\cos\theta
\label{space}
\end{eqnarray}
with $\beta_0=\sqrt{1-4m_e^2/s}$, $s'=(p_2+p_3)^2$ and $V_{2}=- 2p p_{3}
$. The first integration over $\phi_\gamma$ is performed analytically, and
the remaining integrations are tackled numerically. The program 
TOPFIT~\cite{FRW:2002sw} allows one to set cuts on the invariant mass of 
the
top-quark pair, $s'$, and/or the scattering angle $\cos \theta$. The 
lower
hard photon energy (being also the upper soft photon energy) is $\omega =
E_\gamma^{\rm min} $ , the final result is of course independent of
$\omega$.~\footnote{The formulas for the hard photon scattering are rather
lengthy and can be found in~Refs.~\cite{Fleischer:2003kk,Dittmaier:2002ap}.}
\begin{equation}
\frac{{\rm d} \sigma}{{\rm d} \cos \theta}(s,t,m_t, ...) =\frac{{\rm d} \sigma^{soft}}{{\rm d} \cos \theta}(\omega,s,t,m_t, ...) + \frac{{\rm d} \sigma^{hard}}{{\rm d} \cos \theta}(\omega,s,t,m_t, ...) + \frac{{\rm d} \sigma^{virtual}}{{\rm d} \cos \theta}(s,t,m_t, ...)
\end{equation} 
Soft photon terms have to be identified in order to
combine them with virtual corrections for a finite net elastic cross section.  
The four-dimensional integration of the bremsstrahlung contributions
is divergent in the soft-photon part of the phase-space and is treated
in $d$ dimensions. The photonic phase-space part is parametrized 
with Born-like kinematics for the matrix-element squared.
To obtain a soft photon contribution, 
the bremsstrahlung amplitude has to be taken
without $p^0 \equiv E_{\gamma} \leq \omega$ in the numerators. 
In  this limit, $s'$ approaches  $s$ and  the soft contribution
to the differential cross section takes the  form
\begin{eqnarray}
\label{dsoft}
\frac{d \sigma^{soft}}{d\cos\theta} =   
\frac{\alpha}{\pi} 
\left( Q_e^2~ \delta^{soft}_{ini} 
+ Q_e Q_t~ \delta^{soft}_{int} + Q_t^2~ \delta^{soft}_{fin} \right)
\frac{d  \sigma^{Born}}{d\cos\theta}  
\end{eqnarray}
with 
\begin{eqnarray}
\label{delsini}
\delta^{soft}_{ini}(m_e,\omega,\epsilon,\mu) &=&
2~\left(P_{IR}+\ln\frac{2\omega}{\mu}\right)
\left[
-1+\frac{s-2 m_e^2}{s\beta_0}\ln\left(\frac{1+\beta_0}{1-\beta_0}\right)
\right]
+~\frac{1}{\beta_0}\ln\left(\frac{1+\beta_0}{1-\beta_0}\right)
\nonumber \\[2mm]
&-& \!\! \frac{s-2 m_e^2}{s\beta_0}\!\!
\Biggl[
\frac{1}{2}\ln^2\left(\frac{2\beta_0}{1-\beta_0}\right)
+\mbox{Li}_2(1)+\mbox{Li}_2\left(\frac{\beta_0-1}{2\beta_0}\!\!\right)
+\mbox{Li}_2\left(\frac{2\beta_0}{\beta_0+1}\right)
\Biggr] ,
\nonumber\\[3mm]
\\[5mm]
\label{delsofin}
\delta^{soft}_{fin}(m_t,\omega,\epsilon,\mu) &=&
\delta^{soft}_{ini}(m_t,\omega,\epsilon,\mu),
\\[3mm]
\label{delsoint}
\delta^{soft}_{int}(m_e,m_t,\omega,\epsilon,\mu) &=&
2\left(P_{IR}+\ln\frac{2\omega}{\mu}\right)
\left( 
\frac{T}{\sqrt{\lambda_T}}
\ln\frac{T+\sqrt{\lambda_T}}{T-\sqrt{\lambda_T}} 
-\frac{U}{\sqrt{\lambda_U}}
\ln\frac{U+\sqrt{\lambda_U}}{U-\sqrt{\lambda_U}} 
\right)
\nonumber\\
&&
+~\frac{1}{2}\left[ T ~{\cal F}(T) - U ~ {\cal F}(U) \right],
\end{eqnarray}
and
\begin{eqnarray}
\lambda_T &=& T^2-4 m_e^2 m_t^2, \hspace*{0.5cm}
 {\cal F}(T) = - \frac{4}{s} \int_0^1 d\alpha \frac{1}{\beta_T(1-\beta_T^2)}
\ln\frac{1+\beta_T}{1-\beta_T},
\end{eqnarray}
and analogous definitions for $T \leftrightarrow U$.
The infrared-divergent parts proportional to $P_{IR}$ have been shown to 
cancel 
analytically against those obtained from virtual corrections. 
This cancellation is 
indeed true for all orders in perturbation theory.

For the numerical evaluation performed by the code 
TOPFIT~\cite{FRW:2002sw} the following input values~\footnote{ All
observables exhibit very little dependence on the variation of the Higgs
mass.} are 
assumed~\cite{Fleischer:2003kk,Dittmaier:2002ap,Fleischer:2002rn,Fleischer:2002nn,Fleischer:2002kg}:
\begin{equation}
\begin{array}[r]{llll}
  \Gamma_Z = 2.49977 \,{\rm~GeV} \,\,  \, \,,  & \alpha = \frac{e^2}{4\, \pi}
= 1 / 137.03599976 \,, & m_t = 173.8 \,{\rm~GeV} \,\, \,, & m_b = 4.7 \,{\rm~GeV} \,\, \,, \\[2mm]
 M_W = 80.4514958 \,{\rm~GeV} \,\, \, , & M_Z = 91.1867 \,{\rm~GeV} \,\, \,, & M_H = 120 \,{\rm~GeV} \,\, \,.
\end{array}
 \end{equation}

\begin{figure}[htbp] %
\begin{center}
\epsfig{file=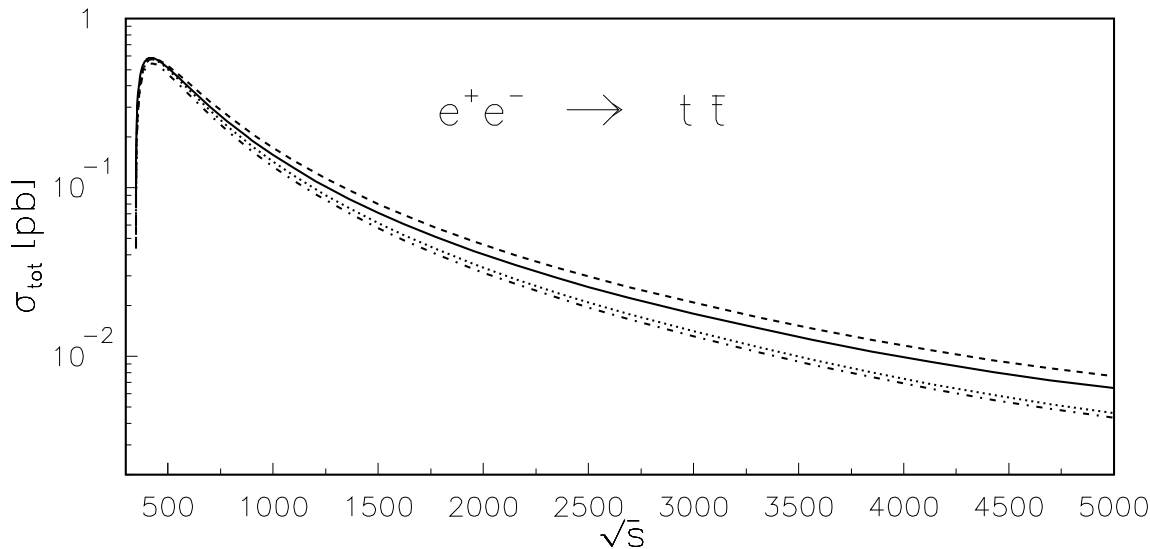,bbllx=25,
bblly=280,bburx=575,bbury=555,width=15cm}
\end{center}
\caption{
Total cross-section for top-pair production as a function of $s$.
Born (solid lines), electroweak (dashed lines), electroweak with
$s'$~=~0.7~$s$-cut (dotted lines)  and electroweak with  $s'$~=~0.7~$s$- and
$\cos\theta$~=~0.95-cut (dashed-dotted lines).}
\label{figuretx}
\end{figure}

\begin{figure}[t] 
\begin{center}
\begin{tabular}{c c}
\epsfig{file=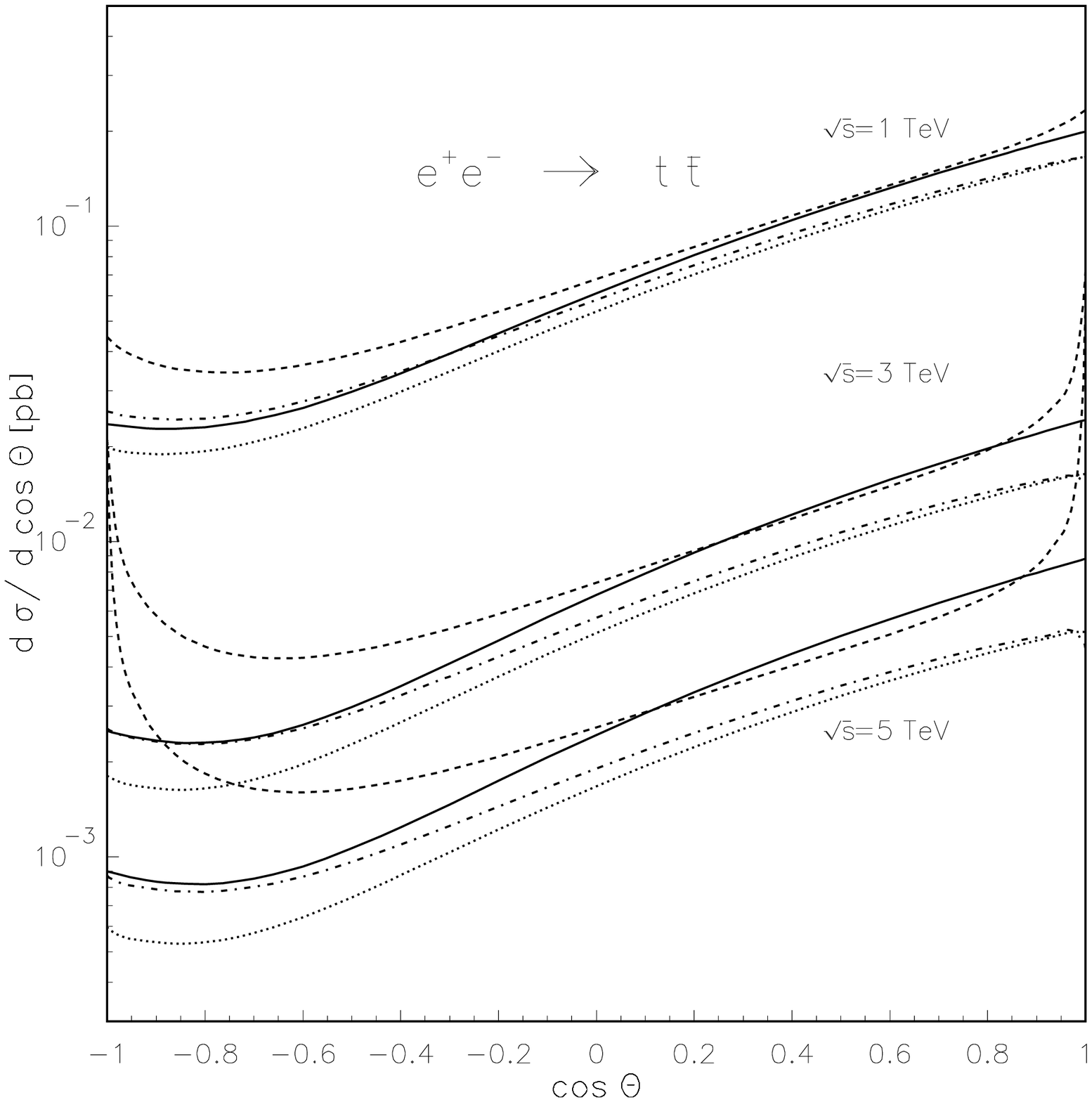,bbllx=30,
bblly=130,bburx=565,bbury=690,width=7.1cm} &
\epsfig{file=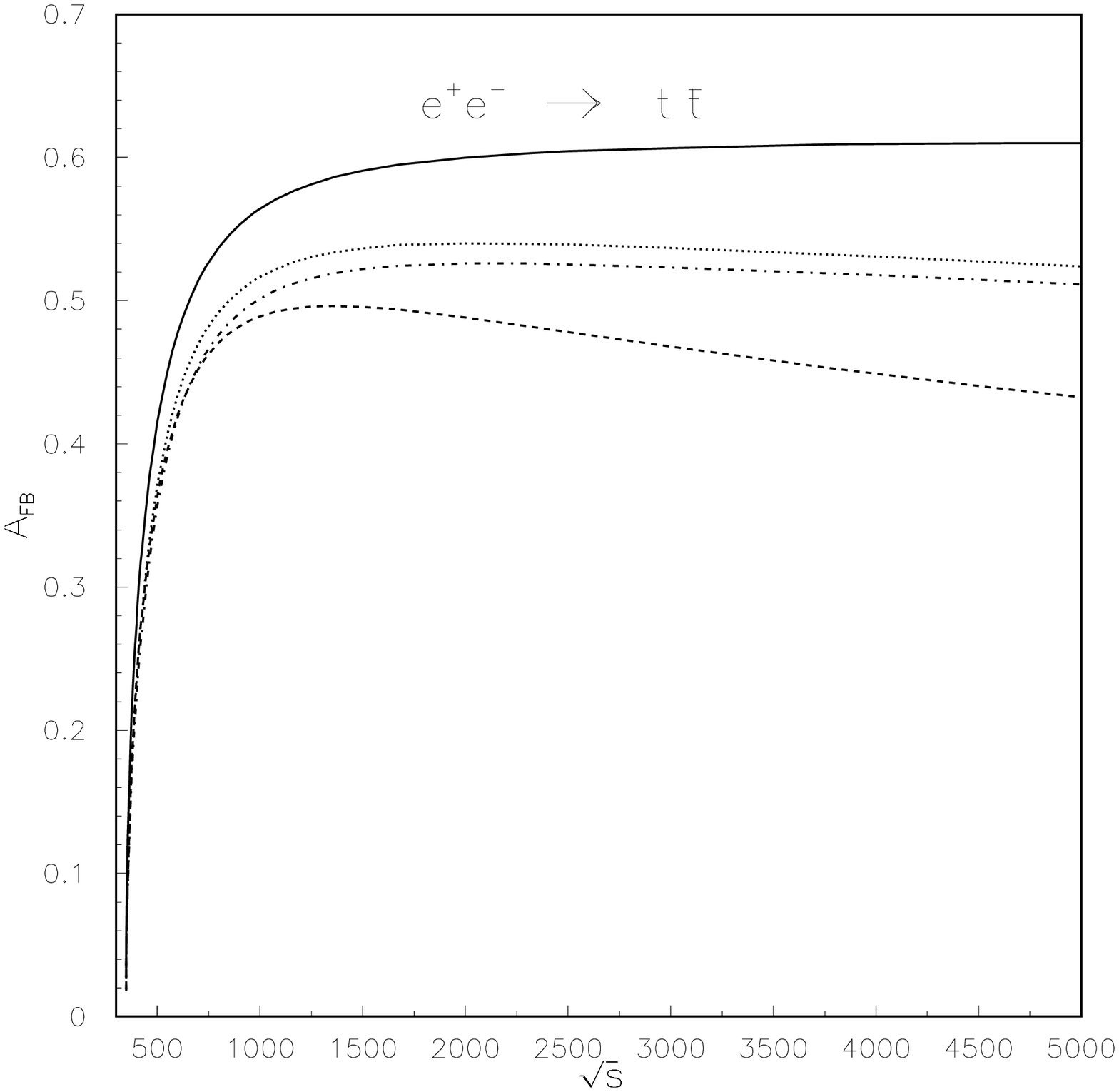,bbllx=20,
bblly=140,bburx=575,bbury=700,width=7.5cm}\\
\end{tabular} 
\end{center}
\caption{\label{eett}
Top-pair production: Differential cross sections (left) and
forward--backward asymmetries for $t \bar{t}$ pair-production in the Born
approximation (solid lines), and with full electroweak corrections (dashed
lines). Cross sections with a $s$~'=~0.7~$s$ cut are shown as dashed--dotted
lines. Asymmetries have $s'$~=~0.7~$s$- and a $\cos\theta$~=~0.95-cut
shown as dashed--dotted lines. The pure weak corrections (dotted lines,
photonic corrections and running of $\alpha$ excluded) for 
$\sqrt{s}$~=~3~TeV and 5~TeV are also shown for the cross sections.}
\end{figure}

\noindent
The  package LoopTools~\cite{Hahn:1998yk} has been  used for the 
numerical evaluation of the loop integrals. To illustrate the
increasingly sizeable effects of electroweak radiative 
corrections, the total cross section, the differential cross section and 
the forward--backward asymmetry are shown 
in~Figs.~\ref{figuretx} and~\ref{eett} for two typical CLIC
centre-of-mass energies, namely $\sqrt{s}$~=~3~TeV and $\sqrt{s}$~=~5~TeV. 

The total cross section of top-pair production decreases by almost one
order of magnitude over the considered energy range. This has the effect
that sizeable {\it relative} radiative corrections at higher energies
result in moderately sized {\it absolute} radiative corrections.
Considering the increasing luminosity at higher energies, the effect is
more or less compensated and again, for reliable predictions, the cross
sections have to be taken into account beyond leading order in $\alpha$.

The differential cross section is, as often, an observable containing more
detailed information about the underlying dynamics of the process.
Radiative corrections in the forward and backward directions have indeed
very different sizes, while the total cross section is mainly sensitive to
the dominating effects in the forward direction (see Fig.~\ref{eett}). In
particular, QED radiative corrections, which dominate the electroweak
correction (dashed lines) in~Fig.~\ref{eett} are the main contributor to
the overall radiative corrections. The clear message in the context of
CLIC physics is that, for all signal processes, a fixed-order calculation
will be insufficient and higher-order QED effects need to be taken into
account.


\newpage
\thispagestyle{empty}
~

\newpage
\chapter{HIGGS PHYSICS} 
\label{chapter:four}

\def\brique{}
\def\Black{}
\def\noir{}
\def\Blue{}
\def\Green{}
\def\rouge{}
\def\bleu{}
\def\vert{}

\newcommand{\beq}{\begin{equation}}
\newcommand{\eeq}{\end{equation}}

\newcommand{\beqn}{\begin{eqnarray}}
\newcommand{\eeqn}{\end{eqnarray}}
\newcommand{\ena}{\end{eqnarray}}
\newcommand{\ra}{\rightarrow}
\newcommand{\susy}{{{\cal SUSY}$\;$}}
\newcommand{\su}{$ SU(2) \times U(1)\,$}

\newcommand{\gag}{$\gamma \gamma$ }
\newcommand{\gagt}{\gamma \gamma }
\newcommand{\gam}{\gamma \gamma }
\def\W{{\mbox{\boldmath $W$}}}
\def\B{{\mbox{\boldmath $B$}}}
\def\V{{\mbox{\boldmath $V$}}}
\newcommand{\np}{Nucl.\,Phys.\,}
\newcommand{\pl}{Phys.\,Lett.\,}
\newcommand{\pr}{Phys.\,Rev.\,}
\newcommand{\prl}{Phys.\,Rev.\,Lett.\,}
\newcommand{\prep}{Phys.\,Rep.\,}
\newcommand{\zp}{Z.\,Phys.\,}
\newcommand{\sovjnp}{{\em Sov.\ J.\ Nucl.\ Phys.\ }}
\newcommand{\nuclinst}{{\em Nucl.\ Instrum.\ Meth.\ }}
\newcommand{\annp}{{\em Ann.\ Phys.\ }}
\newcommand{\intjmp}{{\em Int.\ J.\ of Mod.\  Phys.\ }}

\newcommand{\eps}{\epsilon}
\newcommand{\mw}{M_{W}}
\newcommand{\mww}{M_{W}^{2}}
\newcommand{\mwmw}{M_{W}^{2}}
\newcommand{\mhmh}{M_{H}^2}
\newcommand{\mz}{M_{Z}}
\newcommand{\mzz}{M_{Z}^{2}}

\newcommand{\cw}{\cos\theta_W}
\newcommand{\sw}{\sin\theta_W}
\newcommand{\tw}{\tan\theta_W}
\def\cww{\cos^2\theta_W}
\def\sww{\sin^2\theta_W}
\def\tww{\tan^2\theta_W}

\newcommand{\gev}{\,\, \mathrm{GeV}}
\newcommand{\fh}{{\em FeynHiggs}}
\newcommand{\br}{{\rm BR}}
\newcommand{\He}{h_1}
\newcommand{\Hd}{h_3}

\newcommand{\epm}{$e^{+} e^{-}\;$}
\newcommand{\epemt}{$e^{+} e^{-}\;$}
\newcommand{\epem}{e^{+} e^{-}\;}
\newcommand{\ememt}{$e^{-} e^{-}\;$}
\newcommand{\emem}{e^{-} e^{-}\;}

\newcommand{\lra}{\leftrightarrow}
\newcommand{\tr}{{\rm Tr}}
\def\ls1{{\not l}_1}
\newcommand{\cms}{centre-of-mass\hspace*{.1cm}}


\newcommand{\dkg}{\Delta \kappa_{\gamma}}
\newcommand{\dkz}{\Delta \kappa_{Z}}
\newcommand{\dz}{\delta_{Z}}
\newcommand{\dgz}{\Delta g^{1}_{Z}}
\newcommand{\dgzt}{$\Delta g^{1}_{Z}\;$}
\newcommand{\la}{\lambda}
\newcommand{\lag}{\lambda_{\gamma}}
\newcommand{\lambdae}{\lambda_{e}}
\newcommand{\laz}{\lambda_{Z}}
\newcommand{\lnl}{L_{9L}}
\newcommand{\lnr}{L_{9R}}
\newcommand{\lt}{L_{10}}
\newcommand{\lu}{L_{1}}
\newcommand{\ld}{L_{2}}
\newcommand{\eeww}{e^{+} e^{-} \ra W^+ W^- \;}
\newcommand{\eewwt}{$e^{+} e^{-} \ra W^+ W^- \;$}
\newcommand{\epemww}{e^{+} e^{-} \ra W^+ W^- }
\newcommand{\epemwwt}{$e^{+} e^{-} \ra W^+ W^- \;$}
\newcommand{\eennhht}{$e^{+} e^{-} \ra \nu_e \bar \nu_e HH\;$}
\newcommand{\eennhh}{e^{+} e^{-} \ra \nu_e \bar \nu_e HH\;}
\newcommand{\ppwg}{p p \ra W \gamma}
\newcommand{\wwhh}{W^+ W^- \ra HH\;}
\newcommand{\wwhht}{$W^+ W^- \ra HH\;$}
\newcommand{\ppwz}{pp \ra W Z}
\newcommand{\ppwgt}{$p p \ra W \gamma \;$}
\newcommand{\ppwzt}{$pp \ra W Z \;$}
\newcommand{\gamgamt}{$\gamma \gamma \;$}
\newcommand{\gamgam}{\gamma \gamma \;}
\newcommand{\egamt}{$e \gamma \;$}
\newcommand{\egam}{e \gamma \;}
\newcommand{\gamgamwwt}{$\gamma \gamma \ra W^+ W^- \;$}
\newcommand{\gamgamwwht}{$\gamma \gamma \ra W^+ W^- H \;$}
\newcommand{\gamgamwwh}{\gamma \gamma \ra W^+ W^- H \;}
\newcommand{\gamgamwwhht}{$\gamma \gamma \ra W^+ W^- H H\;$}
\newcommand{\gamgamwwhh}{\gamma \gamma \ra W^+ W^- H H\;}
\newcommand{\ggww}{\gamma \gamma \ra W^+ W^-}
\newcommand{\ggwwt}{$\gamma \gamma \ra W^+ W^- \;$}
\newcommand{\ggwwht}{$\gamma \gamma \ra W^+ W^- H \;$}
\newcommand{\ggwwh}{\gamma \gamma \ra W^+ W^- H \;}
\newcommand{\ggwwhht}{$\gamma \gamma \ra W^+ W^- H H\;$}
\newcommand{\ggwwhh}{\gamma \gamma \ra W^+ W^- H H\;}
\newcommand{\ggwwz}{\gamma \gamma \ra W^+ W^- Z\;}
\newcommand{\ggwwzt}{$\gamma \gamma \ra W^+ W^- Z\;$}

\newcommand{\ptu}{p_{1\bot}}
\newcommand{\vecptu}{\vec{p}_{1\bot}}
\newcommand{\ptd}{p_{2\bot}}
\newcommand{\vecptd}{\vec{p}_{2\bot}}
\newcommand{\ie}{{\em i.e.}}
\newcommand{\cm}{{{\cal M}}}
\newcommand{\cl}{{{\cal L}}}
\newcommand{\cd}{{{\cal D}}}
\newcommand{\cv}{{{\cal V}}}
\def\slashc{c\kern -.400em {/}}
\def\slashp{p\kern -.400em {/}}
\def\slashL{L\kern -.450em {/}}
\def\slashcl{\cl\kern -.600em {/}}
\def\Ww{{\mbox{\boldmath $W$}}}
\def\B{{\mbox{\boldmath $B$}}}
\def\noi{\noindent}
\def\nn{\noindent}
\def\sm{${\cal{S}} {\cal{M}}\;$}
\def\smn{${\cal{S}} {\cal{M}}$}
\def\nph{${\cal{N}} {\cal{P}}\;$}
\def\sb{$ {\cal{S}}  {\cal{B}}\;$}
\def\ssb{${\cal{S}} {\cal{S}}  {\cal{B}}\;$}
\def\ssbe{{\cal{S}} {\cal{S}}  {\cal{B}}}
\def\cviol{${\cal{C}}\;$}
\def\pviol{${\cal{P}}\;$}
\def\cpviol{${\cal{C}} {\cal{P}}\;$}

\newcommand{\lgg}{\lambda_1\lambda_2}
\newcommand{\lww}{\lambda_3\lambda_4}
\newcommand{\ppin}{ P^+_{12}}
\newcommand{\pmin}{ P^-_{12}}
\newcommand{\ppout}{ P^+_{34}}
\newcommand{\pmout}{ P^-_{34}}
\newcommand{\sinsq}{\sin^2\theta}
\newcommand{\cossq}{\cos^2\theta}
\newcommand{\yt}{y_\theta}
\newcommand{\hppll}{++;00}
\newcommand{\hpmll}{+-;00}
\newcommand{\hpplt}{++;\lambda_30}
\newcommand{\hpmlt}{+-;\lambda_30}
\newcommand{\hpptt}{++;\lambda_3\lambda_4}
\newcommand{\hpmtt}{+-;\lambda_3\lambda_4}
\newcommand{\dk}{\Delta\kappa}
\newcommand{\klam}{\Delta\kappa \lambda_\gamma }
\newcommand{\kac}{\Delta\kappa^2 }
\newcommand{\lac}{\lambda_\gamma^2 }
\def\gamgamtzz{$\gamma \gamma \ra ZZ \;$}
\def\gamgamtww{$\gamma \gamma \ra W^+ W^-\;$}
\def\gamgamtwwe{\gamma \gamma \ra W^+ W^-}
\def\sinb{\sin\beta}
\def\cosb{\cos\beta}
\def\sinbb{\sin (2\beta)}
\def\cosbb{\cos (2 \beta)}
\def\tgb{\tan \beta}
\def\tgbt{$\tan \beta\;\;$}
\def\tgbsq{\tan^2 \beta}
\def\sinal{\sin\alpha}
\def\cosal{\cos\alpha}
\def\sinb{\sin\beta}
\def\cosb{\cos\beta}
\def\sinbb{s_ {2\beta}}
\def\cosbb{c_{2 \beta}}
\def\tgb{\tan \beta}
\def\tgbt{$\tan \beta\;\;$}
\def\tgbsq{\tan^2 \beta}
\def\tgbsqt{$\tan^2 \beta\;$}
\def\sinal{\sin\alpha}
\def\cosal{\cos\alpha}
\def\sb{s_\beta}
\def\cb{c_\beta}
\def\tb{t_\beta}
\def\ttb{t_{2 \beta}}
\def\sa{s_\alpha}
\def\ca{c_\alpha}
\def\ta{t_\alpha}
\def\stb{s_{2\beta}}
\def\ctb{c_{2\beta}}
\def\sta{s_{2\alpha}}
\def\cta{c_{2\alpha}}
\def\sbma{s_{\beta-\alpha}}
\def\cbma{c_{\beta-\alpha}}
\def\sbpa{s_{\beta+\alpha}}
\def\cbpa{c_{\beta+\alpha}}
\def\lone{\lambda_1}
\def\ltwo{\lambda_2}
\def\lthree{\lambda_3}
\def\lfour{\lambda_4}
\def\lfive{\lambda_5}
\def\lsix{\lambda_6}
\def\lseven{\lambda_7}
\def\stop{\tilde{t}}
\def\sto{\tilde{t}_1}
\def\stt{\tilde{t}_2}
\def\stl{\tilde{t}_{\rm L}}
\def\str{\tilde{t}_{\rm R}}
\def\msto{m_{\sto}}
\def\mstosq{m_{\sto}^2}
\def\mstt{m_{\stt}}
\def\msttsq{m_{\stt}^2}
\def\mt{m_t}
\def\mtsq{m_t^2}
\def\sint{\sin\theta_{\stop}}
\def\sintt{\sin 2\theta_{\stop}}
\def\cost{\cos\theta_{\stop}}
\def\sintsq{\sin^2\theta_{\stop}}
\def\costsq{\cos^2\theta_{\stop}}
\def\mqtt{\M_{\tilde{Q}_3}^2}
\def\mutt{\M_{\tilde{U}_{3R}}^2}
\def\sbottom{\tilde{b}}
\def\sbo{\tilde{b}_1}
\def\sbt{\tilde{b}_2}
\def\sbl{\tilde{b}_L}
\def\sbr{\tilde{b}_R}
\def\msbo{m_{\sbo}}
\def\msbosq{m_{\sbo}^2}
\def\msbt{m_{\sbt}}
\def\msbtsq{m_{\sbt}^2}
\def\mt{m_t}
\def\mtsq{m_t^2}
\def\selectron{\tilde{e}}
\def\seo{\tilde{e}_1}
\def\set{\tilde{e}_2}
\def\sel{\tilde{e}_L}
\def\ser{\tilde{e}_R}
\def\mseo{m_{\seo}}
\def\mseosq{m_{\seo}^2}
\def\mset{m_{\set}}
\def\msetsq{m_{\set}^2}
\def\msel{m_{\sel}}
\def\mser{m_{\ser}}
\def\me{m_e}
\def\mesq{m_e^2}
\def\snu{\tilde{\nu}}
\def\snue{\tilde{\nu_e}}
\def\set{\tilde{e}_2}
\def\snul{\tilde{\nu}_L}
\def\msnue{m_{\snue}}
\def\msnuesq{m_{\snue}^2}
\def\smuon{\tilde{\mu}}
\def\smul{\tilde{\mu}_L}
\def\smur{\tilde{\mu}_R}
\def\msmul{m_{\smul}}
\def\msmulsq{m_{\smul}^2}
\def\msmur{m_{\smur}}
\def\msmursq{m_{\smur}^2}
\def\stau{\tilde{\tau}}
\def\stauo{\tilde{\tau}_1}
\def\staut{\tilde{\tau}_2}
\def\staul{\tilde{\tau}_L}
\def\staur{\tilde{\tau}_R}
\def\mstauo{m_{\stauo}}
\def\mstauosq{m_{\stauo}^2}
\def\mstaut{m_{\staut}}
\def\mstautsq{m_{\staut}^2}
\def\mtau{m_\tau}
\def\mtausq{m_\tau^2}
\def\gluino{\tilde{g}}
\def\mgluino{m_{\tilde{g}}}
\def\mchi{m_\chi^+}
\def\neuto{\tilde{\chi}_1^0}
\def\mneuto{m_{\tilde{\chi}_1^0}}
\def\neutt{\tilde{\chi}_2^0}
\def\mneutt{m_{\tilde{\chi}_2^0}}
\def\neutth{\tilde{\chi}_3^0}
\def\mneutth{m_{\tilde{\chi}_3^0}}
\def\neutf{\tilde{\chi}_4^0}
\def\mneutf{m_{\tilde{\chi}_4^0}}
\def\chargop{\tilde{\chi}_1^+}
\def\mchargo{m_{\tilde{\chi}_1^+}}
\def\chargtp{\tilde{\chi}_2^+}
\def\mchargt{m_{\tilde{\chi}_2^+}}
\def\chargom{\tilde{\chi}_1^-}
\def\chargtm{\tilde{\chi}_2^-}
\def\bino{\tilde{b}}
\def\wino{\tilde{w}}
\def\photino{\tilde{\gamma}}
\def\zino{tilde{z}}
\def\sdowno{\tilde{d}_1}
\def\sdownt{\tilde{d}_2}
\def\sdownl{\tilde{d}_L}
\def\sdownr{\tilde{d}_R}
\def\supo{\tilde{u}_1}
\def\supt{\tilde{u}_2}
\def\supl{\tilde{u}_L}
\def\supr{\tilde{u}_R}
\def\mh{m_h}
\def\mht{m_h^2}
\def\MH{M_H}
\def\MHt{M_H^2}
\def\MA{M_A}
\def\MAt{M_A^2}
\def\MHp{M_H^+}
\def\MHm{M_H^-}
\def\mqt{\M_{\tilde{Q}_3}}
\def\mut{\M_{\tilde{U}_{3R}}}
\def\mqtz{\M_{\tilde{Q}_3(0)}}
\def\mutz{\M_{\tilde{U}_{3R}(0)}}
\def\mqtzt{\M_{\tilde{Q}_3^2(0)}}
\def\mutzt{\M_{\tilde{U}_{3R}^2(0)}}
\def\stw{s_{2w}}
\def\sbb{s_ {2\beta}}
\def\cbb{c_{2 \beta}}

Understanding the origin of electroweak symmetry breaking and mass
generation will be a central theme of the research programme in particle
physics over the coming decades. The Standard Model (SM), successfully
tested to an unprecedented level of accuracy by the LEP, SLC and Tevatron
experiments, and now also by the $B$ factories, achieves electroweak
symmetry breaking and mass generation via the Higgs
mechanism~\cite{Higgs}.  The primary direct manifestation of the Higgs
mechanism is the existence of at least one Higgs boson, $H^0$. The
observation of a new spin-0 particle would represent the first sign that
the Higgs mechanism of mass generation is realized in nature. Data from 
the direct Higgs search at LEP indicate that it is heavier than 
114.4~GeV~\cite{LEPHiggsWG}, and precision electroweak data 
suggest that it is probably lighter than about
212~GeV~\cite{ichep}. The discovery of the Higgs boson may be made at the
FNAL Tevatron~\cite{TevatronHiggsWG}, or may have to wait for the 
LHC, the CERN
hadron collider currently under construction. Experiments at the LHC
will determine the mass of the Higgs boson and perform a first survey of
its basic properties, measuring a few of its couplings~\cite{LHCHiggs}. A 
TeV-class linear
collider, operating at centre-of-mass energies 350~GeV$< \sqrt{s}
\le$~1~TeV, will provide the accuracy needed to validate further the Higgs
mechanism and to probe the SM or the possible extended nature of the Higgs
sector. It will perform crucial measurements in a model-independent way
and it will also complement the LHC in the search for the heavy
Higgs bosons expected in extensions of the SM~\cite{LCHiggs}. Their 
observation
would provide direct evidence that nature has chosen a route different
from the minimal Higgs sector of the SM.

Neither the precision study of the Higgs profile nor the search for
additional Higgs bosons will be completed at energies below 1~TeV. There
are measurements that will be limited in accuracy or may be not feasible
at all at the LHC or a TeV-class linear collider, owing to limitations
in both the available statistics and the centre-of-mass energy. CLIC
represents a unique opportunity to probe further the Higgs sector in
$e^+e^-$ collisions, complementing the information that the LHC and
a TeV-class linear collider will obtain. Thus, CLIC may complete our
understanding of the origin of electroweak symmetry breaking and mass
generation.

\section{Completing the Light Higgs Boson Profile}

The TeV-class LC will perform highly accurate determinations of the Higgs
profile~\cite{LCHiggs}. However, even at the high design luminosities of 
TESLA, the JLC and the NLC, there are properties of even a light 
Higgs boson that cannot be tested exhaustively.

\begin{figure}[t] 
\begin{center}
\epsfig{file=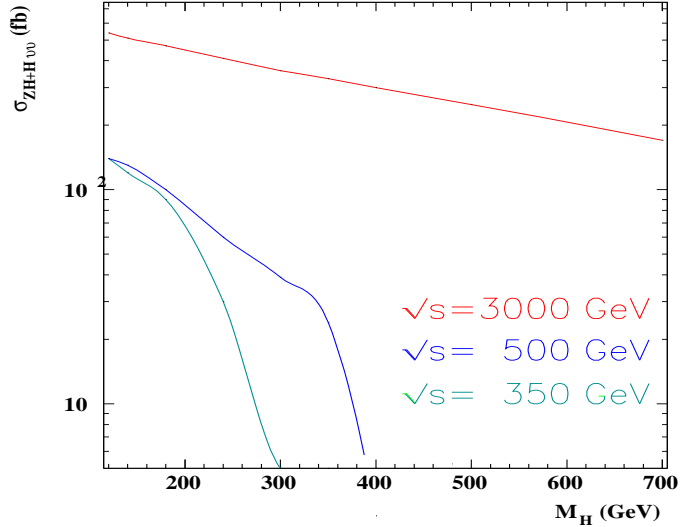,width=9cm,height=7cm,clip}
\end{center} 
\caption{Inclusive Higgs production cross section as a function of the 
Higgs mass $M_H$ for three values of the $e^+e^-$ centre-of-mass
energy $\sqrt{s}$}  
\label{fig:xsec}
\end{figure}

A fundamental test of the Higgs mechanism in the Standard Model is the
predicted scaling of the Higgs couplings to fermions, $g_{Hff}$, with
their masses, $M_f$: $\frac{g_{Hff}}{g_{Hf'f'}} \propto
\frac{M_f}{M_{f'}}$. This must be checked with good accuracy for all
particle species, whatever the Higgs boson mass. A LC with
$\sqrt{s}$~=~350--500~GeV will be able to test exhaustively the Higgs
couplings to gauge bosons and quarks, {\it if} the Higgs boson is light.
To complete this programme for Higgs couplings to leptons, to study the
couplings of intermediate-mass Higgs bosons, and to study Higgs
self-couplings, it is necessary to study rare processes, which need Higgs
samples in excess of 10$^5$ events. As the cross section for $e^+e^-
\rightarrow H^0 \nu \bar \nu$ production increases with energy as
$\log~\frac{s}{M^2_H}$, this process dominates at CLIC energies. The
resulting large Higgs production rate at $\sqrt{s} \ge$~3~TeV seen in 
Fig.~\ref{fig:xsec} yields
samples of the order of (0.5--1)~$\times$~10$^6$ decays of a Standard Model
Higgs boson in 1--2 years at the design luminosity of 
$L$~=~10$^{35}$~cm$^{-2}$s$^{-1}$.

\subsection{\boldmath{$H \to \mu^+\mu^-$} }

\begin{figure}[htbp] 
\begin{center} 
\begin{tabular}{c c}
\epsfig{file=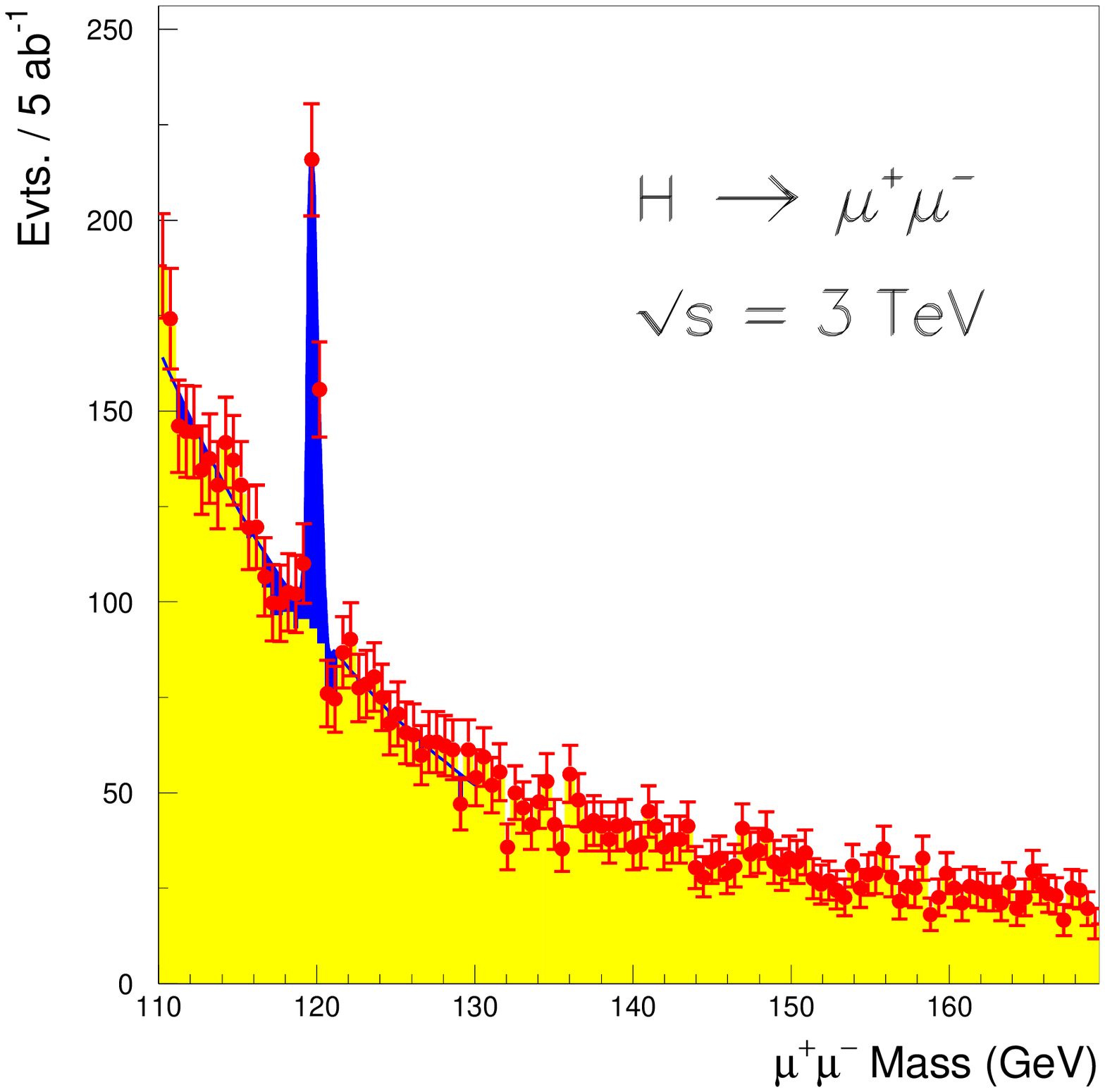,width=7.7cm,height=5.7cm,clip} &
\epsfig{file=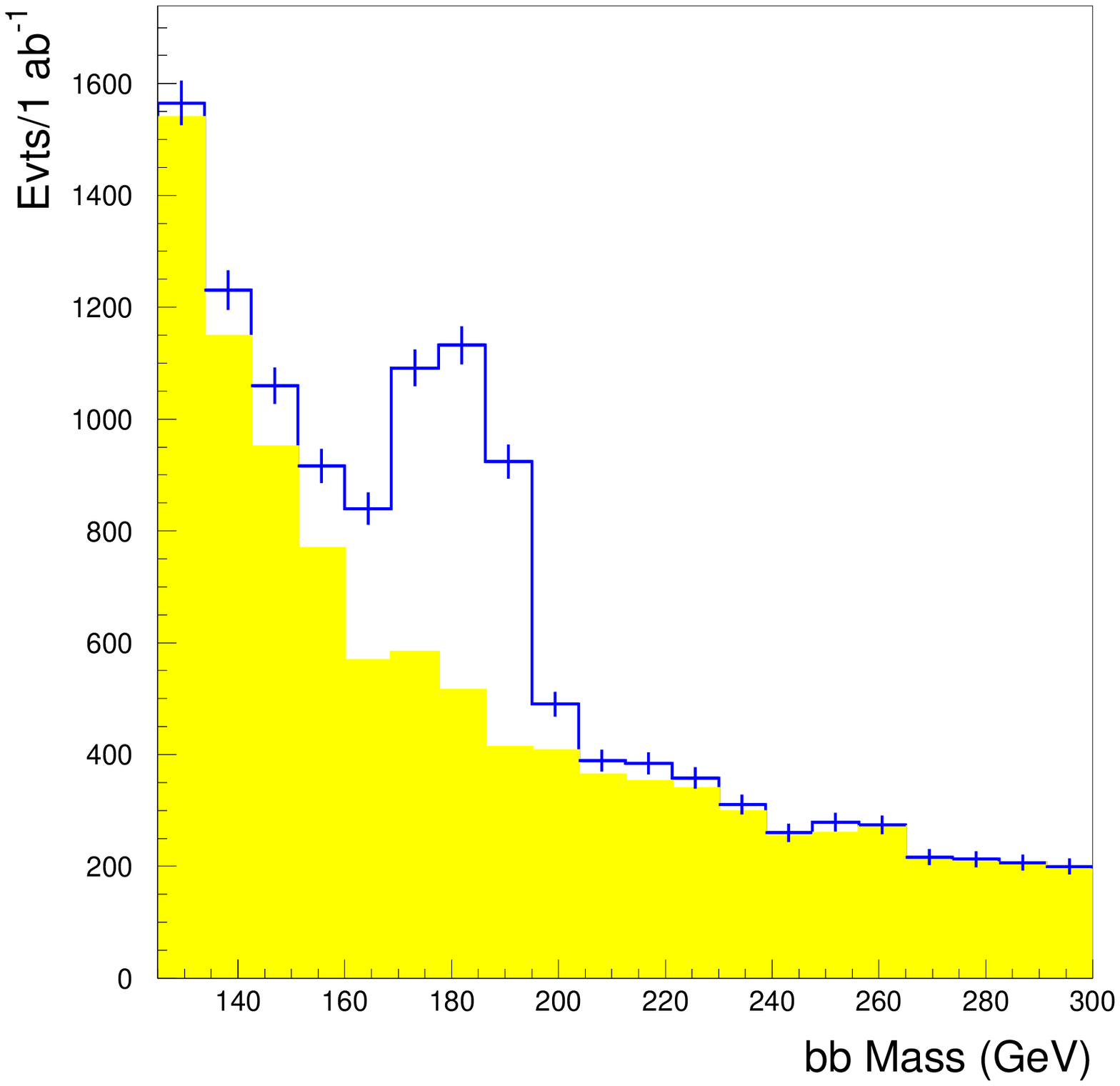,width=7.7cm,height=5.7cm} \\
\end{tabular} 
\end{center} 
\caption{Reconstructed signals for $H^0 \to \mu^+\mu^-$ (left) and
$H^0 \to b\bar{b}$ (right) for $M_H$~=~120~GeV and 200~GeV,
respectively, at $\sqrt{s}$~=~3~TeV} 
\label{fig:rare} 
\end{figure}

Measuring the muon Yukawa coupling by the determination of the $H^0
\rightarrow \mu^+\mu^-$ branching fraction would complete the test of the
scaling of the Higgs couplings to gauge bosons, quarks and leptons
separately, ensuring that the observed Higgs boson is indeed responsible
for the masses of all types of elementary particles. A data set of
3~ab$^{-1}$ at 3~TeV corresponds to 400 $H \rightarrow \mu\mu$ decays, for
$M_H$~=~120~GeV, assuming Standard Model couplings. The main background from
$WW \rightarrow \mu \nu \mu \nu$ can be reduced by cuts on the dimuon
recoil mass and energy. The total background has been estimated including
also the $ZZ\nu\bar\nu$, $WW\nu\bar\nu$ and the inclusive
$\mu\mu\nu\bar\nu$ processes, evaluated without the Higgs contribution.  
The resulting dimuon invariant mass for all particle species is shown in
Fig.~\ref{fig:rare}. A signal significance of more than $5\sigma$ is
obtained up to $M_H \simeq$~155~GeV. The number of signal events is
extracted from a fit to the dimuon invariant mass, where the signal is
modelled by a Gaussian distribution peaked at the nominal Higgs mass and
the background by a polynomial curve fitted on the peak side bands.  The
accuracies in the product of production cross section and $\mu\mu$ decay
branching fraction, derived from the fitted number of signal events, are
summarized in Table~\ref{tab:tab1} for different values of $M_H$.

\begin{table}[htbp]  
\caption{Statistical accuracy on the $g_{H\mu\mu}$ coupling expected
with an integrated luminosity of 3~ab$^{-1}$ at $\sqrt{s}$~=~3~TeV,
for different values of the Higgs boson mass, $M_H$} 
\label{tab:tab1} 

\renewcommand{\arraystretch}{1.5} 
\begin{center}

\begin{tabular}{lccc}\hline \hline \\[-4mm]
\boldmath{$M_H$} & 
$\hspace*{4mm}$ 120~GeV $\hspace*{4mm}$ & 
$\hspace*{4mm}$ 140~GeV $\hspace*{4mm}$ & 
$\hspace*{4mm}$ 150~GeV $\hspace*{4mm}$ \\ 
\boldmath{$\delta g_{H\mu\mu}/g_{H\mu\mu}$} $\hspace*{14mm}$
& 0.042 & 0.065 & 0.110 
\\[3mm] 
\hline \hline
\end{tabular}
\end{center}
\end{table}


These accuracies are comparable to those achievable with a muon
collider operating at the Higgs mass peak~\cite{MC}, and more than a
factor of 2
better than those expected at a very large hadron collider (VLHC) with a 
centre-of-mass energy of 200~TeV~\cite{VLHC}.  

\subsection{\boldmath{$H \to b \bar{b}$} for an Intermediate-Mass Higgs Boson}

The scaling of the Higgs couplings to fermions also needs to be tested for
Higgs bosons of intermediate mass. Beyond the $H \to WW$ threshold, the
branching fractions $H \to f \bar{f}$ fall rapidly with increasing $M_H$
values. The process $e^+e^- \rightarrow \nu \bar \nu H \rightarrow b \bar
b$ at $\sqrt{s} \ge$~1~TeV still offers a favourable signal-to-background
ratio for probing $g_{Hbb}$ for these intermediate-mass Higgs bosons. The
$H^0 \to b \bar{b}$ decay can be measured for masses up to about 240~GeV,
with the accuracies summarized in~Table~\ref{tab:bb}.
%
\begin{table}[htbp]  
\caption{Signal significance and relative accuracy for the
determination of the $b$ Yukawa coupling for different values of the
$M_H$ mass at $\sqrt{s}$~=~3~TeV} 
\label{tab:bb}

\renewcommand{\arraystretch}{1.4} 
\begin{center}

\begin{tabular}{cccc}\hline \hline \\[-4mm]
\boldmath{$M_H$} \textbf{(GeV)} & $\hspace*{3mm}$ &
$\hspace*{5mm}$ \boldmath{$S/\sqrt{B}$}  $\hspace*{5mm}$ & 
$\hspace*{5mm}$ \boldmath{$\delta~g_{Hbb}~/~g_{Hbb}$}  $\hspace*{5mm}$ 
\\[4mm]   

\hline \\[-3mm]
180 & & 40.5 & 0.016 \\
200 & & 25.0 & 0.025 \\
220 & & 18.0 & 0.034
\\[3mm] 
\hline \hline
\end{tabular}
\end{center}
\end{table}


These accuracies are comparable to those expected for a light Higgs at a
300--500~GeV~LC. They will ensure an accurate test of the Yukawa coupling
to quarks for masses up to the limit set by the electroweak data for the
Standard Model Higgs boson.

\subsection{Triple Higgs Coupling and Reconstruction of the Higgs Potential}

Another fundamental test of the Higgs sector with a light Higgs boson,
which would benefit significantly from multi-TeV data, is the 
study of the
Higgs self-couplings and the reconstruction of the Higgs potential. We
recall that symmetry breaking in the Standard Model is completely encoded
in the scalar Higgs potential.  With the assumption of one Higgs doublet
$\Phi$, the most general potential takes the 
form~\cite{ChopinHiggs,Barger:2003rs}:
\beqn
\label{generalpotential2} V_{\ssbe}= \lambda \left\{ \sum_{n=2}
\frac{g^{2(n-2)}}{\Lambda^{2(n-2)}} \frac{a_n}{(n-1)^2} \left[
\Phi^\dagger \Phi - \frac{v^2}{2} \right]^{n} \right\}\,,
\eeqn
in order that spontaneous symmetry breaking ensues with the correct value
of the vacuum expectation value that gives the gauge bosons correct
masses. The leading dimension-4 operator with $n$~=~2 above corresponds to
the minimal Standard Model realization. In particular, this fixes the mass
of the Higgs $M_H$ and its self-couplings in terms of a single parameter,
$\lambda$. Higher-order operators may change the relationship between the
Higgs mass and its self-couplings, as well as the couplings to the
Goldstone bosons. Taking into account dimension-6 operators, and denoting
the Goldstone bosons by $\phi_3$ and $\phi^\pm$ and the Higgs boson with
$H$, Eq.~(\ref{generalpotential2}) leads to:
\beqn
\label{generalpotentiald} V_{\ssbe}&=&\frac{1}{2} M_H^2 \left\{
H^2 + \frac{g}{M_W} H \, \left (\Phi^+ \Phi^-
+\frac{\Phi_3^2}{2} \right ) + \,
\frac{g}{2 M_W} \, h_3 H^3 \, + \right. \nonumber \\
&& \;\;\;\;\;\left. h_4 \left( \frac{g}{4 M_W} \right)^2 H^4 +
h_3' \left( \frac{g}{2 M_W} \right)^2  H^2 \left (\Phi^+ \Phi^-
+ \frac{\Phi_3^2}{2} \right ) ...\right\}  \\
h_3&\equiv& g_{HHH}=1+ a_3 \frac{M_W^2}{\Lambda^2}=1+\delta
h_3\,, \quad h_4\equiv g_{HHHH}=1+ 6\delta h_3\,, \quad h_3'=1+
3\delta h_3\;.
\nonumber
\eeqn
Probing the Higgs potential amounts to measuring the self-couplings of the
Higgs, of which the most accessible is the trilinear coupling.

The triple-Higgs coupling $g_{HHH}$ can be accessed at a TeV-class LC in
the double-Higgs production process 
$e^+e^- \to H H Z$~\cite{Djouadi:1999gv}. 
However, this measurement is made difficult by
the tiny production cross section and by the dilution due to diagrams,
leading to double Higgs production, that are not sensitive to the triple
Higgs vertex.  It has been concluded that a LC operating at $\sqrt{s}$ =
500~GeV can measure the $HHZ$ production cross section to about 15\%
accuracy if the Higgs boson mass is 120~GeV, corresponding to a fractional
accuracy of 23\% in $g_{HHH}$~\cite{Castanier:2001sf}.

\begin{figure}[htbp] 
\begin{center}
\begin{tabular}{c c}
\epsfig{file=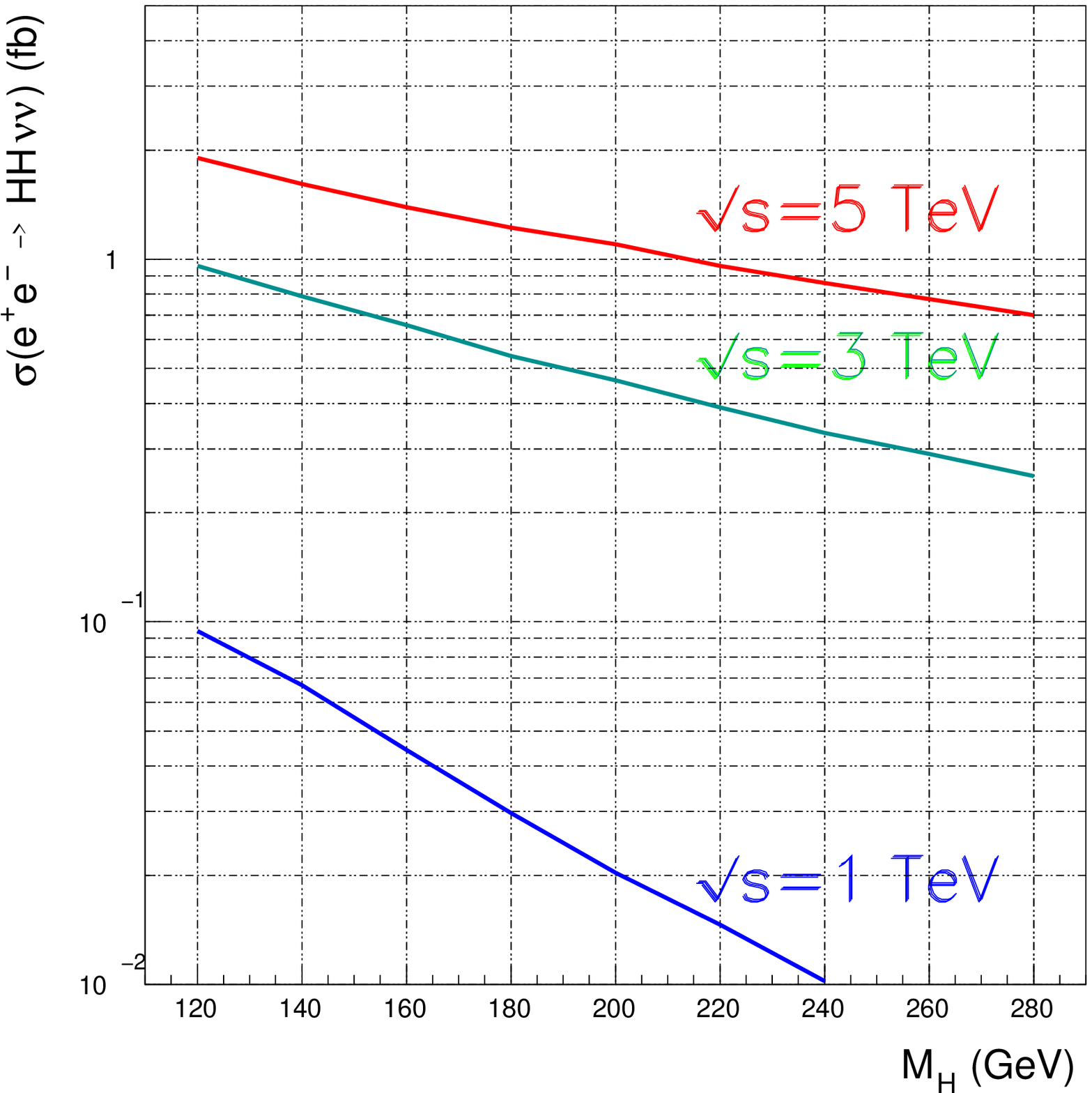,width=7.65cm,height=6.9cm,clip} &
\epsfig{file=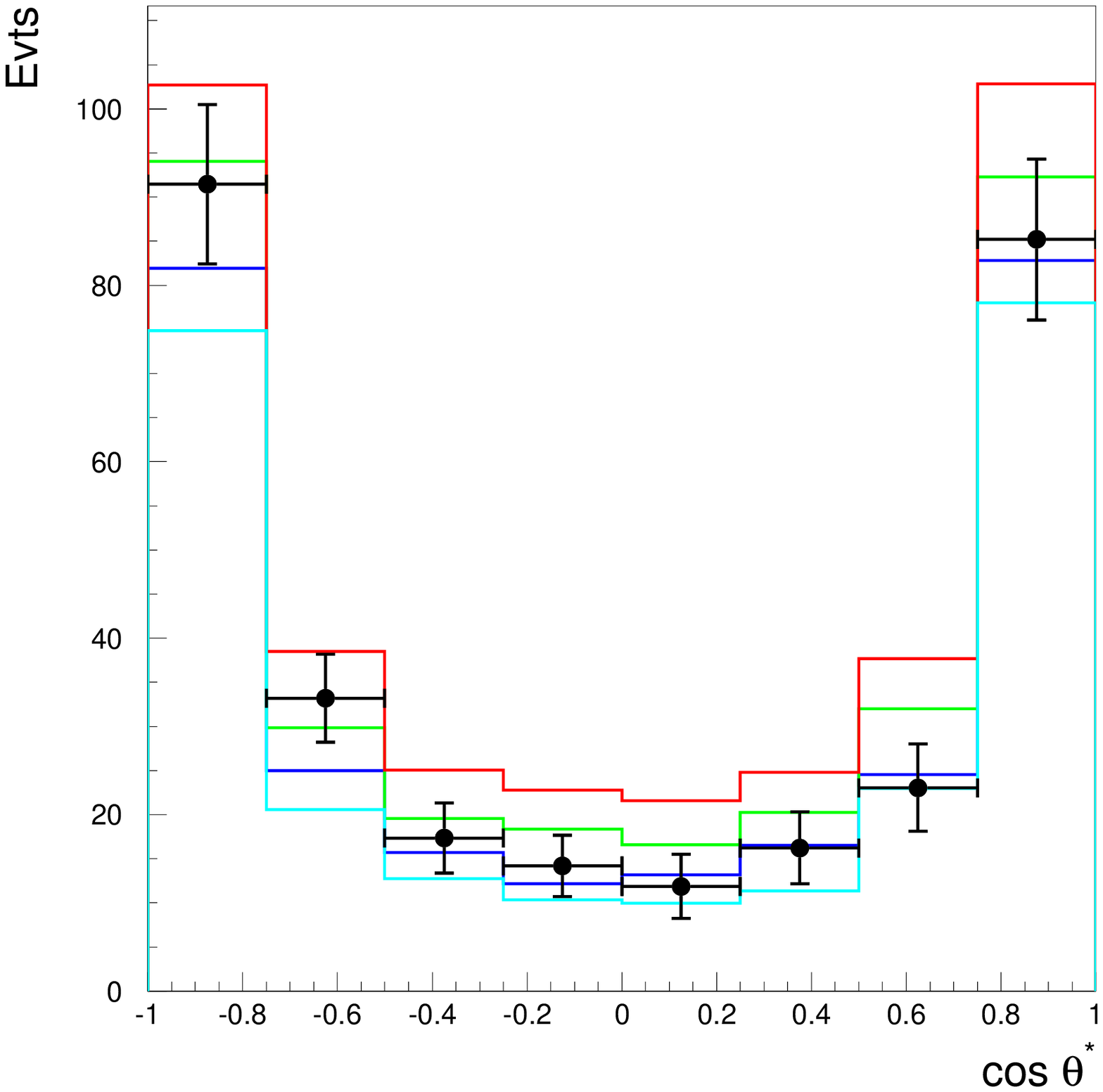,height=7.85cm,width=7.5cm,clip} \\
\end{tabular}
\end{center}
\caption{Double Higgs production at CLIC: cross section for the 
$e^+e^- \to HH \nu \bar{\nu}$ process as a function of the Higgs boson
mass for different centre-of-mass energies (left) and the $|\cos \theta^*|$ 
distributions for different values of the triple Higgs coupling
(right). A combined fit to the cross section and the 
decay angle distribution is used to extract the triple Higgs coupling.}
\label{fig:hhnn}
\end{figure}

This accuracy can be improved by performing the analysis at multi-TeV
energies, as seen in~Fig.~\ref{fig:hhnn}, 
using the process $e^+e^- \to H H \nu \bar{\nu}$ and
introducing observables sensitive to the presence of the triple-Higgs
vertex. The $H^* \to HH$ decay angle $\theta^*$ in the $H^*$ rest frame,
which is sensitive to the scalar nature of the Higgs boson, has been
adopted for the $H H \nu \bar{\nu}$ process. The resulting distribution in
$|\cos \theta^*|$ is flat for the signal, while it is peaked forward for
the other diagrams leading to the $H H \nu \bar{\nu}$ final state. Results
are summarized in Table~\ref{tab:hhh}~\cite{Battaglia:2001nn}.  No
polarization has been included in this study. However, the double Higgs
production for polarized beams is four times larger, indicating a further
potential improvement of the accuracy by a factor of 2.  
Assuming that only some fraction of the time favorable polarized beams
will be used, the precision is on the triple-Higgs coupling will be
in the range of 0.07 to 0.09.
%
\begin{table}[htbp]  
\caption{Relative accuracy for the determination of the 
triple-Higgs coupling $g_{HHH}$ for different values of the $M_H$ mass at 
$\sqrt{s}$~=~3~TeV, assuming unpolarized beams}
\label{tab:hhh}

\renewcommand{\arraystretch}{1.4} 
\begin{center}

\begin{tabular}{cccc}\hline \hline \\[-4mm]
\boldmath{$M_H$} \textbf{(GeV)} & $\hspace*{3mm}$ &
$\hspace*{9mm}$ \textbf{Counting}  $\hspace*{9mm}$ & 
$\hspace*{9mm}$ \textbf{Fit}  $\hspace*{9mm}$ 
\\[4mm]   

\hline \\[-3mm]
120 & & $\pm$~0.131 (stat) & $\pm$~0.093 (stat) \\
180 & & $\pm$~0.191 (stat) & $\pm$~0.115 (stat)
\\[3mm] 
\hline \hline
\end{tabular}
\end{center}
\end{table}

\begin{figure}[!h] 
\begin{center}
\epsfig{file=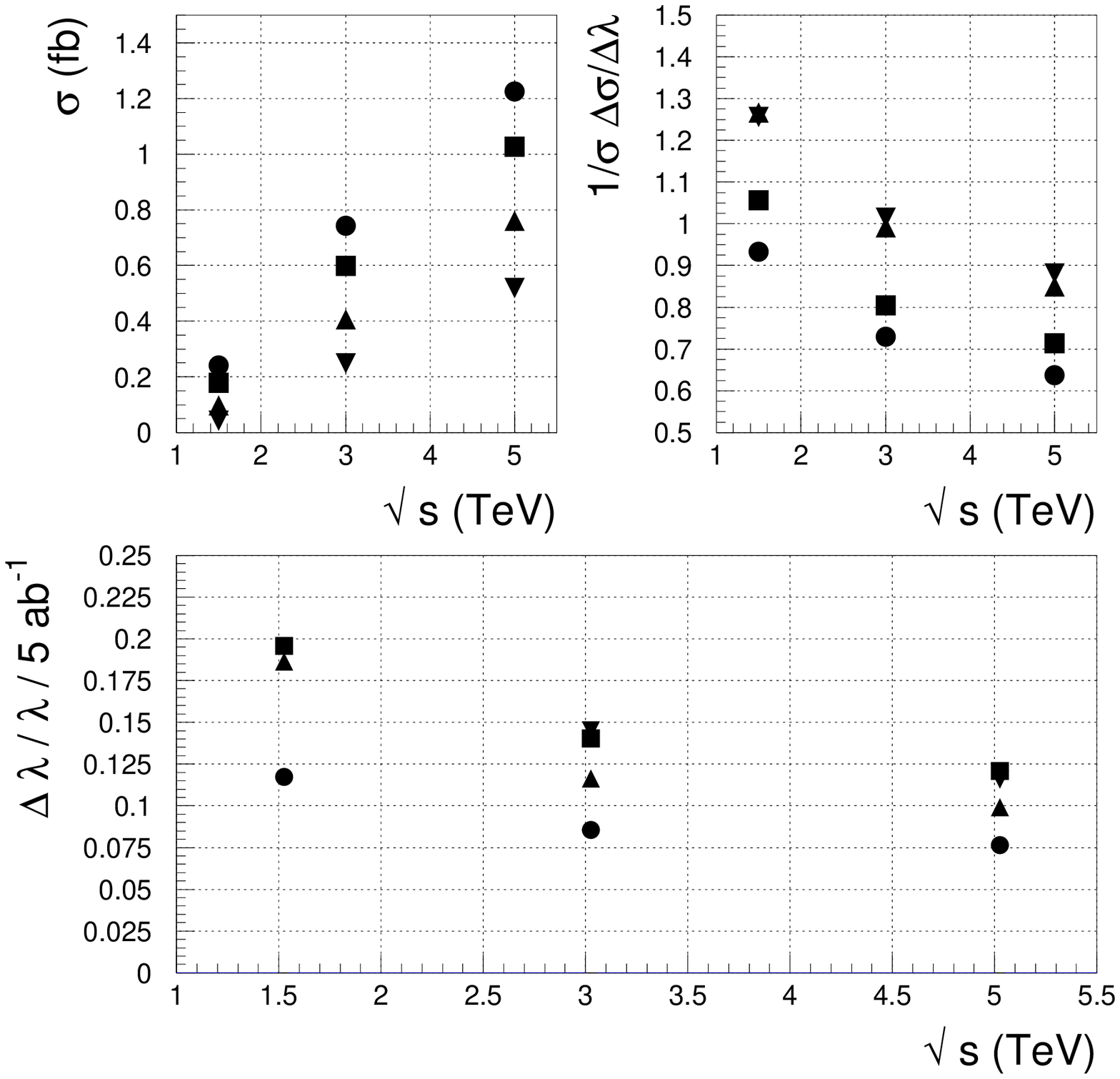,width=8.2cm} 
\end{center}

\vspace*{-5mm}

\caption{The cross section of the processes $e^+e^- \rightarrow H\nu\nu$, 
the sensitivity of the cross section to the triple 
Higgs coupling, and the expected precision with which  the triple
Higgs coupling, for the masses 120~GeV (circle), 140~GeV (box), 180~GeV
(traingle), and 240~GeV (inverse triangle), for 5 ab$^{-1}$}
\label{sec4:HHHall}
\end{figure}

A qualitative study of the channel $e^+e^- \rightarrow H\nu\nu$ 
was further made
at the generator level, taking into account both background and signal,
for a wider range of Higgs masses (120 to 240 GeV) and several 
CMS energies (1.5 to 5~TeV). Figure~\ref{sec4:HHHall} shows the cross section 
of the processes, the sensitivity of the cross section to the triple 
Higgs coupling, and and the expected precision with which  the triple
Higgs coupling can be determined. The calculations include the branching 
ratios into the relevant channels, and a reconstruction efficiency
as determined for 3~TeV, using the fast simulation. The assumed
integrated luminosity is 5 ab$^{-1}$.

The figure shows that for all Higgs masses concerned, a 3~TeV CLIC will 
give a measurement of the triple Higgs coupling 
with a precision of 10--15\%. If the Higgs is relatively 
heavy one can gain in precision by operating CLIC at a higher 
(centre-of-mass system) (CMS) energy. Note that these numbers do not include
beam polarization and do not use a topological selection, as was done for the 
more detailed study reported above. Hence one can gain another 30-50\% in
precision.

On the other hand, the quartic Higgs coupling remains elusive at CLIC,
due to the smallness of the relevant triple-Higgs production cross
sections. Even a 10-TeV collider would be able to produce only about
five such events in one 
year (=~10$^7$s) of operation at a luminosity of 10$^{35}$~cm$^{-2}$s$^{-1}$, 
while the dilution of the quartic coupling sensitivity is large owing to the 
effect of the background diagrams in $HHH\nu\bar{\nu}$, 
as seen in~Table~\ref{tab:h4}.
%
\begin{table}[!h] 
\caption{Production cross sections for $e^+e^- \rightarrow HHH\nu\bar{\nu}$ at 
various centre-of-mass energies for different values of the quartic Higgs 
coupling}
\label{tab:h4}

\renewcommand{\arraystretch}{1.3} 
\begin{center}

\begin{tabular}{ccccc}\hline \hline \\[-4mm]
\boldmath{$\sqrt{s}$} & $\hspace*{2mm}$ &
$\hspace*{3mm}$   \bf\boldmath{ $\frac{ g_{HHHH} }{ g_{HHHH}^{SM}}$~=~0.9} 
$\hspace*{3mm}$ &
$\hspace*{3mm}$ \bf\boldmath{ $\frac{ g_{HHHH} }{ g_{HHHH}^{SM}}$~=~1.0}  
$\hspace*{3mm}$ & 
$\hspace*{3mm}$ \bf\boldmath{$\frac{ g_{HHHH} }{ g_{HHHH}^{SM}}$~=~1.1}  
$\hspace*{3mm}$ 
\\[4mm]   

\hline \\[-3mm]
~3~TeV &  &0.400 & 0.390 & 0.383\\
~5~TeV &  &1.385 & 1.357 & 1.321 \\
10~TeV &  &4.999 & 4.972 & 4.970
\\[3mm] 
\hline \hline
\end{tabular}
\end{center}
\end{table}

\begin{figure}[htbp] 
\begin{center}
\begin{tabular}{cc}
\includegraphics[width=7.5cm,height=7cm]{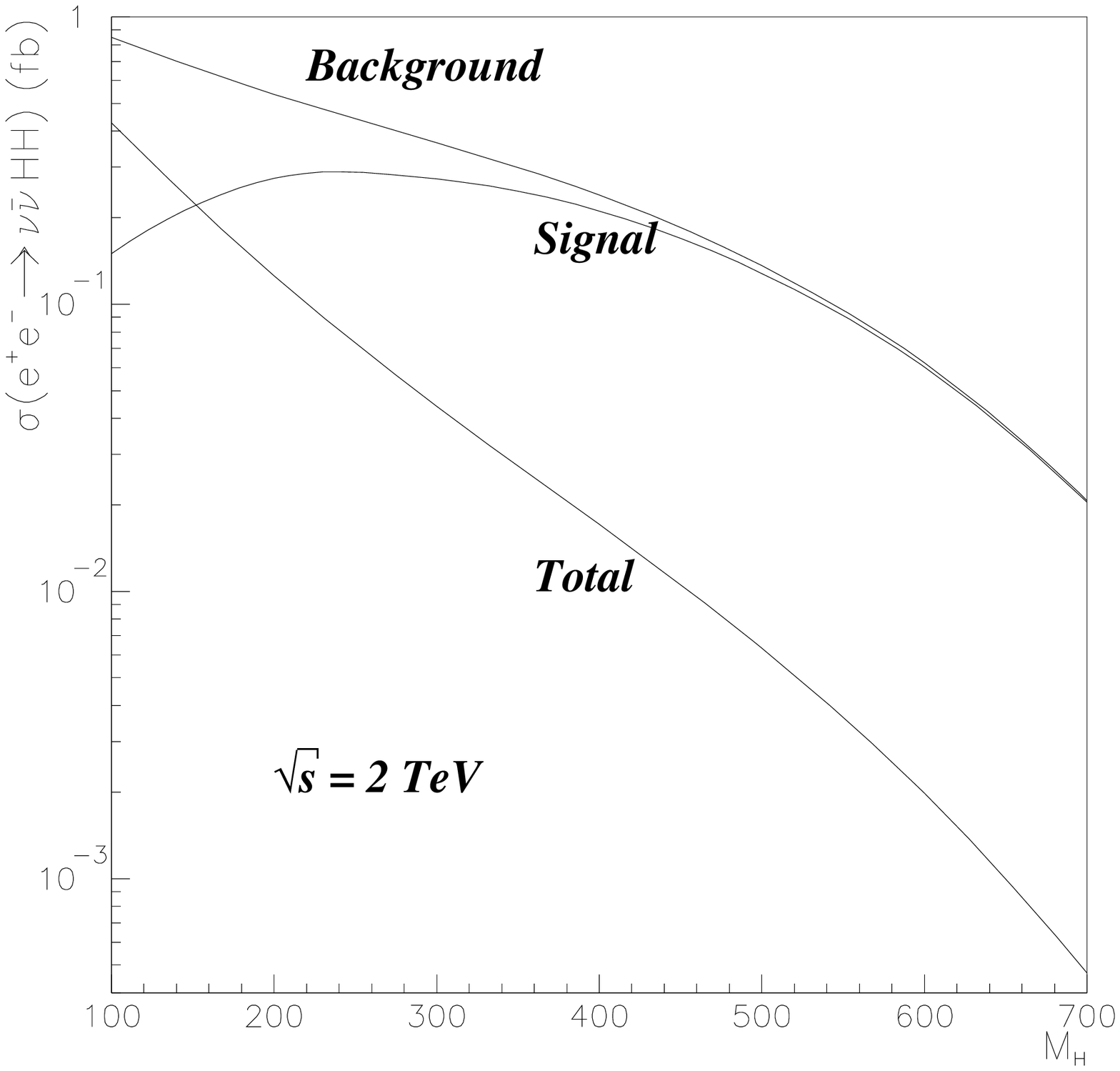} &

\includegraphics[width=7.5cm,height=7cm]{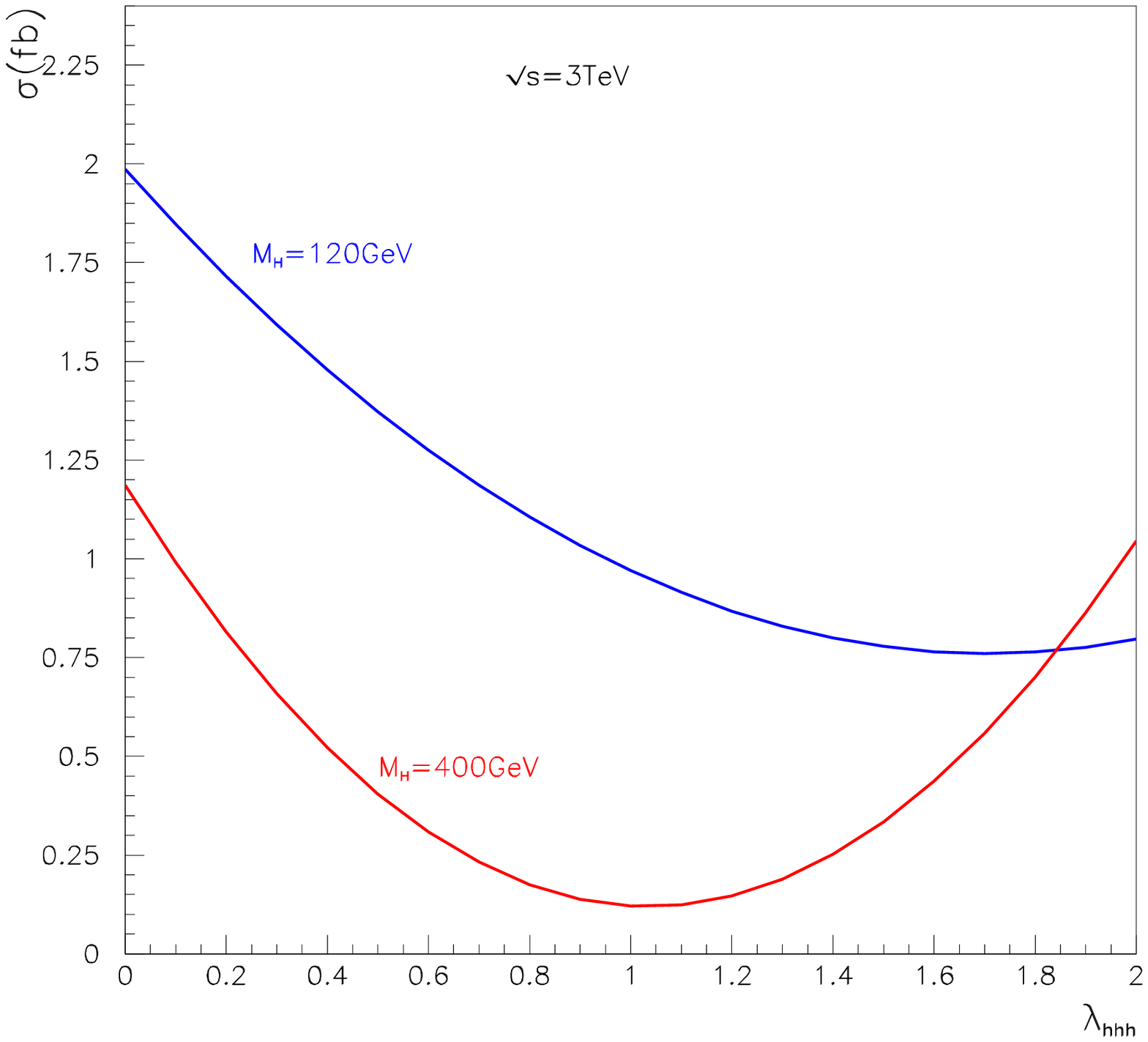} \\
\end{tabular}
\end{center}
\caption{The $\epem \ra \nu \bar \nu HH$ signal and background,
together with the total cross section, as functions of the
Higgs mass for $\sqrt{s}$~=~2~TeV. The second panel shows how the
cross section varies as a function of $g_{HHH}$ for
$M_H$~=~120, 400~GeV at $\sqrt{s}$~=~3~TeV.}
\label{fawh3smfigs}
\end{figure}

A further important advantage of multi-TeV $e^+ e^-$ collisions is that
heavier Higgses can also be scrutinized. Despite the much reduced cross
section for heavier Higgs masses, any deviation in the Higgs
self-couplings gets amplified as the Higgs mass gets higher, and can
easily make up for the reduction in cross section that occurs for a Higgs
with standard self-couplings. This is illustrated in
Fig.~\ref{fawh3smfigs} for $\epem \ra \nu \bar \nu HH$.

\vspace*{4mm}

The signal is identified by the contribution of the diagram containing the
$g_{HHH}$ coupling normalized to its Standard Model value,
$g_{HHH}$~=~1, and the remaining contributions are identified as
background. The signal increases as a function of the Higgs mass up to
about $M_H$~=~350~GeV, before dropping. For higher centre-of-mass energies,
this turnover occurs for even higher masses, and the background decreases
at the same time. As the Higgs mass increases, a subtle cancellation
between these two contributions takes place.  This leads to a precipitous
drop in the total cross section for a Standard Model Higgs boson. This 
type of
cancellation is reminiscent of what occurs in \eewwt between the neutrino
exchange and the triple vector gauge coupling diagrams. If the
$g_{HHH}$ strength deviates slightly from unity, this cancellation
may not be effective, and would increase the cross section, as shown in
the second panel of Fig.~\ref{fawh3smfigs}. A deviation in $g_{HHH}$
has a more drastic effect for $M_H$~=~400~GeV than for $M_H$~=~120~GeV. For
example, for $g_{HHH}$~=~2 the cross section is larger for
$M_H$~=~400~GeV than for~$M_H$~=~120~GeV.

This behaviour can easily be explained, if we recall that in this case
double Higgs production is due to longitudinal $W$ bosons that can be
identified at these high energies with the charged Goldstone modes. The
amplitude for $W_L W_L \ra H H$ may be approximated as
\bea
\tilde{\cal M}_{W_L W_L \ra H H} \ra  \frac{g^2}{4} r (3h_3-2)
+\dots, \quad r=M_H^2/M_W^2 \,.
\ena
Not only does this reveal the enhanced coupling factor
$r=M_H^2/M_W^2$, but it also shows that, if $h_3$ deviates from its
standard value, the cancellation between signal and background is
disrupted. Even in the present absence of a full simulation for the
extraction of the Higgs self-couplings for high Higgs masses, we can
attempt to derive some limit for $M_H$~=~400~GeV, since a naive 3$\sigma$
limit for $M_H$~=~120~GeV reproduces quite closely the result of a more
detailed analysis. For $M_H$~=~400~GeV, both the produced $H$ bosons will
usually decay into a $WW$ pair. We take $\int {\cal L}$~=~5~ab$^{-1}$ and
assume 50\% efficiency for the reconstruction of the $WWWW, WWZZ$ and
$ZZZZ$ final states. A 3$\sigma$ accuracy on the cross section leads to
the constraint 0.88~$< g_{HHH} <$~1.18, i.e., a precision of
about 15\% on the trilinear self-coupling of the Higgs boson.

\subsection{Heavy Higgs Boson}

\begin{figure}[htbp] %
\begin{center}
\begin{tabular}{c c}
\epsfig{file=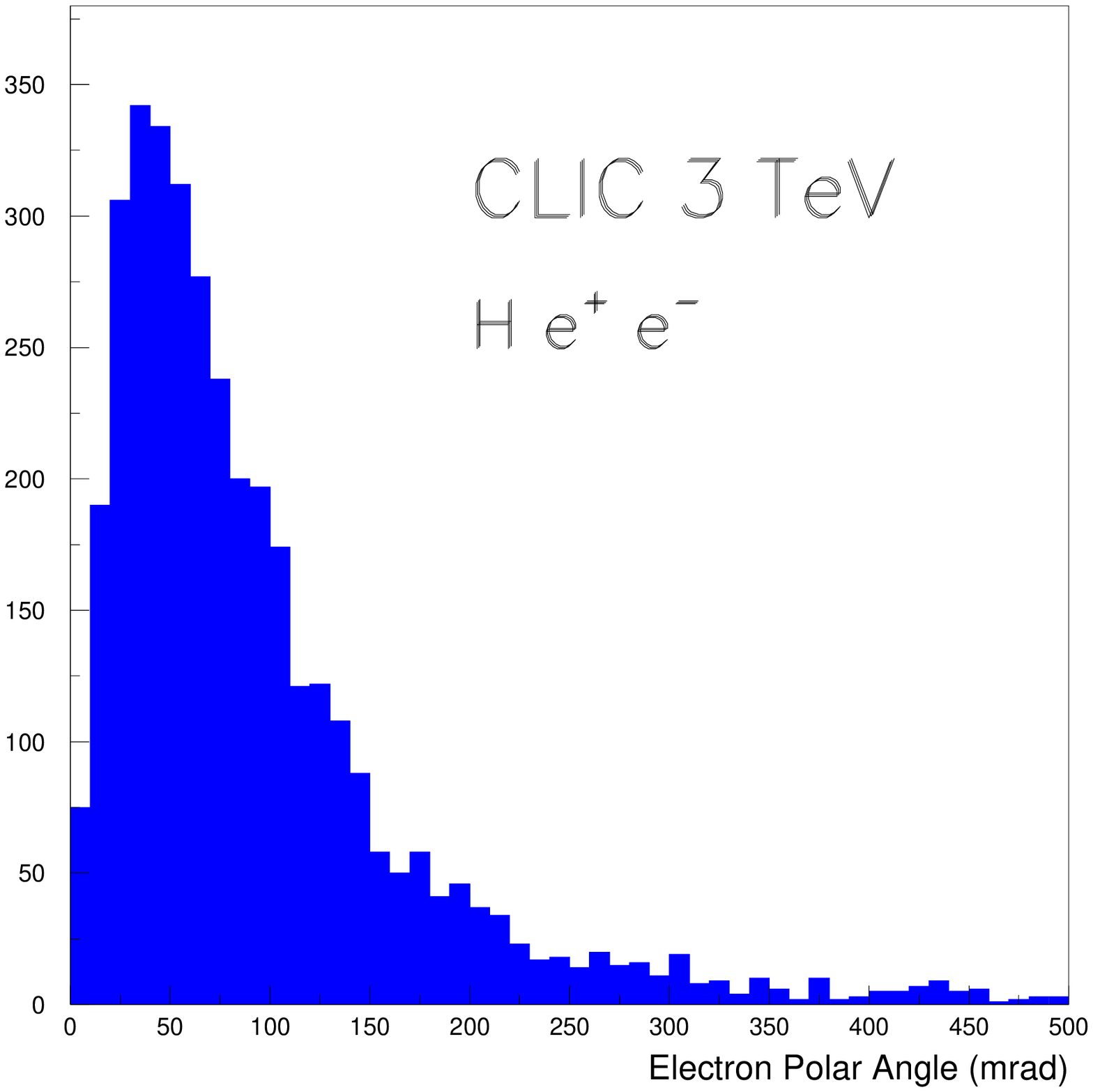, width=7.8cm,height=7.5cm} &
\epsfig{file=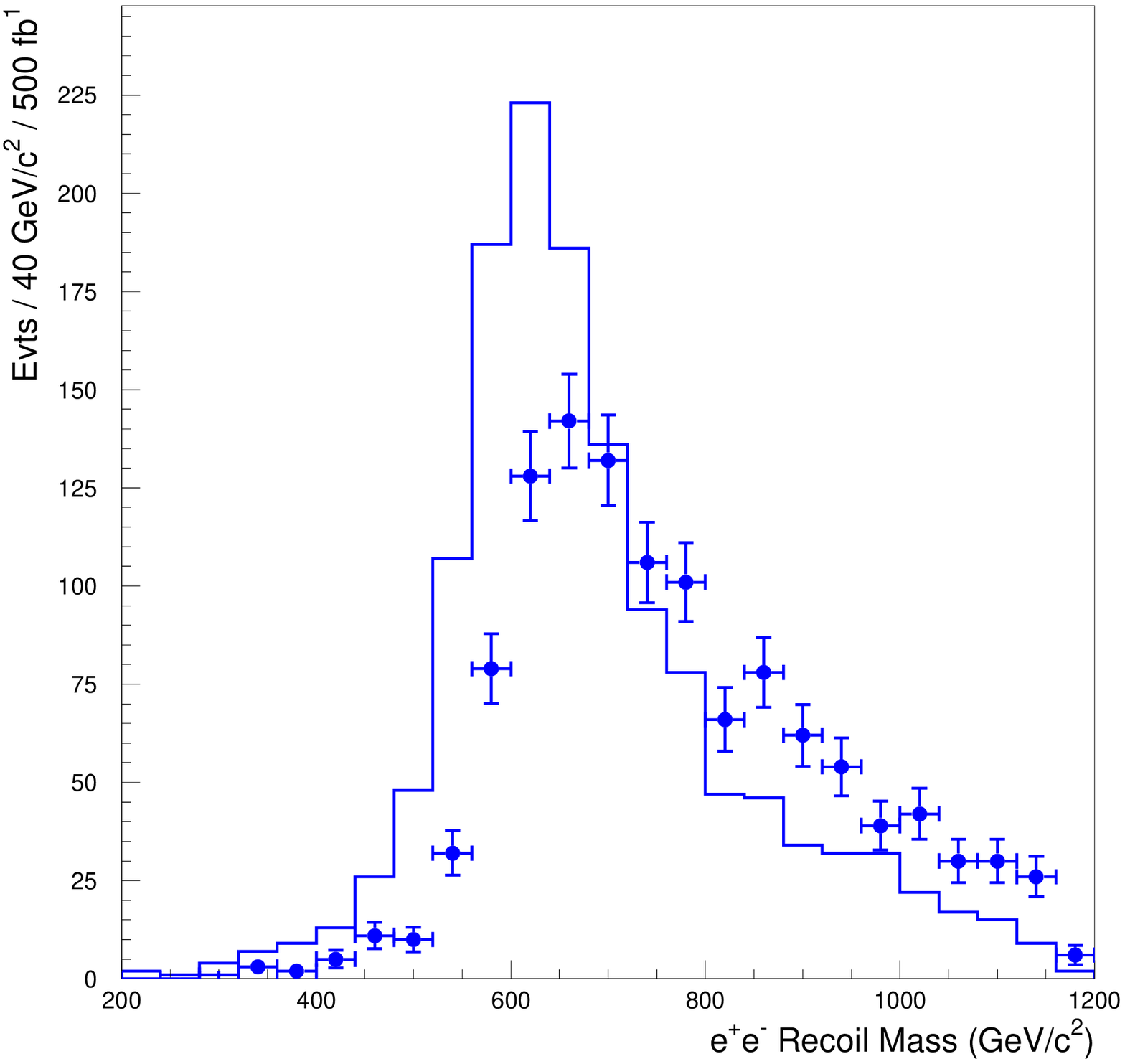, width=7.8cm,height=7.5cm} \\
\end{tabular}
\end{center}
\caption{Left panel: The distribution of electron polar angles in 
$e^+e^- \to H^0 e^+ e^-$ events at CLIC  with $E_{c.m.}$~=~3~TeV. Right 
panel: The distribution of recoil masses $M_X$ in $e^+e^- \to (H^0 \to X) 
e^+ e^-$ events at CLIC  with $E_{c.m.}$~=~3~TeV.}
\label{fig:heavyH}
\end{figure}

The precision electroweak data indicate that the Higgs boson is probably
lighter than about 212~GeV~\cite{LEPEWWG}. However, this limit can be
evaded if new physics cancels the effect of the heavy Higgs boson
mass. In these
scenarios, it is interesting to search for a heavier boson through the
$ZZ$ fusion process $e^+e^- \to H^0 e^+ e^- \to X e^+e^-$ at high
energies. Like the associated $HZ$ production in the
Higgs-strahlung process at lower energies, this channel allows a
model-independent Higgs search, through the tagging of the two forward
electrons and the reconstruction of the mass recoiling against
them~\cite{Battaglia:2001yi}. This analysis needs to identify electrons 
and measure their energies and directions down to $\simeq$~100~mrad, as 
seen in the left panel of Fig.~\ref{fig:heavyH}. This
is close to the bulk of the $\gamma\gamma \to {\mbox{\rm hadrons}}$ and pair
backgrounds, creating a challenge for forward tracking and for the
calorimetric response of the detector. Preliminary results show that a
clean Higgs signal can be extracted at $\sqrt{s}$~=~3~TeV for $M_H
\le$~900~GeV, as seen in the right panel of~Fig.~\ref{fig:heavyH}.

\section{Testing New Physics in the Higgs Sector}

The accurate measurement of Higgs couplings achievable at the LC may
reveal discrepancies with the SM predictions. These might provide us with
a first signal that nature realizes the Higgs mechanism through a scalar
sector, which is more complex than the minimal Higgs doublet of the SM. A
minimal extension of the Higgs sector contains two complex scalar
doublets, in which case the Higgs sector has five physical Higgs bosons:  
two that are CP-even and neutral ($h^0$ and $H^0$), one CP-odd and neutral
($A^0$) and two that are charged ($H^{\pm}$). In addition to the Higgs
masses, there are two more parameters: the ratio ${\rm tan}\beta$ of the
vacuum expectation values of the two fields and a mixing angle $\alpha$.
In the minimal supersymmetric extension of the SM (MSSM) discussed here,
only two of these parameters are independent.

\subsection{Heavy MSSM Higgs Bosons}

\begin{figure}[h!]
\begin{center}
\epsfig{file=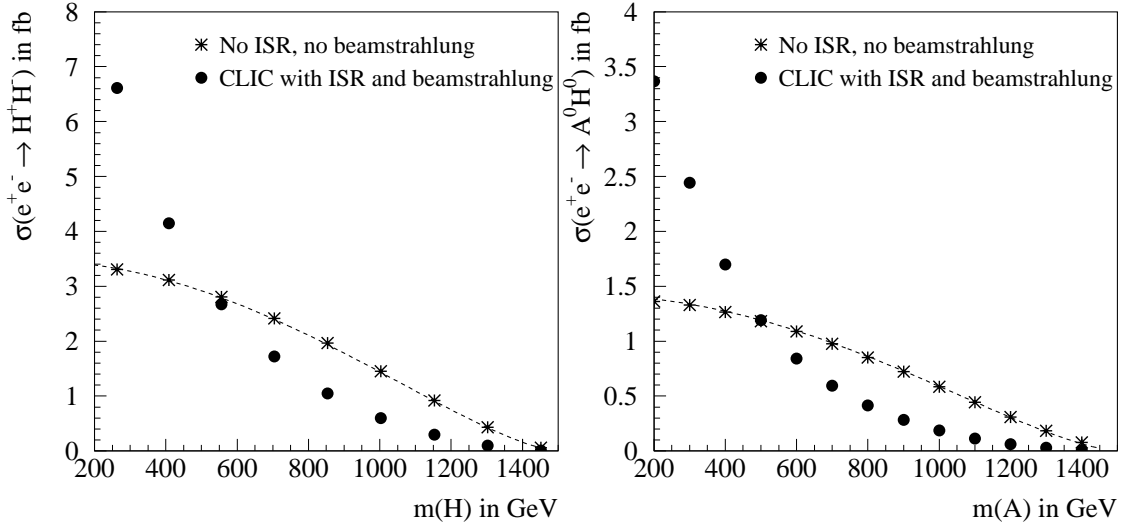,width=15cm} 
\end{center}
\caption{Heavy Higgs production at CLIC: the 
$e^+e^- \rightarrow H^+H^-$ (left) and
$e^+e^- \rightarrow H^0 A^0$ (right) production cross sections for
$\sqrt{s}$~=~3~TeV, as functions of the Higgs mass}
\label{fig:hxs}
\end{figure}

If the heavier Higgs bosons $H^0, A^0$ and $H^{\pm}$ of a non-minimal
Higgs sector are not too heavy, they will be pair-produced directly at a
LC. At tree level, the production cross sections for 
$e^+e^- \rightarrow H^+H^-$ 
and $e^+e^- \rightarrow H^0 A^0$ are independent of ${\rm tan}\beta$. 
In calculating the effective production cross section, initial-state
radiation (ISR) and the beam--beam effects for the CLIC.01 parameters
have been taken into account. As a result, for low values of $M_{H^{\pm}}$
the effective cross section is enhanced with respect to the tree-level cross
section, while close to the kinematical threshold only a smaller fraction
of the energy spectrum is available and the effective cross section is
reduced, as seen in~Fig.~\ref{fig:hxs}.

The sizeable production cross sections for the $H^+H^-$ and $H^0A^0$ pair
production processes provide sensitivity at CLIC for masses up to
about 1~TeV and beyond, for all values of ${\rm tan}\beta$, thus extending the
LHC reach. For large values of $M_A$, the main processes of
interest are $e^+e^- \rightarrow H^+H^- \rightarrow t \bar{b}\bar{t} b$ and 
$e^+e^- \rightarrow H^0 A^0 
\rightarrow b \bar{b} b \bar{b}$ or $t \bar{t}t\bar{t}$. These 
result in highly distinctive, yet challenging, multijet final states
with multiple $b$-quark jets, which must be identified and reconstructed 
efficiently. We now discuss each of these processes in turn.

\subsubsection{\boldmath{$e^+e^- \rightarrow H^+H^-$}}

\begin{figure}[htbp] %
\begin{center}
\begin{tabular}{c c}
\epsfig{file=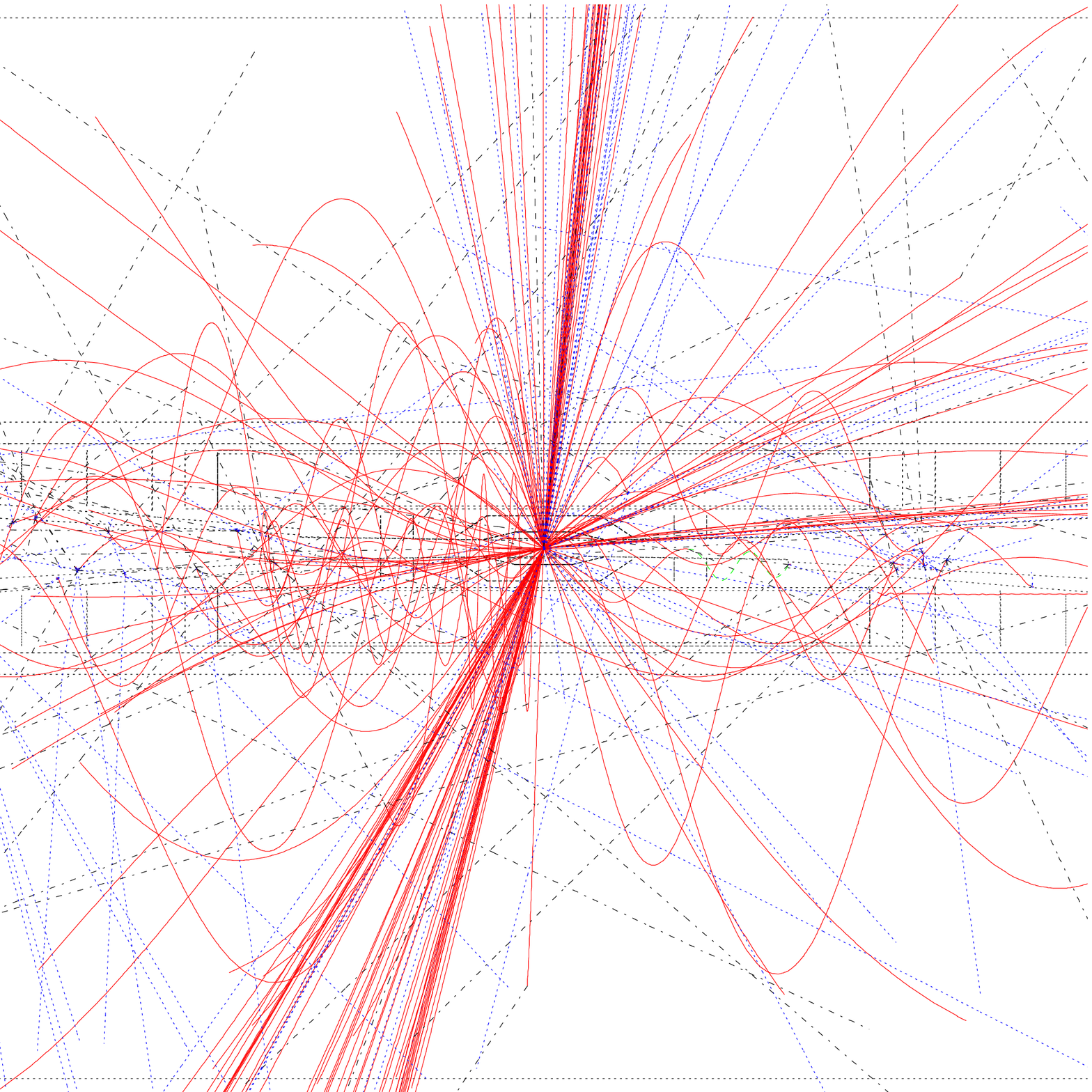, width=7.0cm} &
\epsfig{file=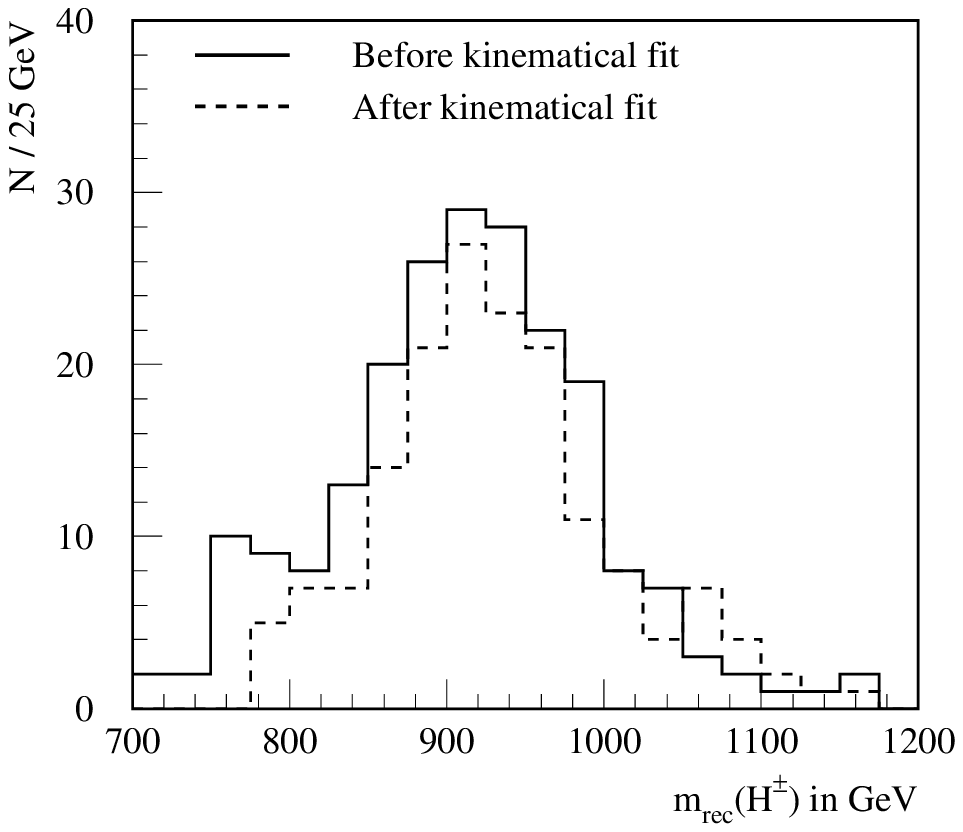, width=7.5cm} 
\end{tabular}
\end{center}
\caption{Charged-Higgs analysis. Left: Display of a 
$e^+e^- \rightarrow H^+H^- \rightarrow t \bar{b} \bar{t} b$ 
event at $\sqrt{s}$~=~3~TeV. The accelerator-induced backgrounds are not
overlaid. Right: $H^{\pm}$ mass signal reconstructed before
(continuous line) and after  
(dashed line) applying a global mass-constrained fit.} 
\label{fig:hpm}
\end{figure}

A detailed analysis has been performed for the reconstruction
of $e^+e^- \rightarrow H^+H^- \rightarrow t \bar{b} \bar{t} b$ 
with $M_H$~=~880~GeV, corresponding to the 
constrained minimal supersymmetric standard model (CMSSM) 
benchmark point~J of
Ref.~\cite{cmssm}. A sample Higgs analysis is 
illustrated in~Fig.~\ref{fig:hpm}, whose left panel shows a simulation of 
a `typical' $e^+ e^- \rightarrow H^+ H^-$ event. Event reconstruction
is based on the identification of the $W bb W bb$ 
final state, and applies a mass-constrained fit to improve the mass
resolution and reject multifermion and other combinatorial
backgrounds~\cite{hpm}, as seen in the right panel of~Fig.~\ref{fig:hpm}. 
Events with no jet combination compatible with the
$W$ or $t$ masses within the observed resolution have been discarded. The
fit uses energy and momentum conservation, the $W$ and $t$ mass
constraints, and imposes equal masses for the $H^\pm$ bosons. The cross
section for the irreducible SM $e^+ e^- \rightarrow t \bar{b} \bar{t} b$
background has been estimated to be 1.5~fb~at~$\sqrt{s}$~=~3~TeV.

The superposition of accelerator-induced $\gamma \gamma \rightarrow
{\mathrm{hadrons}}$ events has also been included.  Here special care has
been taken to make the result robust in the presence of all high-energy
beam--beam effects.  The additional hadrons generated by these $\gamma
\gamma$ collisions, mostly in the forward regions, may either be merged
into the jets coming from the $H^{\pm}$ decay or result in extra jets
being reconstructed. In order to minimize the impact on the event
reconstruction, only the four leading non-$b$ jets have been considered,
together with the $b$-tagged jets. Besause of the significant loss of energy
of the colliding $e^+$ and $e^-$, energy and momentum conservation
constraints cannot be applied on the reconstructed system for the nominal
$\sqrt{s}$. Instead, the kinematical fit allows for an extra particle to
be radiated, but imposes its transverse momentum to be zero. This
kinematical fit improves the Higgs mass resolution by a factor of nearly
2, as seen in~Fig.~\ref{fig:hpm}.

The signal event rate has been estimated including the $\gamma\gamma$ 
background, assuming that the detector integrates
over 15 bunch crossings. 
The fitted signal width is 45~GeV after the mass-constrained fit and 
there are 149 $e^{+}e^{-} \rightarrow H^{+}H^{-} \rightarrow
(t\bar{b})(\bar{t}b)$ signal events with a reconstructed mass in the
signal window, corresponding to an efficiency of 0.05. The $tbtb$ SM
background processes contribute 43 events for 3~ab$^{-1}$. By scaling
these results, the estimated maximum mass reach for a 5$\sigma$ discovery
with 3~ab$^{-1}$ at 3~TeV is 1.25~TeV if the $tb$ mode saturates
the $H$ decay width, or 1.20~TeV if it accounts for only 0.85\% of the
decays.

By reconstructing either both the $H^{\pm}$ or only one, the analysis can 
be optimized in terms of efficiency and resolution for either mass
measurement or an unbiased study of $H^{\pm}$ decays, respectively.
Fractional accuracies of 0.5--1.0\% on the heavy boson mass for 
600 $< M_H <$ 900~GeV and of 5--7\% on the
product of production cross section and $tb$ decay branching fraction are
expected with ${\cal{L}}$~=~3~ab$^{-1}$ at $\sqrt{s}$~=~3~TeV. The results
do not depend critically on the underlying hadronic background.

\subsubsection{\boldmath{$e^+e^- \to H^0 A^0$}}

\begin{figure}[htbp] 
\begin{center}
\begin{tabular}{c c}
\epsfig{file=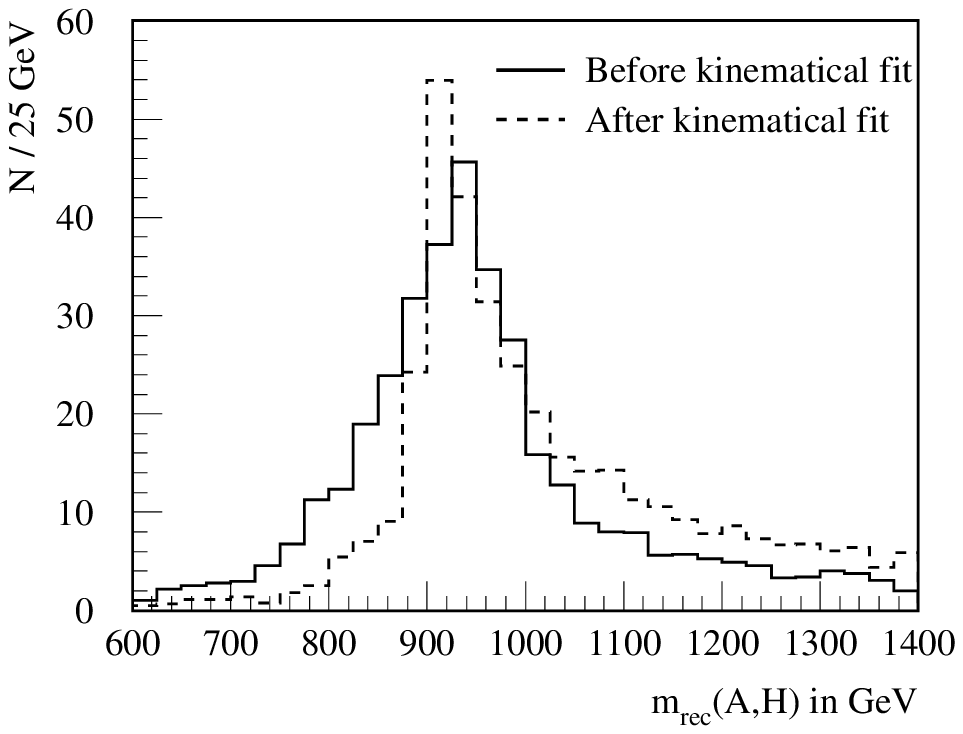,height=7cm} &
\epsfig{file=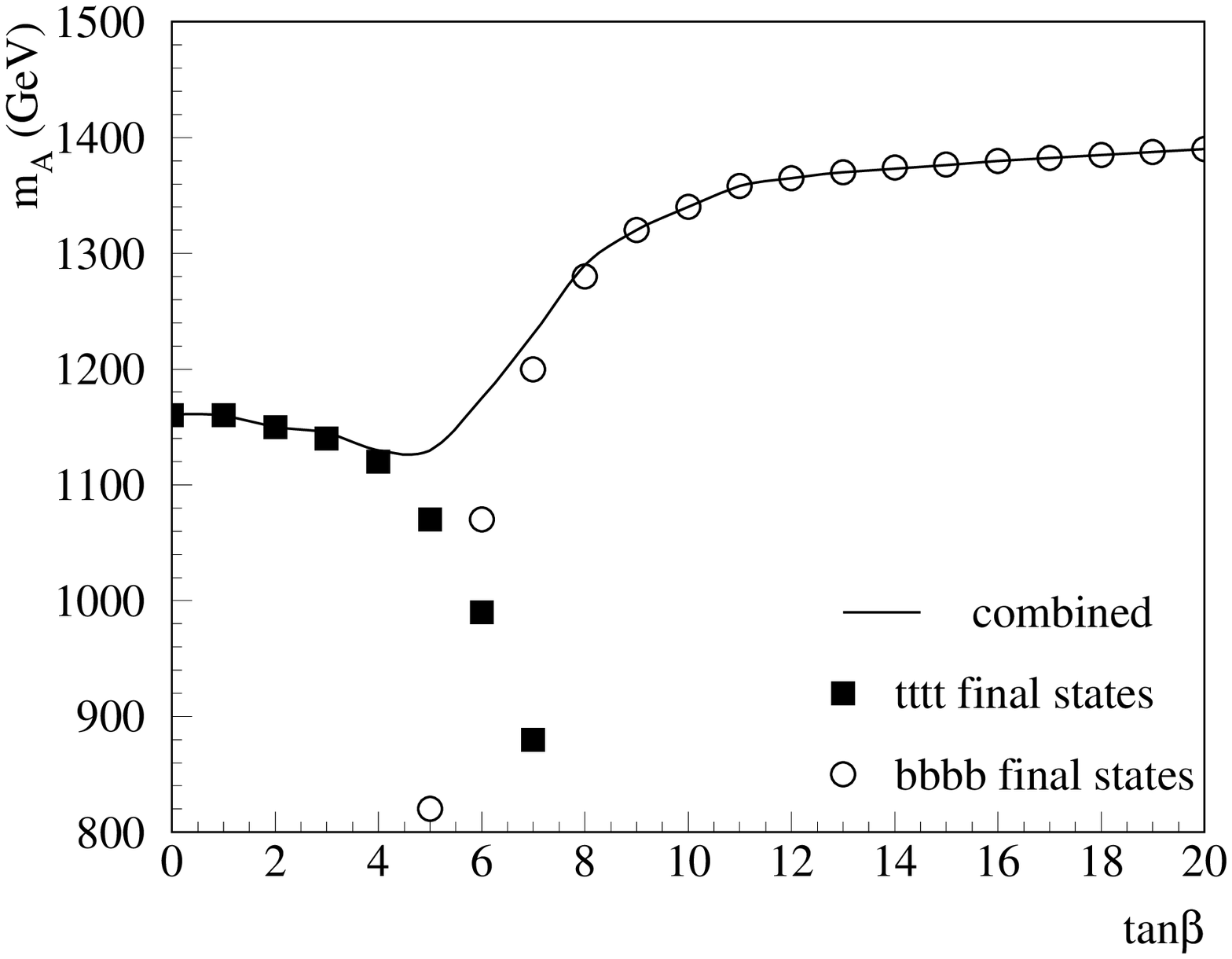,height=6cm} \\
\end{tabular}
\end{center}
\caption{Neutral Higgs analysis. Left: The signal reconstructed in the 
$e^+e^- \to H^0 A^0 \to b \bar{b} \bar{b} b$ for the CMSSM point J. 
Right:~The $H^0A^0$ discovery reach with 3~ab$^{-1}$ of CLIC data 
at 3~TeV as a function of ${\rm tan}\beta$ obtained by summing the
$b \bar {b} \bar{b} b$ and $t \bar{t} \bar{t} t$ channels.}
\label{fig:ha}
\end{figure}

A similar study has been performed for the reconstruction of the 
$e^+e^- \to H^0A^0 \to b \bar{b} b \bar{b}$ process, as illustrated 
in~Fig.~\ref{fig:ha}. The reconstruction has been based 
on the identification of two pairs of $b$-tagged jets with close invariant 
masses. Again, masses have been chosen to
correspond to the CMSSM benchmark point J, which has $M_A$~=~876~GeV.  
The reconstruction is based on the identification of two pairs of
$b$-tagged jets with equal dijet masses. Similarly to the study of
charged Higgs bosons, a kinematical fit has been applied, imposing energy
and momentum conservation and equal masses. Energy loss due to
initial-state radiation and beamstrahlung has also been accounted for, as
described above. Owing to the simpler final state considered for this
analysis, the efficiency for $e^+e^- \to H^0A^0 \to \bar{b} b \bar{b} b$ 
is about seven times larger than that obtained for 
$e^+e^- \to H^+H^- \to t \bar{b} \bar{t} b$.  After reduction of the SM 
$e^{+}e^{-} \rightarrow bbbb$ background, the discovery potential for 
$e^+e^- \to A^0H^0$ in the $bbbb$ final states extends to 
$M_A \le$~1.3~TeV, for an integrated luminosity of 3~ab$^{-1}$. 
The reconstruction yields relative accuracies of 0.004 and 0.062 
for the measurement of the $A^0$ mass and of the signal event rate,
respectively, for the CMSSM benchmark point J which has 
BR($H$, $A \to b \bar{b}$)~=~0.87.

For low values of ${\rm tan}\beta$, the main decay channel is 
$A^0$, $H^0 \to \bar{t} t$. 
The analysis in this channel is based on the reconstruction of the hadronic 
decays of the four $W$ bosons and the identification of the four $b$~jets.
Because of the complex final state, the signal efficiency reaches only 2\% in 
this channel. After reduction of the SM 
$e^{+}e^{-} \rightarrow t \bar{t} \bar{t} t$ 
background, the discovery potential is up to about 1.1~TeV,~for~3~ab$^{-1}$.

The relative accuracies achievable on the mass and production cross sections 
are $\delta M_A/M_A$~=~0.012 and $\delta \sigma_{HA} / \sigma_{HA}$~=~0.075, 
respectively, assuming $m_A$~=~576~GeV and the $\bar{t} t$ channel saturating 
the decay width. The CLIC mass reach for observing the $A^0$ and
$H^0$ bosons as function of ${\rm tan}\beta$ is shown~in~Fig.~\ref{fig:ha}.

\subsubsection{\boldmath{$\gamma\gamma \rightarrow H,A$}}

Photons generated by Compton back-scattering of laser light off the 
incoming $e^-$ beams can reach c.m.~energies of 70--80\% of the 
primary collider
energies and high degrees of polarization \cite{plc}. 
Higgs bosons can be produced as s-channel resonances and thus the full
(photon) beam energy can be used to produce the heavy particles. 
In this way, $\gamma\gamma$ colliders can extend the 
Higgs discovery to higher mass regions than $e^+e^-$ colliders with the 
same centre-of-mass energies. With integrated 
annual luminosities of 100--200~fb$^{-1}$,
sufficiently high signal rates are guaranteed so that photon colliders 
provide a useful instrument for the search of Higgs bosons in 
regions of the parameter space not
accessible elsewhere~\cite{muehldiss}.

\begin{figure}[htbp] %
\begin{center}
\epsfig{figure=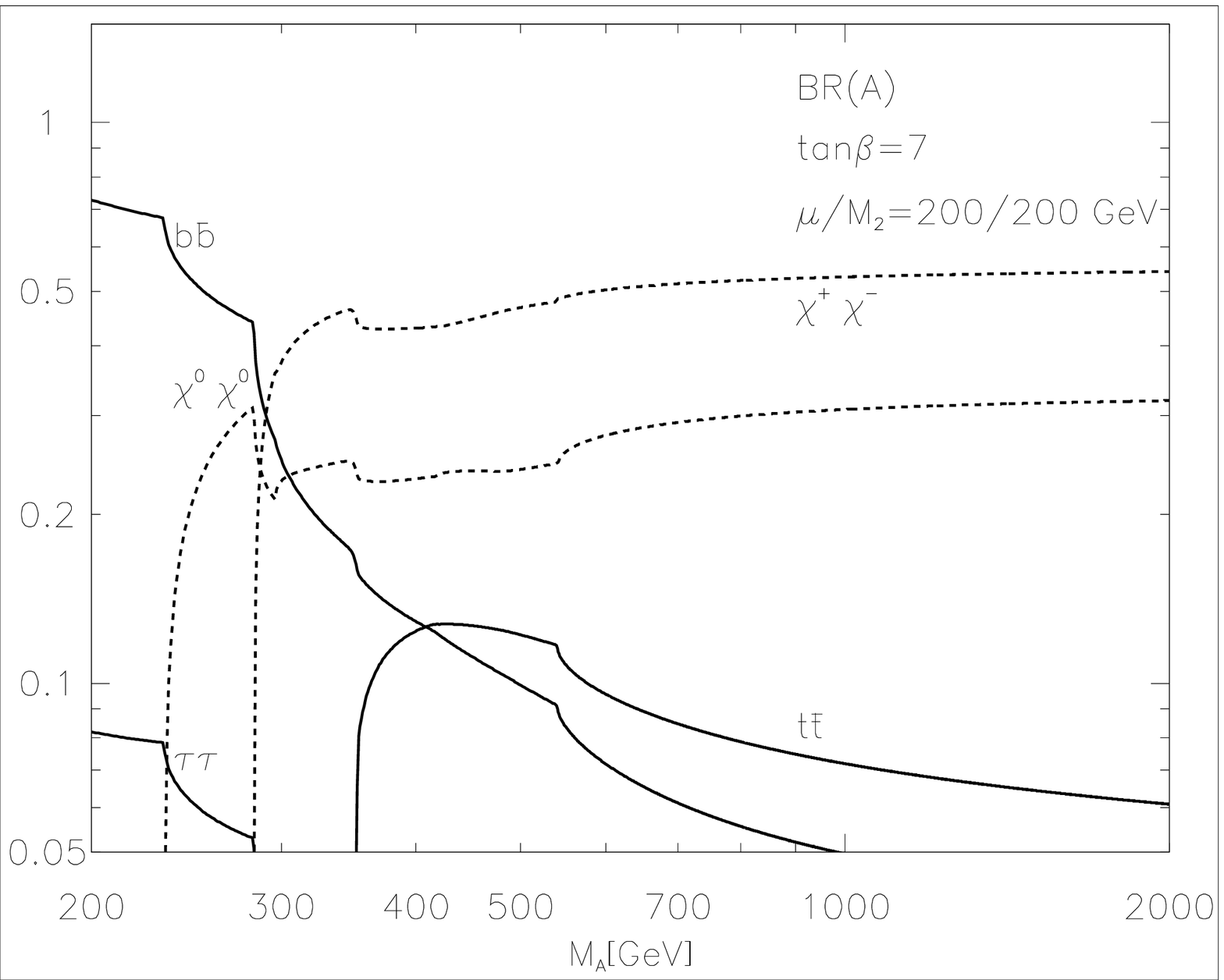,bbllx=3,bblly=3,bburx=560,bbury=450,width=7cm,clip=}
\hspace{0mm}
\epsfig{figure=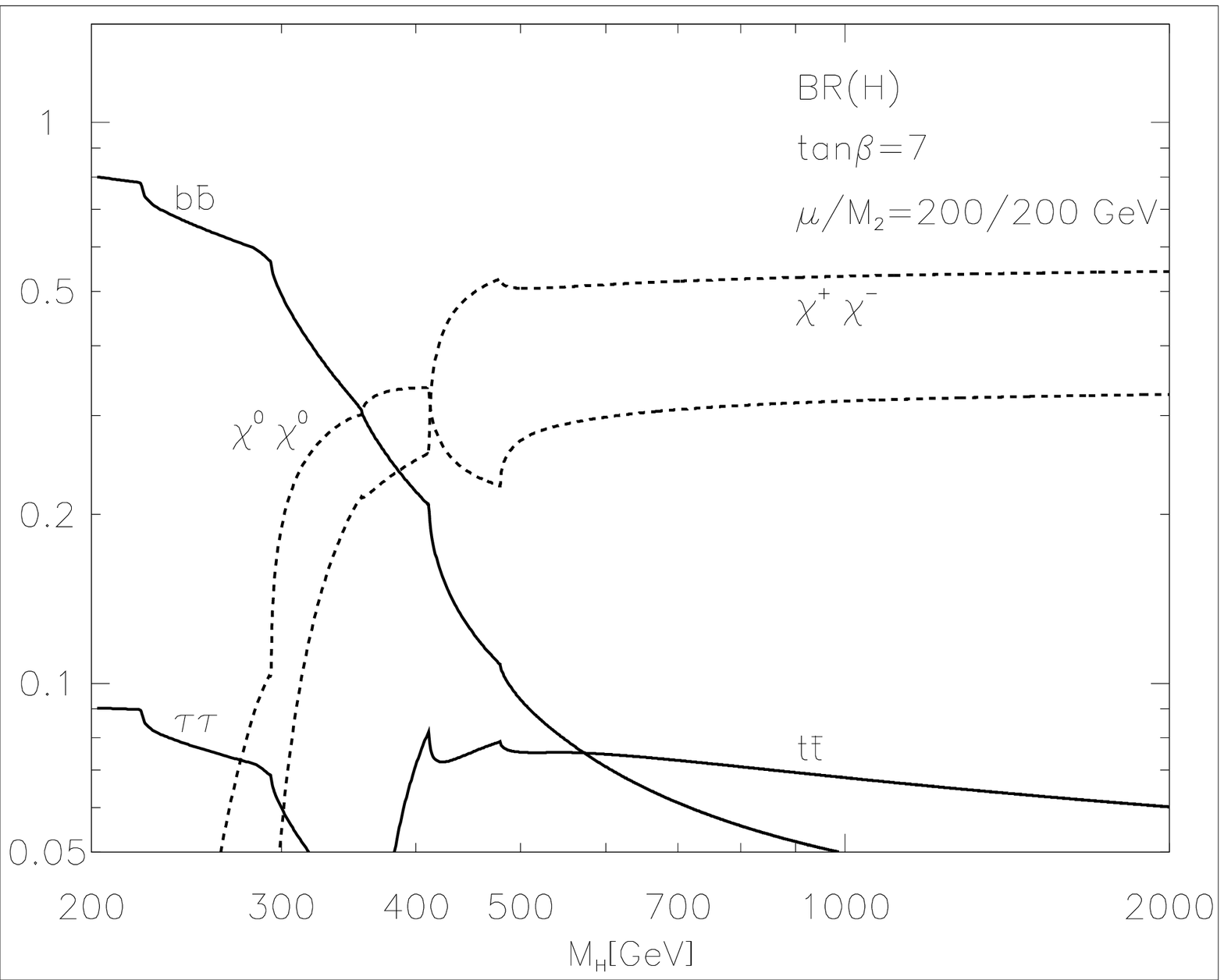,bbllx=3,bblly=3,bburx=560,bbury=450,width=7cm,clip=}
\epsfig{figure=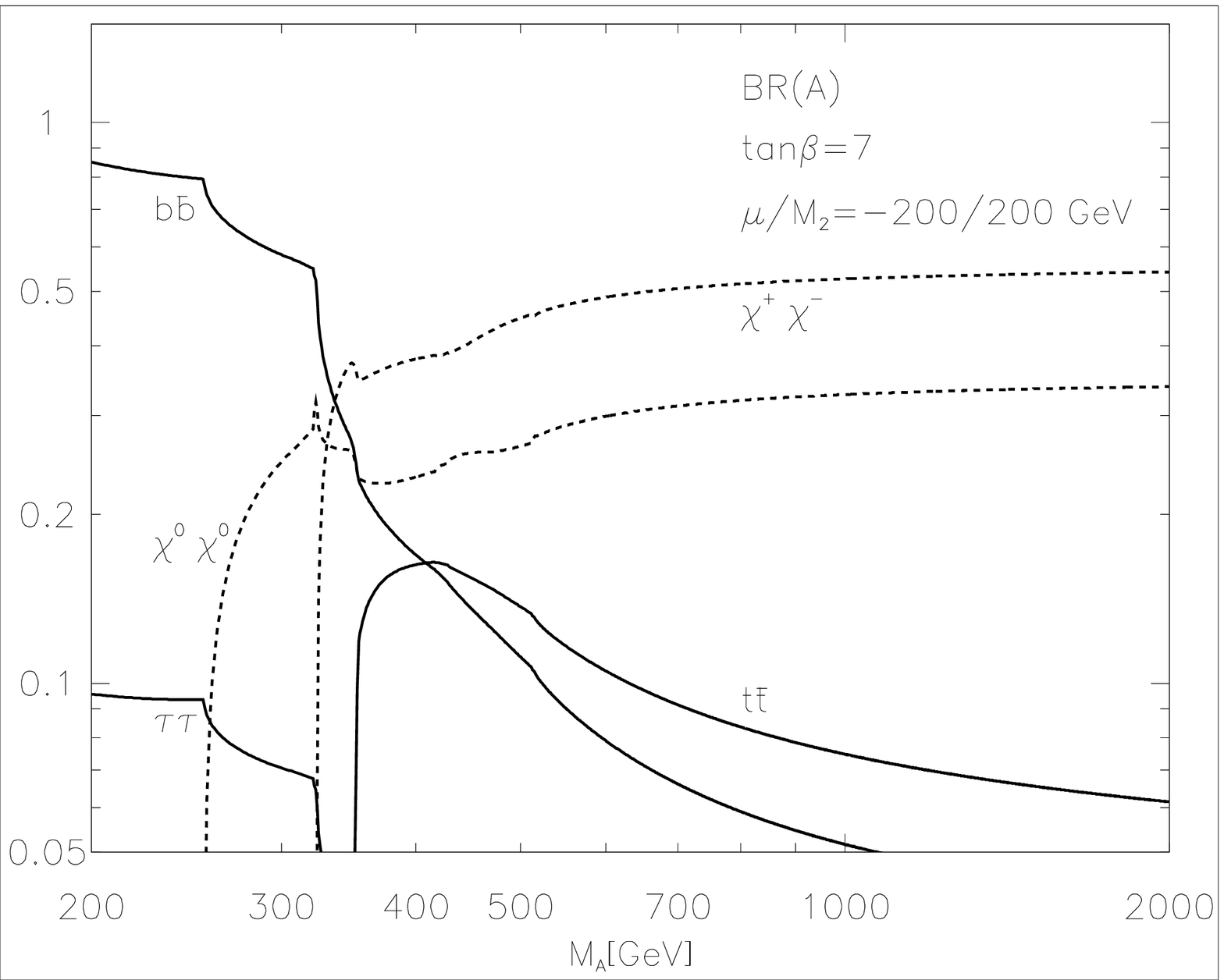,bbllx=3,bblly=3,bburx=560,bbury=450,width=7cm,clip=}
\hspace{0mm}
\epsfig{figure=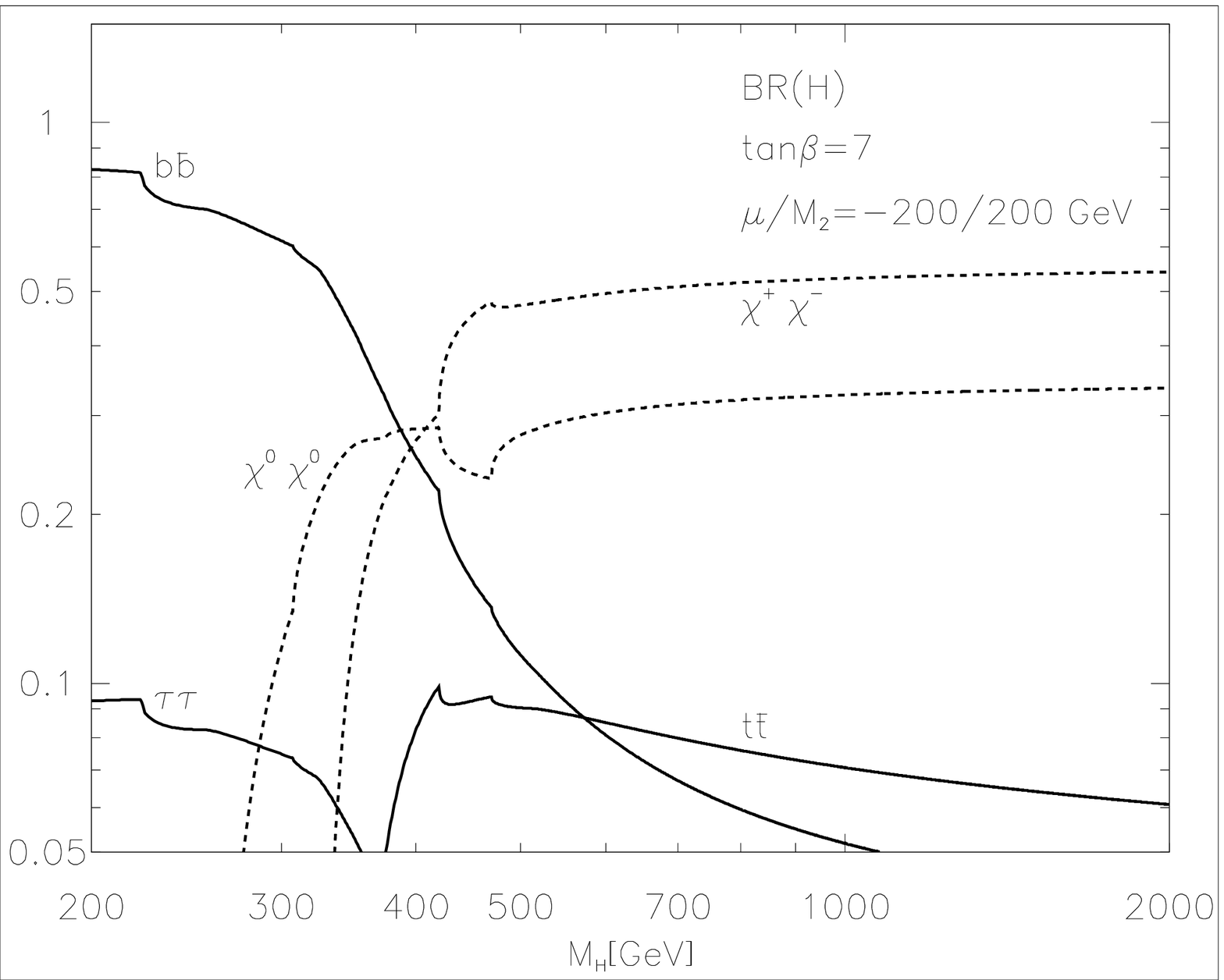,bbllx=3,bblly=3,bburx=560,bbury=450,width=7cm,clip=}
\end{center}
\caption{Branching ratios of the heavy Higgs bosons
$H,A$ as a function of the corresponding Higgs mass: 
$\tilde{\chi}^0\tilde{\chi}^0$ ($\tilde{\chi}^+\tilde{\chi}^-$) represents 
the sum of all neutralinos (charginos) except the lightest neutralino pair}
\label{gamma_higgs1} 
\vspace{-0.5cm}
\end{figure}

The decay channel $H,A \to b \bar{b}$ is particularly promising for the $H$ 
and $A$ discovery. Figure~\ref{gamma_higgs1} shows the branching ratios
of $H,A$~\cite{42,43} for ${\rm tan}\beta$~=~7 and $M_{H,A}>$~200~GeV, a
parameter region not covered by LHC discovery modes. The higgsino and
gaugino MSSM parameters have been set to $|\mu|=M_2$~=~200~GeV, with a
universal gaugino mass at the GUT scale. Squarks and sleptons are
assumed to be so heavy that they do not affect the results significantly.
Results are shown for both signs of the $\mu$ parameter.
Owing to the enhancement of the MSSM Higgs boson couplings to
$b\bar{b}$ at large ${\rm tan}\beta$, the $b\bar{b}$ decay is 
sizeable and dominates for moderate masses.

\begin{figure}[htbp] 
\begin{center}
\epsfig{figure=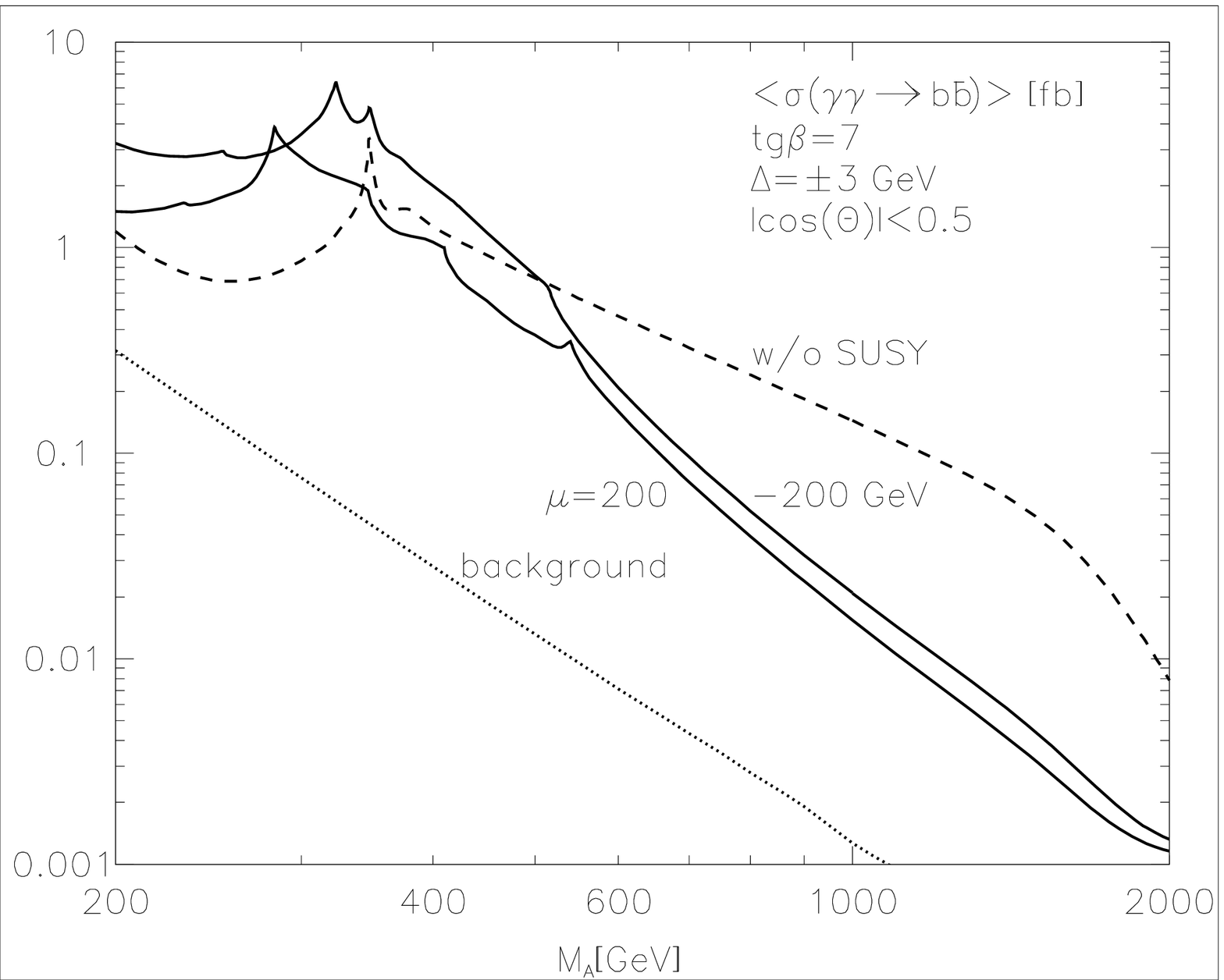,bbllx=3,bblly=3,bburx=560,bbury=450,width=7cm,clip=}
\epsfig{figure=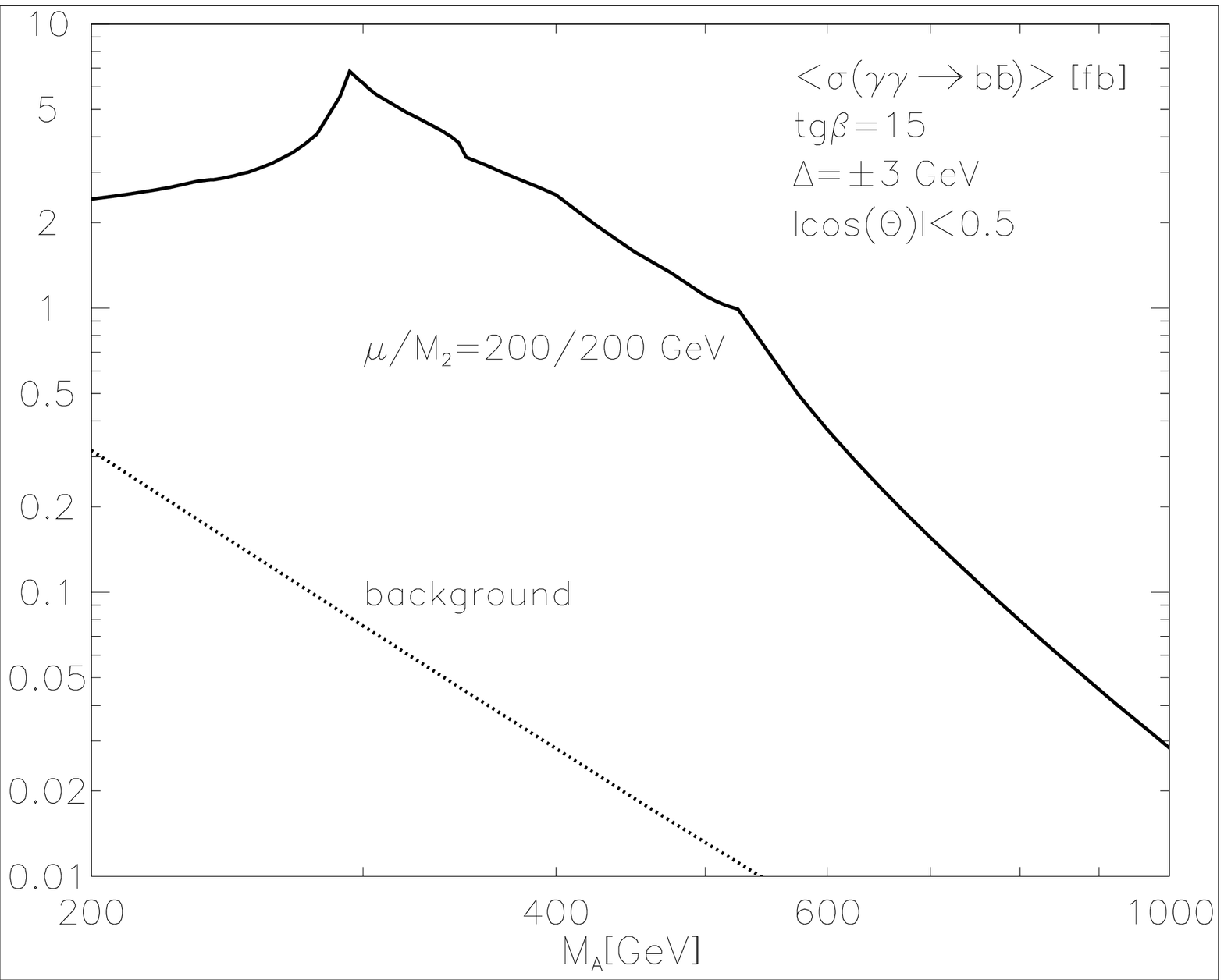,bbllx=3,bblly=3,bburx=560,bbury=450,width=7cm,clip=}
\end{center}
\caption{Left: Sum of the signal, background and interference cross sections
 (full, dashed lines) for the resonant $H,A$ production in $\gamma\gamma$ 
fusion as a function of $M_A$ with final decays into $b\bar{b}$, and the 
corresponding background cross section. The SUSY parameters are chosen as 
${\rm tan}\beta$~=~7, $M_2=\pm\mu$~=~200~GeV. For comparison the cross
section in the case of vanishing supersymmetric-particle contributions
is also shown. Right: The same as on the left, but for ${\rm tan}\beta$~=~15.}
\label{gamma_higgs2}
\end{figure}

Figure~\ref{gamma_higgs2} shows the result for $\gamma\gamma \to 
 b{\bar b}$ and polarized $e^-$ and laser photon beams for 
${\rm tan}\beta$~=~7 
and $M_{H,A}>$~200~GeV. The NLO QCD corrections to the
 signal~\cite{42,hggqcd}, background~\cite{bkgqcd} and interference
 term~\cite{muehldiss} have been included. Since the signals are generated
 for equal photon helicities, the initial beam polarizations have been 
chosen such that this configuration is enhanced. Assuming that evidence 
for a Higgs boson has been found in a preliminary rough scan of the 
$\gamma\gamma$ energy, the maximum of the $\gamma\gamma$ luminosity 
spectrum, which is at 70--80\% of the $e^\pm e^-$ c.m. energy for 
equal photon helicities \cite{plc,kuehn}, has been tuned to the mass 
$M_A$. The analysis can be optimized by applying final-state cuts.
A cut in the production angle of the bottom quarks, $|\cos\theta| < 0.5$,
strongly reduces the background, whereas the signal is affected only
moderately. By collecting $b\bar{b}$ final states with an expected 
resolution $M_A \pm$~3~GeV~\cite{schreiber}, the sensitivity to the 
resonance signal above the background is strongly increased. Furthermore,
in order to suppress gluon radiation, which increases the background at 
NLO in the same-helicity configuration, slim two-jet configurations have
been selected in the final state as defined within the Sterman--Weinberg 
criterion; events with radiated gluon energies above 10\% of the 
$\gamma\gamma$ invariant energy and opening angles between the three 
partons exceeding 20$^\circ$ have been rejected. Higher-order corrections
 beyond NLO, which become important in the two-jet final states, have been 
taken into account by (non-)Sudakov form~factors~\cite{resum}. 

As can be inferred from Fig.~\ref{gamma_higgs2}, 
the background is strongly suppressed (even if the experimental resolution
is less favourable than assumed) and 
the significance of the heavy Higgs boson signals is sufficient for the 
discovery of the Higgs particles up to about 0.8--1~TeV, if about 10--20 
events are needed for the discovery. In order to reach higher masses more 
luminosity needs to be accumulated. The discovery/measurement reach increases
in this channel for higher ${\rm tan}\beta$ values, as
shown~in~Fig.~\ref{gamma_higgs2}(right). 

The production of Higgs bosons $h,H,A$ in $\gamma\gamma$ collisions
may serve a second purpose beyond discovery. The formation of Higgs
bosons in $\tau^+\tau^-$ collisions, with the $\tau$'s generated
in the splitting of the two photons, $\gamma + \gamma \to (\tau^+\tau^-)
(\tau^+\tau^-) \to (\tau^+\tau^-) + h/H/A$, is very sensitive to
${\rm tan}\beta$ for high values, cf.~Ref.~\cite{choi}. The production
cross section rises quadratically with ${\rm tan}\beta$ in this regime, 
$\sigma_{\gamma\gamma} \approx {\rm tan}^2\beta$, so that the mixing
parameter can be measured very well by this method.  
With expected accuracies $\Delta {\rm tan}\beta / {\rm tan}\beta$ at a
level of 5 to 10\%, sufficiently below the kinematical limit
[specifically, 
$\Delta {\rm tan}\beta / {\rm tan}\beta \simeq$~5.7\% [$H + A$] for 
$M_{H,A}$~=~500~GeV, $\sqrt{s_{\gamma\gamma}}$~=~2~TeV and 
$L$~=~500~fb$^{-1}$, this $\gamma\gamma$ method of 
Higgs analyses can improve significantly on complementary measurements in 
channels at the LHC and in the $e^+e^-$ mode of a linear~collider.

\subsection{CP Violation in the Supersymmetric Higgs Sector}

Several of the soft supersymmetry-breaking parameters of the MSSM can have 
a complex phase, inducing new sources of CP violation in addition to the
phase in the Cabbibo--Kobayashi--Maskawa matrix of the SM, even if 
minimal flavour violation is assumed, as here. 
The phenomenology of such an MSSM with complex
parameters (CMSSM) can be substantially different from the case in
which all the soft supersymmetry-breaking parameters are assumed to be 
real (see, for example, Ref.~\cite{CEMPW} for an analysis concerning
LEP constraints  
on the Higgs sector in the CMSSM). The possibly complex parameters 
are the trilinear Higgs--sfermion couplings, $A_q, q~=~t, b, \tau,\ldots$,
the Higgs mixing parameter $\mu$, and the soft supersymmetry-breaking 
gaugino mass parameters $M_1$, $M_2$, $M_3$.

The situation becomes simpler in the constrained version of the MSSM
with universal soft super\-sym\-metry-breaking
scalar masses $m_0$, gaugino masses $m_{1/2}$, and
trilinear couplings $A_0$ (the CMSSM) input at some high scale. In this 
case there are just two new sources of CP violation.
In a suitable convention, these can be chosen as the phases of
the complex parameters $m_{1/2}$ and $A_0$ relative to that of the Higgs
mixing parameter $\mu$. The values of these phases are constrained by the
upper limits on the electric dipole moments (EDMs) of the electron,
neutron~\cite{EDMrecent,APEDM} and mercury atom~\cite{FOPR} and other
measurements, but large CP-violating phases in $m_{1/2}$ and 
in the third-generation trilinear couplings $A_0$ cannot yet be excluded. 
The ongoing probes of CP violation at $B$~factories can test the
consistency of this and other non-minimal CP-violating supersymmetric
scenarios~\cite{CS,DP}.

At the tree level, the Higgs boson sector of the CMSSM is
CP-invariant. However, taking radiative corrections into account,  
all three neutral Higgs bosons, $h$, $H$, and $A$ can
mix with each other~\cite{APLB,PW}. Thus the charged Higgs boson mass
$M_{H^\pm}$ is the best choice as input parameter, contrary to the case 
of the real MSSM, where $M_A$ is usually chosen as input. 
In the approximation of vanishing 
external momentum the mixing can be written as
\begin{equation}
\left( \begin{array}{c} H_1 \\ H_2 \\ H_3 \end{array} \right) \; = \;
\mathbf{O}
\left( \begin{array}{c} h \\ H \\ A \end{array} \right)~, \qquad
\mathbf{O} = \left( \begin{array}{ccc}
O_{11} & O_{12} & O_{13} \\ O_{21} & O_{22} & O_{23} \\
O_{31} & O_{32} & O_{33} \end{array} \right)\,,
\end{equation}
with
\begin{equation}
m_{H_1} \le m_{H_2} \le m_{H_3}\,.
\end{equation}
This mixing is induced by sizeable off-diagonal
scalar--pseudoscalar contributions 
${\cal M}^2_{SP}$~\cite{PW,mhiggsCPV,mhiggsFD,CEPW} 
to the general 3~$\times$~3 Higgs-boson mass matrix.  Each of the
individual CP-violating off-diagonal scalar--pseudoscalar mixing
entries ${\cal M}^2_{\rm SP}$ in the neutral MSSM mass-squared matrix contains 
terms scaling qualitatively as
\begin{equation}
\label{MSP}
{\cal M}^2_{\rm SP}\ \sim\ \frac{m^4_t}{v^2}\,
\frac{{\rm Im}\, (\mu A_t)}{32\pi^2\, M^2_{\rm SUSY}}\
\bigg( 1,\ \frac{|A_t|^2}{M^2_{\rm SUSY}}\,,\
\frac{|\mu|^2}{\tan\beta\,M^2_{\rm SUSY}}\,,\
\frac{2{\rm Re}\, (\mu A_t)}{M^2_{\rm SUSY}}\,\bigg)\, ,
\end{equation}
at the one-loop level, which could be of order $M^2_Z$. The 
gluino phase contributes at two-loop order. In Eq.~(\ref{MSP}),
$M^2_{\rm SUSY}$ denotes a common soft supersymmetry-breaking scale 
defined
by the arithmetic average of the squared stop masses.

Furthermore, the effective couplings of Higgs bosons to bottom and top 
squarks~\cite{deltamb} can be affected by the additional CP 
violation~\cite{PW,CEPW}. These effective couplings are induced by 
loops involving gluinos and higgsinos, as well as top and bottom squarks. 
Although these
effects enter the charged and neutral Higgs-boson masses and couplings
formally at the two-loop level, they can still modify the numerical
predictions or masses and couplings in a significant way, and therefore
have to be included in the~analysis. 

Two of the main  production mechanisms for producing neutral Higgs bosons
in $e^+ e^-$ collisions are the Higgs-strahlung process
and the pair-production reaction: 
\begin{align}
e^+ e^- \to Z H_i  & \qquad {\rm with} \qquad  i = 1,2,3 \\
e^+ e^- \to H_i H_j  & \qquad {\rm with} \qquad  i\neq j, \; i,j = 1,2,3 ~.
\end{align}
The effective $H_iZZ$ and $H_iH_jZ$ couplings are given by
\begin{eqnarray}
  \label{HVV}
{\cal L}_{\rm int} &=& \frac{g_w}{2\cos\theta_w}\, \bigg[\,M_Z\,
\sum\limits_{i=1}^3\, g_{H_iZZ}\, H_i Z_\mu Z^\mu\: +\:
\sum\limits_{j>i=1}^3\, g_{H_iH_jZ}\, ( H_i\, \!\!
\stackrel{\leftrightarrow}{\vspace{2pt}\partial}_{\!\mu} H_j
)\,Z^\mu\, \bigg]\, ,
\end{eqnarray}
where              $\cos\theta_w              \equiv M_W/M_Z$,
$\stackrel{\leftrightarrow}{\vspace{2pt}  \partial}_{\!  \mu}\ \equiv\
\stackrel{\rightarrow}{\vspace{2pt}      \partial}_{\!      \mu}     - 
\stackrel{\leftarrow}{\vspace{2pt} \partial}_{\! \mu}$, and
\begin{eqnarray}
  \label{gHZZ}
g_{H_iZZ} &=& \cos\beta\, O_{1i}\: +\: \sin\beta\, O_{2i}\,,\nonumber\\
 \label{gHHZ}
g_{H_i H_j Z} &=& O_{3i}\, \Big(
\cos\beta\, O_{2j}\, -\, \sin\beta\, O_{1j}  \Big)\ -\
 O_{3j}\, \Big(  \cos\beta\, O_{2i}\,  -\,  \sin\beta\, O_{1i} \Big)\, .
\end{eqnarray}
From the above coupling structure it follows that

\begin{itemize}
\item
the effective couplings  $H_i ZZ$ and $H_i H_j Z$  are related to each
other through 
\begin{equation}
  \label{Orel}
g_{H_k ZZ}\ =\ \varepsilon_{ijk}\, g_{H_i H_j Z}\,, \quad {\rm and}
\end{equation}

\item
unitarity leads to the coupling sum rule~\cite{Alex}
\begin{equation}
\sum\limits_{i=1}^3\, g_{H_i ZZ}^2\ =\ 1\, ,
\end{equation}
which reduces the number of  independent $H_iZZ$ and $H_iH_jZ$
couplings. 

\end{itemize}

It is obvious from Eq.~(\ref{MSP}) that CP-violating effects on the
neutral Higgs-boson mass matrix become significant when the product
${\rm Im}\, (\mu A_t)/ M^2_{\rm SUSY}$ is large. Motivated by this
observation, the following {\it CP-violating benchmark scenario (CPX)}
is defined as~\cite{cpx}:
\begin{eqnarray}
  \label{benchCP}
\widetilde{M}_Q \!&=&\! \widetilde{M}_t\ =\
\widetilde{M}_b\ =\ M_{\rm SUSY}\,,\qquad \qquad 
\mu \ =\ 4 M_{\rm SUSY}\,,\nonumber\\
|A_t| \!&=&\!  |A_b|\ =\ 2M_{\rm SUSY}\,,\qquad \qquad 
\hspace*{-1.3mm} {\rm arg}(A_t)\ =\ 90^\circ\,, \nonumber\\
|m_{\tilde{g}}| \!&=&\! 1~{\rm TeV}\,,\qquad \qquad
\hspace*{15mm} {\rm arg}(m_{\tilde{g}})\ =\ 90^\circ\, ,
\end{eqnarray}
where we follow the notation of~Ref.~\cite{CEPW}.
Without loss  of generality, the  $\mu$ parameter is considered  to be
real. We note  that  the  CP-odd phases  ${\rm  arg}(A_t)$  and  ${\rm
arg}(m_{\tilde{g}})$  are chosen to take their  maximal CP-violating
values. In the following, we also discuss variants of the CPX scenario
with other values of ${\rm arg}(A_t)$ and ${\rm arg}(m_{\tilde{g}})$,
keeping the other quantities fixed at the values in~Eq.~(\ref{benchCP}).

\begin{figure}[htb!]
   \leavevmode
 \begin{center}
   \epsfxsize=11.5cm
    \epsffile[0 0 539 652]{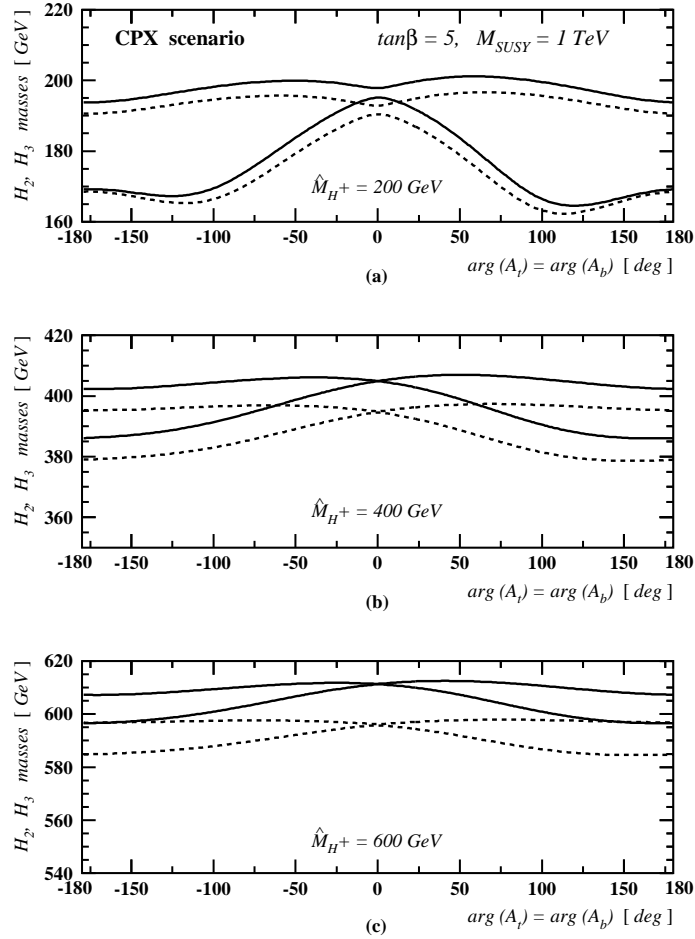}
 \end{center}
\caption{Numerical estimates of the two heaviest $H_2$- and
$H_3$-boson masses versus ${\rm arg}\, (A_t)$ for different charged
Higgs-boson pole masses in a CPX scenario with 
${\rm arg} ( m_{\tilde{g}} )$~=~90$^\circ$ and 
$M_{\rm SUSY}$~=~1~TeV. Effective-potential masses are  
indicated by solid lines and pole masses by dashed ones.}
\label{fig:poleheavy}
\end{figure}

In Fig.~\ref{fig:poleheavy}, we display the masses of the two heaviest
neutral Higgs bosons, $H_2$ and $H_3$, as functions of ${\rm arg}\,
(A_t)$, in the CPX scenario with ${\rm tan}\beta$~=~5, $M_{\rm SUSY}$~=~1~TeV.
Going from the upper to the
lower panel in Fig.~\ref{fig:poleheavy}, we vary the charged Higgs-boson
pole mass: $M_{H^\pm}$~=~200, 400, 600~GeV.  Numerical results
pertaining to the effective-potential masses are indicated by solid lines,
while the results of pole masses are given by the dashed lines. 

We see that the mass difference between the two heavier neutral Higgs
bosons $H_2$ and $H_3$ may be very different from its value when ${\rm
arg}\, (A_t)~=~0$, because of repulsive CP-violating mixing between
the two states.   
Although in general every possible mass difference for CP-violating 
parameters can also be reached in the real MSSM~\cite{markusPhD}, the 
measurement of $m_{H_3} - m_{H_2}$ can give valuable information 
about the complex phases in the CMSSM Higgs sector.
Measuring these mass differences accurately will require an $e^+ e^-$
linear collider, which can only be CLIC if $M_{H^\pm}$ is large. As
shown in the previous section, it should be possible to measure the 
$H^\pm$ and $H_{2,3}$ masses with an accuracy $\sim $~1\% at CLIC,
smaller than the possible mass shifts shown~in~Fig.~\ref{fig:poleheavy}.

If the mass of the charged Higgs boson is large, $M_{H^\pm}
\gg M_Z$, the lightest MSSM Higgs boson becomes an almost pure CP-even
state, whilst the other two neutral Higgs bosons mix through a single
mixing angle $\theta_{23}$ between the CP eigenstates $H$~and $A$. If the
mixing is substantial, which can happen in certain parts of the CMSSM
parameter space~\cite{markusPhD}, determining the value of $\theta_{23}$
will be one of the most important measurements in the Higgs sector.  
Since $\theta_{23}$ can be strongly influenced by ${\rm arg}\,(A_t)$ via the
loop-corrected Higgs boson self-energies and the loop-induced Higgs boson
mixing, one can expect a non-negligible numerical correlation between the
two parameters.

In the limit when $M_{H^\pm} \gg M_Z$, the only mode for producing the
neutral heavy Higgs bosons at CLIC is the associated production mechanism
$e^+ e^- \to Z^* \to H_2 H_3$. Determining the angle $\theta_{23}$ can in
principle be done in two different ways. In the first approach, the decay
products of the $H_2$ and $H_3$ would have to be analysed. If the $H_{2,3}
\to \tau^+\tau^-$ channels have sufficiently large branching ratios, the
$\tau$ polarizations might be used to get a handle on
$\theta_{23}$~\cite{theta23direct}.  However, for $m_{H_{2,3}} \gsim 2
\mt$ the $\tau^+\tau^-$ decay branching ratio becomes strongly suppressed,
and other methods to determine $\theta_{23}$ become more~interesting.

We have analysed the correlation of the branching ratios
of $H_2$ and $H_3$, the angle $\theta_{23}$ and the phase 
${\rm arg}\, (A_t)$ for $M_{H^\pm}$~=~300, 400~GeV.
Assuming heavy charginos, the relevant decay channels for the heavy
Higgs bosons are in this~case
\begin{equation}
H_{2,3} \; \to \; \bar t t,\; \bar b b,\; \tau^+\tau^-,\; 
                  \bar{\tilde t}_1 {\tilde t_1},\; H_1 H_1\,.
\end{equation}
Any of these channels is of course absent if it is kinematically
forbidden. If the $ \bar{\tilde t}_1 {\tilde t_1}$ channel is 
allowed, it usually
dominates the other decays, and hardly any variation of the branching 
ratios with ${\rm arg}\, (A_t)$ can be
observed. Therefore we focus on the more favourable scenarios in
which ${\tilde t_1}$ is too heavy to be produced in Higgs~decays. 

\begin{figure}[htbp]
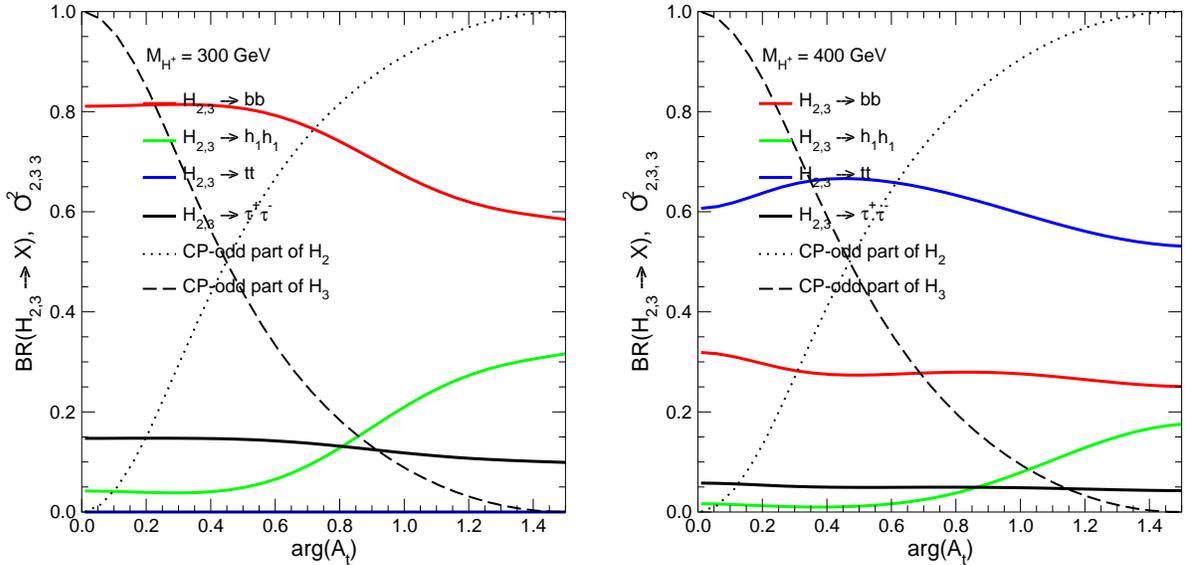
 
\begin{center}
\epsfig{figure=Chap4/clic03c.br.cl.eps, width=7.5cm,height=7.5cm}
\hspace{5mm}
\epsfig{figure=Chap4/clic05c.br.cl.eps, width=7.5cm,height=7.5cm}
\caption{The principal branching ratios (see text) and the CP-odd parts
$\sim \cos^2 \theta_{23}, \sin^2 \theta_{23}$ of $H_2$ and $H_3$ are shown 
as functions of ${\rm arg}\, (A_t)$. In the left (right) plot the
charged Higgs boson mass is set to $M_{H^\pm}$~=~300~(400)~GeV. The
other parameters are specified in  the text.}
\label{fig:h23BRs}
\end{center}
\end{figure}

As representative sample parameters, we choose ${\rm tan}\beta$~=~5, 
$M_{\rm SUSY}$~=~500~GeV, 
$|A_t|~=~A_b~=~A_\tau$~=~1200~GeV, $\mu$~=~800~GeV, $M_2$~=~400~GeV 
and $m_{\tilde g}$~=~500~GeV. We vary ${\rm arg}\, (A_t)$ in the
interval $[ 0, \pi/2 ]$~radians, and note that a measurement of 
${\rm arg}\, (A_t)$ thus
corresponds directly to a determination of $\theta_{23}$. Our evaluations
have been performed with \fh 2.0~\cite{feynhiggs,feynhiggs2.0}, taking
into account corrections up to the two-loop level in the Higgs boson
masses, mixing and decays. The results for $M_{H^\pm}$~=~300, 400~GeV are
shown in the left and right plots of Fig.~\ref{fig:h23BRs}, respectively.
Since it will be nearly impossible to distinguish the decay of the
$H_2$ from that of the $H_3$, we have averaged over the two branching
ratios, i.e. what is shown in Fig.~\ref{fig:h23BRs} is 
$[\br(H_2 \to X) + \br(H_3 \to X)]/2$.  
For $M_{H^\pm}$~=~300~GeV, the channel $H_{2,3} \to \bar t t$ is
absent, whilst for $M_{H^\pm}$~=~400~GeV it is dominant. Thus these two
different choices of $M_{H^\pm}$ represent two distinct cases for the
branching-ratio analysis.

Together with the branching ratios, we also display the CP-odd parts of
$H_2$, scaling like $\cos^2\theta_{23}$, and of $H_3$, scaling like
$\sin^2\theta_{23}$.  The numerical correlations between the branching
ratios, ${\rm arg}\, (A_t)$ and $\{\cos,\sin\}^2\theta_{23}$ 
can be read off from
the plots. Therefore, a precise measurement of ${\rm arg}\, (A_t)$ via the
branching ratios would
allow one to retrieve information about $\theta_{23}$. 
However, one should keep in mind that
we have varied only one parameter and kept the others fixed. This would
correspond to perfect experimental knowledge of the other parameters,
which is of course an unrealistic assumption.  However, the potential for
determining $\theta_{23}$ and ${\rm arg}\, (A_t)$ in this way is
nevertheless demonstrated in this exploratory~study.

Furthermore, the precision with which ${\rm arg}\, (A_t)$ and 
$\theta_{23}$ can be determined 
depends strongly on the achievable precision of the branching-ratio
measurement.  The cross section at CLIC for the 
$e^+ e^- \to Z^* \to H_2 H_3$ 
production process is slightly larger than 1~fb~\cite{eehZhA}, and is
largely independent of the value of $M_{H^\pm}$, which results in about
1000~events for each collected 1~ab$^{-1}$. Unfortunately, no evaluation of the
precision with which the branching ratios of the heavy neutral Higgs
bosons could be measured at CLIC is yet available.

\begin{figure}[htbp] %
\centerline{\includegraphics[width=15.5cm]{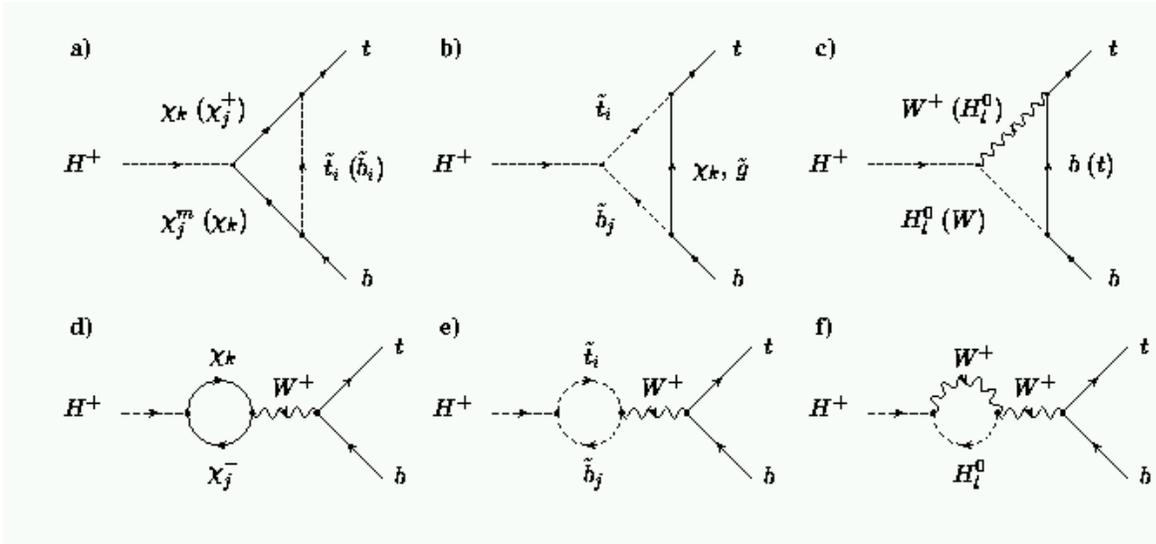}}

\vspace*{2mm}

\caption{Sources of CP violation in $H^+\to t\bar b$ decays
at the one-loop level in the MSSM with complex couplings
($i,j$~=~1,~2; $k$~=~1, ..., 4; $l$~=~1, 2, 3)}
\label{fig:feyngraphs}
\end{figure}

\begin{figure}[htbp] 
\setlength{\unitlength}{1mm}
\begin{center}
\begin{picture}(60,62)
\put(0,0){\mbox{\epsfig{figure=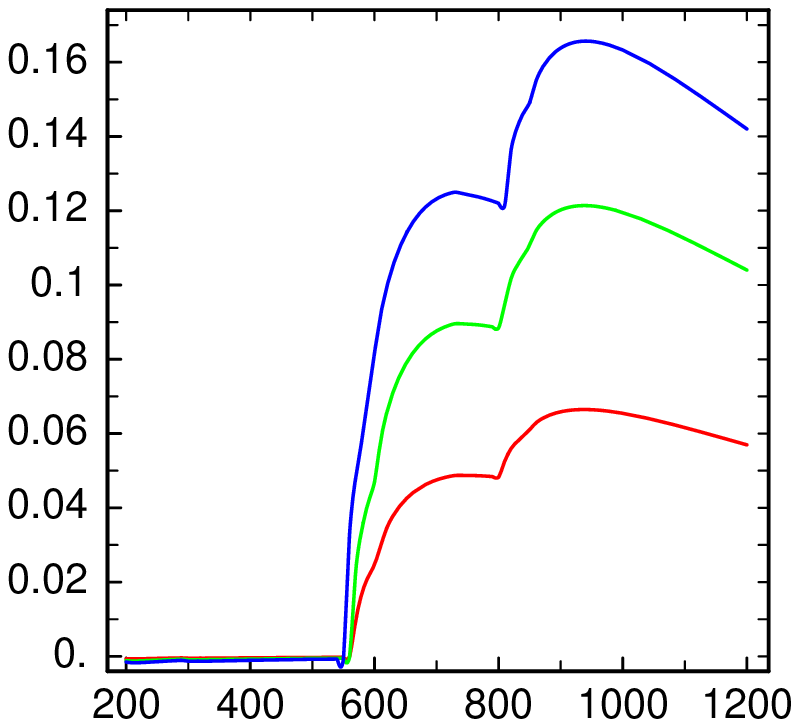,height=61mm}}}
\put(24,49){\footnotesize $\phi_t=\pi/2$}
\put(49,43.5){\footnotesize $\pi/4$}
\put(49,28.5){\footnotesize $\pi/8$}
\put(24,-2){$m_{H^+}$~[GeV]}
\put(-6,28){\rotatebox{90}{$-\delta^{CP}$}}
\put(-6,56){\bf a)}  
\end{picture}
\hspace{16mm}
\begin{picture}(60,62)
\put(0,0){\mbox{\epsfig{figure=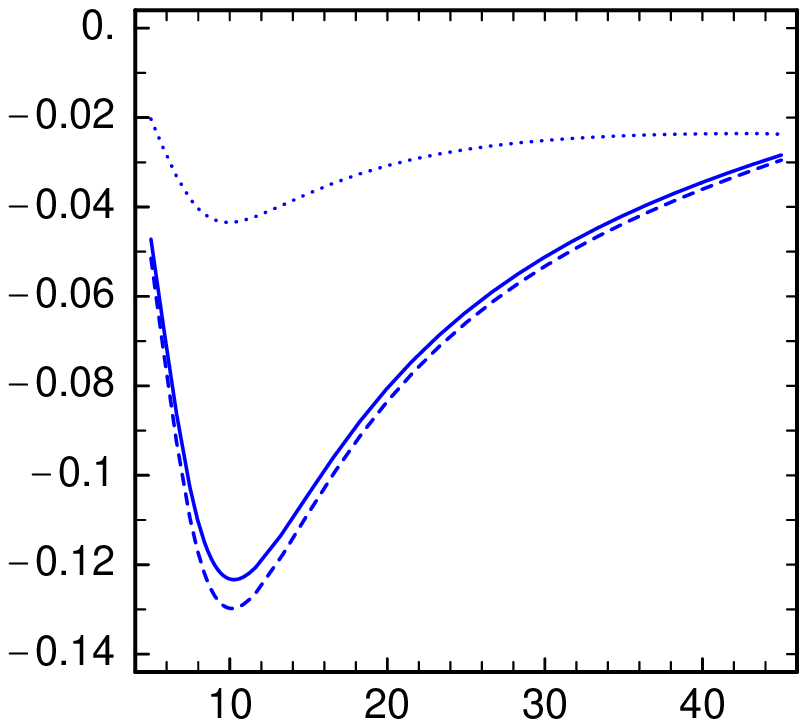,height=63mm}}}
\put(34,51.5){\tiny $\mu=+350$~GeV, $\phi_b=0$}
\put(25,36){\footnotesize $\phi_b=0$}
\put(25.5,16){\footnotesize $\phi_b=\frac{\pi}{2}$}
\put(34,-2){$\tan\beta$}
\put(-5,30){\rotatebox{90}{$\delta^{\rm CP}$}}
\put(-5,56){\bf b)}
\end{picture}
\end{center}
\caption{ CP violation in charged Higgs decays: 
(a) the asymmetry $\delta^{\rm CP}$ as a function of $m_{H^+}$ for 
selected values of $\phi_t$, assuming ${\rm tan}\beta$~=~10 and
$\phi_b~=~0$, and (b) $\delta^{\rm CP}$ as a function of 
${\rm tan}\beta$ for $m_{H^+}$~=~700~GeV and $\phi_t=\pi/2$. The other
parameters are fixed by (\ref{eq:parset}).}
\label{fig:cphpm}
\end{figure}

Finally, another possible observable in the MSSM with complex parameters is 
a difference in the decay rates of $H^+$ and $H^-$~\cite{Christova:2002ke}.
For the $tb$ final state, the rate asymmetry 
\begin{equation}
  \delta^{\rm CP}  \equiv 
  \frac{\Gamma(H^+ \to b\bar{t})-\Gamma(H^- \to t\bar{b})} 
       {\Gamma(H^+ \to b\bar{t})+\Gamma(H^- \to t\bar{b})} 
\end{equation}
can be induced by the one-loop diagrams shown  
in~Fig.~\ref{fig:feyngraphs}~\cite{Christova:2002ke}.
For the numerical analysis of this possibility, we use
\begin{align}
   & M_2=200~{\rm GeV}\,,\quad \mu=-350~{\rm GeV}\,,\quad 
     M_{\tilde Q} = 350~{\rm GeV}\,, \nonumber \\
   & M_{\tilde U} : M_{\tilde Q} : M_{\tilde D} = 0.85 : 1 : 1.05\,, 
     \quad A_t = A_b = -500~{\rm GeV}\,,
\label{eq:parset}
\end{align}
assuming GUT unification between the gaugino masses $M_1, M_2, M_3$, 
and investigate the effect of CP-violating phases $\phi_{t,b}$ in the 
soft trilinear supersymmetry-breaking parameters $A_{t,b}$. 
In Fig.~\ref{fig:cphpm}(a) we show the CP-violating asymmetry
$\delta^{\rm CP}$ as a function of the charged Higgs mass for various
values of $\phi_t$, with $\phi_b$~=~0 and ${\rm tan}\beta$~=~10. In
Fig.~\ref{fig:cphpm}(b) we show  
the ${\rm tan}\beta$ dependence of $\delta^{\rm CP}$ for
$m_{H^+}$~=~700~GeV and $\phi_t=\pi/2$. 
As can be seen, $\delta^{\rm CP}$ could amount to $\simeq$~15\%.  
The leading contribution is due to the diagram of
Fig.~\ref{fig:feyngraphs}(b)  
with $\tilde{t}$, $\tilde{b}$ and $\tilde{g}$ in the loop; 
$\delta^{\rm CP}$ can become large if this diagram has an absorptive
part, i.e. if the decay channel $H^\pm\to \tilde t\tilde b$ is open.

\vspace*{2mm}

With $m_{H^+}$~=~700~GeV, taking the expected statistics of 
$e^+e^- \to H^+H^- \to t \bar{b} \bar{t} b$ at $\sqrt{s}$~=~3~TeV and 
assuming realistic charge tagging performances,
a 3$\sigma$ effect would be observed with ${\cal{L}}$~=~5~ab$^{-1}$, for an
asymmetry $|\delta^{\rm CP}|$~=~0.10.

\section{Summary}

The CLIC multi-TeV $e^+e^-$ linear collider has the potential to
complete the study of the Higgs boson and to investigate an extended Higgs
sector over a wide range of model parameters. Preserving the LC signature
properties of clean events, with well-defined kinematics, will require a
substantial effort of machine parameter optimization, detector design and
data analysis techniques. However, exploratory studies, accounting for
realistic experimental conditions, confirm that CLIC will perform
precision measurements and push its sensitivity up to the kinematical
limits.

While the main motivation for experimentation at a multi-TeV LC arises
from the search for new phenomena beyond the Standard Model, as discussed
in the following Chapters, CLIC's role in studying the Higgs sector
will also be crucial in completing the mapping of the $H^0$ boson profile.
Table~\ref{tab:summary} summarizes the accuracies expected from some {\sc
Clic} measurements of the properties of a Higgs boson in the SM.
Additionally, CLIC would have unique capabilities for studying heavy
Higgs bosons in extended scenarios.  In this way, the multi-TeV
CLIC project will ensure the continuation of a competitive $e^+e^-$ physics
programme via a second-generation project able to address deeper physics
questions in the Higgs~sector.

\newpage 

\begin{table}[!ht] 
\caption{Summary of expected accuracies on Higgs properties at 
CLIC with $\sqrt{s}$~=~3~TeV and 5~ab$^{-1}$ of luminosity}
\label{tab:summary}

\renewcommand{\arraystretch}{1.3} 
\begin{center}

\begin{tabular}{cccc}\hline \hline \\[-4mm]
\textbf{Parameter} & $\hspace*{3mm}$ &
$\hspace*{8mm}$ \boldmath{$M_H$} \textbf{(GeV)} $\hspace*{8mm}$ & 
$\hspace*{8mm}$ \boldmath{$\delta X/X$} $\hspace*{8mm}$ 
\\[4mm]   

\hline \\[-3mm]
$\delta g_{Htt}/g_{Htt}$ & &120--180 & 0.05--0.10  \\
$\delta g_{Hbb}/g_{Hbb}$ & &180--220 & 0.01--0.03 \\
$\delta g_{H\mu\mu}/g_{H\mu\mu}$ & &120--150 &0.03--0.10 \\
$\delta g_{HHH}/g_{HHH}$   & &120--180 & 0.07--0.09  \\
$g_{HHHH}$ & &120 & $\ne$~0 (?) 
\\[3mm] 
\hline \hline
\end{tabular}
\end{center}
\end{table}


\newpage
\chapter{SUPERSYMMETRY}
\label{chapter:five}

\def\b               {\beta}
\def\t               {\theta}

\newcommand{\lsim}{\;\raisebox{-0.9ex}{$\textstyle\stackrel{\textstyle<}
           {\sim}$}\;}

Supersymmetry (SUSY) is one of the best-motivated theories beyond the
Standard Model~\cite{Nilles:1983ge}.  Not only is it theoretically
elegant, providing a unified description of fermions and bosons, including
matter particles and force carriers, but it is also potentially capable of
connecting gravity with the other interactions, and appears an essential
ingredient of string theory.  From the phenomenological point of view, the
most compelling feature of SUSY is that it stabilizes the Higgs mass
against radiative corrections, provided the supersymmetric particles have
masses $\tilde m\leq$~O(1)~TeV~\cite{hierarchy}. The minimal
supersymmetric extension of the Standard Model (MSSM) also predicts a
light Higgs boson with $m_h\lsim$~130~GeV~\cite{mhiggscorr}, 
which is favoured by present precision electroweak
data~\cite{:2001xw,Erler:2000cr}. Furthermore, unification of the Standard
Model gauge couplings and electroweak symmetry breaking occur naturally in
the MSSM as a result of renormalization-group (RG) evolution~\cite{GUTs},
and the lightest supersymmetric particle (LSP) is a good cold dark matter
candidate, if SUSY appears within the TeV energy range~\cite{EHNOS}.

\begin{figure}[htbp] %
\centerline{\includegraphics[width=14.0cm]{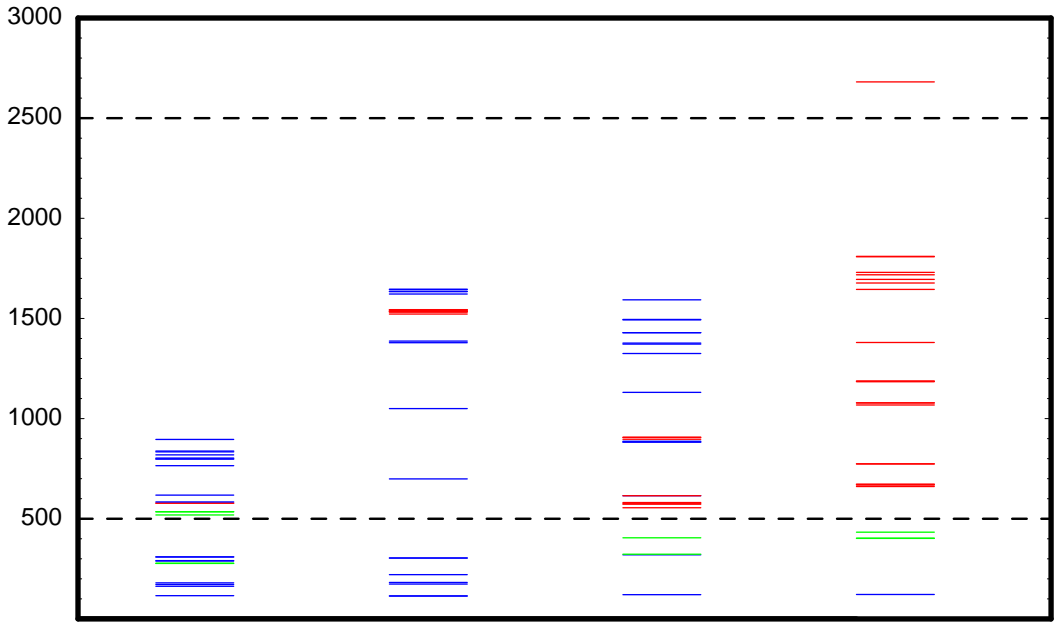}}
\caption{Examples of mass spectra of updated post-LEP benchmark 
points~\cite{bench03}. Sparticles that would be discovered at the LHC,
  a 1-TeV LC and CLIC are shown as blue, green and red lines, respectively. 
  The kinematic reaches of a 1-TeV LC and CLIC at 5~TeV are shown as 
  dashed lines.} 
\label{fig:newpostLEPspectrum}
\end{figure}

In the MSSM, as well as two scalar partners for each 
quark and lepton, there are four neutralinos and two charginos as 
superpartners for the gauge and Higgs bosons. Moreover, there are five 
physical Higgs bosons, $h^0\!$, $H^0\!$, $A^0\!$, and $H^\pm\!$.  
If nature is indeed supersymmetric, it is likely that the LHC will have 
observed part of the spectrum by the time CLIC comes into operation. 
In addition, it is likely that a TeV-scale $e^+e^-$ linear collider (LC) 
will have made 
accurate measurements of some kinematically accessible states, as 
illustrated by the sample spectra in Fig.~\ref{fig:newpostLEPspectrum}.
However, to explore the theory fully, we will need to measure 
accurately the complete sparticle spectrum, just as we need measurements 
of the top quark and the Higgs boson to complete the Standard Model. 
We will need to determine all the MSSM masses, mixing angles, 
couplings, spins, etc., in order to 

\vspace*{1.5mm}

\begin{itemize}
  \item determine all the soft SUSY-breaking parameters,
  \item test their unification at some high scale, 
        which might be the GUT scale or \\
        the SUSY-breaking scale,
  \item pin down the SUSY-breaking mechanism, 
        e.g. whether it is mediated by gravity, \\
        gauge interactions or anomalies,
  \item test the consistency of the model.
\end{itemize}

\vspace*{1.5mm}

\noindent
We also emphasize that the minimality of the MSSM is an 
assumption, which must be tested. 

The LHC has discovery potential for squarks and gluinos for masses up to
about 2.5~TeV. However, its reach for neutralinos, charginos and sleptons
is much lower, namely of the order of 0.5~TeV. We recall that the
questions of unification of SUSY parameters and of the nature of SUSY
breaking require that the measured parameters be extrapolated 
over many orders
of magnitude, from 10$^3$~GeV to 10$^{16}$~GeV.  Very accurate
measurements are required in order to avoid large errors at the
high scale, which needs the most precise measurements possible at an
$e^+e^-$ collider, even of sparticles previously discovered at the LHC.

At CLIC, the clean experimental conditions of $e^+e^-$ annihilation at
$\sqrt{s}$~=~3--5~TeV will enable us to make many such detailed
measurements. Tunable energy and beam polarization will be powerful tools
to disentangle various production channels, enhance signals, and reduce
background processes. Tunable energy also allows for threshold scans, and
polarization is vital to determine the quantum numbers, couplings, and
mixing angles. A high luminosity of 1~ab$^{-1}$ per year is essential for
precision measurements, and even higher energy or luminosity may be
necessary for some particularly difficult scenarios. A multi-TeV $e^+e^-$
linear collider such as CLIC will thus be the ideal machine to complete
the measurements of the LHC and a TeV-scale LC in order to fulfil the
above tasks.

In the following, we discuss the potential of CLIC for studying heavy
charginos, neutralinos, sleptons, and squarks.  To this end, we use for
reference a specific set of benchmark points in the CMSSM, a constrained
version of the MSSM with universal soft SUSY-breaking parameters. We 
concentrate in particular on sparticles with masses beyond the
reach of the LHC and a TeV-scale LC.  We also discuss the determination of
the underlying SUSY-breaking parameters and their extrapolation to the GUT
scale, in order to test unification and to clarify the nature of SUSY
breaking.

\section{Post-LEP Benchmarks and the CLIC Reach}

Benchmark scenarios provide helpful aids for better understanding the
complementarity of different accelerators in the TeV energy range. A set
of benchmark supersymmetric model parameters that are consistent with the
constraints from LEP and other experiments, as well as the cosmology relic
density, have been proposed in~Ref.~\cite{bench01}. They were framed in the
constrained version of the MSSM (CMSSM) with universal soft
symmetry-breaking scalar masses $m_0$, gaugino masses $m_{1/2}$ and
trilinear supersymmetry-breaking parameters $A_0$ at a high input scale,
as expected in a minimal supergravity (mSUGRA) model of soft supersymmetry
breaking. In this framework, the pseudoscalar Higgs mass $m_A$ and the
Higgs mixing parameter $\mu$ (up to a sign) can be derived from the other
MSSM parameters by imposing the electroweak vacuum conditions for any
given value of $\tan\beta$. Thus, given the set of input parameters
determined by $m_{1/2}, m_0, A_0, \tan\beta$ and ${\rm sgn}(\mu)$, the entire
spectrum of sparticles can be derived. For simplicity, the analysis was
restricted to $A_0 = 0$.

These post-LEP benchmark scenarios have recently been
updated~\cite{bench03}, so as to respect the improved restrictions on the
relic density of cold dark matter particles imposed by the WMAP
measurements~\cite{WMAP}. We summarize below some features of 
the updates mandated by WMAP. In the subsequent discussion, we
use the updated post-WMAP benchmarks as far as possible, commenting on
differences from the original set when necessary.

\subsection{Benchmark Points}

Details of the experimental constraints imposed on the CMSSM, the values 
of the parameters chosen as benchmark points, their justifications and the
resulting sparticle spectra may be found in~Refs.\cite{bench01,bench03}.

\begin{figure}[htbp] 
\begin{center}
\begin{tabular}{cc}
\mbox{\epsfig{file=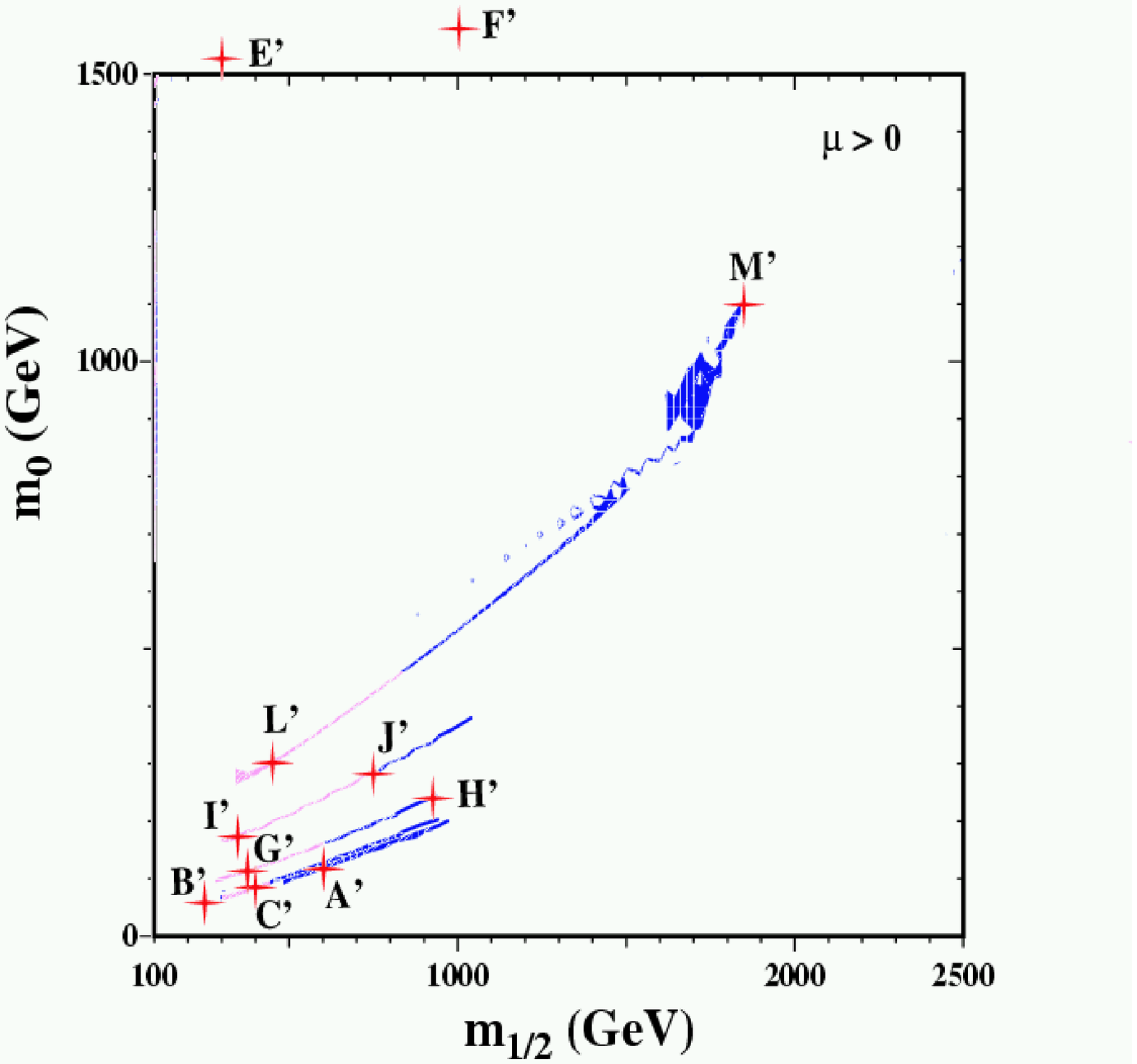,height=7.3cm}} &
\mbox{\epsfig{file=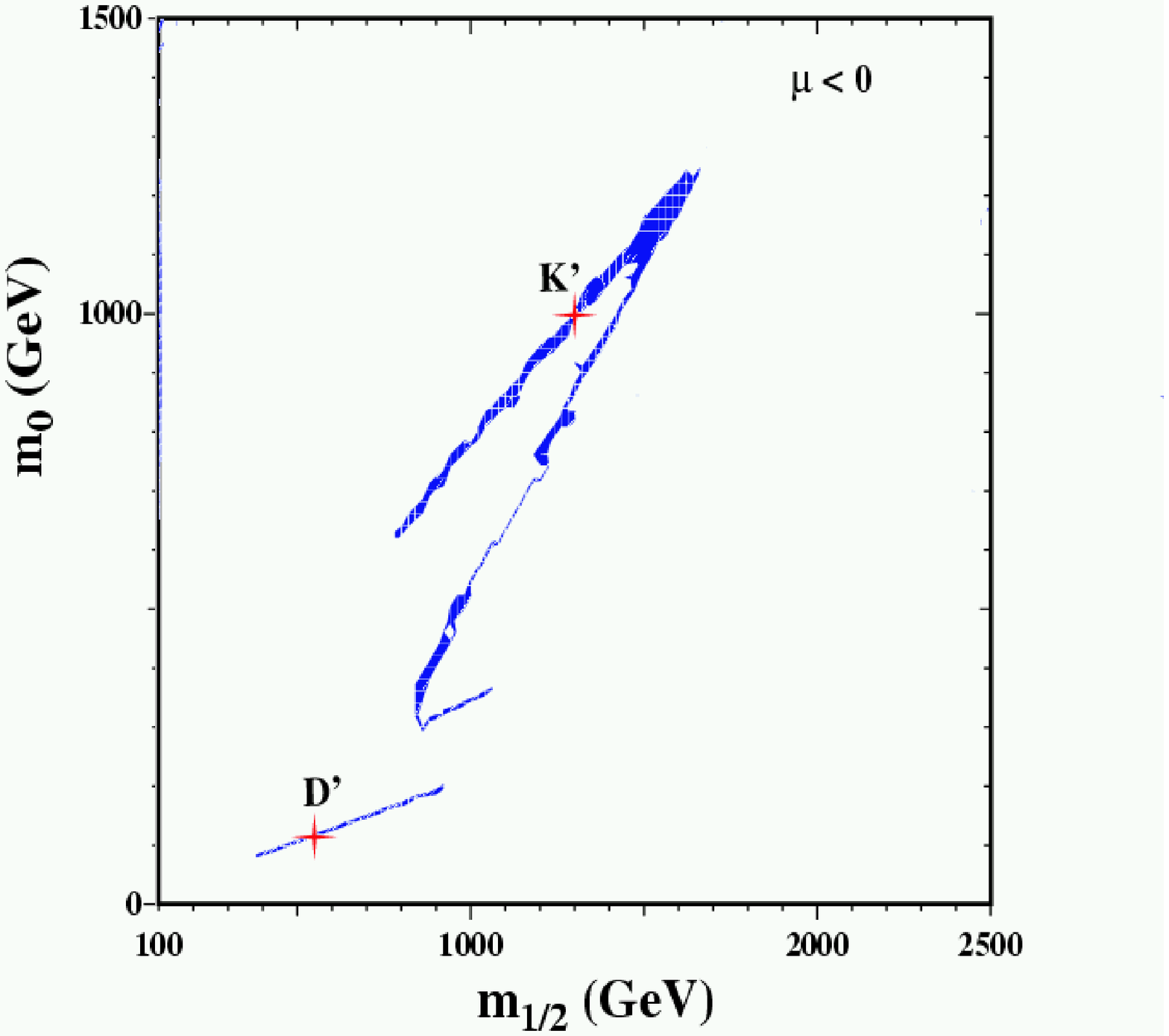,height=7.3cm}} \\
\end{tabular}
\end{center}
\caption{Overview of the updated proposed CMSSM benchmark points in
the $(m_0,m_{1/2})$ planes, superposed on the strips allowed by 
laboratory limits and the relic density constraint, for $\mu >$~0 and
$\tan \beta$~=~5, 10, 20, 35, 50, and for $\mu<$~0 and 
$\tan \beta$~=~10, 35~\cite{bench03}}
\label{fig:WMAPstrips}
\end{figure}

Figure~\ref{fig:WMAPstrips} displays most of the proposed CMSSM
benchmark points, superimposed on the regions of the $(m_{1/2}, m_0)$
plane allowed by laboratory limits, particularly that from LEP on $m_h$,
from $b \rightarrow s \gamma$, and cosmology. The original versions of the
CMSSM benchmark points were chosen with a relic density in the range 
0.1~$< \Omega_\chi h^2 <$~0.3~\cite{bench01}, but WMAP and previous
data now prefer the more limited range 0.094~$< \Omega_\chi h^2
<$~0.129, corresponding to the narrow strips shown in
Fig.~\ref{fig:WMAPstrips}. For most of the benchmark
points, a small reduction in $m_0$ sufficed to relocate them on the WMAP
strip for the corresponding value of $\tan\beta$~\cite{bench03}. However, 
in some cases, notably benchmarks H and M, more substantial changes in 
$m_0$ and/or $m_{1/2}$ were made in order to accommodate the new WMAP 
constraint. Later, where relevant for specific sparticle analyses, we 
comment on the implications of these changes.

The lightest supersymmetric particle would be charged in the bottom right
dark-shaded triangular region, which is therefore excluded. The
experimental constraints on $m_h$ and $b \rightarrow s \gamma$ exert
pressures from the left, which depend on the value of $\tan\beta$ and the
sign of $\mu$.  The indication of a deviation from the Standard Model in
$g_\mu - 2$ disfavours $\mu < 0$ at the 2$\sigma$ level. Large values 
of $m_0$ and $m_{1/2}$
for $\mu > 0$ are disfavoured at the 1$\sigma$ level, as indicated by 
darker shading on parts of the WMAP lines.  
The improved WMAP constraint on the relic density has shrunk the previous
`bulk' region at low $m_0$ and $m_{1/2}$, and narrowed and shortened the
coannihilation `tails' extending to large $m_{1/2}$, which dominate
Fig.~\ref{fig:WMAPstrips}. Not shown is the `focus-point' region at large
$m_0$ near the boundary of the region with proper electroweak symmetry
breaking, where two more benchmark points are located. 

The proposed benchmark points are not intended to sample the allowed CMSSM
parameter space in a statistically unbiased way, but rather to span the
essential range of theoretical possibilities, given our present knowledge.
Estimates of the numbers of CMSSM particles accessible to different
accelerators in the various proposed benchmark scenarios are summarized in
Fig.~\ref{fig:Manhattan} in the general introduction.

\subsection{Detection at the LHC}

\begin{figure}[t!]
\centerline{\epsfxsize = 0.5\textwidth \epsffile{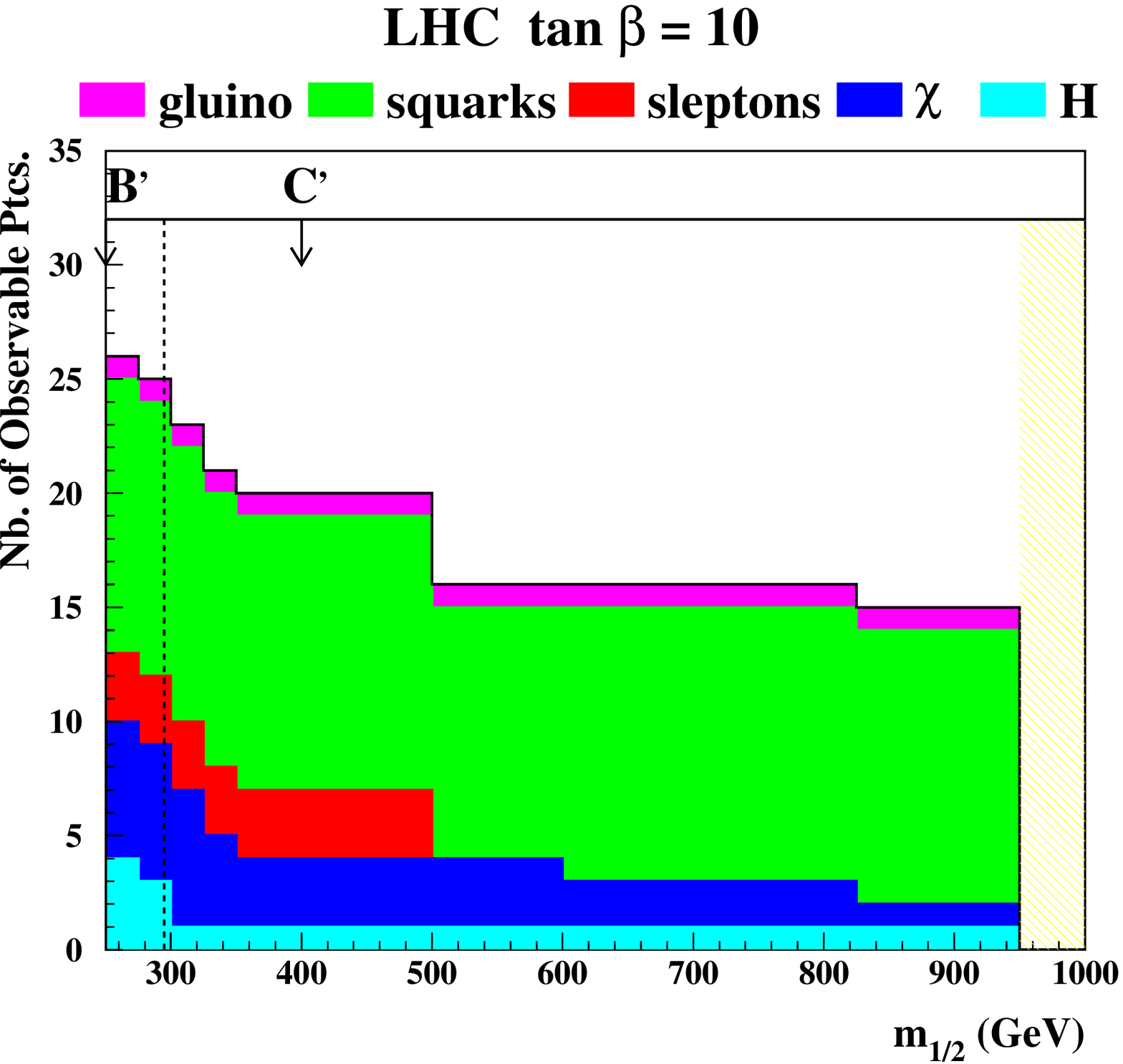}
\hfill \epsfxsize = 0.5\textwidth \epsffile{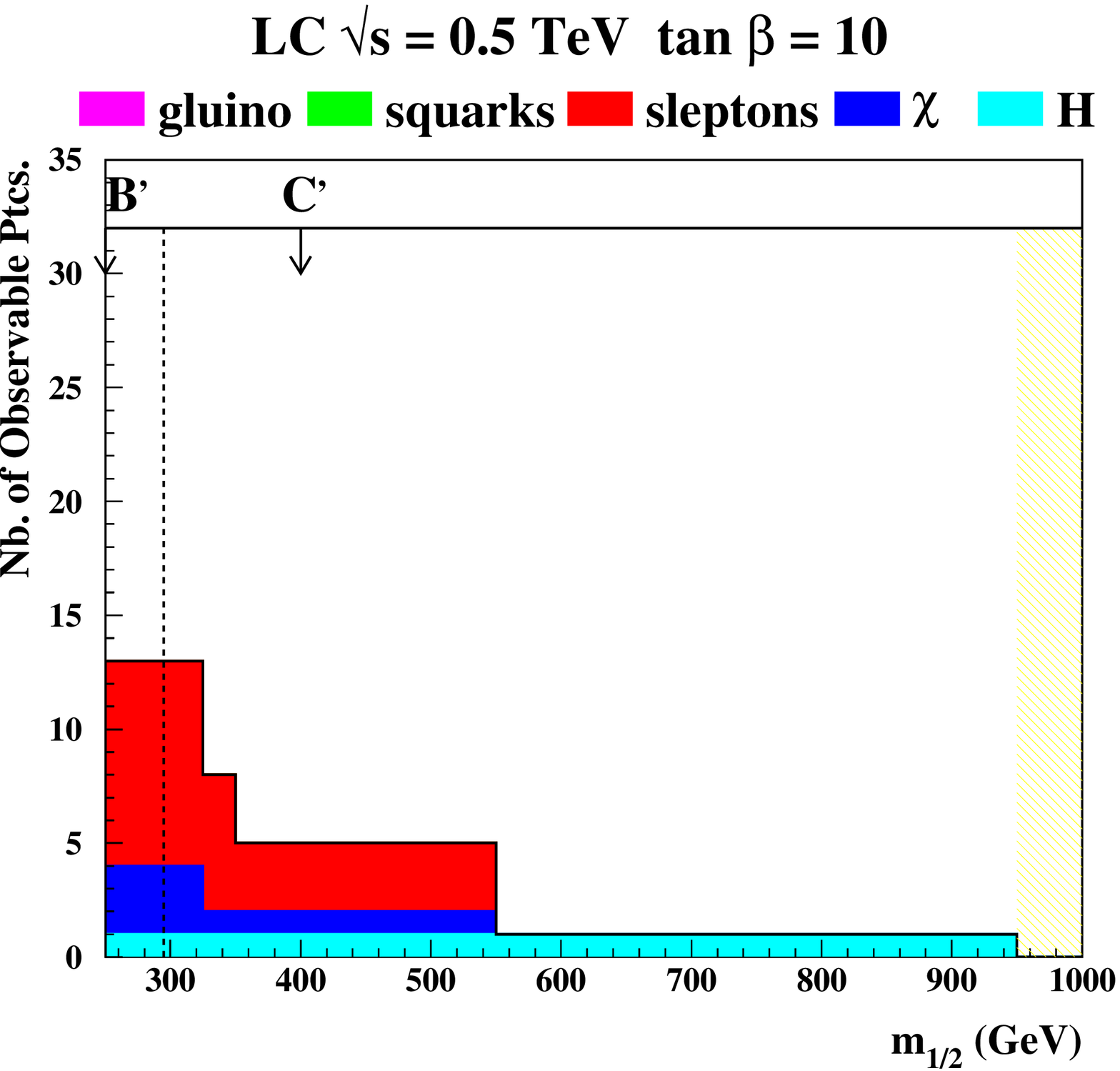}}

\vspace*{3mm}

\centerline{\epsfxsize = 0.5\textwidth \epsffile{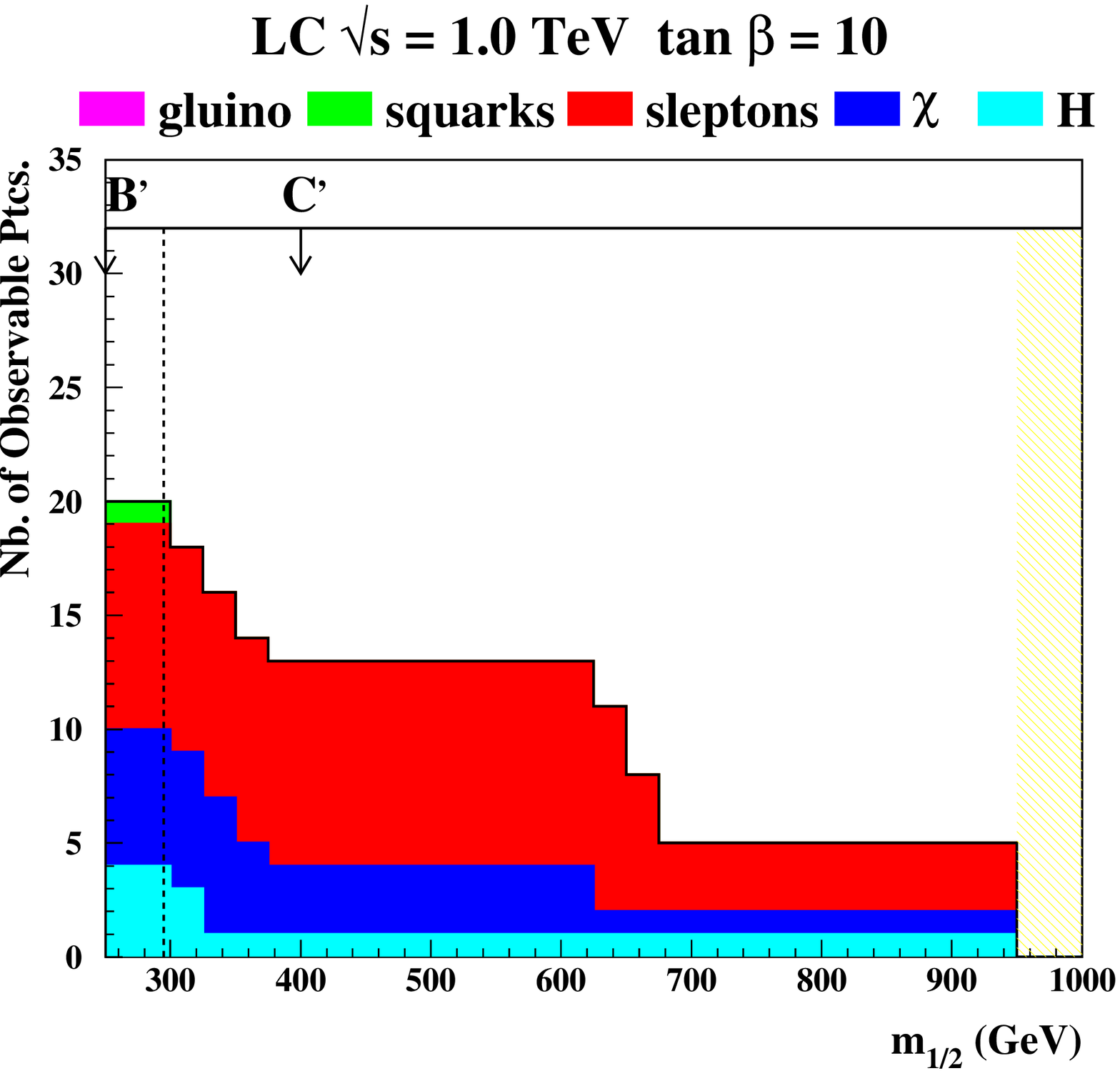}
\hfill \epsfxsize = 0.5\textwidth \epsffile{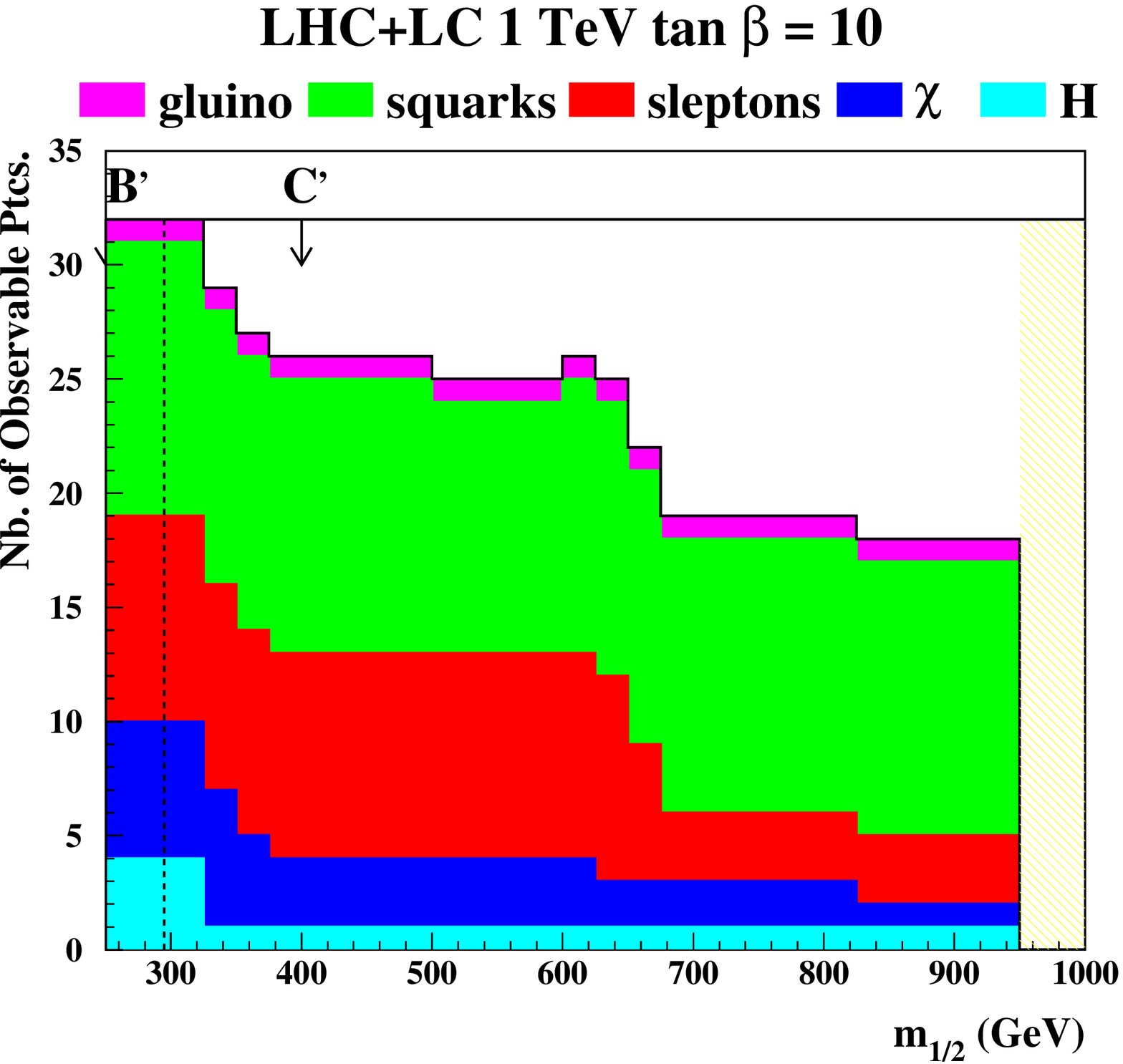}}
\caption{Estimates of the numbers of MSSM particles that may be 
detectable as functions of $m_{1/2}$ along the WMAP line for $\mu >$~0 and 
$\tan \beta $~=~10, for the LHC, a 0.5-TeV $e^+e^-$ linear collider, a 
1-TeV $e^+e^-$ linear collider, and the latter combined with the LHC. The 
complementarity of the LHC and a linear collider emerge clearly. The 
locations of the benchmark points B$^\prime$ and C$^\prime$ along these 
WMAP lines are indicated, as is the nominal lower bound on $m_{1/2}$ 
imposed by $m_h$ (dashed lines)~\cite{bench03}. 
}
\label{fig:lhc_lc_lines}
\end{figure}

In compiling Fig.~\ref{fig:Manhattan}, a preliminary inspection was made
of the LHC potential for these benchmark points, based on the simulation
results summarized in the ATLAS Technical Design
Report~\cite{ATLASTDR} and in the CMS Note~\cite{CMS98006}.  A detailed
study is clearly required before a real assessment of the LHC physics
potential for these benchmarks can be made, and both ATLAS and CMS have
made more detailed studies of these and earlier benchmark scenarios.
However, for this preliminary look, simplified assumptions were adopted to
estimate the discovery potential of the LHC, assuming ATLAS and CMS
combined, together with an integrated luminosity of 300~fb$^{-1}$ per
experiment~\cite{bench01,bench03}. We see in
Fig.~\ref{fig:lhc_lc_lines}(a) how the number of MSSM particles detectable
at the LHC varies along the WMAP line for $\tan\beta$~=~10, including the
points B and C as special cases. For a more complete discussion of
sparticle observability along
WMAP lines, including other values of $\tan\beta$, see~Ref.~\cite{bench03}. We
see that the LHC can detect gluinos and all the squarks, for any allowed
value of $m_{1/2}$, but may be expected to miss many sleptons,
neutralinos, charginos and Higgs bosons, except in the lower part of the
$m_{1/2}$ range.

Thus, in the framework of the CMSSM, the open questions after the LHC 
might include completing the spectrum of electroweakly-interacting 
sparticles, as well as detailed measurements of the gluino and squarks, 
which will not be easy at the LHC.

\subsection{Detection at Linear Colliders}  

Sparticles can generally be produced at any linear $e^+ e^-$ collider 
(LC) if
its centre-of-mass energy is larger than twice the mass of the sparticles,
the pair-production threshold. Exceptions include heavier charginos and
neutralinos, which can be produced in association with lighter charginos
or neutralinos, respectively. Also, for sufficiently light neutralinos and
sneutrinos, observation of the radiative production of otherwise invisible
final states may be experimentally accessible.

Typical supersymmetric signals are multilepton final states and 
multijet ones with large missing transverse energy. Sneutrinos can be
detected at threshold energies if they decay into channels including
charged leptons with a sufficiently large branching ratio. For example, in
some scenarios, the ${\tilde \nu}_{\tau}$ can decay into $\tau {\tilde W}$. 
A counter example to this possibility is the updated benchmark point
B, where all the sneutrino decays are invisible. However, even in such
a case, sneutrinos can be detected at higher energies via the decays
of charginos. 

For $e^+e^-$ linear colliders, data samples of the order of 1000~fb$^{-1}$
or more are expected to be collected over a period of several years. Apart
from the threshold requirement to produce the particles, the product of
the branching ratio times the unpolarized cross section for the detectable
channels was required to be larger than 0.1 fb in order to observe the
sparticle, leading to at least 100 produced sparticles in the total data
sample.

We see in Fig.~\ref{fig:lhc_lc_lines}(b) that a 0.5-TeV $e^+ e^-$ linear
collider would already be able to fill in many of the gaps in the MSSM
spectrum left by the LHC for smaller $m_{1/2}$. The reach of a 1-TeV 
$e^+ e^-$ linear
collider would be correspondingly greater, as
seen in Fig.~\ref{fig:lhc_lc_lines}(c). Such a machine would be largely
complementary to the LHC, as seen in Fig.~\ref{fig:lhc_lc_lines}(d).

\newpage

\subsection{Perspectives}

Assuming that the LHC and an $e^+ e^-$ linear collider in the range
$\lappeq$~1~TeV have been taking data for several years before the start
of a 3~TeV machine like CLIC (CLIC3000), we infer that supersymmetry will
most probably already have been discovered by that time, if it exists. 
Hence the primary r\^ole of CLIC may be to complete the 
sparticle
spectrum, and disentangle and measure more precisely the properties of
sparticles already observed at the LHC and/or a lower-energy $e^+ e^-$
linear collider. However, a machine like CLIC would be needed even for the
direct discovery of supersymmetry in the most problematic cases, such as
benchmark scenarios H and particularly M, where no sparticles may be seen 
at either the LHC or a TeV-class LC, as seen in~Fig.~\ref{fig:Manhattan}.

A few benchmark points emerge as typical of situations that could arise
in the future. 

\begin{itemize}

\item Point C has very low masses, and is representative also of points A,
B, D, G, I, L.  In these cases, the LHC would have discovered the $H^{\pm}$,
as well as seen the $h^0$, and also the gauginos $\chiz_1$, $\chiz_2$ and
$\chipm_1$, the charged sleptons, the squarks and the gluino.  A 1-TeV
linear collider would enable the detailed study of the $h^0$ and of the
same gauginos and sleptons, and it might discover the missing gauginos in
some of the scenarios. However, one would require CLIC, perhaps running
around 2 TeV, to complete the particle spectrum by discovering and
studying the heavy Higgses and the missing gauginos.  CLIC could also
measure more precisely the squarks and in particular disentangle the left-
and right-handed states and, to some extent, the different light squark
flavours.

\item Point J features intermediate masses, much like point K. 
Here, the LHC would have discovered all the Higgs bosons, the squarks and
the gluino, but no gauginos or sleptons.  The 1-TeV $e^+ e^-$ linear
collider would study in detail the $h^0$ and could discover the $\sEl_R$,
$\sMu_R$ and $\sTau_1$, but other sparticles would remain beyond its
kinematic reach.  CLIC3000 could then study in detail the heavy Higgses, 
as discussed in the previous chapter. 
It would also discover and study the gauginos and the missing sleptons,
and even observe in more detail a few of the lighter squarks that had
already been discovered at the LHC. However, to see the remaining
squarks at a linear collider would require CLIC to reach slightly more
than 3~TeV.  

\item Point E has quite distinctive decay characteristics, due to the
existence of heavy sleptons and squarks.  In this situation, the LHC would
have discovered the $h^0$, all squarks and the gluino.  The gauginos are
in principle accessible, but their discovery may be made more difficult
by their predominant decays into jets, contrary to the previous
benchmark points, and sleptons would remain unobserved.  At a 1-TeV $e^+
e^-$ linear collider, the detailed study of the $h^0$ and of the gauginos
could be undertaken.  The discovery of the first slepton, actually a
$\sNu_e$, could be made at CLIC3000, which could also study the three
lightest squarks.  The discovery and analysis of the heavy Higgses would
then require the CLIC energy to reach about 3.5 TeV, which would also
allow the discovery of all sleptons and the observation of all squarks.
A detailed analysis of the accuracy in the determination of the smuon mass 
at $\sqrt{s}$~=~3.8--4.2~TeV is presented later in this chapter.

\item Point H has quite heavy states, as does scenario M.  The LHC would
only discover the $h^0$, all other states being beyond its reach, so the 
LHC might leave the existence of supersymmetry as an open question! At 
point H,
a 1-TeV linear collider would discover the lighter $\tilde \tau$ and the LSP
$\chi$, but no other sparticles. A 1-TeV linear collider would discover no
sparticles at point M. However, CLIC at 3~TeV would be able to discover
most of the gauginos and sleptons.  The CLIC sensitivity to the smuon
mass, using both a muon energy technique and a threshold scan,
is discussed later. On the other hand, to discover all the
squarks, $\ell^+ \ell^-$ collisions in excess of 5~TeV would be needed.
There is currently no $e^+ e^-$ project aiming at such energies, and we
recall that neutrino radiation would become a hazard for a $\mu^+ \mu^-$
collider at such a high energy.

\item Along the lines defined by the WMAP constraints, the reach in
supersymmetric particles for a given collider and the phenomenology of
their decays change significantly. As we discuss later, the CLIC reach 
for the dilepton decay signature of a heavier neutralino, $\chi_2 \to 
\ell^+ \ell^- \chi$ is significantly greater than that of the LHC or a 
1-TeV linear collider. Additionally, we have chosen a point at $m_{1/2}$ 
= 750~GeV and $\tan\beta$~=~10 to study the potential accuracy in the
determination of the mass of the sleptons and of the $\chiz_2$. This
point is located at the limit of the sensitivity of the LHC and of a 1-TeV
linear collider for probing the heavy neutralinos and the slepton sectors,
and represents the limit of the coverage of the full supersymmetric
spectrum at CLIC at 3~TeV.

\item As in the case of a 1-TeV $e^+ e^-$ linear collider, a photon 
collider
option for CLIC would extend the discovery range for heavy Higgs bosons. 
Additionally, it would allow one to discover all four Higgs bosons in
scenarios E, H and M, for a 3-TeV collider, and also in F, for a 5-TeV
collider. The detection of heavier MSSM Higgs bosons at a CLIC-based 
$\gamma \gamma$ collider is discussed in more detail in the previous 
section.

\end{itemize}

\begin{figure}[htbp] 
\begin{center}
\begin{tabular}{cc}
\epsfig{file=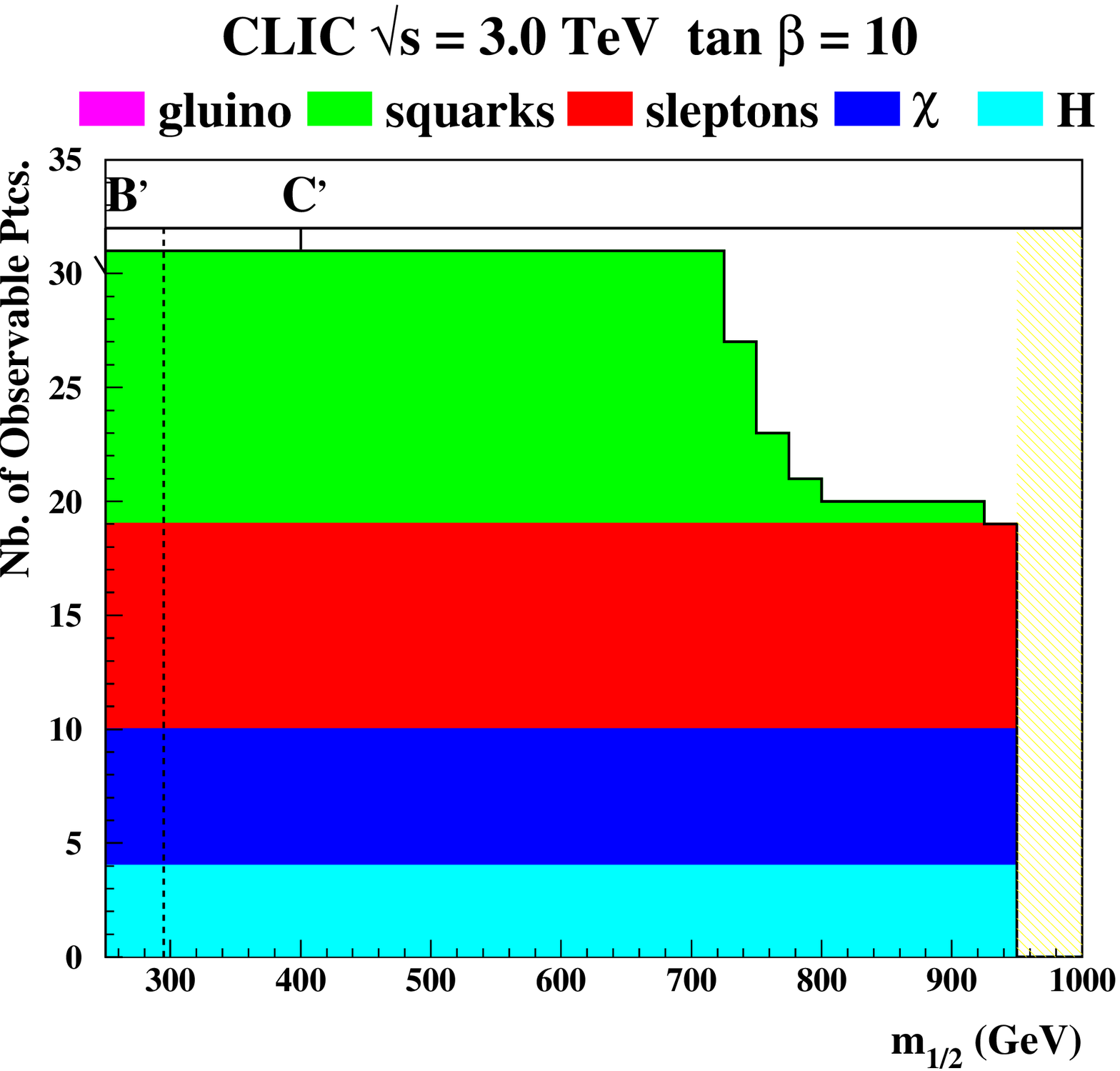,width=7.5cm} &
\epsfig{file=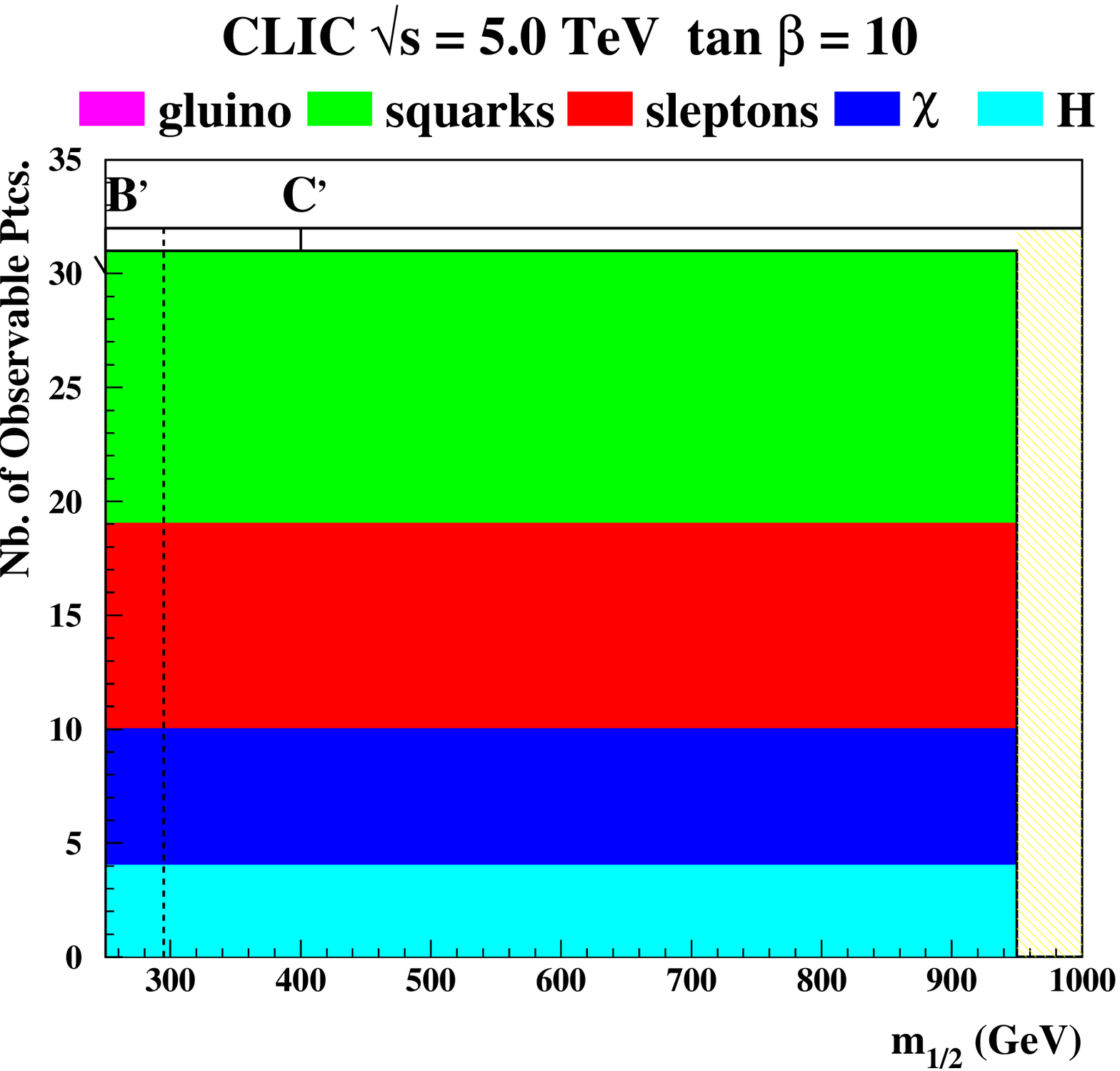,width=7.5cm} \\
\end{tabular}
\end{center}
\caption{Estimates of the numbers of MSSM particles that may be
detectable as functions of $m_{1/2}$ along the WMAP line for $\mu>$~0 and
$\tan \beta$~=~10, for CLIC operating at 3~TeV (left panel) and 
5~TeV (right panel). The locations of the benchmark points B$^\prime$
and C$^\prime$ along these   
WMAP lines are indicated, as is the nominal lower bound on $m_{1/2}$
imposed by $m_h$ (dashed lines)~\cite{bench03}.}
\label{fig:fig5-4}
\end{figure}

The corresponding estimates of the numbers of MSSM particles that can
be detected are shown in~Fig.~\ref{fig:fig5-4}.
Beyond the discovery of sparticles, a crucial issue in the understanding
of the nature of any new physics observed will be the accuracy obtainable
in the determination of the sparticle masses and decays, and also their
quantum numbers and mixing.  
A strong advantage of lepton colliders is the
{\it precision} with which such sparticle properties can be measured. 
%
A strong advantage of lepton colliders is the
{\it precision} with which such sparticle properties can be measured, as 
discussed in the next section of this chapter. 
Typically, the masses of sleptons and gauginos can be determined with a
precision of a few per mille, by threshold
scans and by measuring end-points of two-body decay channel signatures in
inclusive distributions. 
This high precision, even for a limited number
of sparticles, will be of cardinal importance for the reconstruction of
the underlying supersymmetric model and exploration of the 
supersymmetry-breaking mechanism~\cite{Blair:2000gy}. 

The availability of polarized beams at CLIC would, furthermore, provide
additional tools for identifying supersymmetric particles and allow for
additional measurements of parameters of the supersymmetric model, such as
the mixing angles of the sparticles, as discussed later in this chapter.

\subsection{Generalized Supersymmetric Models}
\begin{figure}[th] 
\begin{center} 
\mbox{\epsfig{file=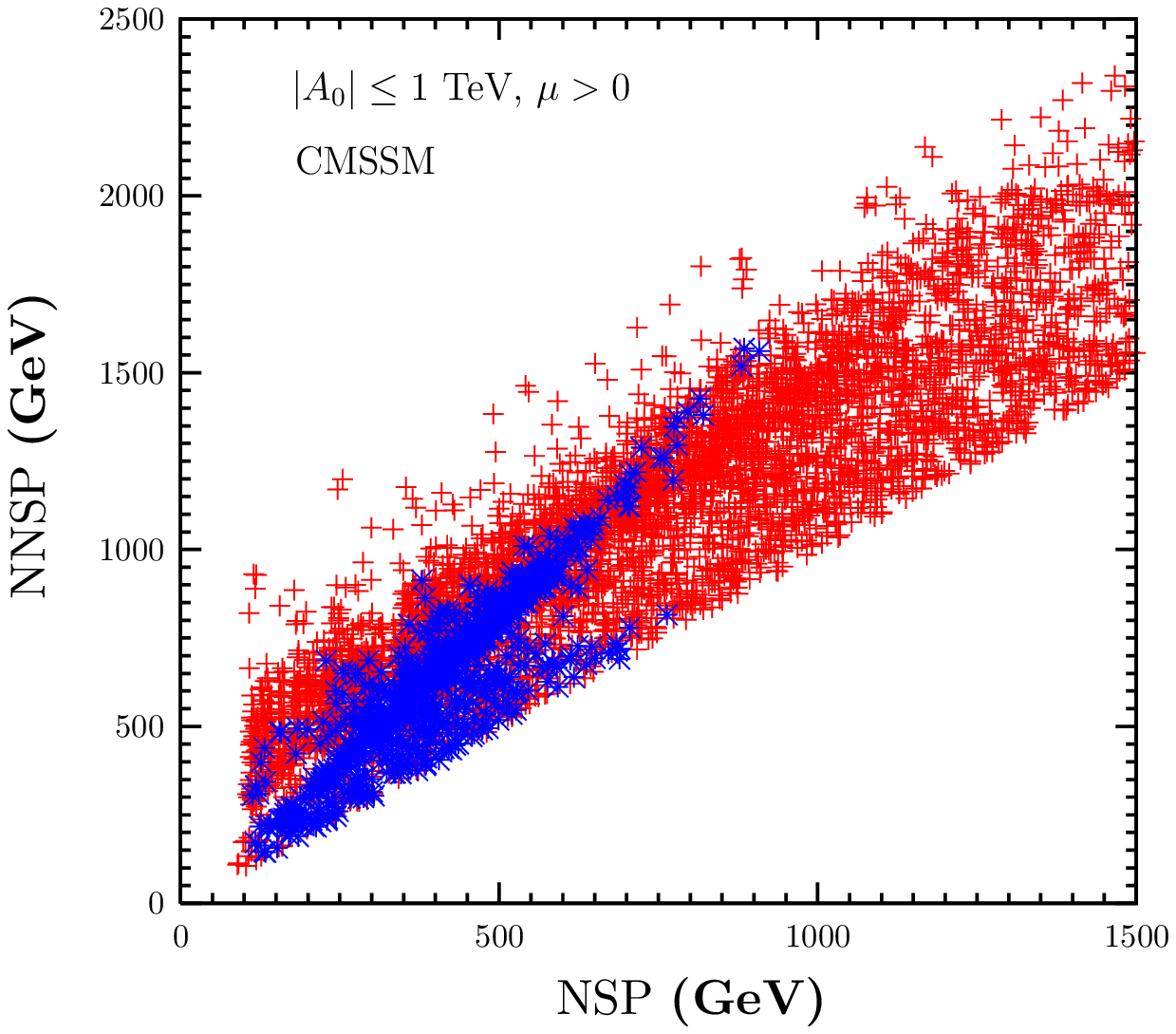,height=6.0cm}}
\mbox{\epsfig{file=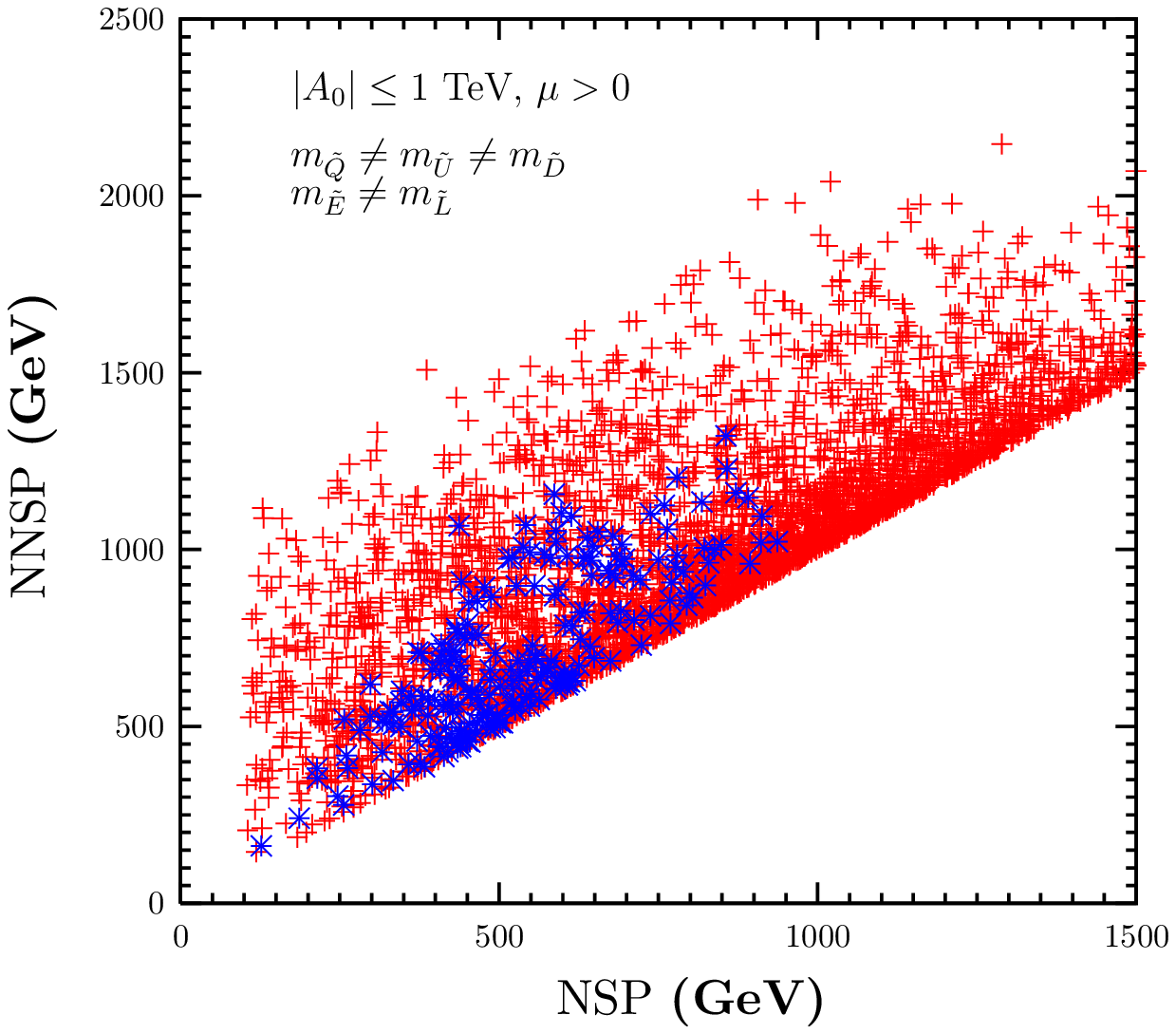,height=6.0cm}}
\end{center}
\begin{center}
\mbox{\epsfig{file=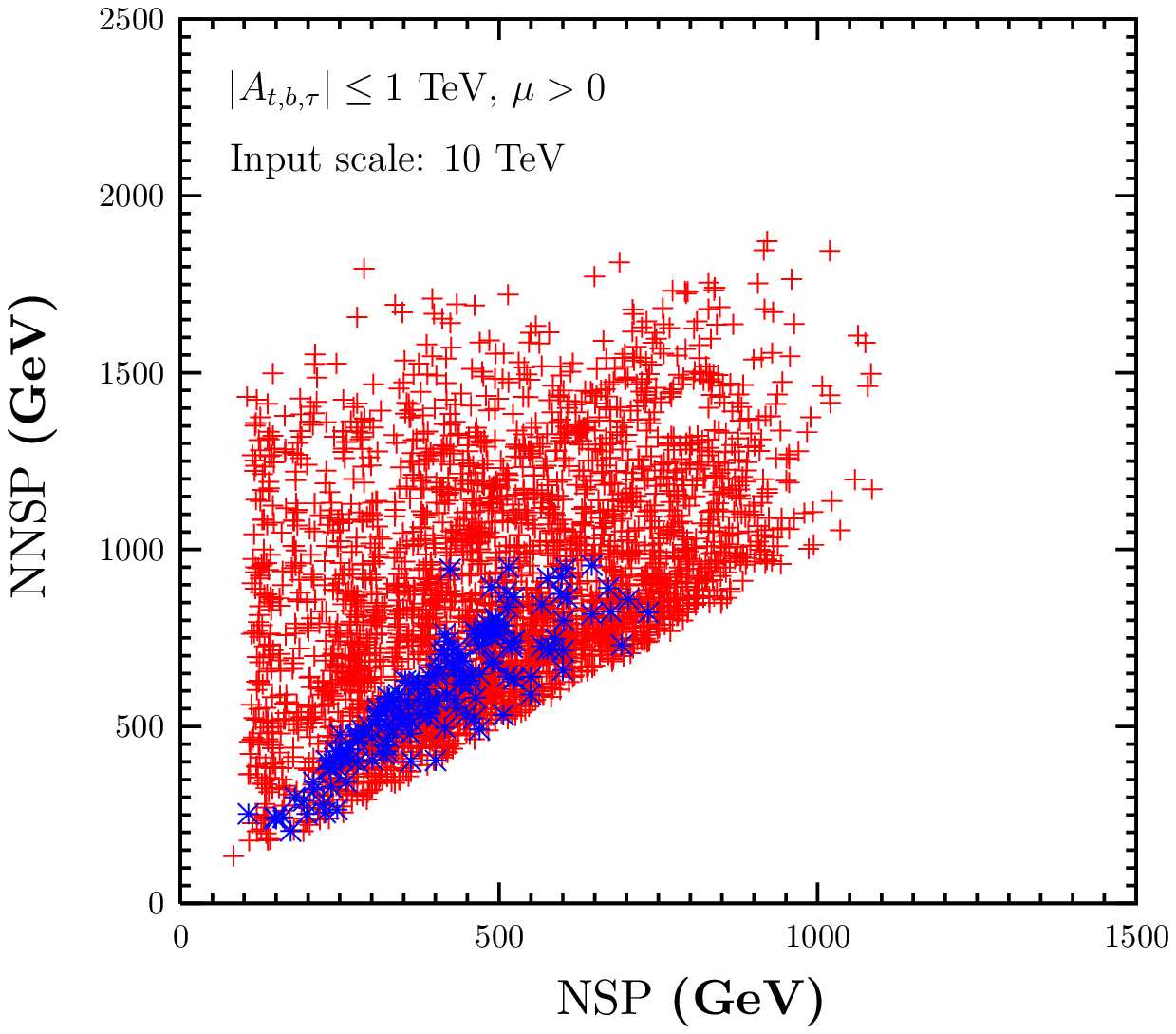,height=6.0cm}}
\mbox{\epsfig{file=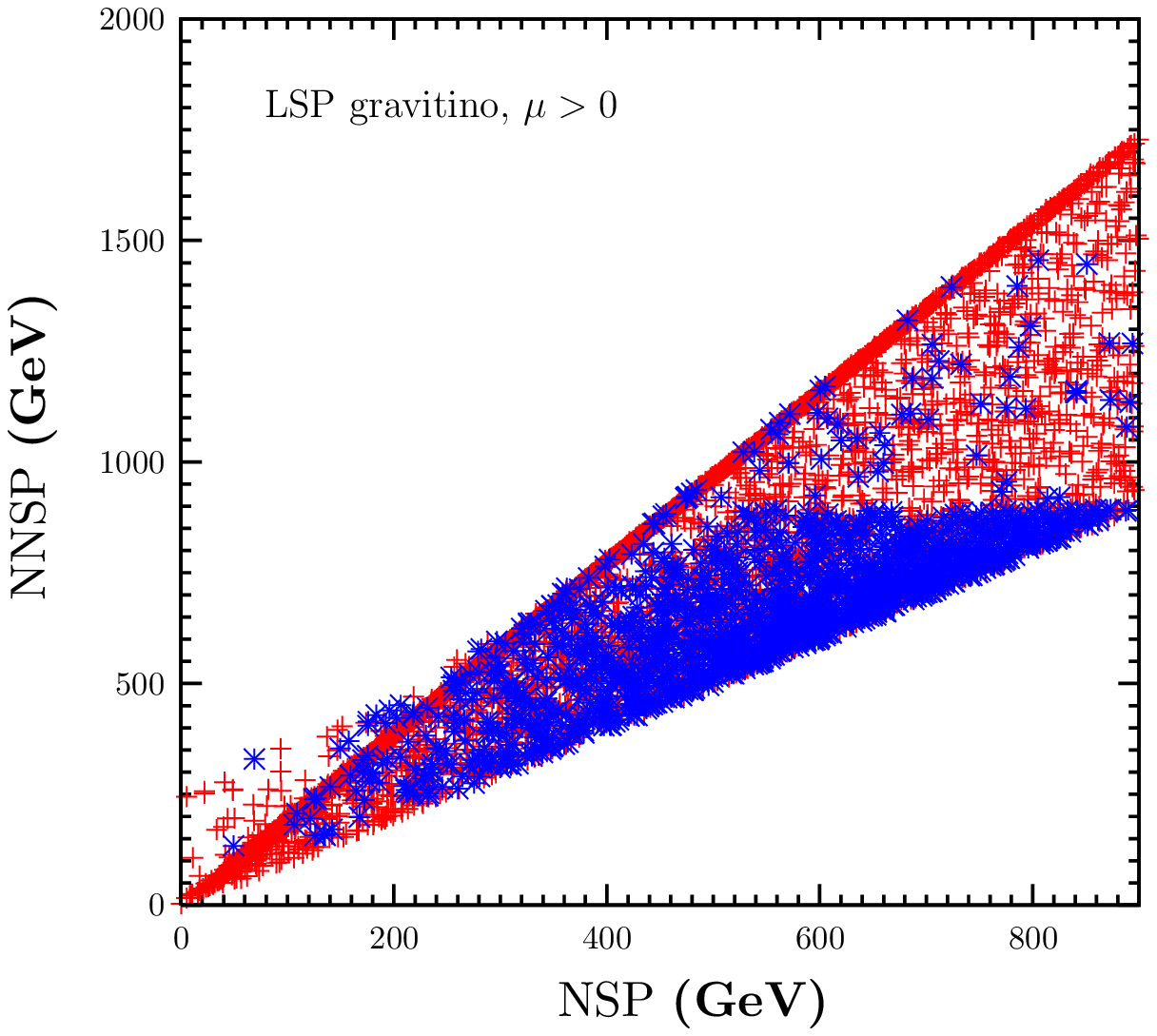,height=6.0cm}}
\end{center}
\caption{
Scatter plots of the masses of the next-to-lightest supersymmetric
particle (NSP)  and the next-to-next-to-lightest supersymmetric particle
(NNSP) for (a) the CMSSM, (b) models without squark and slepton mass
universality, (c) universality for the squark and slepton masses
separately but stability of the effective potential up to 10~TeV only, and
(d) a gravitino LSP. Each scatter plot contains 50,000 points with
$m_{\tilde Q}, m_{\tilde D}, m_{\tilde U}, m_{\tilde L}, m_{\tilde E},
m_{1/2}, m_1, m_2 <$~2~TeV (where $m_{1,2} $ are the soft
supersymmetry-breaking contributions to the Higgs boson masses), 
 -- 1~TeV~$< A_0 <$~1~TeV, $\tan \beta <$~58 and $\mu>$~0.
We thank Vassilis Spanos for supplying these figures}
\label{fig:scatters}
\end{figure}

We complete this introductory discussion by looking beyond specific CMSSM
benchmark scenarios and WMAP lines, considering the prospects for
observing supersymmetric particles at $e^+ e^-$ colliders in more general
scenarios that conserve $R$ parity. For this purpose, we display in
Fig.~\ref{fig:scatters} random samples of supersymmetric models, initially
disregarding the cosmological relic-density constraint (red points) and
subsequently including it (blue points).  The plots feature the masses of
the two lightest observable sparticles, namely the next-to-lightest
supersymmetric particle (NSP)  and the next-to-next-to-lightest
supersymmetric particle (NNSP).

In order to make contact with the previous analysis, we first consider in
panel (a)  of Fig.~\ref{fig:scatters} the CMSSM with parameters chosen to
obey the standard experimental (and cosmological) constraints. We then
consider in panel (b) models in which 
no universality is assumed for the squark
and slepton masses, subject to the supplementary constraint that the
effective potential remain stable when the theory is
extrapolated up to the GUT scale, using the renormalization-group equations
(RGEs). Panel (c) assumes universality for the squark and slepton masses
separately, but stability is required only for RGE evolution up to 10~TeV.
Finally, panel (d) is for models in which the LSP is assumed to be a
gravitino.

In each case, we have generated 50,000 points with $m_{\tilde Q},
m_{\tilde D}, m_{\tilde U}, m_{\tilde L}, m_{\tilde E}, m_{1/2}, m_1, m_2
<$~2~TeV (where $m_{1,2}$ are the soft supersymmetry-breaking contributions
to the Higgs boson masses), 
--~1~TeV $< A_0 <$~1~TeV, $\tan\beta<$~58
and $\mu >$~0 (the plots for $\mu >$~0 are not very different). The 
upper limits on the soft supersymmetry-breaking parameters are somewhat
arbitrary, and one could argue that a varying measure related to the
amount of hierarchical fine-tuning should be applied. However, we do note
that the range considered contains comfortably most points allowed by the
cosmological relic-density constraint.

We see that CLIC with $E_{c.m.}$~=~3~TeV would observe both the NSP and the
NNSP in most cases, and CLIC with $E_{c.m.}$~=~5~TeV would observe them both
in all the cases considered.

\subsection{Alternative Benchmark Scenarios} 
Apart from the points discussed in~Ref.~\cite{bench03}
several other groups have recently proposed benchmark points for SUSY
studies.
For example, in~Ref.~~\cite{weiglein} a number of mSUGRA benchmark
points similar to the ones discussed in~\cite{bench03} are proposed. 
Since these points were conceived at the Snowmass workshop in 2001, these
benchmark points are often referred to as the Snowmass points (SPS points).
In addition, 
points generated with different SUSY-breaking mechanisms are 
proposed: two points from a gauge-mediated SUSY-breaking 
scenario (GMSB), and from an anomaly-mediated SUSY-breaking 
scenario (AMSB).
Both  GMSB points have heavy squarks and gluinos, which are 
beyond the reach of a 1 TeV LC, but within reach of CLIC. 
For the second GMSB point also the heavy Higgses are beyond the reach of 
a 1 TeV LC (and even the LHC).
The AMSB point chosen has many sparticles with a mass around or above 1 TeV:
several gauginos, the heavy Higgses and all coloured sparticles. All these
sparticles are, however, within the reach of a 3~TeV~CLIC.

An alternative set of `string-inspired' benchmark points,
are given in~Ref.~\cite{lykken2}. These points are chosen such that there are
always a few light sparticles which could possibly be detected at the
Tevatron. However, each of these proposed benchmark scenarios always has 
a large number of sparticles in the TeV mass range and can be studied by CLIC.
In some cases the masses are so large that even CLIC will not be able
to see the full sparticle spectrum.
In some of these scenarios only a few gauginos are accessible at a 1-TeV
LC, and the sleptons and Heavy Higgses can only be studied at CLIC.

In all, these benchmark points, in scenarios different from mSUGRA, all 
seem to indicate that there is a role for CLIC to complete the sparticle 
spectrum.

\section{Slepton and Squark Mass Determination}

\subsection{Smuon Mass Determination}

A study has been performed of the reaction 
$e^+e^- \to \sMu_L\sMu_L \to \mu^+\chiz_1 \mu^-\chiz_1$ at CLIC.
The three main sources of background, also leading to 
two muons plus missing energy, are 
i)  $e^+e^- \to W^+W^- \to \mu^+ \mu^- \nu_{\mu} \bar{\nu}_{\mu}$, 
ii) $e^+e^- \to W^+W^- \bar{\nu}\nu 
            \to \mu^+\mu^-\nu_{\mu} \bar{\nu}_{\mu}\nu_e \bar{\nu}_e$ and 
iii) $e^+e^- \to \chiz_1\chiz_2, ~\chiz_2 \chiz_2 
             \to \mu^+\mu^- \nu \bar{\nu}\chiz_1\chiz_1$. 
These backgrounds can be suppressed by 
requiring central production and decay kinematics compatible with
those characteristic  
of smuon pair production. A multidimensional discriminant based on 
$M_{\mu\mu}$, $M_{\rm recoil}$, $E_{\rm missing}$, $\mu \mu$ acolinearity, 
$|\cos \theta_{\rm thrust}|$, $E_t$ and $E_{\rm hem}$ has been
applied. The signal efficiency is flat with the muon energy.

\subsubsection{The energy distribution method for mass determination}

\begin{figure}[htbp] 
\begin{center}
\begin{tabular}{c c}
\epsfig{file=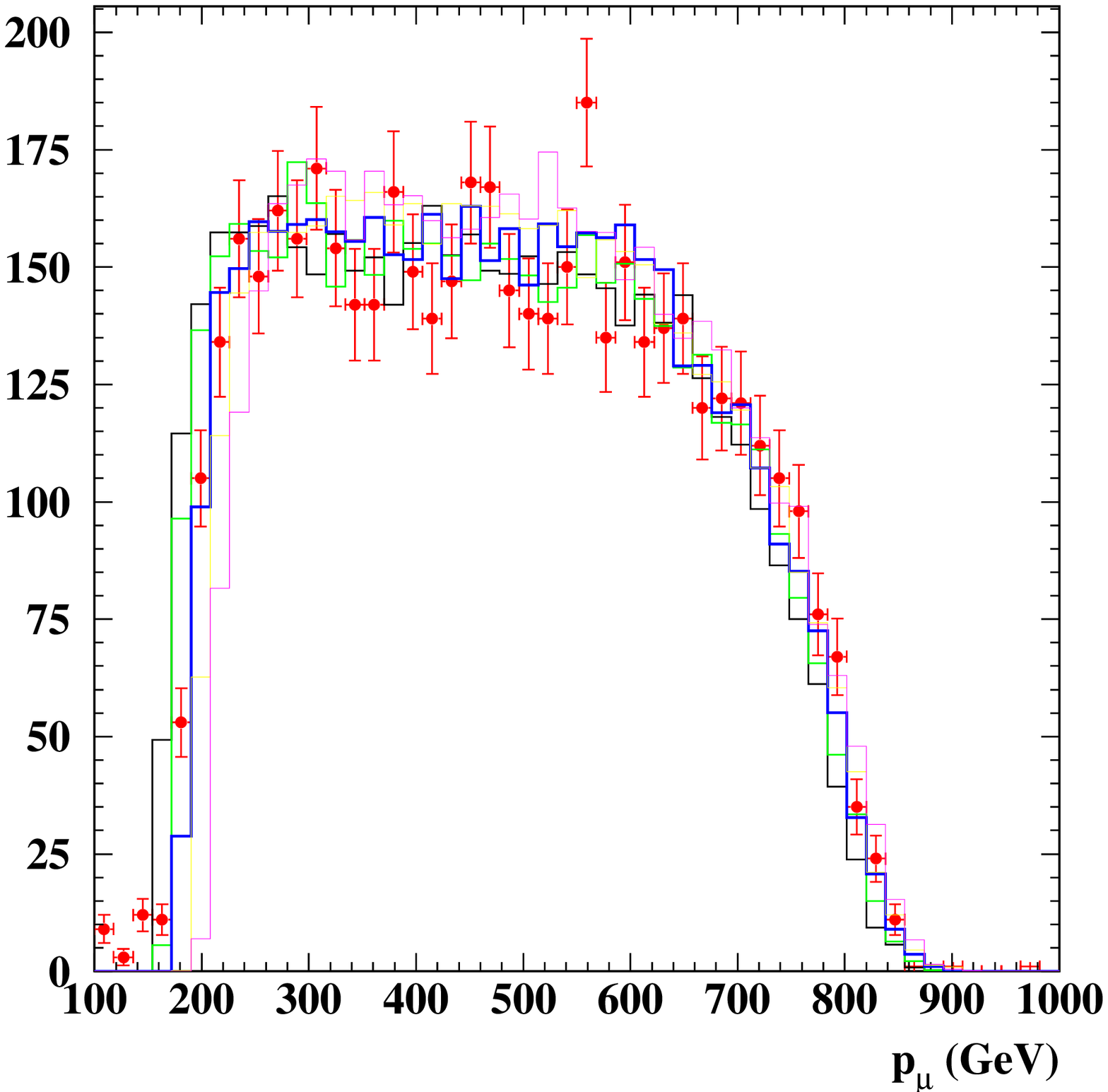,width=7.5cm,height=7cm,clip} &
\epsfig{file=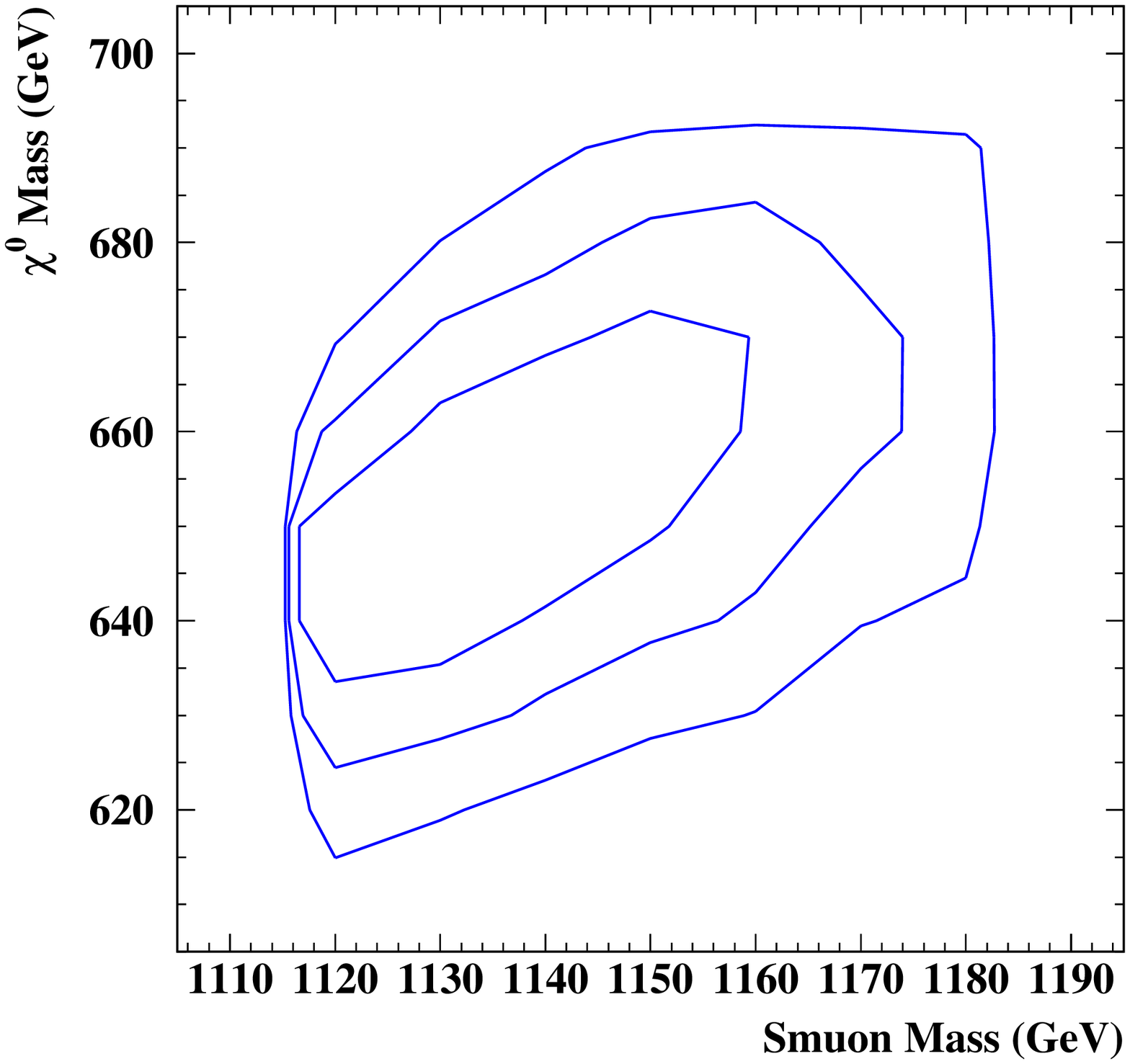,width=7.5cm,height=7cm,clip} \\
\end{tabular}
\end{center}
\caption{Left panel: Muon energy spectrum in the decay $\sMu_L \to \mu 
\chiz_1$ for the benchmark point H, corresponding to 
$M_{\sMu_L}$~=~1150~GeV and $M_{\chiz_1}$~=~660~GeV, as obtained for 
$\sqrt{s}$~=~3~TeV, assuming the baseline CLIC luminosity spectrum.
Right panel: Accuracy in the determination of the $\sMu_L$ and 
$\chiz_1$ masses by a two-parameter fit to the muon energy
distribution. The lines give the contours at 
1$\sigma$, 68\% and 95\% C.L. for 1~ab$^{-1}$ of data at $\sqrt{s}$~=~3~TeV.}
\label{fig:pmu}
\end{figure}

If the centre-of-mass energy $\sqrt{s}$ is significantly larger than 
twice the sparticle mass $M_{\tilde{\mu}}$, the latter can be determined 
by an analysis of the energy spectrum of the muon emitted in the two-body 
$\sMu \rightarrow \chiz_1 \mu$ 
decay, as seen in Fig.~\ref{fig:pmu}.
The two end-points, $E_{\rm min}$ and $E_{\rm max}$, of the spectrum
are related  
to the $\sMu$ and $\chiz_1$ masses and to the $\sMu$ boost by:
\begin{equation}
E_{\rm{max/min}} = 
\frac{M_{\sMu}}{2} \,\left(1 - \frac{M^2_{\chiz_1}}{M^2_{\sMu}}\right) 
\times \left( 1 \pm \sqrt{1 - \frac{M^2_{\sMu}}{E^2_{\rm beam}}} 
\right)
\end{equation}
from which either the smuon mass $M_{\sMu}$ can be extracted, 
if $M_{\chiz_1}$ 
is already known, or both masses can be simultaneously fitted. 

This technique, considered for the determination of squark masses at 
a 500-GeV LC in~Ref.~\cite{feng}, has already been applied to sleptons for 
the LHC~\cite{ATLASTDR,rurua} and a TeV-class LC~\cite{martyn}.  
It is interesting to consider its implications for the required 
momentum resolution in the detector. Two
values of the solenoidal magnetic field $B =$~4 and 6~T have been tested,
corresponding to momentum resolutions $\delta p/p^2$ of 
4.5~$\times$~10$^{-5}$ and 
3.0~$\times$~10$^{-5}$~GeV$^{-1}$ respectively. No appreciable
difference on the resulting mass accuracy has been found for these two
momentum resolutions. This reflects the fact that, at CLIC, the main
issue is the significant beamstrahlung smearing of the luminosity
spectrum, and thus of the effective $E_{\rm beam}$ value. The corresponding
effect has been estimated by assuming both a perfectly well known and
constant beam energy and the smearing corresponding to the baseline 
CLIC parameters at a nominal $\sqrt{s}$~=~3~TeV. Results are summarized in
Table~\ref{tab:res1} for the original version of benchmark point H. Since 
the updated post-WMAP version of point H has smaller $m_{1/2}$ and $m_0$, 
it would present a lesser experimental challenge.
%
\begin{table}[!h] 
\caption{Results of a one-parameter $\chi^2$ fit to the muon energy 
distribution for benchmark point H, obtained under different
assumptions on the $\delta p/p^2$ momentum resolution and the 
beamstrahlung spectrum. Accuracies are given for an integrated
luminosity of  1~ab$^{-1}$.}
\label{tab:res1}

\renewcommand{\arraystretch}{1.3} 
\begin{center}

\begin{tabular}{cccc}\hline \hline \\[-4mm]
\boldmath{$\delta p/p^2$} & $\hspace*{3mm}$ &
$\hspace*{6mm}$ \textbf{Beamstrahlung}  $\hspace*{6mm}$ & 
$\hspace*{6mm}$ \textbf{Fit result (GeV)}  $\hspace*{6mm}$ 
\\[4mm]   

\hline \\[-3mm]
0 & & none & 1150 $\pm$ 10 \\ 
3.0 $\times$~10$^{-5}$ & & none & 1150 $\pm$ 12 \\
4.5 $\times$~10$^{-5}$ & & none & 1151 $\pm$ 12 \\ 
4.5 $\times$~10$^{-5}$ & & standard & 1143 $\pm$ 18 
\\[3mm] 
\hline \hline
\end{tabular}
\end{center}
\end{table}


The smuon mass has been extracted by a $\chi^2$ fit to the muon energy
spectrum by fixing $M_{\chiz_1}$ to its nominal value (see
Table~\ref{tab:res1}).  The fit has been repeated, leaving both masses
free and performing a simultaneous two-parameter fit. The results are
$M_{\sMu_L}$~=~(1145~$\pm$~25)~GeV and 
$M_{\chi^{0}_1}$~=~(652~$\pm$~22)~GeV (see Fig.~\ref{fig:pmu}).

\subsubsection{The threshold scan method for mass determination}

\begin{figure}[htbp] 
\begin{center}
\begin{tabular}{c c}
\epsfig{file=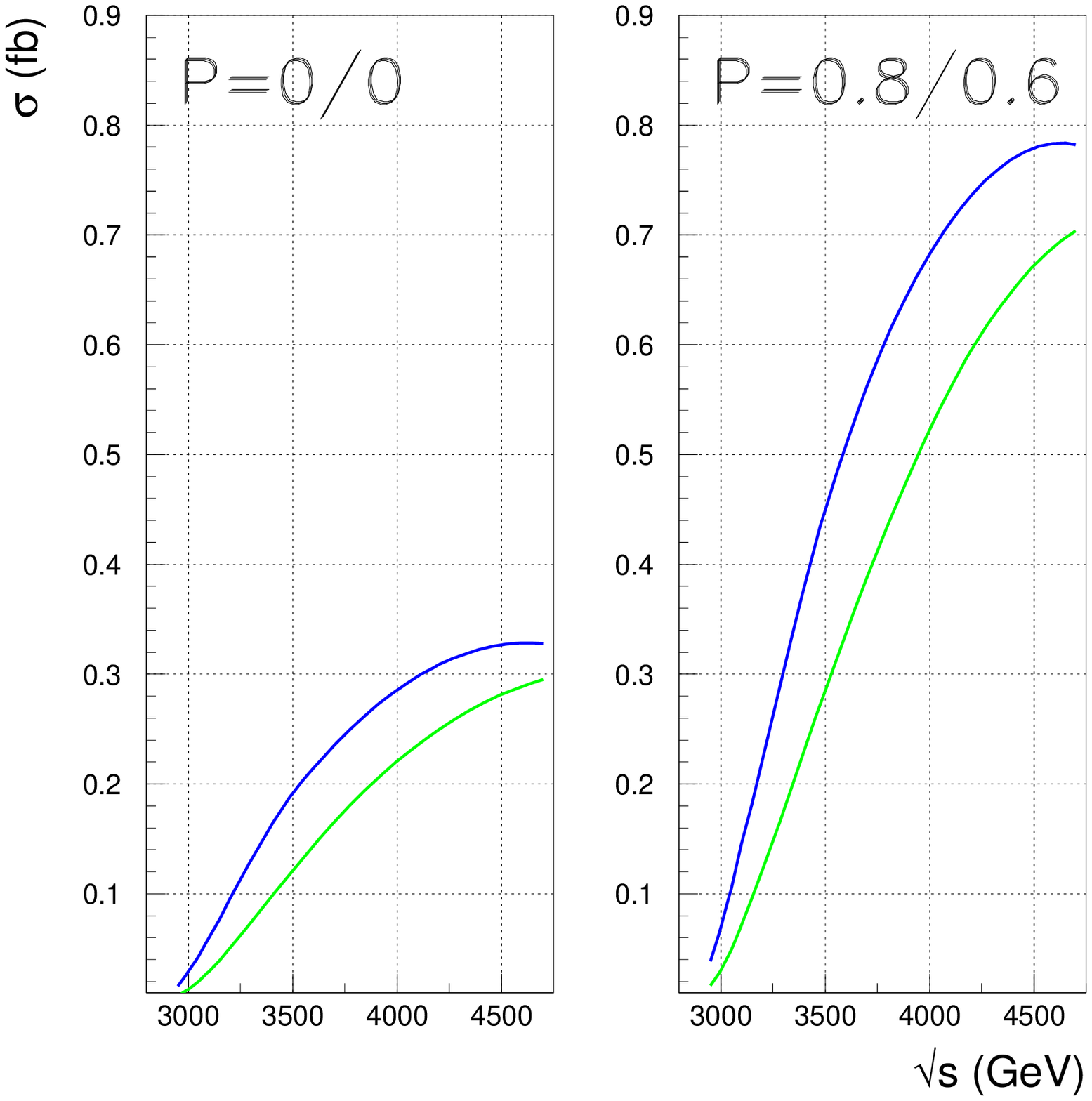,width=7.5cm,clip} &
\epsfig{file=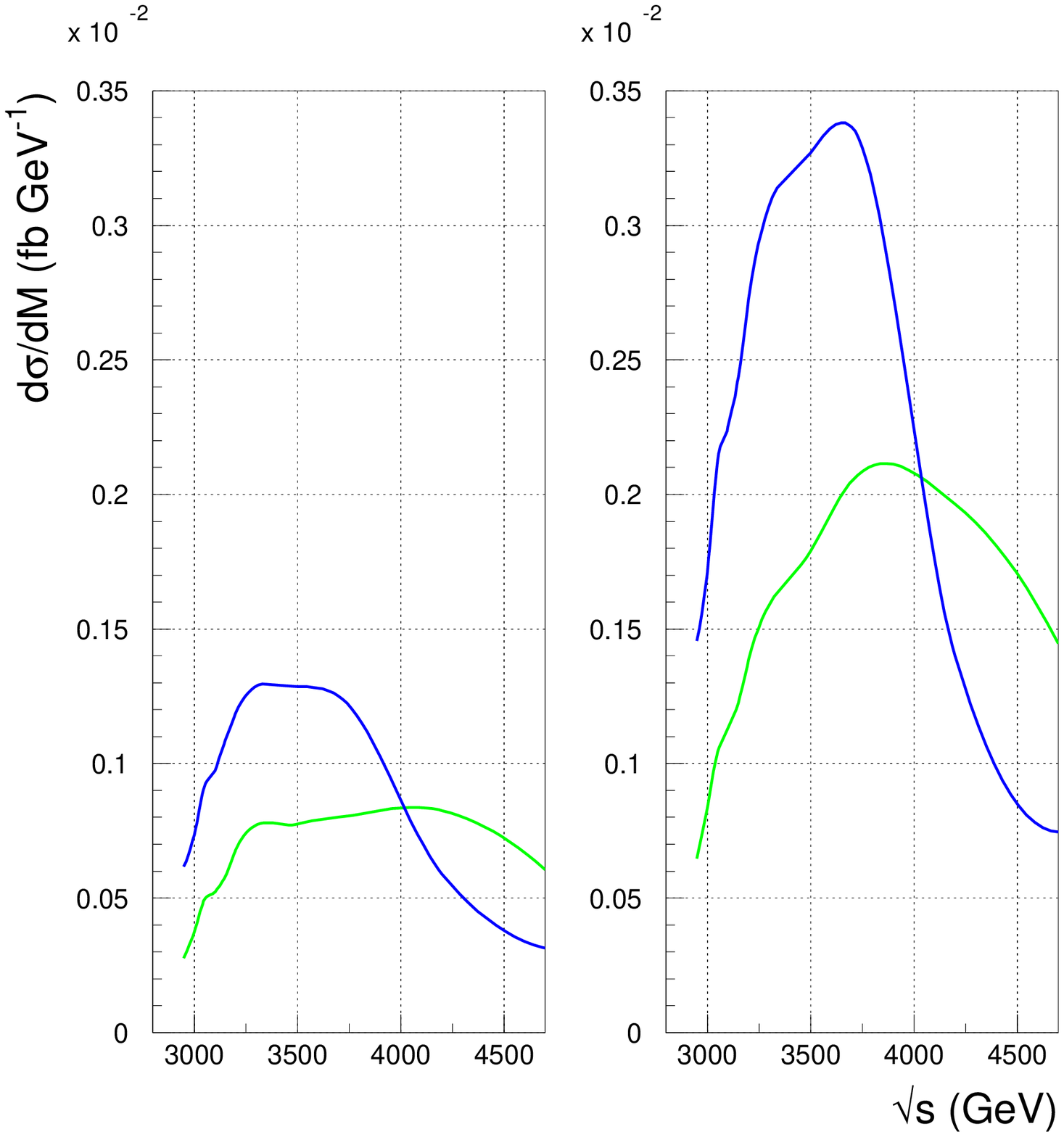,width=7.5cm,clip} \\
\end{tabular}
\end{center}
\caption{Smuon mass determination with threshold scans. 
Left: cross section as a function of $\sqrt{s}$ for smuon pair
production with and without beam 
polarization. The curves indicate the effect of the luminosity
spectrum on the rise of the cross section. Right: cross section
$\sigma$ sensitivity to the smuon mass $M$ indicated in terms of 
$d \sigma/d M$ as a function of centre-of-mass energy $\sqrt{s}$,
showing the optimal 
energy point for the threshold scan. Again the two panels are for no
beam polarization and for 80/60\% polarization, and the two curves show
the effect of the CLIC luminosity spectrum.}
\label{fig:thres}
\end{figure}

An alternative method to determine the $\sMu_L$ mass is an energy
scan of the rise of the $e^+e^- \rightarrow \sMu_L^+ \sMu_L^-$
cross section close to its kinematical threshold. It has been shown that
an optimal scan consists of just two energy points, sharing the total
integrated luminosity in equal fractions and chosen at locations
optimizing the sensitivities to the $\sMu_L$ width and mass,
respectively~\cite{blair}. Including the beamstrahlung effect induces a
shift of the positions of the maxima in mass sensitivity towards higher
nominal $\sqrt{s}$ energies (see~Fig.~\ref{fig:thres}). 
For benchmark point E (considering once 
again the pre-WMAP version), the cross section at
$\sqrt{s}$~=~3~TeV is too small for an accurate measurement. A higher
centre-of-mass energy, 4~TeV, and polarized beams need to be considered.
By properly choosing the beam polarization, not only are the pair-production
cross sections increased (as discussed in a later section of this 
chapter), but also their sensitivities to the smuon
masses. Results are summarized in Table~\ref{tab:res2}. In the case of 
point H~(also in the pre-WMAP version), we see that a mass resolution 
$\delta M$~=~15~GeV is attained, 
which increases to 36~GeV for point E. However, polarized beams would, in 
that case, reduce $\delta M$ to 22~MeV.

%
\begin{table}[htbp] 
\caption{Accuracies in the determinations of the smuon masses for 
the two benchmark scenarios considered in this study, using threshold
scans with different experimental conditions}
\label{tab:res2}

\renewcommand{\arraystretch}{1.3} 
\begin{center}

\begin{tabular}{cccccc}
\hline \hline\\[-4mm]
$\hspace*{3mm}$ \textbf{Point} $\hspace*{3mm}$ 
& $\hspace*{6mm}$ \textbf{Beam-} $\hspace*{6mm}$ 
& $\hspace*{6mm}$ \textbf{Pol.} $\hspace*{6mm}$  
& $\hspace*{6mm}$ \boldmath{$\sqrt{s}$} $\hspace*{6mm}$ 
& $\hspace*{6mm}$ \boldmath{$\int{\cal{L}}$} $\hspace*{6mm}$ 
& $\hspace*{6mm}$ \boldmath{$\delta M$} $\hspace*{6mm}$  \\
 & \textbf{strahlung} & & \textbf{(TeV)} & \textbf{(ab$^{-1}$)} &
\textbf{(GeV)} \\[1.5mm]  \hline\\[-4mm]
H   & none & 0/0 & 3.0--3.5   & 1 & $\pm$ 11 \\
H   & standard & 0/0 & 3.0--3.5   & 1 & $\pm$ 15 \\ 
E   & none & 0/0 & 3.8--4.2 & 1 & $\pm$ 29 \\ 
E   & standard & 0/0 & 3.8--4.2 & 1 & $\pm$ 36 \\ 
E   & none & 80/60 & 3.8--4.2 & 1 & $\pm$ 17 \\ 
E   & standard & 80/60 & 3.8--4.2 & 1 & $\pm$ 22 
\\[2mm] 
\hline \hline
\end{tabular}
\end{center}
\end{table}

\subsection{Analysis of Stop Squarks}
 
The scalar partners of the top quark (stops), $\stL$ and $\stR$, 
are of particular interest among the squarks.
Owing to their large Yukawa coupling, they play  
an important role in the renormalization group (RG) evolution 
and in the radiative corrections to the lightest Higgs mass. 
Moreover, the $\stL$ and $\stR$ mix to mass eigenstates 
$\stone~=~\stL\cos\t_{\sTop} + \stR\sin\t_{\sTop}$ and 
$\sttwo = \stR\cos\t_{\sTop} - \stL\cos\t_{\sTop}$, where
$\t_{\sTop}$ is the stop mixing angle. 
This mixing also leads to a sizeable mass splitting of 
$\stone$ and $\sttwo$, which contrasts with the near-degeneracy 
of the squarks $\sQua_{\rm L,R}^{}$ of the first and second generations. 
The masses of the stops are given by
\begin{equation}
  {\rm diag}(m_{\stone},\,m_{\sttwo}) = R^{\sTop}  
  \left(\!\!\begin{array}{cc} 
      m_{\stL}^2 & (A_t - \mu\cot\b)\,m_t \\
      (A_t - \mu\cot\b)\,m_t & m_{\stR}^2
  \end{array}\!\!\right) (R^{\sTop})^\dagger \,,
\end{equation}
where $A_t$ is the trilinear stop--Higgs coupling,  
$R^{\sTop}$ is the stop mixing matrix
\begin{equation}
  R^{\sTop} = 
  \left(\!\!\begin{array}{rr}
      \cos\t_{\sTop} & \sin\t_{\sTop} \\
     -\sin\t_{\sTop} & \cos\t_{\sTop}
  \end{array}\!\!\right) \,,
\end{equation}
and $m_{\stone}<m_{\sttwo}$ by convention.
As already mentioned, stops play an important role 
in the RG running. For the extrapolation to the GUT scale 
discussed later in this chapter, it is thus essential to know
their parameters as precisely as possible. 
At the LHC, stop masses may be measured with errors of a few per cent, 
but measuring $\t_{\sTop}$ and $A_t$ will hardly be possible.
In $e^+e^-$ collisions, however, measurements of the products of
cross sections with branching ratios 
(especially with polarized beams), threshold scans and 
kinematical distributions may be used to determine 
$m_{\stone}$, $m_{\sttwo}$, $\t_{\sTop}$ and $A_t$ quite precisely. 

\begin{figure}[t]
\unitlength=1mm
\begin{center}
\begin{picture}(62,62)
\put(0,0.5){\epsfig{file=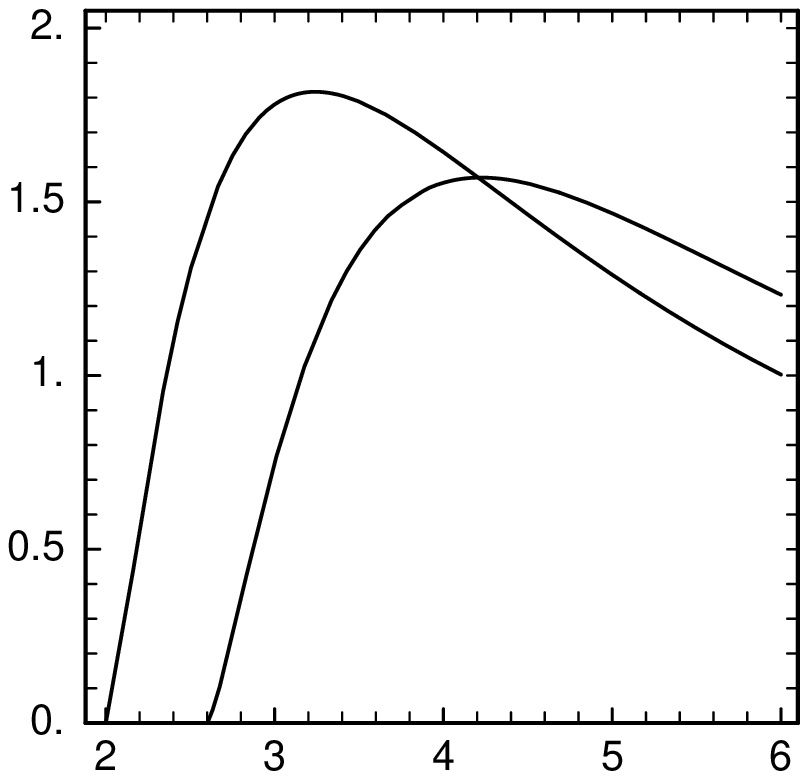, height=6.3cm}}
\put(-6,19){\rotatebox{90}{$\sigma(e^+e^-\to\sTop_i \bar{\sTop_i})$~[fb]}}
\put(28,-3){\mbox{$\sqrt{s}$~[TeV]}}
\put(31,55){\mbox{$\stone \bar{\stone}$}}
\put(48,48){\mbox{$\sttwo \bar{\sttwo}$}}
\put(42,12){\mbox{$\t_{\sTop}$~=~70$^\circ$}}
\end{picture}
\end{center}
\caption{Cross Sections for stop pair production for $m_{\stone}$~=~1~TeV, 
$m_{\sttwo}$~=~1.3~TeV, $\t_{\sTop}$~=~70$^\circ$ and 
$m_{\sGlu}$~=~750~GeV}
\label{fig:stopxsect}
\end{figure}

\begin{figure}[!ht] 
\unitlength=1mm
\begin{center}
\epsfig{file=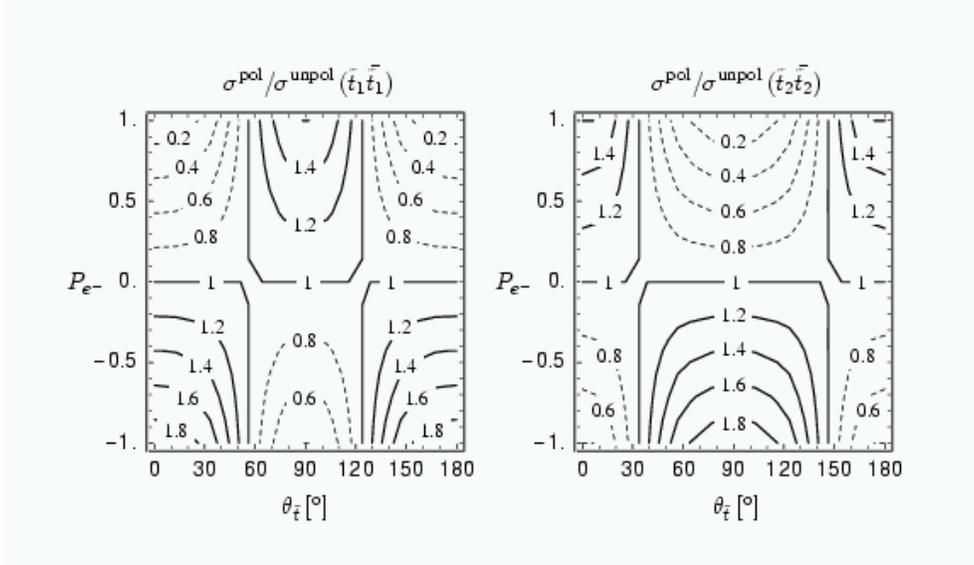,width=13cm}
\end{center}
\vspace{-1cm}
\caption{Polarized cross sections for $\stone \bar{\stone}$ and 
$\sttwo \bar{\sttwo}$ production normalized to the unpolarized ones, 
$\sigma^{\rm pol}/\sigma^{\rm unpol}$, 
as functions of the electron beam polarization $P_{e^-}$ and 
the stop mixing angle $\t_{\sTop}$, 
for $m_{\stone}$~=~1~TeV, $m_{\sttwo}$~=~1.3~TeV and $\sqrt{s}$~=~3~TeV. 
The positron beam polarization has been assumed to vanish.}
\label{fig:stopxsectpol}
\end{figure}

Figure~\ref{fig:stopxsect} shows cross sections for
$e^+e^-\to\sTop_i \bar{\sTop_{i}} \,$ ($i$~=~1,~2) for $m_{\stone}$~=~1~TeV, 
$m_{\sttwo}$~=~1.3~TeV and $\t_{\sTop}$~=~70$^\circ$ as functions of 
$\sqrt{s}$ for unpolarized beams. 
Supersymmetric QCD corrections~\cite{Drees:1990te} and initial-state 
radiation (ISR) are included. 
The cross sections show characteristic dependences~\cite{Kraml:1999qd} 
on the mixing angle and on beam polarization. 
Measurements with different polarizations may thus be used to 
determine $\t_{\sTop}$. 
Figure~\ref{fig:stopxsectpol} shows polarized cross sections for
$\stone\, \bar{\stone}$ and $\sttwo\, \bar{\sttwo}$ production normalized 
to the unpolarized ones, $\sigma^{\rm pol}/\sigma^{\rm unpol}$, 
as functions of the electron beam polarization $P_{e^-}$ and 
the stop mixing angle. Positron polarization is assumed to be zero,
but could be used to enhance further the dependence~on~$\t_{\sTop}$.

If both stop masses and the mixing angle are measured, 
and $\tan\b$ and $\mu$ are known from, for example, the chargino/neutralino 
sector, $A_t$ can be determined by
\begin{equation}
  A_t = \frac{1}{2m_t}\left(m_{\stone}^2-m_{\sttwo}^2\right)
        \sin 2\t_{\sTop} + {\mu \over \tan\b }\,.
\end{equation}
This can then be compared with indirect estimates of $A_t$ 
from precision measurements of $m_h$~\cite{mtprec}.
It is also possible to determine $A_t$ via the associated production of 
stops with Higgs bosons~\cite{Belanger:1998rq} 
(which has, however, a very small cross section at CLIC energies) 
and/or from $\sttwo\to\stone\,(h^0\!,H^0\!,A^0)$ and $\sBot_i H^+$ 
decays~\cite{Bartl:1998xk}, which can have large branching ratios 
if $A_t$ and/or $\mu$ are large.

\section{Neutralino Mass Determination}

\begin{figure}[htbp] %
\begin{center}
\begin{tabular}{cc}
\epsfig{file=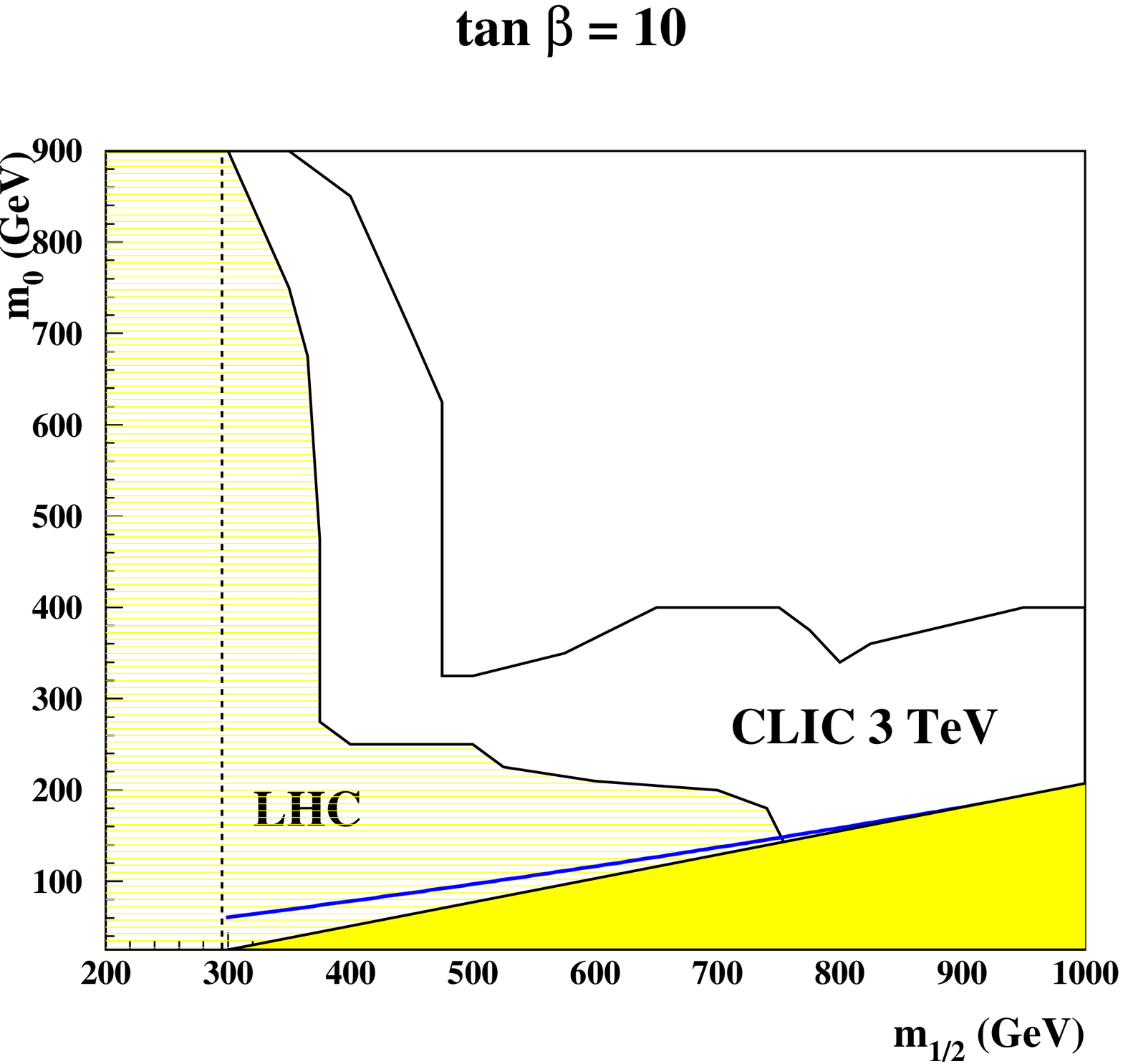,width=8.5cm} &
\epsfig{file=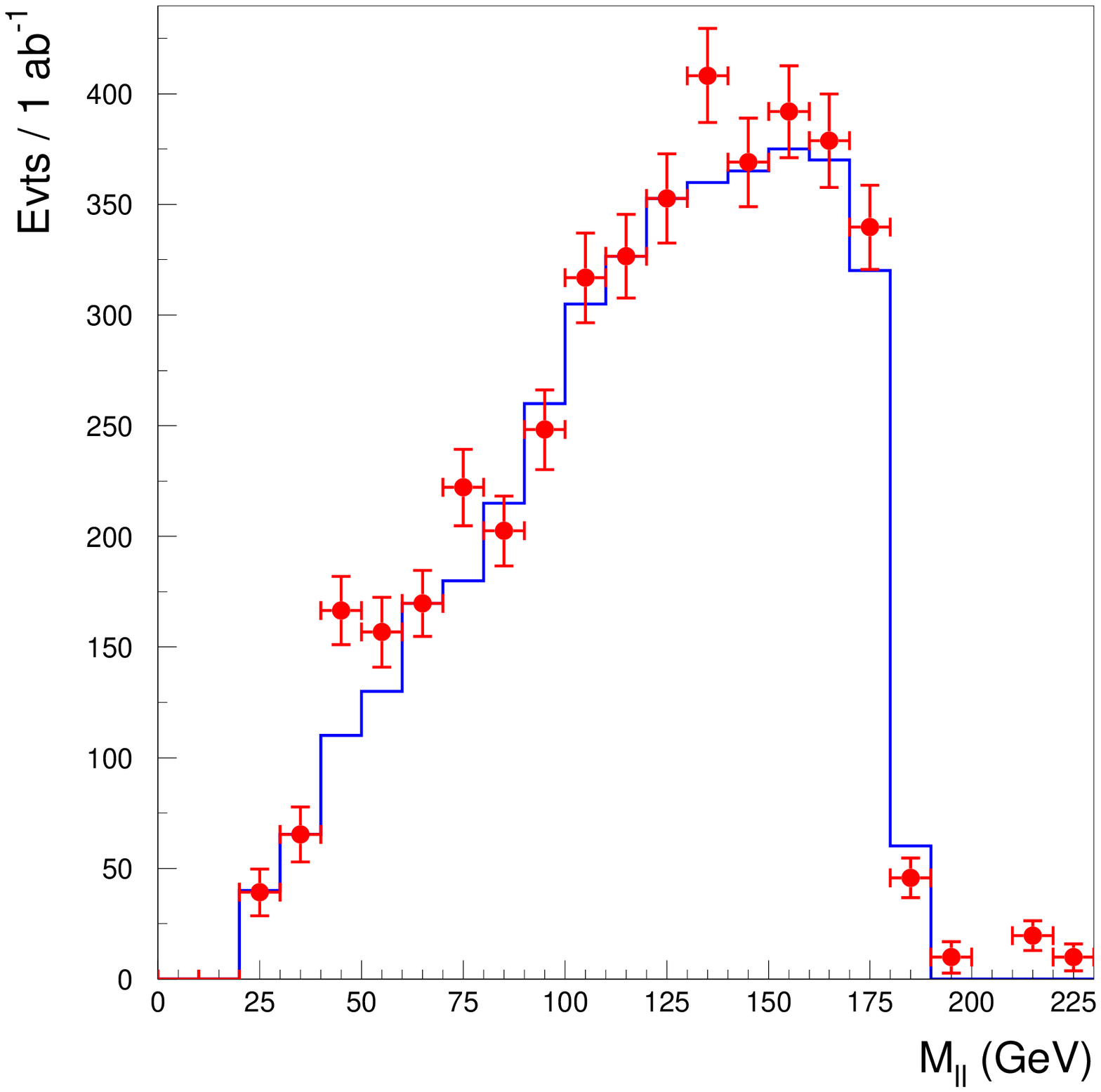,width=7.5cm} \\
\end{tabular}
\end{center}
\caption[]{Left panel: The sensitivity to $\chiz_2$ production at 
CLIC for $\sqrt{s}$~=~3~TeV compared with that of the LHC. 
Right panel: Dimuon spectrum measured at CLIC, assuming $m_0$~=~150~GeV,
$m_{1/2}$~=~700~GeV and $\tan\beta$~=~10, used to derive the particle
masses as discussed in the text.}
\label{fig:sec5_chi20}
\end{figure}

Operating at 3 to 5~TeV with large luminosity, CLIC will be able to
improve significantly on the reach for heavy neutralinos, beyond the
anticipated LHC sensitivity. A general study of the prospective 
CLIC sensitivity has been carried out, including a simulation of the 
detector response parametrized by the SIMDET program. Inclusive
chargino and neutralino pair production have been analysed to determine 
the sensitivity obtainable from the dilepton invariant mass spectrum.  
The decay chains 
$\chiz_j \to \ell^\pm\tilde{\ell}^\mp \to \ell^+\ell^-\chiz_1$ and 
$\chiz_j \to \chiz_i Z^0 \to \chiz_i\ell^+ \ell^-$ give rise to an 
excess of events 
in the $\ell^+\ell^-$ mass spectrum corresponding to the phase space 
for the cascade two-body decay of the first process or to the $Z^0$ 
mass peak in the second. 
A full scan of the $(m_{1/2},\;m_{0})$ plane has been carried out, 
using PYTHIA~6.21 interfaced to ISAJET~7.64.  Events with
at least two leptons and significant missing energy have been selected.
Both SUSY backgrounds involving sleptons and SM gauge-boson pair
production have been considered. Combinatorial backgrounds have been
subtracted by taking the difference of the pair $e^+e^- + \mu^+\mu^-$
events and the mixed $e^{\pm} \mu^{\mp}$ events, and a sliding window has
been used to search for an excess of 5 standard deviations or more of
events in the $M_{\ell\ell}$ mass distribution. Results are shown in
Figure~\ref{fig:sec5_chi20} for $\tan \beta$~=~10, where the extended reach
provided by CLIC is manifest. At larger values of $\tan \beta$,
decays into $\tilde{\tau}$ become more significant, requiring a more
detailed study, which must include $\tau$-lepton reconstruction.

To verify the CLIC capability for measuring the masses of heavy
neutralinos, a representative point has been chosen at $m_0$~=~150~GeV,
$m_{1/2}$~=~700~GeV and $\tan\beta$~=~10, which is compatible with the 
constraint on cold dark matter from the WMAP data and has 
$M_{\chiz_2}$~=~540~GeV, $M_{\chiz_1}$~=~290~GeV  
and~$M_{\ell_L}$~=~490~GeV.  

The dilepton invariant mass distribution shows a clear upper edge at
120~GeV, due to $\chiz_2 \to \mu^\pm \sMu^\mp_L \to \mu^+\mu^- \chiz_1$,
which can be very accurately measured, already with 1~ab$^{-1}$ of data.
However, in order to extract the mass of the $\chiz_2$ state, the masses
of both the $\sMu_L$ and $\chiz_1$ need to be known. As already
discussed, a two-parameter fit to the muon energy distribution yields the 
masses of the $\sMu_L$ and $\chiz_1$ with
accuracies of 3\% and 2.5\% respectively, using 1~ab$^{-1}$ of data.
An improved accuracy can be obtained with higher luminosity and using also a
threshold energy scan. We therefore assume that these masses can be known
to 1.7\% and 1.5\% respectively, which gives an uncertainty of 8~GeV, or
1.6\%, on the $\chiz_2$ mass, when accounting for correlations. This
uncertainty is dominated by that in the $\sMu_L$ and $\chiz_1$
masses. The accuracy in the determination of the end-point would
correspond to a $\chiz_2$ mass determination of 0.3~GeV, fixing all other
masses, and is not sensitive to the details of the beamstrahlung and
accelerator-induced backgrounds.

Last but not least, at CLIC with 3~TeV also the heavier $\chiz_3$ 
and $\chiz_4$ are accessible in this scenario and can be detected via 
the decays $\chiz_3\to\chiz_{1,2}Z^0$ and $\chiz_4\to\chiz_{1,2}h^0$.

\section{Gluino Sensitivity in $\gamma \gamma$ Collisions}

Measurements of the mass and coupling of the gluino pose some
difficulties in $e^+e^-$ collisions, since the gluino couples only to
strongly-interacting particles and is thus produced only at the one-loop
level or in multiparton final states. In a recent publication, gluino
pair production through triangular quark/squark loops in $e^+e^-$
annihilation with energies up to 1 TeV was investigated~\cite{Berge:2002ev}. 
Owing to the large cancellation effects, it was found that promisingly
large cross sections can only be expected for scenarios with large
left-/right-handed 
up-type squark mass splittings, or with large top-squark mixing and for
gluino masses up to 500~GeV. Small gluino masses of 200 GeV might be
measured with a precision of about 5~GeV in centre-of-mass energy scans
with luminosities of 100 fb$^{-1}$ per point.

Here we assume that CLIC can provide electron (positron) beam
polarization of 80\% (60\%) and an integrated  luminosity
of 1000 fb$^{-1}$ per year. We compare the generally small production
rates in $e^+e^-$ annihilation to the larger ones in photon--photon
collisions. Further 
details on gluino pair production in high-energy photon collisions can be
found in~Ref.~\cite{Berge:2003cj}. In the photon--photon collider version of 
CLIC, discussed in Section 2.6, 
100\%-polarized laser photons are backscattered from two electron beams,
whose helicities must be opposite to those of the laser photons in order to
maximize the number of high-energy photons. In the strong fields of
the lasers, the electrons (or high-energy photons) can interact
simultaneously with several laser photons. This non-linear effect
increases the threshold parameter 
for $e^+e^-$ pair production to $X=(2+\sqrt{8})(1+\xi^2)\simeq$~6.5, which
corresponds to laser wavelengths of 4.4~$\mu$m at an $e^+e^-$
centre-of-mass energy of 3 TeV~\cite{Burkhardt:2002vh}. The maximal
fractional energy of the high-energy photons is then $x_{\max}=X/(X+1)
\simeq $~0.87. Since the  low-energy
tail of the photon spectrum is neither useful nor well understood, we use only
the high-energy peak with $x>0.8\,x_{\max}$~=~0.69 and normalize our cross
sections such that the expected number of events can be obtained through simple
multiplication with the envisaged photon--photon luminosity of 100--200
fb$^{-1}$ per year. This requires reconstruction of the total
final-state energy, 
which may be difficult because of the missing energy carried away by the
escaping lightest SUSY particles (LSPs). However, high-energy
photon collisions allow for cuts on the relative longitudinal energy in
addition to the missing-$E_T$ plus multi-jet, top or bottom (s)quark, and/or
like-sign lepton analyses performed at hadron colliders, and 
(in $R$-violating scenarios) sufficiently
long-lived gluinos can be identified by their typical $R$-hadron~signatures.

We adopt the current mass limit $m_{\sGlu}\geq $~200~GeV from 
CDF~\cite{Affolder:2001tc} and D0~\cite{Abachi:1995ng} searches in the 
jets- with-missing-energy channel, relevant to non-mixing squark
masses $m_{\sQ}\geq$~325~GeV and $\tan\beta$~=~3. Since values for the
ratio of the Higgs vacuum expectation values $\tan\beta <$~2.4 are
already excluded by the LEP experiments, and since values between 2.4
and 8.5 are only allowed in a very narrow window of light Higgs boson
masses between 113 and 127~GeV~\cite{lhwg:2001xx}, we employ a safely
high value of $\tan\beta$~=~10. For a conservative comparison of the
$\gamma\gamma$ and $e^+e^-$ options at CLIC, we maximize the $e^+e^-$
cross section by adopting the smallest allowed universal squark mass
$m_{\sQ}\simeq m_{\rm SUSY}$~=~325~GeV and large top-squark 
mixing with $\t_{\sTop}$~=~45.195$^\circ$, $m_{\stone}$~=~110.519~GeV,
and $m_{\sttwo}$~=~505.689~GeV, which can be generated by choosing
appropriate values for the Higgs mass 
parameter, $\mu$~=~--500 GeV, and the trilinear top-squark coupling,
$A_t$~=~648.512~GeV~\cite{Hahn:2001rv}. The SUSY one-loop contributions to the 
$\rho$ parameter and the light top-squark mass $m_{\stone}$ are then,
respectively, still significantly below and
above the LEP limits, $\rho_{\rm SUSY} <$~0.0012$^{+0.0023}_{-0.0014}$ and
$m_{\stone}\geq$~100~GeV~\cite{Hagiwara:pw,lswg:2002xx}. For small
values of $\tan\beta$, mixing in the bottom-squark sector remains
small, and we take $\t_{\sBot}$~=~0$^\circ$. A full set of Feynman
diagrams can be generated and evaluated with the computer algebra
packages {\tt FeynArts}~\cite{Hahn:2000kx} and  
{\tt FormCalc}~\cite{Hahn2:1998yk}.

\begin{figure}[htbp] %
 \centering
 \includegraphics*[width=0.49\columnwidth]{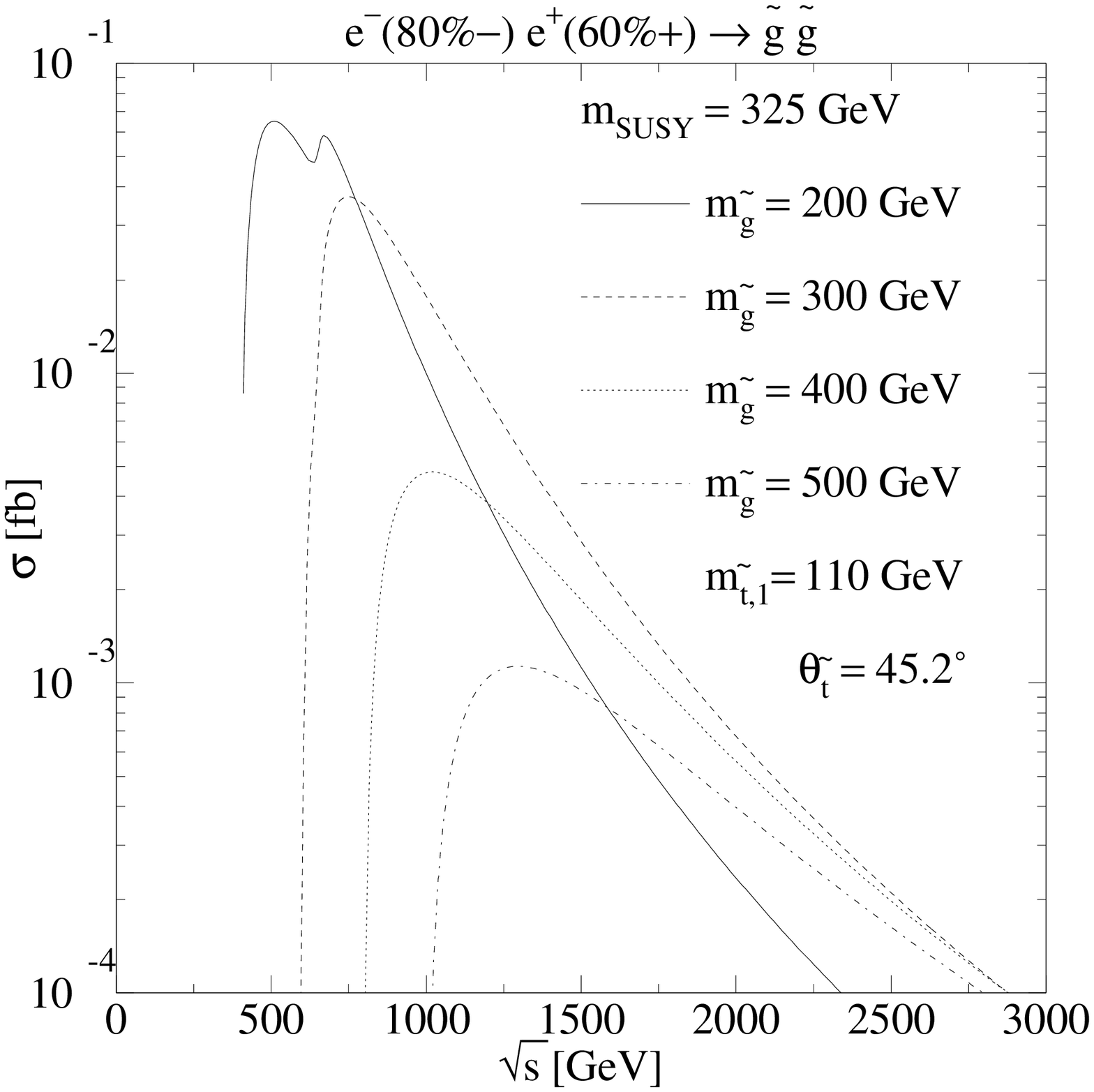}
 \includegraphics*[width=0.49\columnwidth]{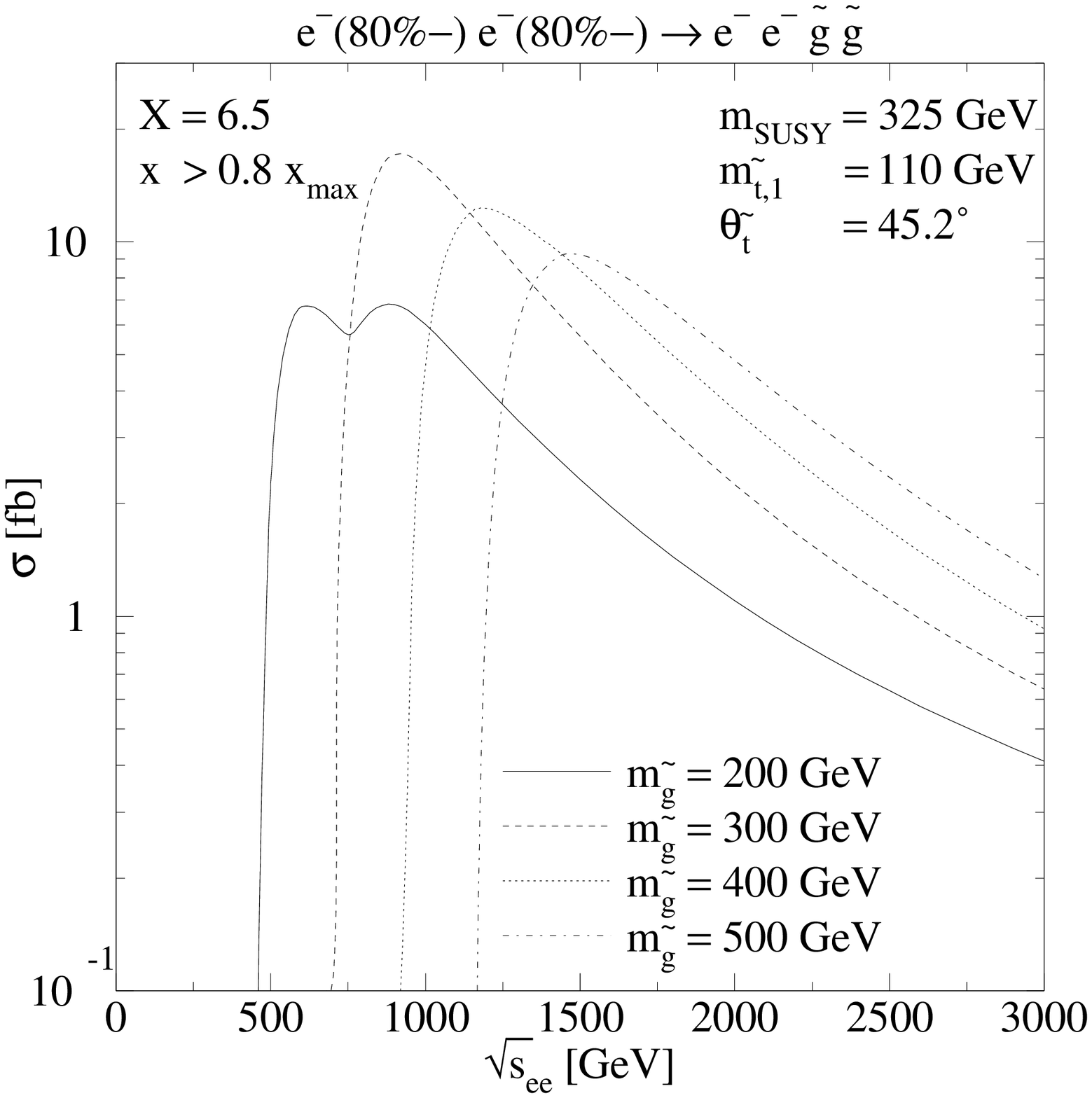}

\vspace*{5mm}

\caption{Gluino pair-production cross sections in $e^+e^-$ annihilation 
  (left) and $\gamma\gamma$ collisions (right) as functions of  
  the $e^\pm e^-$ centre-of-mass energy, for various gluino masses. 
  The photon--photon luminosity has been normalized to unity in the 
  high-energy peak.
\label{gluino:fig1}}
\end{figure}

Figure~\ref{gluino:fig1} shows a comparison of the total cross 
sections expected in $e^+e^-$ annihilation and $\gamma\gamma$ collisions for 
gluino masses between 200 and 500~GeV. Despite its maximization, the 
$e^+e^-$ cross section stays below 0.1 fb and falls steeply with $m_{\sGlu}$, 
so that gluino pair production will be unobservable for $m_{\sGlu}>$~500 GeV, 
irrespective of the collider energy. In contrast, the $\gamma\gamma$ cross 
section reaches around 10~fb for a wide range of $m_{\sGlu}$. In $e^+e^-$ 
annihilation the gluinos are produced in a $P$ wave and the cross section 
rises rather slowly, whereas in $\gamma\gamma$ collisions they can be produced 
as an $S$ wave and the cross section rises much faster.

\begin{figure}[htbp] 
 \centering
 \includegraphics*[width=0.49\columnwidth]{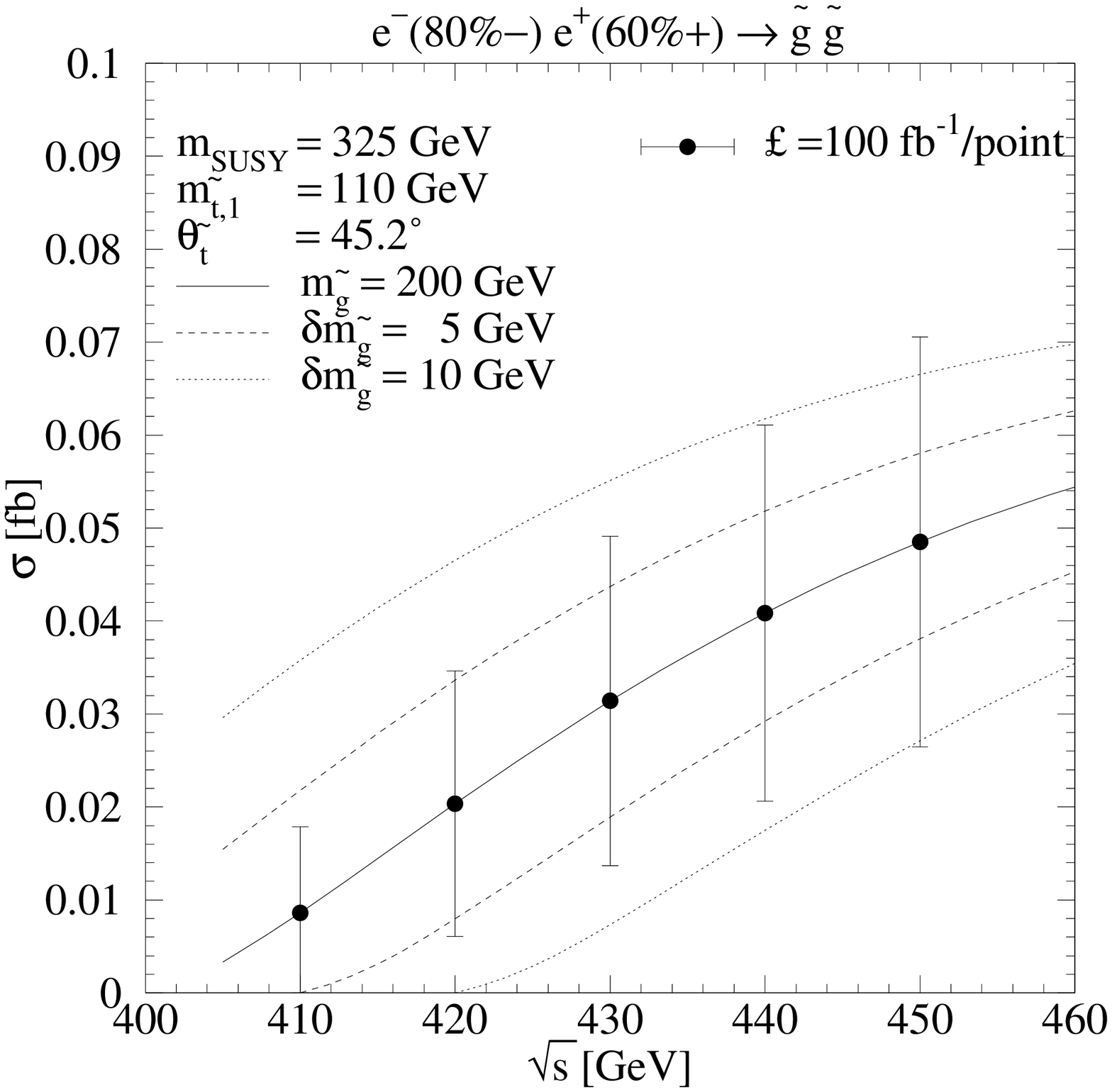}
 \includegraphics*[width=0.49\columnwidth]{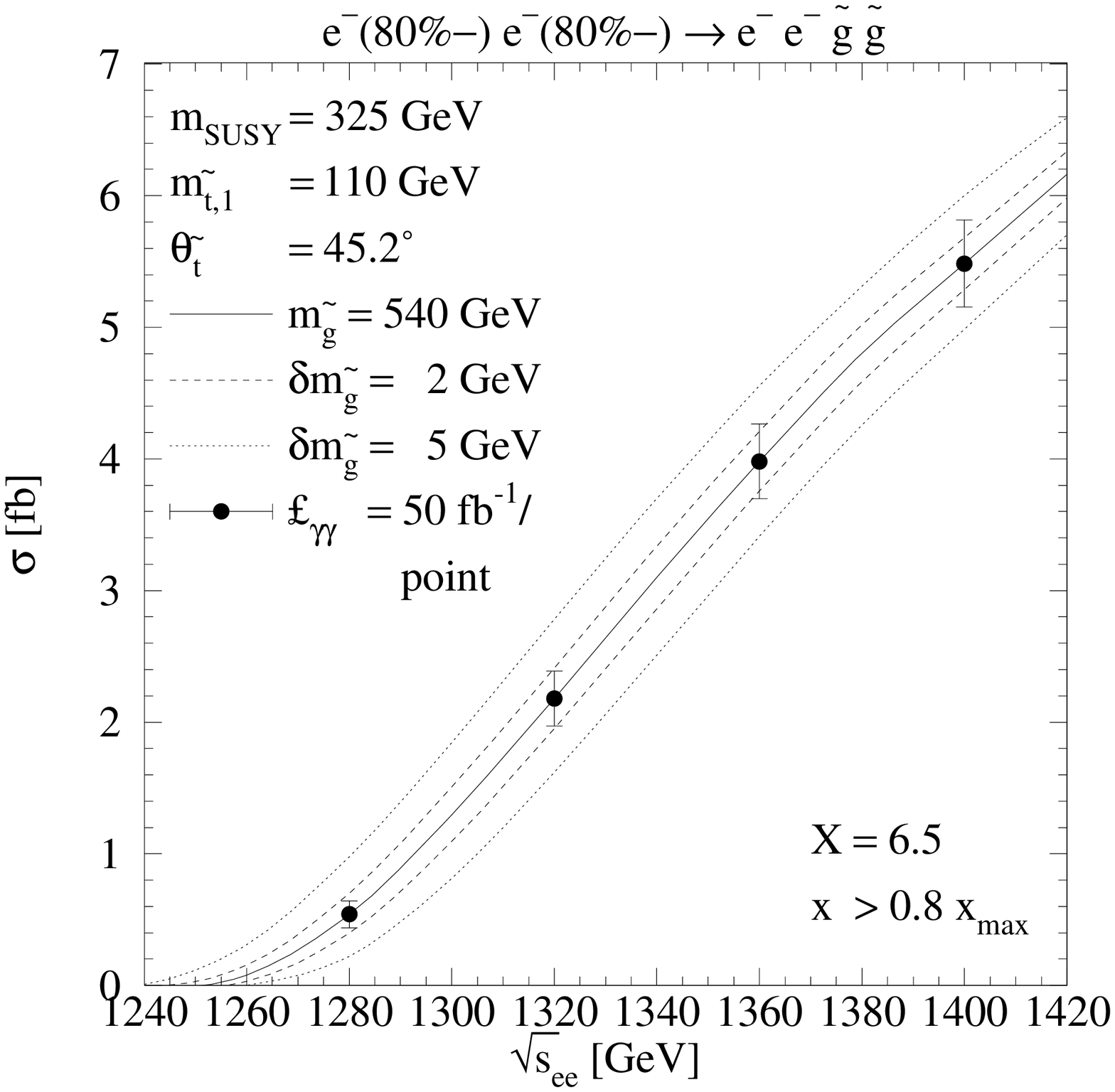}

\vspace*{5mm}

 \caption{Sensitivity of the $e^+e^-$ annihilation (left) and
 $\gamma\gamma$ collision cross section (right) to the mass of the
 pair-produced gluino. The photon--photon luminosity has been normalized to
 unity in the high-energy peak.}
\label{gluino:fig2}
\end{figure}

The different threshold behaviours can be observed even more clearly in 
Fig.~\ref{gluino:fig2}, 
where the sensitivities of $e^+e^-$ and 
$\gamma\gamma$ colliders to the gluino mass are compared. For the  
LHC experiments, a precision of $\pm$~30...60 (12...25)~GeV is 
expected for gluino masses of 540 (1004)~GeV~\cite{:1999fr,Abdullin:1998pm}. 
If the masses and mixing angle(s) of the top (and bottom) squarks are known, 
a statistical precision of $\pm$~5...10~GeV can be achieved in $e^+e^-$ 
annihilation for $m_{\sGlu}$~=~200~GeV for an integrated luminosity of 
100~fb$^{-1}$ per centre-of-mass energy point.
A precision of $\pm$~2...5~GeV may be obtained at a CLIC photon
collider for $m_{\sGlu}$~=~540~GeV and an integrated photon--photon luminosity 
of 50~fb$^{-1}$ per point, provided that the total final-state energy can be
sufficiently well reconstructed. Of course, uncertainties from a realistic
photon spectrum and the detector simulation should be added to the 
statistical error.

\begin{figure}[t] 
 \centering
 \includegraphics*[width=0.9\columnwidth]{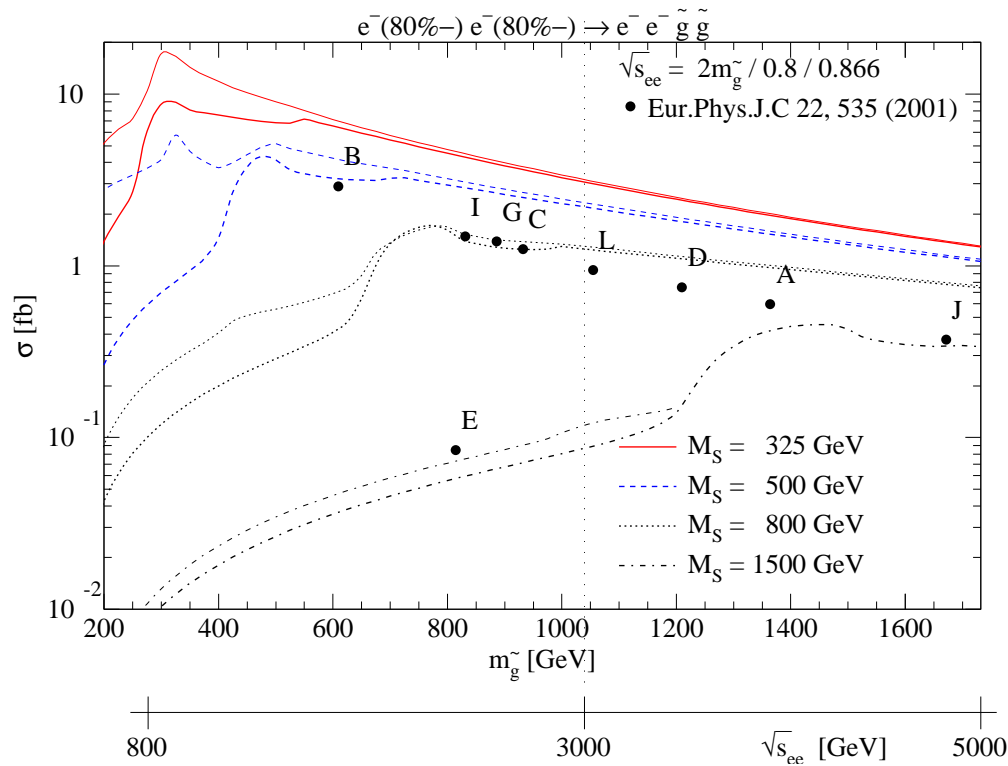}
 \caption{\label{gluino:fig3}
Dependence of the gluino pair-production cross section
in $\gamma\gamma$ collisions on the universal squark mass $m_{\rm SUSY}$ for
no squark mixing (thick curves) and maximal top-squark mixing (thin curves).
The photon--photon luminosity has been normalized to unity in the high-energy
peak. Also shown are the cross sections for the pre-WMAP 
versions of the supersymmetric benchmark 
points of~Ref.~\cite{Battaglia:2001zp}.}
\end{figure}

Finally, we demonstrate in~Fig.~\ref{gluino:fig3} that gluino pair 
production in $\gamma\gamma$ collisions depends only weakly on the 
universal squark mass $m_{\rm SUSY}$, and even less on the top-squark 
mixing. This is in sharp contrast to the results obtained in $e^+e^-$
annihilation~\cite{Berge:2002ev}. In this plot, the $e^+e^-$ or
$e^-e^-$ centre-of-mass energy is chosen close to the threshold for
gluino pair production and is varied simultaneously with $m_{\sGlu}$. 
Also shown in~Fig.~\ref{gluino:fig3} are several of the post-LEP 
supersymmetric benchmark points, considered here in their pre-WMAP 
versions~\cite{Battaglia:2001zp}. We expect similar results for most of 
the post-WMAP versions, since they have similar values of 
$m_{1/2}$~\cite{bench03}. Studies similar to those performed in 
Fig.~\ref{gluino:fig2} show that, with the exception of point E, where 
only about ten events per year are to be expected, the gluino mass can
be determined with a precision of $\pm$~20~GeV (as at point J) or better.

In summary, the 
determination of the gluino mass and coupling will be difficult 
in $e^+e^-$ collisions since the gluino couples only strongly and its
pair-production cross section suffers from large cancellations in the
triangular quark/squark loop diagrams. A photon collider may therefore be
the only way to obtain precise gluino mass determinations and visible gluino
pair production cross sections for general squark masses, and would thus
strongly complement the physics programme feasible in $e^+e^-$ 
annihilation.

\section{Reconstructing High-Scale SUSY Parameters}

If LHC measurements of the MSSM spectrum are complemented by high-precision 
measurements at a prospective LC with sufficiently high 
energy, one can try to reconstruct the original theory at the high 
scale in a model-independent way, as shown
in~Refs.~\cite{Blair:2000gy,Blair2}.  
We first summarize the procedure used, referring 
to~Refs.~\cite{Blair:2000gy,Blair2,spheno} for further details.

We take the masses and cross sections of a particular point in SUSY
parameter space together with their expected experimental errors from the
LHC, a $\sqrt{s}$~=~800~GeV LC of the TESLA design~\cite{tesla}
and a 4-TeV LC of the CLIC design.
We assume that electrons can be polarized to 80\% and positrons to
40\% at the LCs. We then fit the underlying SUSY-breaking
parameters at the electroweak symmetry-breaking scale, $Q_{\rm EWSB} =
\sqrt{m_{\tilde t_1} m_{\tilde t_2}}$, to these observables. An initial 
set of parameters is obtained by inverting tree-level formulas for
masses and cross sections. This set serves as a starting point for the
fit, which is carried out with MINUIT~\cite{James:1975dr} to obtain
the complete correlation matrix. In the fit,  
the complete spectrum is calculated 
at the 1-loop level using the formulae given in~\cite{Pierce:1996zz}. In
the cases of the neutral Higgs bosons and the $\mu$ parameter, the
2-loop corrections given in~\cite{Dedes:2002dy} 
are included. In addition, the cross sections for third-generation
sfermions at a LC are calculated, including the effect of initial-state
radiation~\cite{drees1} and, in the case of squarks, also the 
supersymmetric QCD corrections~\cite{drees1,Eberl:1996wa}). 
The low-scale SUSY-breaking parameters and the errors on them are then run
up to the high-energy (GUT) scale.  In this way one can check the extent 
to which the original theory can be reconstructed. 

As an example we take the pre-WMAP version of benchmark point E, which has
$m_{1/2}$~=~300~GeV, $m_0$~=~1.5~TeV, $A_0$~=~0, $\tanb$~=~10 and 
${\rm sign}(\mu)$~=~1,
to generate the SUSY spectrum -- later `forgetting' this origin of
the masses and cross sections. For the experimental errors, we assume 
that:
(i) the lightest Higgs mass can be measured with a precision of 50~MeV,
(ii) the charginos and neutralinos have been measured at a TeV-class 
linear collider, and we rescale the corresponding errors given 
in~Ref.~\cite{Blair2},
(iii) the LHC has measured the gluino mass with an accuracy of 
3\%~\cite{albert}. 
For the heavy Higgs bosons and the sfermions, we consider three
different scenarios: 
(a) slepton masses can be measured with an accuracy of 2\% and the 
remaining heavy particle masses within 7\%,
(b) slepton masses can be measured with an accuracy of 2\% and the 
remaining heavy particle masses within 3\%,
(c) all the masses of heavy particles can be measured with an 
accuracy of 1\% at CLIC. 
Moreover, we assume that the production cross sections for stops, sbottoms 
and staus, as well as their corresponding branching ratios, can be 
measured with errors that are twice the statistical errors.

In Table~\ref{tab:err}, the best-fit values and the corresponding errors
are given for the three cases described above. The errors on the underlying 
SUSY-breaking parameters of the first-generation sfermions scale roughly like 
the errors on the corresponding physical masses. 
The situation is different for the parameters of the third-generation 
and the Higgs mass parameters since 
(1) the precise knowledge of $m_{h^0}$ restricts in particular 
$M^2_{H,2}$, $M^2_{U,3}$, $M^2_{Q,3}$ and $A_t$, and 
(2) there is additional information from the measurements of cross 
sections and branching ratios in the cases of the sfermions. Note 
that most of the errors are correlated, because all particles
appear in the higher-order corrections to the masses. 
%
\begin{table}[!h]
\caption{Fitted parameters and their errors at the electroweak scale
         for the three cases discussed in the text. In the first
         two columns $\Delta m_{\tilde l}/m_{\tilde l}$~=~0.02.}
\label{tab:err}

\renewcommand{\arraystretch}{1.35} 
\begin{center}

\begin{tabular}{cccc}\hline \hline \\[-4mm]
$\hspace*{16mm}$ & 
$\hspace*{3mm}$ \boldmath{ $\Delta m/m$~=~0.07} $\hspace*{3mm}$ &
$\hspace*{6mm}$ \boldmath{$\Delta m/m$~=~0.03}  $\hspace*{6mm}$ & 
$\hspace*{3mm}$ \boldmath{$\Delta m/m$~=~0.01}  $\hspace*{3mm}$ 
\\[4mm]   

\hline \\[-3mm]
$M_1$ [GeV] & 127.5 $\pm$ 0.3 &  127.5 $\pm$ 0.3 &  127.5 $\pm$ 0.3 \\ 
$M_2$ [GeV] & 234.5 $\pm$ 0.5 & 234.5 $\pm$ 0.5 & 234.5 $\pm$ 0.5   \\
$M_3$ [GeV] & 664   $\pm$ 18 & 664   $\pm$ 17 & 664   $\pm$ 16  
\\[2mm] \hline \\[-3mm]
$M^2_{H,1}$ [GeV$^2$] & (2.197~$\pm$~0.068)~$\times$~10$^6$ & 
                      (2.197~$\pm$~0.055)~$\times$~10$^6$
                      & (2.197~$\pm$~0.025)~$\times$~10$^6$  \\
$M^2_{H,2}$ [GeV$^2$] & (--~1.092~$\pm$~0.005)~$\times$~10$^5$  & 
                       (--~1.092~$\pm$~0.005)~$\times$~10$^5$ 
                     & (--~1.092~$\pm$~0.004)~$\times$~10$^5$  
\\[2mm] \hline \\[-3mm]
$M^2_{E,1}$ [GeV$^2$] & (2.25~$\pm$~0.09)~$\times$~10$^6$ 
& (2.25~$\pm$~0.09)~$\times$~10$^6$
                      & (2.25~$\pm$~0.05)~$\times$~10$^6$ \\
$M^2_{L,1}$ [GeV$^2$] & (2.27~$\pm$~0.06)~$\times$~10$^6$ 
& (2.27~$\pm$~0.06)~$\times$~10$^6$
                      & (2.27~$\pm$~0.03)~$\times$~10$^6$ \\
$M^2_{D,1}$ [GeV$^2$] & (2.53~$\pm$~0.37)~$\times$~10$^6$ 
& (2.49~$\pm$~0.15)~$\times$~10$^6$
                      & (2.49~$\pm$~0.05)~$\times$~10$^6$ \\
$M^2_{U,1}$ [GeV$^2$] & (2.53~$\pm$~0.36)~$\times$~10$^6$ 
& (2.49~$\pm$~0.15)~$\times$~10$^6$
                      & (2.49~$\pm$~0.05)~$\times$~10$^6$ \\
$M^2_{Q,1}$ [GeV$^2$] & (2.49~$\pm$~0.25)~$\times$~10$^6$ 
& (2.52~$\pm$~0.11)~$\times$~10$^6$
                      & (2.52~$\pm$~0.04)~$\times$~10$^6$  
\\[2mm] \hline \\[-3mm]
$M^2_{E,3}$ [GeV$^2$] & (2.21~$\pm$~0.05)~$\times$~10$^6$ 
& (2.21~$\pm$~0.05)~$\times$~10$^6$
                      & (2.21~$\pm$~0.03)~$\times$~10$^6$ \\
$M^2_{L,3}$ [GeV$^2$] & (2.25~$\pm$~0.03)~$\times$~10$^6$ 
& (2.25~$\pm$~0.03)~$\times$~10$^6$
                      & (2.25~$\pm$~0.02)~$\times$~10$^6$ \\
$M^2_{D,3}$ [GeV$^2$] & (2.46~$\pm$~0.08)~$\times$~10$^6$ 
& (2.46~$\pm$~0.07)~$\times$~10$^6$
                      & (2.46~$\pm$~0.04)~$\times$~10$^6$ \\
$M^2_{U,3}$ [GeV$^2$] & (9.55~$\pm$~0.40)~$\times$~10$^5$ 
& (9.55~$\pm$~0.33)~$\times$~10$^5$
                      & (9.55~$\pm$~0.18)~$\times$~10$^5$ \\
$M^2_{Q,3}$ [GeV$^2$] & (1.75~$\pm$~0.03)~$\times$~10$^6$ 
& (1.75~$\pm$~0.03)~$\times$~10$^6$
                      & (1.75~$\pm$~0.02)~$\times$~10$^6$  
\\[2mm] \hline \\[-3mm]
$A_t$~[GeV] & --~519~$\pm$~71  & --~519~$\pm$~69  & --~519~$\pm$~65 
\\[3mm] 
\hline \hline
\end{tabular}
\end{center}
\end{table}

\begin{figure}[htbp] %
\setlength{\unitlength}{1mm}
\begin{picture}(160,70)
\put(0,-80){\mbox{\epsfig{figure=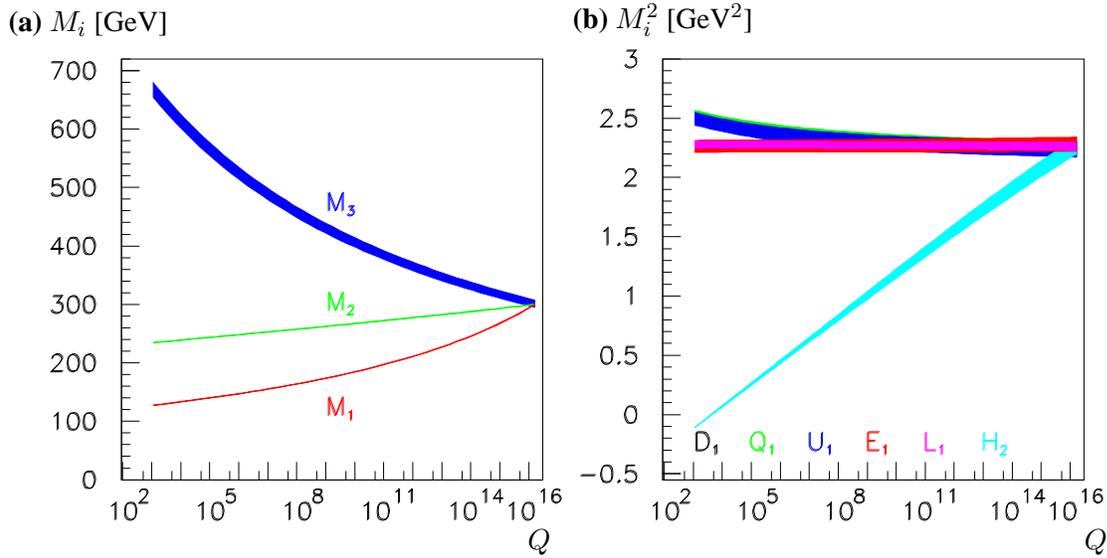,height=16cm,width=16cm}}}
\put(1,67){\makebox(0,0)[bl]{{\bf (a)}  $M_i$~[GeV]}}
\put(70,-2){\makebox(0,0)[bl]{$Q$}}
\put(76,67){\makebox(0,0)[bl]{{\bf (b)} $M^2_i$~[GeV$^2$]}}
\put(144,-2){\makebox(0,0)[bl]{$Q$}}
\end{picture}

\vspace*{5mm}

\caption{Running of (a) gaugino mass parameters and (b)  
         first-generation sfermion mass parameters and $M^2_{H,2}$ assuming
         1\% errors on  sfermion masses and heavy Higgs boson masses.
         The width corresponds to 1$\sigma$ errors.}
\label{fig:run}
\end{figure}
\begin{figure}[htbp] %
\setlength{\unitlength}{1mm}
\begin{picture}(160,60)
\put(-10,0){\mbox{\epsfig{figure=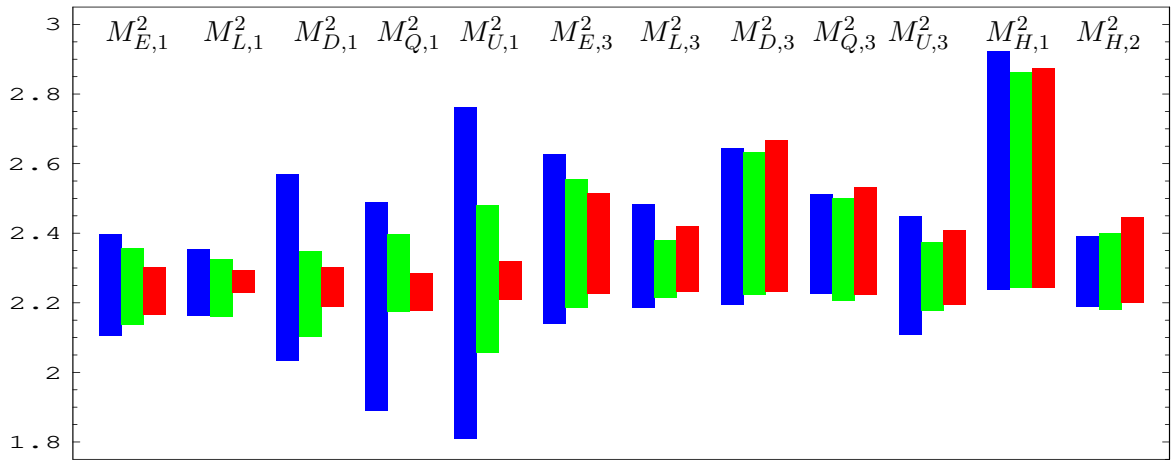,height=6cm,width=18cm}}}
\put(16,54){\makebox(0,0)[bl]{$M^2_{E,1}$}}
\put(29,54){\makebox(0,0)[bl]{$M^2_{L,1}$}}
\put(41,54){\makebox(0,0)[bl]{$M^2_{D,1}$}}
\put(52,54){\makebox(0,0)[bl]{$M^2_{Q,1}$}}
\put(63,54){\makebox(0,0)[bl]{$M^2_{U,1}$}}
\put(75,54){\makebox(0,0)[bl]{$M^2_{E,3}$}}
\put(87,54){\makebox(0,0)[bl]{$M^2_{L,3}$}}
\put(99,54){\makebox(0,0)[bl]{$M^2_{D,3}$}}
\put(110,54){\makebox(0,0)[bl]{$M^2_{Q,3}$}}
\put(120,54){\makebox(0,0)[bl]{$M^2_{U,3}$}}
\put(133,54){\makebox(0,0)[bl]{$M^2_{H,1}$}}
\put(145,54){\makebox(0,0)[bl]{$M^2_{H,2}$}}
\end{picture}

\vspace*{5mm}

\caption{The 1$\sigma$ bands for the sfermion and Higgs mass parameters 
         in TeV$^2$ at $M_{\rm GUT}$.
         The following cases are considered: 
        (dark boxes) slepton masses can be measured with an accuracy of 2\%
           and the remaining particle masses within 7\%;
        (light gray boxes) slepton masses can be measured with an accuracy
          of 2\% and the remaining particle masses within 3\%;
        (dark gray boxes) sfermion and heavy Higgs boson masses can be measured
         with an accuracy of 1\%.}
\label{fig:SfMgut}
\end{figure}

In Fig.~\ref{fig:run} the evolutions of the gaugino mass parameters, 
the squared-mass parameters of the first-generation sfermions and 
$M^2_{H,2}$ are
shown for scenario (c) where 1\% accuracy on all heavy masses has 
been assumed. One clearly sees that the evolution is under excellent
control. In the cases of the third-generation sfermion squared-mass 
parameters and $M^2_{H,1}$, the accuracy at the GUT scale is somewhat 
worse, as can be
seen in Fig~\ref{fig:SfMgut}, where the 1$\sigma$ ranges for the sfermion
mass parameters and the Higgs mass parameters (in TeV$^2$) are given
for the three scenarios with the different mass errors.
As can be seen, there is a clear overlap between all mass parameters,
as expected for universal boundary conditions.

\vspace*{-4mm}

One sees that the mean value of some parameters can be somewhat shifted 
with respect to the ideal value of 2.25~TeV$^2$. The reasons are (i)
the correlations 
between the errors and (ii) the anomaly term $S={\rm Tr}(Y M^2)$ in the RGEs
is not zero any more, because one starts from a slightly shifted minimum.
In the case of the third generation, there is an additional shift due
to the large errors on $A_b$ and $A_\tau$. The use of branching ratios and
the cross sections for sfermion production hardly restrict these parameters,
and the allowed ranges are $\pm$ several TeV.  
These parameters always give positive shifts in the RGEs, because they 
appear squared in the RGEs of the scalar squared-mass parameters.

\begin{figure}[htbp] %
\setlength{\unitlength}{1mm}
\begin{picture}(160,60)
\put(-10,0){\mbox{\epsfig{figure=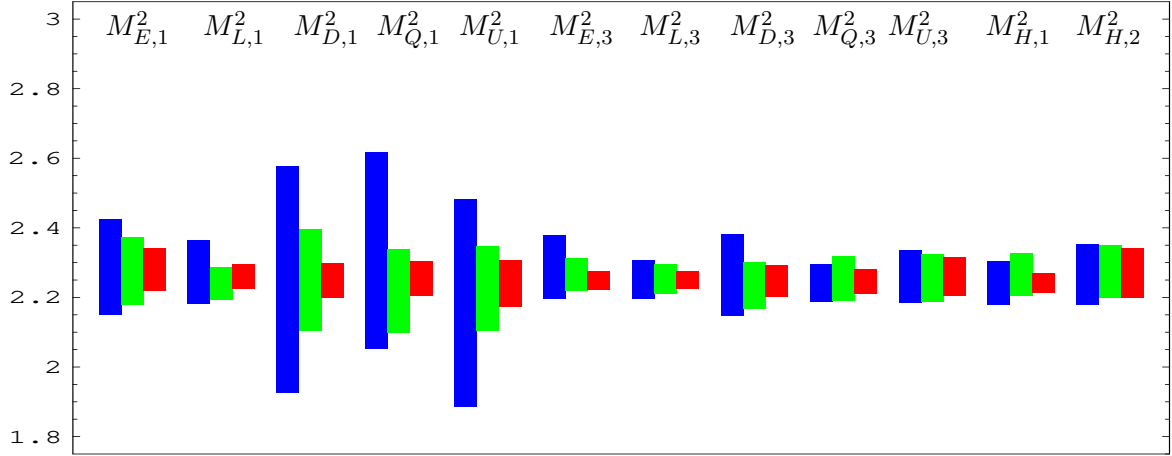,height=6cm,width=18cm}}}
\put(16,54){\makebox(0,0)[bl]{$M^2_{E,1}$}}
\put(29,54){\makebox(0,0)[bl]{$M^2_{L,1}$}}
\put(41,54){\makebox(0,0)[bl]{$M^2_{D,1}$}}
\put(52,54){\makebox(0,0)[bl]{$M^2_{Q,1}$}}
\put(63,54){\makebox(0,0)[bl]{$M^2_{U,1}$}}
\put(75,54){\makebox(0,0)[bl]{$M^2_{E,3}$}}
\put(87,54){\makebox(0,0)[bl]{$M^2_{L,3}$}}
\put(99,54){\makebox(0,0)[bl]{$M^2_{D,3}$}}
\put(110,54){\makebox(0,0)[bl]{$M^2_{Q,3}$}}
\put(120,54){\makebox(0,0)[bl]{$M^2_{U,3}$}}
\put(133,54){\makebox(0,0)[bl]{$M^2_{H,1}$}}
\put(145,54){\makebox(0,0)[bl]{$M^2_{H,2}$}}
\end{picture}
\caption{The 1$\sigma$ bands for the sfermion and Higgs mass parameters at
         $M_{\rm GUT}$.
         The following cases are considered: 
        (dark boxes) slepton masses can be measured with an accuracy of 2\%
           and the remaining particle masses within 7\%;
        (light gray boxes) slepton masses can be measured with an accuracy
          of 2\% and the remaining particle masses within 3\%;
        (dark gray boxes) sfermion and heavy Higgs boson masses can be measured
         with an accuracy of 1\%.
         In addition, it has been assumed that $A_b$ and $A_\tau$ can be 
         measured within 30\%.
\label{fig:SfMguta}}
\end{figure}

The errors on $A_b$ and $A_\tau$ can be reduced significantly
if one can either use the polarization of sfermion decay products~\cite{polar} 
or can measure the triple coupling between sfermions and Higgs 
bosons~\cite{stauexp}. The latter measurement  
would require a LC with at least 6 TeV for this particular example.
Its effect is, however, quite dramatic, as can be seen
in~Fig.~\ref{fig:SfMguta}, where it is assumed that $A_b$ and $A_\tau$
can be measured within 30\% (roughly a factor 2 worse than $A_t$).
Note that such a measurement requires a multi-TeV collider even
if the sfermion and Higgs masses are of the order of a few hundred~GeV.

In summary, the errors on the low-energy parameters roughly scale
as the errors on the corresponding masses. The exceptions are the
third-generation squark parameters, because they are tightly constrained
by the precise measurement of $m_h$, where these parameters appear in
the radiative corrections. The error in the high-scale parameters
of the first generation at $M_{\rm GUT}$ improves by a factor 2 to 7
when the precision in the corresponding masses is improved from 7\% to
1\%. In the case of the third-generation parameters, a determination
of $A_b$ and $A_\tau$ is necessary for reducing the errors at $M_{\rm GUT}$.
This can be done by measuring the trilinear Higgs sfermion couplings,
for example from processes like 
$e^+ e^- \to \tilde b_i \tilde b_j A^0$ or  
$e^+ e^- \to \tilde \tau_i \tilde \tau_j A^0$. If $A_b$ and
$A_\tau$ can be determined within 30\% in such processes, the errors on 
the high-scale parameters can be reduced by an order of magnitude.
This conclusion remains valid for scenarios where most of
the spectrum can be measured precisely at an $e^+ e^-$ collider in
the~TeV~range.

\section{The Role of Beam Polarization}

Polarized beams provide an important analysis tool at any LC, and in the
following we summarize some highlights~\cite{Steiner}.
In particular, for the analysis 
and the precise determination of the underlying structure of New Physics 
(NP), polarized cross sections and asymmetries are superior observables.
One may expect to reach an electron polarization of about 80\% via a
strained photocathode technology as used at the SLC, where
$P(e^-)$~=~(77.34~$\pm$~0.61)\% was reached. One may also
expect to reach a positron polarization of about 40\%--60\%.

For Standard Model processes, it is suitable to 
introduce the effective luminosity
\begin{equation}
  {\cal L}_{\rm eff}/{\cal L}= \frac{1}{2}\, [1-P(e^-)P(e^+)]\,,
  \label{eff-lum}
\end{equation}
as a fraction of the total colliding-beam luminosity,
and the effective polarization
\begin{equation}
  P_{\rm eff}=[P(e^-)-P(e^+) ]/[1-P(e^-)P(e^+)]\,.
  \label{eff-pol}
\end{equation}
Comparing these two numbers (${\cal L}_{\rm eff}/{\cal L}$, 
$P_{\rm eff}$) for the unpolarized case (0.5, 0), the case with only electron
polarization of about $P(e^-)$~=~--80\% (0.5, -- 0.8) and 
the case with $P(e^-)$~=~--80\% and $P(e^+)$~=~+60\% 
(0.74, --0.95), we see that only the latter option can enhance the
effective luminosity. 

It is well known that suitable beam polarization suppresses 
the dominant SM background process $WW$ (as well as $ZZ$):
the use of 80\% electron polarization leads to a reduction by a factor
0.2 (0.76) and $P(e^-)$~=~+80\% and $P(e^+)$~=~--60\% leads to a
scaling factor 0.1 (1.05). This reduction of background processes is
important for all SUSY analyses.

An interesting further option for a LC with both beams polarized is to 
make studies with transversely polarized beams.  One example
was already worked out some time ago for $e+e-\to W^-W^+$ 
in~\cite{Jegerlehner}. At very high energies
$\sqrt{s}\gg $~2$m_W$, longitudinal $W_L$ production dominates. However,
the signal peaks are towards the beam direction and the final-state 
analysis becomes more
difficult. Since transverse beams project out the $W_L$ state 
one gets information about electroweak
symmetry breaking and about deviations from the SM expectation
without the analysis of the final decay products. 

Particularly important, however, are polarized beams for the precise
determination of the structure of NP. As already noted, supersymmetry
is one of the most promising candidates for NP, but it leads, even
in its minimal version, to about 105 new parameters.  In order to
determine the model, e.g. MSSM versus (M+1)SSM, to measure the
fundamental parameters without assuming a specific supersymmetry-breaking
scheme, to measure precisely the mixing angles, e.g. in the stop
sector (see also Section 4.2) or to test whether R-parity is broken or
not, beam polarization is very important, see~Ref.~\cite{Steiner} and
references therein.  
In order to test SUSY decisively, one has to go one step further 
and prove experimentally that the SM particles and their supersymmetric 
partners carry the same internal quantum numbers,
and that the gauge couplings are identical to the gaugino
couplings. 
Beam polarization is particularly needed for this purpose.

\subsection{Slepton Quantum Numbers}
Supersymmetry associates the chiral leptons $e^-_{\rm L,R}$ 
with spinless
particles whose couplings should reflect the chiral quantum numbers of 
their
lepton partners: $e^-_{\rm L,R} \to \sEl^-_{\rm L,R}$ and 
$e^+_{\rm L,R}\to\sEl^+_{\rm L,R}$. These associations could be tested
uniquely by separating the $t$-channel contribution to 
$e^+ e^- \to \sEl^+ \sEl^-$ annihilation, which is controlled by the 
$e\sEl\chiz_i$ vertex, from the $s$-channel contribution.
This separation could in principle be made via threshold scans, 
since the pair-production cross sections for the ${\rm LL}$ and 
${\rm RR}$ states have P-wave thresholds, whereas production of
off-diagonal ${\rm LR}$ and ${\rm RL}$  
pairs exhibit S-wave thresholds~\cite{Soffer}. 
However, since threshold scans require considerable luminosity and time, 
we concentrate here on another procedure~\cite{Bloechinger}: 
selectron pair production occurs with $V$ and $A$ couplings via 
$\gamma$ and $Z$ exchanges in the $s$-channel, and through the scalar 
coupling in the $t$-channel. The beam configurations $e^-_{\rm L}
e^+_{\rm L}$ and $e^-_{\rm R} e^+_{\rm R}$ suppress the $s$-channel
completely, and only the $t$-channel survives.
Since, however, the beams cannot be completely polarized,
one has to study to which extent this test of the sparticle quantum 
numbers can be done with partially polarized beams.

\begin{figure}[htbp] %
\centerline{\includegraphics[width=16.0cm]{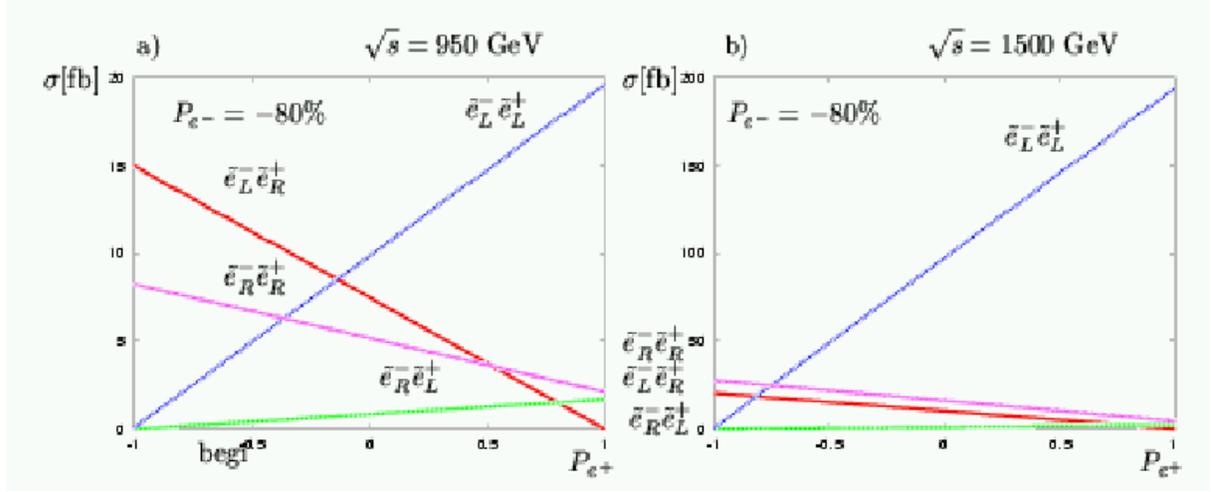}}
\caption{Test of selectron quantum numbers in
$e^+e^-\to\sEl_{\rm L,R}^+\sEl_{\rm L,R}^-$ with fixed electron
polarization $P(e^-)$~=~--80\% and variable positron polarization
$P(e^+)$. The masses are $m_{\sEl_{\rm L}}$~=~448~GeV, 
$m_{\sEl_{\rm R}}$~=~417~GeV
and a) at $\sqrt{s}$~=~950~GeV both the pairs 
$\sEl_{\rm L}^- \sEl_{\rm L}^+$ and $\sEl_{\rm L}^- \sEl_{\rm R}^+$ 
have comparable cross sections for unpolarized beams. 
For  $P(e^+)$~=~--40\% ($P(e^+)$~=~--60\%) the $t$--channel 
pair $\sEl_{\rm L}^- \sEl_{\rm R}^+$  dominates by a factor of about 1.5 
(about 2). b) At $\sqrt{s}$~=~1500~GeV, the pair 
$\sEl_{\rm L}^- \sEl_{\rm L}^+$ dominates kinematically and the
$t$-channel process can not be separated.}
\label{fig_slep}
\end{figure}

In Fig.~\ref{fig_slep}  we show the polarized cross sections for
the processes $e^- e^+\to \sEl^-_i \sEl^+_j$ ($i,j~=~{\rm L,R}$) with
a fixed electron polarization of $P(e^-)$~=~--80\% and variable positron 
polarization, and study which polarization of $P(e^+)$ would be
necessary to separate the pair $\sEl^-_{\rm L}\sEl^+_{\rm R}$. We
have chosen the high $\tan\beta$~=~50 CMSSM scenario of the `Snowmass
Points and Slopes' SPS4~\cite{SPS}, where the slepton masses are 
$m_{\sEl_{\rm L}}$~=~448~GeV and $m_{\sEl_{\rm R}}$~=~417~GeV. We 
first study all four production processes at an energy close to the
threshold $\sqrt{s}$~=~950~GeV, shown in~Fig.~\ref{fig_slep}a. 
With unpolarized positrons, the production of 
$\sEl_{\rm L}^-\sEl^+_{\rm L}$ is 
dominant. When the positrons are left-polarized, the rate of 
$\sEl^-_{\rm L} \sEl^+_{\rm R}$ is increased. 
For $P(e^+)$~=~--40\% it dominates by a factor of about 1.5, and for
$P(e^+)$~=~--60\% by a factor of~about~2. 

This test of quantum numbers is, however, strongly dependent on the 
beam energy. In Fig.~\ref{fig_slep}b it can be seen that at 
$\sqrt{s}$~=~1500~GeV for our particular example the production of the 
off-diagonal pairs $\sEl^\pm_{\rm R} \sEl^\mp_{\rm L}$ is suppressed relative 
to $\sEl^+_{\rm L} \sEl^-_{\rm L}$ production, and the $t$-channel
process cannot be separated clearly, in which case the method would fail.

\subsection{Gaugino Couplings}
At a LC, the parameters of the neutralino/chargino sector and their phases,  
$M_1$, $\Phi_1$, $M_2$, $\mu$, $\Phi_{\mu}$, and $\tan\beta$ 
can be determined via the analysis of masses and polarized cross sections 
without any assumption on the supersymmetry-breaking scheme. 
Even if only the lighter states $\chiz_{1,2}$ and $\chipm_1$ are 
kinematically accessible, a precise determination of these parameters 
could be made~\cite{ckmz}. Once these parameters are known, another 
important check of fundamental supersymmetric relations 
could be made with the help of polarized beams: the test whether 
the SU(2) and U(1) gauge couplings $g$ and $g'$ are identical to 
the gaugino couplings $g_{\tilde{W}}$ and $g_{\tilde{B}}$  from the 
$\ell\tilde{\ell}\tilde{W}$ and $\ell\tilde{\ell}\tilde{B}$
interactions. 
The comparison between the polarized cross sections $\sigma_{\rm L}$ and 
$\sigma_{\rm R}$ for neutralino production with the theoretical prediction 
for variable ratios $g_{\tilde{W}}/g$ and $g_{\tilde{B}}/g'$ provides 
a precise test of the gaugino couplings at the \% level~\cite{ckmz,Ayres}.

\subsection{Distinction between the MSSM and (M+1)SSM}
\begin{figure}[htbp]
\begin{center}
\epsfig{file=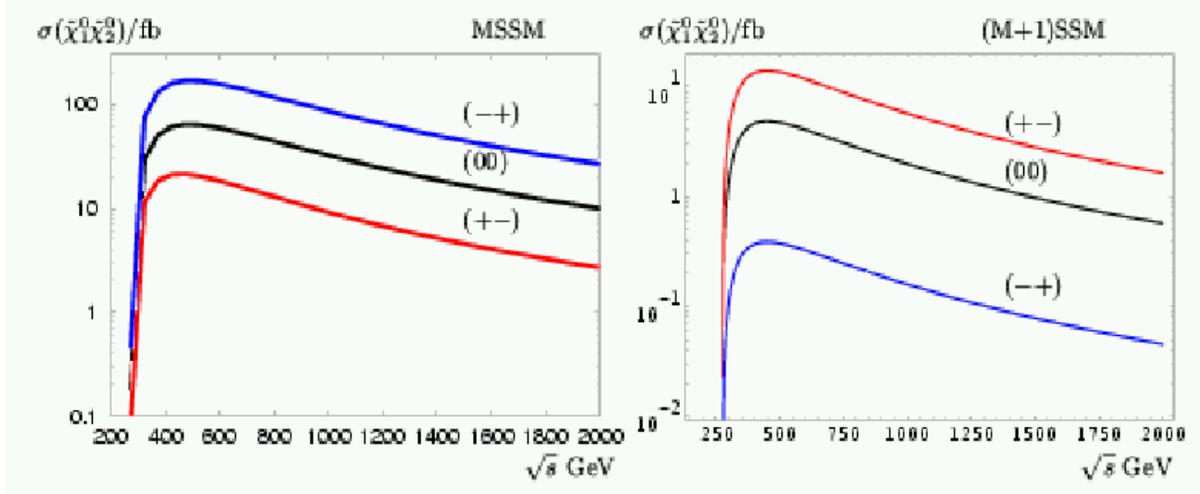,width=16cm}

\vspace*{3mm}

\caption{Cross sections for the process $\sigma(e^+e^-\to
\tilde{\chi}^0_1\tilde{\chi}^0_2)$ with polarized beams 
$(P_{e^-}=\pm$~80\%, $P_{e^+}=\mp$~60\%) for an example in the MSSM and the
(M+1)SSM, where the mass spectra of the light neutralinos are
similar~\cite{Hesselbach}. 
\label{fig_nmssm}}
\end{center}
\end{figure}

Supersymmetric models with additional chiral or vector superfields give 
rise to
extensions of the gaugino/higgsino sector in general, and additional
neutralinos occur. We consider here the (M+1)SSM with one new Higgs
singlet, which leads to an extra singlino state. Since the neutralino 
mass spectra of the lightest four neutralinos in such extended models 
can be similar to those in the MSSM, a distinction between the models 
might be difficult.
However, the coupling structure is different in the two models. Therefore 
they could be distinguished via the polarization dependences of the cross 
sections. In Fig.~\ref{fig_nmssm} we show the
cross section for $e^+e^-\to \tilde{\chi}^0_1\tilde{\chi}^0_2$ for one
example in the MSSM and one in the (M+1)SSM~\cite{Hesselbach},
where the mass spectra of the light particles in the two models are similar.

\begin{figure}[!h] 
\begin{center}
 \epsfxsize=10cm \epsfbox{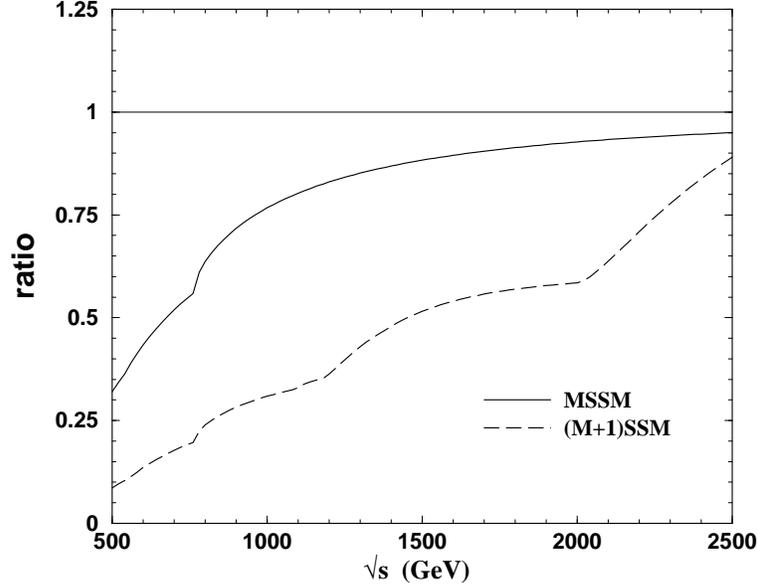}

\vspace*{3mm}

\caption{The energy dependence of the sum of all the
  neutralino--pair-production cross sections normalized to the
  asymptotic form of the summed cross section. The solid line
  represents the exact sum in the MSSM and
   the dashed line the sum of the cross sections
  for the first four neutralino states for a specific parameter set
  of the (M+1)SSM~\cite{ckmz}. }
\label{fig_sum}
\end{center}
\end{figure}

Powerful tests for the closure of the system are evaluations of sum rules. 
The four-state mixing in the MSSM induces sum rules for the couplings that
follow from the unitarity of the diagonalization matrices.
For asymptotic energies, these sum rules can be transformed directly 
into sum rules for the associated cross sections~\cite{ckmz},
and are given in the MSSM by
\begin{equation}
\lim_{s\rightarrow \infty}\,s\,\sum_{i\leq
     j}^4\,\sigma\{ij\}
     =\frac{\pi\alpha^2}{48\,c^4_W s^4_W}\left[64 s^4_W-8 s^2_W+ 5\right]\,.
\end{equation}
In Fig.~\ref{fig_sum} the exact values for
the summed cross sections are shown for one example, normalized to the 
asymptotic values. 
The final
state $\tilde{\chi}^0_1\tilde{\chi}^0_1$ is  invisible in
$R$-invariant theories, and its detection is
difficult. Nevertheless, it can be studied directly  by
photon tagging in the final state
$ \gamma \tilde{\chi}^0_1\tilde{\chi}^0_1$, which can be observed at
the LC. In this example the mass spectra
of the light neutralinos in the two models are similar, and
the asymptotic cross-section value is not affected by the additional 
singlet of the (M+1)SSM. 
However, as can be seen in~Fig.~\ref{fig_sum},
because of the incompleteness of these states
below the thresholds for producing the heavy neutralino, the value 
measurable in the (M+1)SSM
differs significantly from the corresponding sum rule of the MSSM.
Therefore, even if the extended neutralino states are very heavy,
the study of sum rules can shed light on the underlying structure of the
supersymmetric model. In this context, beam polarization is important to 
enhance the rates of even 
those pairs that have extremely small cross~sections.

\vspace*{6mm}

\subsection{Determination of $\tan\beta$ and Trilinear Couplings}
\begin{figure}[htbp] %
\begin{center}
\epsfig{file=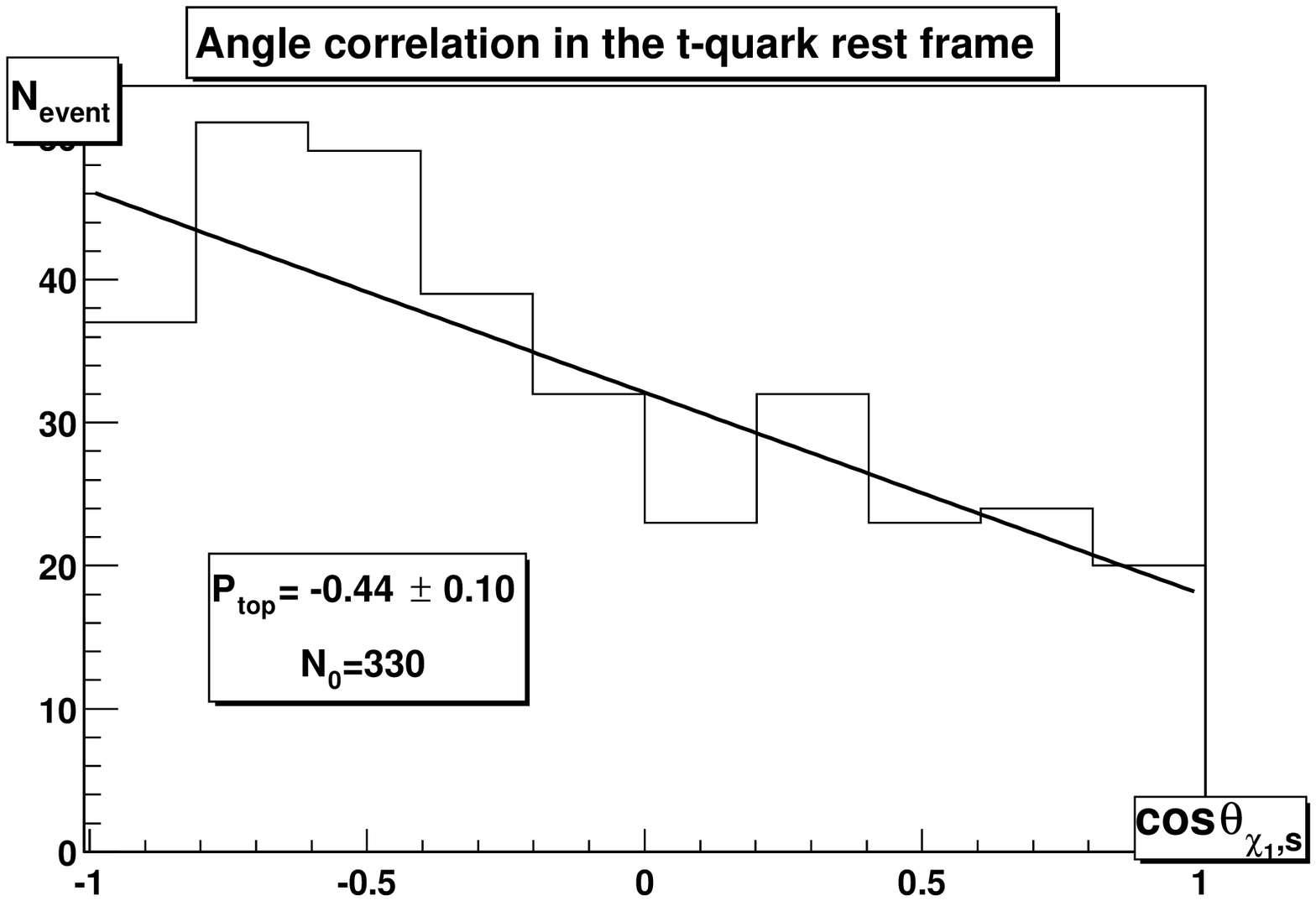,bbllx=5,bblly=5,bburx=563,bbury=384,width=12cm} 
  \caption{Angular distribution in $\cos\theta^{*}_s$, the angle between the
    $\tilde{b}_1$ and $\bar{s}$ partons in the top rest frame of $t\to
    b\,c\bar{s}$ decays from
    $e^+_Le^-_R\to \tilde{b}_1\tilde{b}_1\to t\tilde{\chi}_1^\pm + \tilde{b}_1$
    production
    at $\sqrt{s}$~=~1.9~TeV. The line represents a fit to a top polarization
    of $P_t$~=~--~0.44~$\pm$~0.10~\cite{polar,boos}.}
\label{fig_top}
\end{center}
\end{figure}

Several studies have been made of how $\tan\beta$ can be measured in
the Higgs sector via rates, widths and branching
ratios~\cite{hightb}. However, if $\tan\beta >$~10, its precise
determination will be difficult.
It was shown in~Ref.~\cite{polar,boos} that in this case the $\tau$
and $t$ polarizations are suitable observables for 
determining $\tan\beta$. The procedure is as follows: one studies the 
rates for producing the lighter sparticle pairs $e^+e^-\to 
\tilde{\tau}_1 \tilde{\tau}_1$ or
$\tilde{t}_1 \tilde{t}_1$, and measures the mixing angle of the system
by analysing polarized rates, as already
worked out in earlier studies~\cite{stau}. The masses as well as the 
mixing angle can be measured
with precisions at the per cent level. Ambiguities can be
resolved, e.g. via measurements of the cross sections
with another configuration of the beam polarization~\cite{polar,boos}.  
Determining the polarization of the final decay fermions,
the $\tau$ or $t$ polarization, while 
taking into account the complete neutralino and
chargino mixing, leads to a determination of 
$\tan\beta$ with an accuracy of 10\% 
even at high $\tan\beta$.  In this context, the measurement of the
$t$ polarization from $\tilde{b}_1\to t \tilde{\chi}^{\pm}$ decays
could be very tricky, since the reconstruction of the system is 
difficult because of the
missing transverse energy. However, analysing the
distributions of the quark jets in the hadronic top decays $t\to b+c
\bar{s}$ could lead to a very precise measurement of the top
polarization~\cite{polar,boos} as seen in~Fig.~\ref{fig_top}. 

If the heavier states $\tilde{\tau}_2$, $\tilde{b}_2$ or $\tilde{t}_2$
are also accessible, one can even go a step further, and determine the
trilinear  
couplings $A_f$. If $\tilde{t}_2$ and $\tilde{b}_2$ can be measured at 
the per cent level, $A_t$ and $A_b$ may be extracted with 30--50\% 
precision~\cite{Kraml:1999qd,boos}.
Because of the small $\tau$ mass, 
one cannot expect to determine the parameter $A_{\tau}$ with very
high precision. 

\section{Measuring Neutrino Mixing Angles at CLIC}

Recent neutrino experiments~\cite{skam,sno,kland}
clearly show that neutrinos are massive particles and that they mix.
A possible way to account for the observed neutrino data in supersymmetric
theories is to break lepton number, either spontaneously or explicitly, and 
thus R-parity.
The simplest realization of this idea is to add bilinear terms to the 
superpotential $W$:
\begin{eqnarray}
W = W_{\rm MSSM} + \epsilon_i \hat L_i \hat H_u\,.
\label{eq:model}
\end{eqnarray}
For consistency one also has to add the corresponding 
bilinear terms to soft SUSY-breaking, which induce small vacuum expectation
values (vevs) for the sneutrinos. These vevs induce a mixing between
neutrinos and neutralinos, giving mass to one neutrino. The second neutrino
mass is induced by loop effects 
(see~Ref.~\cite{Diaz:2003as} and references therein). The same parameters
that induce neutrino masses and mixings are also responsible for the
decay of the lightest supersymmetric particle (LSP). This implies that there
are correlations between neutrino physics and LSP decays 
\cite{Mukhopadhyaya:1998xj,Porod:2000hv,Hirsch:2002ys}.

As a specific example we present the relationship between the atmospheric
neutrino mixing angle and the decays of a neutralino into semileptonic
final states $l_i q' \bar{q}$ ($l_i=e,\mu,\tau$). There are several more
examples, which are discussed in~Ref.~\cite{Porod:2000hv}.
In the model specified by Eq.~(\ref{eq:model})
the atmospheric mixing angle is given by
\begin{eqnarray}
& & \tan \theta_{\rm atm} = \frac{\Lambda_2}{\Lambda_3} \\
& & \Lambda_i = \epsilon_i v_d + \mu v_i\,,
\end{eqnarray}
where $v_i$ are the sneutrino vevs and $v_d$ is the vev of $H^0_1$. 
It turns out that the dominant part of the 
$\tilde \chi^0_1$-$W$-$l_i$ coupling $O^L_i$ is given by
\begin{eqnarray}
 O^L_i = \Lambda_i f(M_1,M_2,\mu,\tan \beta, v_d, v_u)\,,
\end{eqnarray}
where the exact form of $f$ can be found in Eq.~(20) 
of Ref.~\cite{Porod:2000hv}. The important point is, that $f$ depends
only on MSSM parameters and not on the R-parity-violating
parameters. Putting everything together one finds:
\begin{eqnarray}
 \tan^2 \theta^2_{\rm atm} = \left| \frac{\Lambda_2}{\Lambda_3} \right|^2
   \simeq \frac{BR(\tilde \chi^0_1 \to \mu^\pm W^\mp)}
               {BR(\tilde \chi^0_1 \to \tau^\pm W^\mp)}
   \simeq  
 \frac{BR(\tilde \chi^0_1 \to \mu^\pm \bar{q} q')}
      {BR(\tilde \chi^0_1 \to \tau^\pm \bar{q} q' )}\,.
\label{eq:corr}
\end{eqnarray}
The restriction to the hadronic final states of the $W$ is necessary
for the identification of the lepton flavour. Depending on the
specific parameters, one gets additional small contributions due to
three-body decays of intermediate sleptons and squarks. Note that
Eq.~(\ref{eq:corr}) is a prediction of the bilinear model independent
of the R-parity-conserving parameters. It has been shown that
under favourable conditions the LHC can measure the ratio of branching
ratios in Eq.~(\ref{eq:corr}) with an accuracy of  
about 3\%~\cite{Porod:2004rb}.

In this note we study an example of such a correlation in a scenario where
the lightest neutralino is rather heavy and where it is 
most likely that the LHC 
will have difficulties to measure the branching ratios accurately.
For the numerical example, we take point H as reference, which is
characterized by $M_{1/2}$~=~1.5~TeV, $M_0$~=~419~GeV, $A_0$~=~0,
$\tan\beta$~=~20 and a positive sign of $\mu$. We vary randomly the 
R-parity-conserving
 parameters around this point so that the masses of the
supersymmetric particles vary within 10\% for the particles that can
be observed at a 5 TeV collider. For the remaining particles we set a
lower cut of 2.5~TeV on the masses.
To these sets of parameters  we add the R-parity-breaking parameters
such that the correct differences between the squared neutrino masses and 
the correct mixing angles are obtained within the experimental allowed range:
1.2~$\times$~10$^{-3}$~eV$^2 < \Delta m^2_{\rm atm}<$~4.8~$\times$~10$^{-3}$~eV$^2$,
5.1~$\times$~10$^{-5}$~eV$^2 < \Delta m^2_{\rm sol}<$~1.9~$\times$~10$^{-4}$~eV$^2$,
0.29~$< \sin^2 \theta_{\rm atm}<$~0.86, $\sin^2 \theta_{13}<$~0.05.
For illustrative purpose only we allow a range for 
$\sin^2 \theta_{\rm atm}$ wider than the
experimentally allowed one: 0.3~$< \sin^2 \theta_{\rm atm} <$~0.7. 
The main LSP decay modes are: 
BR$(\tilde \chi^0_1 \to h^0 \sum_{i~=~1,3} \nu_i) \simeq$~50\%,
BR$(\tilde \chi^0_1 \to W^\pm e^\mp +  W^\pm \mu^\mp + W^\pm \tau^\mp)
 \simeq$~30\%,
BR$(\tilde \chi^0_1 \to Z^0 \sum_{i~=~1,3} \nu_i) \simeq 20$\%.
In addition there are three-body decay modes mainly mediated by virtual
sleptons whose sum of the branching ratios is~about~1--2\%.

\begin{figure}[htbp] %
\setlength{\unitlength}{1mm}
\begin{center}
\begin{picture}(70,70)
\put(0,0){\mbox{\epsfig{figure=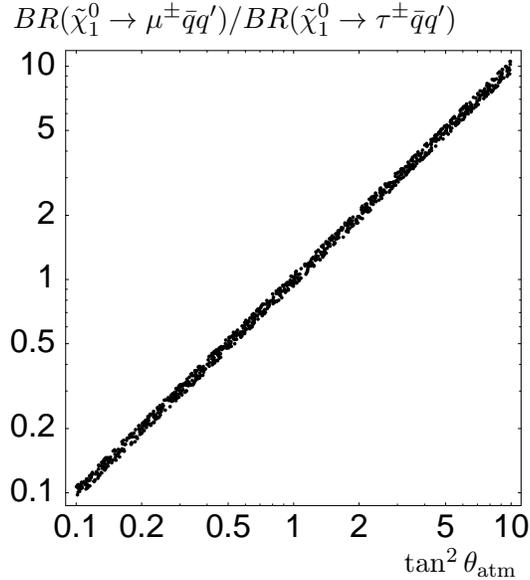,height=7cm,width=7cm}}}
\put(1,70){\makebox(0,0)[bl]{{$BR(\tilde \chi^0_1\to \mu^\pm \bar{q} q')
 / BR(\tilde \chi^0_1 \to \tau^\pm \bar{q} q' )$}}}
\put(53,-2){\makebox(0,0)[bl]{$\tan^2 \theta_{\rm atm}$}}
\end{picture}
\end{center}
\vspace{0.5cm}
\caption{Correlation between atmospheric neutrino mixing angle and
 semileptonic branching ratios of the lightest neutralino.}
\label{fig:corr}
\end{figure}

In Fig.~\ref{fig:corr} we show the correlation between 
$\tan^2 \theta_{\rm atm}$ and 
$BR(\tilde \chi^0_1 \to \mu^\pm \bar{q} q') /
BR(\tilde \chi^0_1 \to \tau^\pm \bar{q} q' )$.
Note that the width
of the band is due to the variation of the parameters as discussed above.
It becomes by a factor of 4--5 smaller if the SUSY masses can be measured
with an accuracy of 2\% instead of the assumed 10\%.
Figure \ref{fig:corr} clearly shows that these
decay modes can be used to relate observables measured at CLIC to
those measured at neutrino experiments. Moreover, they allow for
a consistency test of the theory.
Note that R-parity-violating decay modes into
gauge bosons (Higgs bosons) and leptons are forbidden in 
R-parity-violating models where only trilinear R-parity-breaking
couplings exist. Therefore, the observation of such decays give a
clear proof that bilinear R-parity-breaking is realized~in~nature.

In summary, models where R-parity is broken by bilinear terms 
can explain recent neutrino data and predict at the same time
certain decay properties of the lightest supersymmetric particle.
CLIC is particularly suited to explore the resulting correlations
between neutrino mixing angles and ratios of LSP branching ratios in
scenarios where the LSP is rather heavy.

\section{Summary}

The CLIC multi-TeV linear $e^+ e^-$ collider has important potential for
supplementing the information on supersymmetry that might be provided by
previous colliders such as the LHC and a TeV-class linear collider, as we
have illustrated within the CMSSM. The CMSSM parameter space is restricted
by experimental constraints from LEP and elsewhere, and by cosmology. If
sparticle masses are in the lower part of the range still allowed by these
constraints, much of the supersymmetric spectrum will have been discovered
by the LHC and/or by a TeV-class linear collider, which would also be able
to make some precision measurements. In this case, CLIC would be
invaluable for extending precision measurements to squarks and heavier
neutralinos and charginos, in particular. On the other hand, if sparticle
masses are in the upper part of the allowed range, CLIC might be the first
accelerator to produce many of the SUSY particles.

Using benchmark CMSSM points as examples, we have demonstrated that CLIC
would be able to make detailed measurements of sparticle masses and mixing
angles. These would enable accurate tests of unification hypotheses for
mechanisms of supersymmetry breaking to be made. Electron and positron
beam polarization would be a useful tool for disentangling sparticle
properties, and the $\gamma \gamma$ option would be especially
interesting for gluino
production, in particular. Although most of this discussion referred 
to the CMSSM, we have also shown that CLIC has important capabilities for 
distinguishing this from more complicated supersymmetric models, and for 
exploring scenarios in which R parity is violated.

We conclude that CLIC will be an invaluable tool for exploring
supersymmetry, even if it has been discovered previously at the LHC and
explored at a TeV-class linear collider.


\newpage
\chapter{PROBING NEW THEORIES}
\label{chapter:six}

\setlength{\textheight}{241mm} \setlength{\textwidth}{170mm}
\def\MPL #1 #2 #3 {Mod. Phys. Lett. {\bf#1},\ #2 (#3)}
\def\NPB #1 #2 #3 {Nucl. Phys. {\bf#1},\ #2 (#3)}
\def\PLB #1 #2 #3 {Phys. Lett. {\bf#1},\ #2 (#3)}
\def\PR #1 #2 #3 {Phys. Rep. {\bf#1},\ #2 (#3)}
\def\PRD #1 #2 #3 {Phys. Rev. {\bf#1},\ #2 (#3)}
\def\PRL #1 #2 #3 {Phys. Rev. Lett. {\bf#1},\ #2 (#3)}
\def\RMP #1 #2 #3 {Rev. Mod. Phys. {\bf#1},\ #2 (#3)}
\def\NIM #1 #2 #3 {Nuc. Inst. Meth. {\bf#1},\ #2 (#3)}
\def\ZPC #1 #2 #3 {Z. Phys. {\bf#1},\ #2 (#3)}
\def\EJPC #1 #2 #3 {E. Phys. J. {\bf#1},\ #2 (#3)}
\def\IJMP #1 #2 #3 {Int. J. Mod. Phys. {\bf#1},\ #2 (#3)}
\def\JHEP #1 #2 #3 {J. High En. Phys. {\bf#1},\ #2 (#3)}
\def\f{\frac}
\def\gt{\tilde g}
\def\ct{{\tilde c}_\theta }
\def\st{{\tilde s}_\theta }
\def\sb{s_\beta }
\def\cb{ c_\beta }
\def\eps{\epsilon}

\def\s{s_\theta }
\def\c{c_\theta }

The LHC is expected to probe directly possible new physics (NP)
beyond the Standard Model (SM) up to a scale of a few TeV. While
its data should provide answers to several of the major open
questions in the present picture of elementary particle physics,
it is important to start examining how this sensitivity can be
further extended at a next generation of colliders. 
Today we have a number of indications that NP could be of
supersymmetric nature. 
If this is the case, the LHC will
have a variety of signals to discover these new particles, and a
linear collider will be required to complement the probe of
the supersymmetric spectrum with detailed measurements, as discussed in 
the previous chapter. However, beyond
supersymmetry, there is a wide range of other scenarios invoking
new phenomena at, and beyond, the TeV scale. They are aimed at
explaining the origin of electroweak symmetry breaking, if
there is no light elementary Higgs boson, at stabilizing the SM,
if supersymmetry is not realized in nature, or at embedding the SM in a
theory of grand unification.

\section{Extra Dimensions}

In the last few years there has been considerable interest in the
possibility that new spatial dimensions can be observed at
high-energy colliders. One interesting theoretical option is that
only gravity propagates in the extra dimensions, whilst gauge and matter
fields are confined on a three-dimensional brane~\cite{add,RS}.
This hypothesis is motivated by the hierarchy problem and naturally
realized in string theory. Two different kinds of scenarios have been proposed:
those in which the space metrics are factorizable~\cite{add} and
those in which they are non-factorizable~\cite{RS}. Their relative
phenomenological consequences are quite distinct. When discussing
the experimental signals of these models, one should bear in mind
that the theory is non-renormalizable, with an unknown ultraviolet
completion. It is then necessary to distinguish various kinematical
regimes, which can be tackled with different theoretical tools.

We first consider the {\it cisplanckian region}, in which the centre-of-mass
energy of the collision is much smaller than the fundamental quantum-gravity
scale $M_D$. We can then describe graviton production using an effective
theory obtained by linearizing the Einstein action. In the case of
factorizable extra dimensions, we obtain a nearly-continuous spectrum
of Kaluza--Klein (KK) graviton excitations, which interact very weakly
and give rise to experimental missing-energy signatures.

In the case of non-factorizable metrics, the graviton KK excitations
have mass gaps of the order of the fundamental scale $M_D$ and they
have a sizeable interaction with ordinary matter, leading to
characteristic resonant production at high-energy collisions. A study
of the predictions for CLIC and the comparison with the LHC reach
is presented in Section~1.2.

The effective theory, valid at energies below $M_D$, also contains the
most general set of gauge-invariant higher-dimensional operators.
The coefficients of these operators cannot be computed without knowledge
of the short-distance theory. Therefore, a general experimental study of
contact interactions will be very useful to gain information about
the underlying theory. One can perform a more restrictive (but more
predictive) analysis by considering only those operators generated by
graviton exchange. At tree level, a single dimension-8 operator is
generated. This operator contributes to a variety of processes at CLIC.
The most interesting is probably $e^+e^-\to \gamma \gamma$, since no
other local operator of lower dimension can generate this process.
Therefore, the study of diphoton final states can be of particular 
importance
in probing new gravitational effects. The exchange of gravitons at the 
loop
level generates a single dimension-6 operator, which is the product
of two flavour-universal axial currents. In this case, the relevant
final state is dilepton or dijet. We do not pursue here an analysis
of these processes, but we refer the reader to the later section on 
contact interactions.

At centre-of-mass energies of the order of $M_D$, we are entering the
{\it Planckian region}, where the effective theory approach discussed
above is no longer valid. This is the most interesting region from the
experimental point of view, since new phenomena related to quantum
gravity and string theory will fully manifest themselves. However,
it is also the most
difficult region to tackle theoretically, in the absence of a complete
description of the short-distance behaviour of gravity.
It is however
easy to predict that, in case of a discovery at the LHC, the r\^ole
of a multi-TeV linear collider will be crucial to disentangle
the new physics signals, leading to new understanding of the underlying
theory.

In the {\it transplanckian region}, where $\sqrt{s}\gg M_D$, we can
give a reliable description of the scattering process (at least under
certain kinematical conditions), since semiclassical physics give the
dominant effects. At impact parameters larger than the Schwarzschild
radius, gravitational elastic scattering is computable in
the eikonal approximation. This gives a prediction for Bhabha scattering
at small angles which can be compared with observations at CLIC.
When the impact parameter is of the order of the Schwarzschild
radius or smaller, strong classical-gravity effects complicate the
theoretical description, but there are plausible arguments that
suggest that black holes are formed. If this is the case, any purely
short-distance interactions from quantum gravity will be masked by
the black hole, as the collision energy is increased. We also
briefly discuss some of the implications of transplanckian physics
at~CLIC. 

So far we have considered the case in which gauge and matter fields
are strictly four-dimensional. It is not implausible that, within
the scenarios considered above, also SM fields have extra-dimensional
excitations, although different in nature from those of the gravitons.
We consider such possibility in Section~1.4, where we describe the
present bounds on KK excitations of SM fields and future prospects
for discovery at CLIC. To emphasize the importance of the complementarity
r\^ole of a multi-TeV linear collider with respect to the LHC, we
also discuss in Section~1.4 how studies at CLIC can distinguish between the
surprisingly similar collider signals of KK excitations of SM particles from
those of a supersymmetric model with a nearly degenerate superpartner
spectrum.

\subsection{ADD Type of Extra Dimensions}
 
The first model we consider is that of Arkani-Hamed, Dimopoulos and 
Dvali (ADD)~\cite{add}; we limit our discussion to 
the case of graviton tower exchange in $e^+e^-\to f\bar f$. 
The effect of summing the KK gravitons is to produce a set of effective 
dimension-8 operators of the form $\sim \lambda T^{\mu\nu}T_{\mu\nu}/M_s^4$, 
where $T_{\mu\nu}$ is the stress-energy tensor of the SM matter exchanging 
the tower{\cite {pheno}}. 
This approximation only applies in the limit that the centre-of-mass 
energy of the collision process lies sufficiently below the cut-off 
scale $M_s$, which is of the order of the size of the Planck scale in the 
extra-dimensional space. In the convention used by Hewett~\cite{pheno}
and 
adopted here, the contribution of the spin-2 exchanges can be universally 
expressed in terms of the scale $M_s$ and a sign $\lambda$.
Current experimental constraints 
from LEP and the Tevatron{\cite {greg}} tell us that $M_s \geq 1$ TeV for 
either sign of $\lambda$; values of $M_s$ as 
large as the low 10's of TeV may be contemplated in this scenario.

\begin{figure}[t] 
\centerline{
\includegraphics[width=4.7cm,angle=90]{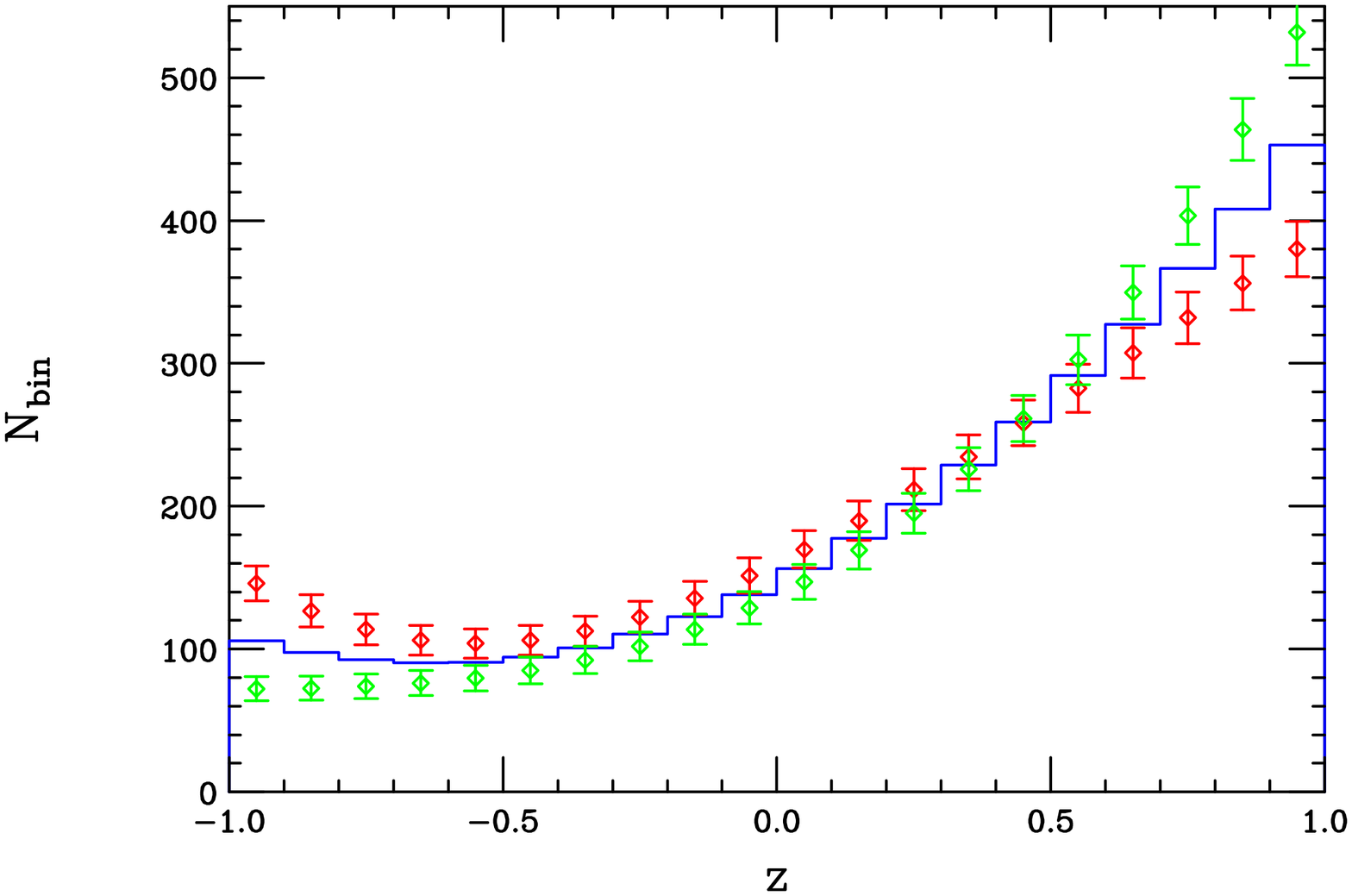}
\hspace*{5mm}
\includegraphics[width=4.7cm,angle=90]{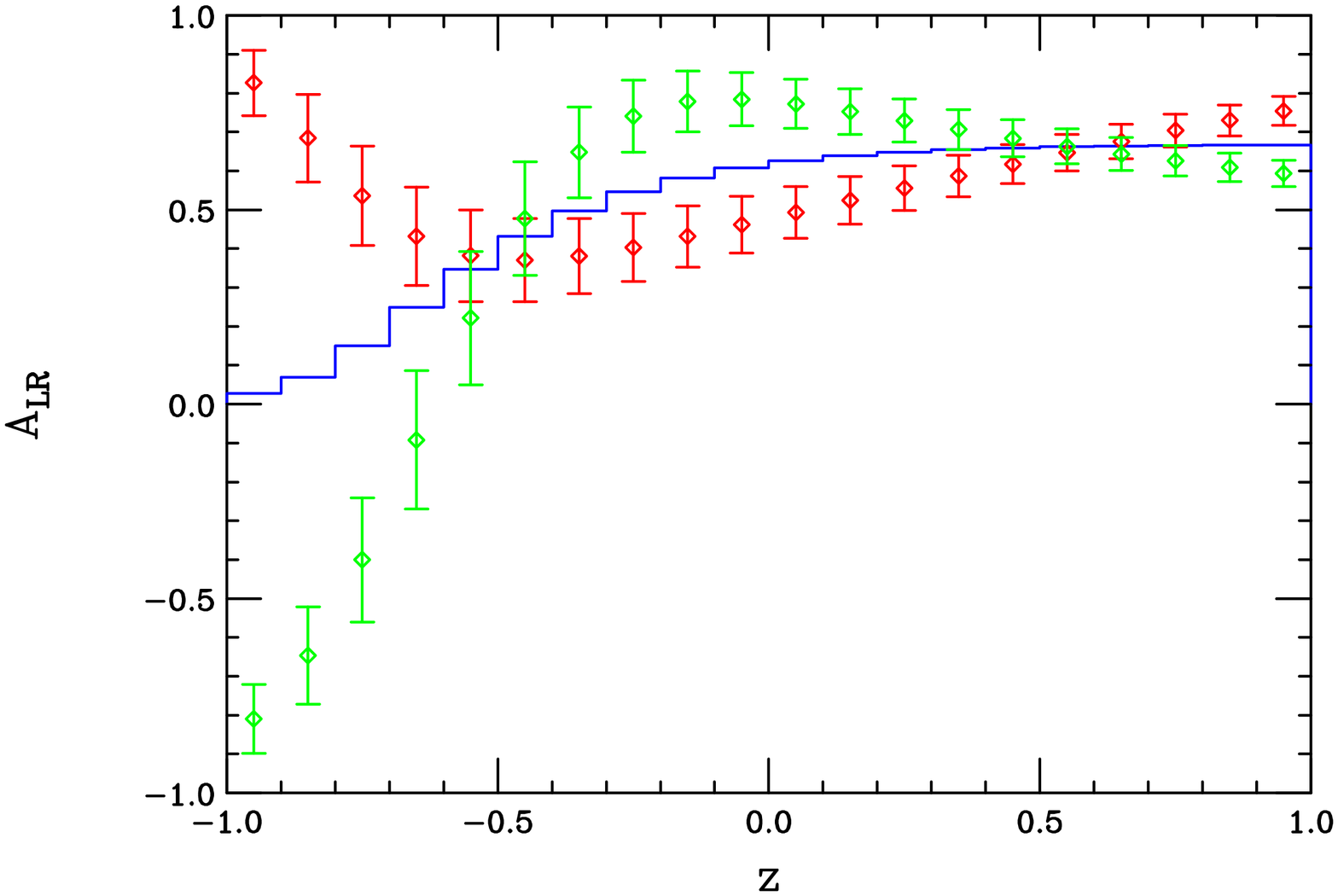}}
\vspace*{-2mm}
\caption{Deviations in the cross section for $\mu$ pairs (left) and 
$A_{\rm LR}$ for $b$ quarks (right) at $\sqrt s$~=~5~TeV for
$M_s$~=~15~TeV in the ADD model for an integrated 
luminosity of 1~${\rm ab}^{-1}$. The SM is represented by the
histogram while the red and green data points show the ADD predictions
with $\lambda=\pm$~1. In both plots $z=\cos \theta$.}
\label{p3-01_fig1}
\end{figure}

\begin{figure}[htbp]
\centerline{
\includegraphics[width=4.7cm,angle=90]{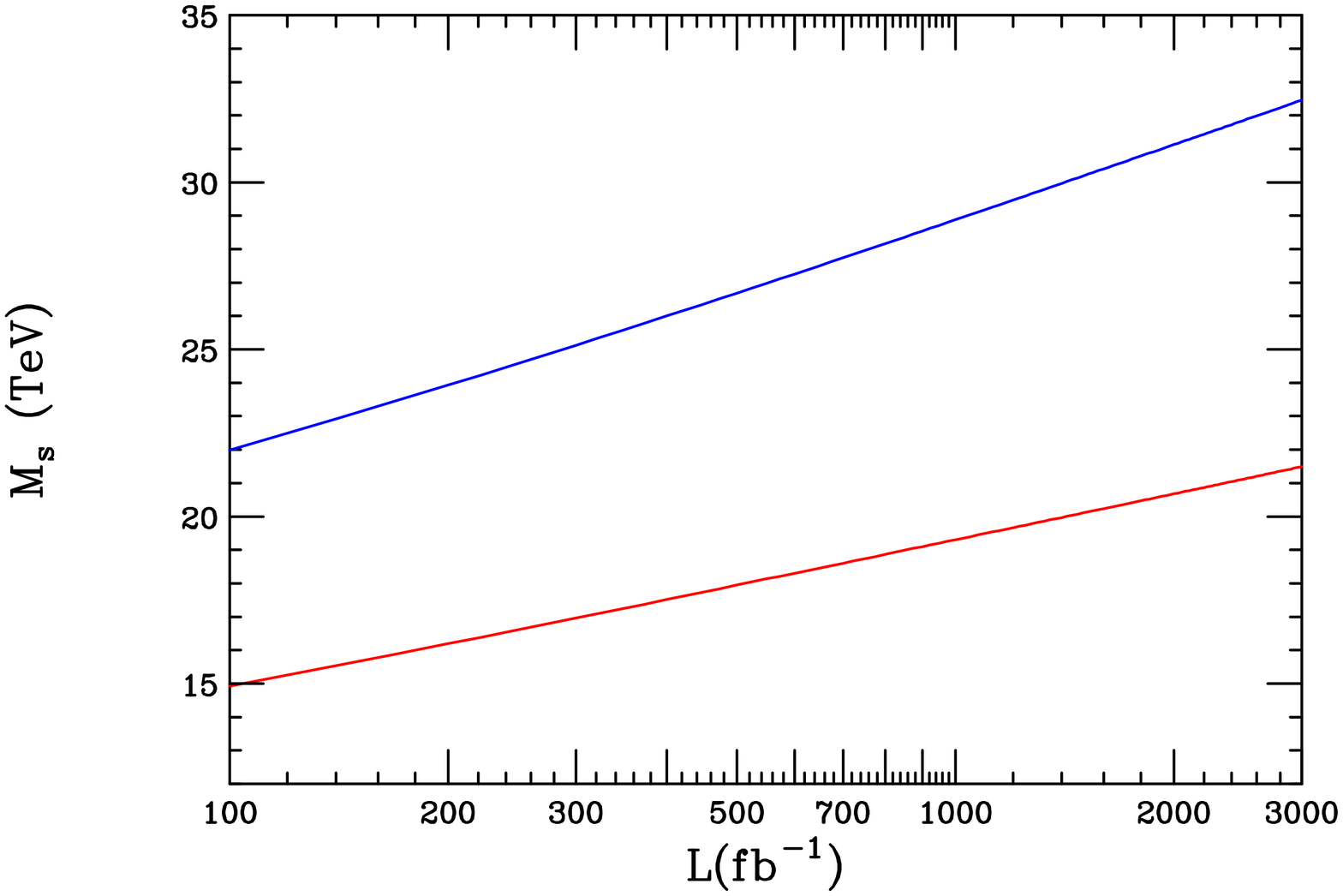}
\hspace*{5mm}
\includegraphics[width=4.7cm,angle=90]{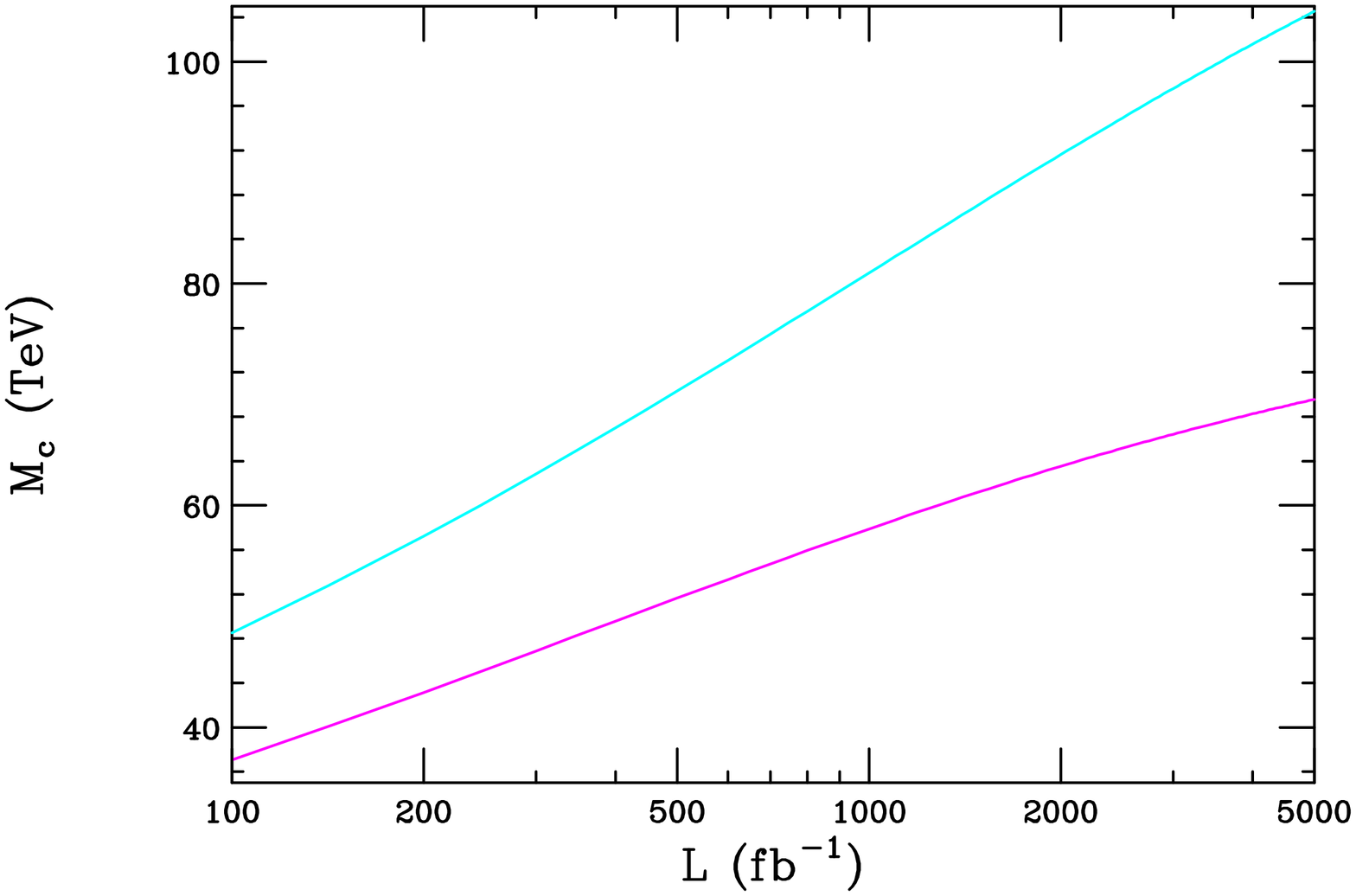}}
\caption{(Left) Search reach for the ADD model scale $M_s$ at CLIC as a 
function of the integrated luminosity from the set of processes $e^+e^-\to 
f\bar f$, assuming $\sqrt s$~=~3 (bottom) or 5 (top)~TeV. Here 
$f=\mu,\tau,b,c$, $t$, etc. (Right) Corresponding reach for the
compactification scale of the KK gauge bosons in the case of one extra
dimension and all fermions localized at the same orbifold fixed point.}
\label{p3-01_fig2}
\end{figure}

In the case of $e^+e^-\to f\bar f$, the addition of KK tower exchange leads to 
significant deviations in differential cross Sections and polarization 
asymmetries from their SM values, which are strongly dependent 
on both the sign of $\lambda$ and the ratio $s/M_s^2$. 
Such shifts are observable in final states of all flavours. In addition, the 
shape of these deviations from the SM with varying energy and 
scattering angle, as shown by Hewett~\cite{pheno}, tells us that the 
underlying physics arises from dimension-8 operators and not, for example, 
$Z'$ exchange. 
Figure~\ref{p3-01_fig1} shows an example of how such deviations from the SM 
might appear at a 5-TeV CLIC in the case that $M_s$~=~15~TeV for either sign 
of $\lambda$. 
The indirect search reach for the scale $M_s$ can be obtained by
combining the data for several of the fermion final states 
($\mu, \tau, c, b, t$,~etc.) in a single overall fit. The result of
this analysis for CLIC is the $\lambda$-independent bound shown in 
Fig.~\ref{p3-01_fig2} as a function of the integrated luminosity for 
$\sqrt s$~=~3 or 5~TeV. For an integrated luminosity of 1 ${\rm ab}^{-1}$ we
see that the reach is $M_s\simeq 6\sqrt s$, which is consistent with
analyses at lower-energy machines~\cite{pheno}. 

\begin{figure}[htbp] 
\begin{center} 
\epsfig{file=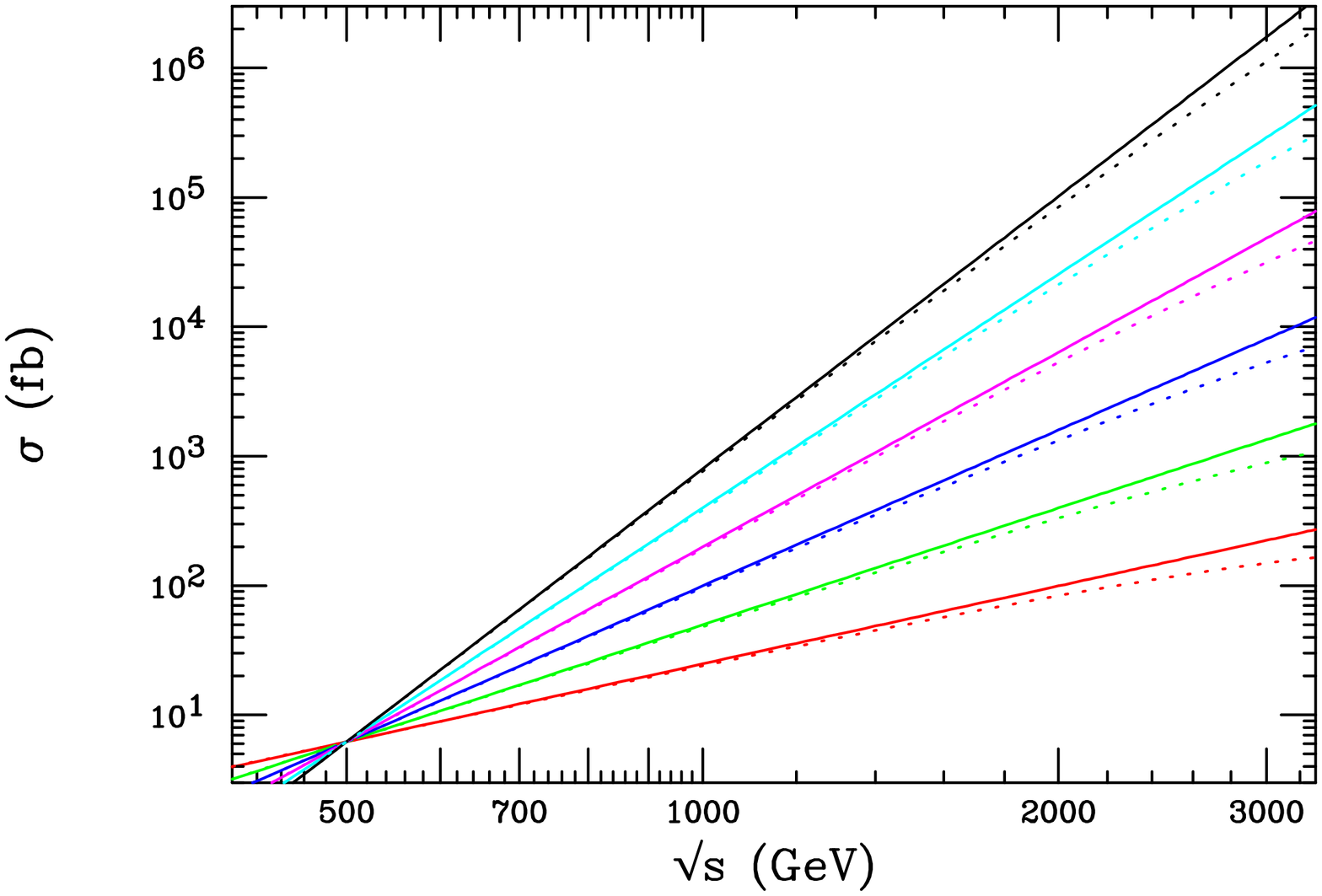,bbllx=0,bblly=100,bburx=540,bbury=700,width=5.8cm,angle=90}
\hspace*{11mm}
\epsfig{file=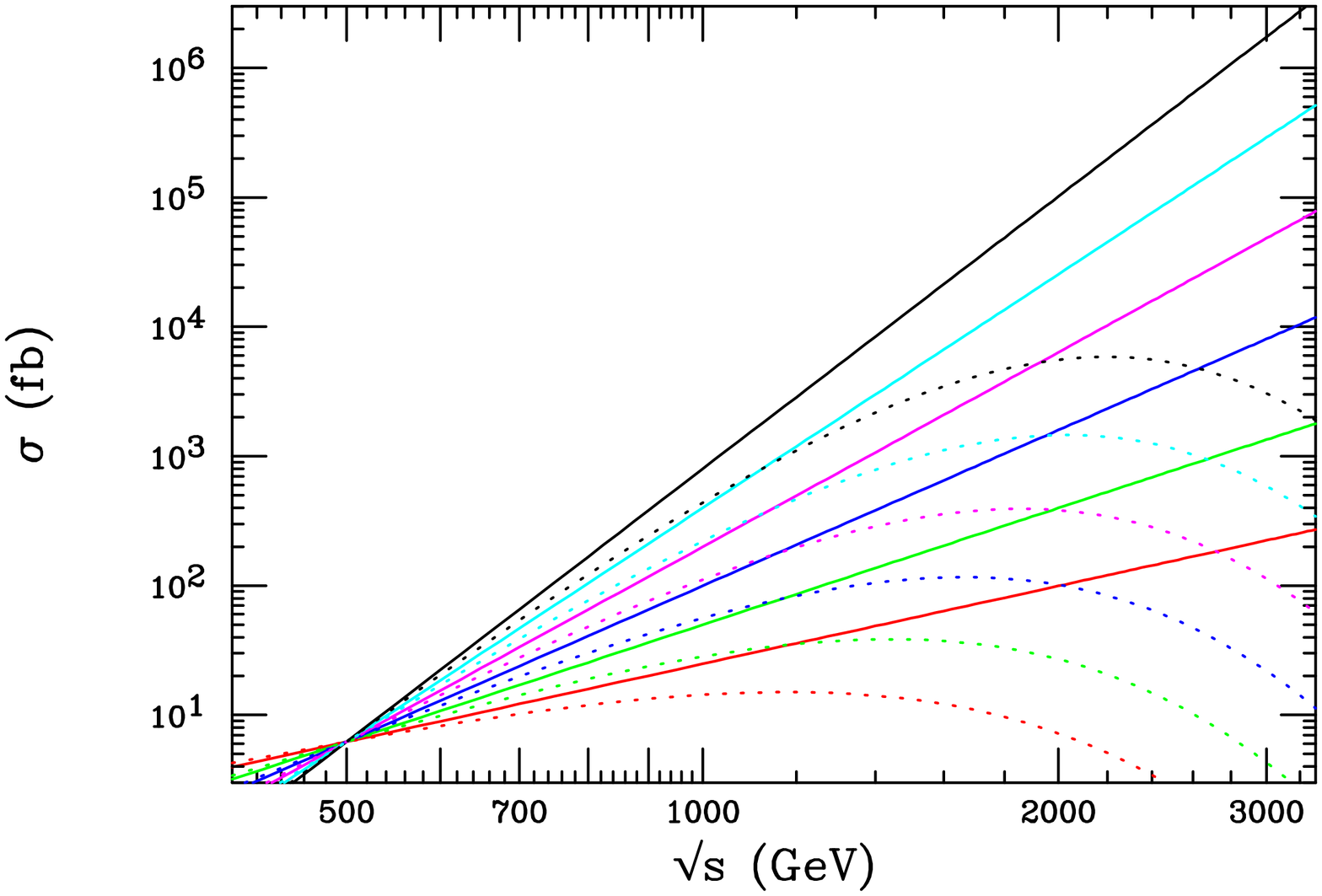,bbllx=0,bblly=100,bburx=540,bbury=700,width=5.8cm,angle=90}

\vspace*{-4mm}

\caption{The cross section for $e^+e^- \rightarrow G\gamma$ with cuts as
described in the text, for rigid (solid lines) and soft (dashed lines)
branes. The different curves correspond, from bottom to top, to different
numbers of extra dimensions $\delta$~=~2, 3, 4, 5, 6 and 7.
Left, for $\Delta$~=~4~TeV, right for $\Delta$~=~1~TeV.}
\label{fig3}
\end{center}
\end{figure}

Scenarios with large extra dimensions have been studied for 
TeV-class linear colliders, e.g.
TESLA~\cite{Aguilar-Saavedra:2001rg}, and are searched for typically
in the channel $e^+e^- \rightarrow \gamma +G$.
One of the acclaimed advantages of a LC with respect to the LHC is that, by 
measuring the $\gamma G$ cross section at different centre-of-mass 
energies, one can disentangle the Planck scale and the number of extra
dimensions $\delta$ simultaneously, as is shown for CLIC 
in~Fig.~\ref{fig3} by the solid lines. The cross sections are
calculated for cuts similar to those in~\cite{Aguilar-Saavedra:2001rg}:
$\sin \theta_{\gamma}>$~0.1, $p^{\gamma}_t >$~0.06~$E_{\rm beam}$
and $x_{\gamma} <$~0.65. The cross sections are normalized in such a
way that 
for each value of $\delta$, the scale $M_D$ is chosen to give the
same cross section at 500~GeV. 

These predictions assume, however, that lower-dimensional physics is 
attached to rigid branes. 
Allowing for flexible branes instead~\cite{wells} introduces a new dependence
on a  parameter $\Delta$, the softening scale, which is  related to the brane 
tension. The dashed lines in Fig.~\ref{fig3} show the effect of $\Delta$ for 
values of 4~TeV  and 1~TeV, respectively.

For a brane tension of 4~TeV, the effect on the cross section is rather 
small.
A collider in the range of 0.5--1~TeV would not be sensitive to
the effect and thus $M_D$ and $\delta $ can still be 
disentangled. However,  at CLIC the 
cross sections are 30--40\% lower  than expected, allowing
one to observe the softness of the brane.
For a brane of 1~TeV tension the effect is more spectacular.
For the example given here, a lower-energy
LC would get fooled when measuring cross sections only
at 0.5~TeV and 1~TeV, and  
extract wrong values of $\delta$ and $M_D$. Extending the range in the 
multi-TeV region will again allow this effect to be observed in its full 
drama. The cross section of the background channel
$e^+ e^- \rightarrow e^+ e^- \gamma$ with the cuts as listed above
is 16~fb at 3~TeV, which sets the scale for the detectability of a signal:
for $\delta $~=~2 or 3 the signal event rate at 3~TeV gets too small
for such a soft brane scenario.
  
\subsection{Graviton Production at CLIC: Randall--Sundrum Model}

In the extra-dimension scenario proposed by
Randall and Sundrum (RS)~\cite{RS}  the hierarchy between the Planck
and the electroweak scale is generated by an exponential function
called the `warp factor'. This model predicts KK graviton
resonances with weak-scale masses and strong couplings to matter.
Hence the production of~TeV-scale graviton resonances is
expected  in two-fermion channels~\cite{dhr}. In its simplest
version, with two branes and one extra dimension, and where all
of the SM fields remain on the brane, the model has two
fundamental parameters: the mass of the first KK state
$m_1$, and the parameter $c= k/M_{\overline{Pl}}$, where $k$ is
related to the curvature of the 5D space and
$M_{\overline{\rm Pl}}$ is the 4D effective Planck scale. The
parameter $c$ controls the effective coupling strength of the
graviton and thus the width of the resonances, and should be less
than 1 but yet not too far away from unity.

\begin{figure}[htbp]
\begin{center}
\epsfig{file=Chap6/e3_3022_fig1a.ps,bbllx=90,bblly=90,
bburx=550,bbury=720,width=6.5cm,angle=90,clip}
\epsfig{file=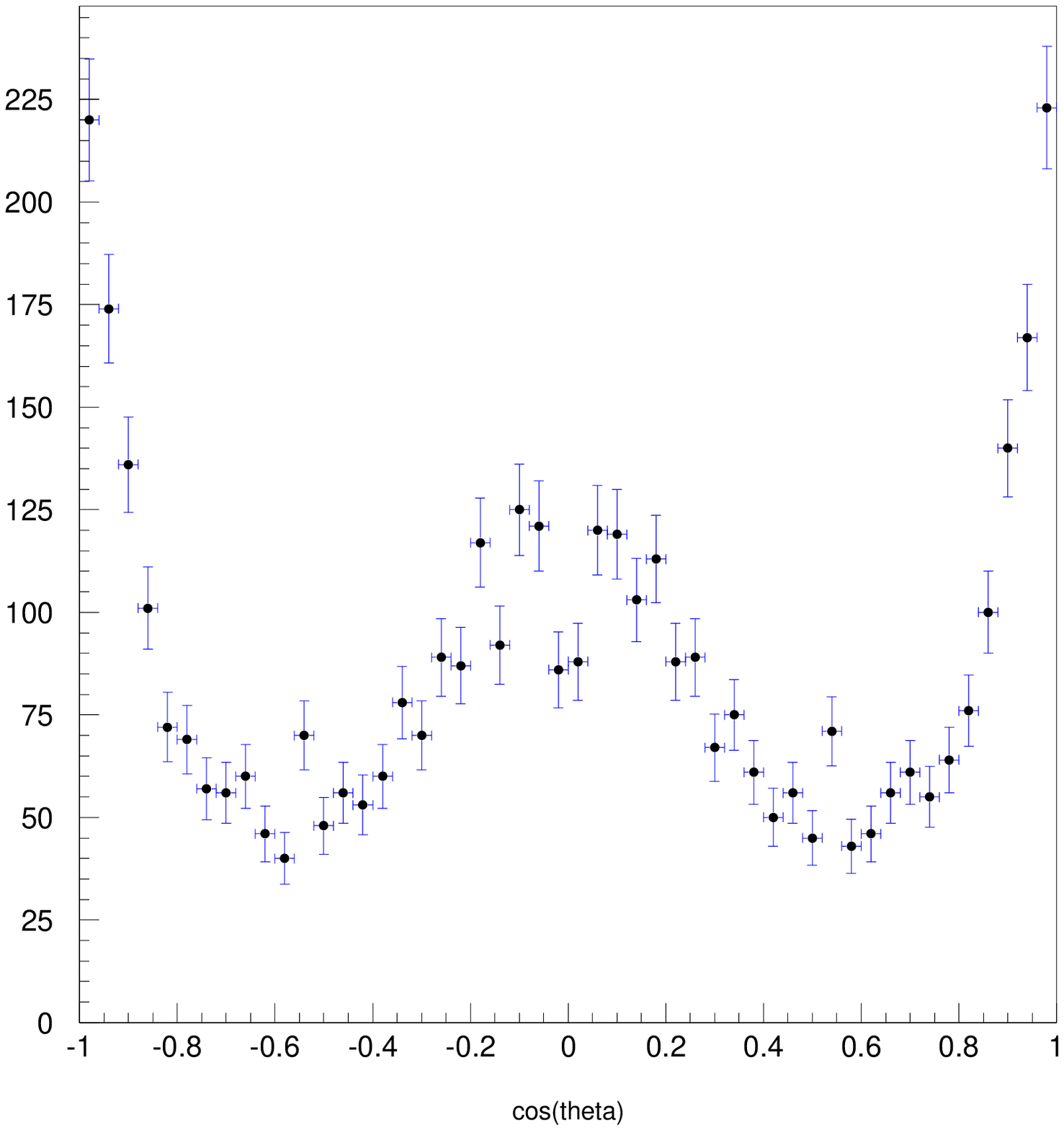,bbllx=0,bblly=20,
bburx=565,bbury=565,width=6.7cm}

\vspace*{2mm}

\caption{ Left: KK graviton excitations in the RS model produced
in the process $e^+e^-\to \mu^+\mu^-$. From the most narrow to
widest resonances, the curves are for 0.01~$<c<$~0.2. Right:
Decay-angle distribution of the muons from
 $G_3$ (3200~GeV) $\to \mu \mu$.} 
\label{fig:kk1}
\end{center}
\end{figure}

The resulting  spectrum for $e^+e^-\rightarrow \mu^+\mu^-$ is
shown in Fig.~\ref{fig:kk1}. The cross sections are huge and the
signal cannot be missed at a LC with sufficient centre-of-mass energy. If
such resonances are observed --- perhaps first by the LHC in the
range of a few~TeV --- it will be important to establish the nature
of these newly produced particles, i.e. to measure their
properties (mass, width and branching ratios) and quantum numbers
(spin). Note that  the mass $m_1$ of the first resonance
determines the resonance pattern: the masses of all higher-mass
resonances are then fixed.

The signal for one KK resonance ($G_1$) is implemented in PYTHIA
6.158~\cite{pythia} via process 41.  For the study,
PYTHIA has been extended to include  two
more resonances ($G_2, G_3$, corresponding to processes 42 and
43) to allow a check on the measurability of the graviton
self-coupling. The decay branching ratios  of these resonances
were modified according to~\cite{dhr,rizzo3}. In particular, the
gravitons can decay into two photons in about 4\% of the cases, a
signature that would distinguish them from, for example, new heavy $Z'$
states~\cite{snow}.

The resonance spectrum was chosen such that the first resonance
$G_1$ has a mass around 1.2~TeV, just outside the reach of a~TeV-class LC, 
and consequently the mass of the third resonance $G_3$
will be around 3.2~TeV, as shown in Fig.~\ref{fig:kk1}. The $\sqrt{s}$
energy for the $e^+e^-$ collisions of CLIC was taken to be 3.2~TeV
in this study. Mainly the muon and photon decay modes of the
graviton have been studied. The  events used  to reconstruct the
$G_3$ resonance signal were selected via either two muons or two
$\gamma$'s with $E>$~1200~GeV and $|\cos\theta| <$~0.97. The
background from overlaid two-photon events --- on average four
events per bunch crossing --- is typically important only for
angles below 120 mrad, i.e. outside the signal search
region considered.

First we study the precision with which one can measure the shape,
i.e. the  $c$ and  $M$ parameters, of  the observed  new
resonance. A scan similar to that of the $Z$ at LEP was made
for an integrated luminosity of  1 ab$^{-1}$. The precision with
which the cross sections are measured allows one to determine $c$ to
0.2\% and $M$ to better than 0.1\%.

Next we determine some key  properties of the new resonance:
the spin and the branching ratios.
The graviton is a spin-2 object, and Fig.~\ref{fig:kk1} shows the
decay angle of the  fermions $G\rightarrow \mu\mu$ for the $G_3$
graviton, obtained using PYTHIA/SIMDET for 1~ab$^{-1}$ of data,
including the CLIC machine background.
The typical spin-2 structure of the decay angle
of the resonance is clearly visible.

For gravitons as proposed in~\cite{dhr,rizzo3} one expects 
$BR(G \rightarrow \gamma\gamma)/BR(G \rightarrow \mu\mu)$~=~2. With
the present SIMDET simulation we get efficiencies in the mass peak
($\pm~200$ ~GeV) of 84\% and 97\% for detecting the muon and
photon decay modes, respectively. With cross sections of
O(1~pb), $\sigma_{\gamma\gamma}$ and $\sigma_{\mu\mu}$
can be determined to better than a per cent. Hence the ratio
$BR(G\rightarrow \gamma\gamma)/BR(G\rightarrow \mu\mu)$ can be
determined to an accuracy of 1\%  or better.

\begin{figure}[t] 
  \begin{center}
    \mbox{ \psfig{file=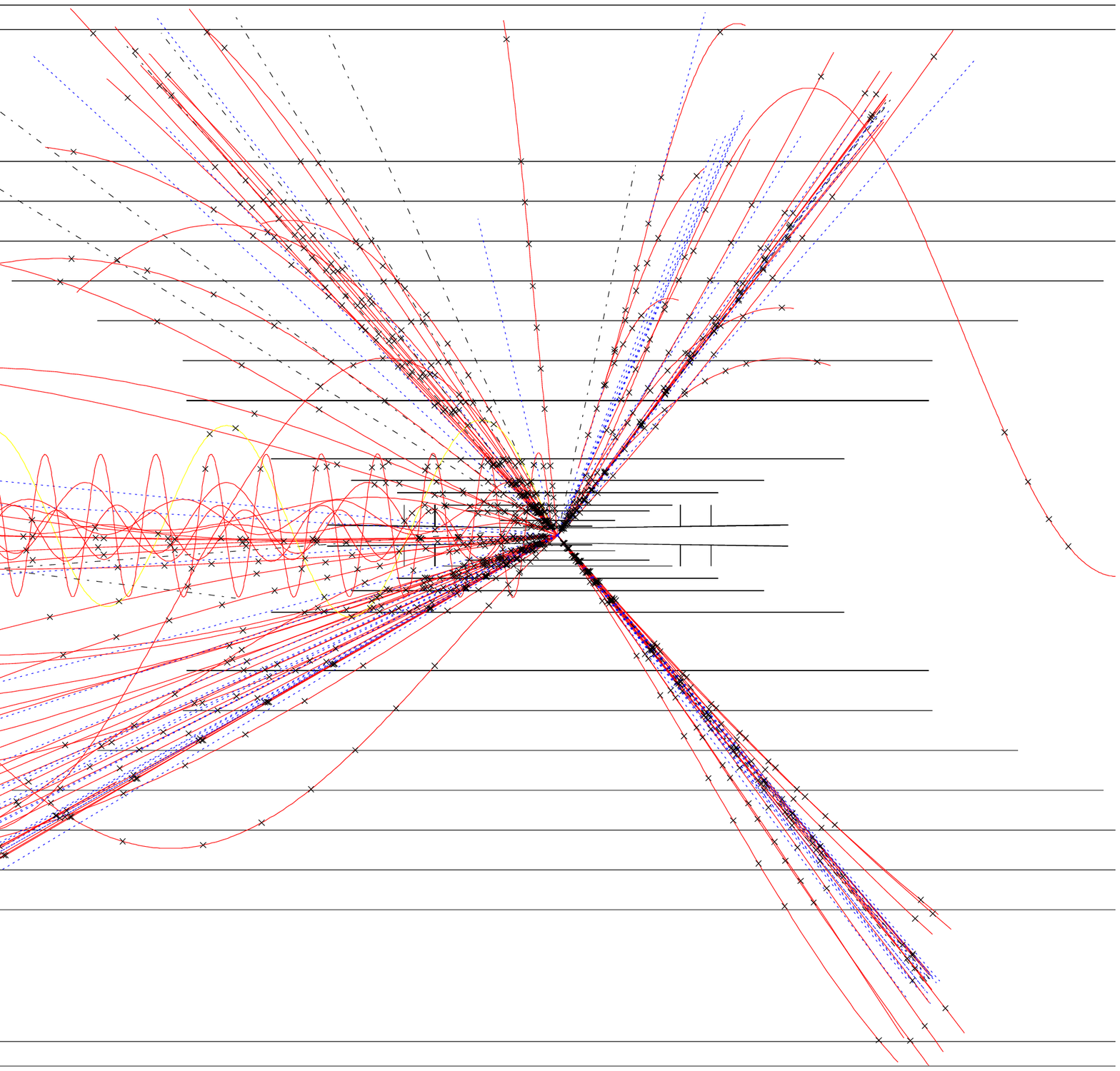,width=0.4\textwidth} 
\hspace*{11mm}
           \psfig{file=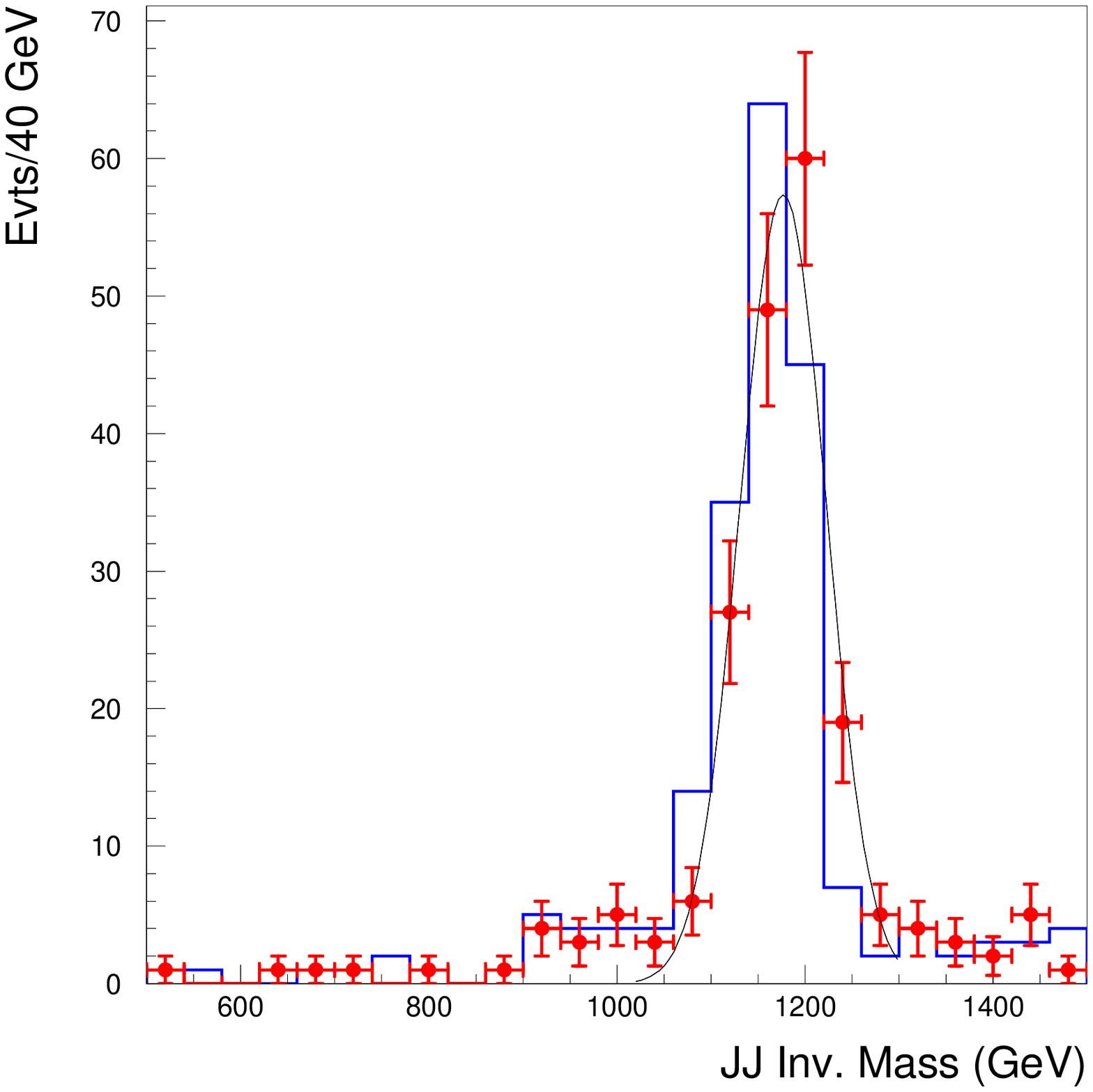,width=0.43\textwidth} }
\caption{Left: Event in a CLIC central detector with the decay
$G_3\rightarrow G_1G_1 \rightarrow$ four jets. Right: Invariant
jet--jet mass  of $G_1$(1200~GeV)
produced in $G_3\rightarrow G_1G_1$ and $G_1  \rightarrow$ two jets.}
\label{fig:kk3}
\end{center}
\end{figure}

Finally, if the centre-of-mass energy of the collider is large enough to
produce the first three resonance states, one has the intriguing
possibility to measure the graviton self-coupling via the
$G_3\rightarrow G_1G_1 $ decay~\cite{rizzo3}. The dominant decay
mode will be $G_1 \rightarrow gg$  or $q\bar q$ giving a two-jet
topology. Figure~\ref{fig:kk3} shows the resulting spectacular
event signature of four jets of about 500~GeV each in the
detector (no background is overlaid). These jets  can be used to
reconstruct $G_1$. Figure~\ref{fig:kk3} shows the reconstructed
$G_1$ invariant mass. The histogram does not include the background,
while the dots include 10 bunch crossings of background overlaid
on the signal events. Hence the mass of  $G_1$ can be well
reconstructed and is not significantly distorted by the
$\gamma\gamma$ background.

In summary, a multi-TeV collider such as
CLIC will allow a precise determination of the shape and 
mass of the new resonance(s), and of its spin. In particular, for
the RS model it was shown that the key discriminating properties
of these resonances can be reconstructed, and the underlying model
parameters can be determined precisely.

\subsection{Transplanckian Scattering}

Elastic collisions in the transplanckian region, where the
centre-of-mass energy $\sqrt{s}$ is much larger than the
fundamental gravity mass scale $M_D$, can be described by
linearized general relativity and known quantum-mechanical
effects, as long as the momentum transfer in the process is
sufficiently small. Therefore, the interesting observable at CLIC
is Bhabha scattering at small deflection angle. The relevant
cross section can be computed using the eikonal approximation, in
the kinematical regime in which the impact parameter is smaller
than the Schwartzschild radius 

$$ 
R_S=\frac{2\sqrt{\pi}}{M_D}\left[ \frac{\Gamma \left(
\frac{\delta +3}{2} \right) }{2\pi^2(\delta
+2)}\frac{\sqrt{s}}{M_D} \right]^{\frac{1}{\delta +1}}. 
$$
At small momentum transfer $t$, the gravitational-scattering
differential cross section is 
$$\frac{d\sigma}{dt}= C_\delta
\frac{\sigma_{BH}}{s}\left(\frac{s}{M_D^2}
\right)^{\frac{(\delta+2)^2}{\delta(\delta +1)}},~~~~\sigma_{BH}=
\pi R_S^2\,, 
$$
where $C_\delta$ is a $\delta$-dependent coefficient. 
With large
gravity-induced cross sections ($\sim$~10$^2$--10$^3$~pb) for all
$M_D/\sqrt{s}\ll$~1, combined with relatively low background, CLIC
could certainly provide very important tests of extra-dimensional
theories. The maximum values of $M_D$ that can be studied at CLIC
in the transplanckian gravitational scattering  are uniquely
determined by the condition for validity of the eikonal
approximation ($\sqrt{s}\gg M_D$). Therefore very high energies
should be achieved to compete with the LHC. However, the cleaner
$e^+e^-$ environment offers several advantages for precision tests
and parameter determinations.

\begin{figure}[t] 
\begin{center}
\includegraphics[height=8.0cm]{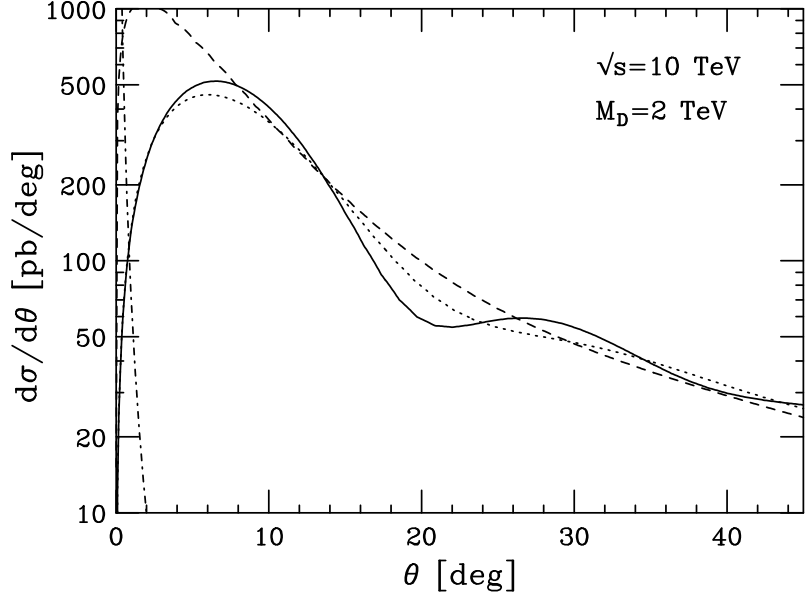}
\end{center}

\vspace*{-4mm}

\caption{Angular distribution of the $e^+e^-\to e^+e^-$ signal for
$M_D$~=~2~TeV and $\sqrt{s}$~=~10~TeV. The solid line is for
$\delta$~=~6, the dotted for $\delta$~=~4 and the dashed for
$\delta$~=~2. The almost-vertical dash-dotted line at the far left of
the figure is the Bhabha scattering rate expected from the Standard
Model.}
\label{clic-theta}
\end{figure}

For our study, we consider centre-of-mass energies as high as
10~TeV, but it is not difficult to rescale our results to
different values of $\sqrt{s}$. In Fig.~\ref{clic-theta} we give
one example of the signal and background distributions in the
scattering angle $\theta$ for $M_D$~=~2~TeV. Here $\theta$ is the
scattering  angle of the electron in the Bhabha process, and
$\theta$~=~0 indicates the electron going down the beam pipe
undeflected by the collision. Notice that the SM background is
completely insignificant as long as we exclude a small region
along the beam direction. This allows experimental studies of the
cross section in a much more forward region than what is possible
at the LHC.

The signal angular distribution is characterized by the peak
structure shown in Fig.~\ref{clic-theta}, with the first peak
approximately given by 
$$\theta^{\rm peak} \simeq \left(
\frac{a_\delta M_D}{\sqrt{s}}\right)^{\frac{\delta+2}{\delta}},
$$
where $a_\delta$ is a numerical coefficient with the values
$a_{2,3,4,5,6}$~=~0.9, 1.2, 1.1, 1.1, 1.0. The $\delta$~=~6
line (solid line) in Fig.~\ref{clic-theta} is the most telling
line, since a careful measurement of the ups and downs of
$d\sigma/d\theta$ would be hard to reproduce in another framework.
Given the good energy and angular resolutions that can be
achieved at a linear collider, CLIC could perform a much more
precise study of the peak structure than what is feasible at the~LHC.

Another advantage of CLIC is that, unlike a $pp$ collider, the
scattering particles are distinguishable.  This means that the
calculable limit of $t\to$~0 (small-angle scattering) is
unambiguously identifiable.  On the other hand, in the $pp$
collider case, the case of partons glancing off each other at
$\hat t\to$~0 is not distinguishable from partons bouncing
backwards at large momentum transfer $\hat t\to -\hat s$.  It is
plausible that these non-calculable large momentum transfer
contributions are negligible, but CLIC can test that assertion.

At large scattering angle, when $t$ approaches $s$, the eikonal 
description breaks down, and non-linear gravitational effects cannot be
neglected. Although a complete calculation has not been performed,
it is plausible to assume that black holes are formed.
If the fundamental scale $M_D$ lies in the~TeV range, the cross section
is expected to be very large, as in the case of the elastic
scattering discussed above: 
$\sigma \simeq \sigma_{\rm BH} \sim$~O(100)~pb~(TeV$/M_D)^2$. 
The lifetime of such a black hole is of order
$\sim$~10$^{-25}$--10$^{-27}$~s, and hence the black hole will
evaporate before it could possibly `attack' any detector material,
certainly causing no safety concern.

\begin{figure}[t] 
\begin{center}
\includegraphics[width=3.0in]{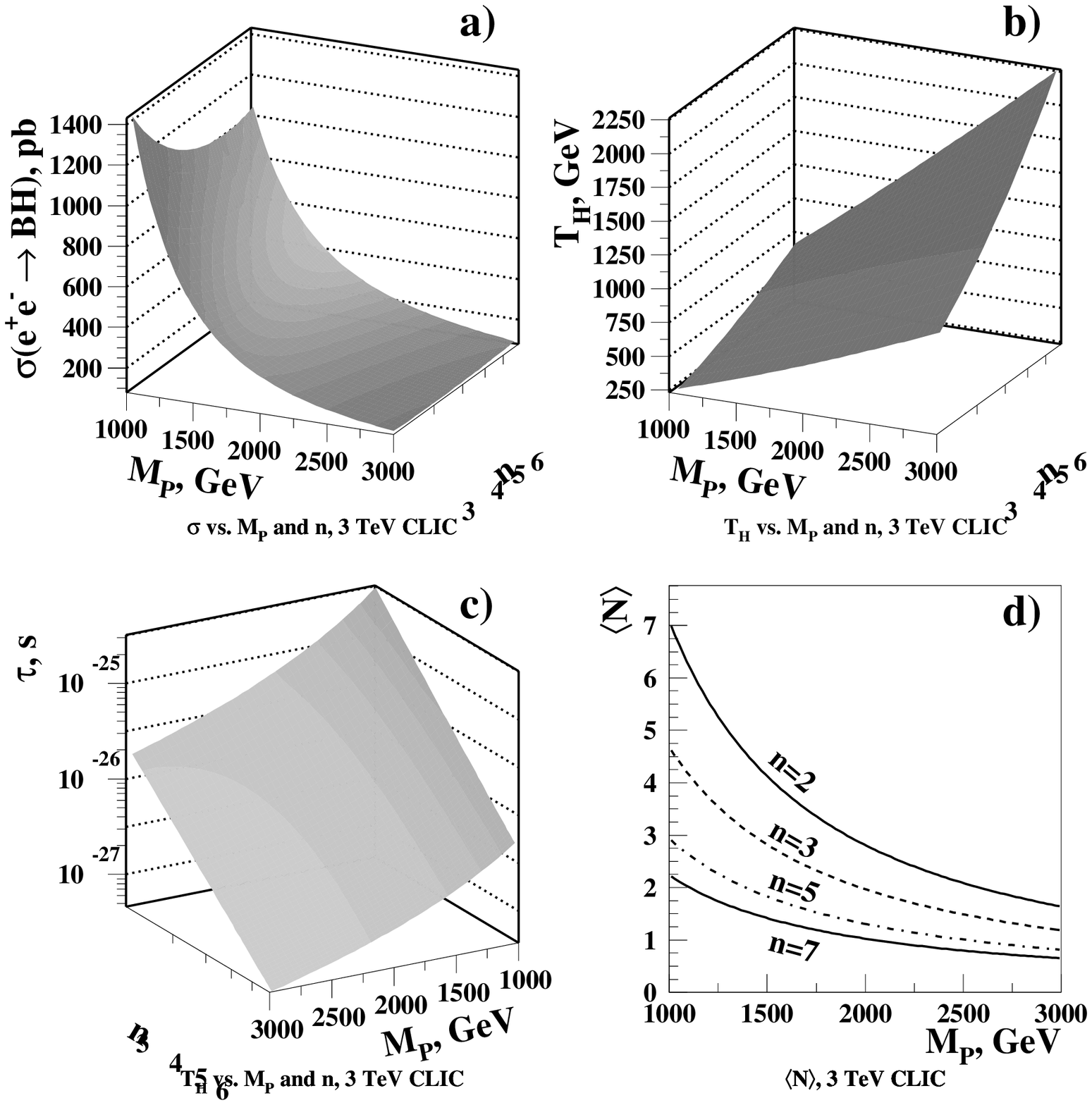}
\includegraphics[width=3.0in]{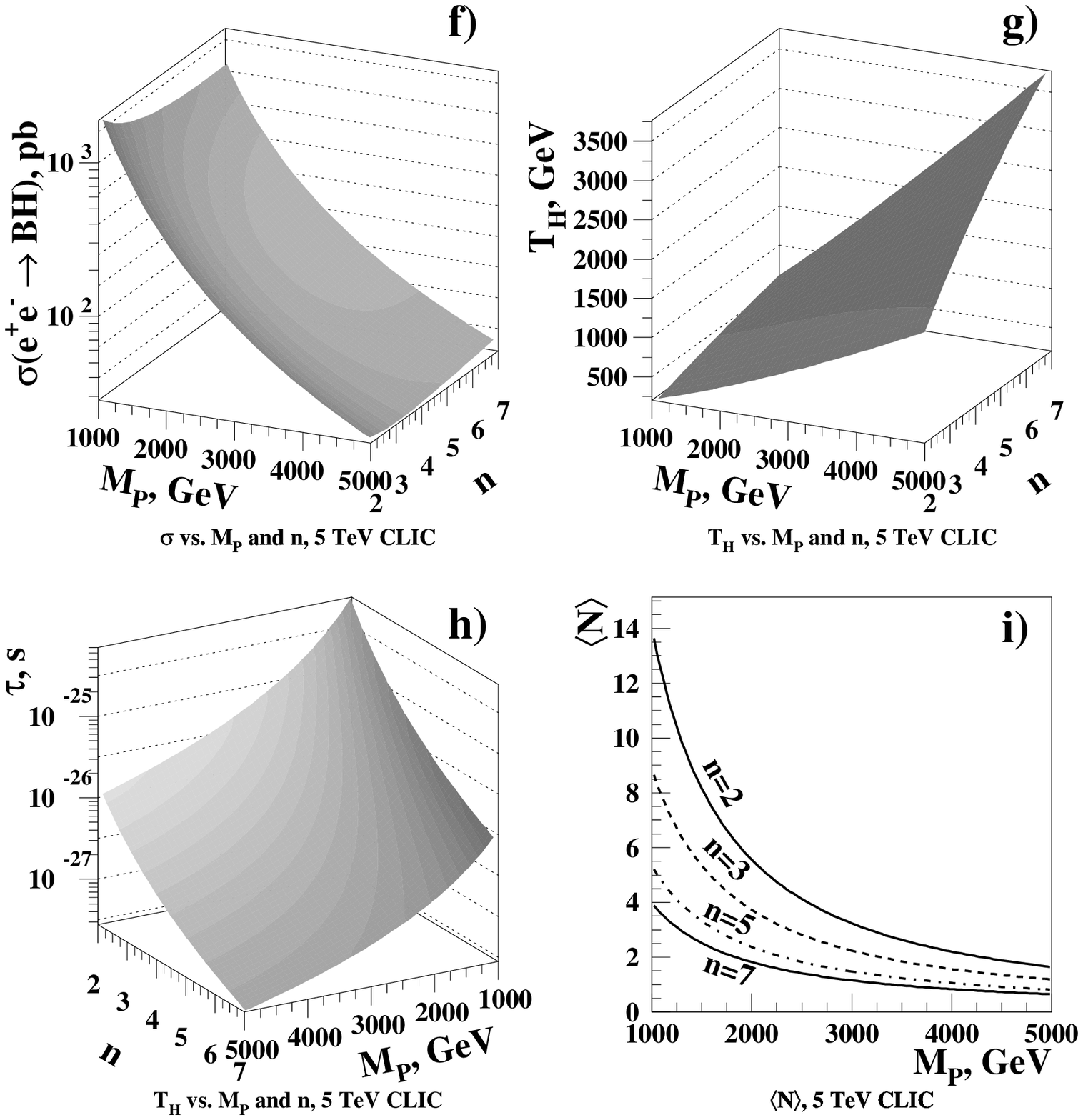}
\medskip
\vspace*{9mm}
\includegraphics[width=3.0in]{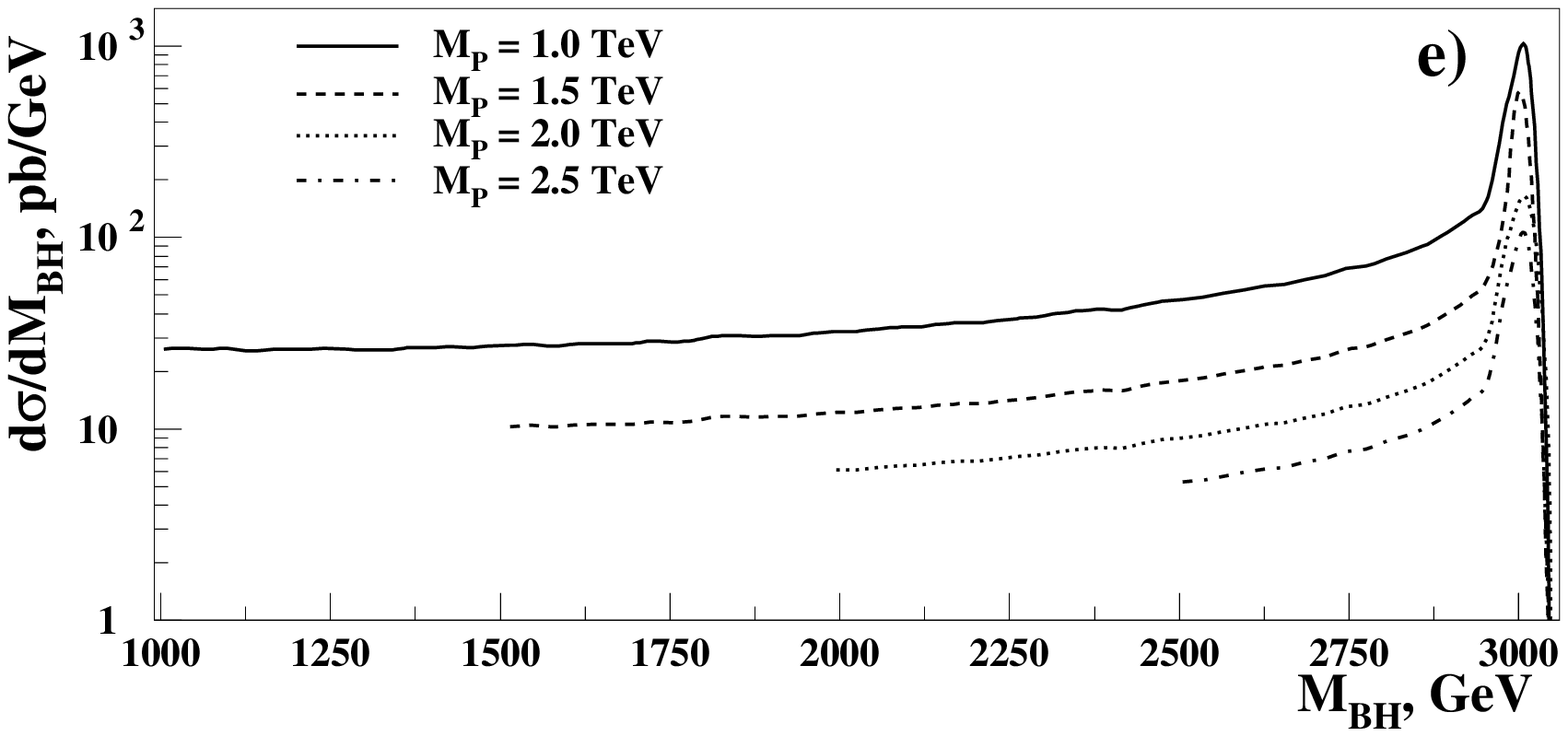}
\includegraphics[width=3.0in]{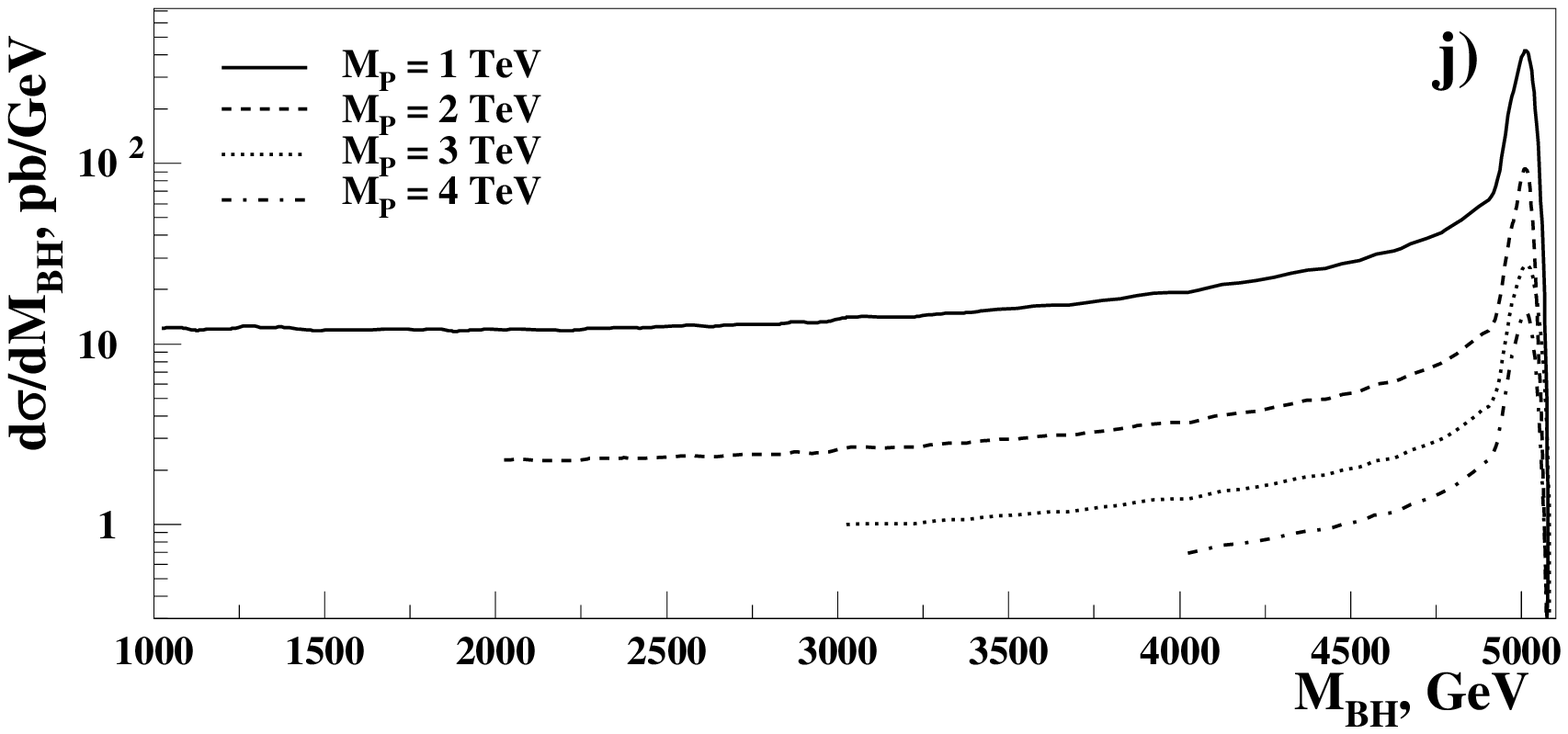}
\end{center}
\caption{Black hole properties at CLIC. Plots a)--d)
and f)--i) correspond to the properties (production cross section,
temperature, lifetime, and average decay multiplicity) of a fixed-mass 3-TeV
and 5-TeV black hole produced at a 3-TeV or 5-TeV machine, respectively.
Plots e), j) show the differential cross section of black-hole production 
for $n$~=~4, as a function of the black-hole mass at a 3-TeV or 5-TeV CLIC 
$e^+e^-$ collider, respectively.}
\label{sec6:landsb1}
\end{figure}

In a first approximation, such a
machine produces black holes of a fixed mass, equal to the energy of the
machine. The total cross section of such a black hole produced at a 3~TeV 
and a 5~TeV machine, as a function of $m_p$ and $n$, is shown in
Fig~\ref{sec6:landsb1}a and Fig.~\ref{sec6:landsb1}f, respectively.
For more elaborated studies of black-hole production at electron 
colliders, one should take into account the machine luminosity
spectrum as discussed in Section 6. Using the beamsstrahlung spectra
for a 3- or 5-TeV CLIC machine,
we show the differential cross section $d\sigma/dM_{\rm BH}$ for black hole
production at a 3- and 5-TeV CLIC machine in Figs.~\ref{sec6:landsb1}e and
Fig.~\ref{sec6:landsb1}j, respectively.

Black hole production at CLIC is complementary to the LHC in many ways, as
the maximum number of black holes produced at CLIC is found at the highest 
accessible
masses. This gives some advantage, as the stringy effects and the
kinematic distortion of the Planck black-body spectrum decrease with the
increase of the black-hole mass. This can be used to extract the 
dimensionality of the extra space, by observing the relationship
between the black-hole horizon temperature ($T_{\rm H}$) vs. the mass
of the black hole, as suggested in~\cite{Dimopoulos:2001hw}.  
This method is less affected at CLIC by unknown stringy and 
kinematic effects. Preliminary studies show
that the statistical sensitivity to the number of extra dimensions and the 
value
of the fundamental Planck scale at CLIC is similar to that at the LHC.

\begin{figure}[t] 
\begin{center}
\includegraphics[height=7.0cm]{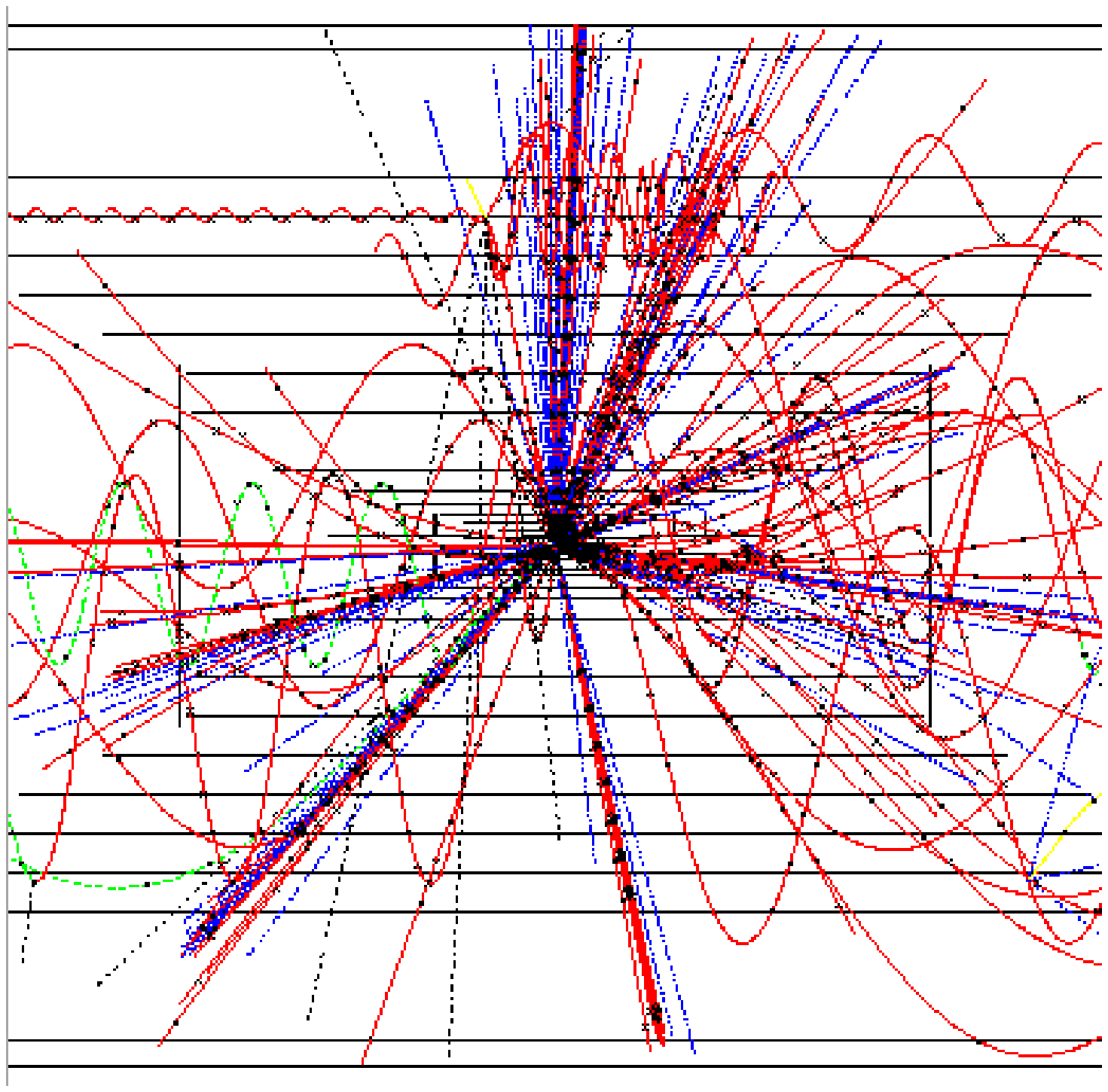}
\hspace*{11mm}
\includegraphics[height=7.0cm]{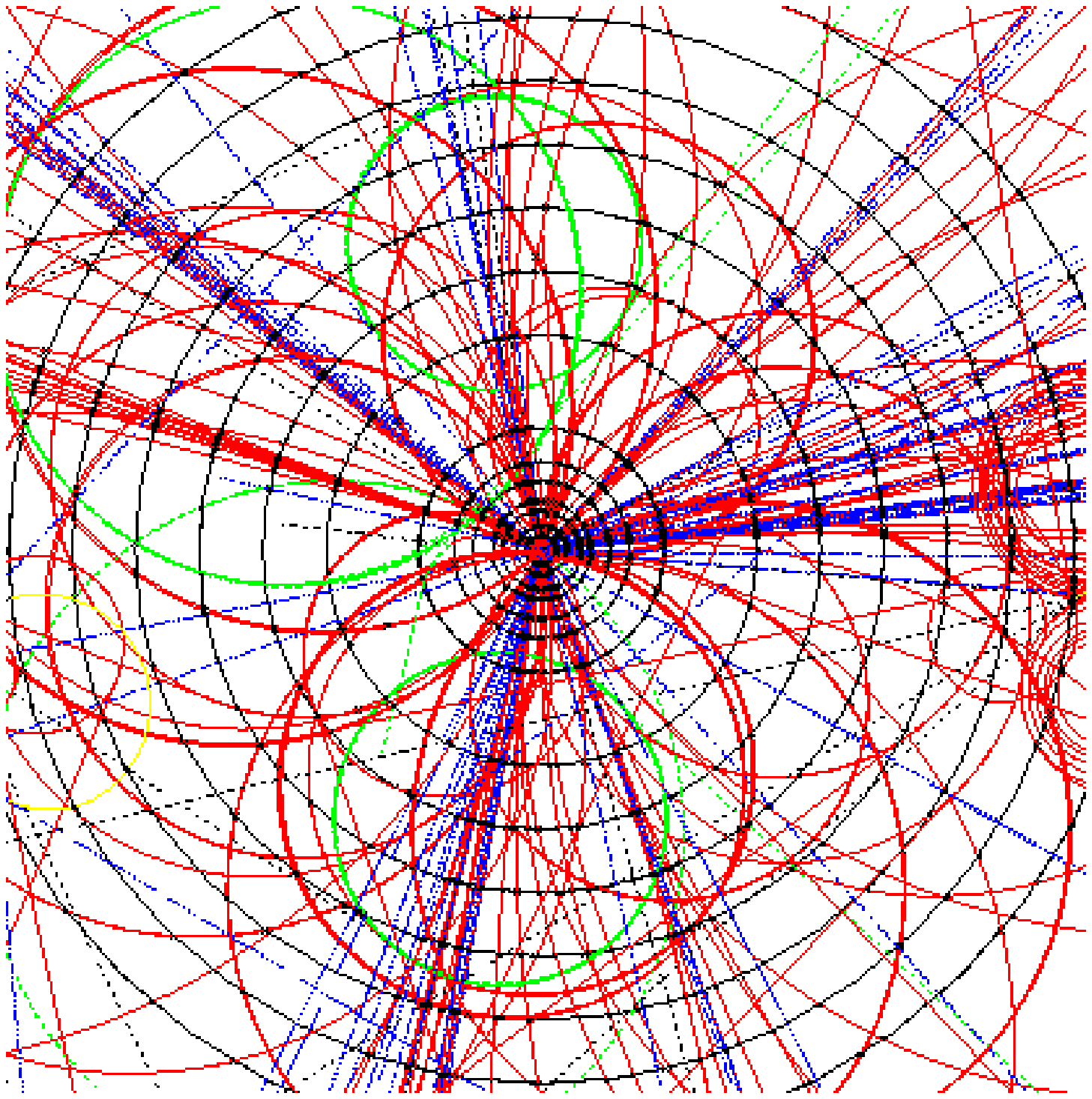}
\end{center}
\caption{Black hole production in a CLIC detector}
\label{fig:bh}
\end{figure}

\begin{figure}[htbp] %
\centerline{\includegraphics[width=7.5cm,height=9.0cm]{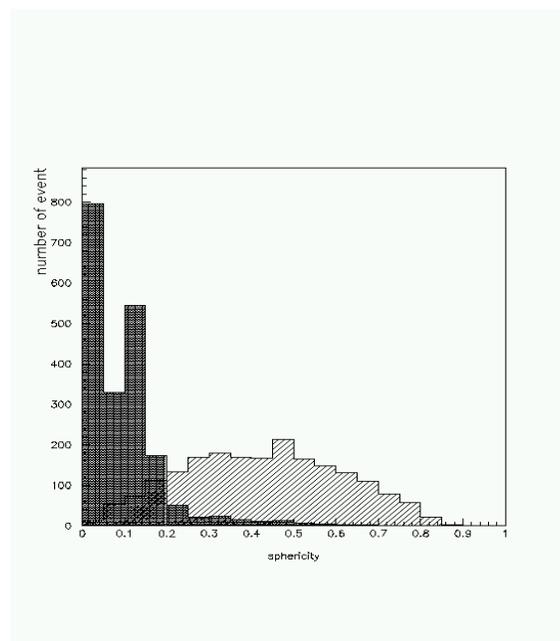}}
\vspace{-1cm}
\caption{Black hole production in a CLIC detector: Sphericity distribution
for 2- and 4-fermion events (full histogram) and black holes
(hatched histogram)}
\label{fig4} 
\end{figure}

The decay of a black hole can be very complex and involves several
stages~\cite{thomas}. If the dominant mode is
Hawking radiation, then all
particles (quarks, gluons, gauge bosons, leptons) are expected to
be produced democratically, with, for instance, a ratio 1/5 between
leptonic and hadronic activity. The multiplicity is expected to
be large. The production and decay processes have been included in
the PYTHIA generator~\cite{landsberg}. Figure~\ref{fig:bh} shows
two black-hole events produced in a detector at CLIC, leading to
spectacular multijet and lepton/photon signals. 
As an example, Fig.~\ref{fig4} shows the sphericity of the events 
measured
after detector simulation and addition of 5 bunch crossings of 
$\gamma\gamma$ 
background for black holes and conventional annihilation events.

If this scenario
is realized in nature, black holes will be produced at high rates
at the LHC. CLIC can be very instrumental in providing precise
measurements.  For example, it may be possible to test Hawking
radiation or get information on the number of underlying extra
dimensions.

\vspace*{-0.5cm}

\vspace{-0.5cm}

\subsection{Kaluza--Klein Excitations in Theories with Extra Dimensions}

Another class of models that leads to a resonance structure in
the energy dependence of the two-fermion cross section has a~TeV-scale 
extra dimension~\cite{anton}. In the simplest
versions of these theories, only the SM gauge fields are in the
bulk, whereas the fermions remain at one of the two orbifold fixed
points; Higgs fields may lie at the fixed points or propagate in
the bulk. In such a model, to a good approximation, the masses of
the KK tower states are given by $M_n=nM$, where $M=R^{-1}$ 
is the compactification scale and $R$ is the compactification radius.

\subsubsection{Kaluza--Klein excitations in two-fermion processes}

\begin{figure}[t] 
\begin{center}
\includegraphics[height=7.5cm]{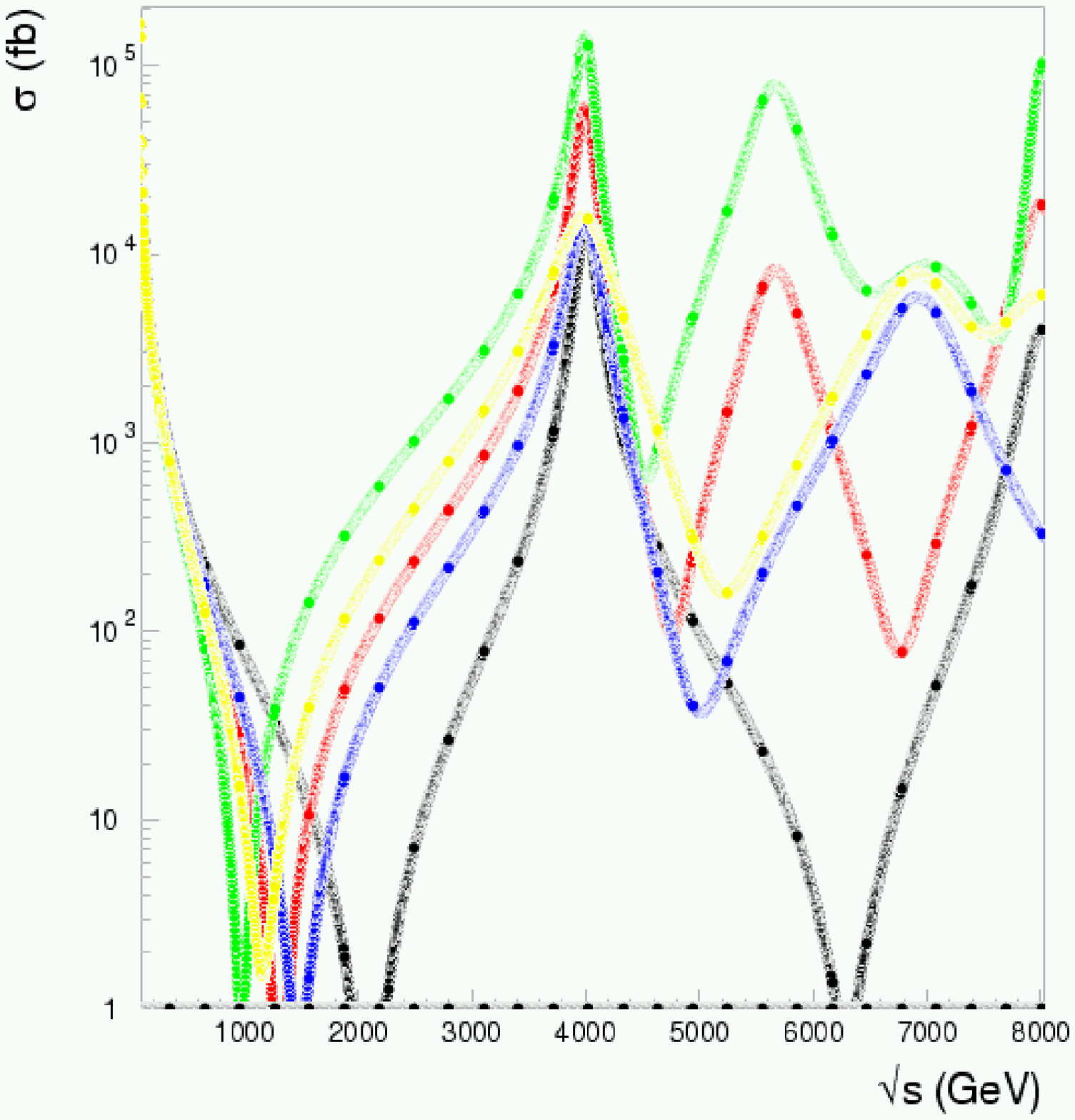}
\hspace*{3mm}
\includegraphics[height=7.5cm]{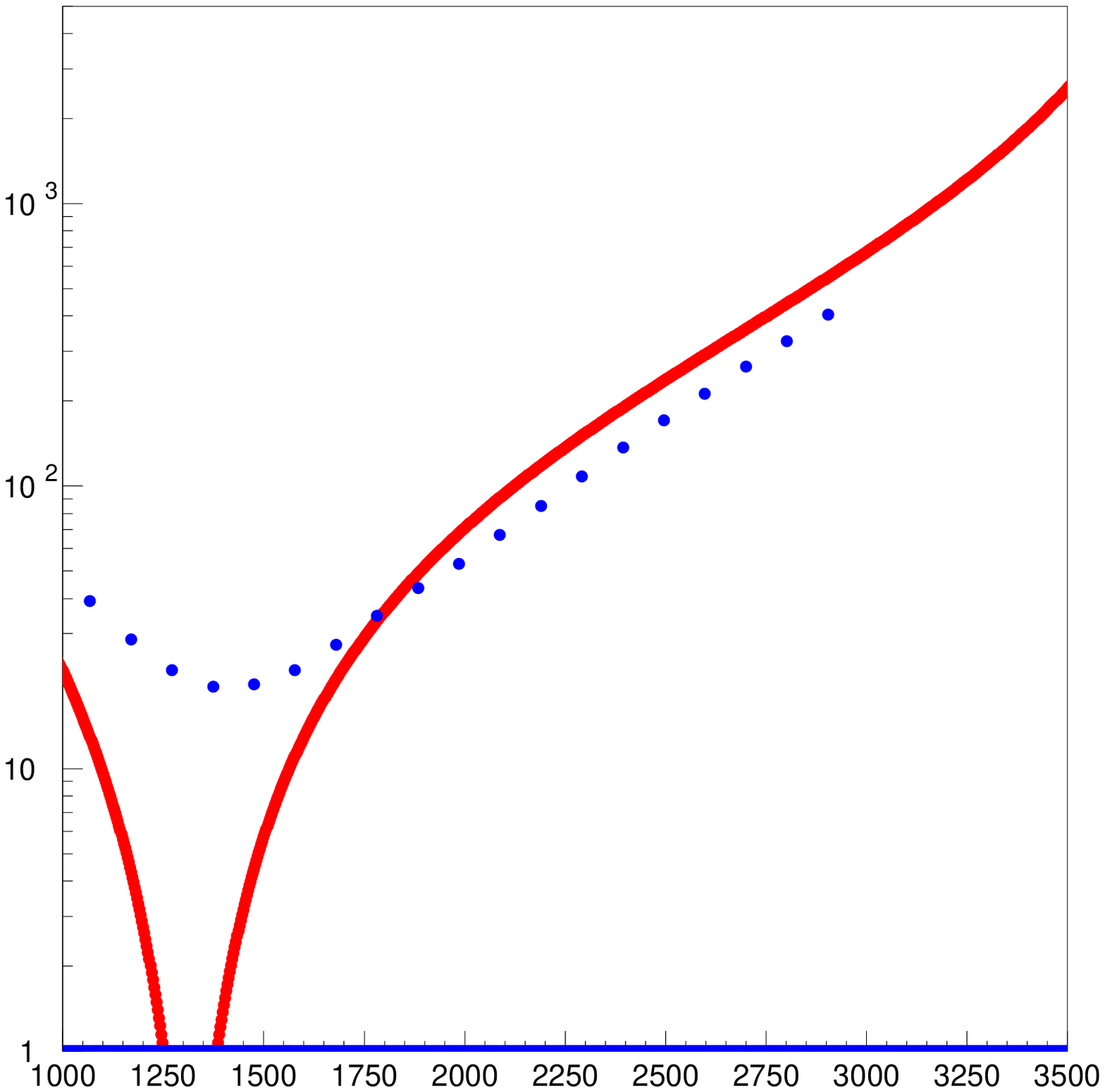}
\end{center}
\caption{ Left: The cross section $\sigma_{\mu\mu}$ for different
models for~TeV-scale extra dimensions. Right: The
cross section $\sigma_{\mu\mu}$ before (solid line) and after
(dotted line)  smearing by the CLIC luminosity spectrum.}
\label{fig:kk5}
\end{figure}

The masses and couplings of the KK excitations are
compactification-scheme-dependent and lead to a rather complex KK
spectrum. Examples of models with one or more extra dimensions,
assuming $M$~=~4~TeV, are shown in Fig.~\ref{fig:kk5}. The  positions
of the peaks and dips and their corresponding cross sections
and widths can be used to identify uniquely the
extra-dimensional model. As an example, one of these models was
taken and the production cross section was folded with the CLIC
luminosity spectrum. The result is shown in
Fig.~\ref{fig:kk5} for the dip position, since the peaks are
likely to be beyond the reach of a 3-TeV collider. The structure
of the dip is largely preserved. It is smeared and somewhat
systematically shifted, but also here the CLIC data will be
sensitive to the model parameters, and will allow one to disentangle
different scenarios.

\begin{figure}[t] 
\begin{center}
\includegraphics[width=8.5cm]{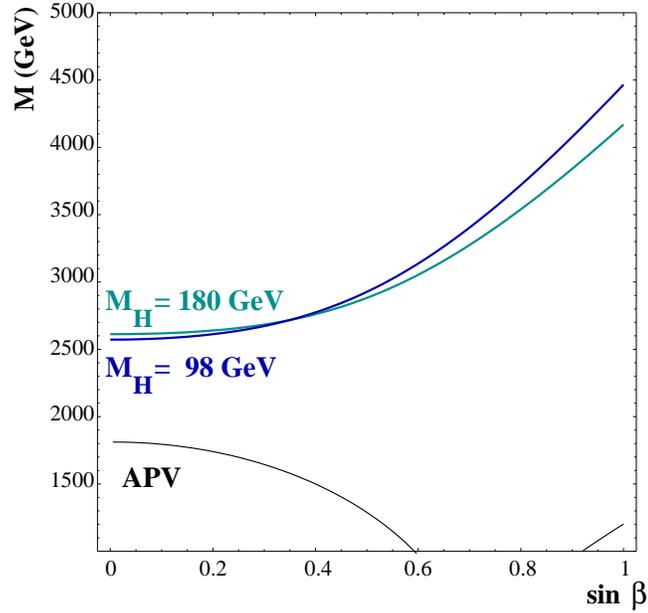}
\end{center}

\vspace*{-3mm}

\caption{95\%~C.L.\ lower bounds on the compactification
scale $M$, as functions of $\sin\beta$, from the high-energy
precision measurements ($\epsilon$ parameters) and from the APV
data. The regions below the lines are excluded.}
\label{kklimits}
\end{figure}

\begin{figure}[!h] 
\begin{center}
\begin{tabular}{c c}
\includegraphics[width=6.5cm]{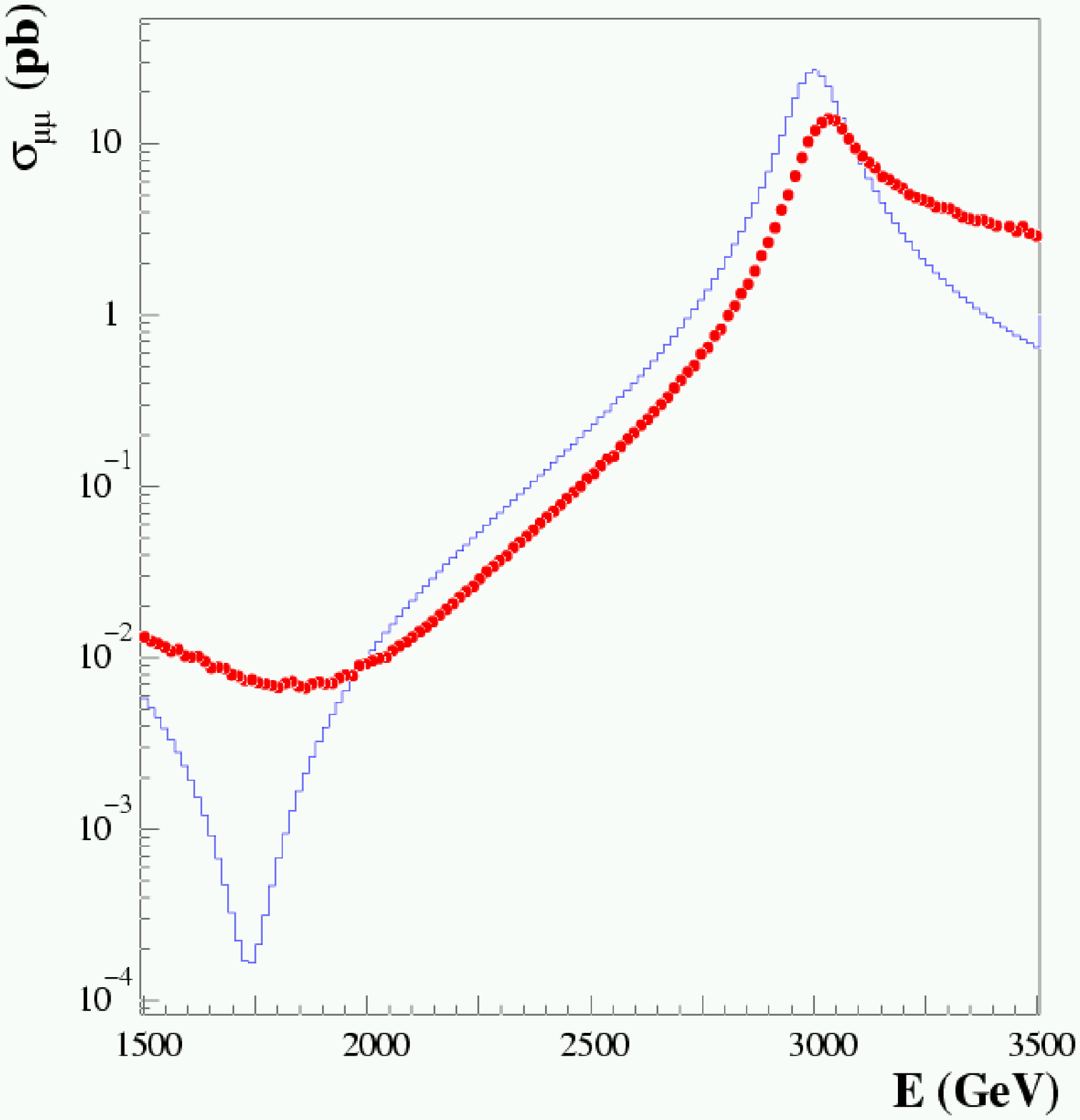} &
\includegraphics[width=6.5cm]{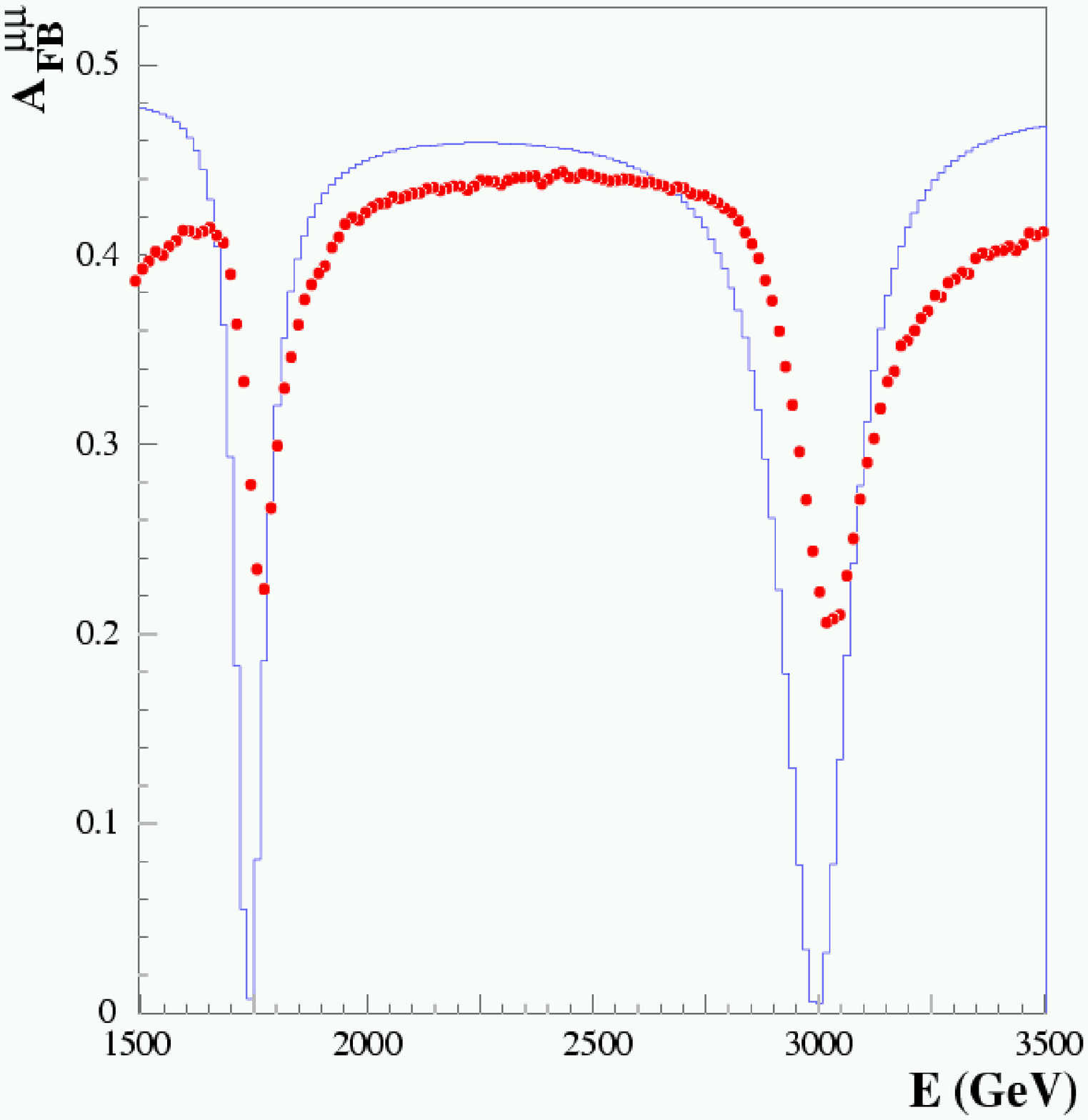}\\
\end{tabular}
\end{center}
\caption{$\mu^+\mu^-$ production cross sections (left) and forward--backward
asymmetry (right) in the 5D SM, including the lower KK 
excitations of the $Z^0$ and $\gamma$ with 
$M_{Z^{(1)}}\sim M_{\gamma^{(1)}}$~=~3~TeV. The continuous lines
represent the Born-level expectations, while the dots include the
effect of the CLIC luminosity spectrum.} 
\label{kksmear}
\end{figure}

Among the models with extra dimensions, we have studied a
5D extension of the SM with fermions on the $y$~=~0
brane of the $S^1/Z_2$ orbifold. This predicts KK excitations of
the SM gauge bosons with fermion couplings a factor of $\sqrt{2}$ larger
than those of the SM \cite{Pomarol:1998sd}. We have
considered two Higgs bosons: one on the $y$~=~0 brane and the other
propagating in the bulk. Indirect limits from electroweak
measurements already exist and are derived by considering the
modifications to the electroweak observables at the $Z^0$ peak
and at low energy~\cite{big}. The masses of the lowest-lying KK
states are constrained to be rather large, i.e. a few~TeV,
depending on the value of $\beta$, which is related to
the ratio between the vev's of the two Higgs
fields and consequently parametrizes the mixing between the KK
excitations and the SM gauge bosons. Bounds can be derived by
considering the latest experimental values of the $\epsilon$
parameters coming from the precision measurement data~\cite{epsi}. The 
95\%~C.L.\ lower bounds on the scale $M$ at
fixed $\sb$, coming from the $\eps$ observables, are given by the
upper curves in Figs.~\ref{kklimits}. The two curves correspond
to $m_t$~=~175.3~GeV with $m_H$~=~98~GeV and 180~GeV. 
The curve for $m_H$~=~120~GeV is very close to the one for
$m_H$~=~98~GeV. Bounds can
be also obtained from low-energy neutral-current experiments, for
example, from the atomic parity-violation (APV) experiments. 
Nevertheless, they lie significantly below the corresponding
high-energy limits, and are shown by the lower curves in
Fig.~\ref{kksmear}. Bounds can be derived from cross sections at
LEP~2 energies. By combining with the precision data, the 95\%~C.L. 
lower limit on $M$ is around  6.8~TeV, while the expected
sensitivity at the  LHC for  100 fb$^{-1}$ is around 
13--15~TeV, depending on the assumed systematics~\cite{cheung}.

At CLIC, the lowest excitations $Z^{(1)}$ and
$\gamma^{(1)}$ could be directly produced~\cite{jhep}. Results
for the $\mu^+\mu^-$ cross sections and forward--backward
asymmetries at the Born level and after folding the effects of
the CLIC.02 beam spectrum are shown in Fig.~\ref{kksmear}.

Hence, if KK excitations appear in the two-fermion processes, cross
Sections in the~TeV range, CLIC will be an ideal tool to study in
detail the properties of these resonances.

\def\lphi{\Lambda_\phi}
\def\gam{\gamma}
\def\epem{e^+e^-}
\def\br{{\mathrm BR}}
\def\mphi{M_\phi}
\def\mpl{M_{Pl}}
\def\gamc{PC}
\def\hsm{h_{SM}}

\subsubsection{Hidden Higgs boson in~TeV$^{-1}$ type of extra dimensions}

In this section the ability of present and future colliders 
to find the lightest Higgs boson in a model that assumes it has a non-trivial 
`location' in a 5D space is explored, in the context of~TeV-scale
extra dimensional models.
The currently missing signal is then due to a suppression 
of the Higgs production cross section at LEP and the Tevatron. 
The model promises a significant enhancement of the signal
at the LHC and possibly at a multi-TeV linear collider (CLIC).
We present results for the cross section of the associated production
of Higgs with gauge bosons at the LC and~LHC. 

First the general features of the  model 
are presented (for a detailed description see~Ref.~\cite{Aranda:2002dz}).
We work with a 5D extension of the SM that contains two Higgs doublets.
The SM fermions and one Higgs doublet ($\Phi_u$) live on a 4D boundary,
the brane, while the gauge bosons and the second Higgs doublet ($\Phi_d$)
are all allowed to propagate in the bulk. The constraints from electroweak
precision data~\cite{Masip:1999mk} show that the compactification scale
can be of O(TeV) (3--4~TeV at 95\% C.L.).
The relevant terms of the 5D $SU(2)\times U(1)$ gauge and Higgs Lagrangian are 
given by
\begin{eqnarray} 
{\cal L}^5 = -\frac{1}{4}\left(F^a_{MN}\right)^2
 -\frac{1}{4}\left(B_{MN}\right)^2
+ |D_M \Phi_d|^2 + |D_{\mu}\Phi_u|^2 \delta(x^5)\,,
\nonumber
\end{eqnarray} 
where the Lorentz indices $M$ and $N$ run from 0 to 4, and $\mu$ runs 
from 0 to 3.

After spontaneous breaking of the electroweak symmetry one obtains the 
following 4D Lagrangian:
\begin{eqnarray} 
\label{sec212interactions}
\nonumber {\cal L}^4 & \supset & \frac{g M_Z}{2c_W}\left(h\sin(\beta-\alpha)
+H\cos(\beta-\alpha)\right)Z_{\mu}Z^{\mu} \\ \nonumber
           &    +    &
\sqrt{2} \frac{g M_Z}{c_W} \left( h \sin\beta\cos\alpha
+H\sin\beta\sin\alpha \right)\sum_{n=1}^{\infty} Z_{\mu}^{(n)}Z^{\mu}
\\ \nonumber
           &    +    & g M_W \left(
h\sin(\beta -\alpha) +H\cos(\beta -\alpha)\right)
W^+_{\mu}W^{-\mu} \\  
           &    +    &
\sqrt{2}gM_W\left(h\sin\beta\cos\alpha + H\sin\beta\sin\alpha\right)
\sum_{n=1}^{\infty}\left(W^+_{\mu}W^{-(n)\mu} + 
W^-_{\mu}W^{+(n)\mu}\right) \,, 
\end{eqnarray}
where $h$ and $H$ are the CP-even Higgses ($m_h <~ m_H$), $\alpha$ is
the mixing angle that appears in the diagonalization of the CP-even
mass matrix, and $\tan\beta$ is the ratio of vev's.

We present results for the associated $h+Z$ production cross section 
at linear colliders and at the~LHC. 

\begin{figure}[t] %
\resizebox*{.30\textwidth}{.25\textheight}
{\includegraphics{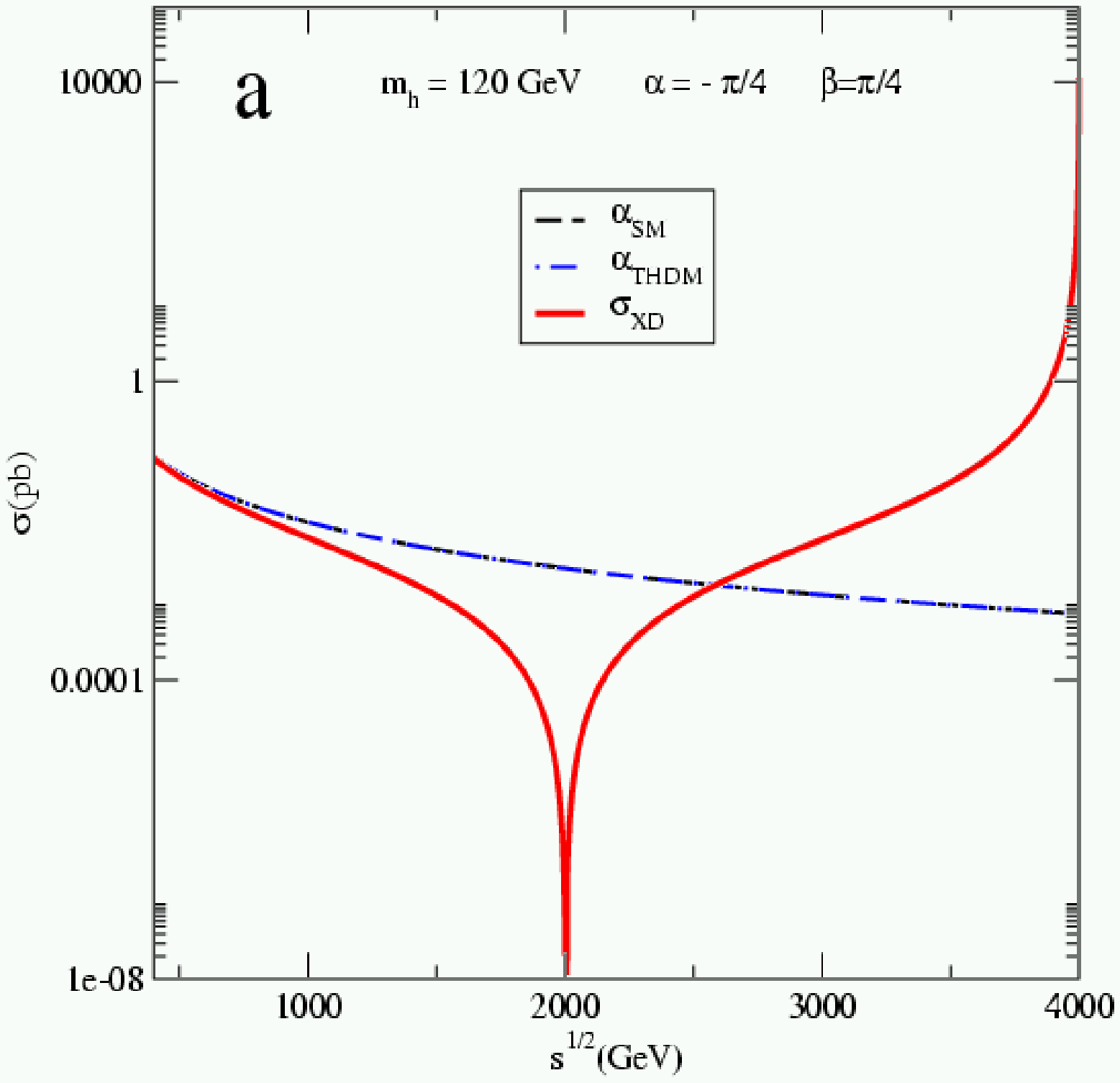}}
\resizebox*{.30\textwidth}{.25\textheight} 
{\includegraphics{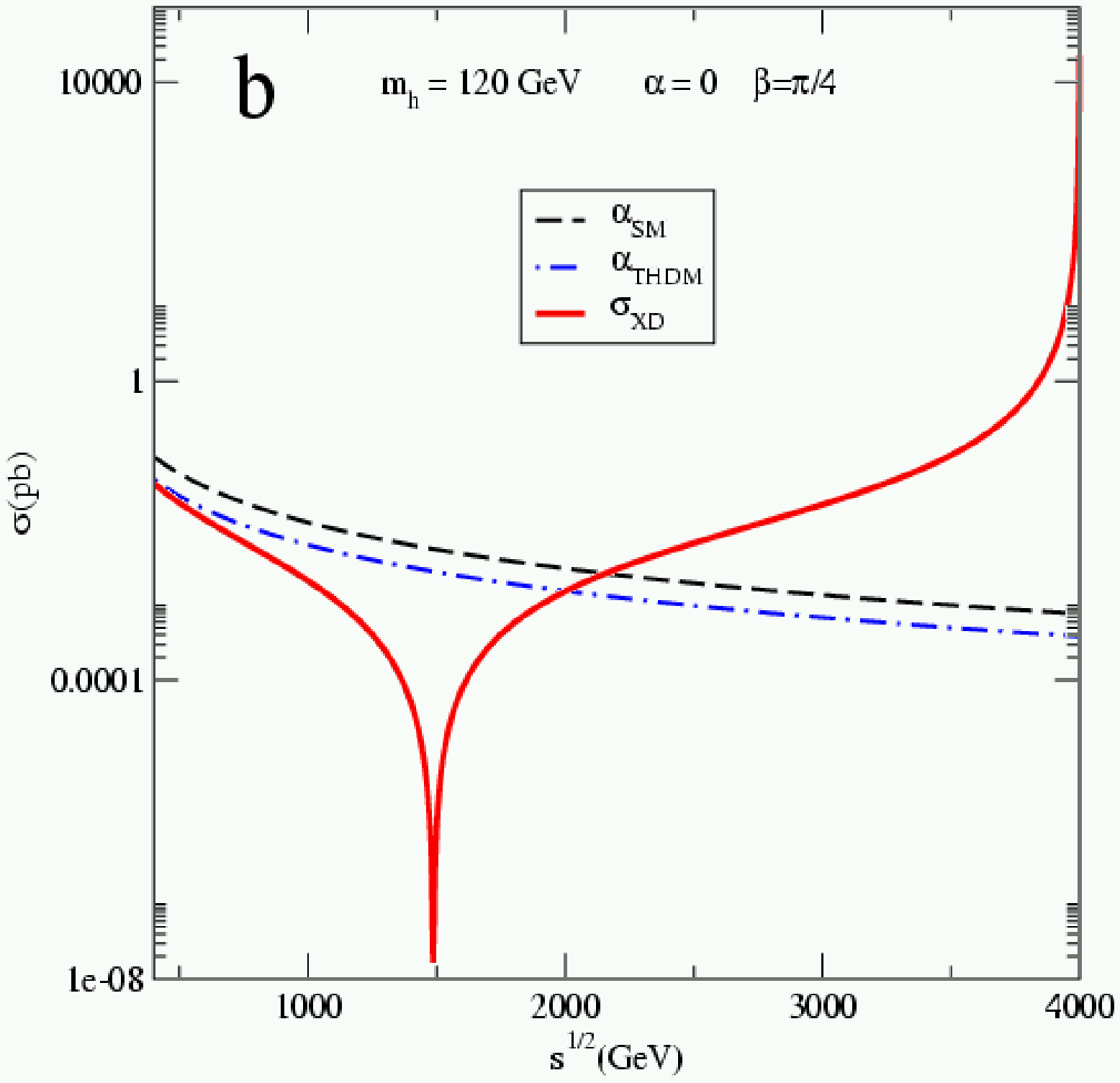}}
\resizebox*{.30\textwidth}{.25\textheight} 
{\includegraphics{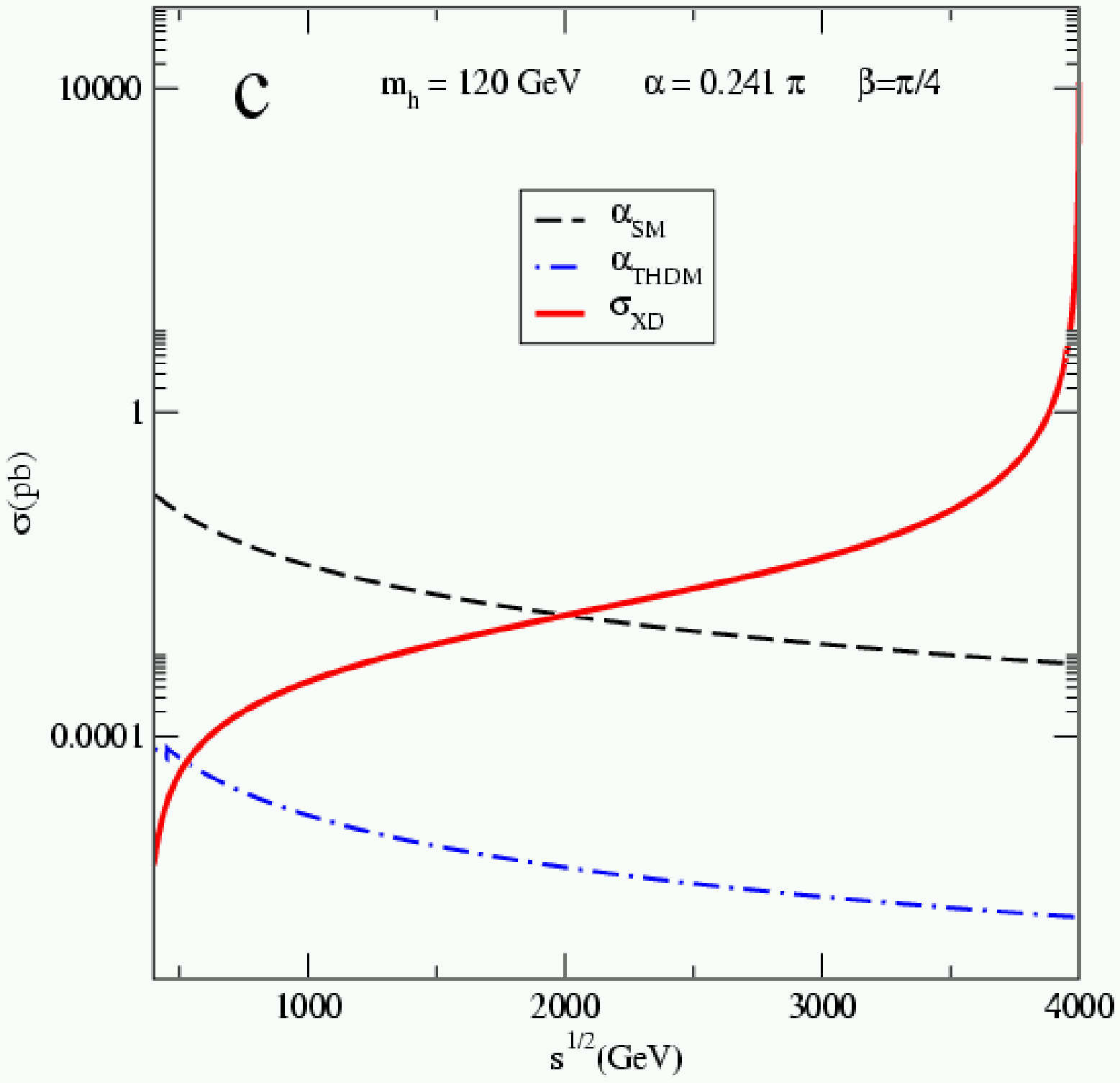}}
\caption{SM, THDM and XD cross Sections for $e^+e^- \rightarrow Zh$. Each plot
corresponds to a different set of values for $\alpha$ and $\beta$, all with
$m_h~=~120$~GeV and with a compactification scale $M_c~=~4$~TeV.}
\label{sec212figure1}
\end{figure}

Figure~\ref{sec212figure1} shows the results for the $e^+e^- \to hZ$
cross section. SM stands for the Standard Model result, THDM
corresponds to the two Higgs doublet model, and the results 
from our model are denoted by XD. The three plots correspond to three
different choices of the parameters $\alpha$ and $\beta$. It can be
observed that the SM cross section dominates in all cases up to
$\sqrt{s}\sim 2$~TeV. This is understood from the fact that the 
heavier KK modes, through their propagators, interfere destructively
with the SM amplitude, thus reducing the XD cross section. Moreover, as
Fig.~\ref{sec212figure1} shows, once the centre-of-mass energy  
approaches the threshold for the production of the first KK state, 
the cross section starts growing. For instance, with $M_c$~=~4~TeV,  
$\sigma_{\rm SM} \simeq \sigma_{\rm XD}$ for $\sqrt{s} \simeq$~2~TeV.
However, one would need higher energies in order to have a cross 
section larger than that of the SM, which may only be possible at~CLIC.

\begin{figure}[!h]
\begin{center}
\includegraphics[width=7.6cm]{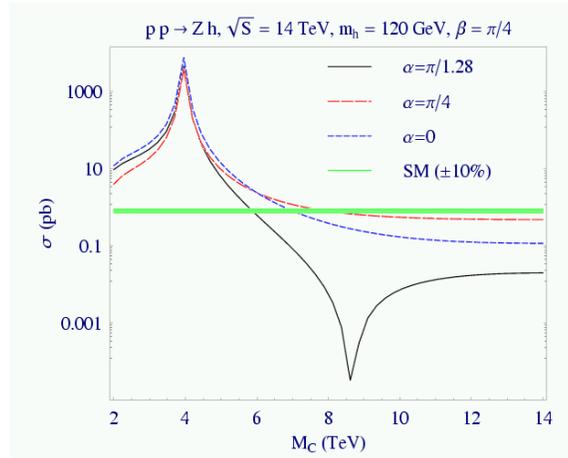}

\vspace*{-4mm}

\caption{Higgs production cross section in association with a $Z$ boson
as a function of the compactification scale for selected values of the mixing
parameters. For reference the SM cross section is shown with a 10\%
uncertainty  band around it.}
\label{sec212figLHC1}
\end{center}
\end{figure}

The Higgs discovery potential in this model is more promising at the LHC. We 
illustrate this in Fig.~\ref{sec212figLHC1} showing the $pp\to hZ$
cross section as a function of the compactification scale $M_c$.
One sees from the figure that depending on the particular values of
$\alpha$ and $\beta$ the enhancement is more or less pronounced. But
independently from the values of $\alpha$, $\beta$ (and the KK width),
the XD production cross section is considerably enhanced with respect to
the SM for $M_c \lesssim$~6~TeV. For certain parameter values there  
is an enhancement even up to $M_c \sim$~8~TeV.

The singularity at $R A(s)$~=~1 is regulated by the width of the KK
mode, which is not included in our calculation. Thus, the peak region
in Fig.~\ref{sec212figLHC1} is only correct in the order of
magnitude. Yet, since the width of the $n^{\rm th}$ KK mode 
$\Gamma_n \sim 2 \alpha_{\rm em} m_n$ is rather small (since
$\alpha_{\rm em}$ is small), the large enhancement is expected to
prevail after the inclusion of the width.
For particular $\alpha$ and $\beta$ values it might also happen that there is
a cancellation among the terms containing the KK state information
(see~Ref.~\cite{Aranda:2002dz}). This cancellation also depends on the
centre-of-mass energy ${\hat s}$ and $M_c$ and happens close to
8.5~TeV in~Fig.~\ref{sec212figLHC1}.

In summary, we presented a model that accounts for the missing Higgs signal. 
Furthermore, the model promises an enhanced signal at the LHC or
CLIC. It assumes one extra dimension and two Higgs doublets, 
one fixed to a brane and the other living in the bulk. The lightest
CP-even scalar may be a mixing of the two and this non-trivial
`location' is the key ingredient in the suppression--enhancement of
the discovery signal. We presented the results for the associated
production cross sections at linear colliders and the LHC.

\vspace{-0.5cm}
\vspace{-0.5cm}

\subsection{Universal Extra Dimensions}

Supersymmetry and extra dimensions offer two different paths to a theory
of new physics beyond the SM. They both address the
hierarchy problem, play a role in a more fundamental theory aimed at
unifying the SM with gravity, and offer a candidate particle for dark
matter, which is compatible with cosmological data. If either supersymmetry 
or extra
dimensions exist at the~TeV scale, signals of new physics are bound to be
found by the ATLAS and CMS experiments at the LHC. 
However, the proper interpretation of such discoveries, namely
the correct identification of the nature of the new physics signals, may
not be straightforward at the LHC. It may have to be complemented by
the data of a multi-TeV $e^+e^-$ colliders,~such
as~CLIC~\cite{Assmann:2000hg}. 

A particularly interesting scenario is offered by Universal Extra
Dimensions (UEDs), as originally proposed in~\cite{Appelquist:2000nn}. It
bears interesting analogies to supersymmetry\footnote{More precisely, the
phenomenology of the first level of KK modes in UEDs is very
similar to that of $N$~=~1 supersymmetric models, with a somewhat degenerate
superpartner spectrum and a stable lightest supersymmetric particle (LSP).  
In what follows, we shall use the term `supersymmetry' in this somewhat
narrower context.}, and has sometimes been referred to as `bosonic
supersymmetry'~\cite{Cheng:2002ab}. In principle, disentangling UED and
supersymmetry may be highly non-trivial at hadron colliders. For each SM
particle, they both predict the existence of a partner (or partners) with
identical interactions. As the masses of these new particles are
model-dependent, they cannot be used to discriminate between the two
theories.

At the LHC, strong processes dominate the production in either case. The
resulting signatures involve relatively soft jets, leptons and missing
transverse energy~\cite{Cheng:2002ab}. The study in~\cite{Cheng:2002ab}
shows that, with 100~fb$^{-1}$ of integrated luminosity, the LHC
experiments will be able to cover all of the cosmologically preferred
parameter space in UED~\cite{Servant:2002aq}. However, the discrimination
between UED and supersymmetry appears very difficult at the LHC.

In fact, there are only three fundamental differences between UED and
supersymmetry. In the case of extra dimensions, there is an infinite tower
of KK partners, labelled by their KK level $n$, while in
supersymmetry, there is just a single superpartner for each SM particle.
However, it is possible that the second and higher KK levels are too heavy
to be observed. The second fundamental distinction between UED and
supersymmetry is the fact that the KK partners have identical spin quantum
numbers, while those of the superpartners differ by $1/2$. Unfortunately,
spin determinations also appear to be difficult at the LHC.
Finally, supersymmetry has an extended Higgs sector and gaugino states,
which have no counterparts in the UED model. However, it is possible that
these heavier Higgs bosons and gauginos are too heavy to be observed.

In its simplest incarnation, the UED model has all the SM particles
propagating in a single extra dimension of size $R$, which is compactified
on an $S_1/Z_2$ orbifold. A peculiar feature of UED is the conservation of
the KK number at tree level, which is a simple consequence of
momentum conservation along the extra dimension. However, bulk and brane
radiative effects~\cite{Georgi:2000ks,vonGersdorff:2002as,Cheng:2002iz}
break the KK number down to a discrete conserved quantity, the so-called
KK parity, (--1)$^n$, where $n$ is the KK level. KK parity ensures that the
lightest KK partners --- those at level one --- are always pair-produced in
collider experiments, similar to the case of supersymmetry models with
conserved $R$-parity. KK-parity conservation also implies that the
contributions to various precisely measured low-energy
observables~\cite{Agashe:2001ra,Agashe:2001xt,Appelquist:2001jz,%
Petriello:2002uu,Appelquist:2002wb,Chakraverty:2002qk,Buras:2002ej} only
arise at the one-loop level and are small. In the minimal UED model, the
boundary terms are assumed to vanish at the cutoff scale $\Lambda$, and
are subsequently generated through RGE evolution to lower scales. Thus the
minimal UED model has only two input parameters: the size of the extra
dimension $R$ and the~cutoff scale $\Lambda$.

In order to study the discrimination of UED signals from supersymmetry
at CLIC, we concentrate on the pair production of level-1
KK muons $e^+e^-\to\mu^+_1\mu^-_1$, and compare it to the analogous
process of smuon pair production in supersymmetry:
$e^+e^-\to\tilde\mu^+\tilde\mu^-$. In UED there are two $n$~=~1 KK
muon Dirac fermions: an $SU(2)_W$ doublet $\mu^D_1$ and an $SU(2)_W$
singlet $\mu^S_1$, both of which contribute. 
In complete analogy, in supersymmetry, there are two smuon
eigenstates, $\tilde\mu_L$ and $\tilde\mu_R$, both of which  
contribute~in~$e^+e^-\to\tilde\mu^+\tilde\mu^-$.

We first fix the UED parameters to $R^{-1}$~=~500~GeV, \, $R\Lambda$~=~20, 
corresponding to the spectrum given in Table~\ref{tab:uedmasses}.

\vspace*{-2mm}

\begin{table}[!h] 
\caption{Masses of the KK excitations for the parameters 
$R^{-1}$~=~500~GeV and $\Lambda R$~=~20 used in the analysis}
\label{tab:uedmasses}

\renewcommand{\arraystretch}{1.2} 
\begin{center}

\begin{tabular}{ccc}\hline \hline \\[-4mm]
$\hspace*{3mm}$ \textbf{Particle} $\hspace*{3mm}$ & $\hspace*{6mm}$ &
$\hspace*{6mm}$ \textbf{Mass}  $\hspace*{6mm}$ \\
 &  & \textbf{(GeV)} 
\\[2mm]   

\hline \\[-3mm]
$\mu_1^D$  &  & 515.0 \\
$\mu_1^S$  &  & 505.4 \\
$\gamma_1$ &  & 500.9
\\[3mm] 
\hline \hline
\end{tabular}
\end{center}
\end{table}


This analysis has backgrounds coming from SM $\mu^+ \mu^- \nu\bar{\nu}$ 
final states, which are mostly due to  gauge-boson pair production 
$W^+W^- \to \mu^+\mu^- \nu_\mu\bar{\nu}_\mu$, 
$Z^0Z^0 \to \mu^+ \mu^- \nu \bar{\nu}$ and from 
$e^+e^- \to W^+W^- \nu_e \bar{\nu}_e$, 
$e^+e^- \to Z^0Z^0 \nu_e \bar{\nu}_e$, followed by muonic decays. 
The total cross Sections are $\simeq 20$~fb and $\simeq 2$~fb, 
respectively. 
In addition to its competitive cross section, this background has leptons 
produced preferentially at small polar angles, therefore biasing
the UED/MSSM discrimination. In order to reduce the background,
a suitable event selection has been applied based on the missing energy, the 
transverse energy and the event sphericity, which provides a factor 
$\simeq$~5 background suppression in the kinematical region of
interest. It does not bias the lepton-momentum distribution. The
estimated background from $\gamma\gamma \to {\mathrm{hadrons}}$
appears to be negligible for muon energies above 2.5~GeV.

\begin{figure}[!h]
\begin{center}
\begin{tabular}{c c c }
\epsfig{file=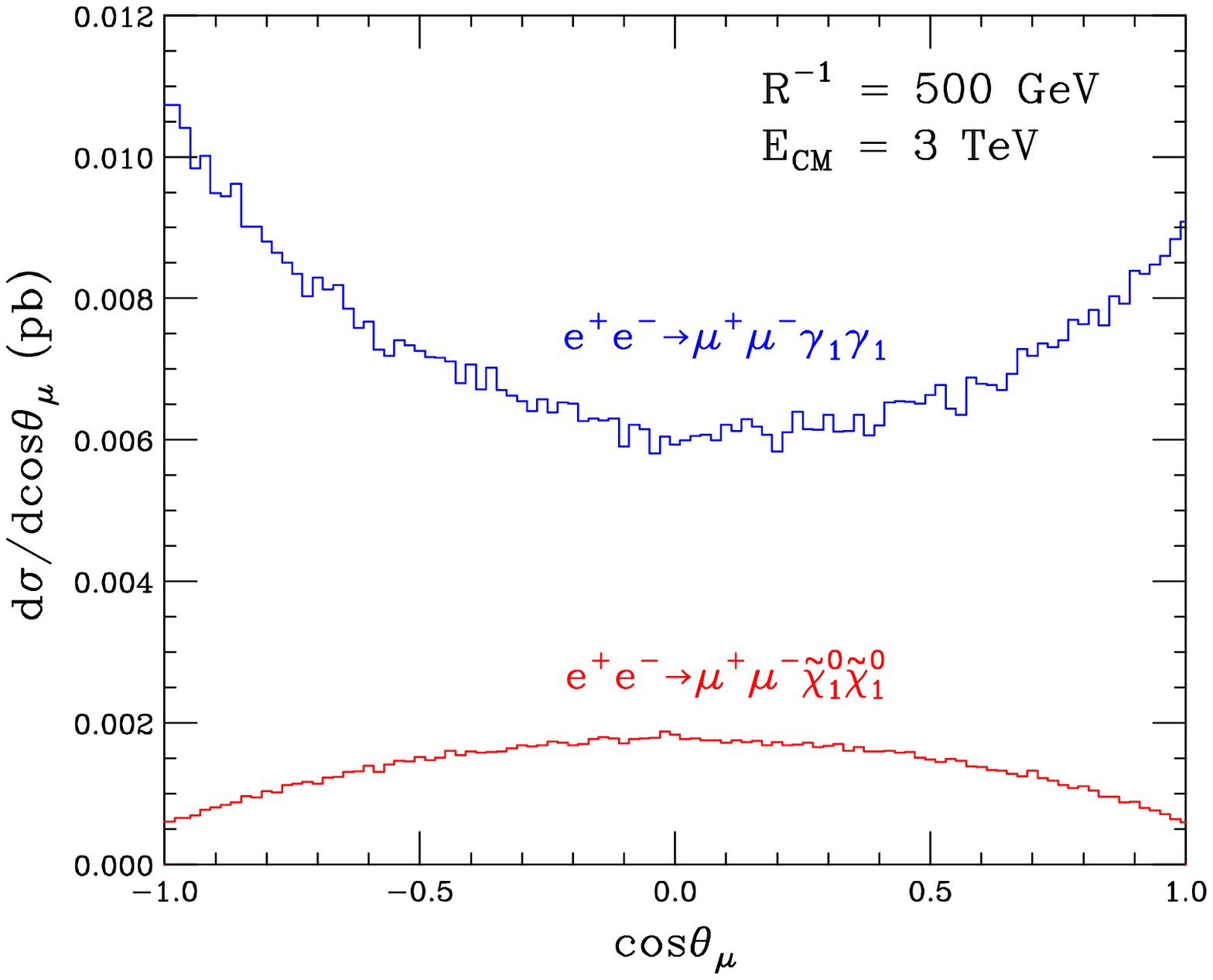,height=5.5cm} & $\hspace{5mm}$
\epsfig{file=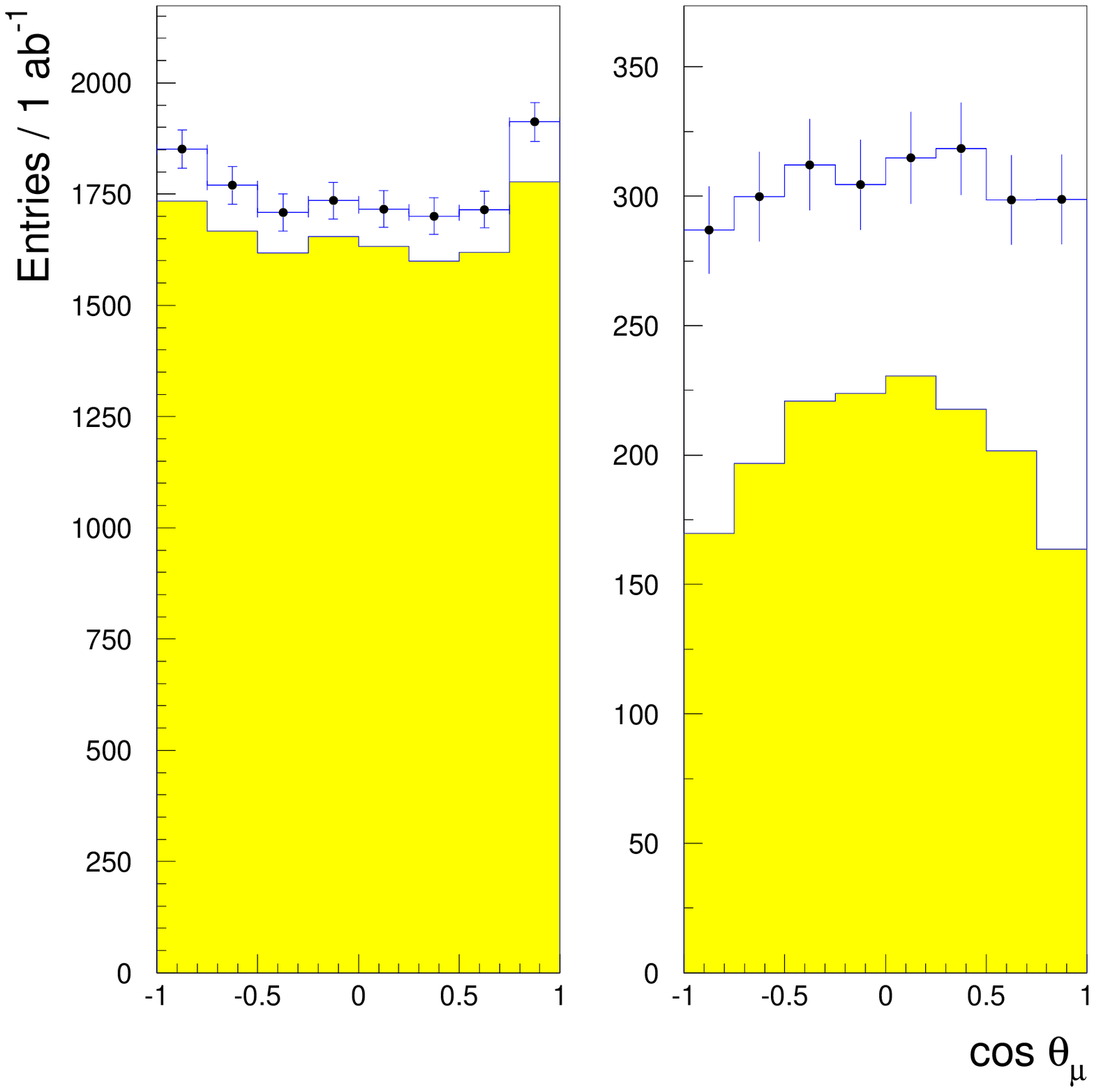,height=6.0cm,width=6.5cm} \\
\end{tabular}
\end{center}
\caption{Differential cross section $d\sigma/d\cos\theta_\mu$ 
for UED (blue, top) and supersymmetry (red, bottom)
as a function of the muon scattering angle $\theta_\mu$.
We have chosen $R^{-1}$~=~500~GeV and $\Lambda R$~=~20
and then adjusted the SUSY-breaking parameters until
we get a perfect spectrum match.
The figure on the left is the ISR-corrected theoretical prediction,
while the result on the right incorporates a detector simulation.}
\label{fig:ang}
\end{figure}

The angular distributions in UED and supersymmetry
are sufficiently distinct to 
discriminate the two cases. However, the polar angles $\theta$ of the 
original KK muons 
and smuons are not directly observable, and the production polar angles
$\theta_\mu$ of the final-state muons are measured instead. As long as 
the mass differences $M_{\mu_1}-M_{\gamma_1}$ and 
$M_{\tilde\mu}-M_{\tilde\chi^0_1}$, respectively, remain small, the
muon directions are well correlated with those of their  
parents~(see Fig.~\ref{fig:ang}a).
%
In Fig.~\ref{fig:ang}b we show the same comparison after detector
simulation and including the SM background. The angular distributions are
still easily distinguishable after accounting for these effects. By 
performing
a $\chi^2$ fit to the normalized polar-angle distribution, the UED
scenario considered here could be distinguished from the MSSM, solely on 
the basis of the distribution shape, with 350~fb$^{-1}$ of data.

Further methods to distinguish between SUSY and UED are the production 
cross section and a threshold scan. The cross section for the UED processes
rises at thresholds $\propto \beta$, while in supersymmetry their threshold
onset is $\propto \beta^3$. Furthermore the cross section for the UED process 
is about 5 times larger compared to the smuon production in
supersymmetry, for these masses.

\begin{figure}[t] 
\begin{center}
\begin{tabular}{c c}
\epsfig{file=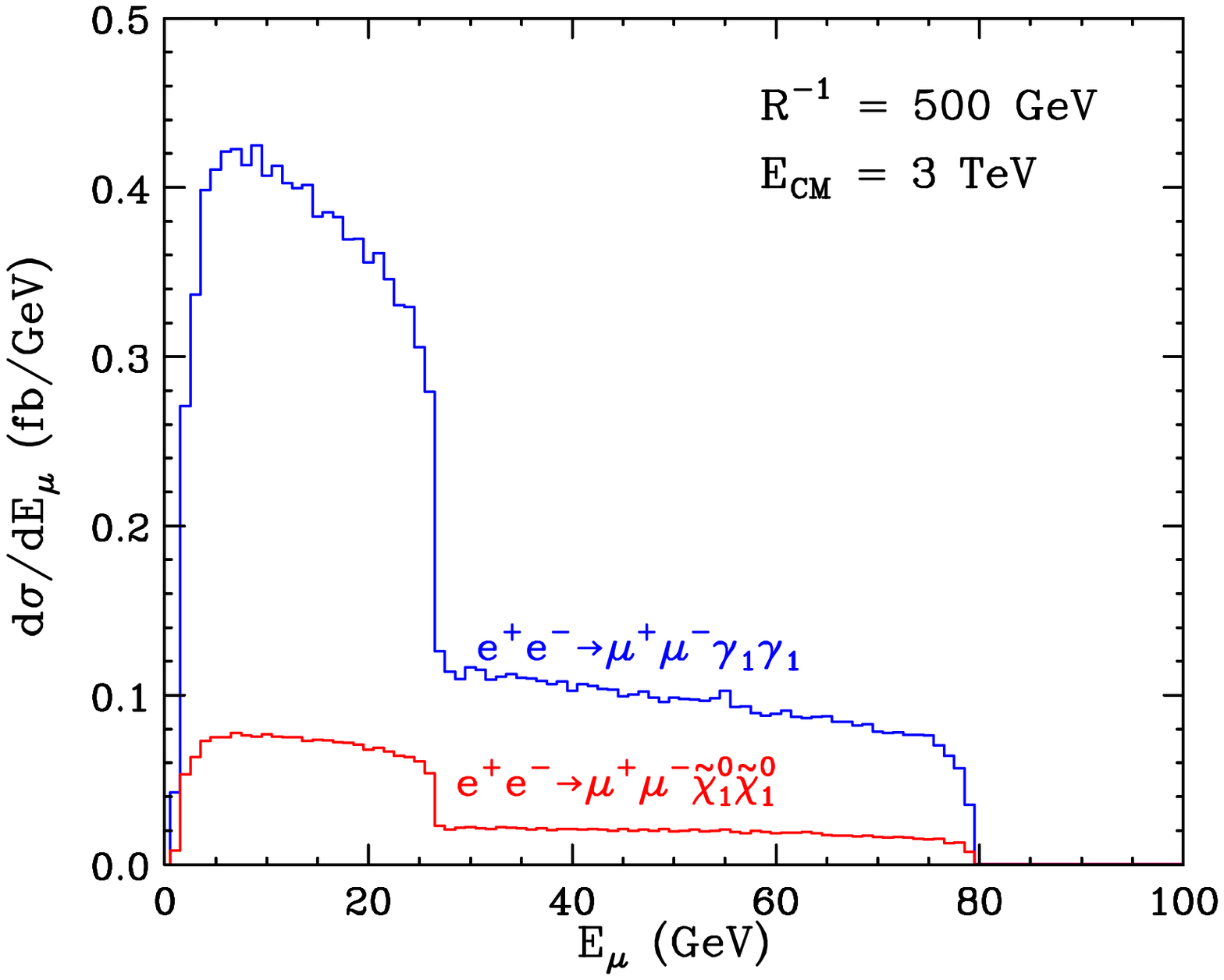,height=5.5cm} &
\epsfig{file=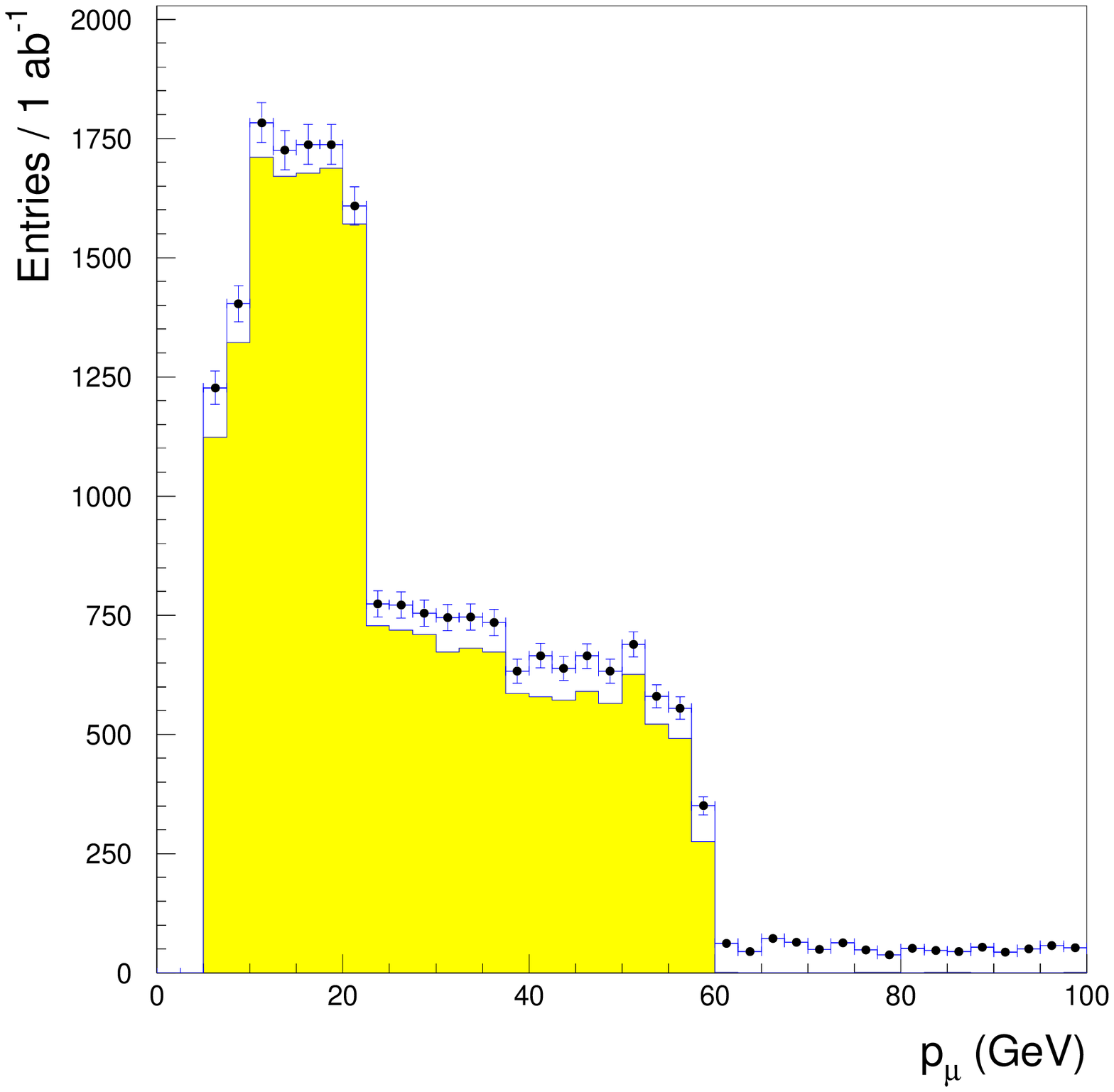,height=6.0cm,width=6.5cm} \\
\end{tabular}
\end{center}
\caption{The muon energy spectrum resulting from KK muon production 
in UED (blue, top curve) and smuon production in supersymmetry
(red, bottom curve), for the same UED and supersymmetry parameters
in~Fig.~\ref{fig:ang}. 
The figure on the left is the ISR-corrected theoretical prediction,
whilst the result on the right shows the distribution after detector 
simulation and including the background contribution.}
\label{fig:emu}
\end{figure}

The muon energy spectrum is completely determined by the kinematics of the
two-body decay, and is therefore equivalent for the case of UED and of
supersymmetry, if the relevant MSSM particle masses are properly tuned. We
show the ISR-corrected expected distributions for the muon energy spectra
at generator level in Fig.~\ref{fig:emu}~(left), using the same
parameters as in 
Fig.~\ref{fig:ang}. We observe that the shape of the $E_\mu$ distribution
in the case of the UED coincides with that for MSSM.

In Fig.~\ref{fig:emu}~(right) we show the muon energy distribution
after detector 
simulation. The end-points can be used to extract the masses by a
three-parameter fit to this distribution. A one-parameter fit, assuming
the mass of the $\gamma_1$ state is known, gives an uncertainty on the
$\mu_1$ mass of $\pm$~0.25~GeV for 1~ab$^{-1}$ of~statistics.

\section{New Vector Resonances}

Many scenarios of NP predict the existence of new particles
that would manifest themselves as rather spectacular resonances in
$e^+e^-$ collisions, if the achievable centre-of-mass energy is
sufficient. A high-energy LC represents an ideal laboratory for
studying this NP. Signals from NP that can be
probed by a collider such as CLIC at
1~TeV~$<\sqrt{s}<$~5~TeV, belong to a rather large domain.
The most striking manifestation of NP in the multi-TeV
region would come from a sudden increase of the $e^+e^-
\rightarrow f \bar f$ cross section indicating the $s$-channel
production of a new particle. The experimental study of such
resonances at a multi-TeV collider will have to accurately
measure their masses, widths, production and decay properties to
determine their nature and identify which kind of NP has
been manifested.
There are several theories that predict the existence of new
vector resonances. A prominent class consists of models with extra
gauge bosons such as a new neutral $Z'$ gauge boson. 

\subsection{Extra-\boldmath{$Z'$} Boson Studies}

One of the simplest extensions of the SM consists of an 
additional $U(1)$  gauge symmetry broken at a  scale close to
the Fermi scale. This extra symmetry is predicted in both
GUT-inspired $E_6$ models and in left--right-symmetric models. For
example, in $E_6$ scenarios we have an additional $U(1)$ current
$J_{Z^\prime\mu}^f = J_{\chi\mu}^f {\cos\theta_6}+
J_{\psi\mu}^f {\sin\theta_6}$
with different models  parametrized by specific values of the
angle $\theta_6$. The values $\theta_6$~=~0, $\theta_6=\pi/2$ and
$\theta_6=-\tan^{-1}\sqrt{5/3}$ are called the  $\chi, \psi$ and
$\eta$ models, respectively.
In the LR models, the new $Z_{\rm LR}$ boson couples to the current
$J_{Z^\prime\mu}=\alpha_{\rm LR}J_{3R\mu}-1/(2\alpha_{\rm LR})
J_{(B-L)\mu}$ with 
$\alpha_{\rm LR}=\sqrt{g_{\rm R}^2/g_{\rm L}^2 \cot^2\theta_W-1}$.
The vector- and axial-vector couplings of the $Z'$ boson to the SM
fermions, for $E_6$-inspired and for LR models, are given in
Table \ref{coupzp}, assuming
$J_{Z^\prime\mu}^f = \bar f\left[\gamma_\mu {v_f^\prime}+
\gamma_\mu \gamma_5 {a_f^\prime}\right] f$
and the parametrization $\theta_2 = \theta_6 + \tan^{-1} \sqrt{5/3}$.
Finally, a useful reference is represented by the so-called
sequential standard model (SSM), which introduces an extra $Z'$
boson with the same couplings as the~SM~$Z^0$~boson.
%
\begin{table}[!h] 
\caption {Vector and axial-vector couplings for the
$E_6$-inspired and LR models, with $s_{\theta} = \sin \theta$,
$s_2=\sin\theta_2$, $c_2=\cos\theta_2$, $c_{2\theta} = \cos 2
\theta$, with $\theta_2=\theta_6+\tan^{-1}\sqrt{5/3}$ and
$\theta\equiv \theta_W$} 
\label{coupzp}

\renewcommand{\arraystretch}{1.4} 
\begin{center}

\begin{tabular}{ccc}\hline \hline \\[-4mm]
$\hspace*{6mm}$ \textbf{Extra}-\boldmath{$U(1)$} $\hspace*{6mm}$ & 
$\hspace*{19mm}$ &
$\hspace*{6mm}$ \textbf{LR} \boldmath{$(g_{\rm L}=g_{\rm R})$} 
$\hspace*{6mm}$ 
\\[4mm]   

\hline \\[-3mm]
 $v_e^\prime=-\frac 1 4 s_\theta\left( c_2+\sqrt{\frac 5 3}
s_2\right)$ & & $  v_e^\prime=\left(-\frac 1 4 +
 s_\theta^2\right)/\sqrt{c_{2\theta}}$\\
 $a_e^\prime=\frac 1 4 s_\theta\left(-\frac 1 3
c_2+\sqrt{\frac 5 3} s_2\right)$& & $a_e^\prime=-\frac 1 4
\sqrt{c_{2\theta}}$\\
$ v_u^\prime=0 $& & $ v_u^\prime=\left(\frac 1 4 - \frac 2 3
s_\theta^2\right)/\sqrt{c_{2\theta}}$\\
 $a_u^\prime=- \frac 1 3 s_\theta c_2 $ & & $  a_u^\prime=\frac 1 4
\sqrt{c_{2\theta}}$\\
 $v_d^\prime=\frac 1 4 s_\theta\left( c_2+\sqrt{\frac 5 3}
s_2\right)$ & & $  v_d^\prime=\left(-\frac 1 4 +
\frac 1 3 s_\theta^2\right)/\sqrt{c_{2\theta}}$\\
 $a_d^\prime=a_e^\prime$& & $a_d^\prime=a_e^\prime$ 
\\[3mm] 
\hline \hline
\end{tabular}
\end{center}
\end{table}


There exist several constraints on the properties of new neutral
vector gauge bosons. Direct searches for a new $Z'$ boson  set
lower limits on the masses~\cite{Abe:1997fd}. These are
summarized in Table~\ref{tab:2} for various models.
An extra $Z'$ naturally mixes with the SM $Z^0$ boson. The
present precision electroweak data constrain  the mixing angle
$\theta_M$ within a few mrad, and the allowed masses 
are shown in~Table~\ref{tab:2}~\cite{Langacker:2001ij,kobel}.
%
\begin{table}[!h] 
\caption{The 95\%~C.L.\ limits on $M_{Z'}$ (GeV) from $\sigma(pp\to Z')
B(Z'\to ll)$ (CDF data) and  from the  average of the four
LEP experiments, for the mixing angle $\theta_M$~=~0}
\label{tab:2}

\renewcommand{\arraystretch}{1.2} 
\begin{center}

\begin{tabular}{ccccccc}\hline \hline \\[-4mm]
$\hspace*{5mm}$  & $\hspace*{5mm}$ &
$\hspace*{2mm}$ \boldmath{$\chi$} $\hspace*{2mm}$ & 
$\hspace*{2mm}$ \boldmath{$\psi$} $\hspace*{2mm}$ & 
$\hspace*{2mm}$ \boldmath{$\eta$} $\hspace*{2mm}$ & 
$\hspace*{2mm}$ \textbf{$LR$} $\hspace*{2mm}$ & 
$\hspace*{2mm}$ \textbf{SSM} $\hspace*{2mm}$ 
\\[4mm]   

\hline \\[-3mm]
CDF &  & 595 & 590 & 620 & 630 & 690\\ 
LEP & & 673 & 481 & 434 & 804 & 1787
\\[3mm] 
\hline \hline
\end{tabular}
\end{center}
\end{table}


A third class of constraints can be derived from the atomic-parity
violation (APV) data~\cite{Casalbuoni:1999yy,jhep}, since models involving
extra neutral vector bosons can modify the $Q_W$ value significantly.  
The present value of $Q_W$ extracted from the caesium data differs by only
$\simeq$~0.8$\sigma$ from the SM prediction. We note that there exist
improved predictions of $Q_W$ in the SM, which account for the effect of
the Breit interaction among electrons and have a refined calculation of
the radiative corrections~\cite{APV}.

Assuming no $Z^0$--$Z'$ mixing, we can evaluate the contribution to the weak
charge due to the direct exchange of the $Z'$ and derive bounds on
$M_{Z'}$. The 95\%~C.L. limit from APV on $M_{Z'}$ for the $\chi$, $\eta$,
LR and SSM models are 627, 476, 665 and 1010~GeV respectively~\cite{jhep},
while no bounds can be set on the $\psi$ model by the APV measurements.
These are less stringent than, or comparable to, those from the LEP
experiments. However, as these bounds are very sensitive to the actual
value of $Q_W$ and its uncertainties, further determinations may improve
the sensitivity.

The LHC will push the direct sensitivity to new vector gauge bosons
past the~TeV threshold. With an integrated luminosity of 100~fb$^{-1}$,
ATLAS and CMS are expected to observe signals from $Z'$ bosons
for masses up to 4--5~TeV, depending on the specific
model~\cite{Godfrey:2002tn}.

Extra-$U(1)$ models can be accurately tested at CLIC. 
With an expected effective production cross section
$\sigma(e^+e^- \to Z'_{SSM})$ of $\simeq$~15~pb, including the
effects of ISR and luminosity spectrum, a $Z'$ resonance will
tower over a $q \bar q$ continuum background of $\simeq$~0.13~pb.
Whilst the observability of such a signal is guaranteed, the accuracy
that can be reached in the study of its properties depends on the
quality of the accelerator beam-energy spectrum and on the
detector response, including accelerator-induced backgrounds. One
of the main characteristics of the CLIC collider is the
large design luminosity, $L$~=~10$^{35}$~cm$^{-2}$~s$^{-1}$ at
$\sqrt{s}$~=~3~TeV for its baseline parameters, obtained in a
regime of strong beamstrahlung effects. The optimization of the
total luminosity and its fraction in the peak has been studied
for the case of a resonance scan. The CLIC luminosity
spectrum has been obtained with a dedicated beam simulation
program~\cite{Schulte:2001kh}, as discussed in Chapter~3, 
for the nominal parameters at $\sqrt{s}$~=~3~TeV. In order to study the 
systematic uncertainties in the knowledge of this spectrum, the modified 
Yokoya--Chen parametrization~\cite{peskin2} has been adopted. In this
formulation, the beam energy spectrum is described in terms of
$N_\gamma$, the number of photons radiated per $e^{\pm}$ in the
bunch, the beam energy spread in the linac $\sigma_p$ and the
fraction $\cal{F}$ of events outside the 0.5\% of the
centre-of-mass energy. Two sets of parameters have been
considered, obtained by modifying the beam size at the
interaction point and therefore the total luminosity and its
fraction in the highest energy region of the spectrum: CLIC.01
with ${\cal{L}}$~=~1.05~$\times$~10$^{35}$~cm$^{-2}$~s$^{-1}$ and
$N_{\gamma}$~=~2.2 and CLIC.02 with
${\cal{L}}$~=~0.40~$\times$~10$^{35}$~cm$^{-2}$~s$^{-1}$ and
$N_{\gamma}$~=~1.2. The $Z'$ mass and width can be determined by
performing either an energy scan, like the $Z^0$ line-shape scan
performed at LEP/SLC, or an autoscan, by tuning the collision
energy just above the top of the resonance and profiting from the
long tail of the luminosity spectrum to probe the resonance peak.

\begin{figure}[t] 
\begin{center}
\mbox{ \epsfig{file=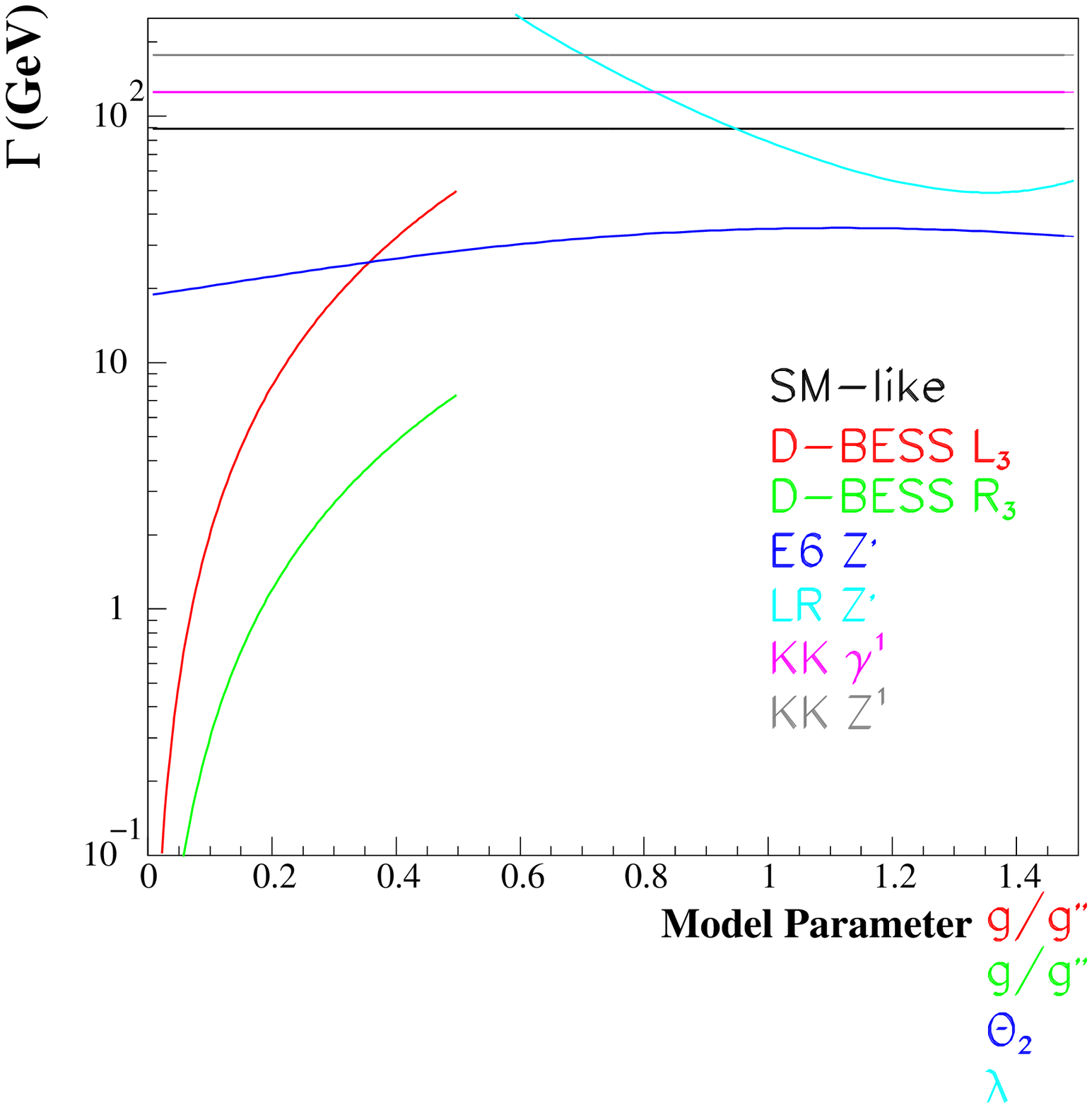,width=7.5cm,height=6cm,clip}
 \epsfig{file=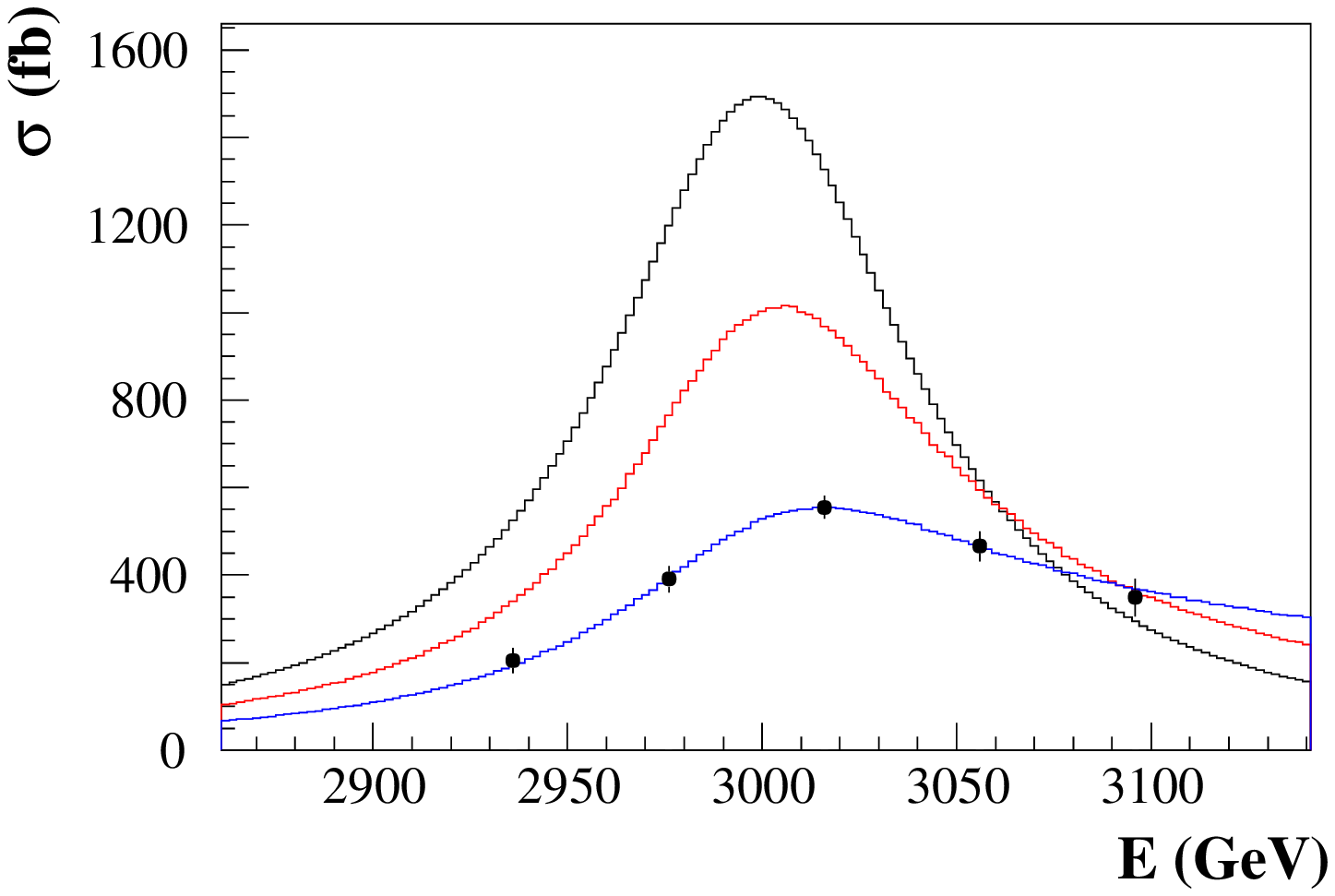,width=8cm,height=6cm,clip}}
\caption{Left: Widths of new gauge vector bosons as functions
of the relevant parameters: $\theta_2$ for $Z'_{E_6}$,
$\lambda=g_{\rm L}/g_{\rm R}$ for $Z'_{\rm LR}$~\cite{alta}, $g/g''$ for
D-BESS. The KK $Z^{(1)}$ width has a negligible dependence on the mixing
angle $\sin\beta$. The $Z'_{E_6}$ and $Z'_{\rm LR}$ widths are
computed by assuming only decays into SM fermions. Right: The
$Z'_{\rm SSM} \to \ell^+ \ell^-$ resonance profile obtained by an energy
scan. The Born production cross section, the cross section with
ISR included and that accounting for the CLIC luminosity
spectrum (CLIC.01) and tagging criteria are shown.}
\label{figwidth}
\end{center}
\end{figure}

For the first method, both dijet and dilepton final states can
be considered, while for the autoscan only $\mu^+ \mu^-$ final
states can provide  the necessary accuracy for the $Z'$
energy. We have generated $e^+e^- \rightarrow Z'$ events for
$M_{Z'}$~=~3~TeV, including the effects of ISR, the luminosity
spectrum and $\gamma \gamma$ backgrounds, assuming SM-like
couplings, corresponding to a total width $\Gamma_{Z'_{SSM}}
\simeq$~90~GeV. The resonance widths for extra-$U(1)$ models as
well as for other SM extensions with additional vector bosons are
shown in~Fig.~\ref{figwidth} as a function of the relevant
model parameters.

A data set of  1~ab$^{-1}$ has been assumed for the CLIC.01 beam
parameters and of 0.4~ab$^{-1}$ for CLIC.02, corresponding to one
year (~=~10$^{7}$~s) of operation at nominal luminosity. This has
been shared in a five-point scan, as seen in Fig.~\ref{figwidth}, and
$M_{Z'}$, $\Gamma_{Z'}/\Gamma_{Z^0}$ and $\sigma_{\rm peak}$ have
been extracted from a $\chi^2$ fit to the predicted cross section
behaviour for different mass and width values (see
Table~\ref{tab:3})~\cite{Battaglia:2001fr}. The dilution of the
analysing power due to the beam energy spread is appreciable, as
can be seen by comparing the statistical accuracy from a fit to
the pure Born cross section after including ISR and beamstrahlung
effects. Still, the relative statistical accuracies are better
than 10$^{-4}$ on the mass and 5~$\times$~10$^{-3}$ on the width.
In the case of wide resonances, there is an advantage in
employing the broader luminosity spectrum of CLIC.01, which offers
larger luminosity. Sources of systematics from the knowledge of
the shape of the luminosity spectrum have also been estimated. In
order to keep $\sigma_{\rm syst} \le \sigma_{\rm stat}$ it is necessary to
control $N_{\gamma}$ to better than 5\% and the fraction
$\cal{F}$ of collisions at $\sqrt{s}<$~0.995~$\sqrt{s_{0}}$ to
about 1\%~\cite{Battaglia:2001dg}.
%
\begin{table}[!h] 
\caption{Results of fits for the cross-Section scan of a
$Z'_{\rm SSM}$ obtained by assuming no radiation and ISR with the
effects of two different choices of the CLIC luminosity
spectrum} 
\label{tab:3}

\renewcommand{\arraystretch}{1.3} 
\begin{center}

\begin{tabular}{cccc}\hline \hline \\[-4mm]
$\hspace*{3mm}$ \textbf{Observable} $\hspace*{3mm}$ &
$\hspace*{5mm}$ \textbf{Breit--Wigner} $\hspace*{5mm}$ &
$\hspace*{5mm}$ \textbf{CLIC.01} $\hspace*{5mm}$ &
$\hspace*{3mm}$ \textbf{CLIC.02} $\hspace*{3mm}$ 
\\[4mm]   

\hline \\[-3mm]
$M_{Z'}$ (GeV) & 3000 $\pm$ .12  & $\pm$ .15 &  $\pm$ .21 \\
$\Gamma_{Z'}/\Gamma_{Z^0}$ & 1. $\pm$ .001 & $\pm$ .003  & $\pm$ .004 \\
$\sigma^{\rm eff}_{\rm peak}$ (fb) & 
1493 $\pm$ 2.0 & 564 $\pm$ 1.7 & 669 $\pm$ 2.9 
\\[3mm] 
\hline \hline
\end{tabular}
\end{center}
\end{table}


\subsection{Heavy Majorana Neutrinos in \boldmath{$Z'$} Decays}

\begin{figure}[t]
\begin{center}
\includegraphics[width=16cm]{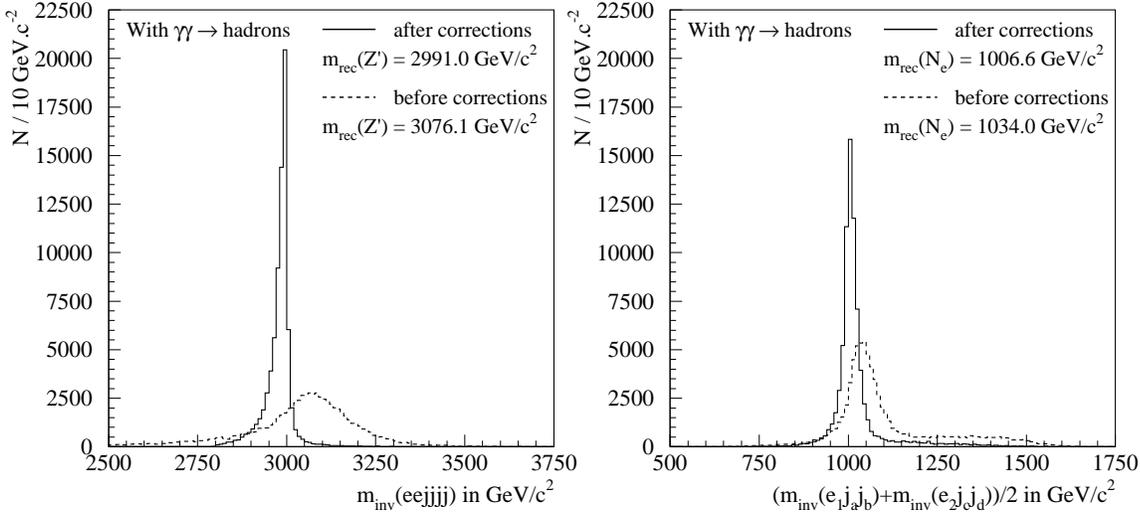}
\caption{Signature of Majorana neutrinos. The left plot shows the 
reconstructed 3~TeV $Z'$ boson before and after the kinematical fit, and the 
right plot shows the reconstructed 1~TeV Majorana neutrino before and 
after the kinematical fit for an integrated luminosity of 1~ab$^{-1}$.} 
\label{arn}
\end{center}
\end{figure}

If neutrinos are Majorana particles, then the left--right-symmetric model
and the see-saw mechanism~\cite{ref4} may provide an explanation for the
lightness of the left-handed neutrinos, by introducing heavy right-handed
neutrinos, $N_{l}$. The LHC is expected to discover these new particles
after  three years of operation at high luminosity, 
mostly through $pp \rightarrow W_R \rightarrow  l N_l$, 
if $m(N_l)$ and $m(W_R)$ are smaller than 4 and 6~TeV, 
respectively~\cite{prd}. At
CLIC, the resonant production of $Z'_{\rm LR} \rightarrow N_{e} N_{e}$ may
lead to signal rates up to about three orders of magnitude higher than for
$pp \rightarrow Z'_{\rm LR}\rightarrow N_{e}N_{e}$ at LHC energies. This
process has been studied using the fast detector
simulation~\cite{ferrari1}. Since $N_{e}$ is expected to decay promptly
into $e^{\pm}$ and a $q_{i}\bar{q}_{j}$ pair, two electrons of equal
charge and four quarks in the final state are the decay signature. The
$\gamma \gamma$ background may induce a shift of the $Z'$ and $N_{e}$
invariant mass peaks towards higher masses. In order to reduce this effect
 a kinematical fit can be applied after
the reconstruction procedure. It requires that the two $N_{e}$ candidates
have equal mass, the system consisting of the two electrons and the four
leading hadronic jets have a zero transverse momentum, the momentum and
energy conservation at $\sqrt{s}$~=~3~TeV. These corrections allow
a good reconstruction of $Z'_{\rm LR}$ and $N_e$, as shown~in~Fig.~\ref{arn}.

\section{Indirect Sensitivity to New Physics}

The direct reach for the mass scales of new phenomena provided by CLIC are
in many cases comparable to those provided by the LHC. However, CLIC would
be able to explore more details of any new processes, possibly allowing
one to determine better the nature of the new physics discovered. With the
centre-of-mass energies and the luminosity envisioned for CLIC, a new era
of precision physics is approaching. Precision electroweak measurements
performed in multi-TeV $e^+e^-$ collisions will push the indirect
sensitivity to new mass scales beyond the 10~TeV frontier.

We review here the sensitivity to new gauge bosons $Z'$, to KK
excitations of the SM gauge bosons and to contact
interactions~\cite{Battaglia:2002sr,Battaglia:2001fr}, as examples of the
anticipated potential of CLIC. These results are based on studies of the
two-fermion ($\mu^+\mu^-$, $b \bar{b}$ and $t \bar{t}$) production cross
sections, $\sigma_{f \bar{f}}$, and forward--backward asymmetries,
$A_{\rm FB}^{f \bar{f}}$.

At CLIC design centre-of-mass energies, the relevant 
$e^+e^- \rightarrow f \bar f$ 
cross sections are significantly reduced with respect to 
those at present-day energies, and the experimental conditions in the
interaction region need to be taken into account in validating the
accuracies on electroweak observables. Since two-fermion cross sections
are of the order of only 10~fb, it is imperative to operate the collider
at high luminosity. This can be achieved only in a regime where beam--beam
effects are important and primary $e^+e^-$ collisions are accompanied by
several $\gamma \gamma \rightarrow hadrons$ interactions. Being
mostly confined in the forward regions, this $\gamma \gamma$ background
reduces the polar-angle acceptance for quark-flavour tagging, and dilutes
the possible jet-charge separation. These experimental conditions require
efficient and robust algorithms to ensure sensitivity to flavour-specific
$f \bar f$ production.

The statistical accuracies for the determination of $\sigma_{f \bar f}$
and $A_{\rm FB}^{f \bar f}$ have been studied using a realistic simulation.
For example, the identification of $b \bar b$ final states was
based on the
sampling of the charged multiplicity in decays of highly-boosted $b$
hadrons at CLIC energies~\cite{Battaglia:2000iw}. Similarly to 
LEP analyses, the forward--backward asymmetry for $b \bar b$ has been
extracted from a fit to the flow of the jet charge $Q^{\rm jet}$, defined as
$Q^{\rm jet} = \frac{\sum_i q_i |\vec p_i\cdot \vec T|^k} {\sum_i |\vec p_i
\cdot \vec T|^k}$, where $q_i$ is the particle charge, $\vec p_i$ its
momentum, $\vec T$ the jet thrust unit vector, $k$ a positive number, and
the sum is extended to all the particles in a given jet.  Here the
presence of additional particles, from the $\gamma \gamma$ background,
causes a broadening of the $Q^{\rm jet}$ distribution and thus a dilution of
the quark charge separation. The track selection and the value of the
power parameter $k$ need to be optimized as a function of the number of
overlaid bunch crossings.  Results for the $e^+e^-\to \bar t t$ channel
have been obtained using a dedicated top tagging algorithm~\cite{laura}.
This uses an explicit reconstruction of the $t\to bW$ decay and also
includes the physics and machine-induced backgrounds. For $t \bar t$
forward--backward asymmetries, the sign of the lepton from the
$W^\pm\to\ell^\pm\nu$ decay has been used. The results are summarized in
terms of the relative statistical accuracies $\delta {\cal{O}}/{\cal{O}}$
in~Table~\ref{acc}.
%
\begin{table}[t] 
\caption{Relative statistical accuracies on electroweak observables
obtainable for 1~ab$^{-1}$ of CLIC data at $\sqrt{s}$~=~3~TeV,
including the effect of the $\gamma \gamma \to hadrons$ background}
\label{acc}

\renewcommand{\arraystretch}{1.4} 
\begin{center}

\begin{tabular}{ccc}
\hline \hline\\[-4mm]
$\hspace*{3mm}$ \textbf{Observable} $\hspace*{3mm}$ & $\hspace*{6mm}$
& $\hspace*{3mm}$ \textbf{Relative stat. accuracy} $\hspace*{3mm}$ \\
 & & \boldmath{$\delta {\cal{O}}/{\cal{O}}$ }
\textbf{for 1~ab}\boldmath{$^{-1}$} 
\\[1.5mm]  \hline\\[-4mm]
$\sigma_{\mu^+\mu^-}$ &  & $\pm$~0.010 \\
$\sigma_{b \bar b}$ &  & $\pm$~0.012 \\
$\sigma_{t \bar t}$ &  & $\pm$~0.014 \\
$A_{\rm FB}^{\mu\mu}$ &  & $\pm$~0.018 \\
$A_{\rm FB}^{bb}$ &  & $\pm$~0.055 \\
$A_{\rm FB}^{tt}$ &  & $\pm$~0.040 
\\[2mm] 
\hline \hline
\end{tabular}
\end{center}
\end{table}

However, it is important to stress that, at the energy scales considered
here, electroweak virtual corrections are strongly enhanced by Sudakov
double logarithms of the type $\log^2 (s/M_W^2)$. Until a complete
two-loop result settles the problem, a theoretical error on the cross
Section of the order of a per cent could be
considered~\cite{Ciafaloni:2000ey}.  We have not included it in the
present analyses. This issue is discussed in detail in the next section.

The indirect sensitivity of a LC to the $Z'$ mass $M_{Z'}$ can be
parametrized in terms of the available integrated luminosity ${\cal{L}}$
and the centre-of-mass energy $\sqrt{s}$. A scaling law for large
$M_{Z'}$ can be obtained by considering the effect of the $Z'$--$\gamma$
interference in the cross section. For $s \ll M_{Z'}^2$ and assuming that
the uncertainties $\delta \sigma$ are statistically dominated, we can
infer the range of mass values that would give a significant difference
from the SM prediction:
\begin{equation}
\frac{|\sigma^{\rm SM} - \sigma^{\rm{SM}+Z'}|}{\delta \sigma} \propto
\frac{1}{M^2_{Z'}}\sqrt{s{\cal{L}}} > \sqrt{\Delta \chi^2},
\end{equation}
and the sensitivity to the $Z'$ mass scales as:
\begin{equation}
M_{Z'} \propto (s {\cal{L}})^{1/4}.
\label{rescaling}
\end{equation}
This relation points to the direct trade-off possible between
$\sqrt{s}$ and ${\cal{L}}$. This needs to be taken into account in the
optimization of the parameters of a high-energy $e^+e^-$ linear collider.

The $\sigma_{f \bar f}$ and $A_{\rm FB}^{f \bar f}$ ($f = \mu,~b,~t$) values
have been computed for 1~TeV~$< \sqrt{s} <$~5~TeV, both in the SM and
including the corrections due to the presence of a $Z'$ boson with 
10~TeV~$< M_{Z'} <$~40~TeV, with couplings defined by the models discussed in
Section~2. Predictions have been obtained by implementing these models in
the {\sc Comphep} program~\cite{comphep}. Relative statistical errors on
the electroweak observables are obtained by rescaling the values of
Table~\ref{acc} for different energies and luminosities. The sensitivity
has been defined as the largest $Z'$ mass, giving a deviation of the actual
values of the observables from their SM predictions corresponding to a SM
probability of less than 5\%.  The SM probability has been defined as the
minimum of the global probability computed for all the observables and
that for each of them taken independently.

\begin{figure}[t] 
\begin{center}
\begin{tabular}{l r}
\includegraphics[width=0.4\textwidth,height=0.45\textwidth]
{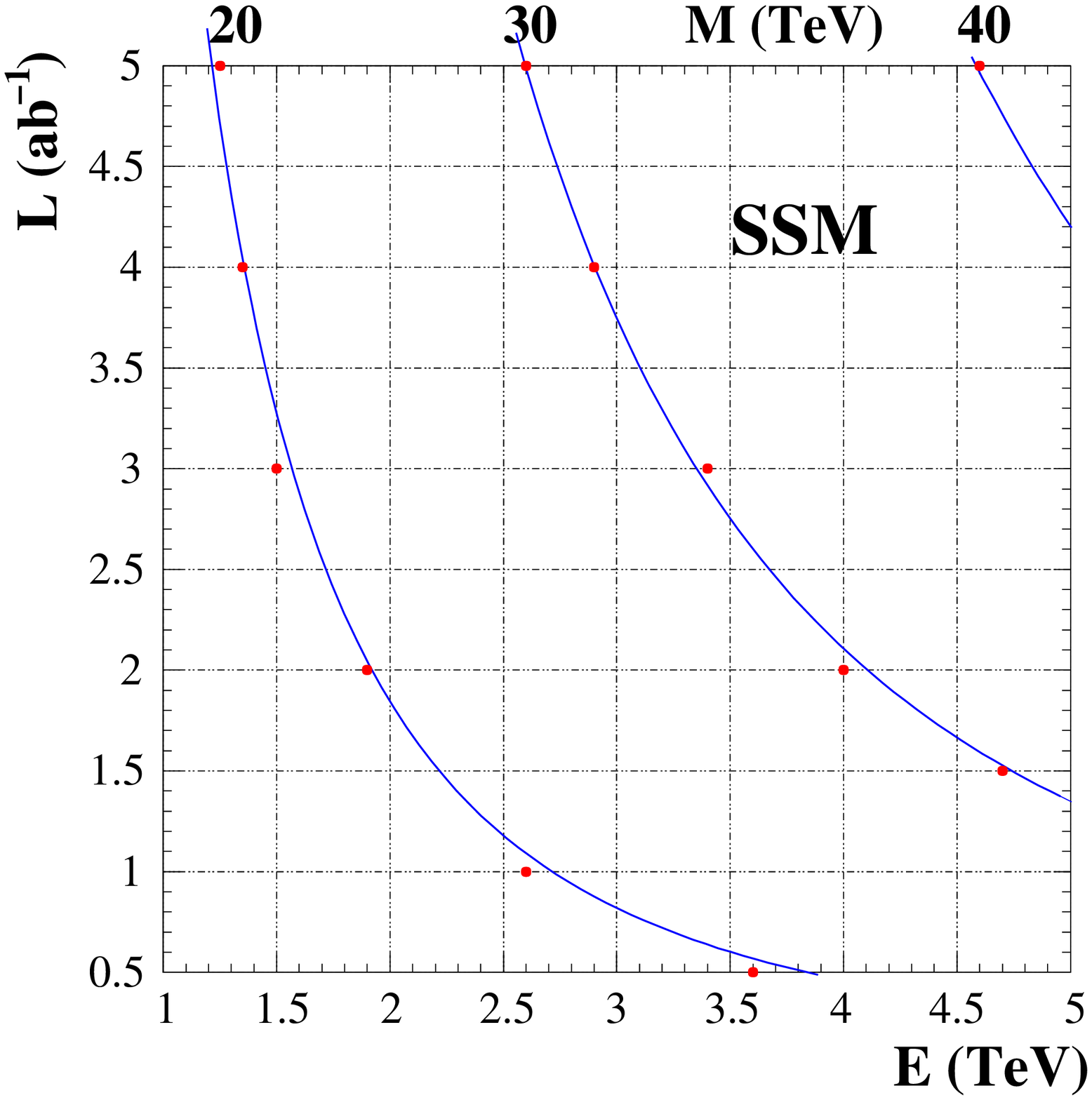}&
\includegraphics[width=0.4\textwidth,height=0.45\textwidth]
{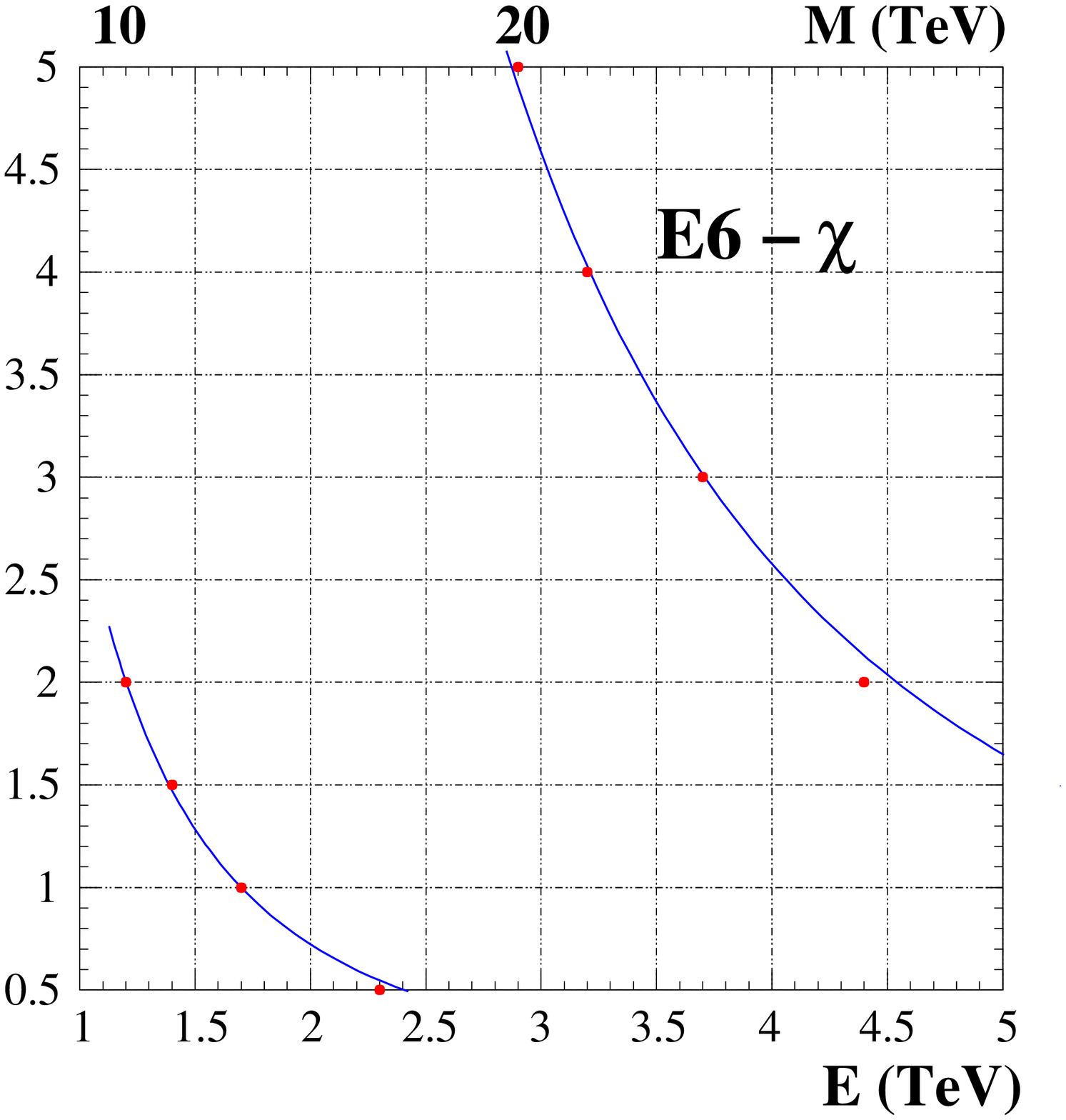}\\
\end{tabular}
\end{center}
\caption{The 95\%~C.L. sensitivity contours in the ${\cal{L}}$ vs.
$\sqrt{s}$ plane for different values of $M_{Z'}$ in the SSM
model (left) and in the $E_6$~$\chi$ model (right).
The points represent the results of the analysis, while the curves
show the behaviour expected from the scaling suggested
in~(\ref{rescaling}).}
\label{fig:zp}
\end{figure}

This sensitivity has been determined, as a function of the $\sqrt{s}$
energy and integrated luminosity ${\cal{L}}$, and compared to the scaling
in (\ref{rescaling}). Results are summarized in~Fig.~\ref{fig:zp}.  For
the $\eta$ model the sensitivity is lower. To reach the sensitivity of
$M_{Z'}$~=~20~TeV, more than 10~ab$^{-1}$ 
of data at $\sqrt{s}$~=~5~TeV are~necessary.

\begin{figure}[htbp] 
\begin{center}
\begin{tabular}{c c}
\includegraphics[width=0.45\textwidth,height=0.46\textwidth]
{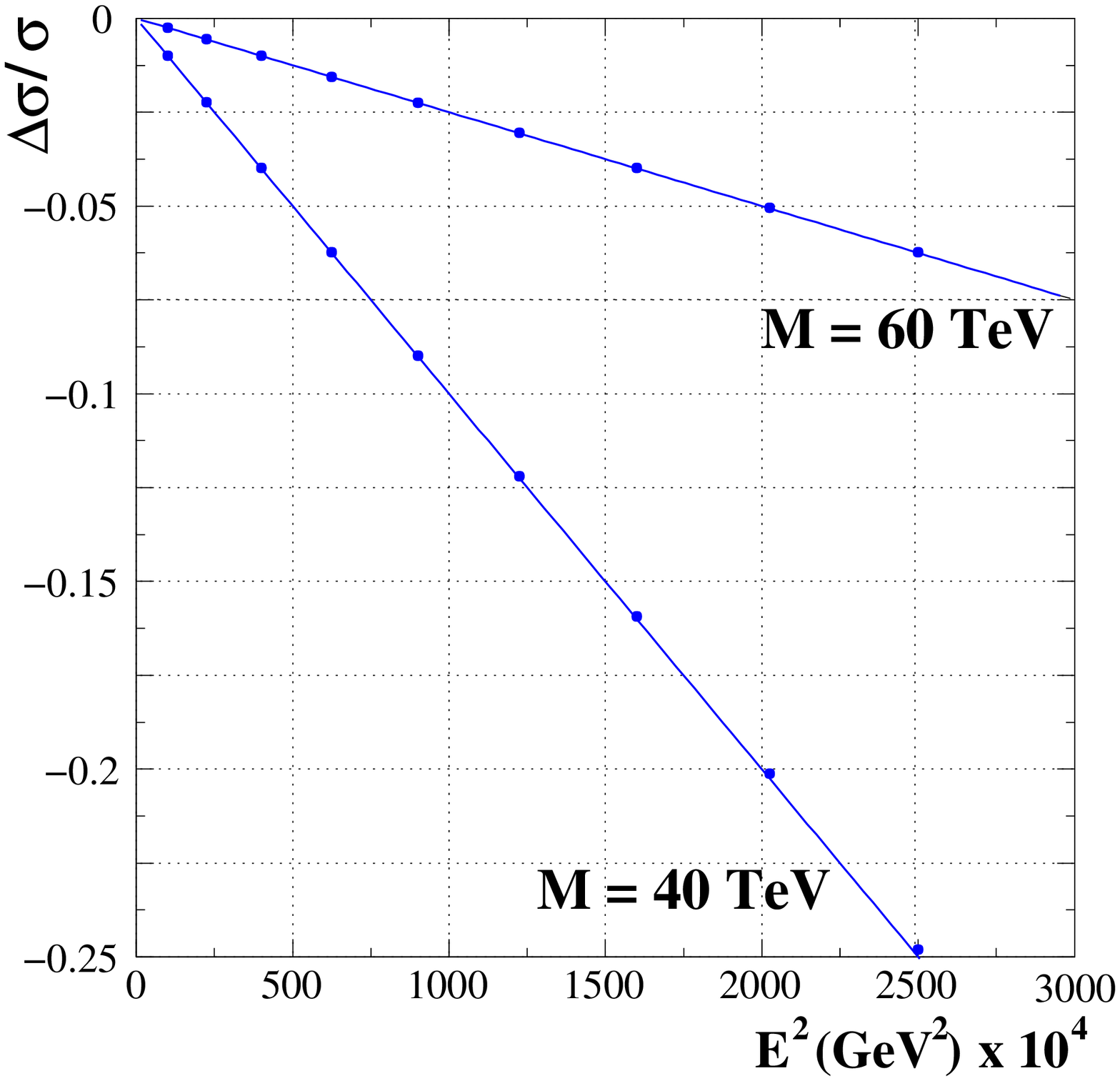}& $\hspace*{3mm}$
\includegraphics[width=0.4\textwidth,height=0.46\textwidth]
{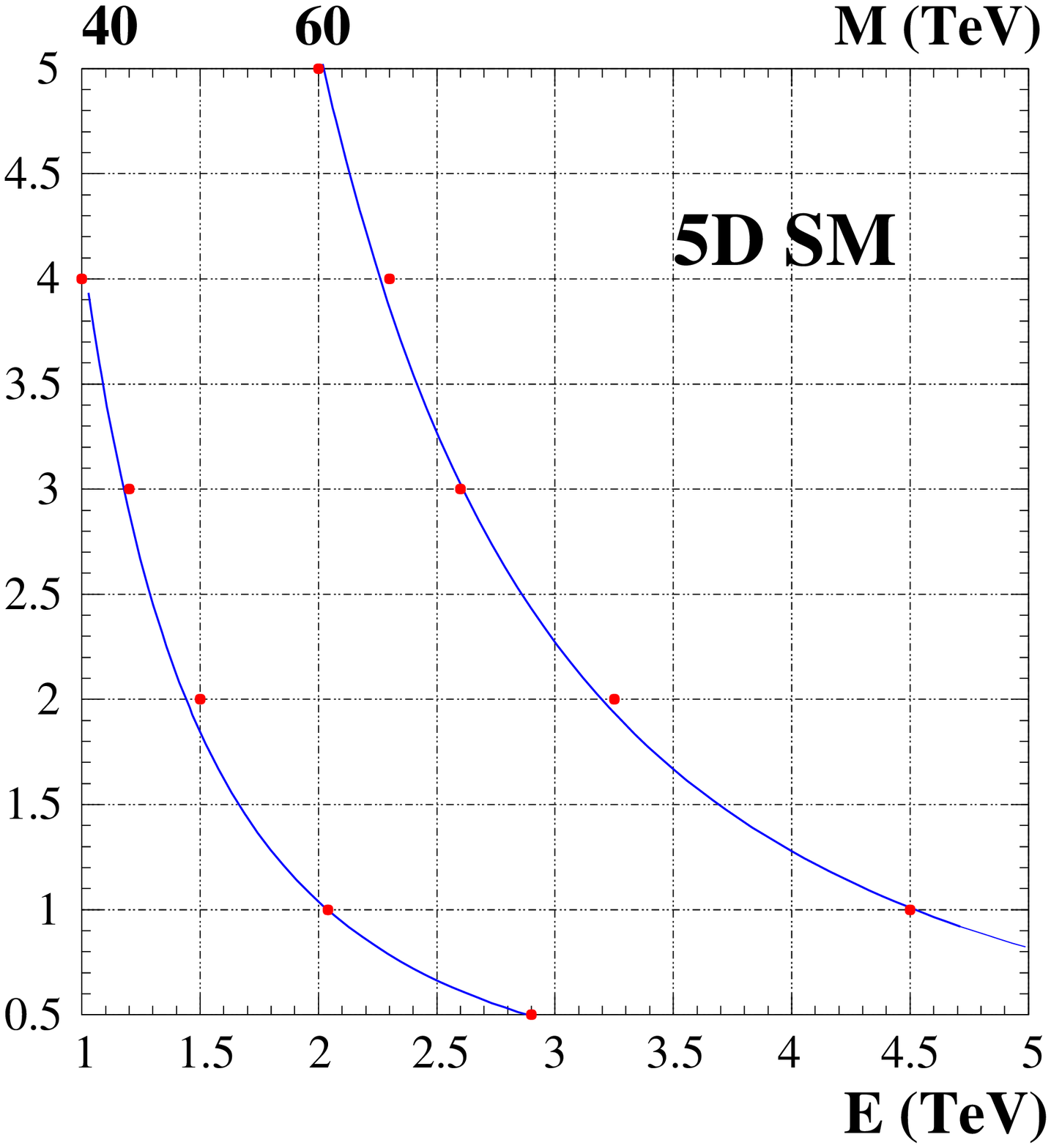}\\
\end{tabular}
\caption{Left: Scaling of the relative change in the $e^+e^- \to b
\bar{b}$ cross section, in the 5D~SM, as a function of the square of the 
centre-of-mass energy, for two values of the compactification scale $M$.
Right: The 95\%~C.L. sensitivity contours in the ${\cal{L}}$ vs.
$\sqrt{s}$ plane for different values of the compactification scale $M$ in the 
5D~SM. The points represent the results of the analysis, while curves show the 
behaviour expected from the scaling suggested in~(\ref{rescaling}).}
\label{scaling} 
\end{center}
\end{figure}

In the case of the 5D~SM, we have included only the effects of the
exchange of the first KK excitations $Z^{(1)}$ and $\gamma^{(1)}$,
neglecting those of the remaining excitations of the towers, which give
only small corrections. The scaling law for the limit on $M$ can be
obtained by considering the interference of the two new nearly degenerate
gauge bosons with the photon in the cross section and taking the 
$s \ll M^2$ limit. 
The result is the same as (\ref{rescaling}). The analysis closely
follows that for the $Z'$ boson discussed above. In Fig.~\ref{scaling}
we give the sensitivity contours as a function of $\sqrt{s}$ for different
values of $M$. We conclude that the achievable sensitivity to the
compactification scale $M$ for an integrated luminosity of 1~ab$^{-1}$ in
$e^+e^-$ collisions at $\sqrt{s}$~=~3--5~TeV is of the order of
40--60~TeV. Results of a similar analysis, including all electroweak
observables, are discussed in~Ref.~\cite{Rizzo:2001gk}.

\begin{figure}[t] 
\begin{center}
\includegraphics[width=0.74\textwidth,height=.74\textwidth]{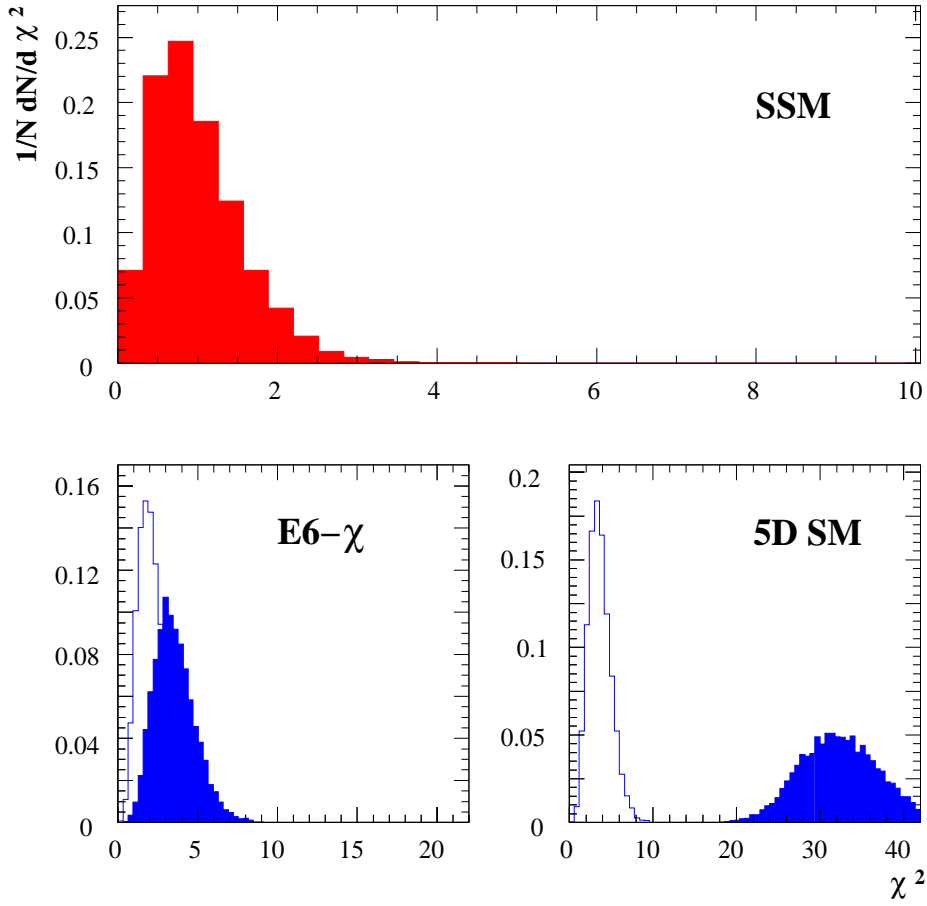}
\caption{The $\chi^2$ distributions obtained for a set of
pseudo-experiments where the SSM is realized with a mass $M_{Z'}$ of 
20~TeV (upper plot). The corresponding distributions for the $E(6)$ $\chi$
and 5D~SM for the same mass scale (full histograms) and for $M$~=~40~TeV 
are also shown for comparison in the lower panels. By integrating these
distributions, one obtains the confidence levels for discriminating 
between these models that are discussed in the text. }
\label{fig:sep}
\end{center}
\end{figure}

An important issue concerns the ability to probe the models, if a
significant discrepancy from the SM predictions were to be observed. Since
the model parameters and the mass scale are {\it a priori} arbitrary, an
unambiguous identification of the realized scenario is difficult. However,
some information can be extracted by testing the compatibility of
different models while varying the mass scale.  
Figure~\ref{fig:sep} shows
an example of such a test. Taking $M$~=~20~TeV, ${\cal{L}}$~=~1~ab$^{-1}$ of
CLIC data at $\sqrt{s}$~=~3~TeV could distinguish the SSM model from
the $E_6$~$\chi$ model at 86\% C.L. 
and from the 5D~SM at 99\%~C.L.  
For a mass scale of 40~TeV, ${\cal{L}}$~=~3~ab$^{-1}$ of CLIC
data at $\sqrt{s}$~=~5~TeV, the corresponding confidence levels become 91\%
and 99\%, respectively. Further sensitivity to the nature of the gauge
bosons could be obtained by studying the polarized forward--backward
asymmetry $A_{\rm FB}^{\rm pol}$ and the left--right asymmetry 
$A_{\rm LR}$ using colliding polarized beams.

The scenarios investigated above address specific models of new physics
beyond the SM. Fermion compositeness or exchanges of very heavy new
particles can be described in all generality by four-fermion contact
interactions~\cite{ref:ci}. These parametrize the interactions beyond the
SM in terms of an effective scale $\Lambda_{ij}$:
\begin{equation}
{\cal L}_{\rm CI} =             \sum_{i,j = \mathrm{L,R}} \eta_{ij}
           \frac{g^2}{\Lambda^2_{ij}}
           (\bar{e_i} \gamma^{\mu} e_i)
           (\bar{f_j} \gamma^{\mu} f_j)\,.
\label{ci_lagr}
\end{equation}
The strength of the interaction is set by convention as $g^2/4\pi$~=~1, and
models can be considered by choosing either $|\eta_{ij}|$~=~1 or
$|\eta_{ij}|$~=~0 as detailed in~Table~\ref{II}. The contact scale $\Lambda$
can be interpreted as an effect of new particles at a mass $M_X$:
$1/\Lambda^2 \propto \lambda^2 / M_X^2$.
%
\begin{table}[!h] 
\caption{Definition of different models of contact interaction}
\label{II}

\renewcommand{\arraystretch}{1.3} 
\begin{center}

\begin{tabular}{ccccccccc}\hline \hline \\[-4mm]
$\hspace*{2mm}$ \textbf{Model} $\hspace*{2mm}$ &
$\hspace*{2mm}$ \textbf{LL} $\hspace*{2mm}$ &
$\hspace*{2mm}$ \textbf{RR} $\hspace*{2mm}$ &
$\hspace*{2mm}$ \textbf{LR} $\hspace*{2mm}$ &
$\hspace*{2mm}$ \textbf{RL} $\hspace*{2mm}$ &
$\hspace*{2mm}$ \textbf{VV} $\hspace*{2mm}$ &
$\hspace*{2mm}$ \textbf{AA} $\hspace*{2mm}$  &
$\hspace*{2mm}$ \textbf{V0} $\hspace*{2mm}$  &
$\hspace*{2mm}$ \textbf{A0} $\hspace*{2mm}$  
\\[4mm]   
\hline \\[-3mm]
$\eta_{\mathrm{LL}}$ & $\pm$~1 & 0 & 0 & 0 & $\pm$~1 & $\pm$~1 
& $\pm$~1 &  0 \\
$\eta_{\mathrm{RR}}$ & 0 & $\pm$~1 & 0 & 0 & $\pm$~1 & $\pm$~1 
& $\pm$~1 & 0 \\
$\eta_{\mathrm{LR}}$ & 0 & 0 & $\pm$~1 & 0 & $\pm$~1 & $\mp$~1 
& 0 & $\pm$~1 \\
$\eta_{\mathrm{RL}}$ & 0 & 0 & 0 & $\pm$~1 & $\pm$~1 & $\mp$~1 
& 0 &$\pm$~1 
\\[3mm] 
\hline \hline
\end{tabular}
\end{center}
\end{table}

\begin{figure}[htbp]
\begin{tabular}{lr}
\includegraphics[width=0.5\textwidth,height=0.55\textwidth]
{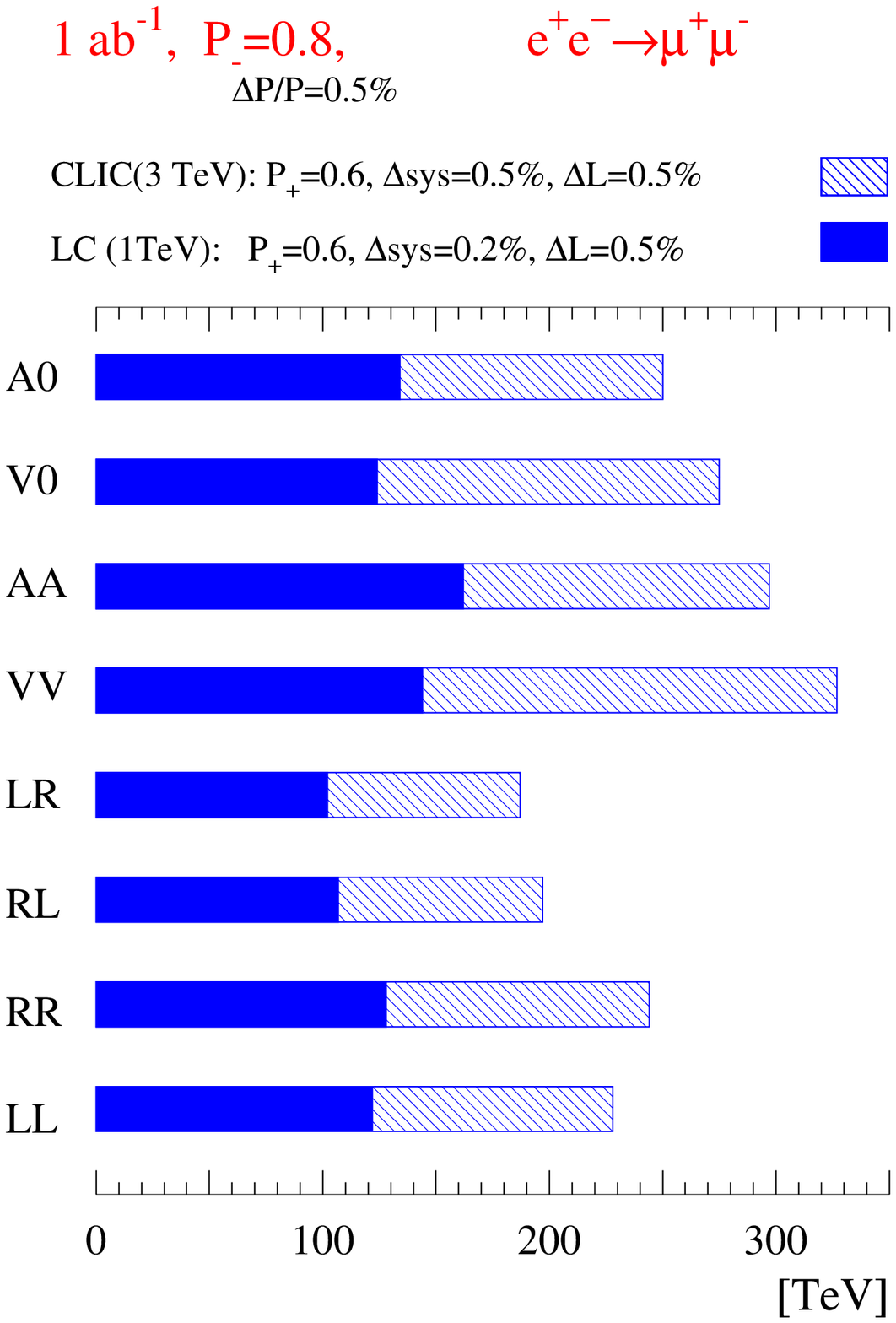} &
\includegraphics[width=0.5\textwidth,height=0.55\textwidth]
{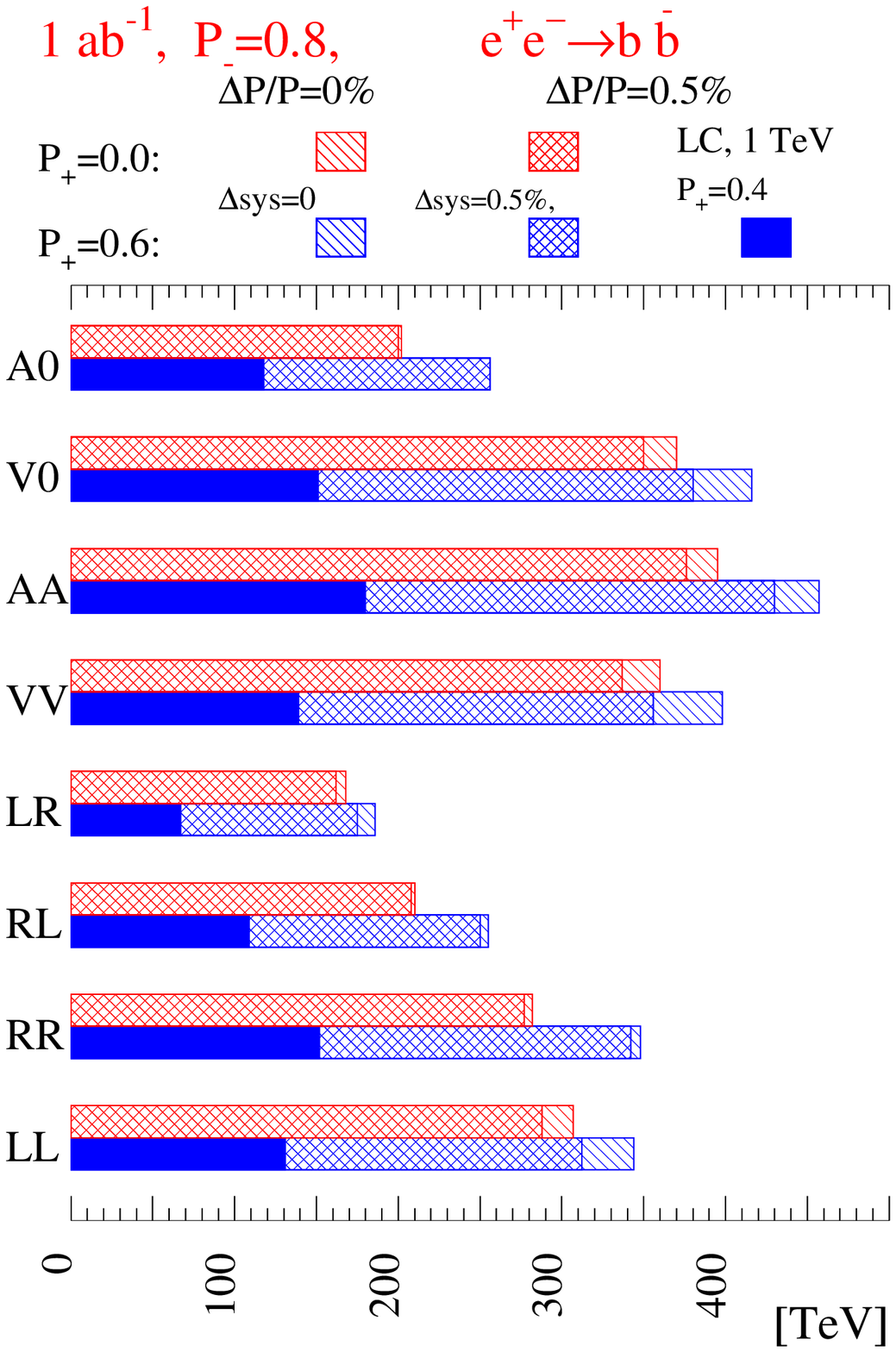} \\
\end{tabular}
\caption{Limits on the scale $\Lambda$ of contact interactions for CLIC
operating at 3~TeV (dashed histogram) compared with a 1~TeV LC (filled
histogram) for different models and the $\mu^+\mu^-$ (left) and $b
\bar{b}$ (right) channels. The polarization of electrons ${\cal{P}}_{-}$
is taken to be 0.8 and that of positrons ${\cal{P}}_{+}$~=~0.6.  For
comparison, the upper bars in the right plot show the sensitivity achieved
without positron polarization. The influence of systematic uncertainties
is also shown.}
\label{fig:ci}
\end{figure}

In order to estimate the sensitivity of electroweak observables to the
contact interaction scale $\Lambda$, the statistical accuracies discussed
in Table~\ref{acc} have been assumed for the $\mu \mu$ and $b \bar{b}$
final states. The assumed systematics of 0.5\% include the contributions
from model prediction uncertainties. Results are given in terms of the
lower limits on $\Lambda$, which can be excluded at 95\% C.L., in
Fig.~\ref{fig:ci}. It has been verified that, for the channels considered
in the present analysis, the bounds for the different $\Lambda_{ij}$ are
consistent. High-luminosity $e^+e^-$ collisions at 3~TeV can probe
$\Lambda$ at scales of 200~TeV and beyond. For comparison, the
corresponding results expected for a LC operating at 1~TeV are also shown.
Beam polarization represents an important tool in these studies.  First,
it improves the sensitivity to new interactions, through the introduction
of the left--right asymmetries $A_{\rm LR}$ and the polarized forward--backward
asymmetries $A_{\rm FB}^{\rm pol}$ in the electroweak fits. If both
beams can be 
polarized to ${\cal{P}}_{-}$ and ${\cal{P}}_{+}$ respectively, the
relevant parameter is the effective polarization defined as ${\cal{P}} =
\displaystyle{ \frac{-{\cal{P}}_{-}+{\cal{P}}_{+}}
{1-{\cal{P}}_{-}+{\cal{P}}_{+}}}$. In addition to the improved
sensitivity, the uncertainty on the effective polarization can be made
smaller than the error on the individual beam polarization measurements.
Secondly, if a significant deviation from the SM prediction were to be
observed, $e^-$ and $e^+$ polarization is greatly beneficial 
to a determination of
the nature of the new interactions. This has been studied in detail for a
LC at 0.5--1.0~TeV~\cite{ref:lc-sr}, and those results also apply,
qualitatively, to a multi-TeV collider.

\begin{figure}[htbp] %
\centerline{\includegraphics[width=9.2cm,angle=90]{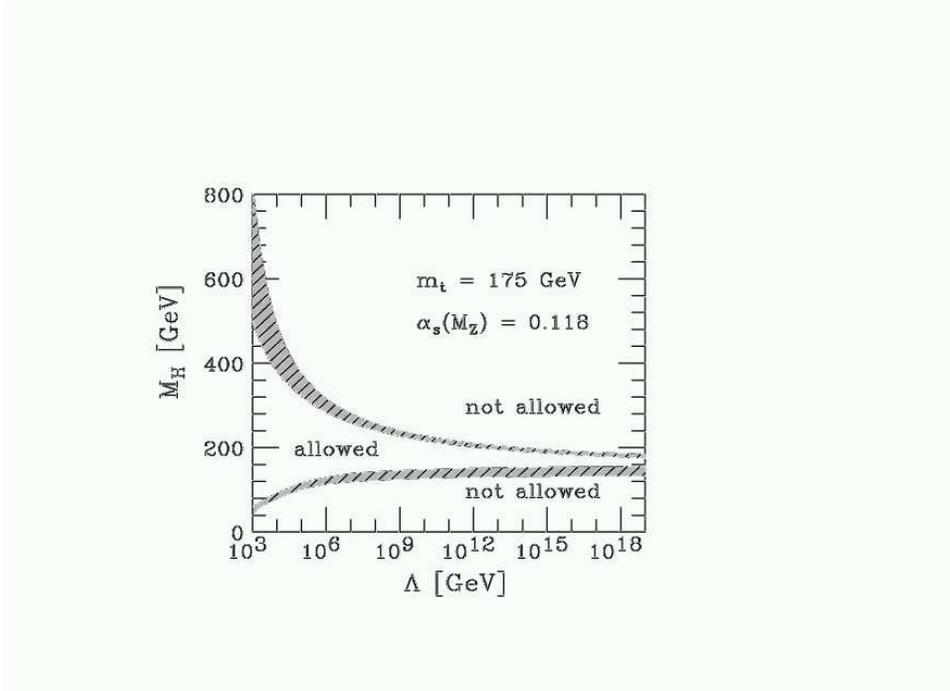}}
\caption{The range allowed for the mass of the Higgs
boson if the SM is to remain valid up to a given scale $\Lambda$.
In the upper part of the plane, the effective potential  blows up,
whereas in the lower part the present electroweak vacuum is
unstable~\protect \cite{Hambye:1997ax}.}
\label{sec6:lighthiggs}
\end{figure}

Using the scaling law, the expected gain in reach on $\Lambda$ for 5
ab$^{-1}$ and a 5~TeV (10~TeV) $e^+e^-$ collider would be 400--800~GeV
(500--1000~GeV). This is a very exciting prospect, {\it if} for the
`doomsday' scenario where in some years from now only a light Higgs has
been discovered, and no sign of other new physics has been revealed by the
LHC or a~TeV-class LC. Indeed, if the Higgs particle is light, i.e. below
150~GeV or so, then the SM cannot be stable up to the GUT or
Planck scale, and a new mechanism is needed to stabilize it, as shown in
Fig.~\ref{sec6:lighthiggs}~\cite{Hambye:1997ax}: only a narrow corridor of
Higgs masses around 180~GeV allow an extrapolation of the SM up to the Planck
scale without introduction of any new physics.  For example, for a Higgs with
a mass in the region of 115--120~GeV, the SM will hit a region of
electroweak unstable vacuum in the range of 100--1000~TeV. Hence, if the
theoretical assessment of Fig.~\ref{sec6:lighthiggs} remains valid, and
the bounds do not change significantly (which could happen following a change
in the top-quark mass from, e.g. new measurements at the Tevatron) {\it
and} the Higgs is as light as 120~GeV, then the signature of new physics
cannot escape precision measurements at CLIC.

Finally, we note that straightforward left--right asymmetry measurements in
M\o ller scattering, as observed in $e^-e^-$ interactions,
 can be used as sensitive probes of new physics effects
due to, say, the existence of higher-mass $Z'$ bosons, doubly-charged
scalars (which might belong to an extended Higgs sector), or the presence
of extra dimensions~\cite{Czarnecki:2000ic}.  The running of
$\sin^2\theta_W$ with $Q^2$ can be measured over a large parameter range
to probe for such novel effects, in a single experiment. The added energy
reach of CLIC will be of major importance for the sensitivity of such
studies. As an example: assuming 90\% polarized
beams at a CLIC energy of 3~TeV, $e^-e^-$ interactions will be sensitive
to interference effects up to a compositeness scale of $\sim 460$~TeV, far
outdistancing the Bhabha scattering sensitivity even if the electron (but
not the positron) is polarized. For the same integrated luminosity, the
sensitivity to $\Lambda$ is about a factor 1.6 larger in $e^-e^-$
scattering, compared with $e^+ e^-$ scattering.

\subsection{Triple-Gauge-Boson Couplings}

Another example of precision measurements is the determination of
triple-gauge-boson couplings (TGCs), studied, e.g. via $e^+e^- \rightarrow
W^+W^-$.  An important feature of the electroweak SM is
the non-Abelian nature of its gauge group, which gives rise to gauge-boson
self-interactions, in particular to TGCs. A precision measurement of these
interactions at high energies will be a crucial test of the validity of
the SM, given that many of its extensions predict deviations, typically
through the effects of new particles and couplings in radiative
corrections: see~Ref.~\cite{Diehl:2002nj}, for example.

\begin{figure}
\begin{center}
\epsfig{file=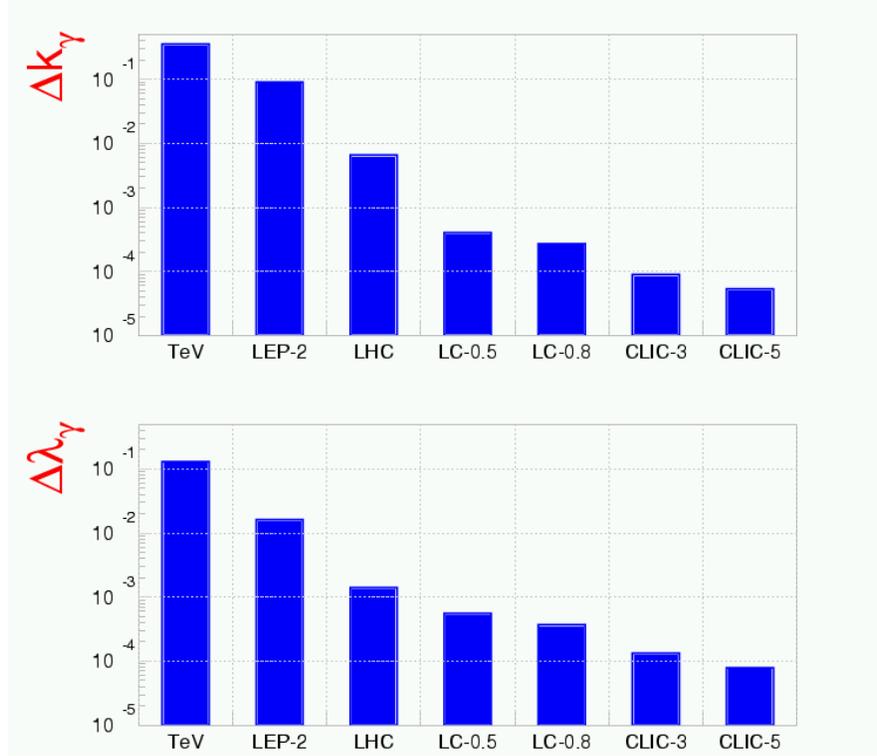,width=11.7cm}
\caption{Precision of triple gauge coupling (TGC) measurements at 
different accelerators, for one year of high-luminosity data
taking~\cite{barklow}}
\label{sec6:tgcs}
\end{center}
\end{figure}

An initial study using~TeV-class LC results for background and detector
effects suggests that the sensitivity to the anomalous $\Delta
\lambda_{\gamma}$ and $\Delta \kappa_{\gamma}$ couplings (which are zero
in the SM) for CLIC at 3 (5)~TeV amounts to roughly 
1.3~$\times$~10$^{-4}$ (0.8~$\times$~10$^{-4}$) 
and 0.9~$\times$~10$^{-4}$ (0.5~$\times$~10$^{-4}$)
respectively, for 1 ab$^{-1}$ of data~\cite{barklow}. A comparative
study is shown in~Fig.~\ref{sec6:tgcs}. The TGC errors are statistical
only, but only $e$ and $\mu$ semileptonic $W W$ events have been included
in the analysis. It has been argued heuristically that the statistical
error from $e$ and $\mu$ semileptonic $W W$ events approximates well the
total statistical and systematic error that is obtained when all $W W$
events are analysed. This was shown to be true at LEP~2, and was also
found to be true in a complete analysis of TGCs for a $\sqrt{s}$~=~500~GeV
analysis. We infer that a multi-TeV collider such as CLIC could probe
these couplings a factor 2--4 times more 
precisely than a~TeV-class~LC.

Allowing for complex couplings, in a form factor approach
the most general $\gamma WW$ and $ZWW$
vertex functions lead to 28 real parameters
altogether~\cite{Hagiwara:1986vm}. All four LEP collaborations have
investigated TGCs and found agreement with the SM. The 68\% C.L.\ errors
for the combined results~\cite{Group:2002mc} are of order 0.02 for $g_1^Z$
and $\lambda_{\gamma}$, and of order 0.06 for $\kappa_{\gamma}$.  The
tightest constraints for the real and imaginary parts of $C$- and/or
$P$-violating
couplings~\cite{Acciarri:1999kn,Abbiendi:2000ei,Heister:2001qt} are of
order 0.1 to 0.6.  All these values correspond to single-parameter
fits. Since only fits with up to three parameters have been performed,
only a small subset of couplings has been considered at a time, thereby
neglecting possible correlations between most of them.  Moreover, many
couplings, notably the imaginary parts of $C$- and $P$-conserving
couplings, have so far been dropped from the experimental analyses.
 
At a future linear $e^+e^-$ collider one will be able to study these
couplings with unprecedented accuracy.  A process particularly suitable
for this is $W$ pair production, where both the $\gamma WW$ and $ZWW$
couplings can be measured at the scale given by the 
centre-of-mass energy~$\sqrt{s}$.

Given the intricacies of a multidimensional parameter space, the full
covariance matrix for the errors on all 28 couplings should be
studied. The high statistics needed for that is available both at a
first-generation LC in the 1-TeV range and at a multi-TeV machine
like CLIC.  With an integrated luminosity of 3~ab$^{-1}$
and unpolarized beams at $\sqrt{s}$~=~3~TeV, about 1.5~million
$W$~pairs are produced before cuts.

In experimental analyses of TGCs and various other processes, optimal
observables~\cite{Atwood:1991ka} have proved to be a useful tool to
extract physics parameters from the event distributions.  These
observables are constructed so as to have the smallest possible statistical
errors. Thanks to this property, they are also a convenient means of
determining the theoretically achievable sensitivity in a given process. In
addition, they take advantage of the discrete symmetries of the cross
Section. In $W$~pair production, the covariance matrix of these
observables consists of four blocks that correspond to CP-even or
CP-odd TGCs and to their real and imaginary parts. For this process it
is convenient to use the following parameters:
\begin{eqnarray}
g_1{\rm ^L} & = & 4\sin^2\theta_W\, g_1^{\gamma} + (2 - 4\sin^2\!\theta_W )\,
\xi\, g_1^Z\,,	\\
g_1^{\rm R} & = & 4\sin^2\theta_W\, g_1^{\gamma} - 4\sin^2\!\theta_W\, \xi\,
g_1^Z\,,		
\nonumber
\end{eqnarray}
and similarly for the other couplings, where $\xi = s/(s-m_Z^2)$.  The
linear combinations are chosen such that the couplings 
$h^{\rm L}$ ($h^{\rm R}$) occur
in the amplitudes of left- (right-) handed $e^-$ and right- (left-) handed
$e^+$.  With unpolarized beams one mainly measures the $h^{\rm L}$, which are
strongly enhanced by the interference with neutrino exchange in the
$t$ channel. Table~\ref{tab:energy} shows the achievable errors on the
real parts of the couplings in the presence of all
TGCs~\cite{Diehl:2002nj}.  Notice the increase in sensitivity for higher
energies, especially for ${\rm Re} \, \Delta \kappa_{\rm L,R}$, ${\rm Re}
\,\lambda_{\rm L,R}$ and ${\rm Re}\, \tilde{\lambda}_{\rm L,R}$.

%
\begin{table*}[!h] 
\caption{Errors on the real parts of CP-even
(top) and -odd (bottom) couplings for different centre-of-mass
energies in units of 10$^{-3}$.  The assumed integrated luminosities
are 500~fb$^{-1}$, 1~ab$^{-1}$ and 3~b$^{-1}$ at
$\sqrt{s}$~=~500~GeV, 800~GeV and 3~TeV,
respectively.  Only those final states are considered where one $W$
decays into a quark--antiquark pair and the other into $e \nu$ or 
$\mu \nu$.  These are the errors that can be obtained in the presence of
all 28 TGCs (the covariance matrices are not shown).}
\label{tab:energy}

\renewcommand{\arraystretch}{1.35} 
\begin{center}

\begin{tabular}{ccccccccc}\hline \hline \\[-4mm]
\boldmath{$\sqrt{s}$}~\textbf{(GeV)}  & 
\textbf{Re}~\boldmath{$\Delta g_1^{\rm L}$} &
  \textbf{Re}~\boldmath{$\Delta \kappa_{\rm L}$} & 
  \textbf{Re}~\boldmath{$\lambda_{\rm L}$} & 
  \textbf{Re}~\boldmath{$g_5^{\rm L}$} & 
  \textbf{Re}~\boldmath{$\Delta g_1^{\rm R}$} & 
  \textbf{Re}~\boldmath{$\Delta\kappa_{\rm R}$}&
  \textbf{Re}~\boldmath{$\lambda_{\rm R}$}&
  \textbf{Re}~\boldmath{$g_5^{\rm R}$}    
\\[4mm]   

\hline \\[-3mm]
500  & 2.6 & 0.85 & 0.59 & 2.0 & 10 & 2.4 & 3.6 & 6.7 \\
800  & 1.6 & 0.35 & 0.24 & 1.4 & 6.2 & 0.92 & 1.8 & 4.8 \\
3000 & 0.93 & 0.051 & 0.036 & 0.88 & 3.1 & 0.12 & 0.36 & 3.2
\\[3mm] 
\hline \hline
\end{tabular}

\vspace*{8mm}

\begin{tabular}{ccccccc}\hline \hline \\[-4mm]
$\hspace*{0.5mm}$ \boldmath{$\sqrt{s}$}~\textbf{(GeV)} $\hspace*{0.5mm}$ & 
$\hspace*{0.5mm}$ \textbf{Re}~\boldmath{$\Delta g_4^{\rm L}$} 
$\hspace*{0.5mm}$ &
$\hspace*{0.5mm}$ \textbf{Re}~\boldmath{$\tilde{\lambda}_{\rm L}$} 
$\hspace*{0.5mm}$ &
$\hspace*{0.5mm}$ \textbf{Re}~\boldmath{$\tilde{\kappa}_{\rm L}$} 
$\hspace*{0.5mm}$ &
$\hspace*{0.5mm}$ \textbf{Re}~\boldmath{$\Delta g_4^{\rm R}$} 
$\hspace*{0.5mm}$ &
$\hspace*{0.5mm}$ \textbf{Re}~\boldmath{$\tilde{\lambda}_{\rm R}$} 
$\hspace*{0.5mm}$ &
$\hspace*{0.5mm}$ \textbf{Re}~\boldmath{$\tilde{\kappa}_{\rm R}$} 
$\hspace*{0.5mm}$ 
\\[4mm]   

\hline \\[-3mm]
500  & 2.5 & 0.60 & 2.7 & 10 & 3.8 & 11 \\
800  & 1.7 & 0.24 & 1.8 & 6.5 & 1.8 & 6.8 \\
3000 & 0.90 & 0.036 & 0.97 & 3.4 & 0.36 & 3.2
\\[3mm] 
\hline \hline
\end{tabular}
\end{center}
\end{table*}


The sensitivity to anomalous TGCs changes considerably in the presence of
initial beam polarization.  With longitudinally-polarized beams, the
strength of the neutrino exchange can in essence be varied freely. In
Table~\ref{tab:polarization} 
we give the errors $\delta h$ on the real
%
\begin{table}[!b] 
\caption{Errors in the real parts of CP-even 
couplings at $\sqrt{s}$~=~3~TeV for different initial beam
polarizations in units of~10$^{-3}$}
\label{tab:polarization}

\renewcommand{\arraystretch}{1.35} 
\begin{center}

\begin{tabular}{cccccccccc}\hline \hline \\[-4mm]
\boldmath{$P^-$} & \boldmath{$P^+$} &
\textbf{Re}~\boldmath{$\Delta g_1^{\rm L}$} &
  \textbf{Re}~\boldmath{$\Delta \kappa_{\rm L}$} & 
  \textbf{Re}~\boldmath{$\lambda_{\rm L}$} & 
  \textbf{Re}~\boldmath{$g_5^{\rm L}$} & 
  \textbf{Re}~\boldmath{$\Delta g_1^{\rm R}$} & 
  \textbf{Re}~\boldmath{$\Delta\kappa_{\rm R}$}&
  \textbf{Re}~\boldmath{$\lambda_{\rm R}$} &
  \textbf{Re}~\boldmath{$g_5^{\rm R}$}    
\\[4mm]   

\hline \\[-3mm]
-- 80\% & +~60\% & 0.54 & 0.028 & 0.021 & 0.50 & 52 & 2.0 & 6.0 & 53 \\
-- 80\% & 0     & 0.68 & 0.036 & 0.026 & 0.64 & 19 & 0.71 & 2.2 & 19 \\
0       & 0     & 0.93 & 0.051 & 0.036 & 0.88 & 3.1 & 0.12 & 0.36 & 3.2 \\
+~80\%  & 0     & 2.4 & 0.14 & 0.089 & 2.2 & 1.0 & 0.040 & 0.12 & 1.1 \\
+~80\%  & -- 60\% & 4.4 & 0.28 & 0.17 & 4.2 & 0.56 & 0.022 & 0.060 & 0.59
\\[3mm] 
\hline \hline
\end{tabular}
\end{center}
\end{table}
%
parts of the CP-conserving couplings (in the presence of all couplings)
for $\sqrt{s}$~=~3~TeV and various combinations of beam
polarizations. For all couplings $h^{\rm L}$ and all couplings 
$h^{\rm R}$ we find 
roughly the following gain or loss in sensitivity, with luminosities given
in the caption of Table~\ref{tab:energy}. Turning on an $e^-$~polarization of
--80\% we gain a factor of 1.5 for $h^{\rm L}$ and lose a factor of 6 for
$h^{\rm R}$. If in addition $P^+$~=~+60\%, we gain a factor of 2 for
$h^{\rm L}$ and lose a factor of 17 for $h^{\rm R}$ compared to
unpolarized beams.  For $P^-$~=~+~80\% we lose a factor of 2.5 for
$h^{\rm L}$ and gain a factor of 3 for $h^{\rm R}$.   
If furthermore $P^+$~=~-- 60\%, we lose a factor of 5 for $h^{\rm L}$
and gain a factor of 5.5 for~$h^{\rm R}$ with respect to unpolarized
beams.  Especially for the right-handed couplings, the gain from
having both beams polarized is thus~appreciable.
%

To see the effects of beam polarization in the full parameter space, it is
advantageous to use a particular basis for the
couplings~\cite{Diehl:1997ft,Diehl:2002nj}. Here one diagonalizes
simultaneously the covariance matrix of the observables and transforms the
part of the cross section that is quadratic in the couplings to the unit
matrix.  In this way one obtains a set of coupling constants that are
naturally normalized for the particular process and---in the limit of
small anomalous couplings---can be measured without statistical
correlations.  The total cross section acquires a particularly simple form
and provides additional~constraints.

Using this method one finds in particular that for unpolarized beams or
longitudinal polarization the process is---in lowest order---insensitive
to the linear combination ${\rm Im}(g_1^{\rm R} + \kappa_R)$ of imaginary
CP-conserving TGCs.  This can be traced back to the analytic expression
of the differential cross section. We remark, however, that this parameter
is measurable with transverse beam polarization, which has been quantified 
in~Ref.~\cite{Diehl:2003qz}.

In contrast to the form-factor approach where everything apart from the
$\gamma WW$ and $ZWW$ vertices is assumed to be SM-like, one may
consider a locally $SU(2)\times U(1)$ invariant effective Lagrangian
with operators of dimension 6 built from SM boson fields, the effective
Lagrangian approach (ELb~approach) of~\cite{Nachtmann:2004ug}. 
 After spontaneous symmetry
breaking some operators lead to new three- and four-gauge-boson interactions,
some contributing to the diagonal and off-diagonal kinetic terms of the gauge
bosons and to the mass terms of the $W$ and $Z$~bosons.  This requires a
renormalization of the gauge-boson fields, which, in turn, modifies the
charged- and neutral-current interactions, although none of the additional
operators contain fermion fields.  Bounds on the anomalous couplings from
electroweak precision measurements at LEP and SLD are correlated with the
Higgs-boson mass~$m_H$.  Rather moderate values of anomalous couplings allow
$m_H$ up to 500~GeV~\cite{Nachtmann:2004ug}.  
The translation of the bounds on the couplings in
the reaction $e^+e^- \rightarrow WW$ at a future LC from the
form-factor approach to the ELb~approach is not straightforward.  
It is done~in~\cite{Nachtmann:2004ug} for this process by defining new
{\em effective} $ZWW$ couplings.

\section{EWSB Without the Higgs Boson}

Present precise electroweak data are consistent with the realization of
the Higgs mechanism with a light elementary Higgs boson. But as the Higgs
boson has so far eluded the direct searches, it remains important to
assess the sensitivity of future colliders to strong electroweak 
symmetry-breaking (SSB) scenarios.

\subsection{\boldmath{$W_L W_L$} Scattering}

In the scenario where no Higgs boson with large gauge boson couplings and a
mass less than about 700~GeV is found, then the $W^{\pm},Z$ bosons are
expected to develop strong interactions at scales of order 1--2~TeV.
Generally one expects an excess of events above SM
expectation, and, possibly, resonance formation. The reaction $e^+e^-
\rightarrow \nu\overline{\nu}W^+_LW^-_L $ at 3~TeV was studied using two
approaches.

The chirally-coupled vector model~\cite{barger} for $W_LW_L$ scattering
describes the low-energy behaviour of a technicolour-type model with a
techni-$\rho$ vector resonance V (spin-1, isospin-1 vector resonance). The
mass of the resonance can be chosen at will, and the cases 
$M\sim$~1.5, 2.0 and 2.5~TeV were~studied. 

In a second approach, the prescription given in~Refs.~\cite{bod,forshaw} using
the electroweak chiral Lagrangian (EHChL) formalism is applied. The Higgs
terms in the SM Langrangian are replaced by terms in the next order of the
chiral expansion
\begin{equation}
{\cal L}~=~{\cal L}^{(2)}
+\alpha_4 \, [\langle (D_{\mu}U) U^\dagger (D^{\nu}U)U^{\dagger} \rangle]^2
+\alpha_5 \, [\langle (D_{\mu}U) U^\dagger(D^{\mu}U) U^{\dagger} \rangle]^2\,,
\label{chL}
\end{equation}
where the matrices $U$ represent chiral boson fields, and the parameters
$\alpha_4$ and $\alpha_5$ quantify our ignorance of the new physics.
Unitarity corrections are important for energies larger than 1~TeV, and
the Pad\'e approach has been used to estimate these. High-mass vector
resonances will be produced in $WW$ and $WZ$ scattering for certain
combinations of $\alpha_4$ and $\alpha_5$.

\begin{figure}[t] 
\begin{center}
\begin{tabular}{c c}
\epsfig{file=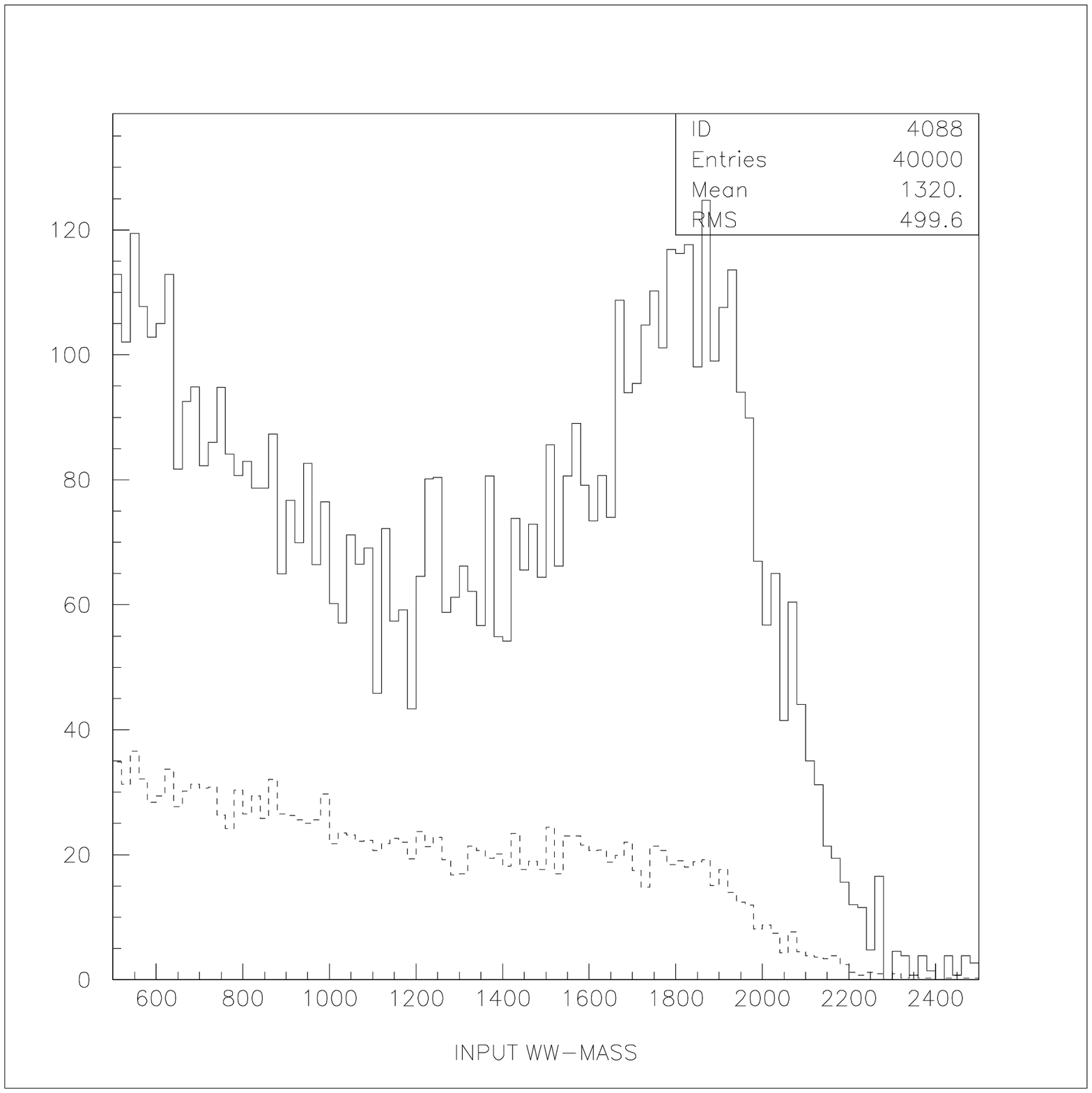,bbllx=30,bblly=150,bburx=560,bbury=680,
width=7.0cm,clip=} &
\epsfig{file=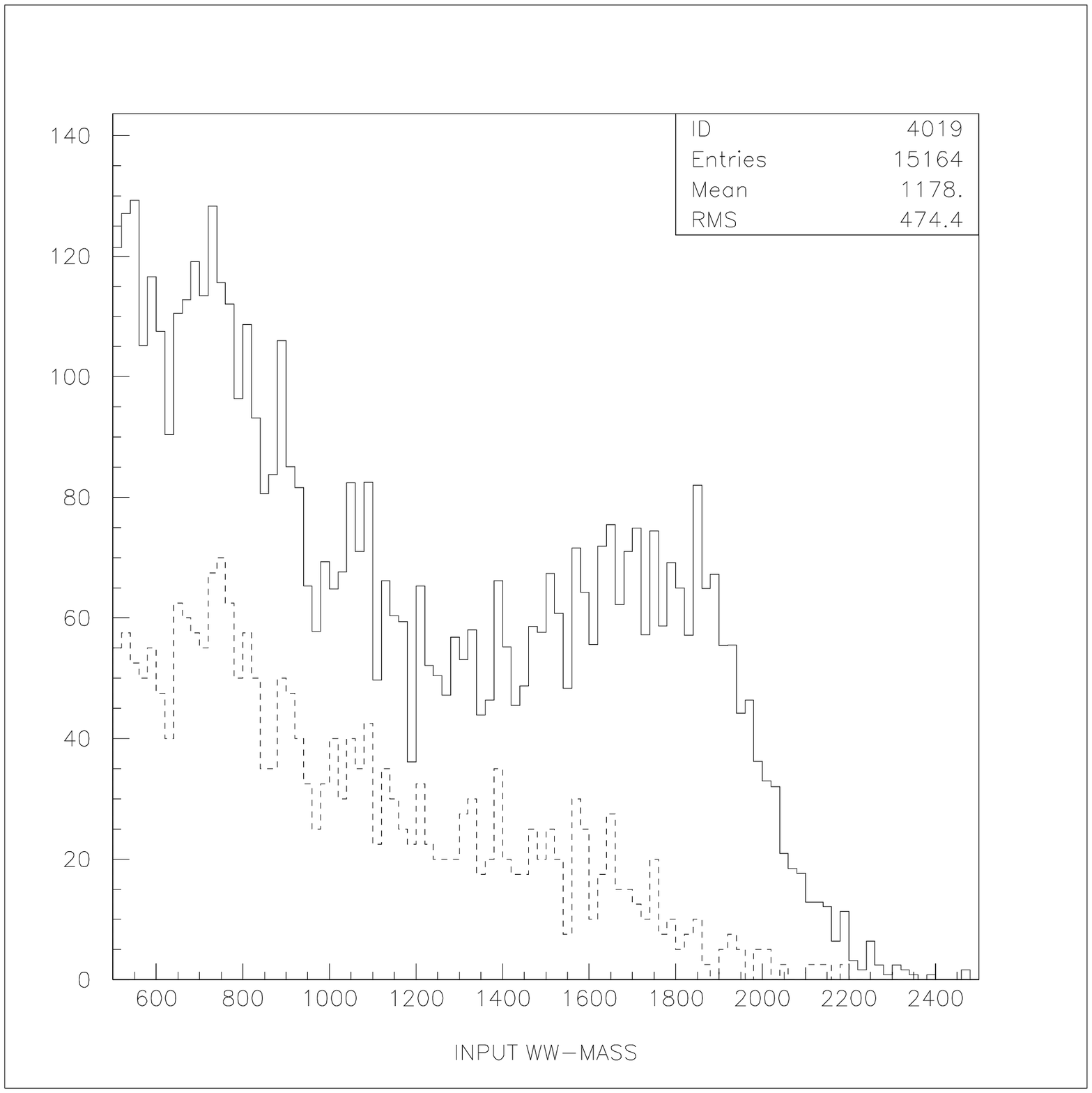,bbllx=30,bblly=150,bburx=560,bbury=680,
width=7.0cm,clip=} \\
\end{tabular}
\caption{Mass spectrum for $WW$ scattering and $WZ$ and $ZZ$
scattering backgrounds for $\alpha_5$~=~-- 0.002, $\alpha_4$~=~0,
before (left) and after (right) detector smearing and addition
of $\gamma\gamma$ background~\protect~\cite{deroeck}}
\label{fig:wwfull}
\end{center}
\end{figure}

The total cross section for $WW\rightarrow WW$ scattering with the values
$\alpha_4$~=~0  and $\alpha_5$~=~-- 0.002, amounts to 12~fb in $e^+e^-$
collisions at 3~TeV, and is measurable at a high luminosity LC. With these
parameters a broad resonance is produced at 2~TeV in the $WW$ invariant
mass, as shown in~Fig.~\ref{fig:wwfull}. A detector study is performed for
this scenario, implemented in the PYTHIA generator~\cite{forshaw}. Events
are selected with the following cuts: 
$p^W_T >$~150~GeV, $|\cos\theta^W|<$~0.8; $M_{W}>$~500~GeV; 
$p_T^{WW}<$~300~GeV; and (200~$< M_{\rm rec}$~(GeV) $<$~1500), 
with $M_{\rm rec}$ the recoil mass. Cross sections including these
cuts for different masses and widths are calculated with the program
of~Ref.~\cite{barger} and are given in Table~\ref{tab:ww}~\cite{BarklowMM}.
%
\begin{table}[!h] 
\caption{Cross sections for vector resonances in $WW$ scattering,
with the cuts given in the text}
\label{tab:ww}

\renewcommand{\arraystretch}{1.3} 
\begin{center}

\begin{tabular}{cccc}\hline \hline \\[-4mm]
$\sqrt{s}$ &  M=1.5 TeV & M=2.0 TeV & M=2.5 TeV\\[1mm]
\\[1mm]   

%
3~TeV &
$\Gamma$~=~35~GeV & $\Gamma$~=~85~GeV & $\Gamma$~=~250~GeV \\[2mm]
\hline \\[-5mm]  
$\sigma$ (fb) & 4.5 & 4.3 & 4.0
\\[3mm] 
\hline \hline
\end{tabular}
\end{center}
\end{table}


\vspace*{-2mm}

\begin{figure}[htbp] 
\begin{center}
\includegraphics[height=5.0cm]{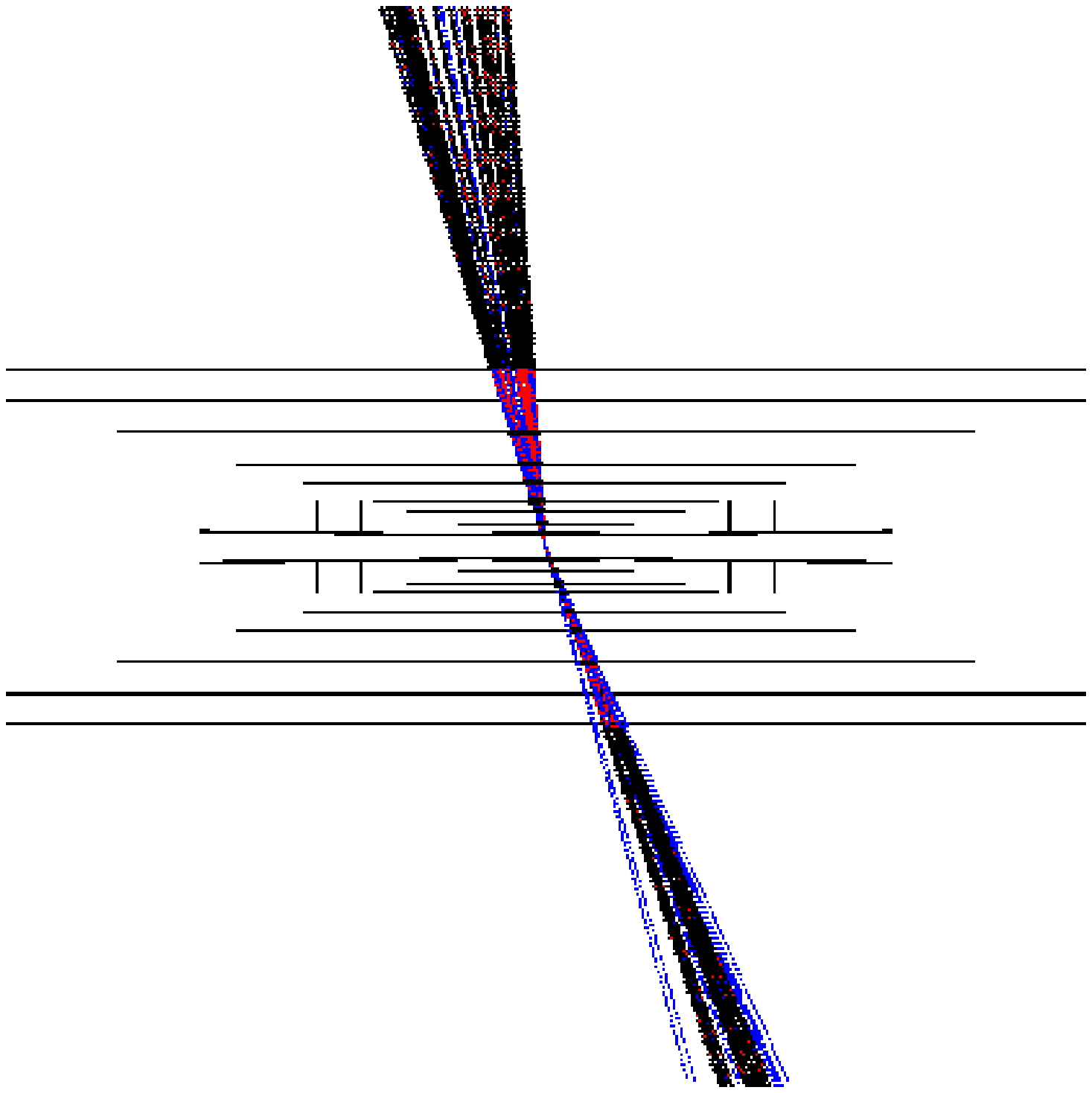}
\includegraphics[height=5.0cm]{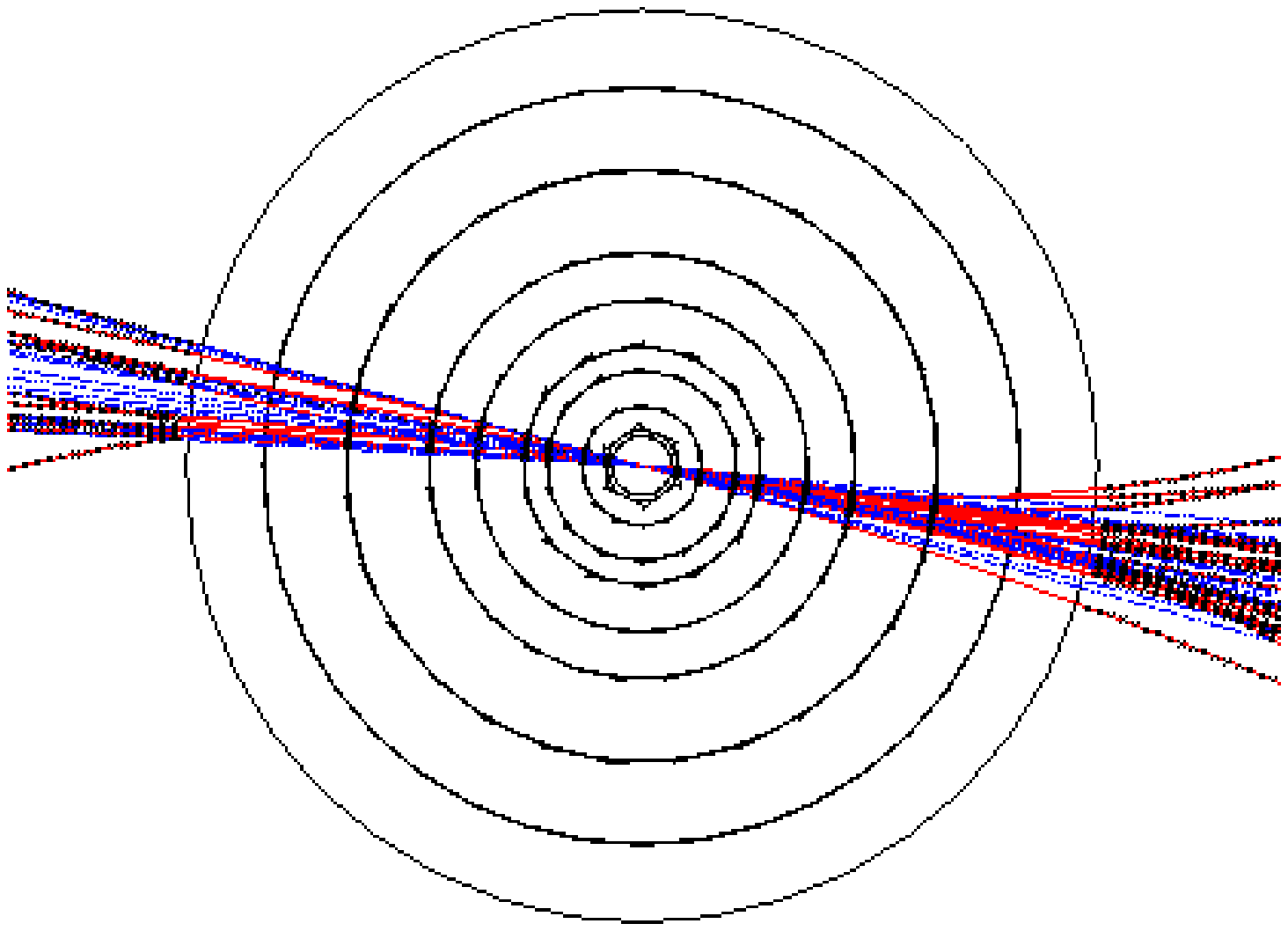}
\caption{Two views of an event in the central detector, of the
type $e^+e^-\to WW\nu\nu\to$ 4 jets $\nu\nu$,
from a resonance with $M_{WW}$~=~2~TeV~\protect~\cite{deroeck}}
\label{fig:wwevent}
\end{center}
\end{figure}

The big advantage of an $e^+e^-$ collider is the clean final state, which
allows the use of the hadronic decay modes of the $W$ bosons in
selecting and reconstructing events. Four jets are produced in the decays
of the two $W$ bosons. However, the boost from the decay of the
heavy resonance makes the two jets of each $W$ very collimated and appear
close to each other, as shown in~Fig.~\ref{fig:wwevent}. With the present
assumptions on the energy flow in SIMDET, the resolution to reconstruct
the $W,Z$ mass is~about~7\%.

A full spectrum, which contains contributions from all the channels
$ZZ\to ZZ, ZZ\to WW, ZW\to ZW, WW\to ZZ$
and $WW\to WW$ is shown for 1.6 ab$^{-1}$ in
Fig.~\ref{fig:wwfull}, before and after detector smearing with parameters
$\alpha_5$~=~-- 0.002, $\alpha_4$~=~0~\cite{deroeck}.

Clearly heavy resonances in $WW$ scattering can be detected at CLIC. The
signal is not heavily distorted by detector resolution and background.
Depending on the mass and width of the signal, about a 1000 events per
year could be fully reconstructed in the four-jet mode at CLIC. Good
energy and track reconstruction will be important to remove
backgrounds and reconstruct the resonance parameters (and hence the
underlying model parameters) accurately.

The particular attractiveness of the study of this possible phenomenon at
CLIC energies using the $e^-e^-$ channel is the availability of highly
polarized electron beams, which either couple to $W^-$ (if the electron is
left-handed ) or decouple (if right-handed). The ease with which the
electron helicity can be inverted means that a real resonance effect in
$W^-W^-$ interactions can trivially be separated from any 
obfuscating~background.

\subsection{Degenerate BESS Model}

Strong symmetry-breaking models are based on low-energy effective
Lagrangians, which provide a phenomenological description of the Goldstone
boson dynamics. Possible new vector resonances
produced by the strong interaction responsible for the electroweak
symmetry-breaking can be introduced in this formalism as gauge bosons of a
hidden symmetry. A description of a new triplet of vector resonances is
obtained by considering an effective Lagrangian based on the symmetry
$SU(2)_L\otimes SU(2)_R\otimes SU(2)_{\rm local}$~\cite{bess}. The new vector
fields are a gauge triplet of the $SU(2)_{\rm local}$. They acquire mass as
the $W^{\pm}$ and the $Z^0$ bosons. By enlarging the symmetry group of the
model, additional vector and axial-vector resonances can be introduced.

The degenerate BESS model (D-BESS)~\cite{dbess} is a realization of
dynamical electroweak symmetry-breaking with decoupling. The D-BESS model
introduces two new triplets of gauge bosons, which are almost degenerate in
mass: ($L^\pm$, $L_3$), ($R^\pm$, $R_3$). The extra parameters are a new
gauge coupling constant $g''$ and a mass parameter $M$, related to the
scale of the underlying symmetry-breaking sector. In the charged sector,
the $R^\pm$ fields are not mixed and $M_{R^\pm}=M$, while
$M_{{L}^\pm}\simeq M (1+x^2)$ for small $x=g/g''$, with $g$ the usual
$SU(2)_W$ gauge coupling constant. The $L_3$, $R_3$ masses are given by
$M_{L_3}\simeq M\left(1+ x^2\right),~~ M_{R_3}\simeq M \left(1+ x^2 \tan^2
\theta\right)$, where $\tan \theta = g'/g$ and $g'$ is the usual $U(1)_Y$
gauge coupling constant. These resonances are narrow, as seen in
Fig.~\ref{figwidth}, and almost degenerate in mass with
$\Gamma_{L_3}/M\simeq$~0.068$x^2$ and 
$\Gamma_{R_3}/M\simeq$~0.01$x^2$,
while the neutral mass splitting is $\Delta M/M=(M_{L_3}-M_{R_3})/M \simeq
\left( 1-\tan^2 \theta \right) x^2\simeq$~0.70$x^2$.

\begin{figure}[t]
\begin{center}
\begin{tabular}{cc}
\includegraphics[width=0.42\textwidth,height=0.42\textwidth]
{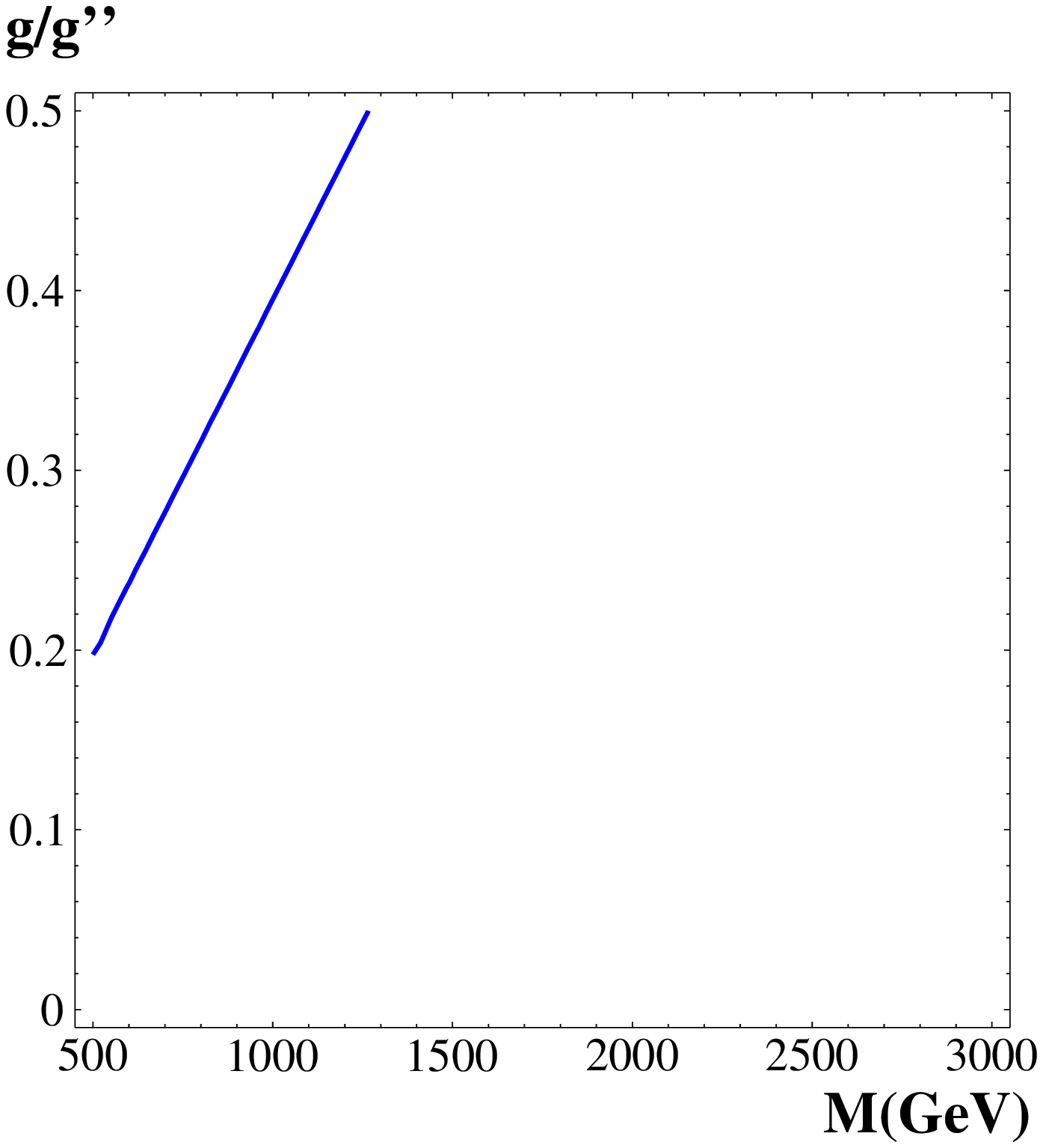} & \hspace*{4mm}
\includegraphics[width=0.42\textwidth,height=0.42\textwidth]
{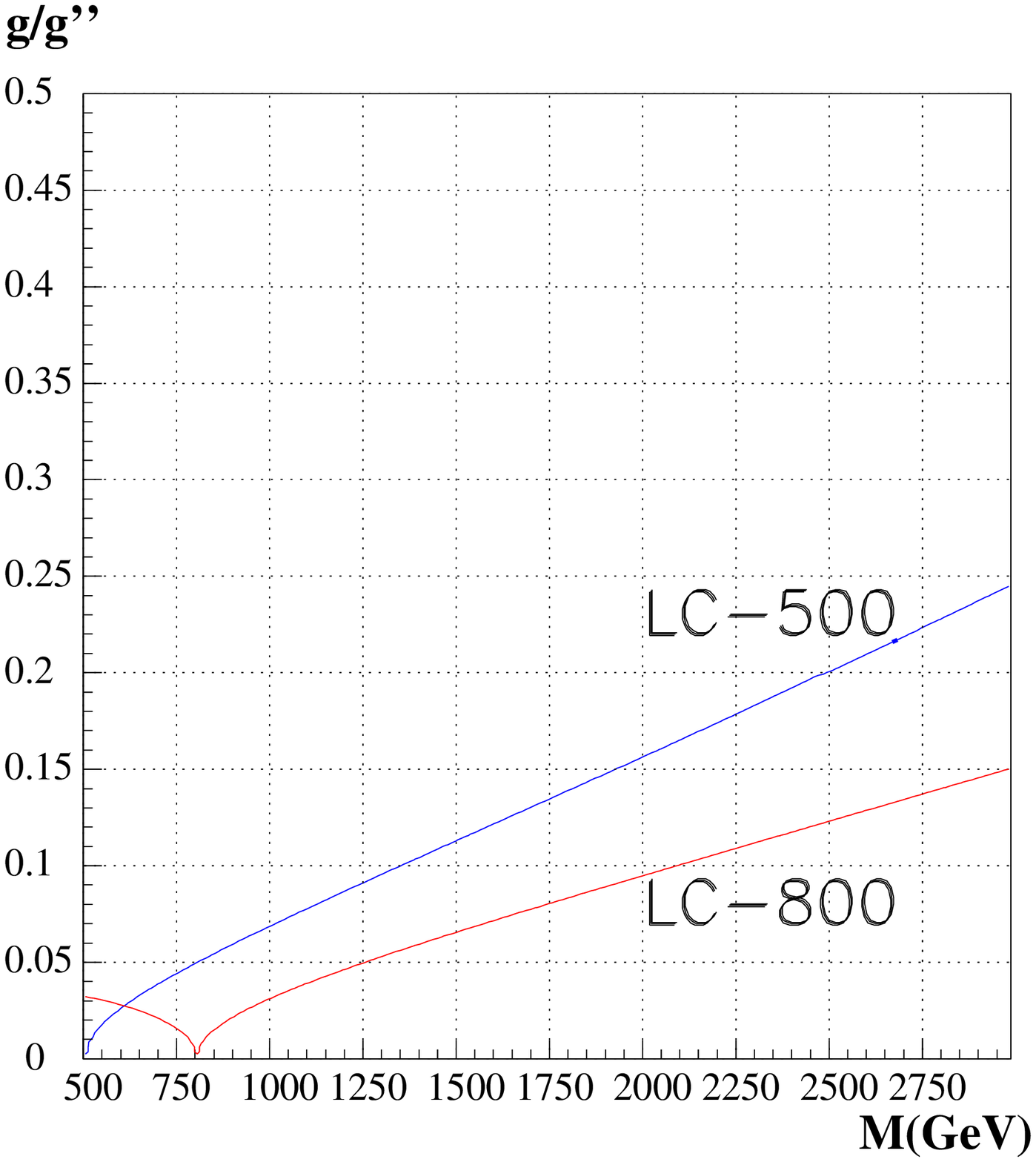}\\
\end{tabular}
\caption {95\%~C.L.\ contours in the plane $(M,~ g/g'')$ from the
present $\epsilon$ measurements (left) and from
measurements of $\sigma_{\mu^+\mu^-}$, $\sigma_{b \bar b}$,
$A_{\rm FB}^{\mu\mu}$, $A_{\rm FB}^{bb}$ at $e^+e^-$ linear colliders
with $\sqrt{s}$~=~500~(800)~GeV  and ${\cal{L}}$~=~1~ab$^{-1}$
(right). The allowed regions are below the curves.}
\label{dom:fig1}
\end{center}
\end{figure}

This model respects the present bounds from electroweak precision data,
since the $S,T,U$ (or $\epsilon_1, \epsilon_2, \epsilon_3$) parameters
vanish at the leading order in the limit of large $M$, because of an
additional custodial symmetry. Therefore, electroweak data set only loose
bounds on the parameter space of the model. We have studied these bounds
by considering the latest experimental values of the $\epsilon$ parameters
coming from the high-energy data~\cite{epsi}: 
$\epsilon_1$~=~(5.4~$\pm$~1.0)~$\times$~10$^{-3}$, 
$\epsilon_2$~=~(--~9.7~$\pm$~1.2)~$\times$~10$^{-3}$, 
$\epsilon_3$~=~(5.4~$\pm$~0.9)~$\times$~10$^{-3}$. 
We have included radiative corrections, taken
to be the same as in the SM, with the Higgs mass as a
cut-off~\cite{dbess}. For $m_t$~=~175.3~GeV and $m_H$~=~1000~GeV, one
has~\cite{epsilon}: $\eps_1^{\rm rad}$~=~3.78~$\times$~10$^{-3}$, 
$\eps_2^{\rm rad}$~=~--~6.66~$\times$~10$^{-3}$, 
$\eps_3^{\rm rad}$~=~6.65~$\times$~10$^{-3}$. The
95\%~C.L.\ bounds on the parameters of the D-BESS model are shown in
Fig.~\ref{dom:fig1}. Comparable bounds come from the direct search at the
Tevatron~\cite{dbess}.

The LHC can discover these new resonances, which are produced via $q
\bar q$ annihilations, through their leptonic decay $q{\bar q'}\to
L^\pm,W^\pm\to (e \nu_e) \mu\nu_\mu$ and $q{\bar q}\to L_3,R_3,Z,\gamma\to
(e^+e^-)\mu^+\mu^-$.

The relevant observables are the dilepton transverse and
invariant masses. The main backgrounds left in these channels
after the lepton isolation cuts are the Drell--Yan processes with
SM gauge bosons exchange in the electron and muon channels. A
study has been performed using a parametric detector
simulation~\cite{redi}. Results are given in Table~\ref{tab:4}
\begin{table}[hb] 
\caption{Sensitivity to production of the ${\rm L}_3$ and ${\rm R}_3$
  D-BESS resonances at the LHC for ${\cal{L}}$~=~100~(500)~fb$^{-1}$
with $M$~=~1.2~(3)~TeV and accuracy on the mass splitting at 
CLIC for ${\cal{L}}$~=~1~ab$^{-1}$.} 
\label{tab:4}

\renewcommand{\arraystretch}{1.2} 
\begin{center}

\begin{tabular}{cccccc}
\hline \hline\\[-4mm]
$\hspace*{3mm}$ \boldmath{$g/g''$} $\hspace*{3mm}$ & 
$\hspace*{4mm}$ \boldmath{$M$} $\hspace*{4mm}$ & 
$\hspace*{4mm}$ \boldmath{$\Gamma_{{\rm L}_3}$} $\hspace*{4mm}$ & 
$\hspace*{4mm}$ \boldmath{$\Gamma_{{\rm R}_3}$} $\hspace*{4mm}$ &
$\hspace*{3mm}$ \boldmath{$S/\sqrt{S+B}$} $\hspace*{3mm}$ & 
$\hspace*{3mm}$ \boldmath{$\Delta M$} $\hspace*{3mm}$ \\
& \textbf{(GeV)} & \textbf{(GeV)} &  \textbf{(GeV)} & 
 \textbf{LHC (\boldmath{$e+\mu$})} &  \textbf{CLIC}
\\[1.5mm]  \hline\\[-4mm]
1000 & 0.7 & 0.1 &17.3 &
\\
0.2 & 1000 & 2.8 & 0.4 & 44.7 &
\\
0.1 & 2000 & 1.4 & 0.2 &3.7 &
\\
0.2 & 2000 & 5.6 & 0.8 & 8.8&
 \\
0.1 & 3000 & 2.0 & 0.3 &(3.4)& 23.20 $\pm$~0.06
\\
0.2 & 3000 & 8.2 & 1.2 &(6.6)& 83.50 $\pm$~0.02
\\[2mm] 
\hline \hline
\end{tabular}
\end{center}
\end{table}
%
for the combined electron and muon channels for
${\cal{L}}$~=~100~fb$^{-1}$, except for $M$~=~3~TeV where
500~fb$^{-1}$ are assumed.
The discovery limit at the LHC, with ${\cal{L}}$~=~100~fb$^{-1}$,
is $M\sim$~2~TeV for $g/g''$~=~0.1. Beyond discovery, the
possibility to disentangle the characteristic double-peak
structure depends strongly on $g/g''$ and smoothly on the mass.

The LC can also probe this multi-TeV region through virtual effects in the
cross sections for 
$e^+e^-\to {{\rm L}_3},{{\rm R}_3},Z,\gamma \to f \bar f $, at 
centre-of-mass energies below the resonances. Thanks to the presence of new
spin-1 resonances, the annihilation channel in $f \bar f$ and $W^+W^-$
has a better sensitivity than the fusion channel. In the case of the
D-BESS model, the ${\rm L}_3$ and ${\rm R}_3$ states are not strongly
coupled to $W$ 
pairs, making the $f\bar f$ final states the most favourable channel for
discovery. The analysis at $\sqrt{s}$~=~500~GeV and $\sqrt{s}$~=~800~GeV is
based on $\sigma_{\mu^+\mu^-}$, $\sigma_{b \bar b}$, $A_{\rm FB}^{\mu\mu}$ and
$A_{\rm FB}^{bb}$. We assume identification efficiencies of
$\epsilon_\mu$~=~95\% and $\epsilon_b$~=~60\% and systematic uncertainties of
$\Delta \epsilon_\mu/\epsilon_\mu$~=~0.5\%,
$\Delta\epsilon_b/\epsilon_b$~=~1\%. The sensitivity contours obtained for
${\cal{L}}$~=~1~ab$^{-1}$ are shown in~Fig.~\ref{dom:fig1}. The 3~TeV LC
indirect reach is lower than that of the LHC or comparable to it.
However, the QCD background rejection essential for the LHC
sensitivity still needs to be validated using a full detector
simulation and pile-up~effects.

\begin{figure}[hb] 
\begin{center}
\hspace{-1cm}
\epsfig{file=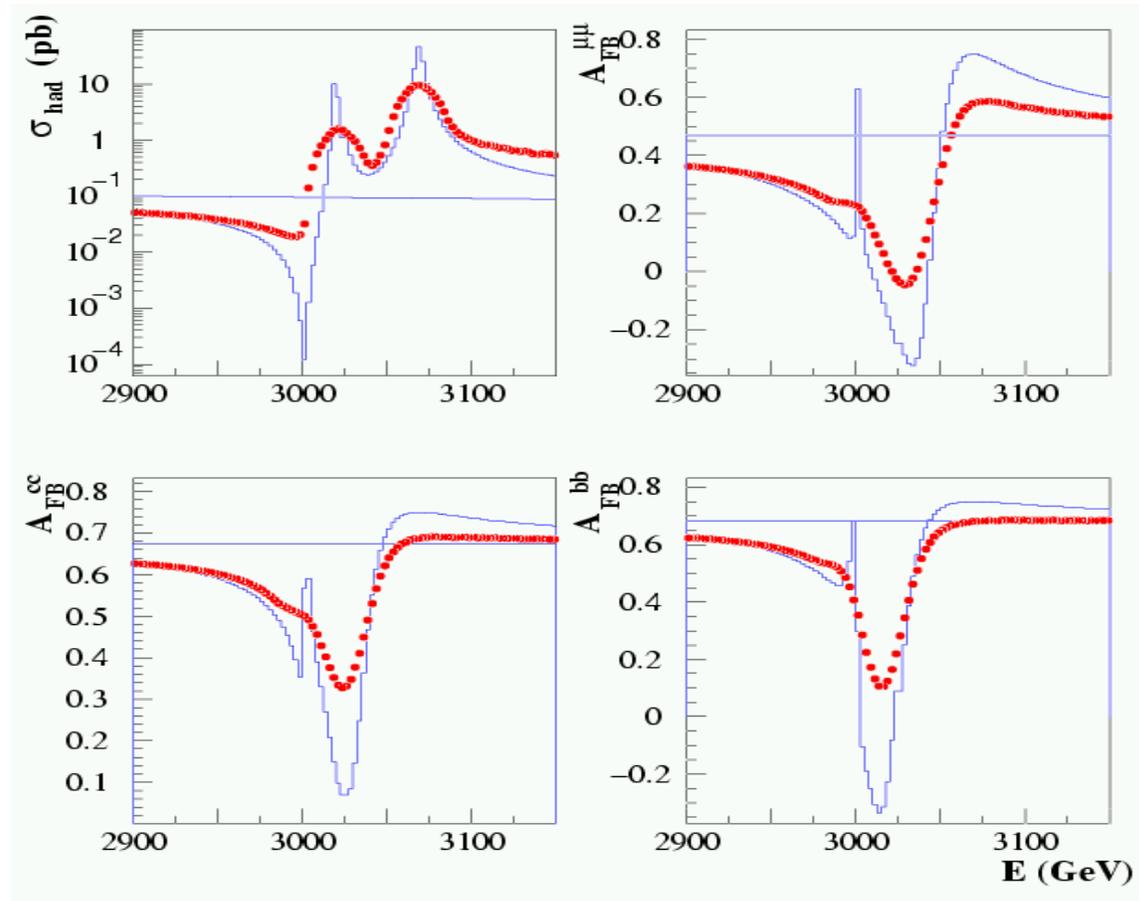,width=15.0cm,height=12cm,clip}
\end{center}
\vspace*{-5mm} 
\caption{Hadronic cross section (upper left)
and $\mu^+\mu^-$ (upper right), $c \bar c$ (lower left) and 
$b \bar b$ (lower right) forward--backward asymmetries at energies
around 3~TeV. The continuous lines represent the predictions for
the D-BESS model with $M$~=~3~TeV and $g/g''$~=~0.15, the flat
lines the SM expectation, and the dots the observable D-BESS signal
after accounting for the CLIC.02 luminosity spectrum.}
\label{fig:afb}
\end{figure}

Assuming a resonant signal to be seen at the LHC or indirect
evidence to be obtained at a lower-energy LC, CLIC could measure the
width and mass of this state and also probe its almost degenerate
structure~\cite{Casalbuoni:1999mm}. This needs to be validated, taking the
luminosity spectrum and accelerator-induced backgrounds into account. The
ability to identify the model's distinctive features has been studied
using the production cross section and the flavour-dependent
forward--backward asymmetries for different values of $g/g''$. The
resulting distributions are shown in~Fig.~\ref{fig:afb} 
for the case of
the narrower CLIC.02 beam parameters~\cite{jhep}. A characteristic feature
of the cross-Section distributions is the presence of a narrow dip, due to
the interference of the ${\rm L}_3$, ${\rm R}_3$ resonances with the
$\gamma$ and $Z^0$, and to cancellations of the ${\rm L}_3$, ${\rm R}_3$ 
contributions. Similar 
considerations hold for the asymmetries. In the case shown in
Fig.~\ref{fig:afb}, the effect is still visible after accounting for the
luminosity spectrum. In this analysis, the beam-energy spread sets the
main limit to the smallest mass splitting observable. With realistic
assumptions and 1~ab$^{-1}$ of data, CLIC will be able to resolve the
two narrow resonances for values of the coupling ratio $g/g''>$~0.08,
corresponding to a mass splitting $\Delta M$~=~13~GeV for $M$~=~3~TeV, and
to determine $\Delta M$ with a statistical accuracy better than 100~MeV,
as seen in Table~\ref{tab:4}.

\section{Further Alternative Theories}

There exist a plethora of other new phenomena that may appear at the~TeV
scale. In the following, we discuss briefly Little Higgs models, a 
possible fourth family of quarks and
leptons, leptoquarks, excited leptons and leptons of finite size, and
finally non-commutative field theories.

\subsection{Heavy Gauge Bosons in Little Higgs Models}

\def\mh{m_h^{}}
\def\gev{\rm GeV}
\def\tev{\rm~TeV}
\def\fbi{\rm fb^{-1}}
\def\abi{\rm ab^{-1}}
\def\ee{e^+e^-}

Within Little Higgs models~\cite{Han:2003wu}, new heavy particles can be
expected in the~TeV range, such as a heavy top quark $T$ and new gauge
bosons. The LHC is sensitive to these particles up to 2.5~TeV for the $T$
quark~\cite{Azuelos:2004dm}, but CLIC will produce these copiously, via
associated ($W',T$) or single ($Z'$) production, allowing for a precise
study of the properties of the particles. 

The existence of the heavy $SU(2)$ gauge bosons $Z_H$ and $W_H$ is
one of the main predictions of the Little Higgs models.  
The masses of $Z_H$ and $W_H$ should be within about a few~TeV in order 
to solve the hierachy problem. 

\begin{figure}[hb!] 
 \centerline{\includegraphics[scale=0.57]{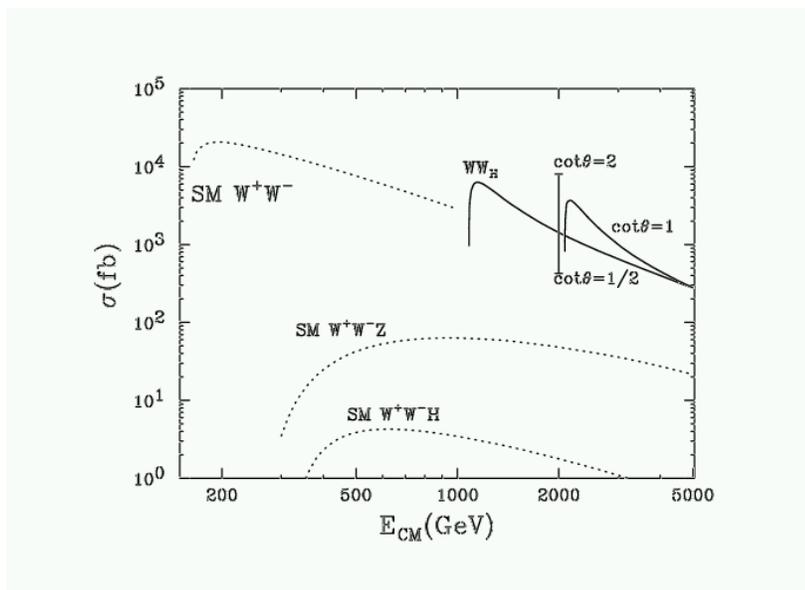}}
 \caption{Total cross section for $\ee\to WW_H$ production versus 
centre-of-mass energy $E_{c.m.}$ for $M_{W_H}$~=~1 and 2~TeV (solid curves).  
Both charge states $W^\pm~W_H^\mp$ have been included.
Some relevant SM processes have also been included 
for comparison (dashed curves).} 
\label{wwh}
\end{figure}

At an $\ee$ linear collider, if the centre-of-mass (c.m.) energy 
can be set at the mass of the vector resonance, 
one is able to reach a substantial production cross section
and perform precision studies of the properties of the particle. Above the
resonance threshold, the dominant production for the heavy gauge bosons 
is through the $WW_H$ final state. 
Fixing $M_{W_H}$~=~1~TeV, we plot in~Fig.~\ref{wwh} (solid curves) 
the total cross section for $WW_H$ production for $M_{W_H}$~=~1 and 2~TeV
as a function of the c.m.~energy.
For a fixed value of $M_{W_H}$, the cross section scales as $\cot^2\theta$,
and the changes for $\cot\theta$~=~1/2 and 2 are indicated by a vertical bar.  
For comparison, we also include some relevant SM processes of
$W^{+}W^{-}$, $WWZ$, and  $WWH$. We see that the signal cross section
for the $WW_H$ final state is large and asymptotically decreases to
the level of $W^{+}W^{-}$.

\begin{figure}[tb!]
\centerline{\includegraphics[scale=0.71]{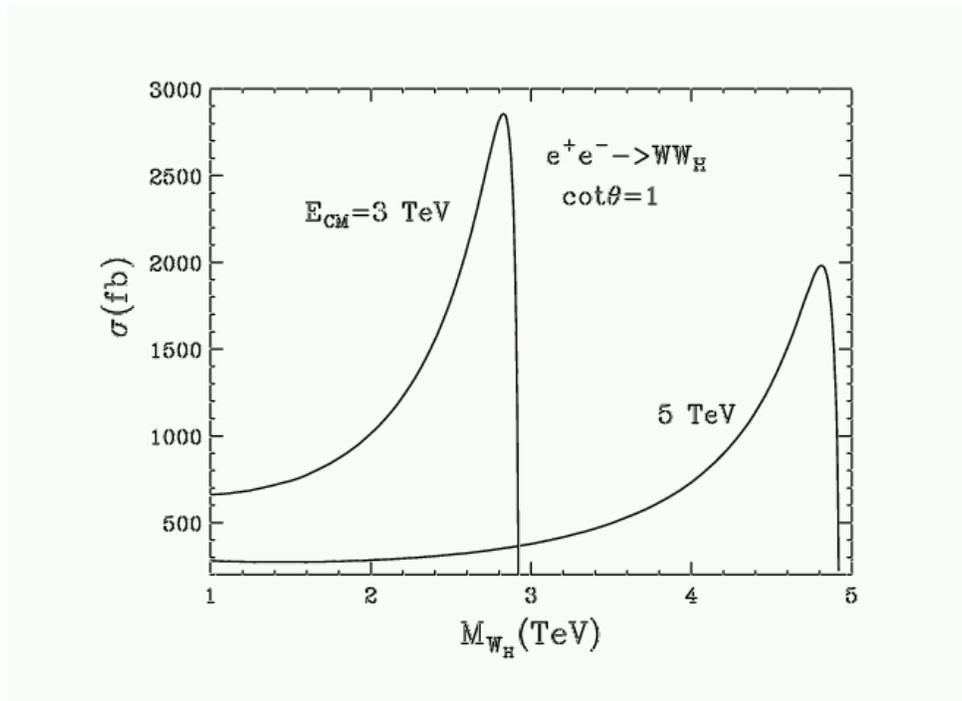}}
\vspace{-1cm}
\caption{Total cross section for $\ee\to WW_H$ production versus its
mass $M_{W_H}$ at the CLIC energies $E_{c.m.}$~=~3 and 5~TeV.
Both charge states $W^\pm~W_H^\mp$ have been included.}
    \label{lh5}
\end{figure}

In Fig.~\ref{lh5}, we show the total cross-Section versus $M_{W_H}$ 
at the CLIC energies, 3 and 5~TeV, for $\cot\theta~=~1$.  
The cross section grows when the mass increases because of the less
and less severe propagator suppression $1/(s-M_{Z_H})^2$, until
it is cut off by the threshold kinematics.

\begin{figure}[hb!]
\centerline{\includegraphics[scale=0.6]{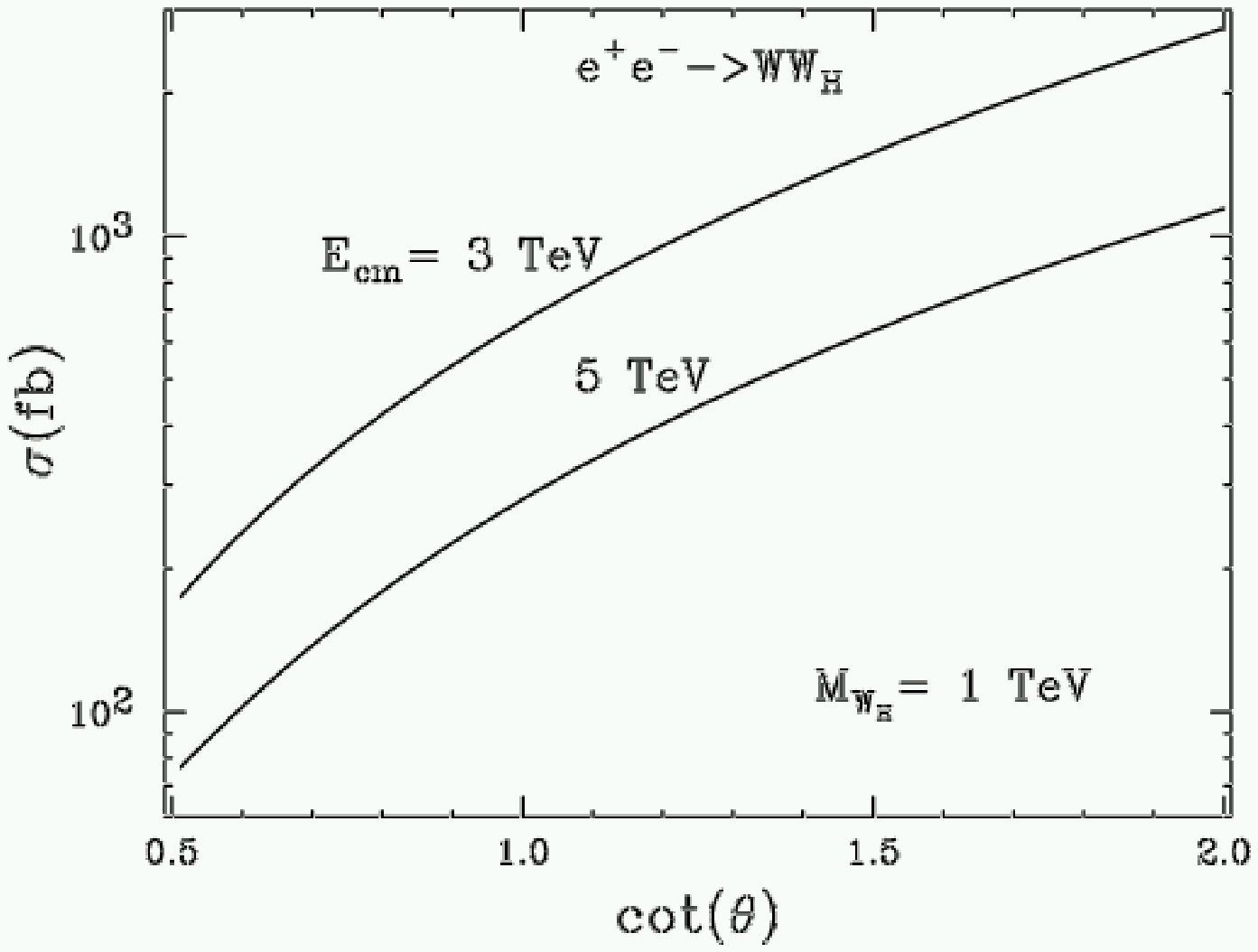}}
\vspace{-1.4cm}
\caption{Total cross section for $\ee\to WW_H$ production as a
function of gauge coupling mixing parameter $\cot\theta$ for
$M_{W_H}$~=~1~TeV} 
\label{cot}
\end{figure}

Generically, the mass and coupling of $W_H$ depend on $\cot\theta$,
the mixing parameter between the SM and new gauge groups 
$SU_L(2)$ and   $SU_H(2).$
Keeping the mass fixed at 1~TeV, we can explore the total cross section 
with respect to $\cot\theta$, as shown in~Fig.~\ref{cot}.  
\vspace{-0.6cm}

\begin{figure}[tb]
\centerline{\includegraphics[scale=0.8]{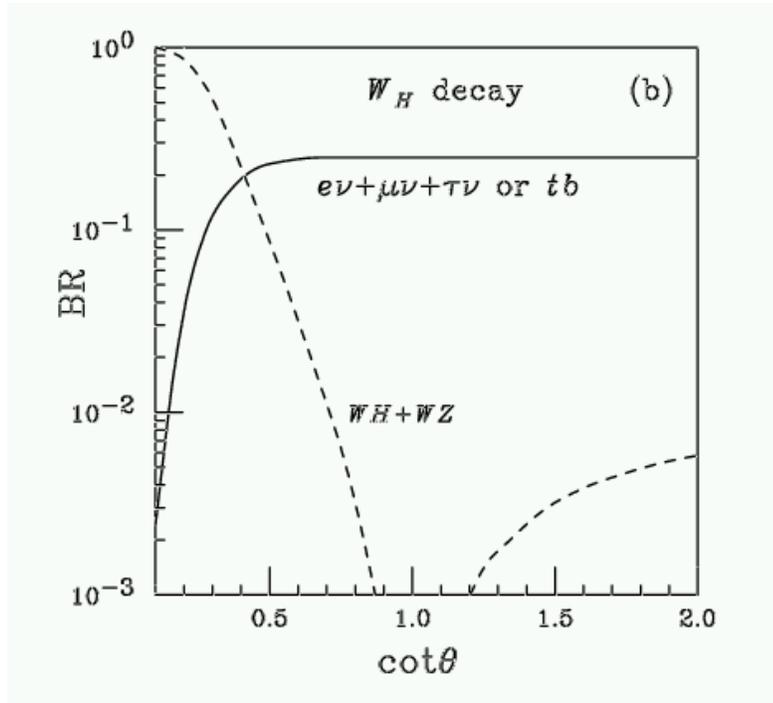}}
    \caption{Branching fractions for $W_H$ decay to SM particles}
    \label{brwh}
\end{figure}

Once $W_H$ is produced, it can decay to SM particles, either 
fermion pairs or $WZ,\ WH$.  The branching fractions are given with
respect to the mixing parameter $\cot\theta$, as in~Fig.~\ref{brwh}.  
The decay channels  to fermions are dominant, asymptotically 1/4 for
the three generations of leptons or one generation of quarks,  
for most of the paramter space.
For a value of $\cot\theta<$~0.4, $W_H\to WH, WZ$ decay modes become more 
important, comparable to fermionic channels, as shown in~Fig.~\ref{brwh}.  
The production rate for the signal is still quite sizeable once above
the kinematical threshold, and likely above the SM backgrounds.

In summary, once above the kinematical threshold, the heavy
gauge boson production at the CLIC energies can be substantial.
The threshold behaviour can determine its mass accurately, and the 
cross section rate will measure the coupling strength $\cot\theta$.

\subsection{Fourth Family}

The mass and mixing patterns of the fundamental fermions are among the
most mysterious aspects of particle physics today. The number of fermion
generations is not fixed by the SM. It is intriguing that
flavour democracy may hint at the existence of a fourth SM family (see
\cite{Sultansoy,Arik} and references therein), and its possible existence
is a fundamental question on which CLIC can cast new light. In order to
avoid violation of partial-wave unitarity, the masses of new quarks and
leptons should be smaller than about 1~TeV~\cite{Chanowitz}. Therefore,
CLIC with 1--3~TeV centre-of-mass energy will give the opportunity to
measure the fourth-SM-family fermions and quarkonia in detail.

Fourth-family quarks, if they exist, will be copiously produced at the
LHC~\cite{Arik2,ATLAS} if their masses are less than $1$~TeV.  Therefore,
any fourth-family quarks might be discovered at the LHC well before CLIC goes
into operation. The same is true for pseudoscalar quarkonia ($\eta_{4}$)
formed by the fourth-family quarks~\cite{Arik3}.  However, the observation of
the fourth-family leptons at the LHC is problematic because of the large
backgrounds from real $W$ and $Z$ boson production, both singly and in
pairs, that hide these heavy lepton signals~\cite{Barger}. Lepton
colliders will therefore be advantageous for the investigation of the
fourth-SM-family leptons and vector ($\psi_{4}$) quarkonia. The potential
of muon colliders was analysed in~Ref.~\cite{Ciftci}. In this section we
consider the production of the fourth-family fermions and quarkonia at CLIC,
starting with pair~production.

First we consider the $e^+e^-$ option of CLIC. The annihilation of
$e^{+}e^{-}$ is a classic channel to produce and study new heavy fermions,
because the cross sections are relatively large with respect to the
backgrounds~\cite{Barger}. The obtained cross section values for pair
production of the fourth-SM-family fermions with $m_{4}$~=~320~(640)~GeV
and the corresponding numbers of events per working year (10$^{7}$~s) are
given in Table~\ref{4_one} (Table~\ref{4_two}). 
Event signatures are
defined by the mass pattern of the fourth family and the 4~$\times$~4
Cabibbo--Kobayashi--Maskawa (CKM) matrix. According to the scenario given
in~Ref.~\cite{Atag}, the dominant decay modes are $u_{4}\to b$ $W^{-}$,
$d_{4}\to t$ $W^{+}$, $l_{4}\to\nu_{\tau}W^{-}$ and
$\nu_{4}\to\tau^{-}W^{+}$. The charged $l_{4}$ lepton will have a
clear signature at CLIC. For example, if the produced $W^{\pm}$ bosons
decay leptonically, one has two acoplanar opposite-charge leptons and
large missing energy. Pair production of the neutral $\nu_{4}$ leptons
will lead to a more complicated event topology. In this case, $\tau$
tagging will be helpful for the identification of the events. Indeed, the
produced $\tau$ leptons will decay at a distance of 1--2~cm from the
interaction point, which can easily be measured by a vertex detector. In
addition, the polarization of electron and positron beams should help to
experimentally determine the axial and vector neutral-current coupling
constants of the fourth-family fermions.
%
\begin{table}[!h] 
\caption{Cross sections and event numbers per year for pair production of
the fourth-SM-family fermions with mass 320~GeV at CLIC 
($\sqrt{s_{ee}}$~=~1~TeV,
$L_{ee}$~=~2.7~$\times$~10$^{34}$cm$^{-2}$s$^{-1}$ and 
$L_{\gamma \gamma}$~=~10$^{34}$cm$^{-2}$s$^{-1}$)}
\label{4_one}

\renewcommand{\arraystretch}{1.35} 
\begin{center}

\begin{tabular}{lccccc}\hline \hline \\[-4mm]
$\hspace*{5mm}$  & 
$\hspace*{5mm}$ &
$\hspace*{3mm}$ \boldmath{$u_{4}\overline{u_{4}}$} $\hspace*{3mm}$ & 
$\hspace*{3mm}$ \boldmath{$d_{4}\overline{d_{4}}$} $\hspace*{3mm}$ & 
$\hspace*{3mm}$ \boldmath{$l_{4}\overline{l_{4}}$} $\hspace*{3mm}$ & 
$\hspace*{3mm}$ \boldmath{$\nu_{4}\overline{\nu_{4}}$} $\hspace*{3mm}$  
\\[4mm]   

\hline \\[-3mm]
$e^{+}e^{-}$ option $\hspace*{9mm}$ & 
$\sigma$ (fb) & 130 & 60 & 86 & 15\\
 & N$_{\rm ev}$/year & 35 000 & 16 000 & 23 000 & 4100 
\\[3mm] \hline\\[-4mm]
$\gamma\gamma$ option & $\sigma$ (fb) & 34 & 2 & 58 & -- \\
& N$_{\rm ev}$/year & 3400 & 200 & 5700 & --
\\[3mm] 
\hline \hline
\end{tabular}
\end{center}
\end{table}

\begin{table}[hb] 
\caption{Cross sections and event numbers per year for pair production of
the fourth-SM-family fermions with mass 640~GeV at CLIC
($\sqrt{s_{ee}}$~=~3~TeV, $L_{ee}$~=~1~$\times$~10$^{35}$cm$^{-2}$s$^{-1}$ and
$L_{\gamma \gamma}$~=~3~$\times$~10$^{34}$cm$^{-2}$s$^{-1}$)}
\label{4_two}

\renewcommand{\arraystretch}{1.35} 
\begin{center}

\begin{tabular}{lccccc}\hline \hline \\[-4mm]
$\hspace*{5mm}$  & 
$\hspace*{5mm}$ &
$\hspace*{3mm}$ \boldmath{$u_{4}\overline{u_{4}}$} $\hspace*{3mm}$ & 
$\hspace*{3mm}$ \boldmath{$d_{4}\overline{d_{4}}$} $\hspace*{3mm}$ & 
$\hspace*{3mm}$ \boldmath{$l_{4}\overline{l_{4}}$} $\hspace*{3mm}$ & 
$\hspace*{3mm}$ \boldmath{$\nu_{4}\overline{\nu_{4}}$} $\hspace*{3mm}$  
\\[4mm]   

\hline \\[-3mm]
$e^{+}e^{-}$ option $\hspace*{9mm}$ & 
$\sigma$ (fb) & 16 & 8 & 10 & 2\\
& N$_{\rm ev}$/year & 16 000 & 8000 & 10 000 & 2000
\\[3mm] \hline\\[-4mm]
$\gamma\gamma$ option & $\sigma$ (fb) & 27 & 2 & 46 & --\\
& N$_{\rm ev}$/year & 8100 & 600 & 14 000 & --
\\[3mm] 
\hline \hline
\end{tabular}
\end{center}
\end{table}


Next we consider the $\gamma\gamma$ option~\cite{Ginzburg1,Ginzburg2,Telnov}
of CLIC. The fourth-SM-family quarks and charged
leptons will also be copiously produced at $\gamma\gamma$ machines. For
$\sqrt{s_{ee}}$~=~1~TeV (3~TeV), which corresponds to 
$\sqrt{s_{\gamma\gamma}}^{\max}$~=~0.83$\sqrt{s_{ee}}$~=~0.83~TeV
(2.5~TeV), the obtained cross section values 
and  number of events per year are presented in Table~\ref{4_one}
(Table~\ref{4_two}). 

Next we discuss quarkonium production. The condition to form 
$(Q\overline{Q})$ quarkonia states with new
heavy quarks is $m_{Q}\leq$~(125~GeV)~$|V_{Qq}|^{-2/3}$,
where $q=d,s,b$ for $Q=u_{4}$ and $q=u,c,t$ for $Q=d_{4}$~\cite{Bigi}.
Unlike the $t$ quark, the fourth-family quarks will form quarkonia 
because $u_{4}$ and $d_{4}$ are almost degenerate and their decays are
suppressed by small CKM mixings~\cite{Celikel,Atag,Sultansoy}. Below,
we consider the resonance
production of $\psi_{4}$ quarkonia at $e^{+}e^{-}$ and $\eta_{4}$ quarkonia
at $\gamma\gamma$ options of CLIC.

Again we first consider the $e^{+}e^{-}$ option for CLIC.
The cross section for the formation of the fourth-family quarkonium is 
given
by the well-known relativistic Breit--Wigner equation
\begin{equation}
\sigma\left(  e^{+}e^{-}\to\left(  Q\overline{Q}\right)  \right)
=\frac{12\pi\left(  s/M^{2}\right)  \Gamma_{ee}\Gamma}{\left(  s-M^{2}\right)
^{2}+M^{2}\Gamma^{2}}\,,
\label{4_bir}
\end{equation}
where $M$ is the mass, $\Gamma_{ee}$ is the partial decay width to $e^{+}%
e^{-}$ and $\Gamma$ is the total decay width of the fourth-family 
quarkonium.
Using corresponding formulae from~\cite{Barger87} in the framework of 
the Coulomb
potential model, we obtain decay widths for the main decay modes of $\psi
_{4}(u_{4}\overline{u_{4}})$ and $\psi_{4}(d_{4}\overline{d_{4}})$, which are
given in Table~\ref{4_three}. One can see that the dominant decay mode for 
both $\psi_{4}(u_{4}\overline{u_{4}})$ and $\psi_{4}(d_{4}\overline{d_{4}})$
quarkonia is $\psi_{4}\to W^{+}W^{-}$. Other important decay modes 
are $\psi_{4}\to\gamma Z$ and $\psi_{4}\to\gamma H$.
%
\begin{table}[hb] 
\caption{Decay widths for main decay modes of $\psi _{4} $ for 
$m_{H}$~=~150~GeV  with $m_{\psi _{4}}\simeq $~1~TeV}
\label{4_three}

\renewcommand{\arraystretch}{1.3} 
\begin{center}

\begin{tabular}{lcc}\hline \hline \\[-4mm]
& 
$\hspace*{5mm}$ \textbf{(\boldmath{$u_{4}\overline{u_{4}}$})} 
$\hspace*{5mm}$ & 
$\hspace*{5mm}$ \textbf{(\boldmath{$d_{4}\overline{d_{4}}$})} 
$\hspace*{5mm}$   
\\[4mm]   

\hline \\[-3mm]
$\Gamma(\psi_{4}\to\ell^{+}\ell^{-})$, 10$^{-3}$ MeV $\hspace*{9mm}$ &
18.9 & 7.3\\
$\Gamma(\psi_{4}\to u\overline{u})$, 10$^{-2}$ MeV &
3.2 & 1.9\\
$\Gamma(\psi_{4}\to d\overline{d})$, 10$^{-2}$ MeV &
1.4 & 1.7\\
$\Gamma(\psi_{4}\to Z\gamma)$, 10$^{-1}$~MeV & 15
& 3.7\\
$\Gamma(\psi_{4}\to ZZ)$, 10$^{-1}$ MeV & 1.7 &
5.4\\
$\Gamma(\psi_{4}\to ZH)$, 10$^{-1}$ MeV & 1.7 &
5.5\\
$\Gamma(\psi_{4}\to\gamma H)$, 10$^{-1}$ MeV &
14.4 & 3.6\\
$\Gamma(\psi_{4}\to W^{+}W^{-})$, MeV & 70.8 &
71.2
\\[3mm] 
\hline \hline
\end{tabular}
\end{center}
\end{table}


In order to estimate the number of produced quarkonium states, one should
take into account the luminosity distribution at CLIC, which is influenced
by the energy spread of electron and positron beams and beamstrahlung. In
our calculations we use the GUINEA-PIG simulation code~\cite{Schulte}. For
illustration, we assume that $m_{\psi_{4}}\simeq$~1~TeV. The estimated
event numbers per year for $\psi_{4}$ production, as well as 
$\psi_{4}\to\gamma H$ and $\psi_{4}\to ZH$ decay channels are
presented in Table~\ref{4_four}. In numerical calculations we use 
$\Delta E/E$~=~10$^{-2}$ for the beam energy spread. 
If $\Delta E/E$~=~10$^{-3}$ can be achieved, the corresponding numbers
will be enlarged by a factor of about 8. 
The $\gamma H$ decay mode of the $\psi_{4}(u_{4}\overline{u_{4}})$
quarkonium is promising for the study of Higgs boson properties,
especially if the energy spread of electron and positron beams of
about 10$^{-2}$ \ could be controlled to about 10$^{-3}$.
%
\begin{table}[htbp]  
\caption{The production event numbers per year for the fourth-SM-family
$\psi_{4}$ quarkonia  at a CLIC 1~TeV option 
with~$m_{\psi _{4}}\simeq$~1~TeV}
\label{4_four}

\renewcommand{\arraystretch}{1.3} 
\begin{center}

\begin{tabular}{lcc}\hline \hline \\[-4mm]
& 
$\hspace*{5mm}$ \textbf{(\boldmath{$u_{4}\overline{u_{4}}$})} 
$\hspace*{5mm}$ & 
$\hspace*{5mm}$ \textbf{(\boldmath{$d_{4}\overline{d_{4}}$})} 
$\hspace*{5mm}$   
\\[4mm]   

\hline \\[-3mm]
$e^{+}e^{-}\to\psi_{4}$ & 26 600 & 10 400\\
$e^{+}e^{-}\to\psi_{4}\to\gamma H$ $\hspace*{11mm}$ & 510 & 50\\
$e^{+}e^{-}\to\psi_{4}\to ZH$ & 60 & 80 
\\[3mm] 
\hline \hline
\end{tabular}
\end{center}
\end{table}


Finally, we discuss again the $\gamma\gamma$ collider.
The pseudoscalar $\eta_{4}$ quarkonia formed by the fourth-SM-family 
quarks will be copiously produced at the LHC~\cite{ATLAS,Arik3}. Decay
widths for the main decay modes of 
$\eta_{4}(u_{4}\overline{u_{4}})$ and 
$\eta_{4}(d_{4}\overline{d_{4}})$ are given in Table~\ref{4_five}. 
From there, it is clear that the dominant decay
mode is $\eta_{4}\to ZH$. One can estimate the $\gamma\gamma
\to\eta_{4}$ production cross section by using 
the following relation~\cite{Gorbunov}:
\begin{equation}
\sigma\approx 50\,{\rm fb}(1+\lambda_{1}\lambda_{2})
\left( \frac{BR_{\gamma\gamma}}{4\times10^{-3}}\right)  
\left( \frac{\Gamma_{\rm tot}}{1\,{\rm MeV}}\right)  
\left( \frac{200\,{\rm GeV}}{M_{\eta}}\right)^{3}\,,
\end{equation}
where $BR_{\gamma\gamma}$ is the branching ratio of 
the $\eta_{4}\to\gamma\gamma$ decay mode, 
$\Gamma_{\rm tot}$ is the\ quarkonium total decay width
and $\lambda_{1,2}$ are helicities of the initial photons. 
Assuming $\lambda_{1}\lambda_{2}$~=~1, we obtain total event numbers
of $\eta_{4}$ produced per year as follows: 
900~$\eta_{4}(u_{4}\overline{u_{4}})$ and 
56~$\eta_{4}(d_{4}\overline{d_{4}})$. 
The corresponding numbers for $\eta_{4}\to ZH$ events are 610 and 38
events, respectively. The advantage of 
$\eta_{4}(u_{4} \overline{u_{4}})$ with respect to
$\eta_{4}(d_{4}\overline{d_{4}})$ is obvious.
%
\begin{table}[htbp] 
\caption{Decay widths for main decay modes of $\eta_{4}$ for $m_H$~=~150
GeV with $m_{\eta_4}$~=~0.75~TeV}
\label{4_five}

\renewcommand{\arraystretch}{1.3} 
\begin{center}

\begin{tabular}{lcc}\hline \hline \\[-4mm]
& 
$\hspace*{5mm}$ \textbf{(\boldmath{$u_{4}\overline{u_{4}}$})} 
$\hspace*{5mm}$ & 
$\hspace*{5mm}$ \textbf{(\boldmath{$d_{4}\overline{d_{4}}$})} 
$\hspace*{5mm}$   
\\[4mm]   

\hline \\[-3mm]
$\Gamma(\eta_{4}\to\gamma\gamma)$, 10$^{-3}$ MeV  &
19.5 & 1.06 \\
$\Gamma(\eta_{4}\to Z\gamma)$, 10$^{-3}$\ MeV &
4.6 & 3.7\\
$\Gamma(\eta_{4}\to ZZ)$, 10$^{-1}$\ MeV & 2.2 &
2.8\\
$\Gamma(\eta_{4}\to gg)$,\ MeV & 5.1 &
5.1\\
$\Gamma(\eta_{4}\to ZH)$, MeV & 47.3 &
47.3\\
$\Gamma(\eta_{4}\to W^{+}W^{-})$, 10$^{-2}$ MeV $\hspace*{11mm}$&
5.7 & 5.7\\
$\Gamma(\eta_{4}\to t\overline{t})$,\ MeV & 16.4 &
16.4\\
$\Gamma(\eta_{4}\to b\overline{b})$, 10$^{-2}$\ MeV &
1.0 & 1.0
\\[3mm] 
\hline \hline
\end{tabular}
\end{center}
\end{table}


\subsection{Leptoquarks}

In the SM of electroweak and colour (QCD)
interactions, quarks and leptons appear as formally independent
components. However, the observed symmetry between the lepton and
quark sectors in the SM could be interpreted as a hint for common
underlying structures. If quarks and leptons are made of
constituents there should appear, at the scale of constituent binding
energies, new interactions among leptons and quarks.
Leptoquarks (LQs) are exotic particles carrying both lepton number
(L) and baryon number (B), which are colour (anti-)triplet, scalar or 
vector
particles that appear naturally in various unifying theories
beyond the SM. The interactions of LQs with the known particles
are usually described by an effective Lagrangian that satisfies
the requirement of baryon- and lepton-number conservation and
respects the $SU(3)_{C}~\times~SU(2)_{W}~\times~U(1)_{Y}$ symmetry
of the SM. There are nine scalar and nine vector leptoquark types
in the Buchm\"uller--R\"uckl--Wyler (BRW) classification~\cite{1}. The
scalar leptoquarks $(S,R)$ can be grouped into singlets
$(S_{0},\widetilde{S}_{0}),$ doublets
$(R_{1/2},\widetilde{R}_{1/2})$ and triplet $(S_{1}).$ Here, we
assume only the mass, and the couplings to right-handed and/or
left-handed leptons, denoted by $g_{\rm R}$ and $g_{\rm L}$, remain as
free parameters~\cite{2}. From the BRW effective interaction
Lagrangian~\cite{1} one can deduce the quantum numbers of scalar
leptoquarks given in Table~\ref{table1}.

The leptoquarks are constrained by different experiments. Direct limits on
leptoquark states are obtained from their production cross sections at
different colliders, while indirect limits are calculated from the bounds
on the leptoquark-induced four-fermion interactions in low-energy
experiments. The mass limits for scalar leptoquarks from single and pair
production, assuming electromagnetic coupling, are
$M_{LQ}>$~200~GeV~\cite{3} and $M_{LQ}>$~225~GeV~\cite{4}, respectively.

We have studied the potential of a CLIC-based $e\gamma$ collider to search
for scalar leptoquarks, taking into account both direct and resolved
photon processes. Scalar leptoquarks can be produced singly in $e\gamma$
collisions, via the process $e\gamma\to Sq$, where $S$ is any type
of scalar leptoquark ($S$ or $R$). At $e\gamma$ colliders, with a photon
beam produced by Compton backscattering~\cite{5}, the maximum $e\gamma$
centre-of-mass energy is about 91\% of the available energy. The projects
of CLIC~\cite{6} colliders will be working at $\sqrt{s_{e^{+}e^{-}}}$~=~1
TeV to $\sqrt{s_{e^{+}e^{-}}}$~=~3~TeV. Corresponding high-energy photon
beams can be obtained from the linacs with energies $\simeq$~415~GeV and
$\simeq$~1245~GeV for CLIC at these energies.

We have implemented the interactions between scalar leptoquarks, leptons
and quarks into the CompHEP program~\cite{comphep} with an internal photon
spectrum. The main contribution to the total cross section comes from the
$u$-channel quark-exchange diagram. For this reason, the total cross
Sections for the production of scalar leptoquarks $R_{1/2}$(--~5/3),
$S_{0}$(--~1/3) and $S_{1}$(--~1/3) practically coincide. This is also true
for $R_{1/2}$(--~2/3), $\widetilde {R}_{1/2}$(--~2/3), 
$\widetilde{S}_{0}$(--~4/3)
and $S_{1}$(--~4/3) type scalar leptoquarks.

The scalar leptoquarks can also be produced in $e\gamma$
collisions by resolved photon processes. In order to produce
leptoquarks in the resonant channel through the quark component of
the photon, we study the following signal for the scalar
leptoquark $S$ (or $R$), $e+q_{\gamma}\to S\to e+q$. 
For CLIC-based $e\gamma$ colliders, the total cross section
for resolved photon contribution is obtained by convoluting with
the back-scattered laser photon distribution and photon structure
function. The photon structure function consists of perturbative
point-like parts and hadron-like parts
$f_{q/\gamma}=f_{q/\gamma}^{\rm PL}+f_{q/\gamma}^{\rm HL}$~\cite{8}.
Since the contribution from the point-like part of the photon
structure function was already taken into account in the
calculation of the direct part, it was subtracted from
$f_{q/\gamma}$ to avoid double 
counting at the leading-logarithmic~level.

\begin{figure}[t]
\includegraphics[height=7cm, width=8cm ]{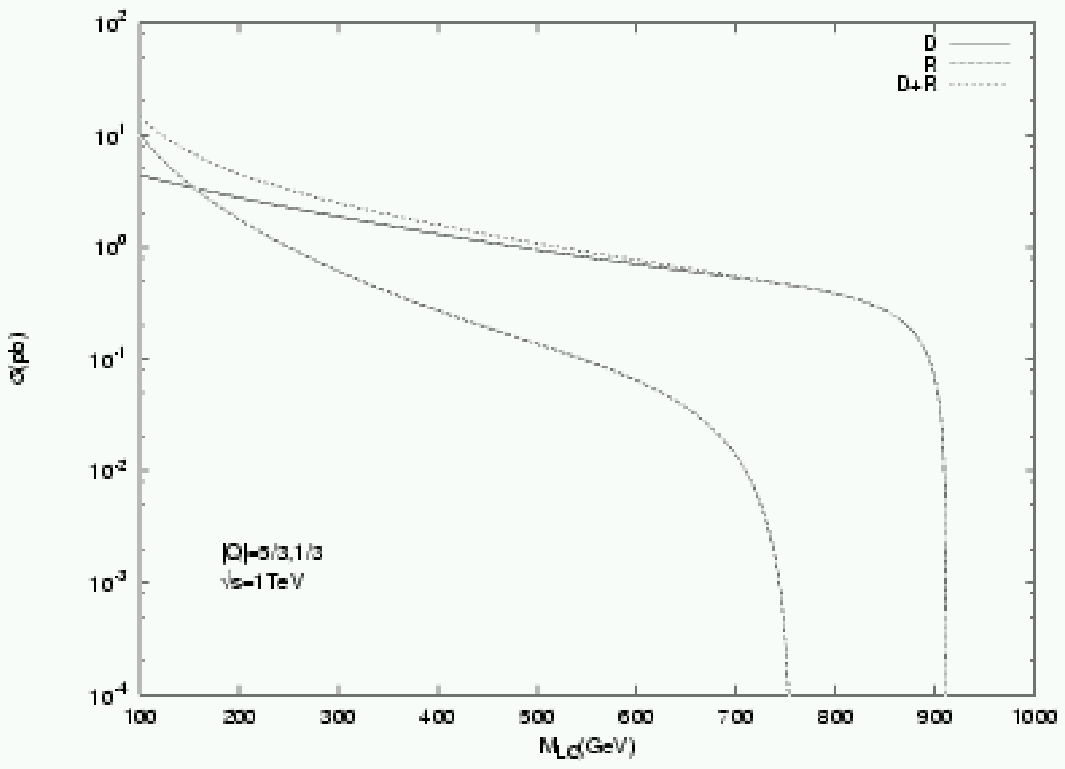}
\includegraphics[height=7cm, width=8cm ]{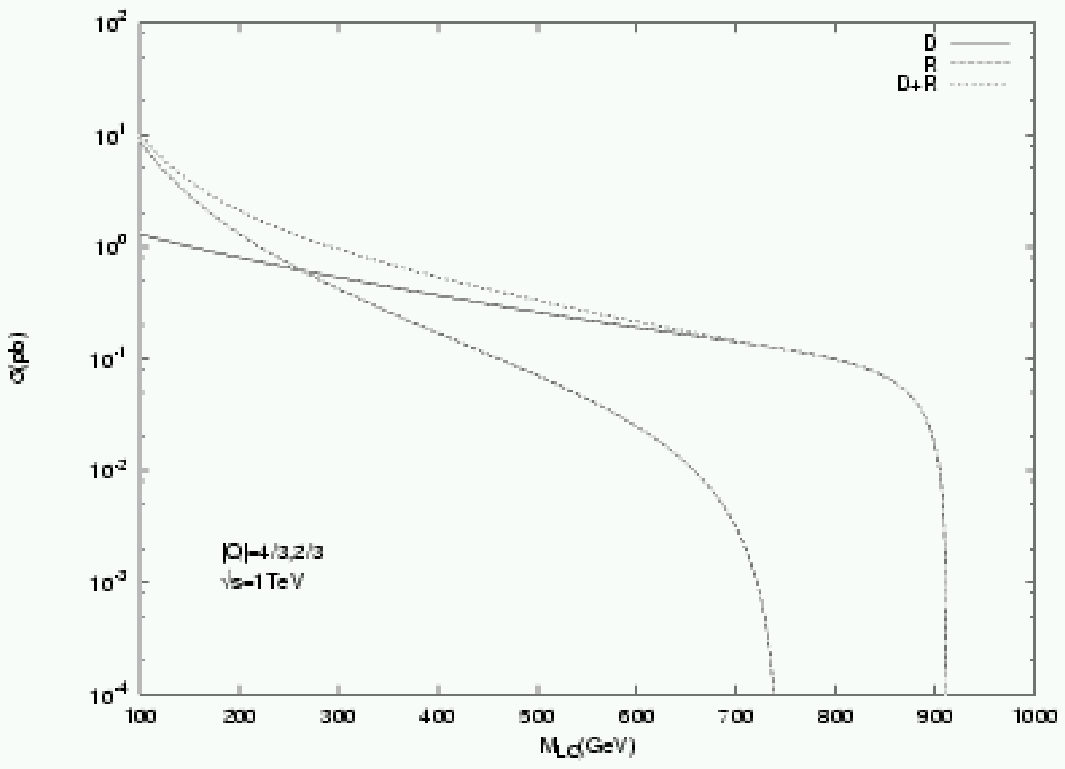}\\
\centerline{(a)
$\qquad\qquad\qquad\qquad\qquad\qquad\qquad\qquad\qquad\qquad$ (b)}
\caption{The direct (D) and resolved (R) contributions to the
cross section depending on scalar leptoquark mass $M_{LQ}$ at
$\sqrt{s_{e^+e^-}}$~=~1~TeV with LQs
charges a) $|Q|$~=~5/3~(1/3) and b) $|Q|$~=~4/3~(2/3).} 
\label{fig1}
\end{figure}

Both schemes can give comparable results for the production cross
Section. Therefore, we add their contribution to form the signal.
In this case, the total cross section for the production of scalar
LQs is $\sigma=\sigma_{\rm D}+\sigma_{\rm R}$, where $\sigma_{\rm D}$ and
$\sigma_{\rm R}$ denote the direct and resolved contributions to the
total cross section, respectively. The total cross sections
including both direct and resolved contributions are plotted in
Fig.~\ref{fig1} for CLIC-based $e\gamma$ colliders with
$\sqrt{s_{e^+e^-}}$~=~1~TeV. As can be seen from this figure, the
contribution from the resolved process is dominant for small
leptoquark masses and the direct contribution is effective for larger
masses up to the kinematical limit. Depending on the centre-of-mass 
energy, the resolved contribution decreases sharply beyond a
leptoquark mass value of about 70\% of the collider energy.
The direct contribution for a scalar leptoquark with $|Q|$~=~5/3 is
larger than the scalar leptoquark with $|Q$|~=~4/3. This can be
explained by the quark-charge dependence of the cross sections
for direct contributions. The coupling for the lepton--quark--leptoquark
vertex can be parametrized as $g_{LQ}=\sqrt{4\pi\alpha\kappa}$,
where $\kappa$ is a parameter.

\newpage

\begin{sideways}
\begin{threeparttable}
\caption{Quantum numbers of scalar leptoquarks according to the BRW
classification. The numbers in the parentheses in the last two columns
denote the values for $g_{\rm L}=g_{\rm R}$.}

\vspace*{-21mm}

\label{table1}

\vspace*{7mm}

\renewcommand{\arraystretch}{1.37} 
%
\large{
\begin{tabular}{cccccccc}
 & & & & & & &   \\
\hline \hline\\[-6mm]
& & & & & & &   \\[-4mm]
$\hspace*{3mm}$ \textbf{Leptoquark} $\hspace*{3mm}$ 
& $\hspace*{4mm}$ \textbf{\it F} $\hspace*{4mm}$ 
& $\hspace*{4mm}$ \textbf{\it I} $\hspace*{4mm}$ 
& $\hspace*{4mm}$ \boldmath{$Q_{\rm em}$} $\hspace*{4mm}$ 
& $\hspace*{5mm}$ \textbf{Decay} $\hspace*{5mm}$ 
& $\hspace*{5mm}$ \textbf{Coupling} $\hspace*{5mm}$
& $\hspace*{3mm}$ \boldmath{$BR(S\to lq)$} $\hspace*{3mm}$ 
& $\hspace*{3mm}$ \boldmath{$BR(S\to\nu q)$} $\hspace*{3mm}$ 
\\[4mm]   

\hline \\[-5mm]
  &  &  &  & $e_{L}u_{L}$ & $g_{0L}$ &  & \\
$S_{0}$ & 2 & 0 & --~1/3 & $e_{R}u_{R}$ & $g_{0R}$ &
$\frac{g_{0L}^{2}+g_{0R}^{2}}{2g_{0L}^{2}+g_{0R}^{2}}\,(\frac{2}{3})$ &
$\frac{g_{0L}^{2}}{2g_{0L}^{2}+g_{0R}^{2}}(\frac{1}{3})$ \\
  &  &  &  & $\nu d_{L}$ & $-g_{0L}$ &  & 
\\[3mm]
\hline \\[-5mm]
$\widetilde{S}_{0}$ & 2 & 0 & -- 4/3 & $e_{R}d_{R}$ & 
$\widetilde{g}_{0R}$ & 1 &
0 \\[3mm]
\hline\\[-5mm]
        &   & 1 & 2/3 & $\nu u_{L}$ & $\sqrt{2}g_{1L}$ & 0 1 \\
$S_{1}$ & 0 & 0 & -- 1/3 & $\nu d_{L},e_{L}u_{L}$ & $-g_{1L}$ & 1/2 &
1/2 \\
     &   & -- 1 & -- 4/3 & $e_{L}d_{L}$ & $-\sqrt{2}g_{1L}$ & 1 & 0 \\[3mm]
\hline \\[-5mm]
  &  & 1/2 & -- 2/3 & $\nu\overline{u}_{R}$ & $g_{1/2L}$ & & \\
  &  & 1/2 & -- 2/3 & $e_{R}\overline{d}_{L}$ & -- $g_{1/2R}$ &
$\frac{g_{1/2R}^{2}}{g_{1/2R}^{2}+g_{1/2L}^{2}}(\frac{1}{2})$ & 
$\frac{g_{1/2L}^{2}}{g_{1/2R}^{2}+g_{1/2L}^{2}}(\frac{1}{2})$\\[-2mm]
$R_{1/2}$ & 0 &  &  &  &  &  & \\[-2mm]
  &  & -- 1/2 & -- 5/3 & $e_{L}\overline{u}_{R}$ & $g_{1/2L}$ & 1 & 0
  \\
  &  & -- 1/2 & -- 5/3 & $e_{R}\overline{u}_{L}$ & $g_{1/2R}$ & & 
\\[3mm]
\hline \\[-5mm]
  &  & 1/2 & 1/3 & $\nu\overline{d}_{R}$ & $\widetilde{g}_{1/2L}$ & 0
  & 1 \\[-2mm]
$\widetilde{R}_{1/2}$ & 0 &  &  &  &  &  & \\[-2mm]
  &  & -- 1/2 & -- 2/3 & $e_{L}\overline{d}_{R}$ &
  $\widetilde{g}_{1/2L}$ & 1 & 0 
\\[2mm] 
\hline \hline
%
\end{tabular}
}
\end{threeparttable}
\end{sideways}

\newpage
\begin{sideways}
\begin{threeparttable}
\caption{The number of events and signal significance for $2j+e$ and
$2j+\not \!p_{T}$ channels in scalar leptoquark decays. The lower and
upper indices on scalar leptoquarks $S$ or $R$ denote weak isospin $I$
and $I_3$, respectively.}

\vspace*{-17mm}

\label{table2}

\vspace*{7mm}

\renewcommand{\arraystretch}{1.7} 
%
\large{
\begin{tabular}{ccccccccccccc}
 & & & & & & & & & & &  &   \\
\hline \hline\\[-6mm]
 & & & & & & & & & & &  &   \\[-4mm]
$\hspace*{2mm}$ \boldmath{$\sqrt{s_{e^{+}e^{-}}}$}\textbf{~=~1~TeV} 
$\hspace*{2mm}$ & 
\multicolumn{2}{c}{ \textbf{Number of events} } & $\hspace*{5mm}$ & 
\multicolumn{5}{c}{\boldmath{$\frac{S}{\sqrt{B}}\,
(e\gamma\to q\overline{q}e)$}} & $\hspace*{5mm}$ &
\multicolumn{3}{c}{\boldmath{$\frac{S}{\sqrt{B}}
(e\gamma\to q\overline{q}^{\prime}\nu)$}} \\[1.5mm]
\boldmath{$L_{\rm int}$\textbf{~=~10}$^5$\textbf{~pb}$^{-1}$} & 
\multicolumn{2}{c}{
\boldmath{$(\sigma_{\rm D}+\sigma_{\rm R})~\times~L_{\rm int}$}} 
& & & 
\\[2.5mm] \hline \\[-3mm]
M$_{\text{\rm LQ}}$(GeV) & 
$\hspace*{1mm}$ $|Q|$~=~5/3 $\hspace*{1mm}$ & 
$\hspace*{1mm}$ $|Q|$~=~4/3 $\hspace*{1mm}$ &  &
$\hspace*{0.5mm}$ $S_{0}$ $\hspace*{0.5mm}$ & 
$\hspace*{0.5mm}$ $S_{1}^{0}$ $\hspace*{0.5mm}$ & 
$\hspace*{0.5mm}$ $R_{1/2}^{-1/2}$ $\hspace*{0.5mm}$ & 
$\hspace*{0.5mm}$ $R_{1/2}^{1/2}$ $\hspace*{0.5mm}$ & 
$\hspace*{0.5mm}$
$\widetilde{R}_{1/2}^{-1/2},\widetilde{S}_{0},S_{1}^{-1}$ 
$\hspace*{0.5mm}$ &  & 
$\hspace*{0.5mm}$ $S_{0}$ $\hspace*{0.5mm}$ & 
$\hspace*{0.5mm}$ $S_{1}^{0}$ $\hspace*{0.5mm}$ & 
$\hspace*{0.5mm}$ $R_{1/2}^{1/2}$ $\hspace*{0.5mm}$ \\[5mm] \hline \\[-3mm]
200 & 452407 & 210085 & & 704 & 556 & 1113 & 258 & 517  & & 89 & 135 & 63\\
300 & 247050 & 94763 & & 401 & 304 & 608 & 117 & 233  & & 49 & 74 & 28\\
400 & 157382 & 53386 & & 255 & 194 & 387 & 66 & 131  & & 31 & 47 & 16\\
500 & 107389 & 32904 & & 174 & 132 & 264 & 40 & 81  & & 21 & 32 & 10\\
600 & 76050 & 21310 & & 123 & 94 & 187 & 26 & 52  & & 15 & 23 & 6\\
700 & 54541 & 14335 & & 88 & 67 & 134 & 18 & 35  & & 11 & 16 & 4\\
800 & 38152 & 9782 & & 62 & 47 & 94 & 12 & 24  & & 7 & 11 & --\\
900 & 6629 & 1660 & & 11 & 8 & 16 & 2 & 4  & & 1 & 2 & --
\\[5mm] 
\hline \hline
%
\end{tabular}
}
\end{threeparttable}
\end{sideways}

\newpage

When a scalar leptoquark is singly produced at an $e\gamma$ collider, the
signal will be two jets and a charged lepton $2j+l$ ($S$ and $R$
leptoquarks), or two jets plus a neutrino $2j+\not\!p_{T}$ ($S$
leptoquarks). Since leptoquarks generate a peak in the invariant ($lj$)
mass distribution, singly-produced leptoquarks are easy to detect up to
mass values close to the kinematical limit. In the case of equal couplings
$g_{\rm L}=g_{\rm R}$, the branching ratios for $S_0$ can be obtained
as 2/3 for $LQ\to lq$ and 1/3 for $LQ\to \nu q$ channels.

In order to calculate the statistical significance $S/\sqrt{B}$ at each
mass value of a scalar leptoquark and for each decay channel, we need to
calculate also the relevant background cross sections. For the background
processes $e\gamma\to W^{-}\nu$ and $e\gamma\to Ze$, we
find the total cross sections 41.20~(49.48)~pb and 2.36~(0.49)~pb
at $\sqrt{s_{e^{+}e^{-}}}$~=~1~(3)~TeV, respectively. We multiply these
cross sections with the branching ratios for the corresponding channels.

For a $e\gamma$ collider with centre-of-mass energy
$\sqrt{s_{e^+e^-}}$~=~1~TeV and total luminosity 
$L$~=~10$^{5}$~pb$^{-1}$, the scalar leptoquarks of types
$S_{0},S_{1}^{0}$ and $R_{1/2}^{-1/2}$ can be
produced up to mass $M_{LQ}\approx$~900~GeV, and 
$R_{1/2}^{1/2},\widetilde{S}_{0},S_{1}^{-1},\widetilde{R}_{1/2}^{-1/2}$ 
up to $M_{LQ}\approx$~850~GeV in
the $2j+e$ channel. However, leptoquarks of types $S_{0},S_{1}^{0}$ can be
produced up to $M_{LQ}\approx$~850~GeV and $R_{1/2}^{1/2}$ up to 
$M_{LQ}\approx$~650~GeV in the $2j+\not \! p_{T}$ channel. For an $e\gamma$
collider with $\sqrt{s_{e^+e^-}}$~=~3~TeV and total
luminosity $L$~=~10$^{5}$~pb$^{-1}$, the scalar leptoquarks of type 
$S_{0},S_{1}^{0},R_{1/2}^{-1/2}$ could be produced up to mass 
$M_{LQ}\approx$~2600~GeV, types 
$\widetilde{R}_{1/2}^{-1/2},\widetilde{S}_{0},S_{1}^{-1}$ up to mass
$M_{LQ}\approx$~2500~GeV, type $R_{1/2}^{1/2}$ up to 
$M_{LQ}\approx$~2100~GeV in
the $2j+e$ channel; and scalar leptoquarks of type $R_{1/2}^{1/2}$ up to 700
GeV, $S_{0}$ up to 900~GeV and $S_{1}^{0}$ up to 1300~GeV in the
$2j+\not \! p_{T}$ channel. The total number of signal events and
statistical significance for these channels
are given in Tables~\ref{table2} and \ref{table3}.

To conclude: scalar leptoquarks can be produced in large numbers at
CLIC-based $e\gamma$ colliders. We have analysed the contributions from
direct and resolved photon processes to the total cross section. We
find that 
the latter contribution is important and cannot be ignored, especially for
small leptoquark masses. Looking at the final-state particles and their
signature in detectors, scalar leptoquarks of various types can be
identified easily.

\newpage
\begin{sideways}
\begin{threeparttable}
\caption{The same as Table~\ref{table2}, but for a CLIC-based
$e\gamma$ collider with $\sqrt{s_{e^+e^-}}$~=~3~TeV} 

\vspace*{-17mm}

\label{table3}

\vspace*{7mm}

\renewcommand{\arraystretch}{1.37} 
%
\large{
\begin{tabular}{ccccccccccccc}
 & & & & & & & & & & &  &   \\
\hline \hline\\[-6mm]
 & & & & & & & & & & &  &   \\[-4mm]
$\hspace*{2mm}$ \boldmath{$\sqrt{s_{e^{+}e^{-}}}$}\textbf{~=~1~TeV} 
$\hspace*{2mm}$ & 
\multicolumn{2}{c}{ \textbf{Number of events} } & $\hspace*{5mm}$ & 
\multicolumn{5}{c}{\boldmath{$\frac{S}{\sqrt{B}}\,
(e\gamma\to q\overline{q}e)$}} & $\hspace*{5mm}$ &
\multicolumn{3}{c}{\boldmath{$\frac{S}{\sqrt{B}}
(e\gamma\to q\overline{q}^{\prime}\nu)$}} \\[1.5mm]
\boldmath{$L_{\rm int}$\textbf{~=~10}$^5$\textbf{~pb}$^{-1}$} & 
\multicolumn{2}{c}{
\boldmath{$(\sigma_{\rm D}+\sigma_{\rm R})~\times~L_{\rm int}$}} 
& & & 
\\[2.5mm] \hline \\[-3mm]
M$_{\text{\rm LQ}}$(GeV) & 
$\hspace*{1mm}$ $|Q|$~=~5/3 $\hspace*{1mm}$ & 
$\hspace*{1mm}$ $|Q|$~=~4/3 $\hspace*{1mm}$ &  &
$\hspace*{0.5mm}$ $S_{0}$ $\hspace*{0.5mm}$ & 
$\hspace*{0.5mm}$ $S_{1}^{0}$ $\hspace*{0.5mm}$ & 
$\hspace*{0.5mm}$ $R_{1/2}^{-1/2}$ $\hspace*{0.5mm}$ & 
$\hspace*{0.5mm}$ $R_{1/2}^{1/2}$ $\hspace*{0.5mm}$ & 
$\hspace*{0.5mm}$
$\widetilde{R}_{1/2}^{-1/2},\widetilde{S}_{0},S_{1}^{-1}$ 
$\hspace*{0.5mm}$ &  & 
$\hspace*{0.5mm}$ $S_{0}$ $\hspace*{0.5mm}$ & 
$\hspace*{0.5mm}$ $S_{1}^{0}$ $\hspace*{0.5mm}$ & 
$\hspace*{0.5mm}$ $R_{1/2}^{1/2}$ $\hspace*{0.5mm}$ \\[5mm] \hline \\[-3mm]
300 & 181523 & 124657 &  & 648 & 491 & 981 & 337 & 674  & & 33 & 49 & 34\\
500 & 73182 & 35515 &  & 261 & 198 & 396 & 99 & 197 &  & 13 & 20 & 10\\
700 & 43722 & 18168 &  & 156 & 118 & 236 & 49 & 98  & & 8 & 12 & 5\\
900 & 30084 & 11135 &  & 107 & 81 & 163 & 30 & 60  & & 5 & 8 & 3\\
1100 & 22141 & 7513 &  & 79 & 60 & 120 & 20 & 41  & & 4 & 6 & --\\
1300 & 16929 & 5335 &  & 60 & 46 & 92 & 14 & 29  & & -- & 5 & --\\
1500 & 13251 & 3906 &  & 47 & 36 & 72 & 11 & 21  & & -- & 4 & --\\
1700 & 10520 & 2922 &  & 38 & 28 & 57 & 8 & 16  & & -- & -- & --\\
1900 & 8406 & 2225 &  & 30 & 23 & 45 & 6 & 12  & & -- & -- & --\\
2100 & 6705 & 1728 &  & 24 & 18 & 36 & 5 & 9  & & -- & -- & --\\
2300 & 5340 & 1377 &  & 19 & 14 & 29 & 4 & 7  & & -- & -- & --\\
2500 & 3842 & 975 &  & 14 & 10 & 21 & -- & 5 & &  -- & -- & --\\
2700 & 824 & 206 &  & 3 & 2 & 4 & -- & 1  & & -- & -- & --
\\[5mm] 
\hline \hline
%
\end{tabular}
}
\end{threeparttable}
\end{sideways}

\newpage

\subsection{Lepton-Size Measurements}

High-energy $e^+e^-$ scattering can also be used to study whether the
leptons have a finite size. Similar studies have been made in the past,
for example at LEP~\cite{Bourilkov:2000ap}. A model inspired by string
theory, which can lead to elastic ministrings, with finite size (dubbed
Nylons), has been proposed in~Ref.~\cite{lykken}.

In the presence of structure, annihilations to fermions $f$ will be
modified to become 
$\frac{d\sigma}{dQ^2}=(\frac{d\sigma}{dQ^2})_{\rm SM}F_e^2(Q^2)
F_f^2(Q^2)$, 
with $ F(Q^2)=1+\frac{1}{6}Q^2R^2$, where $R$
is an effective fermion radius. The current best limit on the radius $R_e$
of the electron comes from LEP, and is 
$R_e<$~2.8~$\times$~10$^{-17}$~cm (95~C.L.). For CLIC, the results for
$ee\to ee$, under the assumption 
that the $t$ channel dominates, for an angular cut on the scattered
electrons of $|\cos\theta|<$~0.9, and for a luminosity of 500~fb$^{-1}$,
are given in Table~\ref{sec6:nylons}. We see that a 5-TeV machine will
probe structures of dimensions a factor of 10 smaller than in the
present results from LEP~\cite{Bourilkov:2000ap}.
%
\begin{table}[!h] 
\caption{ Bounds on the possible electron radius $R_e$ obtainable with 
CLIC operating at various centre-of-mass energies}
\label{sec6:nylons}

\renewcommand{\arraystretch}{1.3} 
\begin{center}

\begin{tabular}{ccc}\hline \hline \\[-4mm]
$\hspace*{3mm}$ \textbf{Radius (cm)} $\hspace*{3mm}$ & $\hspace*{6mm}$ &
$\hspace*{6mm}$ \boldmath{$\sqrt{s}$}~\textbf{(TeV)}  $\hspace*{6mm}$ 
\\[4mm]   

\hline \\[-3mm]
3.0~$\times$~10$^{-18}$ &  &  1 \\
1.2~$\times$~10$^{-18}$ &  &  3 \\
0.9~$\times$~10$^{-18}$ &  &  5
\\[3mm] 
\hline \hline
\end{tabular}
\end{center}
\end{table}


\subsection{Excited Electrons}

The replication of three fermionic generations of known quarks and leptons
suggest, the possibility that they are composite structures made up of more
fundamental constituents. The existence of such quark and lepton
substructure leads one to expect a rich spectrum of new particles with
unusual quantum numbers. A possible signal of excited states of quarks and
leptons~\cite{9} as predicted by composite models~\cite{10} would supply
convincing evidence for a new substructure of matter. All composite models
of fermions have an underlying substructure that may be characterized by
a scale $\Lambda$. The absence of electron and muon electric dipole
moments implies that the excited leptons must have specific chiral
properties. A right-handed excited lepton should couple only to the
left-handed component of the corresponding lepton. Excited leptons may be
classified by $SU(2)~\times~U(1)$ quantum numbers, and they are assumed to
be both left- and right-handed weak isodoublets. In the effective
Lagrangian of~\cite{9}, the parameters $f$ and $f^{\prime }$ associated
with the gauge groups SU(2) and U(1) depend on compositeness dynamics and
describe the effective changes from the SM coupling constants $g$ and
$g^{\prime }$.

Experimental lower limits for the excited electron mass are given as
$m_{e^\star}>$~200~GeV in~\cite{11}, and
$m_{e^\star}>$~306~GeV~in~\cite{12}. The 
higher limits are derived from indirect effects due to $e^\star$ exchange
in the $t$ channel, and depend on a transition magnetic coupling between
$e$ and $e^\star$. Relatively small limits
($m_{\mu^\star,\tau^\star}>$~94.2~GeV) for excited muon ($\mu^\star$) and
excited tau ($\tau^\star$) are given by the L3 experiment
at~LEP~\cite{13}. 

In this work, resonant production of an excited electron in the $s$
channel and its subsequent decay modes $e^\star \to e \gamma$,
$e^\star \to \nu W$, $e^\star \to eZ$ are considered. In
addition, we include the contributions coming from the excited electron in
the $t$ channel. The production cross section and decays of excited
electrons are calculated using the effective Lagrangian of~\cite{9}, which
depends on a compositeness scale $\Lambda$ and on free parameters $f$ and
$f^{\prime}$. We have implemented the interaction vertices into
CompHEP~\cite{comphep} for the excited electron interactions with leptons and
gauge bosons. We choose either $f=f^\prime$ or $f=-f^\prime$ in our
calculations in order to reduce the number of free parameters. For the
case $f=f^\prime$ ($f=-f^\prime$) the coupling of the photon to excited
neutrinos (electrons) vanishes.

When produced, an excited electron decays predominantly into a $W$ boson
and a neutrino, and the branching ratios are insensitive to the excited
electron mass when it is high with respect to $m_W$ or $m_Z$. We obtain the
limiting values for the branching ratios at large $m_{e^\star}$ as 0.28, 0.60
and 0.11 for the coupling $f=f^{\prime}$~=~1, in the photon, $W$ and $Z$
channels, respectively. In the case $f=-f^{\prime}$~=~--~1, the branching
ratio for the $W$ channel does not change, whilst it increases to 0.39
for the $Z$ channel. Taking $f=-f^{\prime}$~=~1 and $\Lambda$~=~1~TeV,
we find the 
total decay width of an excited electron to be 0.83~GeV and 6.9~GeV
for $m_{e^\star}$~=~500~GeV and $m_{e^\star}$~=~1000~GeV, respectively.

Excited electrons can be produced directly via the subprocess $e\gamma
\to e^{\star}\to lV$ ($V=\gamma,Z,W$) and indirectly via a
$t$-channel exchange diagram~\cite{14}. The production cross section of
excited electrons in three modes, taking $f=f^\prime$~=~1 and
$\Lambda=m_{e^\star}$, are given in Table~\ref{table4} 
for CLIC-based $e\gamma$ colliders at the 
centre-of-mass energy $\sqrt{s_{e^+e^-}}$~=~1~TeV
($\sqrt{s_{e^+e^-}}$~=~3~TeV). 
The following information can be read off~Table~\ref{table4} the following
information: the $W$ channel gives higher cross sections than the others,
but there is an ambiguity with the neutrino in this final state.
Therefore, the photon channel gives a more promising result, because of its
simple kinematics. The cross sections and the numbers of signal events are
shown in Table~\ref{table4}. For the signal and background processes we
apply a cut $p^{e,\gamma}_T>$~10~GeV for experimental identification of
final-state particles. The backgrounds to the $W$ and $Z$ decay channels
in the hadronic final states are fairly large. The backgrounds to the
photonic final states are relatively small with respect to the $W$ and
$Z$~channels.

\begin{figure}[htbp] 
\begin{center}
\includegraphics[width=7cm,height=6cm]{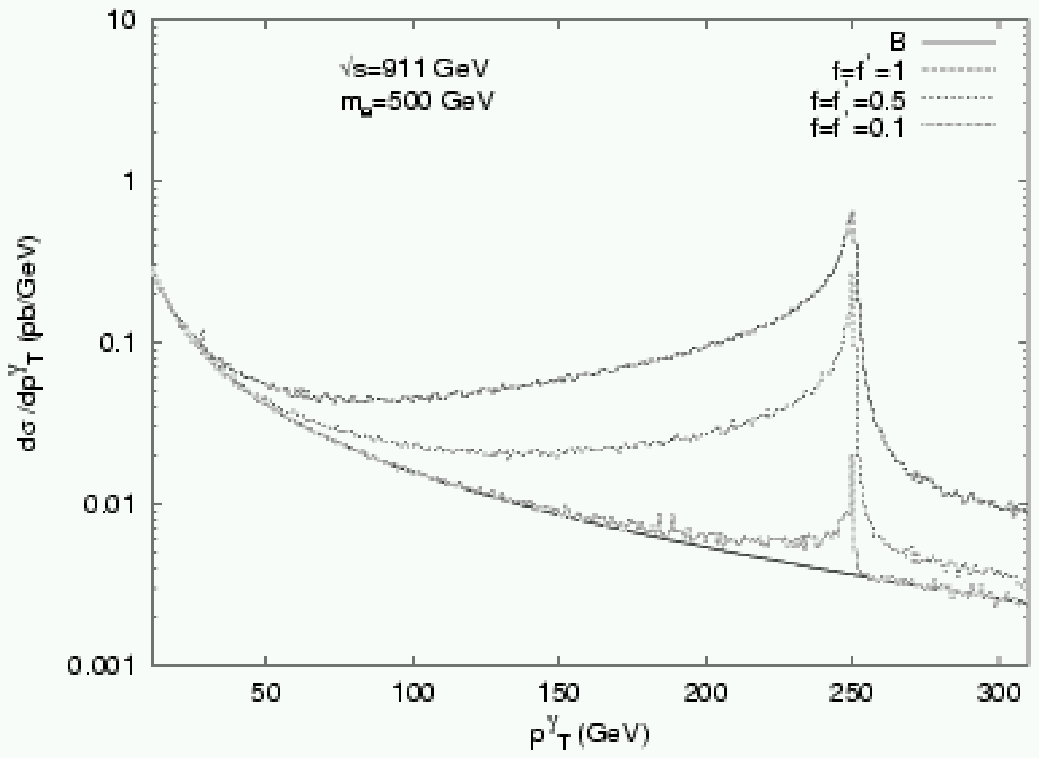}
\includegraphics[width=7cm,height=6cm]{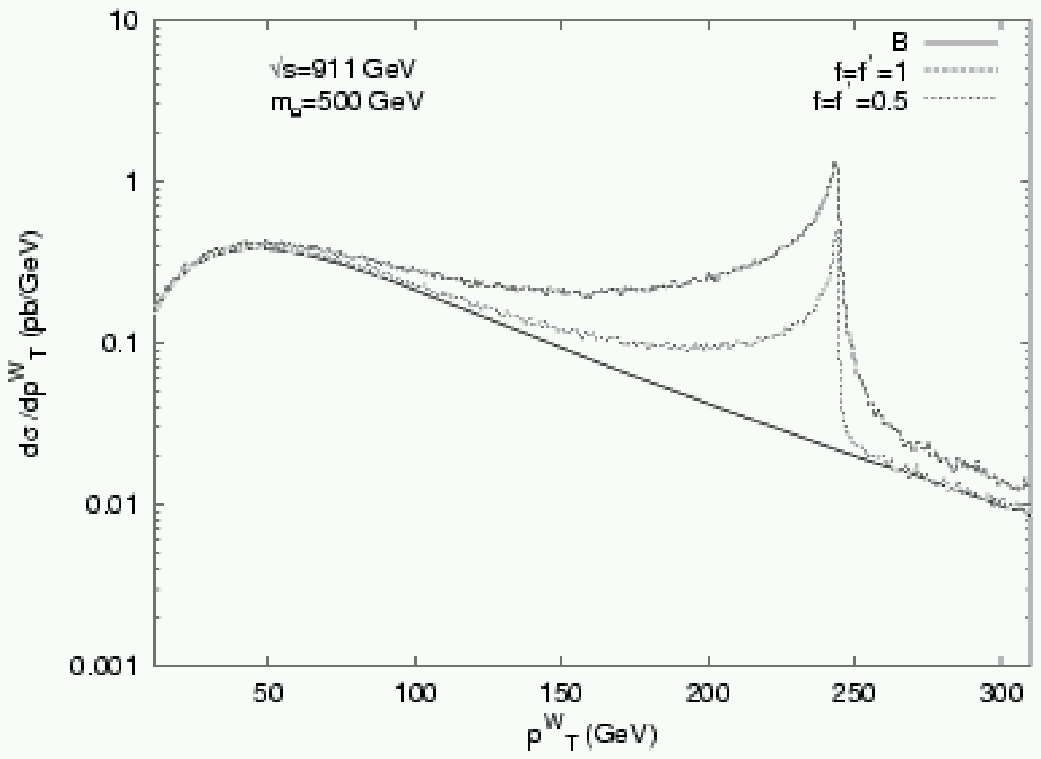}
\includegraphics[width=7cm,height=6cm]{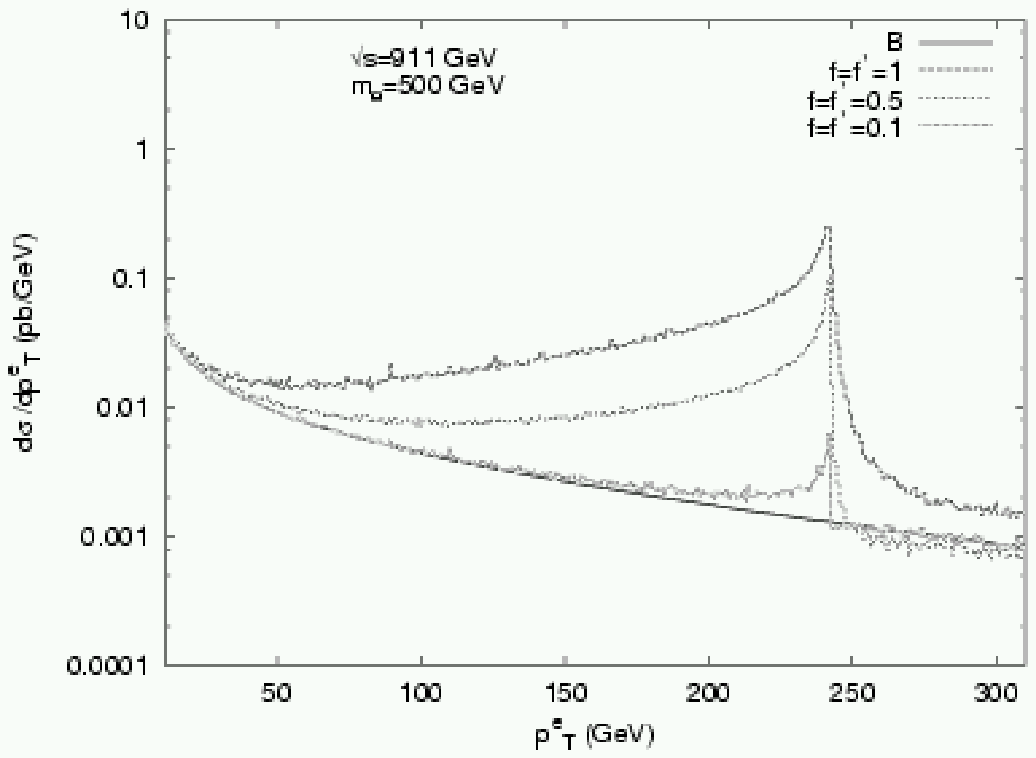}
\caption{Transverse momentum distribution of the photon, $W$ boson and
electron from excited electron production processes at a CLIC-based
$e\gamma$ collider with $\sqrt{s_{e\gamma}^{\rm max}}$~=~911~GeV for
different couplings $f=f^\prime$ and $m_{e^\star}$~=~500~GeV} 
\label{fig2}
\end{center}
\end{figure}

For the signal process {$e\gamma \to e^{\star}\to
e\gamma $}, the transverse momentum $p_{T}$ distribution of the photon or
electron is peaked around the half of the mass value of the  excited
electron, as seen in~Fig.~\ref{fig2}. For the process 
$e\gamma \to e^{\star }\to \nu W \,(eZ)$, the $p_{T}$ distribution of
the $W$ boson (electron) shows a peak around
${m_{e^\star}/2}-{m^{2}_{V}/2m_{e^\star}}$, 
where $m_V$ denotes the $W$ boson
(or $Z$ boson) mass.

In order to estimate the number of events for the signal and background in
a chosen $p_T$ window, we integrate the transverse momentum distribution
around the half the mass of each excited electron
($m_{e^\star}/2-m_V^2/2m_{e^\star}$) in an interval of the transverse momentum
resolution $\Delta p_T$. Here $\Delta p_T$ is approximated as $\approx 10$
GeV for $m_{e^\star}~=~500$~GeV and $\approx 15$~GeV for $m_{e^\star}=
750$~GeV, for
a generic detector. In order to calculate the significance of the signal,
we use integrated luminosities for CLIC-based $e \gamma$ colliders of
$L$~=~10$^5$~pb$^{-1}$. In order to see the potential of CLIC-based $e
\gamma$ colliders to search for an excited electron, we define the
statistical significance $S/\sqrt{B}$, where $S$ and $B$ are the total
number of events for signal and background inside the chosen $p_T$ window,
respectively. We calculate the value of $S/\sqrt{B}$ for different
couplings, assuming $f=f^\prime$ and requiring the condition
$S/\sqrt{B}>$~5 for the signal to be observable. We find from
Table~\ref{table4} that excited electrons can be observed down to
couplings $f\simeq$~0.05 and $f\simeq$~0.1 at CLIC-based $e\gamma$
colliders with $\sqrt{s_{e^+e^-}}$~=~1~TeV and 
$\sqrt{s_{e^+e^-}}$~=~3~TeV,~respectively.

In this study, we have assumed that the excited electrons interact with
the SM particles through the effective Lagrangian~\cite{9}.
This may be a conservative assumption, because it is possible for excited
fermions to couple to ordinary quarks and leptons via contact interactions
originating from the strong constituent dynamics. In this case, the decay
widths can be enhanced~\cite{15}.

To conclude: we have presented results for excited electron production
with subsequent decays into three channels. The excited electrons
can be produced at CLIC-based $e\gamma$ colliders up to the kinematical
limit in each channel, due to the nearly flat photon energy spectrum. We
find that an excited electron with mass 500~(750)~GeV can be probed down
to the coupling $f=f^{\prime}\simeq$~0.05 at $\sqrt{s_{e^+e^-}}$~=~1~TeV. At
a CLIC-based $e\gamma$ collider with $\sqrt{s_{e^+e^-}}$~=~3~TeV the excited
electron can be probed down with couplings as low as 
$f=f^{\prime}\simeq$~0.1.

\newpage
\begin{sideways}
\begin{threeparttable}
\caption{The total cross sections of signal and background inside
the chosen bin and the statistical significance ($S/\sqrt{B}$) values
for different $f=f^\prime$ for $m_{e^\star }$~=~750~GeV at
$\sqrt{s_{e^+e^-}}$~=~1~TeV ($\sqrt{s_{e^+e^-}}$~=~3~TeV) with
$L$~=~10$^5$~pb$^{-1}$.} 

\vspace*{-9mm}

\label{table4}

\vspace*{7mm}

\renewcommand{\arraystretch}{2} 
%
\large{
\begin{tabular}{cccccccccc}
 &  & & & & & &  &  &  \\
\hline \hline \\[-6mm]
 &  & & & & & &  &  &  \\[-10mm]
 & $\hspace*{13mm}$ & 
 \multicolumn{2}{c}{ \boldmath{$e\gamma\to e\gamma$} }  & $\hspace*{13mm}$ 
 & \multicolumn{2}{c}{ \boldmath{$e\gamma\to \nu W$} }  & $\hspace*{13mm}$ 
 & \multicolumn{2}{c}{ \boldmath{$e\gamma\to eZ$} } 
\\[4mm] \hline \\[-5mm]
$f=f^{\prime}$ &  & $\sigma_{\rm tot}$~(pb) & $S/\sqrt{B}$ & & 
$\sigma_{\rm tot}$~(pb) & $S/\sqrt{B}$ &   & 
$\sigma_{\rm tot}$~(pb) & $S/\sqrt{B}$ 
\\[5mm] \hline \\[-3mm]
1.00 &  & 24.70~(4.30) & 1892.4~(963.1)  & &  
80.10~(53.40) & 1869.3~(211.3)  & & 8.90~(1.51)   & 1905.5~(705.4)  \\
0.50 &  & 11.30~(1.80) & 963.1~(183.8)  &  &  
51.40~(50.00) & 466.5~(51.9)  &  &  3.40~(0.53)  & 694.4~(131.2) \\
0.10 &  & 7.04~(1.20)  & 18.9~(6.1)   &    &  
42.13~(48.94) & 13.2~(0.7)  &   &  1.60~(0.32)  & 24.3~(4.1)  \\
0.05 &  & 6.87~(1.18)  & 0.7~(0.8)   &     &  
41.88 & 1.1  &    & 1.52   & 3.9  
\\[5mm] 
\hline \hline
%
\end{tabular}
}
\end{threeparttable}
\end{sideways}

\newpage

\subsection{Non-Commutative Theories}

Recent theoretical studies have demonstrated that non-commutative (NC)
quantum field theories appear naturally in the context 
of string theory in the presence of background 
fields~\cite{seiwit}. In this case, the usual space-time coordinates
no longer commute, and instead obey the relation
\begin{equation}
[\hat x_\mu, \hat x_\nu]=i\theta_{\mu\nu}\,, 
\label{NC1}
\end{equation}
where the $\theta_{\mu\nu}$ are a constant, frame-independent set of six
dimensionful parameters~\cite{rev}. The $\theta_{\mu\nu}$ may be separated
into two classes: (i) space--space non-commutativity with
$\theta_{ij}=c^B_{ij}/\Lambda_{\rm NC}^2$, 
and (ii) space-time non-commutativity
with $\theta_{0i}=c^E_{0i}/\Lambda_{\rm NC}^2$. The quantities $\hat c_{E,B}$
are two, fixed, frame-independent unit vectors associated with the NC
scale $\Lambda_{\rm NC}$.  The NC scale $\Lambda_{\rm NC}$ is most likely to be
near the string or Planck scale, which could be as low as a~TeV; it
characterizes the threshold where NC effects become apparent. Since the
two vectors $\hat c_{E,B}$ are frame-independent, they correspond to
preferred directions in space, related to the directions of the background
fields. Thus, NC theories violate Lorentz invariance, and upper limits on
such effects provide model-dependent constraints on the NC
scale~\cite{rev,sean}. Since experimental probes of NC theories are
sensitive to the directions of $\hat c_{E,B}$, the experiments must employ
astronomical coordinate systems and time-stamp their data so that 
the rotation of the Earth or the Earth's motion around the Sun, for
example, does
not wash out or dilute the effect through time-averaging.  Note that
momenta still commute in the usual way and hence energy and momentum
remain conserved quantities, as does CPT.

It is possible to construct non-commutative analogues of conventional field
theories following two approaches: Weyl--Moyal (WM)~\cite{rev} or
Seiberg--Witten (SW)~\cite{seiwit}, both of which have their own advantages
and disadvantages. In the SW approach, the field theory is expanded in a
power series in $\theta$, which then produces an infinite tower of
additional operators. At any fixed order in $\theta$ the theory can be
shown to be non-renormalizable~\cite{nonr}.  The SW construction can,
however, be applied to any gauge theory with arbitrary matter
representations.  In the WM approach, only $U(N)$ gauge theories are found
to be closed under the group algebra, and the matter content is restricted
to the (anti-)fundamental and adjoint representations.  However, these
theories are at least one-loop renormalizable, and appear to remain so
even when spontaneously broken~\cite{renorm}.  These distinctive
properties of NC gauge theories render it difficult to construct a
satisfactory non-commutative version of the SM~\cite{models}.

However, a NC version of QED is a well-defined theory in the WM approach,
and its implications for very-high-energy $e^+e^-$ colliders may be
explored. This version of NCQED differs from ordinary QED in several ways:
(i) the ordering of products of fields must be preserved through the
introduction of the star product~\cite{rev}, which absorbs the effect of
the commutation relation via a series of Fourier transforms; (ii) all
vertices pick up a Lorentz-violating phase factor that is dependent on the
momenta flowing through the vertex; (iii) NCQED takes on a non-Abelian
nature in that trilinear and quartic photon couplings are generated; and
(iv) only the charges $Q$~=~0, $\pm$~1 are allowed by gauge invariance.  We
note that propagators, however, are not modified since quadratic forms
remain unchanged when one introduces the star product.

NCQED provides a testing ground for the basic ideas behind NC quantum
field theory, and produces striking signatures in QED processes at CLIC.
The NC modifications to pair annihilation, Bhabha, M{\o}ller, and Compton
scattering, as well as $\gamma\gamma\to \gamma\gamma,e^+e^-$ have been
examined by a number of authors~\cite{ncqed}. Some of these processes
receive new diagrammatic contributions due to the non-Abelian couplings,
and all of them acquire a phase dependence due to the relative
interference of the vertex kinematic phases.  The lowest-order NC
contribution to these processes occurs via dimension-8 operators whose
scale is set by $\Lambda_{\rm NC}$.  The most striking result is that an
azimuthal-angle, $\phi$, dependence is induced in $2\to 2$ scattering
processes, since there exist the NC-preferred space-time directions,
providing a unique signature of the Lorentz violation 
inherent in these~theories.

\begin{figure}[t] 
\centerline{
\includegraphics[width=7.7cm,angle=0]{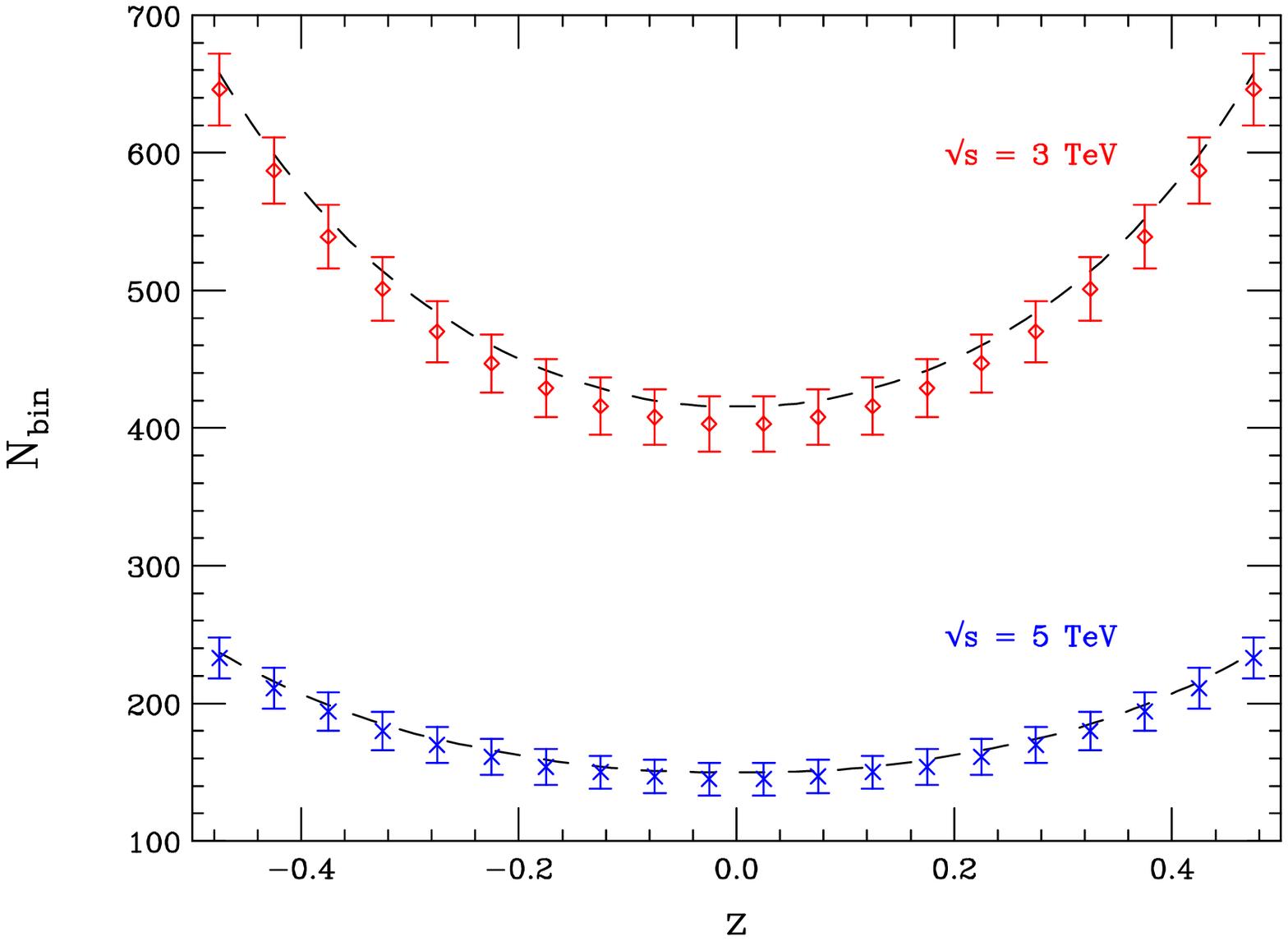}
\hspace*{5mm}
\includegraphics[width=7.7cm,angle=0]{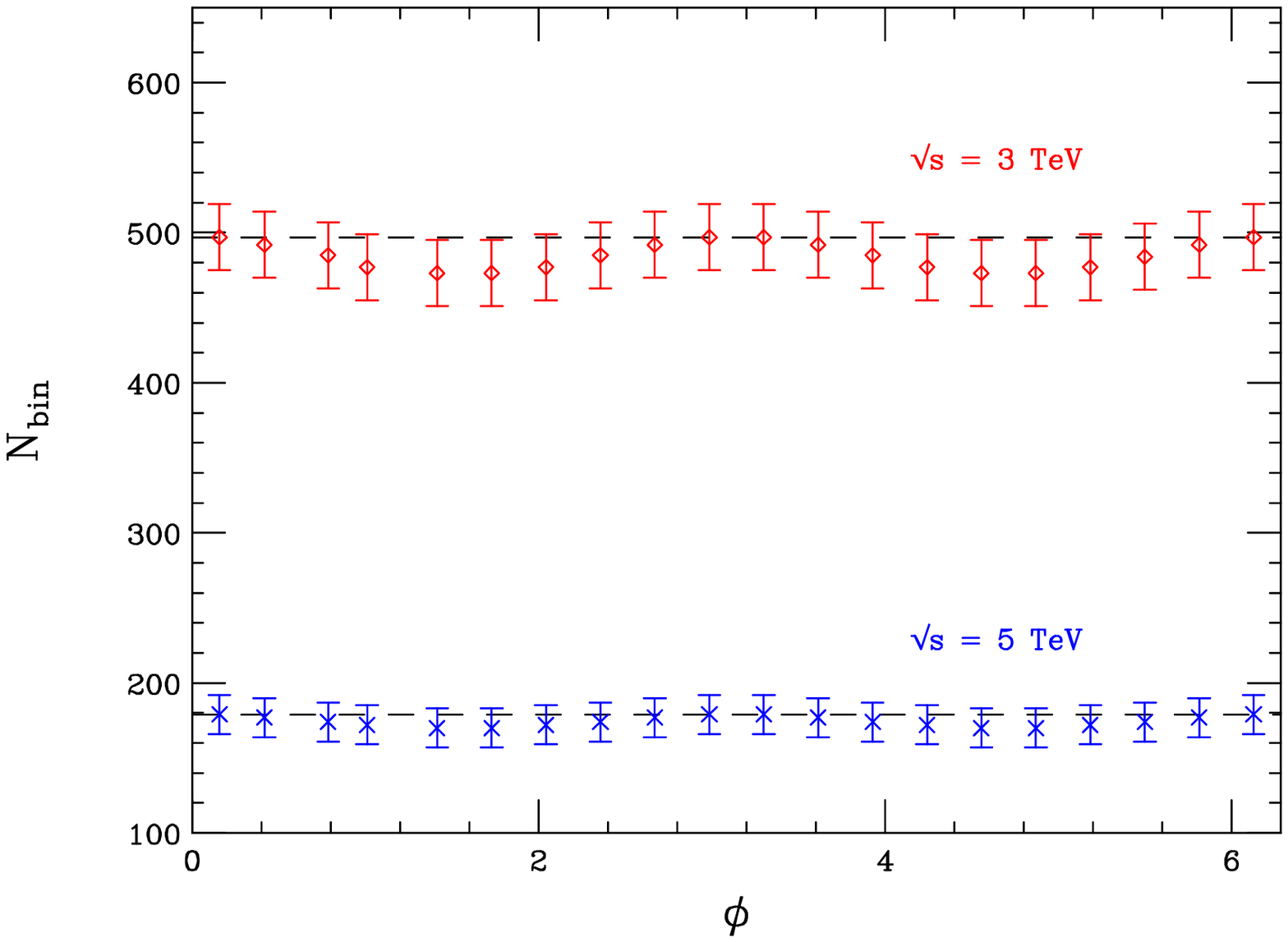}}
\vspace*{5mm}
\caption{Shifts in the $z=\cos \theta$ and $\phi$ distributions for the 
process $e^+e^- \to \gamma \gamma$ at a 3 or 5~TeV CLIC, assuming an
integrated luminosity of 1 ab$^{-1}$. The dashed curves  
show the SM expectations while the `data' assume $c_{02}$~=~1 and 
$\Lambda_{\rm NC}=\sqrt s$. A cut of $|z|<$~0.5 has been applied in
the $\phi$~distribution.}
\label{E3064_fig2}
\end{figure}

\begin{figure}[htbp]
\centerline{
\includegraphics[width=7.7cm,angle=0]{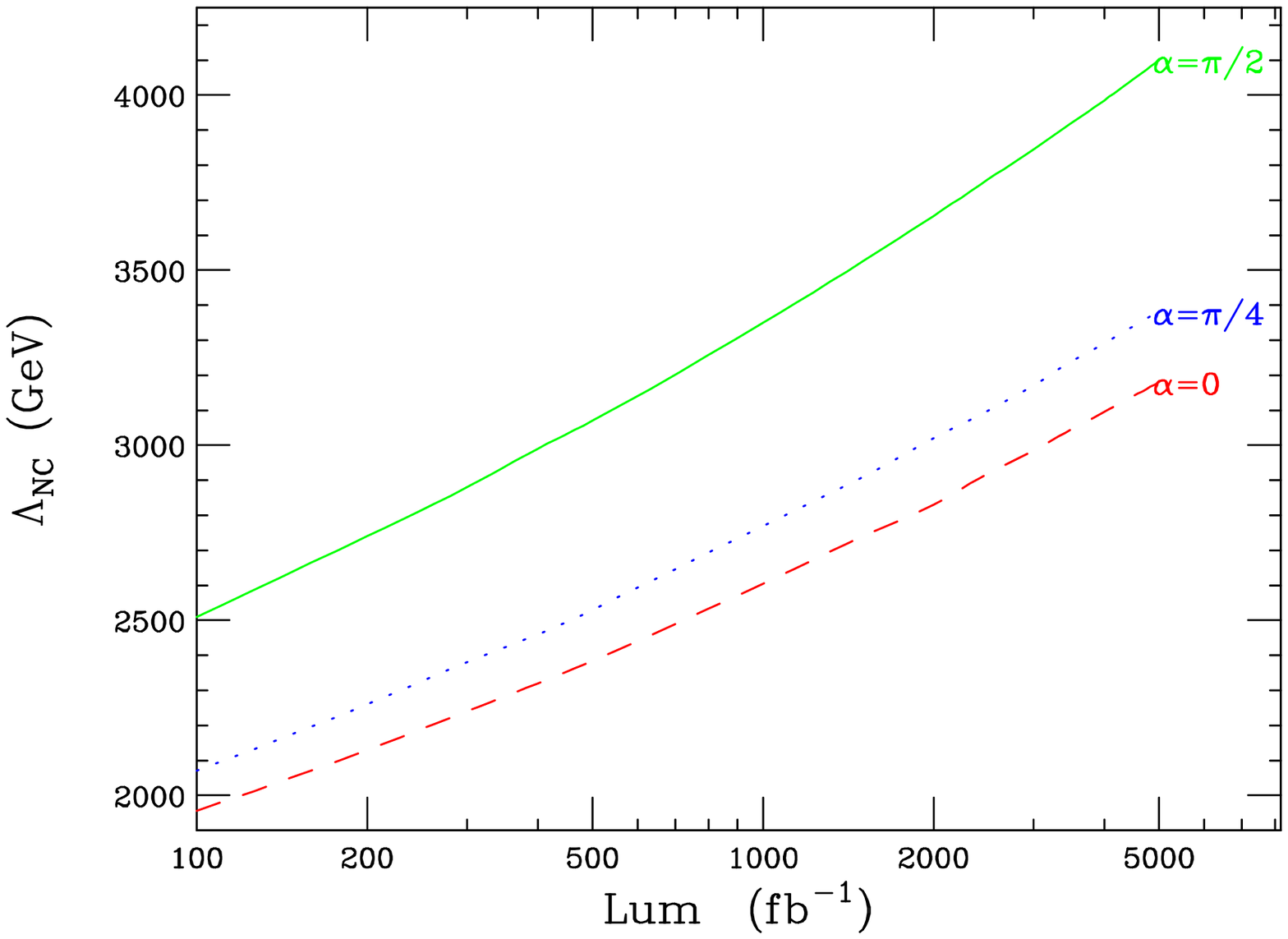}
\hspace*{5mm}
\includegraphics[width=7.7cm,angle=0]{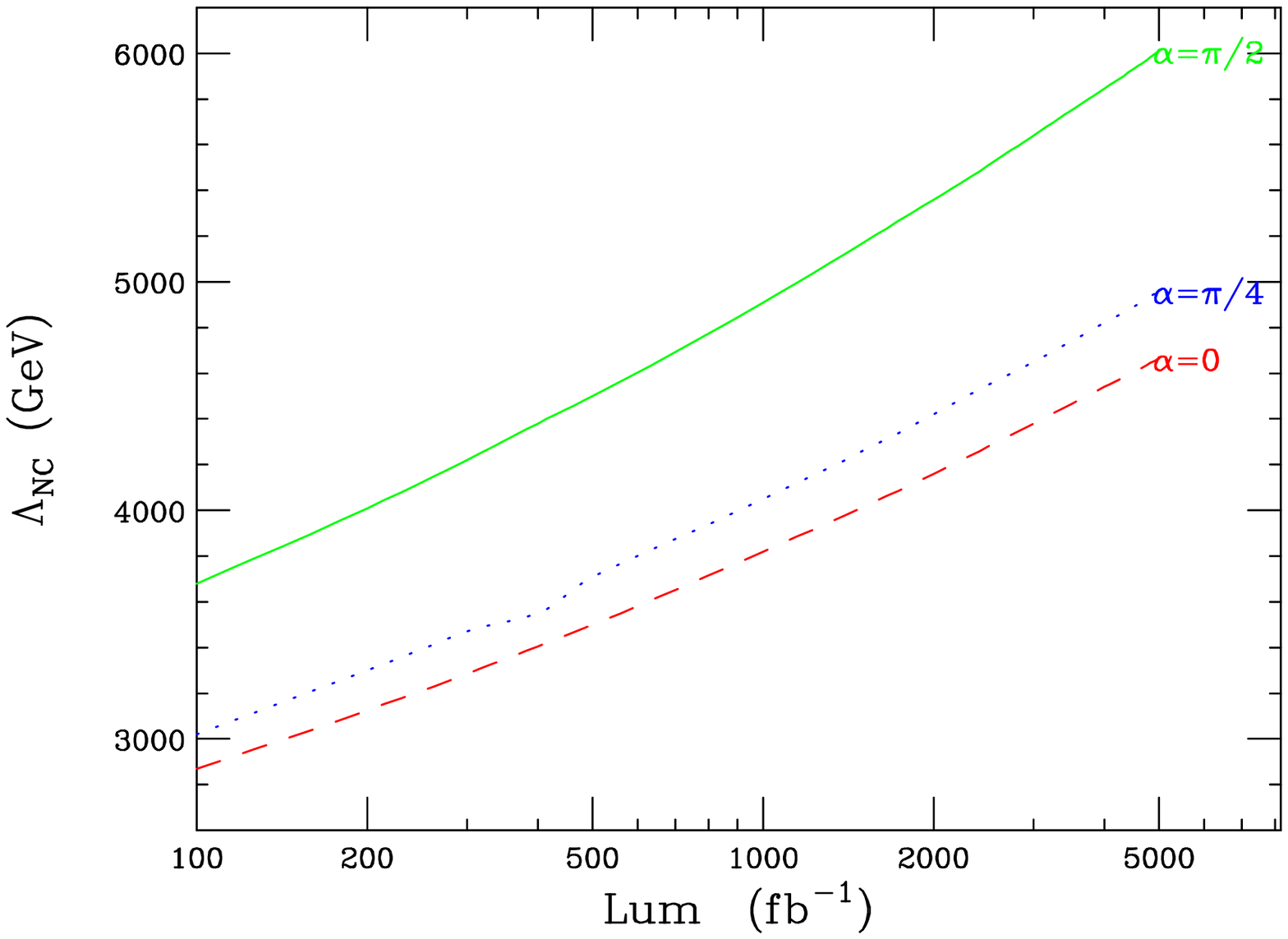}}
\vspace*{5mm}
\caption{Reach for $\Lambda_{\rm NC}$ at a 3~TeV (left) or a 5~TeV 
(right) CLIC as a function of the integrated luminosity for the process 
$e^+e^- \to \gamma \gamma$ following the notation discussed in the text}
\label{E3064_fig3}
\end{figure}

Very-high-energy $e^+e^-$ colliders such as CLIC will allow us to probe
values of $\Lambda_{\rm NC}$ up to several~TeV provided sufficient luminosity,
$\sim$~1~ab$^{-1}$, is available~\cite{jpr}. To demonstrate this claim, we
examine the case of pair annihilation, taking the incoming $e^-$ direction
to be along the $z$-axis.  In addition to the new phases that enter the
$t$- and $u$-channel amplitudes, there is now an additional $s$-channel
photon exchange graph involving the NC-generated three-photon vertex.  
The resulting azimuthal dependence and $\cos\theta$ distribution are shown
in~Fig.~\ref{E3064_fig2} for the illustrative case $c_{02}$~=~1. Writing
$c_{01}=\cos \alpha$, $c_{02}=\sin \alpha \cos \beta$, and 
$c_{03}=\sin\alpha \sin \beta$, Fig.~\ref {E3064_fig3} 
displays the reach for
$\Lambda_{\rm NC}$ for CLIC energies for several values of~$\alpha$.

The 95\% C.L. search reaches for the NC scale in a variety of processes
are summarized in~Table~\ref{summ} for both $\sqrt{s}$~=~500~GeV and CLIC
energies. We see that these machines have a reasonable sensitivity to NC
effects and provide a good probe of such theories.  It turns out that 
M{\o}ller scattering, while sensitive only to space/space
non-commutativity, is a particularly promising process for
experimental investigation: once the 
non-commutative Feynman rules are written down, we can interpret precise
measurements of the M{\o}ller scattering cross-Section.  It may be the
interaction of choice for a possible observation of this new effect.

~~~\\

\vspace*{15mm}

~\\

%

\begin{table}[htbp]  
\caption{Summary of the 95\% C.L. search limits on the NC scale
$\Lambda_{\rm NC}$ from the various processes considered above at
a 500~GeV $e^+e^-$ linear collider with an integrated luminosity
of 500~fb$^{-1}$ or at a 3 or 5~TeV CLIC with an integrated luminosity 
of 1 ab$^{-1}$.  The sensitivities are from the first two papers 
in~Ref.~\cite{ncqed}. The $\gamma\gamma\to e^+e^-$ and $\gamma e\to\gamma e$ 
analyses of Godfrey and Doncheski include an overall 2\% systematic error 
not included by Hewett, Petriello and Rizzo.}
\label{summ}

\vspace*{5mm}

\renewcommand{\arraystretch}{1.4} 
\begin{center}

\begin{tabular}{cccccc}\hline \hline \\[-4mm]
\textbf{Process} & $\hspace*{2mm}$ &
$\hspace*{0.6mm}$ \textbf{Structure probed} $\hspace*{0.6mm}$ & 
$\hspace*{0.6mm}$ \boldmath{$\sqrt s$}\textbf{~=~500~GeV} 
$\hspace*{0.6mm}$ & 
$\hspace*{0.6mm}$ \boldmath{$\sqrt s$}\textbf{~=~3~TeV} $\hspace*{0.6mm}$ & 
$\hspace*{0.6mm}$ \boldmath{$\sqrt s$}\textbf{~=~5~TeV} $\hspace*{0.6mm}$ 
\\[4mm]   

\hline \\[-3mm]
$e^+e^-\to\gamma\gamma$ &  & 
Space-time & 
740--840~GeV & 
2.5--3.5~TeV & 
3.8--5.0~TeV \\
M{\o}ller scattering &  & 
Space--space &  
1700~GeV & 
5.9~TeV & 
8.5~TeV \\ 
Bhabha scattering &  & 
Space-time & 
1050~GeV & 
4.5--5.0~TeV & 
6.6--7.2~TeV \\
$\gamma\gamma\to\gamma\gamma$ &  & 
Space-time & 
700--800~GeV  & & \\ 
&  & Space--space & 
500~GeV & & \\ 
$\gamma\gamma\to e^+e^-$ &  & 
Space-time & 
220--260~GeV &
1.1--1.3~TeV & 
1.8--2.1~TeV \\ 
$\gamma e\to\gamma e$ &  & 
Space-time & 
540--600~GeV & 
3.1--3.4~TeV & 
4.8--5.8~TeV \\ 
&  & Space--space & 
700--720~GeV & 
4.0--4.2~TeV & 
6.3--6.5~TeV
\\[3mm] 
\hline \hline
\end{tabular}
\end{center}
\end{table}


\section{Summary}

A summary of the reach for exotic physics at CLIC, assuming a 
centre-of-mass energy of 5~TeV and an integrated luminosity of 
5~ab$^{-1}$, is given in Table~\ref{exsumm}.

\begin{table}[!h] 
\caption{Summary of the physics reach of CLIC for various examples 
of exotic new physics, assuming $\sqrt{s}$~=~5~TeV and an integrated 
luminosity of~5~ab$^{-1}$}
\label{exsumm}

\vspace*{5mm}

\renewcommand{\arraystretch}{1.35} 
\begin{center}

\begin{tabular}{ccc}\hline \hline \\[-3mm]
$Z'$ (direct) & $\hspace*{30mm}$ & 5~TeV\\
$Z'$ (indirect) & & 30~TeV\\
$l^*,q^*$ & & 5~TeV\\
TGC (95\%) & & 0.00008\\
$\Lambda$ compos. & & 400~TeV\\
$W_{L}W_{L}$ & & $>$~5~TeV \\
ED (ADD) & & 30~TeV  ($e^+e^-$)\\
  & & 55~TeV  ($\gamma\gamma$)\\
ED (RS) & & 18~TeV ($c$~=~0.2)\\
ED (TeV$^{-1}$) & & 80~TeV\\
Resonances & & $\delta M/M, \delta \Gamma/\Gamma \sim$~10$^{-3}$\\
Black holes & & 5~TeV 
\\[3mm] 
\hline \hline
\end{tabular}
\end{center}
\end{table}


As we have shown in this chapter, the particular characteristics of CLIC, 
such as larger beamstrahlung and $\gamma \gamma$ backgrounds than a 
lower-energy LC, are not in general obstacles to full exploitation of 
CLIC's higher centre-of-mass energy. If the new physics is within CLIC's 
kinematic reach, CLIC will find it.

In many cases, CLIC will be able to explore in detail and unravel hints of 
new physics that might be found at the LHC. In other cases, CLIC can reach 
even far beyond the LHC, thanks to its relatively clean experimental 
conditions and its democratic production of new weakly-interacting 
particles.

\newpage
\thispagestyle{empty}
~

\newpage
\chapter{QCD}
\label{chapter:seven}
\catcode`@=11
\def\chkspace{%
  \relax   
  \begingroup\ifhmode\aftergroup\dochksp@ce\fi\endgroup}
\def\dochksp@ce{%
  \unskip              
  \futurelet\chkspct@k\d@chkspc  
}
\def\d@chkspc{%
  \let\nxtsp@ce=\relax
  \ifx\chkspct@k.\else     
    \ifx\chkspct@k,\else
      \ifx\chkspct@k;\else
        \ifx\chkspct@k!\else
          \ifx\chkspct@k?\else
            \ifx\chkspct@k:\else
              \ifx\chkspct@k)\else
              \ifx\chkspct@k(\else
                \ifx\chkspct@k]\else
                  \ifx\chkspct@k-\else
                    \ifx\chkspct@k\egroup\else  
                      \let\nxtsp@ce=\put@space  
                    \fi
                  \fi
                \fi
              \fi
              \fi
            \fi
          \fi
        \fi
      \fi
    \fi
  \fi
  \nxtsp@ce
}
\def\put@space{$\;$}
\catcode`@=12

\def\alpmz{\relax\ifmmode \alpha_s(M_Z)\else $\alpha_s(M_Z)$\fi\chkspace}
\def\z0{\relax\ifmmode Z^0 \else {$Z^0$} \fi\chkspace}
\def\ep{{e$^+$e$^-$}\chkspace}
\def\alp{\relax\ifmmode \alpha_s\else $\alpha_s$\fi\chkspace}

\def\ggx{{$\gamma\gamma$}\chkspace}
\def\sigg{{$\sigma_{\gamma\gamma}^{\rm tot}$}\chkspace}
\def\fg{{$F^{\gamma}_2$}\chkspace}
\def\xg{{$x_{\gamma}$}\chkspace}
\def\syy{{$\sqrt{s}_{\gamma\gamma}$}\chkspace}
\def\gstar{{$\gamma^*\gamma^*$}\chkspace}
\def\be{\begin{equation}}
\def\ee{\end{equation}}
                         \def\bearr{\begin{eqnarray}}
                         \def\eearr{\end{eqnarray}}
\def\benum{\begin{enumerate}}
\def\eenum{\end{enumerate}}
\def\bitem{\begin{itemize}}
\def\eitem{\end{itemize}}
\def\sqeeb{\ifmmode{\sqrt{s_{\protect\bf\mathrm{ee}}}}\else
  {$\sqrt{s_{\protect\bf\mathrm{ee}}}$}\fi}
\def\epem{\ifmmode{\mathrm{e}^{+}\mathrm{e}^{-}}\else
  {$\mathrm{e}^{+}\mathrm{e}^{-}$}\fi}
\def\sqee{\ifmmode{\sqrt{s_\mathrm{ee}}}\else
  {$\sqrt{s_\mathrm{ee}}$}\fi}
\def\kp{{\ifmmode{k_{\perp}}\else{$k_{\perp}$}\fi}}
\def\etmean{\ifmmode{\bar{E}^{\mathrm{jet}}_{\mathrm{T}}}\else
  {$\bar{E}^{\mathrm{jet}}_{\mathrm{T}}$}\fi}
\def\etajmean{\ifmmode{|\bar{\eta}^\mathrm{jet}|}\else 
  {$|\bar{\eta}^\mathrm{jet}|$}\fi}
\def\lxg{\ifmmode{\mathrm{log_{10}}(x_{\gamma})}\else
  {${\mathrm{log_{10}}(x_{\gamma})}$}\fi}
\def\xg{\ifmmode{x_{\gamma}}\else{${x_{\gamma}}$}\fi}
\def\xgp{\ifmmode{x_{\gamma}^+}\else{${x_{\gamma}^+}$}\fi}
\def\xgm{\ifmmode{x_{\gamma}^-}\else{${x_{\gamma}^-}$}\fi}
\def\xgpm{\ifmmode{x_{\gamma}^{\pm}}\else 
  {${x_{\gamma}^{\pm}}$}\fi}
\def\ipb{\ifmmode {\mathrm{pb}^{-1}}\else 
  {$\mathrm{pb}^{-1}$}\fi}
\def\as{\alpha_{\rm s}}
\def\ee{\ifmmode{\mbox{e}^+\mbox{e}^-}\else
  {$\mbox{e}^+\mbox{e}^-$}\fi}
\def\thetamaxp{\ifmmode{\theta_\mathrm{max}'}\else
  {$\theta_\mathrm{max}'$}\fi}
\def\pt{\ifmmode{p_\mathrm{T}}\else{$p_\mathrm{T}$}\fi}
\def\Zzero{\ifmmode{\mathrm{Z}^{0}}\else{$\mathrm{Z}^{0}$}\fi}
\def\ETi{E_{{\rm T}_i}}
\def\etjet{\ifmmode{E^\mathrm{jet}_\mathrm{T}}\else 
  {$E^\mathrm{jet}_\mathrm{T}$}\fi}
\def\ptmiss{\ifmmode{{P}_{\mathrm{T,MISS}}}\else 
  {${P}_{\mathrm{T,MISS}}$}\fi}
\def\mj1h2{\ifmmode{M_{\mathrm{J1H2}}}\else 
  {$M_{\mathrm{J1H2}}$}\fi}
\def\ebeam{\ifmmode{E_{\mathrm{BEAM}}}\else 
  {$E_{\mathrm{BEAM}}$}\fi}
\def\etajet{\ifmmode{|\eta^\mathrm{jet}|}\else 
  {$|\eta^\mathrm{jet}|$}\fi}
\def\etaj{\ifmmode{\eta^\mathrm{jet}}\else 
  {$\eta^\mathrm{jet}$}\fi}
\def\etajdef{\ifmmode{\eta^\mathrm{jet} = 
    -\ln\tan(\theta^\mathrm{jet}/2)}\else{$\eta^\mathrm{jet} = 
    -\ln\tan(\theta^\mathrm{jet}/2)$}\fi}
\def\detajet{\ifmmode{|\Delta\eta^\mathrm{jet}|}\else 
  {$|\Delta\eta^\mathrm{jet}|$}\fi}
\def\costhst{\ifmmode{|\mathrm{cos}\,\Theta^{*}|}\else 
  {$|\mathrm{cos}\,\Theta^{*}|$}\fi}
\def\etajetc{\ifmmode{|\eta^\mathrm{jet}_\mathrm{cntr}|}\else 
  {$|\eta^\mathrm{jet}_\mathrm{cntr}|$}\fi}
\def\etajetf{\ifmmode{|\eta^\mathrm{jet}_\mathrm{fwd}|}\else 
  {$|\eta^\mathrm{jet}_\mathrm{fwd}|$}\fi}
\def\etah{\ifmmode {\hat{\eta}}\else{$\hat{\eta}$}\fi}
\def\dsdetm{\frac{\mathrm{d}{\sigma}_{\mathrm{dijet}}}
  {\mathrm{d}\bar{E}^\mathrm{jet}_\mathrm{T}}}
\def\dsdetml{{\mathrm{d}{\sigma}_{\mathrm{dijet}}}/
  {\mathrm{d}\bar{E}^\mathrm{jet}_\mathrm{T}}}
\def\dsdxg{\frac{\mathrm{d}{\sigma}_{\mathrm{dijet}}}
  {\mathrm{d}x_{\gamma}}}
\def\dsdlxg{\frac{\mathrm{d}{\sigma}_{\mathrm{dijet}}}
  {\mathrm{d log_{10}}\left(x_{\gamma}\right)}}
\def\dsdeta{\ifmmode{\frac{\mathrm{d}{\sigma}_{\mathrm{dijet}}}
  {\mathrm{d}\etajet}}\else
    {$\frac{\mathrm{d}{\sigma}_{\mathrm{dijet}}}
      {\mathrm{d}\etajet}$}\fi}
\def\dsdetac{\frac{\mathrm{d}{\sigma}_{\mathrm{dijet}}}
  {\mathrm{d}\etajetc}}
\def\dsdetaf{\frac{\mathrm{d}{\sigma}_{\mathrm{dijet}}}
  {\mathrm{d}\etajetf}}
\def\dsddeta{\frac{\mathrm{d}{\sigma}_{\mathrm{dijet}}}
  {\mathrm{d}\detajet}}
\def\dscost{\frac{\mathrm{d}{\sigma}_{\mathrm{dijet}}}
  {\mathrm{d}\costhst}}
\def\et{\ifmmode{E_\mathrm{T}}\else{$E_\mathrm{T}$}\fi}
\def\etjone{E^\mathrm{jet}_\mathrm{T,1}}
\def\etjtwo{E^\mathrm{jet}_\mathrm{T,2}}
\def\gg{\ifmmode{\gamma\gamma}\else{$\gamma\gamma$}\fi}
\def\gsg{\ifmmode{\gamma^{\star}\gamma}\else
  {$\gamma^{\star}\gamma$}\fi}
\def\PTMIA{\ifmmode{p_\mathrm{t}^\mathrm{mi}}\else
  {$p_\mathrm{t}^\mathrm{mi}$}\fi}
\def\sas1d{SaS\,1D}
\def\grs{GRS}
\def\grv{GRV}
\def\grvnlo{GRV\,HO}
\def\afgnlo{AFG\,HO}
\def\gs96nlo{GS96\,HO}
\def\lac1{LAC\,1}
\def\hadcor{\ifmmode{(1+\delta_{hadr})}\else{$(1+\delta_{hadr})$}\fi}
%


\section{Introduction}

Strong-interaction measurements at a multi-TeV $e^+e^-$ collider such as
CLIC will constitute an important part of its physics programme. CLIC will
provide unique tests of QCD at high energy scales in a relatively clean
experimental and theoretical environment.  Examples of general QCD
measurements that can be made include the fragmentation and hadronization
of partons at high energies and a precise extraction of $\alpha_s$ 
through multijet rates and event-shape variables, for example. For a
TeV-class linear 
collider, a precision in $\alpha_s$ of the order of 1\% may be achievable
if the relevant perturbative QCD calculations become available with 
sufficiently high
precision~\cite{tesla-tdr}. Also, virtual $\gamma\gamma$ interactions can
be studied at an $e^+e^-$ collider, which will help the understanding of
low-$x$ data at hadron colliders.  A dedicated $\gamma\gamma$ collider
interaction region would allow further detailed measurements of 
the enigmatic
photon structure.

In this chapter, the following items, essentially in the category of
two-photon physics, have been worked out in more detail:

\vspace*{3mm}

\begin{itemize}
\item Measurement of the total $\gamma\gamma$ cross section
  
\item Measurement of the photon structure

\item Measurement of BFKL dynamics
\end{itemize}

\vspace*{3mm}

At lower-energy $e^+ e^-$ colliders, two-photon interactions have traditionally
been studied using bremsstrahlung photons from the electron
beams~\cite{wws}. Similar processes, together with reactions induced by
beamstrahlung photons, will also occur at a future high-energy $e^+ e^-$
collider such as CLIC, and will allow a rich two-photon physics programme.
The energy spectra of the photons emitted in these processes are, however,
peaked at rather low energies.

Since the colliding beams are used only once at linear colliders, other
operation modes become possible, such as a photon ($\gamma\gamma$)
collider (PC)~\cite{ginz,telnov}.  At a PC, the electron and positron
beams of a linear collider are converted into photon beams via Compton
laser backscattering.  This option has been discussed for CLIC in
chapter~2. The advantage of the photon spectrum from laser
backscattering is that it peaks at high energy values, typically 
$\sim $~80\% of the initial electron energy, with a relatively narrow width of
approximately 10\%. Thus, a PC offers the exciting possibility of studying
two-photon interactions at the highest possible energies with high
luminosity.

Finally, we note in passing that, since the overlaid $\gamma\gamma$
collisions form the most important background at CLIC, a good knowledge
and understanding of two-photon processes will be essential for
controlling these background contributions to other processes.

\section{Total Cross Section}

A key example for the study of the properties of the photon is the total
\ggx cross section, which is not yet understood from first principles. The
nature of the photon itself is known to be rather complex. A
high-energy photon can fluctuate into a fermion pair or even into a
hadronic bound state, i.e. a vector meson with the same quantum numbers
as the photon, $J^{PC}$~=~1$^{--}$. These quantum fluctuations lead to the
so-called hadronic structure of the photon.

\begin{figure}
\begin{center}
\epsfig{file=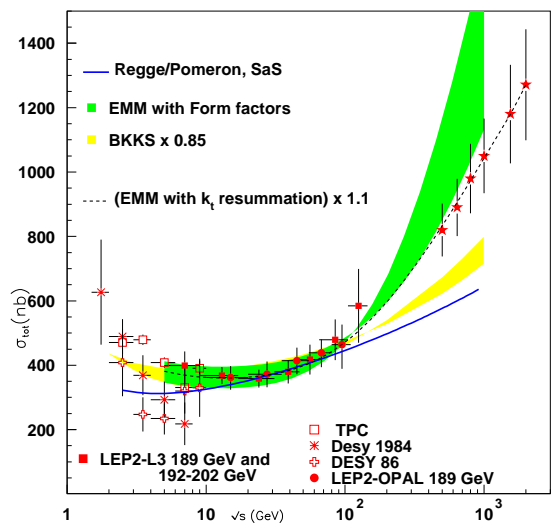,bbllx=20,bblly=150,bburx=560,bbury=670,width=7.8cm}

\vspace*{-2mm}

\caption{
The total \ggx cross section as function of the collision energy, 
compared with model calculations: a band of predictions in the
BKKS model, whose upper and lower limits correspond to different   
photon densities \protect \cite{BKKS};
a proton-like model (solid line \protect~\cite{Donnachie,SAS,aspen});
an eikonal minijet model (EMM) band for the total and 
inelastic cross sections, with different photon densities and 
different minimum jet transverse momenta \protect
\cite{pancheri}). The proton-like and BKKS
models have been normalized to the data, in order to show the 
energy dependence of the cross section.}
\label{sec7:sigtot}
\end{center}
\end{figure}

The data from LEP~\cite{TPC,DESY,L3tot,OPAL} show that the total \ggx
cross section rises with energy. Such a rise was observed previously in
$pp$ collisions, but, intriguingly, the rise in $\gamma\gamma$ collisions
seems to be steeper. Figure~\ref{sec7:sigtot} shows the
data~\cite{TPC,DESY,L3tot,OPAL} together with predictions of various
models.

As seen in Fig.~7.1, models differ in their predictions for
the total $\gamma \gamma$ cross section at CLIC energies. The wide band is
obtained using the eikonal minijet model (EMM), the thinner one using a
model due to BKKS~\cite{BKKS}, and the full curve is an early prediction
by Schuler and Sj\"ostrand~\cite{SAS} using a Regge--Pomeron fit. The EMM,
which embeds QCD minijet cross sections in an eikonal
formalism~\cite{pl,GAP}, can describe the rise observed in present LEP
data quite well, apart from a 10\% normalization uncertainty. The
predictions for $\gamma \gamma$ interactions are obtained with a parameter
set extrapolated from similar fits to the photoproduction cross section at
HERA.  The tuned parameters include the different parton densities,
minimum minijet transverse momenta, the soft cross section and a model for
the transverse momentum distribution of partons in the photon.  It should
be noted, however, that the EMM, with a form-factor model for the parton
transverse momentum distribution and the same set of parameters, does not
reproduce at all the early rise seen in proton--proton scattering, unless
higher-order effects due to soft gluon emission are included~\cite{bn}.
Although at present energies the EMM without soft gluons can describe
the $\gamma \gamma$ data well, within the experimental errors, reliable
predictions at higher energies, like those at CLIC, will require the
proper inclusion of such higher-order contributions. The figure shows the
effect of applying one particular realization of this idea~\cite{bn} to
the $\gamma \gamma$ case.  Predictions for CLIC energies, based on soft
gluon resummation and after fixing the normalization, are shown for a
particular chosen set~of~parameters.

Clearly, new measurements at a higher-energy linear collider will be
indispensable for establishing whether the photon cross section 
indeed rises faster than, for instance, the $pp$ data. A detailed
comparison of  
the predictions of the different models for $\gamma\gamma$ cross sections
reveals that, in order to distinguish between them, the cross sections
should be determined to a precision of better than about
10\%~\cite{pancheri} at a future 0.5--1~TeV  $e^+ e^-$
collider~\cite{tesla-tdr}, though the differences between models get
larger at larger energy. This precision will however be difficult to
achieve at an $e^+ e^-$ collider, since the variable \syy needs to be
reconstructed from the visible hadronic final state in the detector, which
is a problem already at LEP. For example, the hadronic final state in 
the pseudorapidity $\eta = \ln\tan\theta$/2 extends
over the range --~9~$< \eta <$~9
at 3~TeV, while the detector covers roughly the region --~3~$< \eta <$~3.

As already mentioned, at a $\gamma \gamma$ collider the photon beam 
energy can be tuned with a
spread of less than 10\%, so that measurements of \sigg can be made at a
number of `fixed' energies, e.g. in the range 
0.5~$< \sqrt{s_{\gamma \gamma}} <$~2.5~TeV,
as shown in~Fig.~\ref{sec7:sigtot}. 
A detailed study reveals that the
absolute precision with which these cross sections can be measured in the
TeV region ranges from 8\% to 15\%, where the largest contributions to the
errors are due to the modelling of the diffractive component of the cross
section, the choice of the Monte Carlo model used to correct for the event
selection cuts, the knowledge of the absolute luminosity and the shape of
the luminosity spectrum. 
It will be necessary to
constrain the diffractive component using high-energy two-photon data. A
technique to measure diffractive contributions separately, modelled on the
rapidity gap methods used at HERA, has been proposed
in~Ref.~\cite{engel}. The 
potential measurements that can be made at CLIC are shown 
in~Fig.~\ref{sec7:sigtot} with 15\% total errors.

Finally, we can use the above model predictions to update the calculation
of the number of hadronic events per bunch crossing, which is expected to
be significant at a high-energy $e^+e^-$ collider such as CLIC. However, in
this case, it is necessary to add the effect of the beamstrahlung photons
as well.  We do this for CLIC using the spectra of beamstrahlung photons
provided in~Ref.~\cite{schulte_bs}. For the design parameters considered, the
two-photon luminosities per bunch crossing, corresponding to one or both
photons being due to bremsstrahlung, are 
${\cal L}_{eg}^{\gamma \gamma}$~=~5.3589~$\times$~10$^{34}$~m$^{-2}$ or 
${\cal L}_{ee}^{\gamma \gamma}$~=~6.4852~$\times$~10$^{34}$~m$^{-2}$, 
respectively, whereas that due to beamstrahlung
photons alone is 
${\cal L}_{gg}^{\gamma \gamma }$~=~4.9534~$\times$~10$^{34}$~m$^{-2}$.  
The numbers of hadronic events expected per
bunch crossing, using these effective two-photon luminosities per bunch
crossing, are shown in Table~\ref{sec7:tablebs}, 
for three different values of the lower limit on
$s_{\gamma \gamma}$, instead of the fixed value of 50 GeV$^2$
considered~earlier.
\begin{table}[!h] 
\caption{Numbers of $\gamma \gamma$ events per bunch 
crossing expected at CLIC}
\label{sec7:tablebs}

\renewcommand{\arraystretch}{1.3} 
\begin{center}

\begin{tabular}{ccccc}
\hline \hline \\[-3mm]
$\hspace*{3mm}$ \boldmath\textbf{$s_{\rm min}$~GeV$^2$} $\hspace*{3mm}$ &
$\hspace*{5mm}$ \textbf{Aspen~\cite{aspen}} $\hspace*{5mm}$ &
$\hspace*{5mm}$ \textbf{EMM(BN)} $\hspace*{5mm}$ &
$\hspace*{5mm}$ \textbf{BKKS} $\hspace*{5mm}$ &
$\hspace*{3mm}$ \textbf{EMM} $\hspace*{3mm}$ 
\\[4mm]   

\hline \\[-3mm]
5   & 4.0  & 5.5  & 5.7  & 6.3    \\  
25  & 3.4 & 4.7  & 5.0  & 5.5     \\
50  & 3.2 & 4.5  & 4.7  & 5.3 
\\[3mm] 
\hline \hline
\end{tabular}
\end{center}
\end{table}


The number obtained here using the Aspen model~\cite{aspen} is consistent 
with that
obtained in~Ref.~\cite{schulte_had} with the SAS parametrization of
$\sigma_{\rm tot}(\gamma \gamma \to hadrons)$. Thus, we see
that, depending on which theoretical model gives the right high-energy
description, we expect 4 to 7 hadronic events per bunch crossing at CLIC.  
The beamstrahlung photons completely dominate the $\gamma \gamma $
luminosity. Inclusion of the beamstrahlung contribution increases the
expected number of events by a factor of about 10 with respect to that
expected for just the bremsstrahlung photons. About half of these
extra events come from 
the contribution to the $\gamma \gamma$ luminosity from
interactions between bremsstrahlung and beamstrahlung photons.

\section{Photon Structure}

Two-photon interactions have been traditionally used to measure the
structure of the photon. In contrast to the proton, the structure function
of the photon is predicted to rise linearly with the logarithm of the
momentum transfer $Q^2$, and to increase with increasing Bjorken
$x$~\cite{gg_zerwas}. The absolute magnitude of the photon structure
function is asymptotically determined by the strong coupling
constant~\cite{gg_witten}.

The classical way to study the structure of the photon is via 
deep-inelastic electron--photon scattering (DIS), i.e. two-photon interactions
with one quasi-real (virtuality $Q^2 \sim 0$) and one virtual ($Q^2 >$ few
GeV$^2$) photon. The unpolarized e$\gamma$ DIS cross-section is
\begin{equation}
\label{avs1}
  \frac{d \sigma (e \gamma \to eX)}{dQ^2 dx}
  \: = \:\frac{2\pi \alpha^2}{Q^4 x} \,
  \Big[ \big\{ 1 + (1-y)^2 \big\} F_2^{\, \gamma}(x,Q^2) 
  - y^2 F_L^{\, \gamma}(x,Q^2) \Big] \: , 
\end{equation}
where $F_{2,L}^{\, \gamma}(x,Q^2)$ denotes the structure functions of the
real photon. The structure function is given by the quark 
content, i.e.
\begin{equation}
F_2^{\gamma} = \Sigma_q e^2_q( xq^{\gamma}(x,Q^2)+ 
x\overline{q}^{\gamma}(x,Q^2))\,,
\end{equation}
to leading order.

\begin{figure}[!h] 
\begin{center}
\hspace*{-17mm}\epsfig{figure=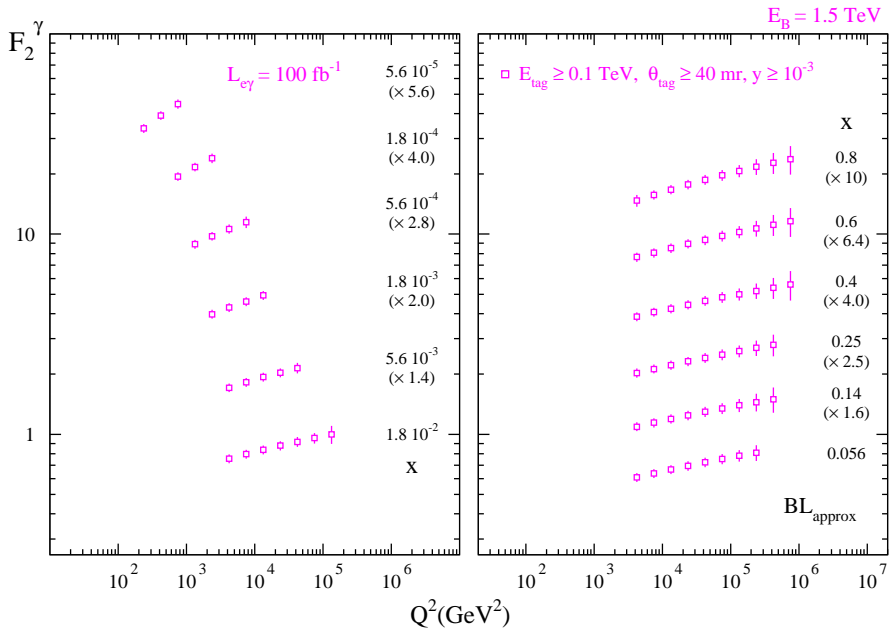,bbllx=80pt,bblly=30pt,bburx=530pt,bbury=720pt,angle=-90,width=12cm}

\vspace*{5mm}

\caption{The kinematic coverage of the measurement of \fg for the 
backscattered $e \gamma$ mode at a 3-TeV linear collider such as 
CLIC~\protect \cite{vogt}}
\label{sec7:f2}
\end{center}
\end{figure}

To measure \fg it is important to detect (tag) the scattered electron
that has emitted the virtual photon. Background studies at CLIC suggest
that these electrons can be detected down to 40 mrad and 100 GeV.  An
important limiting factor for present measurements of \fg at LEP is the
understanding and modelling of the hadronic final state, needed to
reconstruct the kinematics of the events in the $e^+ e^-$ collider mode;
this limitation will become even more severe at higher energies, because of
the increased rapidity span of the hadronic final state. For $e \gamma$
scattering at an $e \gamma$ collider, however, the energy of the probed
quasi-real photon is known (within the beam spread of 10\%) and the
systematic error can be controlled to about 5\%. Figure~\ref{sec7:f2} shows 
the measurement potential for an $e \gamma$ collider~\cite{vogt}. The
measurements are shown with statistical and (5\%) systematical error, for
100~fb$^{-1}$ of $e \gamma$ collider luminosity, i.e. about a year of data
taking. Measurements can be made in the region 
5.6~$\times$~10$^{-5}$~$< x < $~0.56, 
a region similar to the HERA proton structure function
measurements, and in the region 
100~$< Q^2 < $~8~$\times$~10$^{5}$~GeV$^2$.

\vspace*{4mm}

The $Q^2$ evolution of the structure function at large $x$ and $Q^2$ has
often been advocated as a clean measurement of $\alpha_s$. It has 
been shown~\cite{vogt}, however, that a 5\% change on 
$\alpha_s$ results in a 3\%  
change in \fg only, hence such a $\alpha_s$ determination will require
very precise \fg measurements.

At very high values of $Q^2$ $\sim $~10000~GeV$^2$, also $Z$ and $W$ exchange
will become important, the latter leading to charged-current
events~\cite{ridder} with large missing transverse momentum due to the
escaping neutrino. By measuring the electroweak neutral and 
charged-current structure functions, the up- and down-type quark content of the
photon can be determined separately.

A linear collider can also provide circularly-polarized photon beams,
either from the polarized beams of the $e^+ e^-$ collider directly, or via
polarized laser beams scattered on the polarized $e^+ e^-$ drive beam. This
offers a unique opportunity to study the polarized parton distributions of
the photon, for which no experimental data are available to date.

While $e \gamma$ scattering allows one to measure the quark distributions
inside the photon, it constrains only weakly the gluon distribution via
the QCD evolution of the structure functions. Direct information on the
gluon in the photon can however be obtained from measurements of
jet~\cite{wengler}, open charm~\cite{jankowski}, and
$J/\psi$~\cite{indumathi} production in $\gamma\gamma$ interactions at an
$e^+ e^-$ and $\gamma\gamma$ collider.

Dijet production in $\gamma\gamma$ interactions has been studied at CLIC
energies. The two jets can be used to estimate the fraction of the photon
momentum participating in the hard interaction, which is a sensitive probe
of the structure of the photon. The transverse energy of the jets provides
a hard scale, which allows such processes to be calculated in perturbative
QCD. Fixed-order calculations at the next-to-leading order (NLO) in the strong
coupling constant $\as$ for dijet production are available and can be
compared with the data, providing tests of the theory. Measurements of this
kind are available from TRISTAN ~\cite{bib-amy,bib-topaz} and
LEP~\cite{bib-opalgg,bib-djopold, bib-newopal}.  The {\kp}-clustering
algorithm~{\cite{bib-ktclus}} is used to define jets because of the
advantages of this algorithm in the comparison to theoretical
calculations~{\cite{bib-ktisbest}}. The study is done using
PHOJET~\cite{bib-phojet} and PYTHIA~\cite{bib-pythia} for an $e^+ e^-$ 
centre-of-mass energy of 3~TeV with SIMDET modified for CLIC to simulate the
detector response. HADES is used to provide additional {\gg} interactions.

For studies of the photon structure, a pair of variables,
{\xgp} and {\xgm}, is defined~\cite{bib-LEP2}. They estimate
the fraction of the photon's momentum participating in the hard
scattering process:
\begin{equation}
\xgp \equiv \frac{\displaystyle{\sum_{\rm jets=1,2}
 (E^\mathrm{jet}+p_z^\mathrm{jet})}}
 {{\displaystyle\sum_{\rm hfs}(E+p_z)}} \qquad \mbox{and} \qquad
\xgm \equiv \frac{\displaystyle{\sum_{\rm jets=1,2}
 (E^\mathrm{jet}-p_z^\mathrm{jet})}}
{\displaystyle{\sum_{\rm hfs}(E-p_z)}}\,,
\label{eq-xgpm}
\end{equation}
where $p_z$ is the momentum component along the $z$ axis of the
detector and $E$ is the energy of the jets or other hard objects in the 
hadronic
final state (hfs). In leading order, for direct events, all the energy of 
the event
is contained in two jets, i.e. ${\xgp}$~=~1 and ${\xgm}$~=~1, whereas for
single-resolved or double-resolved events one or both values are
smaller than~1. The dijet differential cross section as a function of
{\xg} is therefore particularly well suited to study the structure of
the photon, since it separates predominantly-direct events at high
{\xg} ($\xg>$~0.75) from predominantly-resolved events at low {\xg}
($\xg<$~0.75). However, as is evident from the definition of
{\xg}, it is crucial to identify the two jets from the same
{\gg} interaction if one is to gain access 
to the structure of the~photon.

\begin{figure}[t] 
\begin{center}
\epsfig{figure=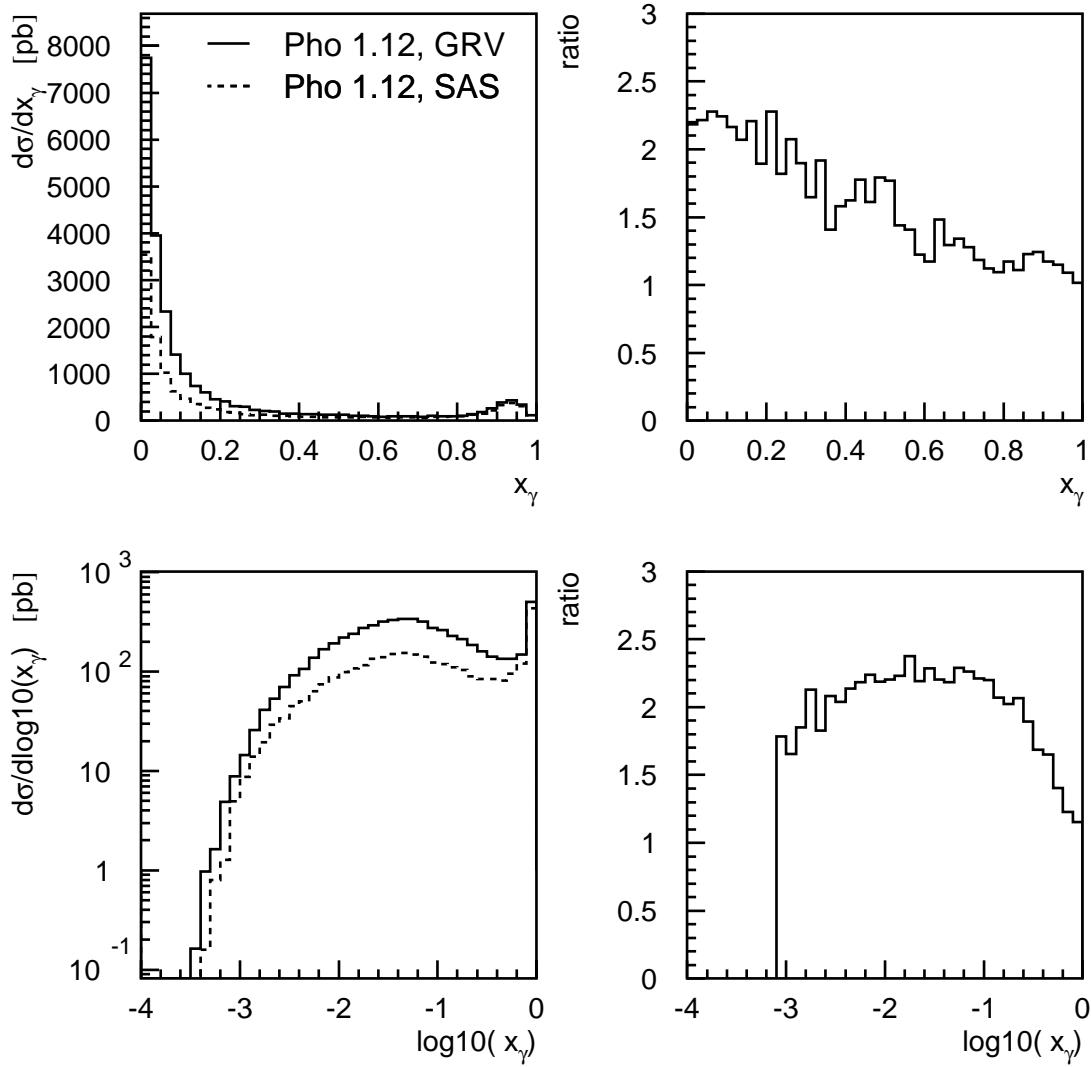,bbllx=0,bblly=0,bburx=530,bbury=520,clip,width=15cm}
\end{center}

\vspace*{-5mm}

\caption{The {\xg} and {\lxg} cross sections for a single {\gg}
collision, for different photon structure functions on a linear
(top) and a logarithmic scale (bottom).
The figures to the right show the ratio of the distributions of the 
two structure functions.}
\label{sec7:dijets}
\end{figure}

Figure~\ref{sec7:dijets} shows a generator study of the distribution of
$\xg$, for events in $e^+e^-$ collisions selected by having two jets with
transverse energies above thresholds $E_T^{\rm jet1} >$~10~GeV and 
$E_T^{jet2}>$~8~GeV both in the region $|\eta| <$~2.5, predicted 
for two different 
photon parametrizations. The measurements can go down to 
$\xg \sim $~5~$\times$~10$^{-4}$. 
The difference between the two parton distributions gives a
factor of 2 difference in the dijet cross sections in this kinematic
region.

This measurement will have to be made at reduced luminosities. For the
nominal high-luminosity scenario, it is expected that each bunch crossing
will yield on average four {\gg} interactions. In addition, several bunch
crossings may be overlaid; these, being genuine {\gg} interactions,
are they part of the~signal. 

\begin{figure}[ht]
\begin{center}
\includegraphics[width=0.95\textwidth]{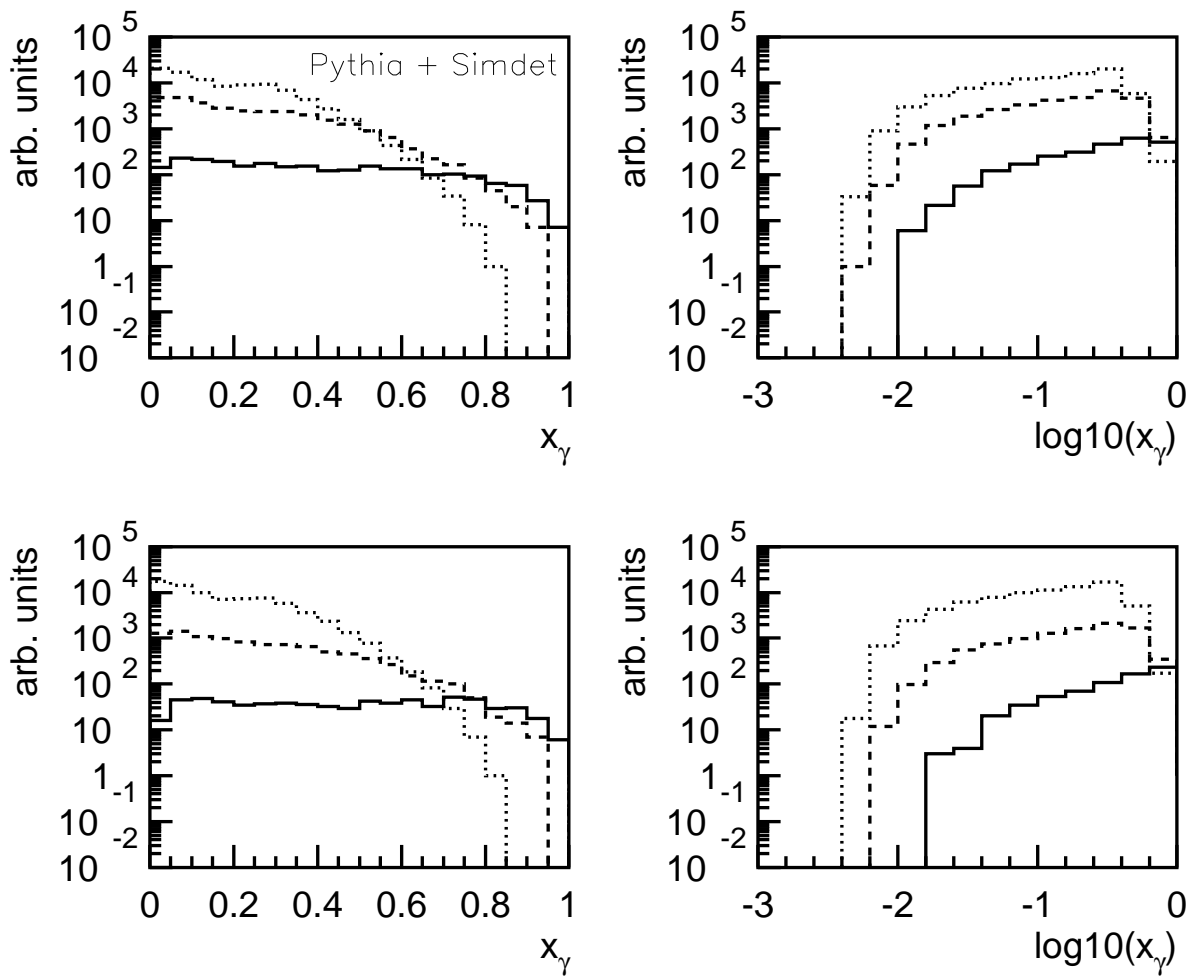}
\end{center}

\vspace*{-3mm}

\caption{The {\xg} and {\lxg} distributions for a single {\gg}
collision compared with the case where one or two bunch crossings of
{\gg} collisions are overlaid. The upper two plots require the
first and second jets in the event to be above 7~GeV and 5~GeV
transverse momentum. In the lower two plots the thresholds have
been raised to 10~GeV and 8~GeV, respectively.}
\label{fig:xg}
\end{figure}

However, overlaid interactions would spoil the measurement, both by
smearing the energy of the jets and by adding additional jets to the
event, making it increasingly unlikely that the leading two jets
originate from the same hard interaction. In addition, the extra energy
added outside the two leading jets leads to a distortion of the $\xg$
distribution towards lower values, as can be seen in~Fig.~\ref{fig:xg}.

\begin{figure}[ht] 
\begin{center}
\includegraphics[width=0.75\textwidth]{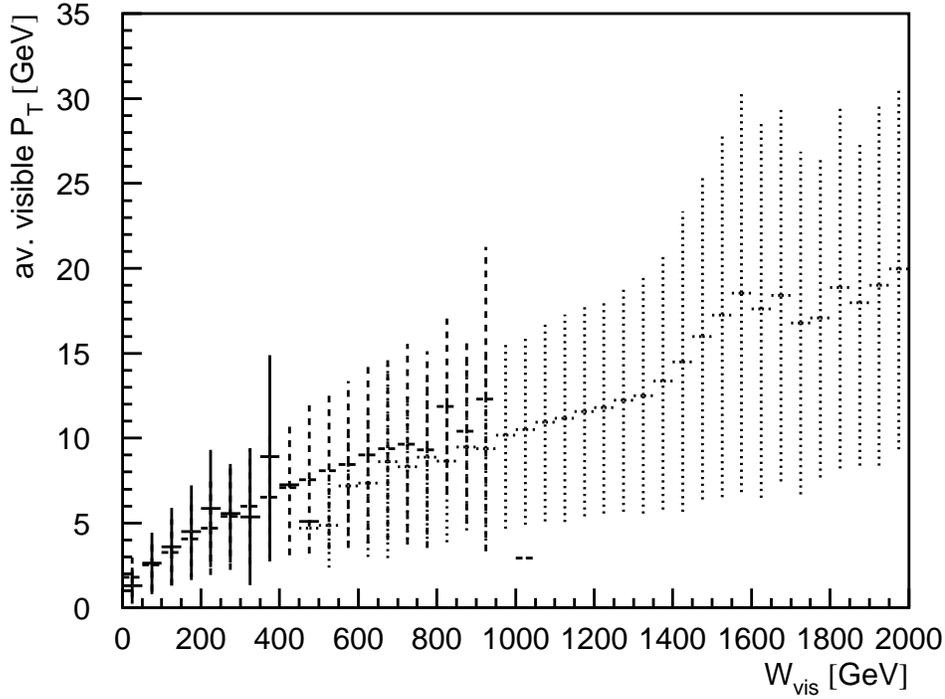}
\caption{Visible transverse momentum of the hadronic final state
contributed on average by background {\gg} collisions in
one (solid), two (dashed) and eight (dotted) bunch crossings,
as a function of the invariant mass of the hadronic
final state. The vertical error bars indicate the spread.}
\label{fig:vispt}
\end{center}
\end{figure}

In addition to the type of physics studies mentioned above, the presence
of many {\gg} collisions in each event has potentially severe implications
for many other analyses. To investigate these effects, a dedicated
study was carried out using PYTHIA for $Z$-pair production, where both $Z$
bosons decay exclusively to neutrinos. This process does not deposit any
signal in the detector and enables us to study the effect of the hadronic
background as simulated by HADES. This study demonstrates that even if the
invariant mass of the hadronic system is required to be above 200~GeV,
between 5 and 10~GeV per unit rapidity are added to the event, per bunch
crossing. Furthermore, the visible transverse momentum of the hadronic
final state rises as a function of the number of {\gg} collisions present.
Figure~\ref{fig:vispt} shows that, already for two bunch crossings,
between 5 and 10~GeV of visible transverse momentum will be 
added to each~event.

\section{Tests of BFKL Dynamics}

One of the important open questions in high-energy QCD is the existence of
BFKL effects. The BFKL equation~\cite{bfkl,Mueller} resums multiple gluon
radiation of the gluon exchanged in the $t$ channel, corresponding to a
resummation of $\alpha_s\ln 1/x$ terms. It predicts a power increase of
the cross section. HERA, the Tevatron and even LEP have been searching for
BFKL effects in the data. Currently the situation is that in some corners
of the phase space the NLO calculations undershoot the measured QCD
activity, e.g. for forward jets and neutral pions at HERA, but the BFKL
dynamics has not yet been established.

The BFKL dynamics can be tested at future high-energy $e^+e^-$ linear
colliders. In this section, the total $\gamma^*\gamma^*$ cross section
derived in the leading-logarithmic QCD dipole picture of BFKL dynamics
is calculated, with higher-order corrections. 
%
%
The advantage of $\gamma^*\gamma^*$
scattering is that it is a process without non-perturbative couplings.

Defining 
$y_1$ ($y_2$) and $Q_1^2$ ($Q_2^2$) to be the rapidities and
the squared transferred energies for the two virtual photons, 
one obtains~\cite{brl,gamma}
\begin{eqnarray}
\label{eeBFKLa}
&&\hspace{-1.5cm}d\sigma_{e^+e^-}
 (Q_1^2,Q_2^2;y_1,y_2) = \frac{4}{9} 
\left(\frac{\alpha_{\rm em}^2}{16}\right)^2  \, 
 \alpha_s^2 \, \pi^2 \sqrt{\pi} \,
\frac{d Q_1^2}{Q_1^2}  \,\frac{d Q_2^2}{Q_2^2} \, \frac{d y_1}{y_1}  \,
\frac{d y_2}{y_2} \,
\frac{1}{Q_1 \, Q_2}
\, \nonumber \\
&&\frac{
\exp \left ( \frac{4\alpha_s N_c}{\pi}\, Y \ln 2 \right )
}
{\sqrt{ \frac{14 \alpha_s N_c}{\pi} Y \zeta(3)}}  
\times \, 
\exp \left (-\frac{\ln^2 \frac{Q_1^2}{Q_2^2}}{\frac{56 
\alpha_s N_c}{\pi} Y \zeta(3)} \right )\,
%
 \left[2 l_1  +
9 t_1 \right] \, \left[2 l_2  +
9 t_2 \right] \,,
\end{eqnarray}
for the leading-order BFKL cross section, where
$t_1 \equiv \frac{1}{2} (1 + (1-y_1)^2), \quad l_1 \equiv 1-y_1$,
$t_2$ and $l_2$ are defined analogously, and  
$Y \equiv \ln \left (s y_1 y_2 /\sqrt{Q^2_1 Q^2_2} \right )$.

The two-gluon-exchange cross section has been calculated exactly in 
the high-energy approximation, and reads
\begin{eqnarray}
\label{2g}
\hspace{-1.2cm}d\sigma_{e^+e^-}
 (Q_1^2,Q_2^2;y_1,y_2) &=& \frac{d Q_1^2}{Q_1^2}  \,\frac{d Q_2^2}{Q_2^2} 
 \,\frac{d y_1}{y_1} \, \frac{d y_2}{y_2} 
 \, \frac{64 \,(\alpha_{\rm em}^2 \alpha_s)^2}{243 \pi^3} \,\frac{1}{Q_1^2} 
\nonumber \\
&~& \left[ t_1 t_2 \ln^3 \frac{Q_1^2}{Q_2^2} + \left( 7 t_1 t_2 
+ 3 t_1 l_2 + 3 t_2 l_1 \right) \ln^2 \frac{Q_1^2}{Q_2^2} \right. \nonumber
\\ &~&  \left.
+ \left( \left ( \frac{119}{2} - 2 \pi^2 \right ) t_1 t_2
+ 5 (t_1 l_2 +t_2 l_1) + 6 l_1 l_2 \right) \ln \frac{Q_1^2}{Q_2^2} \right.
\nonumber
\\ &~& \left.
+ \left( \frac{1063}{9} - \frac{14}{3} \pi^2 \right) t_1 t_2
+ (46 - 2 \pi^2) (t_1 l_2 + t_2 l_1) - 4 l_1 l_2 \right] \,.
\end{eqnarray} 
It is shown in~Ref.~\cite{gamma} that the two-gluon cross section almost always
dominates over the DGLAP one in the double-leading-logarithmic (DLL)
approximation, and that the difference between the BFKL and two-gluon
cross-sections increases with $Y$.

The comparison between the DGLAP-DLL and two-gluon cross sections in the 
leading-order (LO)
approximation shows that they are similar when $Q_1$ and $Q_2$ are not too
different, precisely in the kinematical domain where the BFKL cross
section is expected to dominate. However, when $Q_1^2/Q_2^2$ is further
away from 1, the LO two-gluon cross section is lower than the DGLAP one,
especially at large~$Y$. This suggests that the two-gluon cross section
could be a good approximation to the DGLAP one if both were restricted to
the region where $Q_1^2/Q_2^2$ is close to 1. In this section we use the
exact two-gluon cross-section to evaluate the effect of the non-BFKL
background, since the two-gluon term appears to constitute the dominant part
of the DGLAP cross section in the region 0.5~$ < Q_1^2/Q_2^2 <$~2.

The results given here are
based on these calculations for a future Linear Collider
(LC) with 500~GeV in the centre of mass, and either 3 or 5~TeV for CLIC.
In general, $\gamma^* \gamma^*$ interactions are selected at $e^+e^-$
colliders by detecting in forward calorimeters scattered electrons that
leave the beam pipe.  For the LC, it has been argued~\cite{brl} that angles
as low as 20 mrad should be reached. Currently, angles down to 40 mrad are
foreseen to be instrumented for a generic detector at~CLIC.

Details of the calculation are discussed in~Refs.~\cite{gamma,royon}.
Higher-order (HO) effects are added to the LO BFKL calculation in a
phenomenological way. The results of the BFKL and two-gluon cross sections
are given in Table~\ref{sec7:tab1}, assuming that tagging of scattered
leptons can be done down to 40 mrad and to 50~(100)~GeV at LC (CLIC). The
ratio between HO BFKL and two-gluon cross sections is lower at CLIC than at
LC, owing to the difference in phase space. We find that about 97 (28) events
can be expected per year at CLIC operating at 3 (5)~TeV.
%
\begin{table}[!h] 
\caption{Final cross sections (in fb), for event selections described in 
the text}
\label{sec7:tab1}

\renewcommand{\arraystretch}{1.3} 
\begin{center}

\begin{tabular}{ccccc}
\hline \hline \\[-4mm]
$\hspace*{15mm}$ & $\hspace*{5mm}$ &
$\hspace*{5mm}$ \boldmath\textbf{BFKL$_{\rm HO}$} $\hspace*{5mm}$ &
$\hspace*{5mm}$ \textbf{two-gluon} $\hspace*{5mm}$ &
$\hspace*{3mm}$ \textbf{ratio} $\hspace*{3mm}$ 
\\[4mm]   

\hline \\[-3mm]
 CLIC (3~TeV) &  & 9.7E--2       & 3.7E--2  & 2.6     \\
 CLIC (5~TeV) &  & 2.8E--2       & 1.1E--3  & 2.5     \\
 LC (500~GeV) &  & 8.7    & 2.6  & 3.3  
\\[3mm] 
\hline \hline
\end{tabular}
\end{center}
\end{table}

\begin{figure}[t] 
\begin{center}
\psfig{figure=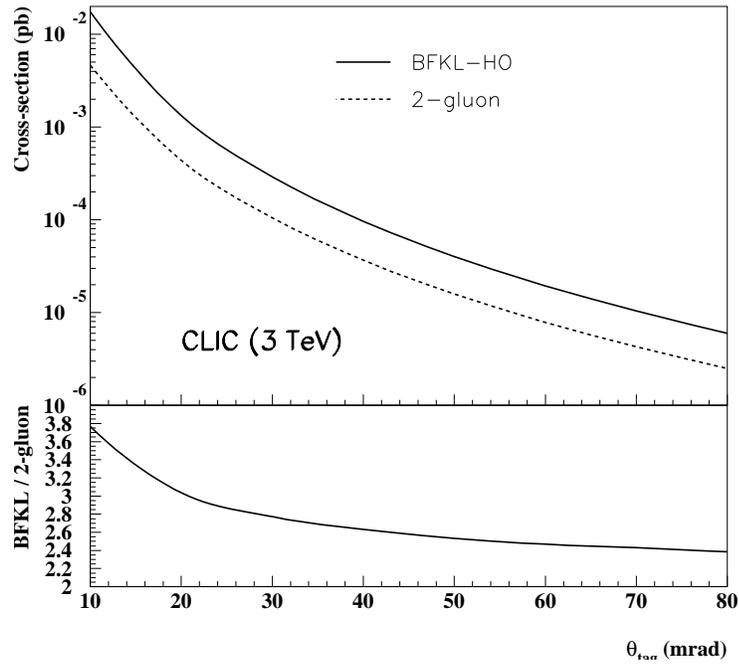,height=3.8in}
\end{center}
\caption{The HO BFKL and two-gluon cross sections and their 
ratio as functions of the tagging angle for CLIC at 3~TeV. Leptons are
tagged beyond 100~GeV.}
\label{sec7:bfkl1}
\end{figure}

The HO BFKL and two-gluon cross sections for a 3~TeV CLIC are plotted in
Fig.~\ref{sec7:bfkl1}, as well as their ratio for different values of the
tagging angle. As was already noted in~Ref.~\cite{gamma}, it is important
to decrease the tagging angle to obtain a larger ratio and large values of
the cross sections. The conclusions are similar for a 5~TeV machine.
%
\begin{figure}[!h] 
\begin{center}
\psfig{figure=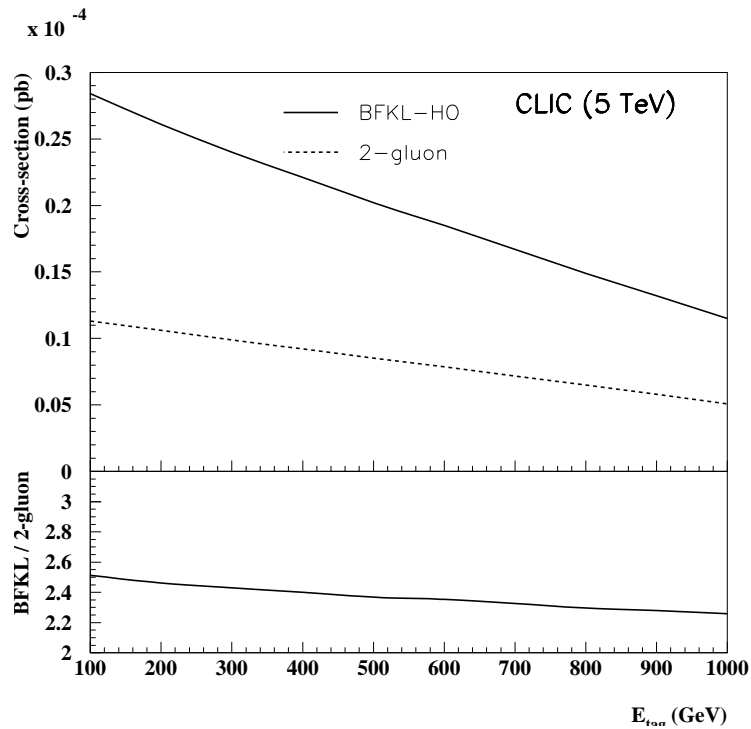,height=3.8in}
\end{center}
\caption{The HO BFKL and two-gluon cross sections and their ratio as
functions of the tagging energy for CLIC at 5~TeV. Leptons are tagged 
beyond 40 mrad.}
\label{sec7:bfkl2}
\end{figure}

The dependence on the tagging energy is studied in~Fig.~\ref{sec7:bfkl2}.
The ratio between the HO BFKL and two-gluon cross sections remains almost
constant, but lowering the tagging energy has an important effect on the
cross-section value, and on the number of events expected per year.  
Another strategy to identify BFKL effects in the data is to study the
energy or $Y$ dependence of the cross sections. To illustrate this, we
show calculations of the HO BFKL and two-gluon cross sections, as well as
their ratio, for given cuts on rapidity $Y$ in Table~\ref{sec7:tab2}. 
We note that one could reach up to a factor 6 difference, while keeping a
cross section measurable at CLIC, if one could tag leptons down to
25~mrad. If this is not possible, a cut on $Y$ at 8 would probably be
sufficient to see BFKL effects.

\begin{table}[htb] 
\caption{Final cross sections (in fb), for event selections described in 
the text, after different cuts on $Y$}
\label{sec7:tab2}

\renewcommand{\arraystretch}{1.3} 
\begin{center}

\begin{tabular}{ccccccc}
\hline \hline\\[-4mm]
$\hspace*{0.6mm}$ \boldmath{$Y$}~\textbf{cut} $\hspace*{0.6mm}$ & 
$\hspace*{0.6mm}$ \boldmath\textbf{BFKL$_{\rm\bf NLO}$} $\hspace*{0.6mm}$ & 
$\hspace*{0.6mm}$ \textbf{Two-gluon} $\hspace*{0.6mm}$ & 
$\hspace*{0.6mm}$ \textbf{Ratio} $\hspace*{0.6mm}$ & 
$\hspace*{0.6mm}$ \boldmath\textbf{BFKL$_{\rm NLO}$} $\hspace*{0.6mm}$ & 
$\hspace*{0.6mm}$ \textbf{Two-gluon} $\hspace*{0.6mm}$ & 
$\hspace*{0.6mm}$ \textbf{Ratio} $\hspace*{0.6mm}$ \\
 & 
\textbf{3~TeV} & 
\textbf{3~TeV} & 
\textbf{3~TeV} & 
\textbf{5~TeV} & 
\textbf{5~TeV} & 
\textbf{5~TeV} 
\\[1.5mm]  \hline\\[-4mm]
 No cut & 9.7E--2   & 3.7E--2  & 2.6 & 2.8E--2  & 1.1E--3  & 2.5  \\
 $Y \ge $ 8. & 7.8E--3  & 1.6E--3  & 4.8 & 2.8E--3  & 6.2E--4  & 4.5 \\
 $Y \ge $ 9. & 1.9E--3  & 3.1E--4  & 6.2 & 9.3E--4  & 1.6E--4  & 5.7 \\
 $Y \ge $ 9. & 3.3E--2  & 4.4E--3  & 7.5 & 1.1E--2  & 1.8E--3  & 6.2 \\
 $\theta \ge $ 25 mrad &  &  & & & &
\\[2mm] 
\hline \hline
\end{tabular}
\end{center}
\end{table}

The $Y$ dependence of the cross section remains a powerful tool to
increase this ratio, and is more sensitive to BFKL effects, even in the
presence of large HO corrections. The uncertainty in the BFKL
cross section after HO corrections is still quite large.  Thus,
measurements performed at a LC or CLIC should be compared with
the precise calculation of the two-gluon cross section after the kinematical
cuts described above, and any difference would be a sign of BFKL effects.
A fit of these cross sections would then be made to determine the BFKL
pomeron intercept after HO corrections as done, for instance, for LEP or
HERA data~\cite{robi}.

Closely related to the \gstar measurement is vector-meson production,
e.g. \ggx $\to J/\psi J/\psi$ or (at large $t$) \ggx $\to
\rho\rho$, where the hard scale in the process is given by the $J/\psi$
mass or the momentum transfer~$t$. The $J/\psi$ particles can be detected 
via
their decays into leptons, and separated from the background through their
peaks in the dilepton invariant mass spectra. Approximately 100 fully 
reconstructed
4-muon events are expected for 200~fb$^{-1}$ of luminosity for a 500~GeV
$e^+ e^-$ collider~\cite{KWADR}.  For this channel it is crucial that the decay
muons and/or electrons can be measured to angles below 10 degrees in the
experiment. Further processes, which are strongly sensitive to BFKL effects,
include $e \gamma$ scattering with associated jet
production~\cite{contreras}, and $e^+e^- \to e^+e^-\gamma X$ and
$\gamma\gamma \to \gamma X$~\cite{evanson}.

In summary, the study of these various processes will provide fundamental
new insight into small-$x$ QCD physics, going beyond that obtainable from 
present accelerators and the LC.

\newpage
\chapter{SUMMARY}
\label{chapter:eight}

The LHC will provide unique physics at the energy frontier in the TeV
energy range, for many years after its commissioning. However, scenarios
for physics in the TeV range generally have aspects that the LHC is unable
to test. Electron--positron linear colliders can complement the LHC by
producing directly new weakly-interacting particles and making possible
precision studies. These are the core motivations for a linear collider
with centre-of-mass energy in the TeV range. However, a complete
understanding of physics in the TeV range may require a multi-TeV linear
$e^+ e^-$ collider, for which the only available candidate is CLIC.

Research and development work on the two-beam acceleration concept to be
used in CLIC has been maturing over a number of years, with the CLIC Test
Facilities 1 and 2 demonstrating many key aspects of the required
technologies. There have, in particular, been significant advances in
demonstrating the feasibility of the high frequency and high
accelerating gradient central to the CLIC design, as well as the high
geometrical stability that it would require. Based on these studies,
parameter sets have been proposed for a nominal CLIC centre-of-mass energy
of 3~TeV and for a possible upgrade to 5~TeV, as well as for lower-energy
linear colliders based on CLIC technologies.

The high-energy CLIC parameter sets imply a large amount of beamstrahlung,
which produces important photon and electron--positron pair-production
backgrounds, as well as a relatively broad spectrum of beam collision
energies. This study has shown that these are not insuperable barriers to
experimentation at CLIC. There are designs for the beam-delivery system,
interaction region and detector that reduce the beam-induced backgrounds
sufficiently for all the essential physics measurements to be made. The
concepts proposed for detectors at lower-energy linear colliders may be
adapted for experiments at CLIC.

Likewise, studies for lower-energy colliders have developed tools for 
simulating physics events and backgrounds that serve also for 
experiments at CLIC. These tools include generators for Standard Model 
events and code suitable for modelling new physics processes. We note in 
particular that the electroweak radiative corrections to interesting 
physics processes are well understood, and can be calculated with 
sufficient accuracy to enable CLIC data to be interpreted reliably.

A key topic for experiments at CLIC will be the completion of the
phenomenological profile of the Higgs boson. Within the Standard Model,
CLIC will make it possible to determine whether a light candidate Higgs
boson is indeed responsible for the muon mass. The higher centre-of-mass
energy of CLIC will enable the triple-Higgs coupling of a light Higgs
boson to be measured much more accurately than at lower-energy linear
colliders. If there is a heavy SM Higgs boson, CLIC will
enable it to be studied closely. If nature is described by supersymmetry,
there will be more than one physical Higgs boson. CLIC will provide
unparalleled reach for the heavier supersymmetric Higgs bosons, opening
new possibilities for probing CP violation.

The relatively high centre-of-mass energy of CLIC will provide unique
kinematic reach for the supersymmetric particles themselves. The
luminosity offered by CLIC will be ample to study their masses and decays
in detail. The strong beamstrahlung at CLIC will smear the centre-of-mass
energy and complicate missing-energy analyses. However, studies of the
production and decay of several species of supersymmetric particles,
including smuons, selectrons, charginos and stop squarks demonstrate that
these complications do not impede accurate measurements of sparticle
masses and decays. In some supersymmetric scenarios, CLIC will produce
many types of sparticle that are too heavy to be produced at a
lower-energy linear collider. In other supersymmetric scenarios, CLIC will
make possible precise measurements of strongly-interacting squarks and
gluinos that cannot be performed at the LHC.

CLIC will likewise provide a unique kinematic reach for probing other
theories of physics beyond the Standard Model, such as extra dimensions.
In each of the cases studied, the experimental conditions at CLIC are no
obstacle to exploiting fully the broad kinematic reach of CLIC.

There will also be unique opportunities at CLIC to study QCD. In addition
to jet production and fragmentation studies in electron--positron
annihilation events, CLIC will make possible detailed probes of QCD
predictions in the collisions of real and virtual photons. In particular,
the high CLIC centre-of-mass energy will enable BFKL predictions for 
high-energy scattering to be tested.

This exploratory study has demonstrated that experiments at CLIC will be
able to exploit fully its high centre-of-mass energy for tests of the
Standard Model as well as unique probes of ideas for new physics beyond
the Standard Model. CLIC will take physics at the energy frontier to a new
scale and level of accuracy.

\end{document}